%% file: hwo_start.tex
\PassOptionsToPackage{svgnames, table}{xcolor}
\documentclass[twocolumn,letterpaper]{aastex701}
\usepackage[T1]{fontenc}
\usepackage{comment}
\usepackage{multirow}
\usepackage{booktabs}
\usepackage{tabularx}
\usepackage{array}
\newcommand{\sleads}[1]{\textcolor{violet}{Science Leads: #1}}
\newcommand{\breakthrough}{\textcolor{teal}{\underline{\textbf{\emph{Breakthrough Science:}}}} }
\newcommand{\enabling}{\textcolor{ForestGreen}{\underline{\textbf{\emph{Enabling Science:}}}} }

\newcommand{\ngg}{15} %20 observations
\newcommand{\nee}{13} %same as \neeall; 19 observations
\newcommand{\nssic}{32} %77 observations
\newcommand{\nlw}{10} %same as \nlwall; 24 observations
\newcommand{\ncases}{70}
\newcommand{\nobs}{140} %number of observing programs within the science cases

\newcommand{\nggall}{18} %3 unofficial GG cases (GG-1, GG-8, GG-15)
\newcommand{\nssicall}{33} % SSiC-13 unofficial
\newcommand{\ncasesall}{74}

\newcommand{\ndecadalanswered}{27}%number of Astro2020 questions addressed by these science cases

\newlength{\cw} %width for capabilities column
\newcommand{\riff}{\vspace{6pt}\newline}
\newcommand{\harder}{\textbf{Strengthened:}\ }
\newcommand{\easier}{\riff\textbf{Softened:}\ }

\usepackage{makecell}

\definecolor{colorGG}{HTML}{E6E2FC} % Soft Lavender
\definecolor{colorEE}{HTML}{F7E2D1} % Pale Peach
\definecolor{colorSSiC}{HTML}{E0F1FB} % Fresh Ice Blue
\definecolor{colorLW}{HTML}{D1EDE6} % Soft Seafoam Mint

\definecolor{colorTextGG}{HTML}{785EF0} % Rich Purple
\definecolor{colorTextEE}{HTML}{D55E00} % Vermillion Orange
\definecolor{colorTextSSiC}{HTML}{56B4E9} % Sky Blue
\definecolor{colorTextLW}{HTML}{009E73} % Mint Green

\definecolor{panelCOS}{HTML}{8F5EB4}
\definecolor{panelGAL}{HTML}{A75E78}
\definecolor{panelCOEP}{HTML}{BE5E3C}
\definecolor{panelISP}{HTML}{41AFCC}
\definecolor{panelSSP}{HTML}{2BA9AE}
\definecolor{panelEAS}{HTML}{16A491}

\newcommand{\pref}[3][]{%
 \mbox{\textcolor{colorText#2}{\textbf{#2#1}}~(\ref{#3})}%
}

\begin{document}

\title[HWO Science]{Early Exploration of the Scientific Discovery Space for the Habitable Worlds Observatory}

\suppressAffiliations

\author[0000-0001-8189-0233]{Courtney D. Dressing}
\affiliation{Department of Astronomy, University of California, Berkeley 501 Campbell Hall \#3411, Berkeley, CA 94720, USA}
\email{dressing@berkeley.edu}

\include{latex_author_list_10august2026}

\begin{abstract} 
The Habitable Worlds Observatory (HWO) is a future NASA flagship mission concept identified by the Astro2020 Decadal Survey as the highest priority for large space missions. HWO should conduct ``transformative astrophysics'' and search for biosignatures in the atmospheres of approximately 25 potentially Earth-like planets. To further the early-stage development of HWO, NASA formed the Science, Technology, Architecture Review Team (START). In turn, START invited the scientific community to join working groups to explore the potential discovery space. In this paper, we present \ncases{} science cases that resulted from this process. The cases address four scientific pillars: growth of galaxies (\ngg~cases), evolution of the elements (\nee~cases), solar systems in context (\nssic~cases), and living worlds (\nlw~cases). Combined, they would address \ndecadalanswered{}~of the 30 science questions and discovery areas identified by Astro2020. The \nobs~observing programs needed for the \ncases~investigations encompass a rich variety of spectroscopic (for 87\% of science cases) and photometric (for 30\%) observations extending from the UV to the NIR. Additionally, high-contrast and polarimetric capabilities would be needed for 34\% and 27\% of science cases, respectively. Access to UV wavelengths is critical: 83\% of science cases need data at wavelengths $<400$~nm, and 26\% extend to $<100$~nm. In the NIR, 26\% of science cases need observations at wavelengths $\ge2000$~nm. Pursuing the full portfolio of science would also necessitate precise astrometry for planet mass measurement, rapid response capabilities, a large instantaneous field of regard, non-sidereal tracking, saturation mitigation strategies, and high dynamic range. 
\end{abstract}


\section{Introduction} 
\label{sec:intro}

In this paper, we describe the results of early exploration of the scientific discovery space that could be explored by a future IR/O/UV facility like the one prioritized by \citet{astro2020}. In Astro2020, the observatory is described as a space-based telescope with an estimated inscribed diameter of approximately 6~meters and broad wavelength coverage from the ultraviolet to the near-infrared. Astro2020 stated that the combination of a ``large, stable telescope'' and an ``advanced coronagraph'' would make IR/O/UV ``capable of surveying a hundred or more nearby Sun-like stars to discover their planetary systems and determine their orbits and basic properties.'' In addition to conducting a census of nearby planetary systems, IR/O/UV would obtain spectra at wavelengths ranging from the UV to the NIR ``to identify multiple atmospheric components that could serve as biomarkers'' in the atmospheres of the ``most exciting $\sim$25 planets.'' 

Due to the incredible power of IR/O/UV for detecting and characterizing potentially habitable worlds, IR/O/UV was later renamed the Habitable Worlds Observatory (HWO). While the first two words in this name emphasize the ability to image temperate terrestrial planets, the final word (``Observatory'') is a reminder that HWO would be able to conduct a broad range of investigations. By having access to the same wavelength range as the Hubble Space Telescope but a much larger aperture, drastically higher angular resolution, and significantly more advanced instruments, Astro2020 noted that the IR/O/UV mission ``would be capable of achieving breakthrough discoveries across nearly all of astrophysics.'' Notably, \citet{astro2020} declared that observations with IR/O/UV ``would directly address two-thirds of
 the 24 key science questions identified [by Astro2020] and will contribute to addressing many of the others.'' A facility like HWO would also have the potential to address many of the questions identified by the 2023 Planetary Science and Astrobiology Decadal Survey \citep{planetary_decadal2023} and address the themes prioritized by the European Space Agency for the next large-class science mission \citep{esa_voyage2050}.

\subsection{Historical Context} 
\label{ssec:history}
One of the Frontier Projects prioritized by the Astro2020 Decadal Survey \citep{astro2020} was a ``future large infrared/optical/ultraviolet telescope optimized for observing habitable exoplanets and general astrophysics.'' As detailed by \citet{postman+stapelfeldt_hwo25}, the observatory envisioned by Astro2020 was the latest in a long string of concepts 
that included the Space Interferometry Mission \citep[SIM]{shao1998, unwin_et_al2008}, the Terrestrial Planet Finder \citep[TPF]{angel+woolf1997}, and the Advanced Technology Large-Aperture Space Telescope \citep[ATLAST]{postman_et_al2009, postman_et_al2010}. The most recent ancestors of HWO are the Habitable Exoplanet Observatory \citep[HabEx,]{gaudi_et_al2020} and the Large UV/Optical/Infrared Surveyor \citep[LUVOIR]{luvoir2019}.

Astro2020 \citep{astro2020} designated the large IR/O/UV space telescope as the ``top priority'' for the new ``Great Observatories Mission and Technology Maturation Program.'' Astro2020 recommended that NASA establish such a program to facilitate development of flagship missions because while there is a tremendous scientific advantage to having a simultaneously operating suite of large-scale observatories covering the electromagnetic spectrum, deploying a fleet of powerful, complementary flagship observatories necessitates shortening the wait between flagship launches and therefore speeding up and/or overlapping the development cycles. Building on past studies such as \citet{bitten_et_al2019} and \citet{armus_et_al2021}, Astro2020 therefore recommended re-phasing the mission development process so that mission concepts could be studied in more detail earlier in the design cycle. This change would allow for more accurate and precise estimates of costs, budget profiles, scientific capabilities, and technical challenges before presenting future missions to Congress for pre-implementation approval. 

Crucially, early concept maturation would involve identifying and exploring cost ``breakpoints'' at which the funding needed for the mission changes dramatically. Astro2020 further recommended that the early studies map science breakpoints at which the scientific return of the mission is dramatically increased or decreased; determine key technologies and their associated maturation risks and requirements; and identify possible architecture trades. The work of the concept study teams and the alignment of mission objectives with decadal priorities would then be reviewed by future Decadal Surveys and Mid-Decadal Surveys, which could recommend certain mission concepts for further development through the Great Observatories Mission and Technology Maturation Program. Once a mission concept finished its maturation phase and was ready for implementation, the cost of a future mission would transition from the Great Observatories Mission and Technology Maturation Program to its own specific mission development line. 

\subsection{Role \& Scope of Paper}
The purpose of this paper is to present an introductory exploration of the diverse range of scientific investigations that could be possible with a future facility like the HWO mission concept. We concentrate on the science cases developed by the Science, Technology, Architecture Review Team (START; see Section~\ref{sec:start}) and the Science Working Groups (SWGs; see Section~\ref{sec:wg}) initiated by START. These science cases are only a subset of the rich range of programs that could be completed by an HWO-like facility and were intended to provide early input to help inform possible investments in technology maturation, architecture trades, and precursor observations. Just as previous flagship missions like the Hubble Space Telescope \citep[e.g.,][]{spitzer1968, bahcall+odell1979, spitzer1990, williams_et_al1996, lallo2012} and JWST \citep[e.g.,][]{stockman1997, gardner_et_al2006, rigby_et_al2023} have conducted observations that would have seemed out of reach before launch, HWO would similarly be used in novel and unexpected ways to deepen our understanding of the universe. The notional programs described in this paper are only an early preview of the wide range of science that could be accomplished by HWO. 

Importantly, this paper should not be interpreted as a description of or a prescription for the future Habitable Worlds Observatory. Rather, this document describes, summarizes, and compares a set of science questions identified by the community as scientifically compelling. While some of these scientific objectives would be achievable with HWO, others would demand different facilities including extremely large ground-based telescopes and future flagship missions operating over a range of wavelengths. Accordingly, the observational capabilities requested for each science case may extend beyond the range of notional capabilities for a facility like HWO. Throughout the document, any mention of scientific ``requirements'' should be interpreted as the observational capabilities needed to conduct a specific scientific study rather than design specifications for HWO. Future mission requirements would be defined by the HWO Project Office by comparing the needs for scientific investigations such as those presented herein to the capabilities of potential observatory architectures. For a summary of current and ongoing engineering work, see \citet{feinberg_et_al2026}.

\input{roadmap}

\subsection{Structure of Paper}
The paper is structured as follows. In Section~\ref{sec:start} and Section~\ref{sec:wg}, we describe the formation, composition, and aims of the START and the SWGs, respectively. In Section~\ref{sec:scdds}, we present and discuss \ncases~science cases identified by the SWGs. We then identify crucial datasets for each SWG in Section~\ref{sec:swg_obs} and highlight similarities across SWGs in Section~\ref{sec:xswg}. Next, in Section~\ref{sec:drivers}, we describe and compare the observational capabilities indicated by each science case and identify the science cases that place the strongest demands on mission design. In Section~\ref{sec:astro2020}, we map the science cases considered in this document to the questions identified by the Astro2020 Decadal Survey \citep{astro2020}, thereby validating the claim that a facility like HWO would enable a broad range of scientifically exciting investigations. We then motivate the need for HWO in particular in Section~\ref{sec:need_hwo}, and discuss observations that would benefit the mission if they were to be conducted prior to the launch of HWO in Section~\ref{sec:precursor}. Finally, we summarize this paper and consider the next phase of work on HWO in Section~\ref{sec:conc}. For reference, we provide an overview of the paper structure and content in Table~\ref{tab:roadmap}.

\section{The Science, Technology, Architecture Review Team}
\label{sec:start}
In 2023, NASA announced that a Science, Technology, Architecture Review Team (START) and Technical Assessment Group (TAG) would be assembled to help advance maturation of the HWO mission concept. For more details about the construction, membership, deliverables, and conclusion of START, see Appendix~\ref{app:start}. 

According to the Terms of Reference\footnote{\url{science.nasa.gov/wp-content/uploads/2023/12/tor-hwo-start-final-fy23-signed.pdf}}, START was tasked with ``quantitatively describ[ing] the relationships between Astro2020 HWO science goals and its needed observatory/instrument properties.'' Specifically, START was asked to ``defin[e] the high-level scope of the mission;'' ``identify science goals for HWO, pulling from the Astro2020 report's lists of ``Science Questions'' and tables of associated capabilities;'' and ``map these goals onto HWO science objectives that are specific, feasible [for] HWO.'' Furthermore, START was tasked with ``quantify[ing] the relationships between HWO performance and the ability of HWO to realize its science objectives'' as well as the relationships between HWO performance and ``top-level architecture/instrument parameters.'' While START was not tasked with conducting trades or making selections, START was asked to ``assess the fidelity of models needed in the future to execute those future coupled trades'' and given permission to ``identify what additional knowledge/research is needed to quantify these relationships.''

\section{Working Groups}
\label{sec:wg}
\begin{figure*}
 \centering
 \includegraphics[width=1\linewidth]{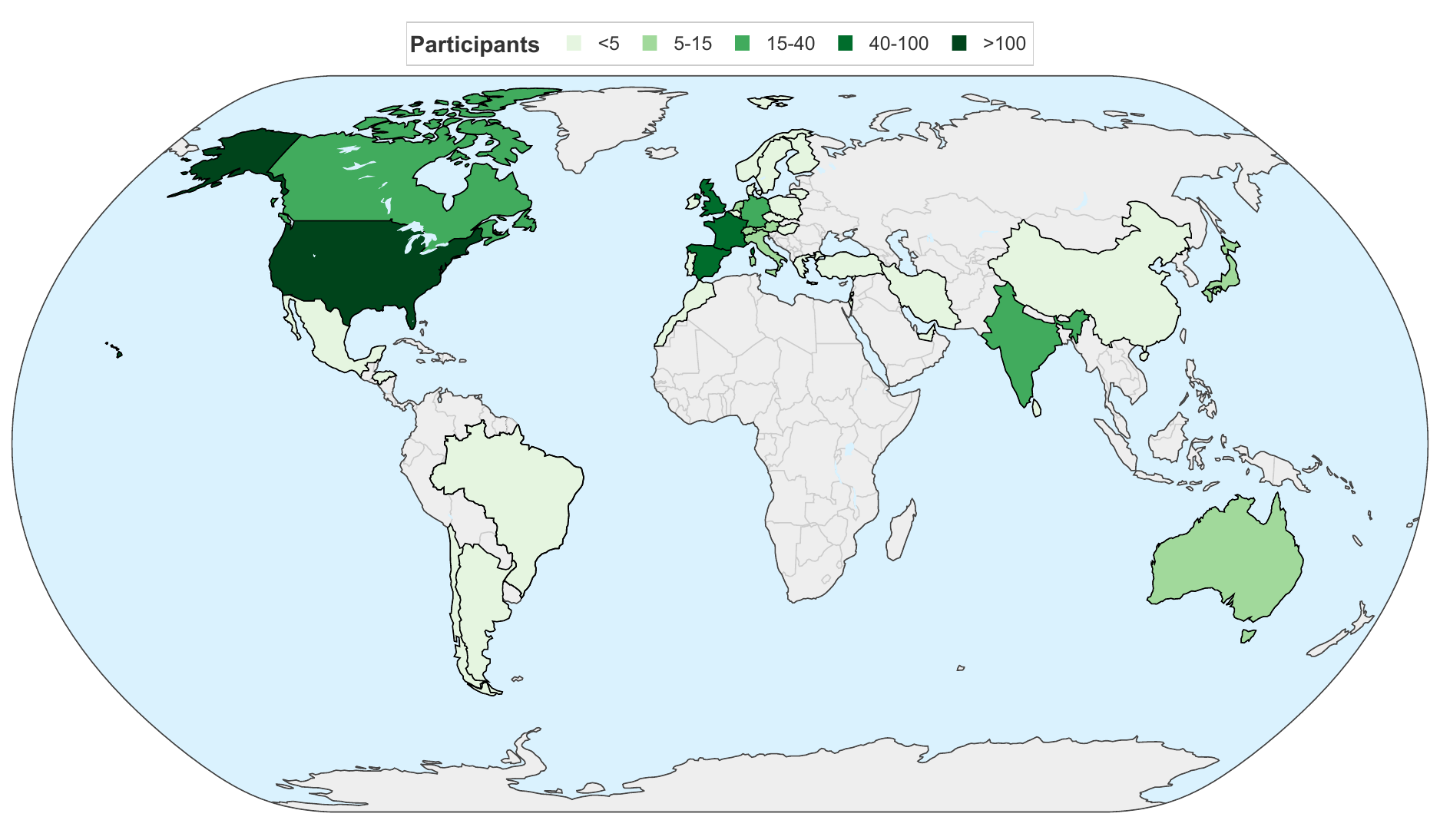}
 \includegraphics[width=1\linewidth]{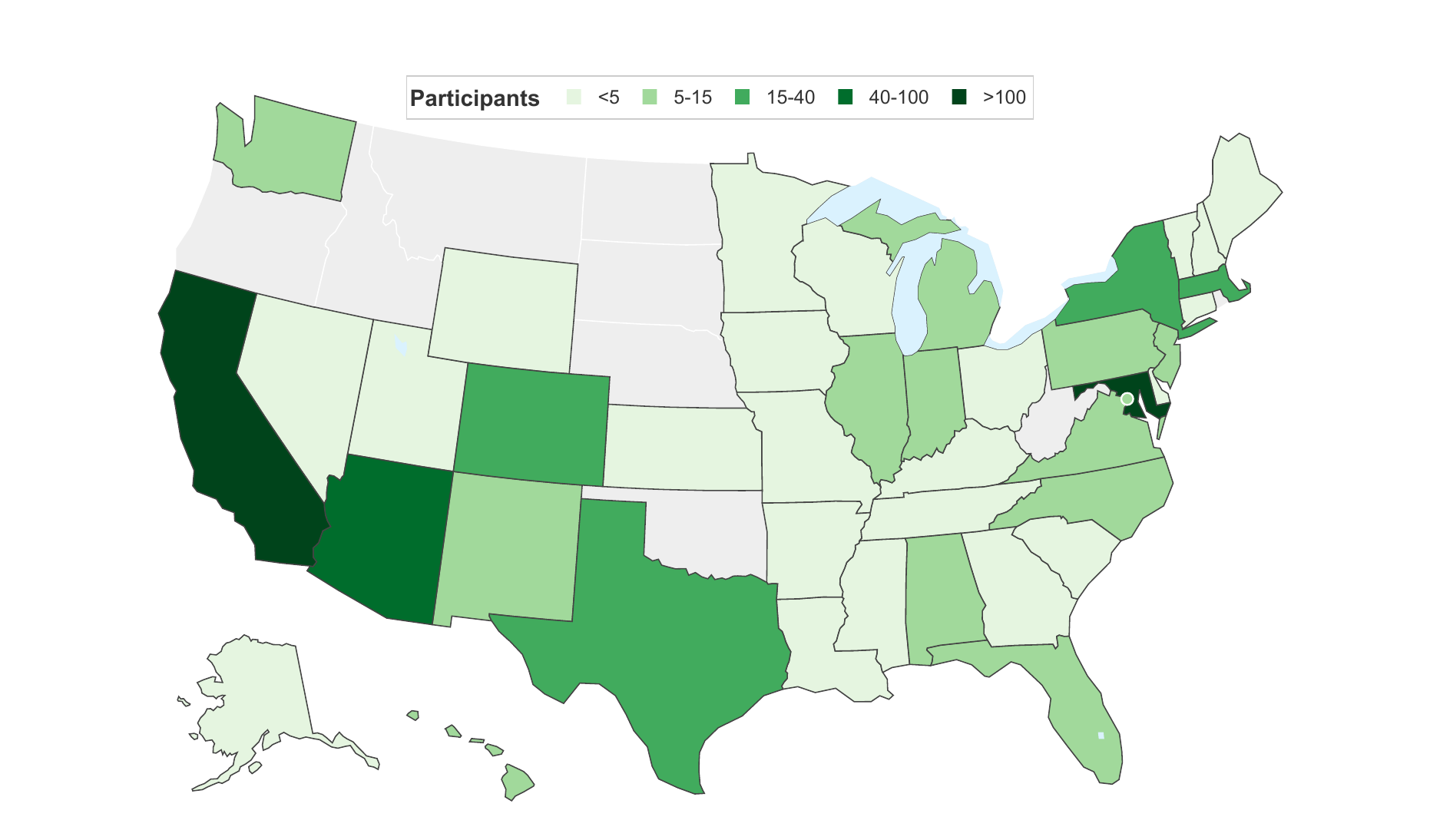}
 \caption{Global (top) and US national (bottom) participation in HWO. The countries and states are shaded to indicate the total number of people from each region who co-authored this paper or filled out the Google form to join working groups. The colorscale is the same for both panels. Individuals with multiple affiliations are counted by their primary affiliations.}
 \label{fig:map}
\end{figure*}

The START and TAG leadership formed working groups focused on specific areas of HWO development. These working groups were open to scientists and engineers around the world, and over 1000~people signed up. In Figure~\ref{fig:map}, we show the global and national participation in the HWO working groups and the authorship of this paper. In total, 38~countries were represented. Within the US, participation spanned 41~states, Puerto Rico, and the District of Columbia. 

The goals of the working groups were two-fold. First, we wanted to ensure that community members with a diverse range of perspectives were able to participate in the discussion of the HWO mission concept. Second, thoroughly considering the observational capabilities needed to address the scientific priorities identified by Astro2020 required input from many experts in various subfields. We describe the working group infrastructure in Section~\ref{ssec:wg_structure} and the process used by the working groups to develop science cases in Section~\ref{ssec:scdd_process}.

\subsection{Working Group Structure}
\label{ssec:wg_structure}
The START and TAG formed four categories of working groups:
\begin{enumerate}
\item{\textbf{Science Working Groups (SWGs)} were focused on identifying and developing the science cases for the observatory.}
\item{\textbf{Joint Working Groups (JWGs)} were tasked with managing the interface between scientific objectives and engineering demands.}
\item{\textbf{Technology Working Groups (TWGs)} were responsible for identifying and maturing the technologies needed to develop HWO.}
\item{\textbf{Community Working Groups (CWGs)} were asked to foster the development of a supportive and stable community and consider the environment in which HWO would operate.}
\end{enumerate}

Arranging them from large scales to small scales, the four SWGs and their associated subgroups were
\begin{enumerate}
\item{\textbf{Growth of Galaxies (GG)} explored how galaxies, their constituents, and their environments evolve over the history of the HWO-observable universe. The working group included four subgroups:}
\begin{enumerate}
\item{\emph{AGN Over Cosmic Time} studied the central engines of galaxies and their impacts on galaxy evolution in imaging and spectroscopy at multiple scales.}
\item{\emph{Ionizing Photons and their History} aimed to understand the galaxies and their stars that drove reionization by observing their analogues at lower redshift in the UV/O/IR.}
\item{\emph{Intergalactic and Circumgalactic Medium} explored the evolution of galaxies and the universe in context of the circumgalactic medium (CGM) and intergalactic medium (IGM) in absorption and emission.}
\item{\emph{The Dark Sector} explored the nature of dark matter and dark energy via their impacts on galaxies and large scale structure, revealed primarily through gravitational lensing.}
\end{enumerate}
\item{\textbf{Evolution of the Elements (EE)} considered the enrichment of the universe over cosmic time by studying the formation, distribution, and evolution of stars. The working group included three subgroups:}
\begin{enumerate}
\item{\emph{Star Formation and the Interstellar Medium} investigated UV/O/IR spectroscopy and imaging of star forming regions.}
\item{\emph{Cosmic Explosions} addressed studies of supernovae, merger-driven stellar and stellar remnant explosions, and sources of gravitational wave events.}
\item{\emph{Stellar Populations and Evolution} considered UV/O/IR spectroscopy and imaging of stars from individuals in the Milky Way, to populations in the Local Group, to stellar clusters across the universe.
}
\end{enumerate}
\item{\textbf{Solar Systems in Context (SSiC)} explored the observations needed to further our understanding of planetary systems and place our own solar system within the broader context of planetary systems orbiting other stars. The working group included four subgroups:}
\begin{enumerate}
\item{\emph{Characterizing Exoplanets} addressed observations and characterization of exoplanets that are observed by HWO directly or indirectly (i.e., phase curve, transits, eclipses).}
\item{\emph{Solar System Observations} considered remote sensing, often at high cadence, of solar system planets, their moons, and small bodies using high spatial resolution imaging and UV/O/IR spectroscopy.}
\item{\emph{Demographics and Architectures} synthesized current knowledge of exoplanet occurrence rates and system architectures for the types of stars that HWO would target and assessed the sensitivity and accessibility needed to constrain system architectures.}
\item{\emph{Birth and Evolution} considered the observational capabilities necessary to advance understanding of the formation of planetary systems during embedded, protoplanetary, and debris disk stages.
}
\end{enumerate}
\item{\textbf{Living Worlds (LW)} investigated the observations needed to detect life on other planets. The working group included three subgroups:}
\begin{enumerate}
\item{\emph{Target Stars and Systems} built on previous work to assemble current knowledge of likely HWO target stars, identify knowledge gaps, and consider the ability of precursor observations, contemporaneous observations with other facilities, and HWO observations to constrain important properties of host stars and their planetary systems.
}
\item{\emph{Biosignature Possibilities} considered the wide variety of biosignatures (e.g., biogenic gases, aerosols, surface biosignatures, technosignatures) that could be detectable with HWO, the conditions under which they might occur, and the associated measurement needs.
}
\item{\emph{Biosignature Interpretation} explored how potential biosignatures could be assessed and considered the additional information about the planet and planetary system needed to interpret biosignatures and rule out false positives.}
\end{enumerate}
\end{enumerate}

The working groups considered past work such as the HabEx mission concept report \citep{gaudi_et_al2020}, the LUVOIR mission concept report \citep{luvoir2019}, the Exoplanet Science Strategy Report \citep{exoplanet_science_strategy2018}, an Astrobiological Science Strategy for the Search for Life in the Universe \citep{astrobio_strategy2019}, the United States 2020 Decadal Survey on Astronomy \& Astrophysics \citep{astro2020}, the United States Planetary Science and Astrobiology Decadal Survey $2023 - 2032$ \citep{planetary_decadal2023}, and the European Space Agency's Voyage 2050 report \citep{esa_voyage2050}. Accordingly, the science cases developed by the SWGs strongly align with the priorities identified in those past studies. 

Outside of the SWGs, the WGs most relevant to this paper were the \emph{Astrophysics in the 2030s/2040s from Space Working Group} and \emph{Astrophysics in the 2030s/2040s from the Ground Working Group}, both of which were CWGs. Both of these groups were assigned to identify facilities that were likely to be operating in the next two decades and assess their observational capabilities. The information gathered by the Astro 2030s/2040s WGs would then be used to determine the areas of parameter space that could be best explored by a facility like HWO and the regions of parameter space that would be better studied by other facilities. Comparing the anticipated capabilities of an HWO-like facility to those of other facilities would then enable the HWO Technology Maturation Project Office to prioritize science cases that could be uniquely accomplished with an observatory like HWO. 

\subsection{Science Case Development Process}
\label{ssec:scdd_process}
The core output of the science working groups was the construction of a set of science cases describing investigations that could be done with a facility like HWO. Working groups presented their findings in the form of Science Case Development Documents (SCDDs) or Super Simple Science Case Development Documents (SSSCDDs). Both SCDDs and SSSCDDs motivated and explained potential observations, but they differed in terms of length and detail: SCDDs were longer documents that included more extensive background material and were more likely to incorporate preliminary calculations while SSSCDDs were more concise and quickly narrowed in on how a certain investigation might potentially push technical capabilities. The more compact SSSCDD template was designed to lower the barrier to science case submission while still providing enough information to prioritize parameter studies of how engineering decisions would impact scientific return. 

\begin{figure*}
 \centering
 \includegraphics[width=1\linewidth]{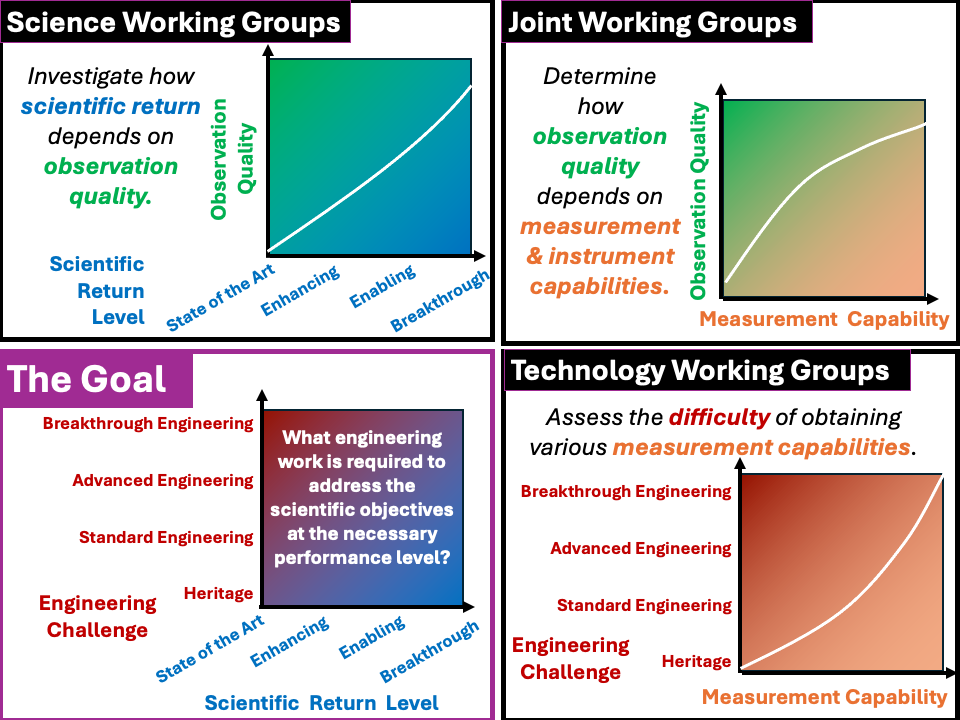}
 \caption{Diagram showing how science working groups, technical working groups, and joint working groups collaborated to investigate how engineering capabilities connect to future science with HWO. Inspired by charts made by John Ziemer.}
 \label{fig:flow}
\end{figure*}

When developing science cases, working groups were instructed to consider multiple levels of scientific progress to better support trade studies. By considering a range of scientific capabilities rather than a single point value, the science working groups generated a mapping between observation quality and science return levels. In parallel, the technology working groups established by the TAG considered the mapping between engineering capabilities and observatory capability while the joint working groups investigated the relationship between observatory capability and observation quality. As shown in Figure~\ref{fig:flow}, combining the work of all three groups would therefore produce a mapping between engineering challenge and science return that could be used to guide trade studies.

Throughout the science case development process, we categorized science return levels using this classification scheme:
\begin{enumerate}
 \item \textbf{State of the Art:} This level represented the scientific understanding that already existed at the time the science case was written or would be almost certainly guaranteed to exist at the time of potential HWO observations. 
 \item \textbf{Incremental Progress (Enhancing):} This level represented a meaningful improvement over the state of the art generated by upgrading one or two capabilities.
 \item \textbf{Substantial Progress (Enabling):} This level would use new capabilities to conduct science at a level significantly higher than the state of the art. At this level, it would be possible to answer the key science questions posed in the science case. 
 \item \textbf{Major Progress (Breakthrough):} This level represented exceptional improvement over the state of the art and would meet expectations for a ``flagship observatory.'' At the breakthrough level, the science questions posed by the science case would be answered robustly and the observations would enable addressing even more sophisticated questions. 
\end{enumerate}

When assessing the corresponding engineering challenge levels,\footnote{Definitions from presentation ``Milestones and Year-Long Plan: CML3 Deliverables, Intermediate Milestones, and Updated Schedule
'' by John Ziemer and Lee Feinberg on 12 March 2024. Slides posted at \url{https://drive.google.com/file/d/1-tR6r6CdMyB8LJ876U9f0q7HUXH6AC9T/view}.} we used this scheme:
\begin{enumerate}
 \item \textbf{Heritage:} ``space-flight experience at the sub-system and component level, including software; parts are still available with validated performance models and on-orbit test data. No new technology. Low risk.''
 \item \textbf{Standard Engineering:} ``custom components or sub-systems that can be designed using known and publicly available, industry standard tools and processes; components are available, manufacturable, with validated performance models that include the scale of sub-systems. No new technology. Low risk.'' 
 \item \textbf{Advanced Engineering:} ``a small number of new technologies or software may be required to meet key performance parameters at the component or sub-system level; design process is understood using custom tools, if necessary (often combined with standard engineering practices) and validated models provide reliable performance predictions. Any technologies typically at TRL\footnote{Technology Readiness Level (TRL) refers to technological maturity. Levels range from 1 (lowest) to 9 (highest). TRL$=5$ is defined by NASA as, ``Component and/or breadboard validation in relevant environment'' (\url{https://www.nasa.gov/directorates/somd/space-communications-navigation-program/technology-readiness-levels/}).} $>5$. Medium risk.''
 \item \textbf{Breakthrough Engineering:} ``end-to-end subsystem or even system that includes multiple new technologies and software; design process is customized, and models need to be developed and validated in parallel with design process and technology development. Laboratory measurements and environmental testing are necessary to validate hardware and models. Technologies TRL$<5$. High Risk.''

\end{enumerate}

\section{Potential Science Cases}
\label{sec:scdds}
A facility like HWO would be capable of accomplishing far more science than can be described in a single manuscript. Here, we summarize the science cases developed by the Science Working Groups described in Section~\ref{ssec:wg_structure} following the process explained in Section~\ref{ssec:scdd_process}. We begin the discussion for each working group with a high-level summary of the associated science cases. We then discuss each science case in a separate subsection. As shown in Table~\ref{tab:refs}, some of these science cases have been described in more detail in standalone peer-reviewed publications. Additionally, most of the science cases were included in the conference proceedings of the ``Towards the Habitable Worlds Observatory: Visionary Science and Transformational Technology'' meeting (henceforth ``HWO25''), which was held in Washington, D.C. in July 2025 \citep{HWO25_proceedings, HWO25_proceedings_part2}. In this section, we discuss all \ncasesall~science cases shown in Table~\ref{tab:refs}, but beginning in Section~\ref{sec:swg_obs} we consider only the \ncases~science cases that were published or posted in the SCDD Portal.

The purpose of this document is to assist in maximizing the potential science return of a facility like HWO by describing how scientific progress would scale with observational capabilities, thereby helping to identify future work that could generate the highest scientific return per engineering investment. In accordance with that emphasis on cutting-edge science, this document focuses on the observational needs for breakthrough and enabling science (levels 3 and 4 defined in Section~\ref{ssec:scdd_process}). For additional information about the samples and measurements needed for incremental progress and the current state of the art, see the individual SCDD references listed in Table~\ref{tab:refs}.

\subsection{Growth of Galaxies}
\label{ssec:gg_scdds}
The Growth of Galaxies (GG) working group addressed how galaxies evolve over time and interact with their environments. Of the \nggall~science cases presented in this section, \ngg~were subsequently posted in the SCDD Portal or published. 

Drawing inspiration from past investigations and simulations of the circumgalactic medium \citep[e.g.,][]{werk_et_al2014, somerville+dave2015, tumlinson_et_al2017}, multiple GG science cases consider the cycle of matter within galaxies and between galaxies and the surrounding intergalactic medium (Sections~\ref{sssec:agn_outflow}, \ref{sssec:resolve_reionization}, \ref{sssec:lyman_escape}, \ref{sssec:green_peas}, \ref{sssec:lyman_indirect}, \ref{sssec:cgm_map}, \ref{sssec:disk_cgm}, \ref{sssec:cgm_elm}, and \ref{sssec:agn_feedback}). Additionally, numerous GG science cases investigate the fundamental questions of how galaxies contributed to the reionization of the universe and the efficiency with which modern galaxies continue to produce ionizing photons (e.g., Sections~\ref{sssec:resolve_reionization}, \ref{sssec:lyman_escape}, \ref{sssec:green_peas}, \ref{sssec:lyman_indirect}, and \ref{sssec:ionizing_lf}). Other GG science cases are focused on the properties of black holes, how supermassive black holes interact with their host galaxies, and the speed of black hole mergers (Sections~\ref{sssec:bh_quiescent}, \ref{sssec:bh_mass_spin}, \ref{sssec:bh_energy}, \ref{sssec:bh_torus}, \ref{sssec:smbh_mergers}, and~\ref{sssec:smbh_pol}). Finally, additional science cases consider the nature of dark matter (Sections~\ref{sssec:dm_power} and \ref{sssec:dm_lensing}) and the effects of dark matter on the growth of structure in the universe (Sections~\ref{sssec:dm_lensing} and \ref{sssec:agn_feedback}).

\subsubsection{Deciphering The Launching of Multi-phase AGN-driven Outflows and Their (Spatially Resolved) Multi-scale Impact (SCDD-GG-5)}
\label{sssec:agn_outflow}
\sleads{Lulu Zhang, Gagandeep Kaur, Tianmu Gao, \'{A}lvaro Labiano, Erin K. S. Hicks, Vivian U, Chris Packham, Missagh Mehdipour, Travis Fischer, Thaisa Storchi Bergmann, Namrata Roy, Isabel M\'{a}rquez, and Christiaan Boersma}

As described by \citet{zhang_hwo25}, this science case aims to explore how feedback from Active Galactic Nuclei (AGN) drives the formation and evolution of galaxies. Specifically, by mapping AGN-driven inflows and outflows and determining their ionization states, temperatures, energies, and relationships with AGN activity, HWO observations would help reveal how AGN affect the interstellar medium and therefore star formation. High spatial resolution (i.e., $<10$ pc) is essential for obtaining the mass and kinetic measurements needed to understand the complex interplay between inflows, outflows, and the ISM. Specifically, Lyman-alpha emission in the UV spectrum can be used to map
ionized gas outflows \citep[e.g.,][]{Heckman_et_al2011, arrigoni_battaia_et_al2016} while CO and H$_2$ features in the NIR spectrum can be used to identify cold/
warm molecular gas outflows \citep[e.g.,][]{feruglio_et_al2010, cicone_et_al2014, cicone_et_al2017}, which are potentially linked to star formation activities that can be traced by infrared emission features of polycyclic aromatic hydrocarbons \citep[e.g.,][]{kim_et_al2012, shipley_et_al2016}.

Space-based observations are essential for this science case because ground-based observations at the desired UV and NIR wavelengths are severely affected by telluric absorption. Accordingly, current instruments on large telescopes and upcoming instruments planned for ELTs will not observe the key wavelength ranges needed for this science case. JWST is sensitive to the desired NIR wavelength range, but the relatively modest FOV of NIRSpec ($3" \times 3"$ for NIRSpec) means that mapping spatially resolved nearby galaxies requires many pointings and is therefore inefficient. Furthermore, fully understanding the role of AGN feedback also requires spatially-resolved UV observations, and such instrumentation is not currently available \citep{james_et_al2019}. The upcoming, high-efficiency survey telescope UVEX \citep{kulkarni_et_al2021} is promising for selecting samples for future HWO IFS observations. 

\breakthrough Achieving breakthrough science would require NIR IFS spectroscopy of 2000~galaxies at spatial resolution $0\farcs03$ at 1~$\mu$m and spectral resolution $R > 5000$ at $1-5\mu$m. Within that sample, 50~galaxies should also have UV/VIS IFS spectra at spatial resolution of $0\farcs003$ at 100~nm and spectral resolution $R > 50,000$ at $100 - 900$ nm. The NIR IFS and UV IFS should have field-of-views of $10\arcsec \times 10\arcsec$. These fields would be mosaicked to cover each galaxy, with a typical target requiring mosaic coverage of $4 \times 10\arcsec \times 10\arcsec$.
The observations should have a limiting sensitivity of $10^{-20}$~erg~s$^{-1}$~cm$^{-2}$ to permit detection of AGN with luminosities $L_{\rm AGN}\sim10^{40}-10^{48}$ erg~s$^{-1}$. 

\enabling For substantial progress, the sample size and resolution could be decreased by a factor of a few. At NIR wavelengths, 1000 galaxies should be observed using an IFS with a field-of-view of $6\arcsec \times 6\arcsec$, spatial resolution $0\farcs04$ at 1~$\mu$m, and spectral resolution $R > 3000$ at $1 - 3 \mu$m. Twenty of those galaxies should also be observed with a UV IFS with the same $6\arcsec \times 6\arcsec$ field-of-view, spatial resolution $0\farcs004$ at 100~nm, and spectral resolution $R > 30000$ at $100-400$~nm. These observations should have a limiting sensitivity of $10^{-19}$~erg~s$^{-1}$~cm$^{-2}$, which would increase the minimum detectable AGN luminosity to $L_{\rm AGN}\sim 10^{41}$~ergs~s$^{-1}$. 

\subsubsection{Exploring the Quiescent Black Hole Population of Nearby Dwarf Galaxies with the Habitable World Observatory (SCDD-GG-4)}
\label{sssec:bh_quiescent}
\sleads{Fabio Pacucci}

\begin{figure}
 \centering
 \includegraphics[width=1\linewidth]{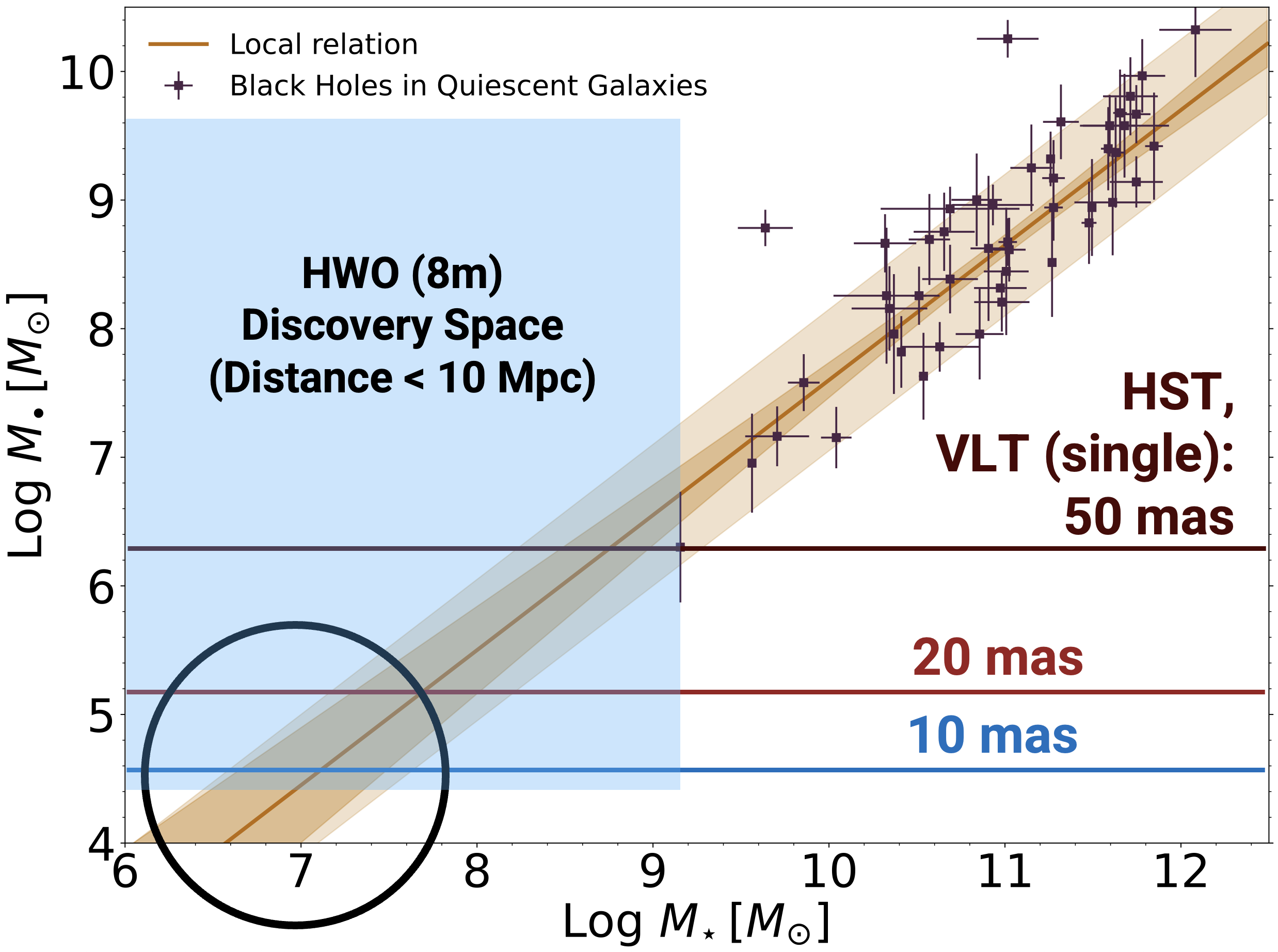}
 \caption{Black hole mass versus galaxy mass showing the currently known population of black holes in quiescent galaxies (points with errors) as reported by \citet{bennert_et_al2021}, representative sensitivities for galaxies within 10~Mpc, and the discovery space for HWO (blue rectangle). The black circle marks the region where HWO could potentially detect MBHs in local dwarf galaxies, thereby testing whether those MBHs also follow the local relation (orange line with shaded uncertainty regions). Figure from \citet{pacucci_hwo25} reproduced with permission.}
 \label{fig:quiescent}
\end{figure}

As discussed by \citet{pacucci_hwo25} and illustrated in Figure~\ref{fig:quiescent}, this science case aims to extend studies of the relationship between central supermassive black holes (SMBHs) and their galaxies to the lowest mass galaxies. The sensitivity and resolution possible with HWO would enable detection of quiescent black holes with masses of $10^5 M_\odot$ (substantial progress) or even $ < 10^5 M_\odot$ (breakthrough progress). Such observations would require MOS UV/VIS spectroscopy of dwarf galaxies at distances of $10-30$~Mpc. HWO could also reveal the effects of BHs on nearby stars via measurements of the stellar velocity dispersion. 

Extending studies of the BH-galaxy correlation to this low-mass regime is essential for addressing long-standing questions such how black holes were seeded in the early universe, which was prioritized by Astro2020 \citep[B-Q4;][]{astro2020}. As summarized in the review by \citet{greene_et_al2020}, three leading formation channels are collapse of Population~III stars, direct collapse, and gravitational runaway. 

Black holes formed by the collapse of Pop~III stars are likely to have masses of roughly 100~$M_\odot$ \citep[e.g.,][]{fryer_et_al2001}, well below the detection limit for HWO, and much smaller than the masses at which central black holes are typically observed. Reaching typical SMBH masses would require very efficient BH growth \citep[i.e., at or near the Eddington limit,][]{haiman+loeb2001, madau_et_al2014} which is challenging but not theoretically impossible \citep{jiang_et_al2019}.

In the direct collapse model, ``massive seed'' black holes are formed from collapsing clouds of gas without first completing the full cycle of stellar evolution. Theoretical predictions suggest that resulting BHs would have masses of $10^4 - 10^6 M_\odot$ and that specific conditions are needed to prevent the gas from cooling and fragmenting into multiple stars rather than forming black holes. Despite the ``direct'' name, these black holes likely experience a brief, intermediate phase of existence as rapidly accreting supermassive stars ($10^5 - 10^6 M_\odot$) before fully collapsing to black holes \citep{inayoshi_et_al2020}. 

In the gravitational runaway model, BHs with masses of $10^3 - 10^4 M_\odot$ are formed in dense stellar clusters through either a slow process or a fast process. In the slow model \citep{miller+hamilton2002}, the $50 M_\odot$ BH remnant of a massive star gradually accretes low-mass BHs and eventually reaches a mass of $10^3 M_\odot$. A relatively massive seed BH is necessary because the timescale for dynamical ejection of lighter BHs is shorter than the timescale for the onset of runaway accretion for seeds with masses $< 20 M_\odot$. The possibility of forming the requisite $50 M_\odot$ seed BHs is debated \citep[e.g.][]{woosley2017}. 

In the fast scenario \citep{portegies_zwart+mcmillan2002}, stellar mergers occur before BH formation. In this case, supermassive stars are formed by mergers in the cores of dense clusters and then those supermassive stars collapse to form intermediate-mass black holes. This process may be metallicity-dependent due to the relationship between stellar metallicity, stellar winds, and mass loss. BH formation via fast gravitational runaway may be more likely in very metal-poor, very dense clusters formed at high redshift, but the existence of such clusters, while predicted in simulations \citep[e.g.,][]{devecchi+volonteri2009}, has not yet been observationally proven \citep{greene_et_al2020}.

As depicted in Figure~1 of \citet{greene_et_al2020}, if the dominant BH formation channel is direct collapse rather than gravitational runaway or the deaths of Population III stars, theory predicts a flatter positive correlation between black hole mass and galaxy mass, a steeper positive correlation between the fraction of galaxies with central black holes and galaxy mass, and a more top-heavy distribution of black hole masses. Conversely, if gravitational runaway is the dominant formation channel, galaxy mass would be expected to display positive correlations with black hole mass and black hole fraction while the distribution of black hole masses would be relatively flat at low masses. Finally, formation via deaths of Population III stars predicts a bottom-heavy black hole mass distribution, a flat relationship between black hole fraction and galaxy mass, and a steep positive correlation between black hole mass and galaxy mass. 

\breakthrough The detection of quiescent black holes with masses $<10^5 M_\odot$ would constitute breakthrough progress. Specifically, \citet{pacucci_hwo25} note that a volume-limited survey of roughly 100~nearby dwarf galaxies sensitive to black hole masses $\gtrsim 10^{4.5} M_\odot$ is needed to test models of black hole seeding. The galaxies should be within $10-30$~Mpc and cover a mass range of $10^6-10^9 M_\odot$. Target selection should favor galaxies for which HWO observations have the potential to resolve the gravitational sphere of influence surrounding black holes with masses $10^{4.5}-10^6 M_\odot$ and robustly measure stellar kinematics. Practically, this requirement may necessitate rejecting targets with very low surface brightness, nearby neighbors, unfavorable inclination angles, and very dusty nuclei. The selected targets should be observed with MOS UV/VIS spectroscopy at 100-1000~nm at high spatial resolution (10~mas). The spectral resolution should be $R\sim 8000$ to permit achieving the required $<30$~km/s velocity precision needed for breakthrough studies of the effects of black holes on the kinematics of nearby stars. 

\enabling Substantial progress also requires MOS UV/VIS spectroscopy of a volume-limited sample of local galaxies within $10-30$~Mpc that are amenable to the detection of low-mass black holes and measurements of the kinematics nearby stars. However, the survey sensitivity could be reduced to black hole masses $\geq 10^5 M_\odot$ rather than $<10^5 M_\odot$. The angular resolution requirement could be relaxed to 20~mas and the velocity precision could be degraded to 30~km/s. This velocity precision is still substantially higher than the current state-of-the-art measurement precision of $50-80$~km/s.

\subsubsection{The Formation and Evolution of Supermassive Black Holes: IMBH Mass and Spin Functions (SCDD-GG-2)}
\label{sssec:bh_mass_spin}
\sleads{Jenna M. Cann, Krista Lynne Smith, Francesca Civano, Satyapriya Das, Erin K. S. Hicks, Gagandeep Kaur, Stephanie LaMassa, Jeffrey D. McKaig, Missagh Mehdipour, Gopika SM}

This science case, which is presented in \citet{cann_hwo25}, also probes the origin of supermassive black holes and their effects on their host galaxies. The three objectives of this investigation are (1) determine the prevalence of massive black holes in dwarf galaxies and compare to the prevalence observed for larger galaxies like the Milky Way; (2) investigate the possible environmental-dependence of the mass and spin distributions of IMBHs; and (3) measure the spin distribution of AGN to assess the relative importance of mergers and accretion in their growth history. As discussed in detail in Section~\ref{sssec:bh_quiescent}, the processes by which SMBH seeds form and evolve are key research areas prioritized by Astro2020 \citep[B-Q4;][]{astro2020}. Constraining the initial mass function of SMBH seeds will also inform studies of dark matter in the early universe because the likelihood of formation and mass distributions of Pop~III stars and supermassive stellar remnants are affected by the masses of dark matter minihalos \cite[e.g.,][]{johnson_et_al2013}. 

To investigate possible differences between IMBHs in nuclear environments (i.e., at the centers of galaxies) and non-nuclear environments (e.g., in globular clusters), the science case proposes observing the central 10~kpc of low-metallicity dwarf galaxies with masses $<10^{8.5} M_\odot$ as well as globular clusters in the Milky Way and nearby galaxies. 
 
The rationale for favoring low-metallicity galaxies is due to their relatively quiescent cosmic history compared to their more metal-rich counterparts, which have likely undergone several epochs of star formation and merger activity, allowing us insight into a more ``pristine" population by which to investigate SMBH seeds. The choice of 10~kpc is motivated by theoretical simulations showing that the majority of accretive off-nuclear IMBHs will lie within that distance \citep{bellovary_et_al2019}. Multiwavelength UV/O/IR observations are needed to distinguish IMBHs from very massive neutron stars. 

For the IMBHs in globular clusters, mass measurements will be essential for distinguishing between the various formation channels discussed at length in \citet{greene_et_al2020} and summarized in Section~\ref{sssec:bh_quiescent}. The science case proposes both a kinematic search to detect quiescient IMBHs and an optical/IR spectroscopic program to detect actively accreting IMBHs displaying nebular lines. 

The current state-of-the-art is the detection of roughly 20 highly accreting BHs with estimated masses of $10^4 - 10^5 M_\odot$ residing in galaxies with masses of $10^9 - 10^{10} M_\odot$. Increasing the sample to 200 BHs (enabling) or 500 - 1200 (breakthrough) would enable more comprehensive studies of BH demographics and formation models. Higher sensitivity could extend the samples to much lower BH masses of $10^3 M_\odot$ (enabling \& breakthrough). For breakthrough science, detecting a large number of such light BHs ($M_{\rm BH}\sim 10^3 M_\odot$) would provide powerful constraints on BH formation channels and refine relationships between coronal line emission and BH mass (breakthrough). Breakthrough observations also have the potential to increase the number of BHs with measured spins by two orders of magnitude from roughly 10 to 1000, thereby providing further insight into their formation and merger history. 

Observationally, the ideal instrument for this science case is a highly sensitive NUV/VIS IFS capable of observing a large field of view at high spatial resolution. If that combination is not realistic, observations at high spatial resolution over smaller FOVs could be mosaicked to sample the full galaxy. However, observations of a larger FOV obtained a lower spectral resolution would be much less helpful because of the increased contamination from star formation. Accordingly, this science case favors spatial resolution over FOV, and will benefit significantly from the increased PSF stability of space-based observations. In addition, spin measurements obtained via continuum-fitting require UV spectra that cannot be obtained from the ground. These spectra should have resolution $R > 1500$ and extend from 90~nm to 800~nm or even redder if possible. Observations with multiple instruments are acceptable as long as the instrumental offsets are well-characterized. 

While some instruments on large ground based telescopes can achieve the necessary spatial and spectral resolution (e.g., Keck/OSIRIS), the limited FOV and decreased PSF stability of ground-based instruments renders ground-based observations less compelling. NIR observations could potentially be done by ground-based facilities after source locations have been well-constrained, but observations at shorter wavelengths require space-based observations from a telescope like HWO. Also collecting NIR observations from HWO is preferable because the less well understood and more variable PSFs of ground-based facilities would hinder precise comparisons of features across the two datasets, particularly given extended time between observations and potential intrinsic variability in the emission signatures.

\breakthrough In general, breakthrough progress for this science case requires detecting $500-1200$ accreting black holes with masses $\geq 10^3~M_\odot$ including some in host galaxies with masses $<10^6 M_\odot$. Breakthrough science also requires measuring the relationship between galaxy mass and black hole mass for galaxy masses $<10^6 M_\odot$. Additionally, breakthrough progress demands determining accretion rate diagnostics for a variety of black hole environments and measuring spins for 1000~black holes out to redshift $z\sim5$. 

Achieving these goals would require IFS observations with sensitivity to line fluxes as low as $10^{-20} \rm \, erg \, cm^{-2} \, s^{-1}$ at $R>3000$ spectral resolution\footnote{Note: \citet{cann_hwo25} quote $R>1500$ in the body of the science case, but their Table~2 lists $R\sim 2000-3000$ for enabling science. Accordingly, we adopt a higher value of $R>3000$ here.} and 1~mas/px spatial resolution\footnote{Note: \citet{cann_hwo25} acknowledge that the listed spatial resolution of 1~mas/px is not practical for HWO but they chose to include this value because it illustrates the resolution needed for kinematic detection of an IMBH at a distance of roughly 10~Mpc. In other tables, we relax this requirement to 50~mas/px.} over regions roughly 1' in diameter spanning 90~nm to 2.5~$\mu$m. Spectra could be obtained in multiple pointings with one or more instruments. If there is tension between spatial resolution and field of view, higher spatial resolution is strongly preferred because mosaics could be used to compensate for a reduced field of view. 

Additionally, detecting coronal lines would require filters spanning $330~\rm nm - 2.5 \mu m$ (e.g., [Si VI] at $1.96\mu$m and [Ne V] at 330~nm and 340~nm). [Si VI] is the brightest NIR coronal line, but the red wavelength limits detectability for objects at higher redshifts. The shorter rest wavelengths of [Ne V] are much more accessible, but those features are over an order of magnitude fainter than NIR lines \citep{mckaig_et_al2024}, thereby requiring higher sensitivity to detect bluer coronal lines. 

\enabling The requirements for enabling science are largely the same as for breakthrough science, but the target sample could be reduced to 200~black holes and low end of the host galaxy mass distribution could be increased to $10^6 - 10^7 M_\odot$ rather than $<10^6 M_\odot$. Additionally, while observations should have the potential to detect black holes with masses as low as $10^3 M_\odot$ and better constrain black hole accretion rates, enabling science does not require detection of a sample of $10^3 M_\odot$ black holes or empirical determination of accretion rate diagnostics. 

Observationally, the line flux sensitivity could be relaxed to $10^{-19} \rm \, erg \, cm^{-2} \, s^{-1}$, and the spectral resolution could be reduced to $R\sim2000-3000$ While an IFS is highly preferred, enabling science could still be achieved by using initial wide-field UV imaging at $100-200$~nm to identify candidate black holes by their high UV emission and point-like appearance and then obtaining MSA spectra of those candidates. In addition to being less efficient, this two-stage strategy also increases the likelihood of observing false positives rather than true black holes. 

Additionally, the wavelength range needed for coronal lines could be narrowed to 330 nm - 2.0 $\mu$m. This range has a bluer cut-off than that needed for breakthrough progress while still including [Si VI] at $1.96\mu$m. However, the redder cut-off will substantially reduce the range of redshifts over which [Si VI] can be observed from $z<0.1$ for $\lambda \leq 2.5~\mu$m to $z<0.01$ for $\lambda \leq 2~\mu$m. Consequently, the sample of possible target galaxies will be reduced from roughly 40,000 to only 6000. If redder observations are not possible, an alternative way to increase the target sample would be to prioritize sensitivity to the [Ne V] coronal line at rest wavelengths of 334.5~nm and 342.5~nm. However, current data suggests that these [Ne V] lines are less common than [Si VI], so increased sensitivity to [Ne~V] may not be sufficient compensation for the reduction in [Ne~V] targets. 

\subsubsection{Probing Energy Extraction from Black Holes with HWO (SCDD-GG-1)}
\label{sssec:bh_energy}
\sleads{Mainak Singha, Peter Senchyna}

This science case was considered during the START process but was not included in the STScI science case Portal or the HWO25 conference proceedings. The science case aims to investigate the relationship between SMBHs and their host galaxies and how SMBHs affect their surroundings. These topics are related to Astro2020 questions B-Q2, B-Q3 and D-Q3 as well as discovery area B-DA \citep{astro2020}. The core of this science case is determining the launching point and driving mechanisms behind black hole jets in active galaxies at $z \lesssim 0.06$. Observations of active galaxies with HWO could measure the line-of-sight velocity of ionized gas and the magnetic field strength along the jet-axis to determine whether the jet is launched from the accretion disk or from the event horizon. 

Theory predicts that jets launched from the accretion disk should have constant line-of-sight gas velocity along the jet axis \citep{fukumura_et_al2010} and magnetic field strength that decreases with increasing distance from the nucleus \citep{dihingia_et_al2021}. Measurements of the line-of-sight velocity and the wind pressure profile are also useful for determining whether magnetic field pressure or winds drive collimation. In the magnetic field case, the jets will likely incorporate a turbulent cocoon \citep[e.g.,][]{levinson+begelman2013} while the combination of high line-of-sight velocities ($> 500$ km s$^{-1}$) and high pressure ($> 10^{-3}$ dyne) would suggest wind-driven collimation \citep[e.g.,][]{fukumura_et_al2014}. 

The current state-of-the-art observations for this science case are NIR observations with JWST/NIRSpec at spatial resolution of 100~mas and spectral resolution of $R\sim 1000$. 
The UV data will reveal absorption features indicative of disk accretion winds and shock structures. Observing highly ionized gases such as O VI (103~nm) is particularly valuable and drives the need for UV data. The needed spatial resolution is set by the size of the black hole sphere of influence and set to half of the value for a black hole at $z\sim0.04$ (i.e., 12~mas given that $z\sim0.04$ black holes are expected to have gravitational spheres of influence spanning 24~mas). 

The request for NIR IFS data is motivated by the desire to detect and trace jet cocoons, which could be formed as a result of interactions between the jet and the surrounding medium. Observations of [Fe II] are particularly informative for this science case. NIR data would also permit measurements of emission line ratios that would reveal ionization conditions and detection of coronal lines from disk-driven winds as in Section~\ref{sssec:bh_mass_spin}. 

Additionally, measurements of the angle and degree of jet polarization could distinguish between spin-driven jets and disk-driven jets. Compared to disk-driven jets, spin-driven jets are expected to have higher polarization ($1\%$ versus $0.1 - 0.5\%$) and narrower polarization angles ($\pm 10^\circ$ versus $\pm 20^\circ$). 

\breakthrough Breakthrough progress would require FUV and NIR IFS observations of galaxies at high spatial and spectral resolution (12~mas, $R > 5000$). The wavelength range should extend blueward to at least $100$~nm to ensure detection of O~VI and other highly ionized lines. The target sample should include nearby radio-loud galaxies at $z<0.01$ (e.g., M87, NGC~1068), radio-loud galaxies at $z<0.06$ from the MOJAVE survey, and radio-quiet galaxies from the BASS survey at $z<0.06$. Overall, the sample should span radio luminosities of $10^{21} - 10^{28}$~W~Hz$^{-1}$ and include some sources with super-Eddington accretion rates ($\lambda_{\rm Edd} = 10^{-4} - 10$). The observations should be able to detect spatially-resolved broadband emission at $7\sigma$ confidence at 12~mas resolution. Observations of [Fe~II] emission would be needed to characterize shocks while measurements of the [Si~VI] and [S~VIII] coronal lines would be required to investigate disk-driven winds. In addition, distinguishing between disk-driven jets and spin-driven jets at 10$\sigma$ would require UV polarization measurements capable of measuring polarization angles to $5^\circ$ and polarization fractions to $0.01\%$

\enabling For substantial progress, the requirements for IFS data would be relaxed to 25~mas spatial resolution and $R>2700$ spectral resolution. Additionally, the range of radio luminosities and accretion rates covered by the target sample could be reduced to $10^{22}-10^{28}$~W~Hz$^{-1}$ and $\lambda_{\rm Edd} =10^{-4} - 1$, respectively. These changes would remove both radio-quiet AGN and super-Eddington sources.

Instead of distinguishing between disk-driven and spin-driven jets, enabling science would require detecting winds at $3\sigma$ confidence. The corresponding observational requirement is the ability to measure the polarization fraction to a reduced precision of 0.03\% rather than the 0.01\% level required for breakthrough progress. 

\subsubsection{Imaging the Dusty Torus Around Supermassive Black Holes
 (SCDD-GG-3)}
\label{sssec:bh_torus}
\sleads{Varoujan Gorjian, Chris Packham, Erin Hicks, Stephanie La Massa}

This science case is discussed by 
\citet{gorjian_hwo25}. Like Section~\ref{sssec:bh_quiescent}, this science case also investigates the growth of SMBHs and how SMBHs affect their host galaxies, but it focuses on the properties and possible mediating effects of the dusty tori surrounding SMBHs. The size and opacity of the torus will affect the portion of the galaxy that is exposed to radiation from the SMBH accretion disk and the spectrum of the escaping radiation. Tori are expected to be small \citep[diameters of 25-130~pc; ][]{garcia-burillo_et_al2021} and therefore are challenging to resolve. 

High spatial resolution UV imaging with HWO could potentially measure the thickness of tori, thereby constraining the fraction of the central AGN accretion disk that is unshielded. For detected tori, UV spectroscopy or multiwavelength imaging could reveal the dust absorption characteristics, thereby improving models. Observations of the dust-induced extinction bump at 217.5~nm are particularly important for understanding dusty tori and will not be possible with ground-based facilities. Furthermore, UV observations are essential for probing how dusty tori affect light emitted by AGN at the peak of the AGN emission spectra. While ground-based observatories could potentially observe dusty tori at optical wavelengths, achieving the requisite spatial resolution will require adaptive optics which in turn requires nearby bright natural guide stars for tip/tilt correction even when laser guide stars are used for most corrections. Accordingly, large ground-based telescope may not be able to achieve the spatial resolution needed to sufficiently study these systems even if the required angular resolution is much larger than their theoretical diffraction limits.

In the Unified Scheme of AGN structure \citep{antonucci1993}, torus geometry should affect the relative frequencies of AGN that exhibit both broad lines and narrow lines (i.e., Type 1 Seyfert galaxies) and those that should exhibit only narrow lines (i.e., Type II Seyfert galaxies). Current observations suggest a 2:1 ratio of Type~II to Type~I Seyfert galaxies, which would be consistent with torus edge thicknesses of 43-225~pc assuming an angular thickness of $120^\circ$. If HWO observations reveal that the torus covering fraction is smaller than anticipated, then a significant fraction of AGN may lack broad line regions, which would have important implications for interactions between AGN and their host galaxies. 

\breakthrough Breakthrough science would require UV/VIS observations of $>10$ nearby ($<65$~Mpc) Seyfert galaxies at high angular resolution (5~mas). The observations should cover a wavelength range of 90-700~nm at a spectral resolution $R>2700$. 

\enabling For enabling science, a smaller sample of five Seyfert galaxies should be observed at a lower spatial resolution of 22~mas over a reduced spectral range of 90 - 400~nm. The spectral resolution ($R>2700$) and sample distance limit ($<65$~Mpc) would be unchanged from the breakthrough case.

\subsubsection{Spatially Resolving the Fundamental Elements of Reionization in Galaxies (SCDD-GG-6)}
\label{sssec:resolve_reionization}
\sleads{Xinfeng Xu, Stephan McCandliss, Allison Strom, Hsiao-Wen Chen, Yumi Choi, Annalisa Citro, H\r{a}kon Dahle, Matthew J. Hayes, Anne Jaskot, Logan Jones, Gagandeep Kaur, Themiya Nanayakkara, Alexandra Le Reste}

As discussed in \citet{xu_reionization_hwo25}, this science case aims to investigate how galaxies, and Lyman continuum star clusters (LyC clusters) in particular, ionize the intergalactic medium. Due to the small spatial extent of LyC clusters (10 - 100~pc), they have thus far been detected primarily in gravitationally lensed galaxies \citep[e.g.,][]{vanzella_et_al2022, pascale_et_al2023}. A notable exception is the extreme dwarf starburst galaxy Haro~11, for which \citet{komarova_et_al2024} obtained COS spectra. The higher UV sensitivity and angular resolution of HWO has the potential to advance studies of ionization by spatially resolving LyC clusters and the interstellar medium (ISM) clouds that surround them. In addition to dramatically expanding the sample of accessible unlensed galaxies, the sensitivity and resolution of HWO would also significantly advance studies of lensed galaxies, which should be detected in large numbers by Euclid and Roman. Observations with HWO could measure the rate at which LyC escapes from the cluster, properties of cluster members (e.g., mass, age, metallicity), and the conditions of the surrounding clouds (e.g., gas abundance, electron density, temperature, pressure, and extinction). 

These observations would also provide valuable insight into how radiation from the most massive members of LyC clusters (i.e., stellar feedback) affects galactic outflows and the distribution of ionized, neutral, and metal-rich gas within galaxies. Current observational and numerical studies of feedback within LyC lack the angular resolution needed to distinguish between various feedback models \citep[e.g.,][]{trebitsch_et_al2017, barrow_et_al2020, kimm_et_al2019, carr_et_al2025a}, but the increased spatial resolution and sensitivity of HWO would allow future observers to resolve LyC clusters and explore their effects on cosmic reionization. 

The number of galaxies targeted is set by assuming that each galaxy harbors $10-100$ LyC clusters. The requested spectral resolution is driven by the need to measure the velocity dispersion of cool clouds in the ISM \citep{chen_et_al2023, carr_et_al2025a}. The wavelength range is set by the need to directly detect LyC (90~nm rest-frame) and other rest-frame UV continuum and lines (i.e., 100 - 200~nm). Observing targets at higher redshifts will shift the necessary wavelength range redward and require increased spatial resolution. 

\breakthrough Breakthrough progress on this science case requires UV observations of roughly 10,000~unlensed galaxies and $10^5-10^6$ LyC clusters at redshifts $z<1$. The observations should cover a $10''\times 10''$ field-of-view at a spatial resolution $0\farcs01/$spaxel (i.e., 20 - 80 pc/spaxel) and spectral resolution $R\sim 30,000$ over $50-200$~nm. The sensitivity should be high enough to measure LyC leakage to $5\%$ at $2\sigma$ confidence when considering a 2~nm region centered at rest frame 90~nm. For breakthrough science, the observations should be sensitive to average flux density $\geq 1 \times 10^{-19}$~erg~s$^{-1}$~cm$^{-2}$~\AA$^{-1}$ at wavelengths corresponding to 90~nm rest-frame. 

\enabling Substantial progress requires a smaller sample of approximately 1000 unlensed galaxies and $10^4 - 10^5$ LyC clusters. The observational requirements would be relaxed to a $5''\times 5''$ field-of-view, spatial resolution $0\farcs05$/spaxel, spectral resolution $R\sim 10,000$, and wavelength range $90-200$~nm. Additionally, the sensitivity limit would be softened to $\geq 5 \times 10^{-19}$~erg~s$^{-1}$~cm$^{-2}$~\AA$^{-1}$ at 90~nm rest-frame.

\subsubsection{How Do Ionizing Photons Escape from Star-Forming Galaxies? (SCDD-GG-7)}
\label{sssec:lyman_escape}
\sleads{Cody Carr, Renyue Cen, Sophia Flury, M. S. Oey, Stephan McCandliss, Allison Strom, Claudia Scarlata}

This science case investigates how ionizing radiation (i.e., Lyman continuum (LyC) photons) escapes from galaxies and was presented in \citet{carr_hwo25, carr_jatis_25}. The overarching goal is to determine the processes and the timescale over which the universe became ionized. Although it might seem logical to explore Epoch of Reionization (EoR) by observing young galaxies at high redshift, any ionizing radiation escaping from those galaxies would be absorbed by the neutral IGM well before reaching Earth \citep{inoue_et_al2014}. 

An alternative approach is to instead study the physics of LyC escape using simulations and observations of local analogs of high redshift galaxies \citep[e.g.,][]{izotov_et_al2016, xu_et_al2016, flury_et_al2022a, flury_et_al2022b, choustikov_et_al2024a, choustikov_et_al2024b, jaskot_et_al2024a, jaskot_et_al2024b}. The key quantity of interest is the Lyman continuum
(LyC) escape fraction ($f_{\rm esc}$), which represents the fraction of LyC photons that escape the interstellar and circumgalactic medium (ISM and CGM) of a galaxy.

\begin{figure*}
 \centering
 \includegraphics[width=\linewidth]{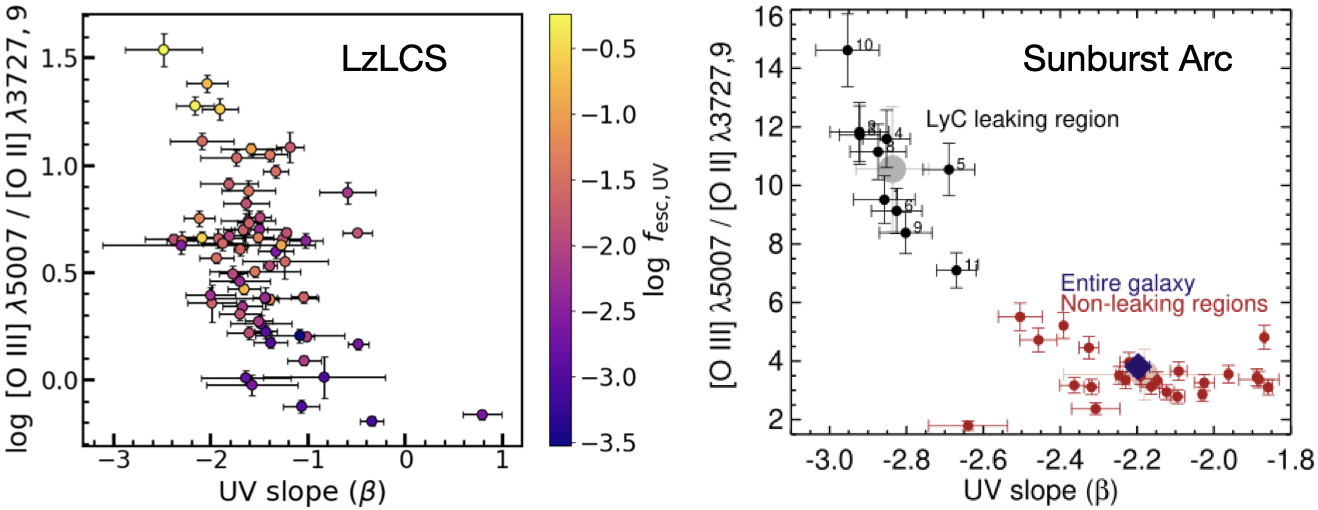}
 \caption{LyC escape according to the O32 = [O III] 5007 / [O II] 3726,9 ratio and the slope of the FUV continuum slope ($\beta$). The left panel shows values derived from HST COS integrated spectra for entire galaxies selected from LzLCS. Data were taken from \citet{flury_et_al2022a, flury_et_al2024}. The right panel shows values drawn from individual regions of the Sunburst Arc galaxy, figure taken from \citet{kim_et_al2023}. The black points correspond to a region with LyC emission and the red points to regions without. The blue point represents the average value measured over the whole galaxy. This sunburst arc demonstrates that $f_{\rm esc}$ and its relation to O32 and likely vary dramatically across the surface of a galaxy, potentially explaining the variation observed in LzLCS.}
 \label{fig:gg7_diagnostics}
\end{figure*}

By using observations of low-redshift galaxies to determine which other parameters correlate with $f_{\rm esc}$, astronomers can identify diagnostics to infer $f_{\rm esc}$ for high-redshift galaxies. Known indicators of high $f_{\rm esc}$ include enhanced [O~III] 500.7~nm/[O~II] 372.69~nm (O32) ratios \citep{zackrisson_et_al2013, nakajima+ouchi2014}, high star formation rate surface density ($\Sigma_{UV}$), and bluer slopes of the FUV continuum \citep[$\beta_{UV}$;][]{chisholm_et_al2022}. However, observations show that the relationship between $f_{\rm esc}$ and these metrics is complex, meaning no single indicator can perfectly predict $f_{\rm esc}$ \citep[e.g.,][]{flury_et_al2022b, jaskot_et_al2024b}. More detailed studies are necessary to understand how the different diagnostics relate to LyC escape.

Gravitational lensing provides a rare opportunity to zoom-in on the small-scale features of LyC emitting galaxies \citep[e.g.,][]{welch_et_al2025}. For example, \citet{kim_et_al2023} measured O32, $\beta_{UV}$, and $f_{\rm esc}$ at various regions within the gravitationally lensed Sunburst Arc, finding that LyC leaking (i.e., high $f_{\rm esc}$) and LyC non-leaking (i.e, low $f_{\rm esc}$) regions of the galaxy had very different relationships between O32 and $\beta_{UV}$. Accordingly, galaxy-integrated observations of O32 and $\beta_{UV}$ are insufficient to estimate $f_{\rm esc}$. The results of \citet{kim_et_al2023} are shown in Figure~\ref{fig:gg7_diagnostics} alongside values inferred from integrated spectroscopy in the Low-z Lyman Continuum Survey \citep[LzLCS;][]{flury_et_al2022b}. While the results of LzLCS show clear differences between the strongest and weakest LyC emitting galaxies in terms of $f_{\rm esc}$, there is not a perfect correlation, reflecting the averaging over the surface of the galaxy.
 
\begin{figure*}
 \centering
 \includegraphics[width=\linewidth]{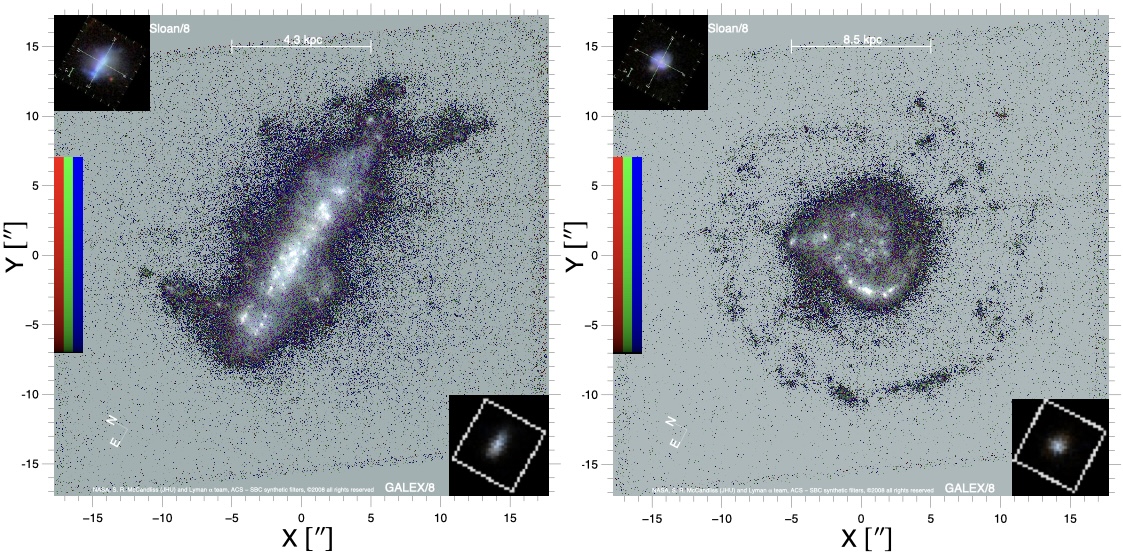}
 \caption{Hubble Space Telescope Advanced Camera for Surveys/Solar Blind Channel (ACS/SBC) synthetic filter images of two nearby star-forming galaxies \citep{mccandliss2009}. Blue indicates regions of diffuse Ly$\alpha$ emission. We count roughly 20 (left panel)
and 30 (right panel) distinct star clusters in each galaxy. Spatially and spectrally resolving these
clusters will be critical for determining which stellar populations contribute to LyC escape and
how stellar feedback clears pathways through the CGM to facilitate LyC escape.}
 \label{fig:gg7_clusters}
\end{figure*}

To better understand the relationship between $f_{\rm esc}$ and various galaxy diagnostics, galaxies must be spatially resolved on the scale of super star clusters (SSCs), which host the
massive O- and B-type stars that produce most LyC photons in star-forming galaxies.
Examples of SSCs in star-forming galaxies are shown in Figure~\ref{fig:gg7_clusters}.

Feedback from SSCs plays a critical role in creating escape pathways for LyC photons. The type of feedback can be inferred from the age of the stellar population: young, metal-poor systems are dominated by radiation feedback, while more evolved systems experience supernova-driven feedback. LyC escape is generally categorized into three regimes: (1) radiation-dominated, where winds and ionization from massive stars clear low-density
channels \citep[e.g.,][]{flury_et_al2024, carr_et_al2025a}; (2) “two-stage burst” scenarios, in which supernovae initiate clearing that is enhanced by radiation pressure \citep[e.g.,][] {martin_et_al2024, flury_et_al2024}; and (3) supernova-dominated escape, where blast waves alone carve out escape routes \citep[][]{kimm+cen2014, cen2020, rosdahl_et_al2022, choustikov_et_al2024a}.

Spatially resolved observations—ideally with integral field spectrographs (IFSs)—are essential to determine how LyC photons escape from galaxies. As demonstrated by \citet{yuan_et_al2023}, spatially resolved spectroscopy can constrain outflow mass to within a factor of two. However, existing telescopes lack the resolution to probe SSC-scale structures ($\sim$1–100 pc) in unlensed systems. The Habitable Worlds Observatory would overcome this limitation.

The observations should target Ly$\alpha$ (rest-frame 121.6~nm), higher-order Lyman lines (e.g., Ly$\beta$ at 102.6~nm, Ly$\gamma$ at 97.3~nm, Ly$\delta$ at 95.0~nm), and the Lyman limit at 91.2~nm, as well as low- and high-ionization metal lines. Low-ionization transitions like Si II (119.0~nm, 119.3~nm, 126.5~nm, 139.3~nm, and 152.7~nm) would provide information about the distribution of cold clouds and the kinematics of galactic winds. 

Lines from higher ionization states such as SiIII (120.6~nm) and SiIV (139.3~nm, and 140.3~nm) would constrain the ionization state, informing total and neutral hydrogen column densities \citep{xu_et_al2022, huberty_et_al2024}. Observations of OVI (103.2~nm and 103.8~nm) would probe the warm-hot CGM and metal recycling processes. Finally, measurements of the far-UV spectral energy distribution could be used to estimate the ages of stellar populations \citep{saldana-lopez_et_al2022}. 

\breakthrough Breakthrough progress requires high-resolution, spatially resolved UV spectra ($30,000 < R < 100,000$, $0.3-30$~mas, $95 - 200$~nm) over a field of view no smaller than $3'' \times 3''$. A larger field-of-view (e.g., $10'' \times 10''$) would be advantageous for covering the full circumgalactic medium in a single pointing. 

\enabling For substantial progress, the spectra should cover the same wavelength range ($95-200$~nm) but the spectral resolution could be reduced $R>15,000$ rather than $R=30,000 -100,000$. Additionally, the spatial resolution could be relaxed to $30-50$~mas.

\subsubsection{Tracking Cosmic Reionization via Green Pea Galaxies with HWO (SCDD-GG-8)}
\label{sssec:green_peas}
\sleads{Mainak Singha, Kristen Garofali, Annalisa Citro, Sophia Flury, Cody Carr, Xinfeng Xu, Themiya Nanayakkara, Karla Z. Arellano-Córdova, Stephan McCandliss}

This science case was developed during the START era and not listed on the STScI Science Case Portal or included in the HWO25 Conference Proceedings. Like the case presented in Section~\ref{sssec:lyman_escape}, this science case also investigates reionization and the escape of ionizing radiation from host galaxies. As discussed in Section~\ref{sssec:resolve_reionization}, IGM absorption prevents direct observations of escaping ionizing radiation from high-redshift galaxies, but observations of local, low-redshift galaxies can be used to explore the physics of LyC escape. This science case focuses specifically on observing $z < 1$``Green Pea'' galaxies, which are small and appear green due to strong emission at 500.7~nm from [O~III] \citep{cardamone_et_al2009}. 

Despite their low masses ($10^{8.5} - 10^{10} M_\odot)$, Green Peas are active sites of star formation (star formation rates $\sim 10 M_\odot \rm yr^{-1}$; specific star formation rates $\lesssim 10^{-8} \rm yr^{-1}$) and appear to be a local analog to UV bright galaxies observed at high redshift \citep[e.g.,][]{cardamone_et_al2009, rhoads_et_al2023}. LyC escape has been detected from multiple Green Pea galaxies, indicating that their high-redshift counterparts may have played a key role in reionization \citep{malkan+malkan2021, izotov_et_al2021, flury_et_al2022a, flury_et_al2022b}. Observations of Green Peas at higher spatial resolution ($< 100$~pc; ideally with an IFS) would better reveal the processes by which ionizing radiation travels through and out of these galaxies. 

The observational goals for this science case are to map how LyC escape fraction $f_{\rm esc}$ varies within Green Peas and determine the drivers of ionization. Measuring and comparing the strengths of various UV~lines will reveal stellar populations at different ages and the contributions of stellar feedback \citep[e.g.,][]{flury_et_al2022a}, constrain the contribution of binary stars to the observed $f_{\rm esc}$ \citep[e.g.,][]{ma_et_al2016, rosdahl_et_al2018}, and map out black hole accretion disks, shocks, and star-forming regions \citep{castellano_et_al2024}. Determining the shape of the UV SED could also reveal the extent of H~I nebular contributions to the observed LyC \citep[e.g.,][]{inoue2010, simmonds_et_al2024}.

In addition, measurements of gas kinematics will support studies of the possible role of turbulence in facilitating LyC escape by altering porosity \citep{seon2009, elmegreen+efremov1997}. If $f_{\rm esc}$ is higher in more turbulent regions, then that would indicate that increased turbulence contributes to escape. Conversely, the absence of a corelation between turbulence and $f_{\rm esc}$ would suggest that ISM inhomogeneities drive escape \citep[e.g.,][]{gazagnes_et_al2020, saldana-lopez_et_al2022}. Alternatively, if regions of high $f_{\rm esc}$ tend to be located in regions affected by stellar winds or supernovae, then stellar feedback may be responsible for carving escape routes through the ISM \citep[e.g., ][see also Section~\ref{sssec:resolve_reionization}]{alexandroff_et_al2015, chisholm_et_al2017, jaskot_et_al2017, komarova_et_al2021, amorin_et_al2024}. The high spatial and medium-high spectral resolution possible with an IFS are essential for distinguishing between or determining the contributions of each of these pathways to LyC escape and reionization. 

\breakthrough Breakthrough progress would expanding the sample of green peas to $500-1000$ galaxies and obtaining high spatial resolution (12~mas/spaxel) and high spectral resolution ($R=8000-10,000$ observations with an IFS. The observations should cover $90 - 400$~nm and span a field of view of $2'' \times 2''$. Coverage of rest-frame LyC (91.2~nm) is essential. The high spatial resolution is driven by the need to resolve the specific regions of Green Pea galaxies responsible for LyC escape. 

\enabling For substantial progress, the sample size would be reduced to $200-500$~green peas and the wavelength range would be narrowed to $90-350$~nm. Additionally, the spatial and spectral resolution requirements could be relaxed to 24~mas/spaxel and $R>6000$, respectively.

\subsubsection{Identifying the Contributors to Cosmic Reionization with the Habitable World Observatory: LyC Escape Calibration in Faint Galaxies (SCDD-GG-9)}
\label{sssec:lyman_indirect}
\sleads{Annalisa Citro, Claudia M. Scarlata, Cody A. Carr, Yumi Choi, Sophia R. Flury, Matthew J. Hayes, Anne Jaskot, Gagandeep Kaur, Alexandra Le Reste, Matilde Mingozzi, Themiya Nanayakkara, Sally Oey}

As discussed by \citet{citro_hwo25}, this science case aims to determine the main contributors to the reionization of the Universe. The specific goal is to determine the fraction of ionizing photons ($f_{\rm esc}$) that are able to escape from galaxies into the diffuse IGM. Observations of lensed galaxies with HST and JWST have significantly improved estimates of the volume density and production rate of ionizing photons, indicating that fainter galaxies are particularly important for reionization \citep[e.g.,][]{atek_et_al2015, livermore_et_al2017, lin_et_al2024, simmonds_et_al2024}, but the increasing opacity of the IGM at high redshifts prevents direct measurements of the escape fraction at $z > 4$.

An alternative path to directly measuring the escape fraction in the high-redshift Epoch of Reionization (EoR) galaxies that reionized the universe is to measure escape fractions from low redshift galaxies that are local analogs to EoR galaxies and use those insights to construct relationships between escape fraction and quantities that can be directly measured for EoR galaxies (e.g., dust content, gas ionization state). Past work has largely been restricted to galaxies brighter than $M_{\rm UV} \sim -18$, but theoretical work suggests that fainter galaxies should have higher escape fractions and are therefore likely to play a key role in reionization \citep{wise_et_al2014, finkelstein_et_al2019}. Accordingly, measuring escape rates from fainter, less massive galaxies and extending the local-EoR calibrations to the faint galaxy regime is essential for understanding the primary sources of the high energy photons that reionized the universe.

The data needed for this science case are photometric and spectroscopic observations of faint, low-redshift galaxies covering both rest-frame UV and optical wavelengths. UV photometry at wavelengths $<100$~nm is needed to measure LyC flux and constrain $f_{\rm esc}$ while optical photometry is needed to measure galaxy compactness and star formation rate surface density. Moderate resolution spectroscopic observations in the UV and blue optical ($R \gtrsim 10,000$, $100~\rm nm \lesssim \lambda \lesssim 510~\rm nm$) will disentangle stellar and ISM absorption features, thereby permitting assessments of the gas covering fraction. 

Combined, the photometric and spectroscopic data will enable measurements of dust attenuation, ISM ionization state, resonant scattering, neutral gas covering fraction, and galaxy compactness, all of which can be used as indirect estimators of $f_{\rm esc}$. The broad spectral coverage and exquisite sensitivity of HWO would provide an unprecedented opportunity to collect the full dataset from a single facility, but it may also be possible to address this science case by combining Roman Space Telescope observations of indirect LyC indicators with observations by ground-based extremely large telescopes of direct LyC indicators. 

\breakthrough Breakthrough progress requires observing a sample of 40,000 galaxies as faint as $m_{\rm UV}=25$. Imaging should be sensitive enough to detect 32.5~ABmag LyC sources at $<100$~nm. Spectra should cover rest-frame wavelengths of $100-510$~nm at resolution $R\gtrsim10,000$ and detect the continuum at S/N=100 at 160~nm. High spectral resolution and sensitivity are required to separate ISM absorption lines from nearby stellar lines and to detect weak features. 

\enabling For enabling science, the sample would be reduced to 2000~galaxies as faint as $m_{\rm UV}=22.5$. Due to the focus on brighter targets, the imaging depth could be relaxed to 30~ABmag at $<100$~nm. Additionally, spectra should cover the same $100-510$~nm wavelength range, but the spectral resolution could be decreased to $R>2000$ rather than $R\gtrsim10,000$ and the sensitivity could be reduced by a factor of two to S/N=50 on the continuum at 160~nm. 

\subsubsection{Evolution of the Ionizing Photon Luminosity Function (SCDD-GG-11)}
\label{sssec:ionizing_lf}
\sleads{Stephan McCandliss, Swara Ravindranath, Sangeeta Malhotra, Chris Packham, Sophia Flury, Alexandra Le Reste, Allison Strom, Marc Postman, John O'Meara
}

As described in \citet{mccandliss_hwo25}, this science case aims to investigate re-ionization by measuring how the number and ionizing luminosity of galaxies has varied over the history of the universe. Quantifying the level of ionizing radiation from a broad sample of galaxies with a range of sizes and environments will allow researchers to construct models of how the spectrum of ionizing radiation has changed over cosmic time and facilitate searches for the modern, low-z counterparts to the high redshift galaxies that ionized the universe. This science case is complementary to SCDD-GG-9 (Section~\ref{sssec:lyman_indirect}), which aims to identify the sources of Lyman continuum radiation that reionized the universe and focuses particularly on galaxies on the faint end of the luminosity function. 

Building on past work with GALEX and upcoming work with UVEX \citep{kulkarni_et_al2021}, this investigation would measure Lyman continuum emission from low-redshift galaxies. Importantly, observations with HWO could constrain the shape of the ionizing continuum at wavelengths shorter than 90~nm, thereby circumventing the challenges of estimating the LyC escape fraction in regions of the spectrum affected by the Lyman bump \citep[i.e., just redward of 90~nm;][]{inoue2010}. 

\breakthrough The desired dataset for this science case consists of UV multi-object spectroscopy of a large number of galaxies in fifteen fields within a region of the sky spanning roughly one square degree to mitigate the effects of cosmic variance. For breakthrough science, the sample should include 500~galaxies per $\Delta z = 0.2$ redshift bin for $0.2 \leq z \lesssim 1.2$. The spectra should be sensitive to LyC luminosities as low as $2 \times 10^{37}$~erg~s$^{-1}$~\AA$^{-1}$ and Lyman continuum escape fractions $f_{\rm esc} \leq 0.01$. The corresponding flux and magnitude limits at rest-frame 90~nm (i.e., observed wavelength $90(1+z)$~nm) should be $10^{-20}$~erg~s$^{-1}$~cm$^{-2}$~\AA$^{-1}$ and 32, respectively. The ideal instrument configuration would be a MOS with three modules, each with $736 \times 384$ shutters. The requested configuration of three 736$\times$384~microshutter arrays covering three distinct 4~arcmin$^{2}$ fields corresponds to a spatial resolution of 225~mas/shutter and a total field-of-view of 12~arcmin$^2$. The MOS should have a spectral resolution of $R\sim1000$ over a wavelength range of 100-180~nm.

\enabling For enabling science, a smaller sample size, coarser redshift binning, and a reduced redshift range would be sufficient (i.e., 200 galaxies per $\Delta z = 0.3$ redshift bin for $0.2 \leq z \lesssim 1.0$). In this case, the MOS would consist of a only a single shutter module, still with $736 \times 384$ shutters, and therefore cover a smaller instantaneous field-of-view of four square arcminutes. The wavelength range would be unchanged (100 - 180~nm) but the spectral resolution would be decreased to $R\sim500$. The flux and magnitude limits at rest-frame 90~nm would be relaxed to $10^{-19}$~erg~s$^{-1}$~cm$^{-2}$~\AA$^{-1}$ and 30, respectively.

\subsubsection{Dark Matter through Dwarf Satellite Galaxies in Milky-Way Analogue Galaxies in the Local Universe
 (SCDD-GG-12)}
\label{sssec:dm_power}
\sleads{Jessica E. Doppel, Ethan O. Nadler, Leonidas Moustakas, Mathilde Jauzac, Richard Massey}

This investigation is described in \citet{doppel_hwo25} and is motivated by the desire to determine the nature of dark matter. The current paradigm of $\Lambda$CDM adequately describes the large-scale structure of the universe, but it is an open question whether $\Lambda$CDM satisfactorily describes the behavior of the universe on small scales \citep[e.g.,][]{sales_et_al2022}. This science case aims to address that question by measuring the dark matter halos of low-mass dwarf galaxies. 

If dark matter undergoes self-interactions, the specific nature of those interactions would affect whether dark matter subhalos of dwarf galaxies are predominantly cuspy or cored \citep[e.g.,][]{turner_et_al2021a, yang_et_al2023}. 
In addition, measuring the number of dwarf satellites in the dark matter subhalos of Milky-Way-mass host galaxies has the potential to distinguish between various models of cold dark matter (CDM), self-interacting dark matter (SIDM), and warm dark matter (WDM). Measurements of the frequency of satellite galaxies with masses below $10^6 M_\odot$ will be particularly useful for ruling out various CDM and WDM models, but satellite counts alone will likely be insufficient for assessing the performance of SIDM models.

The proposed dataset for this science case consists of high-resolution, photometric observations of Milky-Way-mass galaxies (i.e., galaxies with stellar masses of $10^{10} M_\odot - 10^{11} M_\odot$). Ideally, HWO would image the complete virial volume around each target host galaxy, but such observations are likely to be extremely time-consuming. Instead, HWO could image portions of the virial volume of each host galaxy and the resulting satellite galaxy counts could be extrapolated to estimate the full mass distribution of satellite subdwarf galaxies. Observations in at least two filters are required to confirm the association of RGB stars and satellite galaxies. \citet{doppel_hwo25} focused on $g$-band observations because RGB stars emit the most flux at visible wavelengths and therefore $g$-band is a well-motivated choice for one of the observation wavelengths. 

\breakthrough Breakthrough progress in this science case demands sufficient sensitivity to place a lower limit on the warm dark matter particle mass of $\ge 13.0$~keV at $3\sigma$ confidence. Observationally, achieving this goal requires observing $>10$ Milky-Way-mass hosts over $4' \times 3'$ fields-of-view at a sensitivity sufficient to detect ultra-faint dwarf galaxies \citep[UFD galaxies;e.g.,][]{willman_et_al2005, simon2019} with masses as low as $10^{3.5} M_\odot$. Based on simulations, such observations are expected to reveal approximately 1000~UFD galaxies with the most distant galaxies more than 10~Mpc away. UFD galaxy masses would be estimated by counting the number of RGB stars resolved in each satellite, which requires resolving RGB stars as faint as $g$-band $<32.5$~mag. The observations should cover visible and near-infrared wavelengths at a spatial resolution of $0\farcs02$. As mentioned previously, observations in at least two filters are required; $g$-band observations are recommended for one filter to sample the peak of the RGB flux distributions. 

\enabling For enabling science, the observations should be able to limit the warm dark matter particle mass to $\ge10.6$~keV at $3\sigma$ confidence. The corresponding observational requirement is surveying 10~Milky-Way-mass host galaxies with sensitivity sufficient to detect UFD galaxies with masses $\geq 10^4 M_\odot$. As a result, the anticipated yield would be reduced by a factor of two to approximately 500~UFD~galaxies, all of which would be at distances $<10$~Mpc. Additionally, the required sensitivity in $g$-band would be relaxed to $<31$~mag. Like breakthrough progress, enabling progress would also require spatial resolution of $0\farcs02$ and observations in two or more filters. 
 
\subsubsection{Habitable Worlds Observatory: Strong Lensing Constraints on Dark Matter Halo Mass Function down to \texorpdfstring{$10^7 M_\odot$}{10,000,000 M\_sun} (SCDD-GG-13)}
\label{sssec:dm_lensing}
\sleads{Qiuhan He, David Lagattuta, Mathilde Jauzac, Tansu Daylan, Richard Massey, Jason Rhodes, Jessica E. Doppel, Leonidas Moustakas, Alina Kiessling, Russell Smith, Stephane Werner, James Nightingale, Maximilian von Wietersheim-Kramsta, Satyapriya Das, Georgios N. Vassilakis}

As discussed by \citet{he_hwo25}, the goal of this science case is to determine the nature of dark matter and the impact of dark matter on the evolution of the universe. Specifically, the science case aims to determine the nature of dark matter (warm or cold) and study the formation of dark matter structure on small scales to assess whether hierarchical structure extends to the low-mass regime (i.e., $< 10^7 M_\odot$). While the conventional cold dark matter (CDM) paradigm predicts that dark matter particles interact solely via gravity \citep[e.g.,][]{peebles1982}, some alternative frameworks include interactions with photons and stronger interactions between dark matter particles \citep[e.g.,][]{blumenthal_et_al1984}. 

The planned observational strategy is to search for dark matter halos by conducting high-resolution imaging and spectroscopic observations of strong lensing systems. The various theories under current consideration can all describe the observed structure on large scales, but they differ in their predictions for dark matter halos at masses below current detection limits. For instance, if dark matter carriers are keV sterile neutrinos, the minimum dark matter halo mass should be $10^6 - 10^{10} M_\odot$ \citep[e.g.,][]{boyarsky_et_al2019}. The detection of dark matter halos below that mass range would therefore help rule out theories involving keV sterile neutrinos. Observations with HWO have the potential to distinguish between various dark matter theories by measuring the frequency and masses of dark matter objects on sub-galactic scales. 

\breakthrough For breakthrough science, observations of 2000 lensing systems at a spatial resolution\footnote{Note: The spatial resolutions needed for this science case are sharper than anticipated for HWO, but these numbers are useful for mapping the relationship between observational capabilities and possible scientific performance.} of $0\farcs011$ would be sensitive to subhalo masses $> 1 \times 10^7 M_\odot$ and therefore enable a $3\sigma$ lower limit of $\geq12.6$~keV on the mass of warm dark matter particles. Images at UV and visible wavelengths (200 - 500~nm) should cover $4 \times 4$ square arcminutes (obtained using multiple tiled pointings) and be sensitive to targets as faint as 35$^{th}$ magnitude. The pixel scale should be $0\farcs005-0\farcs015$. 

Breakthrough progress also requires IFS spectroscopy over the same UV and visible wavelength range ($200-500$~nm). The data should have angular resolution $0\farcs02$ and spectral resolution $R \sim 10,000$. The IFS data should cover a $20" \times 20"$ region (obtained using multiple tiled pointings) and be sensitive to sources as faint as 30$^{th}$ magnitude. 

\enabling For enabling progress, observations of 800 lensing systems at spatial resolution of $0\farcs016$ would be sufficient. While these observations would be sensitive to subhalo masses $>2 \times 10^7 M_\odot$, the 50\% decrease in the sample size decreases the $3\sigma$ lower limit for the mass of warm dark matter particles to $\ge 9.3$~keV. Further softening the requirements to 200~lensing systems observed at spatial resolution $0\farcs023$ would generate an ``enhancing'' dataset sensitive to subhalo masses $> 5 \times 10^7 M_\odot$ and could exclude at $3\sigma$ dark matter particles with masses $\leq 6.4$~keV. 

For both imaging and spectroscopy, enabling progress requires observations over UV, visible, and NIR wavelengths ($350-2500$~nm). In the red optical and NIR, the data could come from ground-based extremely large telescopes rather than a space-based facility like HWO. Using mosaics as needed, imaging with HWO should cover $3 \times 3$ square arcminutes at a pixel scale of $0\farcs01$. As for breakthrough progress, the images and IFS spectra should be sensitive to targets as faint as 35$^{th}$ and 30$^{th}$ magnitude, respectively. The IFS data should have a spatial resolution of $0\farcs04$ per spaxel. The required field-of-view and spectral resolution for enabling science are not readily discernible from Table~2 of \citet{he_hwo25} because the science case references the capabilities of other upcoming facilities when defining enabling science. They provide two examples: coverage of a $3'' \times 4''$ field at $R=3200-1700$ with ELT/IFU and coverage of $3' \times 3'$ at $R\sim 3500$ with WST/IFU.\footnote{https://wstelescope.eu/}

\subsubsection{Measuring SMBH Merger Timescales with HWO (SCDD-GG-14)}
\label{sssec:smbh_mergers}
\sleads{James Nightingale}

This science case was posted on the STScI SCDD Portal\footnote{\url{https://docs.google.com/document/d/14ZGS_XTkN8vtiaMKtJylhcOS-2EsEL0q6AuBMRA-o_A/edit?tab=t.0}} but not included in the HWO25 Conference proceedings. The goal is to investigate the speed of mergers between supermassive black holes (SMBHs). By focusing on the early stages of SMBH mergers when the two SMBHs are gravitationally unbound and widely separated (i.e., $10^2 - 10^5$~pc), observations with HWO could reveal whether mergers are fast (i.e., $< 100 $ million years) or slow (i.e., $> 1$ billion years). Specifically, HWO images of strong lensing systems could constrain SMBH merger timescales because the rate at which strong lenses show perturbations attributable to SMBHs should correlate with the merger timescale. If mergers are rapid, then a low SMBH detection rate is expected because binary SMBHs would quickly sink to the center of the lens galaxy and merge. Conversely, if mergers are slow, then 5-20\% of lens galaxies would be expected to have detectable SMBHs because the SMBHs would require billions of years to decrease their separation and merge. 

The expected rate of SMBH detections for the slow merger case depends on both the merger timescale and the angular resolution of images, with longer timescales and better angular resolutions leading towards higher expected detection rates. For detected SMBH, the resulting mass estimates will provide valuable insight into the SMBH merger mass function. For non-detections, ``sensitivity mapping'' will be used to determine the population of SMBHs that could have been detected in the image, thereby providing additional constraints on the SMBH mass function.

Given that the merger timescale is currently unconstrained, a sample of 1000 strong lenses is desirable for robustly measuring the merger speed. 
For breakthrough science with 1000 lenses, finding zero SMBHs would point towards fast merger timescales ($\leq 10^8$~yr) because roughly 10 detections would be expected for slow merger timescales ($\sim 10^9$~yr). Reducing the sample to 300~lens would shrink the number of expected detections to only 3 SMBH, thereby leading to an upper limit of $10^9$ years on the merger timescale in the case of no detections. Accordingly, a 300-lens sample is insufficient for determining whether SMBH mergers are fast ($\leq 10^8$~yr) or slow ($\sim 10^9$~yr).

The best lenses for this science case will be symmetric, have large lensed source area, and, across the sample, provide coverage of spatial scales between 100 pc and 10~kpc. The HWO target sample of 1000~lenses could be selected from the sample of roughly 100,000~strong lenses that is expected to be revealed by Euclid, Rubin, and Roman, roughly 10\% of which should be spectroscopically characterized. Due to the nature of their discovery, Euclid-detected lenses will be readily observable with HWO; current estimates suggest that roughly 100~hours of HWO imaging would be sufficient to search 1000~lenses for SMBHs with masses $\geq 10^7 M_\odot$. These calculations assume that HWO imaging would occur at 650~nm (intentionally selected to fall within the $550 - 900$~nm wavelength range of Euclid VIS's imager) and have a spatial resolution of approximately $0.02''$. 

JWST is not able to accomplish this science case due to its lower spatial resolution, which would limit the minimum detectable SMBH mass to $10^8 M_\odot$ (an order of magnitude higher than possible with HWO), and its smaller aperture, which would prevent achieving the necessary SNR on very faint high-redshift strong lenses. Future extremely-large ground based telescopes will have larger apertures and therefore increased sensitivity to faint strong lenses, but atmospheric turbulence may prevent achieving the image stability needed to detect the tiny perturbations caused by SMBHs and determine that they are not due to systematic effects. 

In contrast, space-based observations with HWO would have the advantage of exquisite PSF stability. Depending on the performance of advanced adaptive optics systems on next-generation ground-based telescopes, HWO may be the only facility capable of achieving this science case. Accordingly, this science case may be one of the most significant non-exoplanet drivers pushing the PSF stability of HWO. 

\breakthrough Breakthrough progress would require surveying 1000 lenses to measure the density of SMBHs. If the merger rate is fast, this dataset would constrain the merger timescale to $<10^8$~years. If the rate is slow, roughly 10~merging SMBHs should be detected. The lens should be imaged at an angular resolution of $0\farcs02$, which would permit the detection of SMBHs with masses $\geq 10^{7.5} M_\odot$. Given that the target lenses will be selected from the set detected by Euclid, it would be advantageous to conduct HWO observations at a wavelength within Euclid's $550-900$~nm wavelength range. For this SCDD, Nightingale assumes an observation wavelength of 650~nm. 

\enabling For enabling progress, a smaller sample of 300~lenses should be imaged at an angular resolution of $0\farcs03$. As for breakthrough progress, an observation wavelength of 650~nm or elsewhere in the Euclid bandpass ($550-900$~nm) would be preferred. These observations could constrain the merger timescale to $<10^9$~years in the case of fast mergers or result in the detection of about 3~merging SMBHs if mergers are slow.

\subsubsection{A High Spatial and Spectral Resolution Absorption Map of the Inner CGM Enabled by HWO (SCDD-GG-15)}
\label{sssec:cgm_map}
\sleads{Joseph N. Burchett, Sanchayeeta Borthakur, Rongmon Bordoloi, 
Deborah Lokhorst}

This science case was largely superseded by SCDD-GG-16 (Section~\ref{sssec:disk_cgm}) and was not included in the STScI SCDD Portal or the HWO25 Conference Proceedings. However, the science case is included here for completeness. The goal is to explore how the movement of matter into and out of galaxies affects galaxy growth. A key way to probe these processes is to study the circumgalactic medium (CGM), which is by far the dominant constituent of galaxies by volume. Historically, the CGM has been studied in absorption by looking at the absorption features imposed by the CGM on light from bright background sources. Studies of the CGM are difficult because most CGM features are at UV wavelengths (due to the physical conditions of the CGM, i.e., mostly ionized gas at temperatures of $10^3 - 10^6$~K). At high redshift, these UV lines are shifted to optical wavelengths where they are more easily detected, but probing the CGM of nearby galaxies is more challenging. 

Observations with HWO have the potential to address key mysteries in CGM science such as the structure, physical conditions, and kinematics of the CGM, galactic inflows, and galactic outflows; the apportionment of gases and metals inside and outside galaxies; and the processes by which galaxies quench and the impact on their CGM. Answering these questions requires sensitive IFS spectroscopy covering a range of (rest-frame) FUV and EUV features at the spatial resolution needed to resolve the morphology of inflows and outflows. Detecting spectral lines from gases at different ionization states is crucial for accurately and precisely measuring gas phase abundances and total masses. Covering a wide field of view (in multiple pointings if needed) is necessary to account for heterogeneity and asymmetries in the CGM. 

Due to the uncontrolled positioning of bright background sources, absorption-line spectroscopy of the CGM has been limited by the finite number of available sightlines. As shown by \citep{peeples_et_al2019} and \citep{koplitz_et_al2023}, untangling complex kinematics is tricky when using a limited number of sightlines and the relatively low spectral resolution of HST/COS ($R \sim 20,000$). The increased sensitivity and resolution of HWO would enable more comprehensive studies of the CGM by increasing both the number of galaxies observed with absorption-line spectroscopy and the number of absorption-line sightlines per galaxy. At low redshift ($0.005 \lesssim z \lesssim 0.02$), HWO could observe up to six sightlines in each of 100+ galaxies to study galactic flows on scales of 10~kpc. These galaxies would have large angular sizes, thereby increasing the number of potential QSO sightlines. The lower limit on redshift at $z \sim 0.005$ would prevent Milky Way absorption from interfering with the ability to measure Ly$\alpha$ absorption in the CGM of the target galaxy. The upper limit at $z \sim 0.02$ is driven by the desire to image a large enough sample to yield meaningful insights on kinematics. 

This science case can also be advanced by observing many galaxies with similar properties and analyzing the data assuming that the CGM behaves similarly in galaxies with nearly equivalent masses, star formation rates, and gas content. Furthermore, sightlines to background galaxies can be used to produce high spatial resolution maps of the CGM to complement the high spectral resolution mapping possible with QSOs. Bright background galaxies have a much lower sky density than bright QSOs, but they provide the valuable opportunity to sample a whole region of the CGM rather than a single pencil-beam. 

\breakthrough For breakthrough progress, this science case requires multi-object spectroscopy over 36~arcmin$^2$ at wavelengths of $90 - 320$~nm and spectral resolution $R \sim 100,000$ sensitive to FUV magnitudes $\leq 22.0$. 

\enabling Enabling progress requires a less aggressive wavelength range ($95 - 320$~nm), lower spectral resolution ($R \sim 50,000$), and reduced sensitivity (FUV magnitude $\leq 21.5$) over the same field of view (36~arcmin$^2$). 

\subsubsection{Characterizing Gas Flows Through Observations of the Disk-Circumgalactic Medium Interface with the Habitable Worlds Observatory (SCDD-GG-16)}
\label{sssec:disk_cgm}
\sleads{Sanchayeeta Borthakur, Joseph N. Burchett, Frances Cashman, Andrew J. Fox, Yong Zheng, David M. French, Rongmon Bordoloi, Brad Koplitz}

Continuing the theme of studying how gas flows in and out of galaxies, this science case focuses on the behavior of gas and metals at the disk-CGM interface and addresses questions D-Q2b and D-Q2c identified by Astro2020 \citep{astro2020}. As discussed in detail by \citet{borthakur_et_al2025} and \citet{borthakur_hwo25}, this science case would determine the positions, kinematics, baryonic mass, and metal mass of gas clouds beyond the gas disks of Milky-Way-type galaxies (i.e., $20 - 30$~kpc from the centers of galaxies). These observations would also measure the mass of gas in inflows and outflows in order probe how gas is obtained to feed star formation (D-Q2b) and ejected via feedback. The dataset would also reveal how metals cycle through the stellar disk (D-Q2b) and provide insight into feedback processes by constraining gas kinematics and energetics at the boundary between the disk and the CGM (D-Q2c). The FUV coverage of HWO would be extremely valuable for detecting molecular hydrogen, which is a significant fraction of the baryon mass and yet has been inaccessible at UV wavelengths since the FUSE era \citep[e.g.,][]{bluhm+de_boer2001, tumlinson_et_al2002, hoopes_et_al2004, shull_et_al2021}. While infrared observations are sensitive to the rotational and vibrational bands of cold H$_2$, FUV observations are needed to study high-energy processes. 

Measuring the extent, gas column density, composition, and kinematics of gas disks will clarify the importance of various gas inflow pathways such as spiraling inward from the CGM to the extended disk to the stellar disk or fountain flow in which gas is deposited along the minor axis. In addition, the data will advance studies of why some galaxies exhibit large, extended co-rotating disks while others do not \citep[e.g.,][]{kacprzak_et_al2010, ho_et_al2017, martin_et_al2019, diamond-stanic_et_al2016, nateghi_et_al2024}. The proposed observations will also probe the frequency, sizes, structures, kinematics, and metal content of the high-velocity clouds that may play a key role in star formation by providing a significant amount of inflowing gas \citep{wakker1991}. HWO UV absorption line spectroscopy to QSOs could trace outflows from galaxies and possibly reveal outflow structures similar to the Milky Way's Fermi Bubbles and eROSITA Bubbles, thereby enabling estimates of the frequency, scales, and energetics of bubbles and insight into their involvement in feedback. Combining QSO-absorption line spectra with more conventional ``down-the-barrel'' spectroscopy in which emission from the target galaxy is used as the background source for illuminating the CGM, HWO could study outflows at scales ranging from a few pc to tens of kpc. High spectral resolution ($R \geq 50,000$; ideally $R \sim 100,000$) MOS spectroscopy could constrain outflow rates, metallicities, and kinematics from thousands of 100-pc sized regions such as stellar clusters and associations. 

This science case requires determining the column density, covering fraction, kinematics, ionization state, metallicity, size, and physical density of gas clouds within the Milky Way and other nearby galaxies. The primary dataset for this science case consists of QSO-absorption line spectra. To further constrain CGM properties in the breakthrough science case, lower resolution ($R \geq 5000$) emission line spectra should be obtained near regions exhibiting particularly strong absorption to probe features on larger scales (10's of kpc). For some analyses, the galaxies will be studied individually, but for others investigations, observations of similar galaxies will be stacked to probe trends that may be too small to detect in single galaxies. Notably, all of the observations obtained to study the CGM of other galaxies will also contain information about the disk-CGM interface of the Milky Way. Combined, those observations will enable studies of how the Milky Way's disk-CGM interface varies on kpc (and perhaps sub-kpc) scales. 

\breakthrough Breakthrough progress would require the following observational capabilities: 
\begin{enumerate}
 \item Sensitivity to HI column densities as low as $10^{13}$~cm$^{-2}$ separated by 5~km~s$^{-1}$ along 4-6 sightlines in 25~galaxies within 30~kpc to determine the gas cloud covering fraction of nearby galaxies
 \item Spectral resolution sufficient to measure 3-5~km~s$^{-1}$ kinematics for point and extended sources to explore the motion of gas clouds at the disk-CGM interface 
 \item Ability to obtain UV~spectra covering the Lyman series, H$_2$, O~VI, C~IV, Si~IV, Si~II, Si~III, Mg~II, Zn~II, and Cr~II to constrain the ionization state and composition of the gas
 \item Spatial resolution sufficient to map sub-kpc scales to determine the sizes and covering fractions of gas clouds in target galaxies via extended source absorption spectroscopy using $2-3$ background galaxies per target galaxy
 \item Ability to obtain emission and absorption spectra along similar sightlines across $30-40\%$ of the sky to determine the density of the circumgalactic medium surrounding the Milky Way.
 \item Sensitivity to measure the absorption and emission of tens of extragalactic targets to probe CGM density in other galaxies
\end{enumerate}
 
In summary, breakthrough science would require spectroscopy at high spectral resolution ($R \sim 100,000$, corresponding to $\sim 3$~km~s$^{-1}$ velocity resolution). The spectra should cover $94 - 350$~nm and be sensitive enough to detect the continuum from background sources (GALEX FUV mag = 21) at $S/N \gtrsim 10$ in 4~hours. The observations should be obtained over a $6'\times 6'$ field of view with a multiobject spectrograph with $>1000$ slitlets and spatial resolution $<1''$. 

\enabling For enabling science, the requirements would be softened to the sensitivity required to detect H~I column densities of $10^{13-13.5}$~cm$^{-2}$ separated by 10~km~s$^{-1}$ along $\geq 3$ sightlines in roughly 25~galaxies within 30~kpc and the spectral resolution needed to probe kinematics at 5-7~km~s$^{-1}$. In addition, the required spatial resolution for mapping clouds would be relaxed from sub-kpc scales to $<10$~kpc scales, thereby requiring only 6~sightlines per galaxy rather than the numerous sightlines enabled by using extended background sources rather than point sources. Finally, fewer sightlines would be used for combined emission and absorption line spectroscopy to probe the CGM of the Milky Way (a few dozen sightlines rather than 30-40\% of the sky) and extragalactic targets (4-6 targets rather than tens of targets). 

Accordingly, the wavelength range would be narrowed to $97 - 300$~nm\footnote{For enabling science, we adopt the broader wavelength range repeatedly described as the ``minimum'' wavelength range in the text of \citet{borthakur_hwo25} rather than the narrower range of $100-280$~nm given in their Table~2.}, the spectral resolution would be relaxed to $R\sim 50,000$ (6~km~s$^{-1}$), and the sensitivity would be reduced to GALEX FUV mag $< 20.5$ at $S/N \gtrsim 10$ in 4~hours. A multiobject spectrograph with 1'' spatial resolution, a smaller FOV ($3' \times 3'$), and fewer slitlets (100's) would be sufficient. 

\subsubsection{Mapping Galactic Winds and Small-scale Structure in the Circumgalactic Medium with Habitable Worlds Observatory (SCDD-GG-17)}
\label{sssec:cgm_elm}
\sleads{Joseph N. Burchett, Deborah M. Lokhorst, Yakov Faerman, Kevin France, Kate H. R. Rubin, David S. N. Rupke, and Sanchayeeta Borthakur
}

As detailed in \citet{burchett_et_al2025} and \citet{burchett_hwo25}, this science case aims to better understand the processes by which galaxies exchange matter and energy by using emission line spectroscopy to map CGM features on kpc and sub-kpc scales. The vast majority of past CGM studies have utilized QSO-absorption line spectroscopy (e.g., Section~\ref{sssec:disk_cgm}) because the faintness and diffusivity of the CGM render emission spectroscopy extremely challenging except in unusual cases \citep[e.g.,][]{burchett_et_al2021,nielsen_et_al2024}. Unlike absorption line observations, which are limited to sightlines with serendipitous bright background objects, emission line observations provide more information about spatial variations within the CGM. The combination of absorption-line spectroscopy and emission line spectroscopy for the same galaxies is necessary to measure the gas volume density, which cannot be accomplished with either method alone and is essential for tracking the baryon cycle. Moreover, building a sample of galaxies with coverage in both emission and absorption will allow researchers to develop benchmarks that can then be applied to the much larger sample of galaxies that have been observed with only absorption spectroscopy. 

By improving our understanding of how metals, dust, and gas travel within and across the boundaries of galaxies, this science case would address at least three of the questions identified by Astro2020 \citep[D-Q2a, D-Q2b, and D-Q2d; ][]{astro2020}. While top-of-the-line IFSs on 8-10-m ground-based telescopes have detected the CGM in emission, those facilities must contend with significant absorption from the Earth's atmosphere and are therefore pushed towards higher redshift galaxies whose rest-frame UV emission has shifted to the optical. For instance, detection of $10^4$~K gas using ground-based IFSs is limited to galaxies at $z > 2$ for observations using H I Ly$\alpha$ \citep[e.g.,][]{cantalupo_et_al2014, cai_et_al2019} and to galaxies at $z > 0.3$ for observations using Mg~II \citep[e.g.,][]{burchett_et_al2021, zabl_et_al2021}. Future space-based X-ray observatories may be able to detect hotter gas ($T > 10^6$~K), but those observations will not have the resolution needed to resolve gas kinematics. UV observations such as those proposed for this science case are essential for detecting gas with temperatures of $10^4 - 10^6$~K and providing a comprehensive map of the multiphase CGM within galaxies. Compared to absorption-line spectroscopy, which provides information along particular sightlines, emission line spectra are better suited to detecting large outflow structures such as the ``Makani'' superwind \citep{rupke_et_al2019} and the extended nebulae detected around starburst galaxies \citep{perrotta_et_al2024}. Theoretical simulations predict that extragalactic equivalents to the Milky Way's Fermi Bubbles should be common \citep{pillepich_et_al2021} and readily detectable with HWO. 

In order to comprehensively explore the relationship between host galaxies and the diffuse CGM, HWO should map galaxies with a range of properties. The notional sample is a set of 12 galaxies divided evenly into three mass bins ($10^9 - 10^{10}M_\odot$, $10^{10} - 10^{11} M_\odot$, and $> 10^{11}M_\odot$). Six of the galaxies should be quiescent while the other six should be actively forming stars. The star-forming sample should include one or more nearly edge-on galaxies to clearly separate galactic and circumgalactic material as well as one or more nearly face-on galaxies for ``down-the-barrel'' studies of emission \citep[e.g., ][also see Section~\ref{sssec:disk_cgm}]{chisholm_et_al2016}. 

The second objective of this science case is to determine the clumpiness and small-scale structure of the cold gas in the CGM. Although cold gas is currently detectable (and detected) in absorption line spectroscopy, the current data lack the spatial resolution needed to test whether analytic theory correctly describes the substructure of the cold gas. Specifically, while analytic calculations suggest that cold gas should be clumpy on scales of a few pc, these tiny scales are not resolvable with current numerical simulations, let alone current observations. The sensitivity and spatial resolution of HWO would permit detailed mapping of cold gas, thereby testing the proposal by \citet{mccourt_et_al2018} and other theoreticians that ``shattering'' causes cold gas to form small ``cloudlets'' surrounding galaxies. The observations will also reveal whether ``disruption'' due to processes like turbulence causes the cold gas cloudlets to fragment into even smaller pieces or whether magnetic fields or other processes provide ``shielding'' that hinders fragmentation. Conclusively measuring the filling factor and sizes of cold gas clumps will provide valuable insights into the cooling of the CGM and thereby the accretion of matter onto galaxies.

\breakthrough For this science case, breakthrough progress requires resolving the morphology of the CGM by Ly$\alpha$ emission mapping of hydrogen gas down to surface brightness of $10^{-21}$ erg s$^{-1}$ cm$^{-2}$arcsec$^{-2}$ out to the virial radius. These observations should be sensitive to accretion flows with hydrogen number densities as low as $n_H \sim 10^{-5.5}$~cm$^{-3}$. Additionally, HWO should generate Ly$\alpha$, low ion, and high ion maps of the halos of low-redshift galaxies on kpc scales to search for superwinds and potential biconical structures. To investigate the phase structure of the CGM, HWO should obtain emission maps of the tracers of 10$^4$~K and $10^{5-6}$~K gas (e.g., Ly$\alpha$, Si~II/III/IV, C~III, O~VI). HWO should also obtain emission maps in Mg~II for cross-comparison with observations by ground-based telescopes. Finally, HWO should obtain extremely high resolution images (i.e., physical scales of 10~pc) to map the CGM on small scales. 

The large-scale observations could be accomplished using a multi-object spectrograph with a field-of-view of $8' \times 8'$, an angular resolution of $0\farcs5$, a spectral resolution of $R \sim 20,000$, and a wavelength range of $97.5 - 160$~nm. For the small-scale mapping on pc-scales, a better approach is to use an integral field spectrograph (IFS). The IFS should have a field of view of $10'' \times 10''$, an angular resolution of $0\farcs05$, a spectral resolution of $R \sim 10,000$, and the same wavelength range as the MOS ($97.5 - 160$~nm). Ideally, the MOS and IFS should also observe $279-290$~nm to map Mg~II emission for comparison with ground-based observations of galaxies at $z \lesssim 1$. The observations should be sensitive to targets as faint as $10^{-21}$~erg s$^{-1}$ cm$^{-2}$arcsec$^{-2}$. 

\enabling For enabling science, the MOS would have the same angular resolution of $0\farcs5$, but the field-of-view would be smaller ($6' \times 6'$), the spectral resolution would be substantially lower ($R \sim 4,000$), and the wavelength range would be narrower ($102 - 160$~nm)\footnote{Note that \citet{burchett_hwo25} expresses the required wavelength range for enabling science as $102-140$~nm and $150-160$~nm. For simplicity, we adopt the full range $102-160$~nm.}. For the IFS, the necessary capabilities would be softened to a field of view of $5'' \times 5''$, an angular resolution of $0\farcs10$ (to detect features on scales of a few 10s of pc rather than 10-pc scales), a spectral resolution of $R \sim 4,000$, and the same reduced wavelength range as the MOS ($102-160$~nm). The observations should be sensitive to targets as faint as $10^{-20}$~erg s$^{-1}$ cm$^{-2}$arcsec$^{-2}$. 

\subsubsection{Active Galactic Nuclei Feedback Effects on the Intergalactic Medium (SCDD-GG-18)}
\label{sssec:agn_feedback}
\sleads{Megan T. Tillman, Joseph N. Burchett, Blakesley Burkhart, Vikram Khaire, Sanchayeeta Borthakur}

The science case presented by \citet{tillman_hwo25, tillman_et_al2025b} aims to investigate the relative effects of dark matter and galactic feedback on the growth of structure in the universe. The strategy is to measure how the Lyman-$\alpha$ forest evolves with redshift, paying particular attention to the redshifts $z< 1.8$ where limited data exist. These HWO observations would complement work by cosmological surveys like DESI and eBOSS at higher redshifts \citep[e.g.,][]{desi_et_al2016, dawson_et_al2016}. 

As shown by \citet{tillman_et_al2025}, including AGN feedback changes the predicted shape of the 1-D power spectrum of the Lyman-$\alpha$ forest. Measuring the 1-D power spectrum to a precision of $5-10\%$ could reveal the influence of various AGN feedback mechanisms such as radiative AGN feedback, AGN jet feedback, and AGN X-ray feedback. The difference between the models is most pronounced at $z < 0.5$, but the models differ by roughly 10\% out to redshifts $z \sim 1.4$. The shape of the power spectrum at low $k$ ($< 0.003$~s~km$^{-1}$) will reveal the role of AGN X-ray feedback while the shape at high $k$ ($> 0.07$~s~km$^{-1}$) will illustrate the importance of jet feedback. The data needed for Lyman-$\alpha$ forest measurements are high-resolution, UV spectra towards quasars. Detecting Lyman-$\alpha$ at $z < 0.7$ requires FUV data, but the transition is redshifted into the NUV for $0.7 < z < 1.8$. 

\breakthrough Breakthrough science requires observations of roughly 750 QSO sightlines at $z = 0.7$ and 85-250~QSO sightlines at $z = 1.6$. The spectra should cover $121.5 - 340.2$~nm at spectral resolution of $R \sim 40,000$ with $S/N \geq 20-25$ per pixel and sensitivity to sources with FUV magnitude $\lesssim 17.6$ and NUV magnitude $\lesssim 18$. High resolution is necessary for observing low b-values (i.e., low Doppler broadening parameters indicative of smaller velocity dispersions), and covering a wide range of b-values ($10 - 15$~km~s$^{-1}$) is important for assessing the temperature and velocity distribution of the absorbers along the line of sight as well as the thermal structure of the IGM. 

\enabling For enabling science, the wavelength range and spectral resolution would be unchanged, but the target sample would be reduced to roughly 250 QSOs at $z = 0.7$ and 30-85~QSOs at $z = 1.6$. The observations should have $S/N \geq 10-20$ per pixel and be sensitive to sources with FUV magnitude $\lesssim17.3$ and NUV magnitude $\lesssim 17.8$.

\subsubsection{Unveiling supermassive black hole binaries with FUV-to-NIR spectropolarimetry (GG-19)}
\label{sssec:smbh_pol}
\sleads{Fr\'{e}d\'{e}ric Marin, Julie Biedermann, Thibault Barnouin}

This science case was not included in the HWO25 Conference Proceedings, but it was described by \citet{marin_et_al2025} and posted on the STScI SCDD Portal\footnote{\url{https://docs.google.com/document/d/1WJ9Cw5DVZk33IP3kr_q2bGwm02nbnv--9fXcdVA-o2I/edit?tab=t.0}}. The focus of the science case is the detection and characterization of supermassive black hole (SMBH) binaries. Gravitational wave detections have revealed that such binaries exist and merge \citep[e.g.,][]{abbott_et_al2016, abbott_et_al2019}, but there are still many open questions about the nature of these objects and how they influence their surroundings. Indeed, the Astro2020 Panel on Compact Objects and Energetic Phenomena (COEP) identified ``What are the mass and spin distributions of neutron
stars and stellar mass black holes?'' (B-Q1) and ``What seeds supermassive black holes and how do they
grow?'' (B-Q4) as two of the key questions that should be addressed over the next decade \citep{astro2020}. In addition, electromagnetic observations of SMBHs like those that would be possible with a facility like HWO have the potential to directly address Discovery Area B-DA: ``Transforming our View of the Universe by Combining New Information from Light, Particles, and Gravitational Waves'' \citep{astro2020}.

The purpose of this science case is to directly detect the interaction of binary SMBHs and their accretion disks using polarimetry. Unlike many indirect signatures of binary SMBHs \citep[e.g., strange spectral features or periodic flux variations][]{zheng_et_al2023, pasham_et_al2024}, the spectropolarimetric signatures expected from a binary SMBH cannot be produced by other astrophysical phenomena such as single SMBHs \citep[e.g.,][]{vaughan_et_al2016}. Indicators of asymmetric accretion in SMBH binaries include abrupt variations in UV/VIS flux caused by gaps in the accretion disk around the primary SMBH, a bimodal flux distribution showing a secondary peak at FUV wavelengths, and a complicated dependence of polarization phase angle on wavelength suggesting multiple scattering regions \citep[e.g.,][]{savic_et_al2019, dotti_et_al2022, marin_et_al2023}.

The goal of this science case is to search for binary SMBHs via spectropolarimetry. Each candidate binary SMBH would be observed repeatedly at a cadence selected based on the periodicity observed in the polarized SED and the inferred binary properties. A large space-based observatory is essential for this science case because the FUV peak of the polarized light SED would not be detectable from the ground and smaller telescopes would lack the sensitivity needed to robustly detect small polarization fractions. 

\breakthrough Breakthrough progress would require $R>50,000$ spectropolarimetry of $\geq 300$~candidate binary SMBHs with UV magnitudes $<22$ at $S/N \geq 500$. The broad wavelength range from the FUV (100~nm) to the NIR ($2\mu$m) must be observed simultaneously and the entire AGN core should fit within the instrument aperture. The polarization should be measured to within 0.01\%. The observations should result in full FUV-NIR spectra of polarized light for 100~AGN, detections of SED dips for 10 AGN, detections of intriguing polarization peaks and wavelength-dependent scattering for 10~AGN, and a measurement of polarized periodicity for at least one AGN. 

\enabling Enabling science would require the same target sample and S/N, but the instrumental requirements would be reduced to spectral resolution $R>40,000$, wavelength range 100~nm - $1.5\mu$m, and polarization error $<0.05\%$. The observations should result in 50 AGN with polarized FUV-NIR spectra, 5 AGN with SED dips, and 5 AGN with polarization peaks and wavelength-dependent scattering. Like breakthrough progress, enabling progress also requires measurement of polarized periodicity for at least one AGN.

\subsection{Evolution of the Elements}
\label{ssec:ee_scdds}
The Evolution of the Elements working group focused on studying the chemical enrichment of the universe over cosmic time. All \nee~of the science cases presented in this section were later published or posted to the SCDD Portal. Multiple EE science cases focus on detailed spectroscopic analyses of specific stellar populations (Sections~\ref{sssec:massive_stars_lowZ}, \ref{sssec:vms}, \ref{sssec:first_stars}, \ref{sssec:r_process_nature}, \ref{sssec:r_process_el}, \ref{sssec:flash_ccsne}). Such studies are expected to provide valuable insights into the lives, nucleosynthesis, and explosive deaths of the first generation of stars because they formed from the material added to the ISM by the first stars \citep[e.g.,][]{beers+christlieb2005, karlsson_et_al2013}. These studies could also provide a window into origins and formation history of galaxies and their halos \citep[e.g.,][]{frebel2010}. Additional EE science cases investigate the stellar magnetic fields (Section~\ref{sssec:mag_stars}); the compositions of exoplanets accreted by white dwarfs and possible variations in the values of fundamental physical constants (Section~\ref{sssec:wds}); the expansion rate of the universe (Section~\ref{sssec:distance_ladder}); the compositions of dust grains in the Milky Way and other galaxies (Sections~\ref{sssec:dust_extinction} \& \ref{sssec:dust_uv}); and the morphologies, assembly, star formation histories, and elemental distributions of galaxies (Sections~\ref{sssec:resolved_stellar_pops} \& \ref{sssec:ism_uv}).
 
\subsubsection{Echoes of the First Stars: Massive Star Evolution in Extremely Metal-Poor Environments with the Habitable Worlds Observatory (SCDD-EE-1)}
\label{sssec:massive_stars_lowZ}
\sleads{Peter Senchyna, Calum Hawcroft, Miriam Garcia, Aida Wofford, Janice C. Lee, Chris Evans, Ryan J. Rickards Vaught, Macarena G. del Valle-Espinosa, Grace Telford, Fabrice Martins, G\"{o}ran Ostlin, M\'ed\'eric Boquien, Paul Scowen, John Ziemer, Jane Rigby}

As explained by \citet{senchyna_hwo25}, the purpose of this science case is to understand how very metal-poor massive stars irradiated and enriched the universe. These stars differ significantly from the higher metallicity massive stars found in the Milky Way, but HWO would have the sensitivity necessary to detect and analyze faint, extremely metal-poor massive stars in other galaxies that are more similar to the highly influential massive stars formed in the early universe. Recent JWST observations of high-redshift galaxies are not well-fit by stellar population models \citep{bunker_et_al2023, topping_et_al2024, topping_et_al2025, castellano_et_al2024}, suggesting that models of their integrated light do not sufficiently account for the most massive stars. In addition, some high-redshift galaxies contain highly-ionized gas with unusual abundance patterns that may be the result of fusion processes in low-metallicity, massive stars \citep{cameron_et_al2023, senchyna_et_al2024, charbonnel_et_al2023, marques-chaves_et_al2024, topping_et_al2024}. 

Thoroughly understanding the spectra of low-metallicity, massive stars and their effects on their environments requires spatially resolved observations that isolate the contributions of individual massive stars across the electromagnetic spectrum. Observations at (rest-frame) UV wavelengths are particularly important for revealing how massive stars ionize their surroundings and contribute to feedback. Specifically, this science case aims to determine how ancient, metal-poor massive stars evolved and influenced their host galaxies. More generally, this science case also explores how low-metallicity massive stars differ from their higher-metallicity counterparts. In particular, the science case seeks characterizing how radiation-driven winds depend on metallicity in the low-metallicity regime \citep[e.g.,][]{kudritzki_et_al2002, bouret_et_al2003}. Mass removal by the wind, or by other processes such as eruptions or binary interactions, alters the evolution of massive stars, their life-long feedback, the type of supernova they will experience, or whether their death will lead to the formation of black hole and eventually a gravitaional wave event \citep{eldridge+stanway2009, gotberg_et_al2017, gotberg_et_al2019, klencki_et_al2022, smith2014, langer2012, langer_et_al2020}. 

Studying the effects of massive stars on early galaxies will contribute to addressing Astro2020 question F-Q3c, which questions the universality of the stellar initial mass function \citep{astro2020}. The proposed HWO observations would measure the FUV emission of all massive unobscured stars ($> 30 - 50 M_\odot$) in the target region and have the potential to detect stars with masses above $100 M_\odot$, thereby providing valuable information about the massive end of the initial mass function in extreme environments. 

Additionally, early, extremely metal-poor, massive stars are an ``extreme'' population and the proposed observations have the potential to advance understanding of their multiplicity rate, so this science case is also relevant to Astro2020 questions G-Q1 (extreme stars and stellar populations) and G-Q2 (effects of multiplicity on stellar evolution) \citep{astro2020}. As massive stars can produce black holes, this science case will also inform investigations of the origin, prevalence, and binarity of stellar-mass black holes (e.g., Astro2020 questions B-Q1, B-Q1b) as well as the ionization history of the universe and the cycling of material through galaxies (Astro2020 questions DQ-1, D-Q1b, D-Q1c, D-Q2, D-Q2a, D-Q2b, D-Q4d, DF-Q1c).

The dataset needed for this science case consists of spatially-resolved UV/optical spectroscopy of massive stars ($> 30 M_\odot$) and their surroundings in regions with very low metallicities ($\ll 10\%$ solar). UV coverage is essential for detecting extremely hot stars that are sometimes extremely faint at longer wavelengths \citep[e.g.,][]{gotberg_et_al2017}; estimating mass loss rates in stellar winds \citep[e.g.,][]{zsargo_et_al2008}; determining stellar properties, photospheric features, and feedback \citep[e.g.,]{vink2022}; and identifying FUV gas absorption by galactic outflows in front of emission sources. 

Data at shorter wavelengths ($100 - 180$~nm) would reveal wind properties while optical ($500-700$~nm), or even better, UV/optical ($300 - 700$~nm) spectra would enable measurements of stellar properties. UV/optical spectra of the regions around massive stars are essential for tracing ionization and mapping the effects of stellar feedback. The push towards FUV wavelengths is motivated by the unique sensitivity to stellar-wind diagnostic lines, and to dense, highly ionized gas\footnote{Key FUV lines include N IV] 1483,1486, C IV 1548,1550, He II 1640, O III] 1661,1666, [C III], C III] 1907,1909.} and the previous detection of strong FUV lines in both high-redshift galaxies and nearby low-metallicity galaxies. 

Current facilities lack the requisite sensitivity and angular resolution to resolve and spectroscopically characterize individual massive, metal-poor stars in galaxies 5-10~Mpc away from Earth. Closer metal-poor galaxies such as Sextans~A \citep{lorenzo_et_al2022}, SagDIG \citep{garcia_et_al2018}, Leo~A \citep{gull_et_al2022, urbaneja_et_al2023}, and Leo~P \citep{evans_et_al2019, telford_et_al2021} are not suitable targets for this science case because their relatively low star formation rates do not sample the rare, short-lived, but particularly impactful massive stars ($\gtrsim 50 M_\odot$) that likely dominate ionization and feedback in the early Universe ($M_\star \lesssim 40 M_\odot$ for Sextans~A and $M_\star \lesssim 30 M_\odot$ for the others). HWO would provide the valuable opportunity to spatially resolve the FUV lines and untangle the spatial relationship between highly-ionized, dense gas and the extremely massive stars responsible for most of the ionizing radiation. Although ground-based extremely large telescopes equipped with adaptive optics will be able to resolve the same spatial scales, they will be unable to obtain the critical UV/blue-optical spectra needed to characterize hot stars and analyze feedback \citep[e.g., ][]{evans_et_al2018, garcia_et_al2021}.

For this science case, the most compelling target is IZw18 \citep{aloisi_et_al2007, annibali_et_al2013}, an extremely metal-poor galaxy ($1/32~Z_\odot$) at a distance of 18.9~Mpc. IZw18 has high star formation rate ($1 M_\odot/\rm yr$) and is expected to harbor hundreds or even thousands of O~stars that could be used to decipher the evolution and winds of metal-poor massive stars, for the first time, with HWO. Interestingly, IZw18 shows the same nebular emission in CIV and He~II that is frequently detected in high-redshift galaxies \citep[e.g.,][]{mingozzi_et_al2022}, indicating that spatially resolved observations of IZw18 will be useful for better understanding the influence of ancient massive stars on their host galaxies. Furthermore, UV and optical spectroscopy of IZw18 also display features reminiscent of broad stellar wind emission from Wolf Rayet stars \citep[e.g.,][]{izotov_et_al1997, legrand_et_al1997, brown_et_al2002}. This emission is noteworthy because the only Wolf Rayet stars expected to exist as single stars in very low metallicity environments are those with very high masses and luminosities extremely close to the Eddington limit \citep[e.g.,][]{crowther+hadfield2006, vink2018}. HWO would be able to take spectra of these unusual stars, thereby constraining their nature and formation history. 

If HWO is capable of resolving individual massive stars in IZw18 (i.e., spectroscopic aperture $\lesssim 20$~mas, ideally $\lesssim 15$~mas), it will also have the resolution and sensitivity needed to search for and characterize massive stars in multiple closer galaxies that may or may not host such populations. Designing HWO so that observations of IZw18 are feasible is essential for guaranteeing that HWO would be able to obtained spatially-resolved spectra of massive stars in low-metallicity galaxies.

This science case would significantly benefit from an IFS. If an IFS is not available, a MOS with high spatial resolution ($\lesssim 20$~mas) is a plausible alternative but would require all unobscured stars with masses $\gtrsim 20 M_\odot$ to first be detected via deep precursor FUV/NUV/optical imaging before the MOS could be configured to obtain their spectra. In contrast, an IFS with a FOV $\gtrsim 5''$ could obtain all of the data needed for this science case in a single deep observation of $\lesssim 100$~hours\footnote{Exposure time estimated using the UV spectroscopic exposure time calculator assuming target stars brighter than FUV (GALEX) magnitude = 26~AB and an 8-m telescope aperture.} Additionally, IFS observations have the added advantage of obtaining the data needed to characterize the gas surrounding massive stars while simultaneously acquiring the data needed to characterize the massive stars themselves. For MOS observations, separate observations would be needed to obtain the nebular spectroscopy dataset. Regardless of the specific instrument, high spatial resolution is crucial for this science case because of the need to isolate the spectra of massive stars. 

A spectral resolution of $R > 2000$ would be sufficient for distinguishing stellar and interstellar absorption in P-Cygni profiles and determining terminal wind velocities, but a higher spectroscopic resolution of $R > 4000$ would allow for more robust identification and characterization of the faint wind lines and slow terminal velocities expected for extremely metal-poor stars. In the blue-optical ($300 - 500$~nm), spectra with resolution $R>4000$ (ideally $R >8000$) are necessary for precisely determining stellar temperatures and surface gravities \citep[e.g.,][]{garcia_et_al2021, evans_et_al2023}.

\breakthrough Breakthrough science would entail resolving and observing $\gg10$ stars with masses $>30-50~M_\odot$ and metallicities as low as $0.1 Z_\odot$; measuring winds from extremely massive, low-metallicity stars ($> 30 M_\odot$, $<0.1Z_\odot$); and characterizing extremely metal-poor massive stars ($> 30 M_\odot$, $\leq 1/30 Z_\odot$) using spatially-resolved FUV/optical spectroscopy. Gas characterization requires FUV/optical spectroscopic observations ($120 - 500$~nm, $R > 4000$) of low-metallicity dwarf galaxies at a spatial resolution high enough to probe the relationships between gas properties and stellar properties. The spectral resolution is set by the need to resolve velocities at the virial scale ($\gtrsim 50-100$~km/s) and ascertain the sources of various UV lines. The wavelength range is driven by the desire to probe high-density hot gas and detect unusual enrichment patterns, which requires UV capabilities to capture high-ionization lines such as C~III], C~IV, O~III], N~III], N~IV], and He~II. 

The observations should be acquired using a near-diffraction-limited UV/optical IFS with a wide field of view ($3-10''$), spectral resolution $R \gtrsim 5000$, and access to $90 - 200$~nm and $300-500$~nm. The lowest acceptable spatial resolution for breakthrough science is 20~mas/spaxel (i.e., physical scales of 2~pc/spaxel for I Zw~18), but resolving smaller spatial scales would dramatically increase the number of accessible targets and therefore the scientific reach of the proposed observations. Indeed, \citet{senchyna_hwo25} included a super-breakthrough level of scientific return titled ``The Dream'' that would require 5~mas/spaxel spatial resolution and continuous spectroscopic coverage from 90~nm to 500~nm.

\enabling For enabling science, the ideal instrument is a UV IFS but a MOS is a possible backup option with the caveats discussed above. The IFS should cover 90-200~nm at spectral resolution $R\sim 3000$ and spatial resolution of $30$~mas/spaxel. For I Zw~18, this spatial resolution would correspond to a physical resolution of 3~pc, which is at least 50\% larger than the $\lesssim 2$~pc resolution required for breakthrough progress. 

\subsubsection{Very Massive Stars with the Habitable Worlds Observatory (SCDD-EE-5)}
\label{sssec:vms}
\sleads{Fabrice Martins, Aida Wofford, Miriam Garcia, Peter Senchyna, Janice Lee Paul A. Scowen}

As discussed in \citet{martins_hwo25}, the goal of this science case is to increase the sample of very massive stars (i.e., stars with masses $> 100 M_\odot$) and learn more about their powerful influence on their environments. As of 2025, there are roughly 20 known VMS stars, and only a handful of those are more massive than $150 M_\odot$ \citep[e.g.,][]{crowther+dessart1998, martins_et_al2008, crowther_et_al2010, brands_et_al2022}. The difficulty in detecting very massive stars is two-fold: first, due to the nature of the initial mass function, very massive stars are exceedingly rare; second, due to the overwhelming luminosity of very massive stars ($\gtrsim 10^6 L_\odot$), those that do exist have extremely short lifetimes \citep[$2-3$~Myr;][]{yusof_et_al2013, kohler_et_al2015, grafener_et_al2021, martins+palacios2021}. Despite their scarcity and short-lived nature, very massive stars have an outsized influence on their environments because they dramatically outshine less massive stars, especially at short wavelengths \citep{crowther_et_al2016}; produce key elements; produce powerful winds that may be key sources of feedback \citep{martins_et_al2008, grafener+hamann2008, bestenlehner_et_al2014}; and may end their lives as massive black holes that grow to become supermassive black holes. 

Due to their rarity, very massive stars are most likely found at distances where their light is blended with that of other nearby stars. Even so, regions likely to harbor very massive stars can be identified spectroscopically through characteristic features such as HeII~1640 emission, which models suggest is produced only by stars more massive than $100~M_\odot$ (unless they have entered the Wolf-Rayet phase, which happens after a few Myr of evolution and only for a short period of time) \citep{martins_et_al2023}. 

In addition, very massive stars could be identified by looking for CIV 5802-5812 emission as a narrow doublet rather than broad emission as would be expected for Wolf-Rayet stars. One path to advance studies of very massive stars would therefore be to use archival spectra and data from other facilities to identify likely hosts of very massive stars based on UV and optical spectroscopic features. There are already several hundred intriguing targets \citep[e.g.,][]{hadfield_et_al2005, whitmore_et_al2010, wofford_et_al2014, smith_et_al2016, leitherer_et_al2018, gomez-gonzalez_et_al2021, della_bruna_et_al2022, martins_et_al2023, berg_et_al2024} and even more will be known when HWO begins operations. 

By obtaining spatially-resolved UV-optical observations of a subset of likely host clusters, HWO could characterize dozens of very massive stars by measuring their effective temperatures, luminosities, masses, and wind parameters. Crucially, HWO would be able to investigate how the mass loss rates of very massive stars depend on metallicity. The role of metallicity in shaping mass loss for very massive stars is currently unconstrained, but it is crucially important for understanding the evolution and end stages of very massive stars, and therefore the effects of very massive stars on their host galaxies.

HWO is essential for this science case because of the need for high spatial resolution and a small diffraction limit to resolve very massive stars in crowded fields and the need for UV spectra to measure mass loss rates and metallicity \citep[mostly through Fe lines in the far-UV; e.g., ][]{bouret_et_al2015}. Measurements of iron, oxygen, and carbon lines in UV spectra can also be used to determine effective temperatures of very massive stars, but the traditional method is to use measurements of helium and nitrogen lines in optical spectra \citep[e.g.,][]{massey_et_al2004}. The combination of UV and optical spectra data will constrain stellar luminosities (and therefore stellar masses) as well as surface abundances and stellar mixing \citep[e.g.,][]{hunter_et_al2007, martins_et_al2008, bouret_et_al2013, bestenlehner_et_al2020, brands_et_al2022}. Ground-based ELTs will likely be capable of obtaining resolved images at NIR wavelengths. While the NIR data will constrain stellar luminosities, the combination of UV, optical, and NIR data will result in better determination of the spectral energy distribution and therefore improved measurements of stellar properties. 

\breakthrough For this science case, breakthrough progress would consist of characterizing at least 500 very massive stars at distances of $0.8 - 15$~Mpc. The sample should include stars with metallicities $<0.1 Z_\odot$ to facilitate studies of how the mass loss rates of very massive stars depend on metallicity. These stars should be observed at high spatial resolution (ideally 5~mas, beyond the expected reach of HWO) and moderate spectral resolution ($R > 5000$) at wavelengths of $100 - 700$~nm. The ideal instrument is a UV IFS that is diffraction-limited at UV wavelengths and has 5-mas spatial resolution. The IFS should also have a field-of-view\footnote{\citet{martins_hwo25} does not include field of view among the instrument requirements listed in their Table~2, but in their Section~4.2, they state that a range of $1\farcs5\times 1\farcs5$ to $3'' \times 3''$ would be appropriate for the majority of their targets. We adopt the smaller field for enabling progress and the larger field for breakthrough science.} of $3'' \times 3''$

\enabling For enabling science, the sample would decrease to roughly 100 stars at closer distances of $0.8 - 3.5$~Mpc. The target metallicity range would be reduced to $ 0.1 Z_\odot < Z < 1.5 Z_\odot$ due to the change in the distance range of the sample. A UV IFS is still desired, but the specifications are reduced to be diffraction-limited in the optical rather than the UV. In addition, the desired spectral resolution would be lower ($R > 2000$), the wavelength range would be narrower ($120 - 600$~nm), and the spatial resolution would be relaxed to 15~mas, which is within the range currently considered for HWO. An IFS field of view of $1.5'' \times 1.5''$ would be acceptable.

\subsubsection{Resolving Individual Stars in Nearby Large Galaxies with the Habitable Worlds Observatory (SCDD-EE-6)}
\label{sssec:resolved_stellar_pops}
\sleads{Adam Smercina, Tara Fetherolf, Eric W. Koch, Silvia Martocchia, Chris Mihos, Martin M. Roth, Benjamin F. Williams}

As discussed in \citet{smercina_hwo25}, this science case investigates how the evolution of other galaxies compares to that of the Milky Way. Specifically, the science case is concerned with the interaction, assembly, star formation, and enrichment history of galaxies and how they contribute to the observed diversity in galaxy morphology. The science case also probes the universality of the initial mass function. 

For all of these questions, the science case uses stars as a historical record. Depending on the depth at which any particular galaxy is observed, HWO could address different fractions of the overall science case. For this science case, galaxies are divided into three groups based on imaging depth. For all three depths, high spatial resolution and high sensitivity are essential for achieving the needed data quality in crowded fields. 

In the deepest case (Depth~1), HWO would observe roughly 100 galaxies with sufficiently high sensitivity to detect stars as faint as 1~magnitude below the oldest main sequence turn-off (oMSTO). The target galaxies would be at distances $< 5$~Mpc, have masses of $10^8 - 10^{11} M_\odot$, and cover a range of metallicities and star formation properties. The resulting HWO photometry would be used to construct detailed star formation histories of local group galaxies out to a distance of 5~Mpc. In addition, HWO would be able to resolve thousands of young star clusters, thereby enabling studies of how the stellar initial mass function varies among and within galaxies. Focusing on young star clusters is essential for capturing the short-lived, extremely massive end of the IMF. 

In the intermediate case (Depth~2), HWO would observe hundreds of galaxies within 20~Mpc at a depth sufficient to detect stars as faint as 1~magnitude below the Red Clump. Detecting the Red Clump will permit the use of color-magnitude diagram fitting to establish resolved star formation histories covering the last 5 billion years. Importantly, the increased sensitivity and spatial resolution of HWO would enable investigations of the star formation and metal enrichment histories of galaxies $4-20$ times more distant than possible with JWST or HST, respectively. In addition to increasing the overall number of galaxies for which star formation histories can be studied, this increase in study volume will allow HWO to explore star formation histories for a much broader diversity of galaxies. For example, the HWO sample could include massive elliptical galaxies, galaxies in clusters, and galaxies in the late stages of mergers. As a result, HWO would be able to investigate the effects of mergers on star formation and explore the spatial relationship between galactic structure and star formation. 

In the shallowest case (Depth~3), HWO would be sensitive to stars as faint as 2~magnitudes below the Tip of the Red Giant Branch. By detecting and resolving the most luminous stars (e.g., massive main-sequence stars, core Helium burning Red Giant Branch stars, Asymptotic Giant Branch stars), HWO would support investigations of how the locations of those stars are connected to dust \citep[e.g.,][]{dalcanton_et_al2015, dalcanton_et_al2023}, star formation \citep[e.g.,][]{lazzarini_et_al2021}, and morphological features such as bars and spirals \citep[e.g.,][]{smercina_et_al2023}. HST has begun to construct maps like these for the very closest galaxies \citep[e.g.,][]{williams_et_al2023, dalcanton_et_al2015} and JWST will extend that work to galaxies within 10~Mpc, but HWO would dramatically advance this field by mapping thousands of galaxies within 50~Mpc at UV, optical, and NIR wavelengths. The increased sensitivity and distance reach of HWO would significantly expand the sample of possible targets (e.g., $100\times$ and $30\times$ increase in the number of accessible MW-mass galaxies and dwarf galaxies, respectively), thereby providing a much more comprehensive view of galaxy evolution and an improved map of the 3-D distribution of galaxies within 50~Mpc. 

\breakthrough Breakthrough progress would require NUV/optical/NIR photometry covering a 0.007 deg$^2$ field at high spatial resolution ($0\farcs015$ resolution with a pixel scale of $0\farcs01$/pixel). The imaging should be sensitive to sources as faint as $UV < 32.5$, $B < 34$, and $I < 33$. These magnitudes correspond to 1 magnitude fainter than the oldest Main Sequence Turnoff point for targets at 5~Mpc, 1 magnitude fainter than the red clump for targets at 20~Mpc, and 2 magnitudes fainter than the tip of the red giant branch for targets at 50~Mpc.

\enabling Enabling progress would demand a field-of-view of 0.0031 deg$^2$, a pixel scale of $0\farcs02$/pixel, $0\farcs025$ spatial resolution, and sensitivity to targets with magnitudes $UV < 30.5$, $B < 32$, and $I < 31$. These magnitudes are roughly equivalent to the brightness of the red clump in targets at a distance of 15~Mpc.

\subsubsection{White Dwarfs as Probes of Extrasolar Planet Compositions and Fundamental Astrophysics (SCDD-EE-7)}
\label{sssec:wds}
\sleads{Siyi Xu, Martin Barstow, Andy Buchan, \'{E}rika Le Bourdais, Patrick Dufour}

As discussed by \citet{xu_siyi_hwo25}, this science case uses observations of white dwarfs to probe two very different research areas: exoplanet compositions and fundamental physics. For exoplanet science, spectroscopic characterization of ``polluted'' white dwarfs can reveal the compositions of planets that have been accreted onto the white dwarf \citep[e.g.,][]{jura+young2014}. This technique works because the strong gravitational fields of white dwarfs cause rapid density segregation, meaning that any heavy elements detectable in the photosphere must have been recently acquired, perhaps from infalling planetary debris. White dwarf spectra can also be used to probe the composition of circumstellar material \citep{xu_et_al2024}. UV spectra of polluted white dwarfs are particularly valuable because several of the elements that are key ingredients for terrestrial planets and life as we know it (i.e., C, N, O) have their strongest features in the UV. Additional abundances could be constrained by combining the UV spectra obtained by HWO with optical and NIR spectra obtained by high-resolution spectrographs on ground-based telescopes. 

By determining the composition of material accreted by white dwarfs, HWO could yield constraints on planetary water content as well as the amount and identity of ``light'' elements in iron-dominated planetary cores. Ideally, HWO would measure abundances of six elements that are important building blocks for terrestrial planets: sulfur (S), carbon (C), oxygen (O), magnesium (Mg), iron (Fe), and silicon (Si). To increase the likelihood of successful abundance measurements, the program would observe white dwarfs whose spectra have already been found to show signs of S, C, O, Fe, and Si. This target selection requirement necessitates a pre-screening stage in which thousands of white dwarf spectra will be examined to identify the desired target sample of 45 polluted white dwarfs. 

For fundamental physics, the strong gravitational fields of white dwarfs also provide a test bed for searching for deviations from standard general relativity. White dwarf spectra are an ideal environment for such tests because the gravity of white dwarfs is strong enough that deviations may be measurable as small changes in the positions of their narrow absorption lines \citep[e.g.,][]{flambaum+shuryak2008} but not so strong that gravitational broadening would hinder precise measurement of line centers, as is the case for neutron stars. For example, \citet{berengut_et_al2013} measured Fe~V and Ni~V lines in white dwarf spectra to investigate possible variations in the fine structure constant $\alpha$ between Earth gravity and white dwarf gravity. In another study, \citet{bagdonaite_et_al2014} searched for variations in the ratio $\mu$ of the proton mass to the electron mass using H$_2$ Lyman lines in the spectrum of a white dwarf. 

For this science case, HWO observations obtained at extremely high spectral resolution would be used to constrain the fundamental physics parameters $\alpha$ and $\mu$ via measurement of the wavelengths of H$_2$ and highly ionized Fe and Ni lines (e.g., Fe~IV/V/VI and Ni~IV/V/VI). Note that coverage of both types of features is important because $\alpha$-driven changes will be more noticeable for the highly ionized atomic lines while $\mu$-driven changes will be more detectable for the H$_2$ lines. In their Table~4, \citet{xu_siyi_hwo25} note that there are numerous H$_2$ features at $90-160$~nm and thousands of absorption lines from highly ionized Fe and Ni at $100-160$~nm.

Conveniently, the white dwarfs observed for the fundamental physics case could be drawn from the sample surveyed to investigate planetary compositions. However, as noted below, the spectral resolution required for investigations of fundamental physics is much higher than that needed for studies of exoplanet compositions via white dwarf pollution. For polluted white dwarfs, the spectral resolution is set by the need to distinguish photospheric features from circumstellar or interstellar features. For fundamental physics, the higher spectral resolution results from the need to measure the tiny wavelength shifts expected from changes in $\alpha$ and $\mu$. In addition, the extreme sensitivity of HWO is essential for detecting spectral features at the high S/N needed to precisely measure minuscule shifts in line positions.

\breakthrough Breakthrough progress in white dwarf pollution requires measuring precise abundances (0.1~dex) of S, C, O, Mg, Fe, and Si for 45 polluted white dwarfs. The 45~selected targets should have prior detections of S, C, O, Fe, and Si, so constructing the target sample will likely require observing a total sample of 3200~polluted white dwarfs. An even larger sample of approximately 9500 polluted white dwarfs would be needed to identify a subsample of 155~polluted white dwarfs with detections of both O and Si and abundances measured to 0.1~dex that would be used to investigate the water mass fraction of polluters. The targets are expected to have GALEX FUV magnitudes $\leq20$. Once the targets are identified, measuring stellar abundances will require $R\sim 60,000$ spectra at wavelengths $>90$~nm. 

For breakthrough science in fundamental physics, HWO should obtain $R\sim 200,000$ spectra at wavelengths $>90$~nm\footnote{In their Section 2.2, \citet{xu_siyi_hwo25} state that observations at $100-200$~nm are needed for fundamental physics. However, in their Section 4.2 and Table~3, they adopt a unified wavelength constraint of $>90$~nm for both the planetary science and fundamental physics science goals. We follow suit and also list the $>90$~nm requirement.} for 50~white dwarfs with GALEX FUV magnitudes $\leq20$. To robustly detect or rule out variations in $\alpha$ and $\mu$, the sample should cover the full two orders of magnitude of surface gravity ($7 < \log g < 9$). Additionally, the wavelengths of H$_2$ and high ionization Fe/Ni lines should be measured to a precision of 0.0006~nm. 

\enabling For enabling progress in white dwarf pollution, the same elemental abundances should be measured for the same number of targets (i.e., sulfur, carbon, oxygen, magnesium, iron, and silicon abundances of 45 polluted white dwarfs), but the precision of the abundances would be reduced to 0.2~dex. As for breakthrough science, identifying this sample of white dwarfs polluted by S, C, O, Fe, and Si would require first observing thousands of polluted white dwarfs with GALEX FUV magnitudes $\leq 20$. Once the targets have been identified, abundance measurements could be obtained at a lower spectral resolution of $R\sim50,000$ over a redder wavelength range ($>114$~nm).

For enabling progress in fundamental physics, the wavelengths of H$_2$ and high ionization Fe/Ni lines to the same precision (0.0006~nm) for a smaller set of 25~white dwarfs. The target magnitudes and spectral resolution would be unchanged, but the blue wavelength cut-off would be increased to 114~nm as for the white dwarf pollution case.

\subsubsection{Habitable Worlds Observatory: The Nature of the First Stars (SCDD-EE-8)}
\label{sssec:first_stars}
\sleads{Ian U. Roederer, Rana Ezzeddine, Jennifer S. Sobeck
}

As discussed in \citet{roederer_nature_hwo25}, this science case aims to determine the nature of the first stars \citep[i.e., Population III stars;][]{bromm+larson2004} by observing candidate Population~III stars identified by large stellar surveys and by measuring metal abundances in very low-mass second-generation stars that would have inherited their metals from the first stars \citep[e.g.][]{wise_et_al2012, jeon_et_al2017, hartwig_et_al2018}. UV spectra are essential for both objectives because low abundances of many elements are easier to constrain from UV spectra than from optical or NIR spectra. While Na, Ca, and Ti have stronger optical lines than UV lines \citep[e.g.,][]{starkenburg_et_al2017}, the inverse is true for C, Mg, Al, Si, P, S, Sc, V, Cr, Mn, Fe, Co, Ni, Cu, and Zn. Accordingly, a non-detection of the latter set of elements in optical spectra does not yield a sufficiently low upper limit on abundance in order to place meaningful constraints on their abundance in the first generation of stars. Additionally, the abundances of those metals in second generation stars are so low (e.g., \mbox{[Fe/H] $< -4$}) that they are very challenging to detect in optical spectra. 

Access to the UV is essential for measuring abundances of elements like Co, Ni, and Zn, which are expected to provide valuable information about the energy and morphology of the supernovae that occurred as the first generation of stars died and shaped the metal enrichment of the universe \citep[e.g.,][]{bromm2013}. High-resolution UV spectroscopy with HWO would be sensitive to hundreds of metal lines, providing an unprecedented ability to measure abundances for extremely metal-poor stars. Importantly, metal-poor stars have not yet been observed at $140 - 180$~nm; high-resolution spectroscopy in that wavelength range may provide access to additional elements. 

\breakthrough Breakthrough progress would require measuring abundances of at least 20~elements per star for a sample of 50~stars with [Fe/H]$<-4$ and $V<18$. These measurements would be determined from high-resolution UV spectra obtained with SNR=100 and $R \sim 100,000$ over a wavelength range of $140-310$~nm. A spatial resolution $<100$~mas is required to properly isolate these sources.

\enabling Substantial progress would require measuring 15~elemental abundances per star for a smaller sample of 10~stars with [Fe/H]$<-4$ and $V<14$. The spectra needed for enabling science would have the same spectral resolution, spatial resolution, and signal-to-noise as those needed for breakthrough science, but the wavelength range would be narrowed to $180-310$~nm.

\subsubsection{Habitable Worlds Observatory: The Nature of the Astrophysical r-process (SCDD-EE-9)}
\label{sssec:r_process_nature}
\sleads{Ian U. Roederer, Rana Ezzeddine}

As presented in \citet{roederer_rprocess_hwo25}, this science case probes the physics behind the formation of rapid neutron capture process (r-process) elements in stars. As shown by \citet{roederer_et_al2022}, adding UV data to optical and NIR data expands by 50\% the number of r-process elements that can be measured in the spectra of cool stars. While many s-process peak elements 
(e.g., Sr, Y, \& Zr at the first peak; Ba, La, Ce, Pr, \& Nd at the second peak; Pb at the third peak) have strong features at optical wavelengths \citep[e.g.,][]{merrill1926, st_john+moore1928, aller+greenstein1960, van_eck_et_al2001}, UV data are essential for measuring the abundance patterns of elements at the three analogous r-process peaks in stars enriched in r-process elements. At the temperatures experienced in the atmospheres of metal-poor stars, roughly 20 r-process elements, including many of the r-process peak elements, have their strongest transitions in the UV between 170~nm and 310~nm. For reference, the r-process peak elements are Se, Br, \& Kr (first peak), Te, I, \& Xe (second peak), and Os, Ir, Pt, \& Au (third peak). Other r-process elements have strong transitions in both the optical and the UV. 

As discussed in \citet{roederer_et_al2022}, UV HST/STIS spectra were essential for detecting Se (first peak), Te (second peak), Pt (third peak), and Au (third peak) and for providing improved constraints on Os (third peak) and Ir (third peak) for the r-process-enhanced star HD~222925. While 200 similar stars have been detected, nearly all are too faint for detailed spectroscopy with HST (estimated limit $V < 10$) and only HD~222925 is bright enough for HST to study all 42 r-process elements \citep{roederer_et_al2022}. 

A facility like HWO could significantly enhance studies of r-process physics by providing the UV sensitivity needed to measure the abundances of numerous r-process elements. UV measurements could be combined with optical spectroscopy (from ground-based facilities) of another 20~elements. The stars are unlikely to be in crowded fields, so angular resolution is not a concern for this science case. Still, angular resolution $<100$~mas is recommended to isolate the sources. 

\breakthrough Breakthrough progress would require measuring abundances of 45~r-process elements in 100~FGK stars with $V < 17$. These measurements could be determined from high-resolution ($R \sim 100,000$) UV spectra covering $150-310$~nm at SNR=40. Notably, extending the spectra to 150~nm provides access to Sb II (160.6~nm), Br I (154.0~nm), and In II (158.6~nm).

\enabling Enabling progress would require measuring the same number of abundances per star but for a smaller and brighter sample of 20~stars with $V<14$. The spectra needed for these measurements would cover a narrower wavelength range of $170-310$~nm at the same spectral resolution, spatial resolution, and SNR as the spectra needed for breakthrough science. This narrower and redder spectral range eliminates the opportunity to observe the Sb~II, Br~I, and In~II features highlighted in the breakthrough science case.

\subsubsection{Envisioning the Distance Ladder in the Era of the Habitable Worlds Observatory (SCDD-EE-10)}
\label{sssec:distance_ladder}
\sleads{Gagandeep Anand, Meredith Durbin, Rachael Beaton, Joseph B. Jensen, Adam Riess}

As discussed by \citet{anand_hwo25}, the goal of this science case is to improve measurements of extragalactic distances, and thereby the Hubble Constant, by revamping the cosmic distance ladder. These observations have the potential to resolve the ``Hubble tension,'' which refers to the discrepancy between Hubble constant measurements based on Type Ia~Supernovae \citep[$H_0 = 73.04 \pm 1.04$~km~s$^{-1}$~Mpc$^{-1}$,][]{riess_et_al2022} and those based on Planck observations of the cosmic microwave background \citep[$H_0= 67.4 \pm 0.5$~km~s$^{-1}$~Mpc$^{-1}$,][]{planck_et_al2020}.

Currently, the Hubble Constant is measured via a three-stage process in which the closest distances are measured using a geometric foundation (e.g., trigonometric parallax), intermediate distances are measured using ``primary indicators'' for which apparent magnitude can be compared to absolute magnitude to measure distance (e.g., Cepheid variable stars), and the largest distances are measured using Type Ia Supernovae (SNe Ia). This three-tier approach is necessary because current facilities lack the sensitivity necessary to detect and measure the brightness of primary indicators out to the desired distance. In contrast, the large aperture and sensitivity of HWO would allow astronomers to measure the flux from Cepheids at both peak light and minimum light at distances up to 100~Mpc, thereby eliminating the need for a third stage based on SNe Ia. Using Cepheids rather than SNe Ia is advantageous because Cepheids are much better understood than SNe Ia and therefore distance measurements based on Cepheids would have smaller systematic errors. 

While Cepheid-based distances would be a significant improvement over SNe Ia-based distances, Cepheids alone are not a complete solution to the cosmic distance ladder. Importantly, Cepheids are unlikely to be useful for determining distances to elliptical galaxies because they are young stars ($20 - 200$~Myr) and are therefore associated with recent star formation. In addition, because Cepheids are young, they tend to be found in crowded environments where dust and gas could hinder precise photometry. Furthermore, stellar metallicity affects the relationship between Cepheid luminosity and period, but determining the metallicities of distant Cepheids will require additional spectroscopic data, likely from next-generation ground-based telescopes.

For older galaxies, distances could be determined using the tip of the red giant branch (TRGB) or the J-region asymptotic giant branch (J-AGB), both of which are prominent features on the color-magnitude diagram (CMD). TRGB is more established, and refers to the CMD location at which evolved low-mass stars ($M_\star \lesssim 2 M_\odot$) experience the helium flash and helium core fusion ignites. The helium flash occurs at a fixed core mass \citep[$\pm 0.02 M_\odot$,][]{serenelli_et_al2017}, which has the convenient consequence that the luminosity of TRGB stars is also constant and can be used as a standard candle as along as wavelength-dependent corrections are applied to account for variations in metallicity or measurements are conducted in I-band where metallicity-dependent systematics are reduced \citep[e.g.,][]{hatt_et_al2018, jang_et_al2018, beaton_et_al2018, wu_et_al2023}. Because TRGB stars are old ($>5$~Gyr), they can be detected in all galaxies with enough stars to fill out that region of the CMD \citep[i.e., $>5 \times 10^6 M_\odot$][]{mcquinn_et_al2019}. TRGB stars in galactic halos are ideal targets because they are in less crowded regions where their measured fluxes will be less affected by dust and stellar contamination \citep[e.g.,][]{beaton_et_al2016, beaton_et_al2019, jang_et_al2021, anand_et_al2022}. 

Measuring distances using J-AGB stars is a newer method in which the luminosities of carbon-rich AGB stars are used as standard candles \citep{madore+freedman2020}. These stars have intermediate ages (300 Myr -- 1 Gyr) and form a horizontal plume on the CMD. Although less well-characterized than the TRGB method, the J-AGB method has been applied to JWST data \citep{lee_et_al2024, lee_et_al2025, li_et_al2024} and the resulting distances appear consistent with Cepheid- and TRGB-based distances \citep{riess_et_al2024}. J-AGB stars are best observed in the outer regions of disks.

The differences between the locations and ages of Cepheids, TRGB, and J-AGB stars mean that the three samples will have different systematics, thereby providing an extremely useful dataset for cross-comparison and model validation. Current results are mixed as to the level of agreement between various indicators \citep[e.g.,][]{riess_et_al2024, freedman_et_al2025}.

For the oldest and faintest galaxies ($M_V \gtrsim -8$), the only viable primary distance indicator may be RR~Lyrae stars because the TRGB will be sparsely populated \citep{madore+freedman1995}, and the even smaller number of Cepheids and J-AGB stars that existed in the past will have already moved on to latter stages of stellar evolution. While not necessarily needed for improving measurements of $H_0$, RR-Lyrae-based distances will be useful for studying faint, low-mass dwarf galaxies that are relics of the epoch of reionization and for mapping streams of ancient stars in galactic halos \citep[e.g.,][]{catelan2009, monelli+fiorentino2022}. 

Theoretically, applying a two-stage distance ladder method (i.e., a first rung of geometric measurements and a second rung of primary distance indicators) to a sample of 24 Cepheid host galaxies could yield a $1\%$ measurement of $H_0$. To ensure that the resulting measurement is not biased by large-scale structure, the measurement could be repeated four times using 24~Cepheid host galaxies in each quadrant of the sky. A sample of 24~Cepheid hosts would therefore be needed for enabling science while a larger sample of roughly 100 Cepheid host galaxies at distances up to 100~Mpc is needed for breakthrough science. When using a two-rung distance ladder, the Cepheid host galaxies do not need to contain SNe Ia, which broadens the sample of possible targets. 

For distant galaxies, the ability to observe through a long-pass ``white-light'' filter would significantly improve the efficiency of variable star detection, but the lack of such a filter would not preclude the acquisition of the needed data. An instrument with a wide field of view and dual-band imaging capabilities would further improve efficiency by enabling simultaneous optical and infrared imaging and reducing the number of pointings needed to sample a galaxy. Dual-band capabilities would be particularly impactful for the breakthrough case of measuring distances with the J-AGB as well as Cepheids and the TRGB because the NIR data needed for J-AGB detection would be obtained at the same time as the optical data needed for Cepheid and TRGB detection. 

To refine the distance measurements for the anchor galaxies, spectroscopic observations will be needed to measure radial velocities of faint, detached eclipsing binaries. The radial velocity information will be combined with photometric observations of the eclipses to determine the physical radii of the two stars, which, in tandem with a radius-surface brightness relation, will reveal the distance to the system. These spectroscopic data would most likely be obtained by ground-based extremely large telescopes, but HWO may also be capable of obtaining spectra with the necessary precision and spectral resolution. 

Ground-based extremely large telescopes are poorly suited to obtaining the photometry needed to detect variable stars in the anchor and second-rung samples for several reasons. First, ground-based AO systems perform best in the NIR while the variations of Cepheids and RR~Lyrae stars are best detected at bluer wavelengths. Second, AO-equipped ELT images will have smaller fields of view ($\mathcal{O}~1')$, thereby requiring multiple pointings per nearby target and drastically reducing observational efficiency. Third, AO-equipped ELT images will lack the stability needed to measure flux variations to the necessary precision. 

\breakthrough At present, the distance ladder includes four geometric ``anchors'' on the first rung of the ladder: the Milky Way, the Large Magellanic Cloud, the Small Magellanic Cloud, and NGC~4258, which is at a distance of 7~Mpc and was observed by JWST programs GO-1685, GO-1995, and GO-2875. Breakthrough progress would require expanding to ten geometric anchors, each of which should have distances known to an improved precision of $0.3\%$. Those anchors could be the current sample of four~anchors supplemented by M31, M33, NGC~6822, IC1613, WLM, and IC10. For the second rung of the ladder, distances should be measured using Cepheids, TRGB, and J-AGB stars for $\sim 100$~Cepheid hosts out to distances of 100~Mpc and the results from the different methods should be consistent. 

Obtaining these measurements will require the ability to precisely measure the fluxes of Cepheids in bright, nearby galaxies to calibrate the distance ladder. Due to the brightness of these targets ($I \sim 12$ for those in the Magellenic Cloud and $I \sim 7$ for those in the Milky Way) and the anticipated high sensitivity of HWO, this goal will likely require special capabilities to observe targets brighter than the nominal saturation limit. Possible solutions include dispersing or diffusing light, spatial scanning \citep[e.g.,][]{riess_et_al2018}, and quickly reading out subarrays \citep[e.g.,][]{breuval_et_al2024}. 

Breakthrough science requires high-resolution optical/NIR imaging that is diffraction-limited at 500~nm, Nyquist sampled in $V$ (ie., a pixel scale of $0.01''$/pixel in optical/NIR), and has an absolute flux calibration uncertainty of $0.5\%$. The imaging should cover a field of view of 36 arcmin$^2$ down to depths of $V < 33$ and $I < 32$. 

\enabling For enabling science, there should be at least six geometric anchors on the first rung of the ladder (i.e., the current four plus M31 and M33) and distances to those anchors should be known to $0.5\%$. For the second rung of the ladder, distances to 24~Cepheid host galaxies within 100~Mpc should be measured using both Cepheids and TRGB stars. The pixel scale would be relaxed to $0.02 - 0.025''$/pixel, thereby undersampling the stars, the field of view would be reduced to 25 arcmin$^2$, and the depth would be decreased to $V < 32$ and $I < 31$. In addition, the ability to observe bright targets would be relaxed to $I > 12$ to enable observations of Cepheids in the Magellanic clouds rather than the brighter Cepheids in the Milky Way ($I > 7$). 

\subsubsection{Dust Extinction Beyond the Milky Way (SCDD-EE-2)}
\label{sssec:dust_extinction}
\sleads{R. Paladini, S. Salim, M. Boquien, Y. Choi, M. Hayes, B.Hensley, V. Lebouteiller, J. Roman-Duval}

As presented by \citet{paladini_hwo25}, the goal of this science case is to characterize the properties and possible differences of extragalactic dust grains. While dust is only a small component of the mass budget of the ISM \citep[$1\%$,][]{demyk2011}, it has profound effects on the paths of light through the universe and is deeply connected to the formation of molecules, stars, and planets. Due to the challenges of characterizing dust grains, our current knowledge of dust properties is primarily restricted to the grains that reside in the diffuse ISM of the Milky Way. HWO provides the opportunity to investigate how dust grains in denser Milky Way environments and in other galaxies might differ from the relatively well-characterized dust grains in the Milky Way's diffuse ISM. Past studies of various sightlines within the Milky Way have revealed UV/optical dust extinction curves with a remarkable diversity of wavelength dependencies \citep[e.g.,][]{massa_et_al1983, witt_et_al1984, valencic_et_al2004, fitzpatrick+massa2007, gordon_et_al2009}. While most of these wavelength-dependent shapes are well described by a relationship between total extinction and selective extinction of $R(V) = A(V)/E(B-V) \sim 3.1$ \citep{cardelli_et_al1989}, other sightlines have total-to-selective-extinction ratios as low as $R(V) \sim 2$ or as high as $R(V) \sim 5$. 

The discovery that parameterizing extinction in terms of $R(V)$ explains the majority of directional differences in the extinction curves towards diffuse regions of the Milky Way indicates that the population of dust grains interacting with UV light are somehow connected to those interacting with light at optical and NIR wavelengths. The leading theory is that distribution of dust grain sizes scales with $R(V)$ such that regions with higher $R(V)$ have larger grains \citep{weingartner+draine2001}.

As the sample of well-measured extinction curves has increased, more discrepancies have been identified between observed Milky Way extinction curves and the curves predicted by the $R(V)$ scaling \citep{mathis+cardelli1992, clayton_et_al2003, valencic_et_al2004}. These discrepancies are even more pronounced for sightlines towards the Small and Large Magellanic Clouds, which display sharper declines in the UV \citep{gordon_et_al2003}. In addition, some Small Magellanic Cloud extinction curves do not appear to contain a 217.5~nm absorption feature that is prominent in Milky Way extinction curves \citep{gordon_et_al2024}. 

While the stellar spectra obtained for the three-year HST ULLYSES project \citep{roman-duval_et_al2025} may provide some additional insights into dust extinction in the Magellanic Clouds as well as a few low-metallicity dwarf galaxies (NGC~3109, Sextans~A, WLM, IC~1613, Leo~A, Leo~P), a facility like HWO has the potential to dramatically advance dust studies by increasing the number of observations as well as the observation quality. Enhanced sensitivity and spatial resolution would also provide the valuable opportunity to study how dust properties in diffuse regions vary locally within other galaxies due to localized phenomena such as supernova explosions and related shocks.

For the diffuse ISM, dust grains are expected to be small ($\sim 0.1$~micron). In denser regions, grains have icy veneers that facilitate sticking, coagulation, and grain growth. As a result, the grain sizes tend to be larger and the dust extinction curve becomes noticeably flatter in the near-infrared. The short-wavelength behavior of extinction curves in dense regions is not well characterized, and there are only a small number of NIR extinction curves for dense regions. HWO has the potential to obtain UV/optical/NIR extinction curves for many dense sightlines within the Milky Way and extend studies of extinction in dense regions to the Magellanic Clouds and perhaps even Local Group galaxies. These observations could be possible with a facility like HWO because the extreme sensitivity of the telescope would permit the detection of bright targets even through many magnitudes of extinction. 

This proposed census of dust properties would be accomplished by obtaining UV/O/IR spectra towards various regions within and beyond the Milky Way to determine dust extinction curves. The science objectives are split into Objective~1, which probes diffuse ISM sightlines (extinction $A_V < 3$ magnitudes) out to Local Volume galaxies at distances of 10~Mpc, and Objective~2, which probes dense sightlines ($A_V > 3$ magnitudes) primarily within the Milky Way. 

For Objective 1, the primary targets are bright OB stars ($V = 16$ at 1~Mpc, $V=21$ at 10~Mpc) whose relatively featureless spectra facilitate the detection of interstellar absorption lines. (Section~\ref{sssec:dust_uv} also selects OB~stars as targets for the same reasons.) For Objective~2, the preferred targets are blue A/F/G giants because they are both abundant and bright at UV/O/IR wavelengths. For NIR observations, redder K/M giants will also be suitable targets, but those targets are too faint at blue wavelengths to probe absorption in the UV and optical. 

\breakthrough For breakthrough progress, 22,515 diffuse sightlines and 126~dense sightlines should be observed for Objectives~1 and 2, respectively. The diffuse sightlines should include 20,000 sightlines towards the Milky Way, 2,000 towards the Large Magellanic Cloud, 400 towards the Small Magellanic Cloud, and 100 towards M31. These sightlines should be selected to sample the $A(V)$ range between 0.1~mag and 3~mag (i.e., densities of $\sim 10^{20} - 5 \times 10^{21} \rm cm^{-2}$) with roughly equal numbers of sightlines in each of 10 $A(v)$ bins with widths $\sim0.3$ mag. In addition, the observations should include diffuse sightlines towards roughly ten local group galaxies and five local volume galaxies; those observations will permit mapping the extinction curve for those galaxies. 

The dense sightlines should include 100 through the Milky Way, 10 towards to the LMC, 10 towards the SMC, and five towards M31. Additionally, the dataset should include sightlines towards dense regions in at least one local group galaxy. For both the diffuse and dense sightlines, observations should be obtained at a signal-to-noise ratio $\geq5$. The instrument should be capable of observing targets as faint as $\leq21$~AB mag (i.e., OB stars at a distance of 10~Mpc). 

The ideal instrument to obtain these observations is a UV/VIS/NIR MOS with spectral resolution $R\sim3000$, spatial resolution $0\farcs01$/spaxel, and wavelength coverage of 100 -- 2500~nm. Earlier versions of this science case mentioned a desired field of view $3''-4'' \times 3''-4''$, but \citet{paladini_hwo25} does not specify a field of view. A MOS is preferable to an IFS because this program aims to target bright stars that are likely to be spatially grouped within their host galaxies.

The minimum angular resolution is set by the desire to probe galaxies out to distances of 10~Mpc. The short wavelength limit of 100~nm is motivated by the desire to capture the shape of the extinction curve at FUV wavelengths and the presence or absence of the 217.5~nm bump that is present in Milky Way extinction curves and missing in some Small Magellanic Cloud extinction curves \citep{gordon_et_al2024}. Extended red coverage would capture the 3.3~micron aromatic feature and 3.4~micron aliphatic feature, which have not yet been confirmed in ISM sightlines. At minimum, NIR coverage at $1-2\mu$m would be necessary for assessing the flattening of the NIR region of extinction curves in dense regions. 

\enabling The number of sightlines required for enabling progress is significantly smaller. The diffuse sightlines obtained for Objective~1 should include 1000 towards the Milky Way, 100 towards the Large Magellanic Cloud, 20~towards the Small Magellanic Cloud, and five~towards M31 as well as diffuse sightlines towards five Local Group galaxies and two local volume galaxies. For Objective~2, the dense sightlines should include 50 towards the Milky Way, five towards the Large Magellanic Cloud, five towards the Small Magellanic Cloud, and one towards M31. No dense sightlines towards Local Group or Local Volume galaxies would be required for enabling progress. 

In this case, the spectral resolution ($R\sim3000$) would be unchanged, but the wavelength coverage would be narrowed to 100--1000~nm. In addition, the sensitivity requirement would be softened to $>14$~AB mag. Due to the increased focus on sightlines towards the Milky Way and Magellanic Clouds, an IFS would be sufficient for enabling science even though a MOS would be more efficient for making breakthrough progress by observing sightlines towards more distant Local Group and Local Volume galaxies.

\subsubsection{Probing the Variations of Interstellar Dust Abundance and Properties Within and Between Galaxies with HWO UV Spectroscopy in the Local Volume (SCDD-EE-11)}
\label{sssec:dust_uv}
\sleads{Julia Roman-Duval, Mederic Boquien, Yumi Choi}

As explained in \citet{roman-duval_hwo25}, the goal of this science case is to investigate how the composition and properties of interstellar dust varies within and among galaxies in the Local Volume (distances $< 10$~Mpc). Dust is intimately connected to star formation, stellar evolution, and the cycles of metal within galaxies and among galaxies. Accordingly, this science case is related to Astro2020 questions D-Q4c, which compares the ISM and other properties of local galaxies and high-redshift galaxies, and F-Q1b, which focuses on the formation and environmental interaction of molecular clouds \citep{astro2020}. 

While FIR emission spectroscopy can be used to detect dust, inferences based on FIR data are severely limited by the relative lack of knowledge about dust opacity \citep[e.g.,][]{roman-duval_et_al2014, roman-duval_et_al2017, roman-duval_et_al2022}. In contrast to FIR emission spectroscopy, UV absorption line spectroscopy does not require knowledge of dust opacity. 

When determining dust composition via UV absorption spectroscopy, the reasoning is that the elements with low abundances in the gas phase must instead be in the form of dust. The abundance of the material in the dust phase can then be determined from the gas phase abundance and the total ISM metallicity (i.e., the metallicity considering both the gas and the dust). The total metallicity can be inferred by either using the measured photospheric abundances of young stars as a proxy for the total abundances within the ISM \citep[e.g.,][]{jenkins2009, roman-duval_et_al2021} or by modeling the dust-depletion sequence \citep[e.g.,][]{ritchey_et_al2023, de_cia_et_al2024, konstantopoulou_et_al2024}. For this HWO science case, the most likely scenario is that dust abundances will be determined by combining the gas phase abundances measured with HWO absorption line spectroscopy with photospheric abundances determined from spectroscopy obtained by ground-based extremely large telescopes. 

This investigation has two aims. The first objective is to quantify the amount of various refractory (Fe, Si, Mg, Ni, Cr, Mn) and volatile (Zn, S, P, Kr, Cl, N) elements found in neutral gas and dust along the line of sight to multiple groups of massive stars in the Local Volume. Targeting O and B stars with few photospheric absorption lines provides a fairly uniform canvas for detection of absorption features generated by neutral ISM gas along the line of sight. The target samples would be selected to probe a variety of metallicities ($1 - 100\%$ solar) and gas densities ($20\rm~cm^{-2} < \log N(H) < 22 \rm~cm^{-2}$), thereby mapping how dust properties vary with environmental conditions. The resulting observations will permit assessments of the dust-to-metal mass ratio (D/M) and the dust-to-gas mass ratio (D/G) within and among galaxies. In addition to providing valuable insight into the life cycle of metals, galaxy evolution, and the chemical enrichment of the universe, measurements of D/M and D/G are needed to reduce dust-related systematics for other investigations such as studies of the chemical enrichment history of the universe using damped Ly$\alpha$ systems (see Section~\ref{sssec:agn_feedback}). 

The second objective is to conduct a more local survey of C and O, which contribute the largest reservoir of metals available for dust growth in the ISM, within the Milky Way and Magellanic Clouds. This objective is restricted to much smaller distances because the difficulty in measuring abundances and dust depletions for C and O. Specifically, measuring C and O abundances requires analyzing both highly saturated lines (C~II at 133.5~nm, O~I at 130.2~nm) and extremely weak lines (C~II] at 232.5~nm, O~I at 135.5~nm). This science case cannot be accomplished with HST because fixed pattern noise on COS prevents achieving the necessary S/N and STIS does not have the needed sensitivity. This science case may therefore demand careful attention to HWO detector noise properties. 

\breakthrough Breakthrough progress would require significantly increasing the number of sightlines with robust depletion measurements. Ideally, depletions of C, O, Fe, Si, Mg, Cr, Ni, and other important dust consituents should be determined for 100~sightlines each towards the Milky Way, Large Magellanic Cloud, and Small Magellanic Clouds. Depletions for elements other than carbon and oxygen should also be measured along 100 sightlines per galaxy for other galaxies within 10~Mpc such as IC~1613 and Sextans~A. The sightlines should be selected to sample a range of galaxy properties and span the density range $N(H) = 10^{20}-10^{22} \rm \, cm^{-2}$. Ideally, the sightlines would be selected so that there are $1-3$ sightlines in each 0.1~dex bin in column density and metallicity, but realistically the number of low-metallicity sightlines may be restricted by the number of viable targets (i.e., very low metallicity galaxies within 10~Mpc). 

For this science case, a MOS would be preferable to an IFS given the need to obtain $>100$ spectra over a relatively wide field-of-view ($\geq 2' \times 2'$). The spectra obtained along each sightline should cover the EUV, FUV, and NUV ($95 - 315$~nm) with roughly uniform throughput (i.e., without a sharp decrease at 110~nm). The short wavelength limit is set by the need to capture important features from H$_2$ (i.e., the Lyman-Werner bands at 95-110~nm) and P~II for Objective~1. For Objective~2, the wavelength range could be narrowed to $115 - 315$~nm. Spectral resolution $R\sim 50,000$ would be sufficient for Objective~1, but higher spectral resolution ($R> 100,000$) is needed for Objective~2 because of the weakness of the 135.5~nm O~I and 232.5~nm C~II features that must be measured to determine oxygen and carbon depletions. 

The sensitivity should be high enough to obtain $S/N > 20/{\rm pixel}$ spectra in less than 10~hours for targets with FUV and NUV fluxes $\gtrsim 5 \times 10^{-16}$~ergs/cm$^{2}$/s/\r{A}) when observing at $R\sim 100,000$. For observations at $R\sim 50,000$, $S/N > 20/{\rm pixel}$ should be achievable in $<10$~hours for targets with FUV fluxes $\gtrsim 3 \times 10^{-17}$~ergs/cm$^{2}$/s/\r{A}) and NUV fluxes $\gtrsim 1.6 \times 10^{-17}$~ergs/cm$^{2}$/s/\r{A}). Fully understanding the composition and depletions of dust at low metallicities will require $S/N > 100$. 

\enabling For enabling science, depletions of major dust constituents should still be measured along 100~sightlines towards the Milky Way, Large Magellanic Cloud, and Small Magellanic Cloud but fewer other sightlines would be sampled. Specifically, the distance limit for the other galaxies would be reduced from 10~Mpc to a few MPc and only 30~sightlines per galaxy would be measured. In addition, while carbon and oxygen depletions should still be measured along 100 Milky Way sightlines, those depletions would be measured along only 10 sightlines each for the Large and Small Magellanic Clouds. 

The required wavelength range and spectral resolution would be unchanged from the breakthrough case, but the sensitivity requirements would be relaxed by an order of magnitude. Enabling progress would require reaching $S/N > 20/{\rm pixel}$ spectra in $<10$~hours for targets with FUV and NUV fluxes $\gtrsim 5 \times 10^{-15}$~ergs/cm$^{2}$/s/\r{A}) when observing at $R\sim 100,000$. For observations at $R\sim 50,000$, $S/N > 20/{\rm pixel}$ should be achievable in $<10$~hours for targets with FUV fluxes $\gtrsim 2 \times 10^{-16}$~ergs/cm$^{2}$/s/\r{A}) and NUV fluxes $\gtrsim 3 \times 10^{-16}$~ergs/cm$^{2}$/s/\r{A}). 

\subsubsection{Probing the Full Depth of ISM Properties with a UV-IFU (SCDD-EE-12)}
\label{sssec:ism_uv}
\sleads{Bethan James, Danielle Berg}

As described by \citet{james_hwo25}, the aim of this science case is to determine the distribution of elements within galaxies and how that might depend on galactic properties and cosmic time. This investigation would be subdivided into four objectives, each of which requires particular observations. The first objective is to determine how those materials spread and mix within galaxies on both radial and azimuthal scales. This objective requires contiguous mapping of the interstellar medium across galaxies. The second objective is to investigate how the formation, evolution, and death of stars affects the distribution of elements within galaxies. This investigation requires mapping gas in and near regions where stars are being formed. The third objective is to determine the temporal scales over which elemental abundances change within galaxies. This objective requires comparing multi-phase elemental abundances determined for clusters of various ages. The fourth objective is to evaluate the influence of environmental conditions on elemental abundances. For this objective, abundances should be measured and compared for interacting and isolated star-forming regions. 

For all four objectives, high spatial resolution is essential to connect measured abundance ratios with the local conditions. Although ancient, high-redshift galaxies are too distant to resolve the small spatial scales needed to assess the causes of unusual abundance patterns, observations of nearby local analogs can be used as proxies to better understand the enrichment history and abundance patterns of high-redshift galaxies. 

This science case requires spatially-resolved UV spectra to determine the physical and chemical conditions of the ISM throughout local galaxies. UV emission spectra are needed to determine the density and temperature of electrons as well as the hardness of the ionizing field. The combination of UV emission spectra and UV absorption spectra is needed to measure abundances and column densities for gas and dust. Access to the UV is essential for detecting critical elements such as O, C, Si, and N. For instance, strong emission from SiIII] (rest wavelengths 188.3~nm and 189.2~nm) and CIII] (rest wavelengths 190.7~nm and 190.9~nm) has been detected in redshifted spectra of high-redshift galaxies \citep[e.g.,][]{garnett_et_al1995, berg_et_al2019}, but both C and Si lack emission lines at visible wavelengths. Accordingly, detecting C and Si in the local analogs to high-redshift galaxies requires UV spectroscopy. In addition, the spectroscopy must be spatially resolved in order to properly calculate abundances \citep{yuan_et_al2013, mast_et_al2014, james_et_al2016}. For example, observations at high spatial resolution have revealed significant variation in oxygen abundance on scales of $10 - 100$~pc \citep[e.g.,][]{james_et_al2020, williams_et_al2022}, indicating that inferences based on unresolved observations may be erroneous. 

Deep, sensitive UV IFS spectroscopy has the potential to reveal the mechanisms by which galaxies mix their ISM and the resulting variations of elemental abundances on a range of scales. UV data could be combined with spatially-resolved optical data obtained by ground-based telescopes to determine electron temperatures, ionic abundances, and reddening corrections. Mapping elemental abundances near regions of ongoing star-formation will provide new insights into the impact of mechanical and radiative feedback on the distribution of metals within galaxies and within outgoing galactic winds that enrich the intergalactic medium. Determining multi-phase abundances near and within older stellar clusters will provide insight into the chemical enrichment and mixing history of galaxies because the abundance patterns observed for different clusters can be compared to the ages of the associated O and B stars, which can be dated using photospheric and stellar wind features \citep[e.g.,][]{chisholm_et_al2019}. Furthermore, comparing abundance patterns for isolated star-forming systems to those measured for interacting star-forming systems will reveal how infalling gas mixes with local gas and potentially spurs or impedes star formation. 

Achieving those scientific objectives requires measuring spatially-resolved, multi-phase abundances of C, N, and Si. Accurately and precisely measuring gas abundances, requires direct sensitivity to all of the most important ionization levels of each element rather than detecting only a few ionization levels and estimating the abundances of other ionization states using poorly constrained ionization correction factors. Critically, the data must be spatially contiguous to ensure adequate sampling of spatial variation in abundances.

Characterizing the radiative environment and investigating the effects of stellar feedback requires detecting the stellar continuum and spectral features of the extremely hot stars responsible for the bulk of the stellar ionizing photons. Stellar spectra are also needed to determine the ages and metallicities of various stellar populations. Furthermore, measurements of various ionized species such as C~III], He~II, C~IV, and O~III] on small spatial scales ($\sim 10$~pc) are needed to determine the relative importance of different ionization channels including stars, shocks, and AGN \citep[e.g.,][]{mingozzi_et_al2022}. Gas kinematics determined from emission line spectra will provide additional insight into the sources of ionization and the role of mechanical feedback. 

The ideal sample for this science case consists of nearby galaxies ($z < 0.05$, $D < 20$~Mpc) that have low metallicities ($12 + \log (\rm O/H) < 8.5$), high star-formation rates ($> 1 M_\odot/\rm {yr}$), and low masses ($M < 10^{10} M_\odot$) reminiscent of high-redshift galaxies. Using previous optical observations from other facilities including ground-based telescopes, specific targets should be selected to span a range of potentially important properties such as star formation rate, metallicity, and environment.

\breakthrough Breakthrough progress requires mapping galaxy properties on scales $<10$~pc for a diverse sample of $\sim 100$ nearby galaxies ($z< 0.025$; $D<100$~Mpc). This angular resolution would be high enough to resolve structures within the interstellar medium and therefore prevent inferred galactic properties from being biased by unresolved inhomogeneities. For breakthrough science, the galaxies should be observed at $S/N \geq 5$ per spaxel using an IFS with a contiguous field of view of at least $10'' \times 10''$, wavelength coverage of $95 - 285$~nm, spaxels $\leq 0.1'' \times 0.1''$, and spectral resolution $R > 10,000$. This spectral resolution corresponds to a velocity resolution $\Delta v < 30 \, \rm km \, s^{-1}$. For reference, a higher velocity resolution of $\Delta v 5-10 \, \rm km \, s^{-1}$ would enable resolving the intrinsic widths of emission features and ISM outflows in very low mass galaxies. 

\enabling For enabling science, $\sim50$ galaxies should be mapped on scales of $\sim 100$~pc (i.e., cluster scales). These observations could be obtained by efficient stepping of a MOS with a field of view of $4'' \times 4''$ and $\leq 0.2'' \times 0.2''$ slit extractions. The MOS should cover $115 - 285$~nm \citep[the same wavelength range as UVEX but with an order of magnitude higher throughput;][]{kulkarni_et_al2021} at spectral resolution $R = 10,000$.

\subsubsection{The Heavy Element Enrichment History of the Universe (SCDD-EE-4)}
\label{sssec:r_process_el}
\sleads{Eric Burns, Jen Andrews}

Like Section~\ref{sssec:r_process_nature}, the science case described by \citet{burns_hwo25} focuses on the origin of r-process elements. However, while Section~\ref{sssec:r_process_nature} aims to study r-process nucleosynthesis by measuring abundance patterns in quiescent stars, this science case proposes to obtain rapid observations of kilonovae produced by neutron star mergers \citep[e.g.,][]{burns2020}. These data could allow researchers to assess the nature of the explosions and the contributions from remnant object, jets, and ejecta to the resulting emission \citep[e.g.,][]{arcavi2018, metzger_et_al2018}. Understanding neutron star mergers was identified as a top scientific priority by the Astro2020 Decadal Survey \citep{astro2020}, which identified ``New Messengers and New Physics'' as a key scientific challenge and ``New Windows on the Dynamic Universe'' as a priority area within that theme. By investigating the origins of elements, this science case would also address one of the ``Eleven Questions for the New Century'' raised by the National Academies Committee on the Physics of the Universe \citep{nrc2003}: ``How Were the Elements from Iron to Uranium Made?'' 

A facility like HWO is important for this science case because HST lacks the rapid response capabilities needed for this science case (i.e., data within hours of explosion) and UVEX \citep{kulkarni_et_al2021} will focus on the nearby universe. Theory suggests that the rate of neutron star merges peaked $5-8$ billion years ago and that 80-90\% of mergers occurred 2-11 billion years ago \citep{burns2020}; an observatory like HWO would be uniquely able to study the UV evolution of these kilonovae. 

Kilonovae light curves evolve over time, and this science case aims to acquire data as early as possible to better understand the nature of the explosion. Accordingly, the ability to rapidly schedule and obtain observations is critical for this science case. This science case also drives demand for on-board decision making to assess whether to adjust exposure times, acquire photometric data at particular bands, or obtain spectroscopy. If such decisions can be made autonomously by a space-based observatory, then multi-band, multi-epoch observations could be acquired much more efficiently than if the observations needed to be transmitted to Earth and analyzed there (by humans and/or algorithms) before requesting and obtaining additional data. UVEX is posed to advance kilonovae science, but a facility like HWO would provide a significant enhancement over UVEX. 

\breakthrough Breakthrough progress would require characterizing 80~objects by obtaining UV spectroscopy for nearby objects or UV imaging for distant objects. The velocities will be so high ($>0.1c$) that spectral features will be broadened. Accordingly, low spectral resolution will be sufficient. 

Imaging should be sensitive to objects as faint as UV magnitude = 30 and have 1'' spatial accuracy. The objects should be first observed as soon as possible after explosion (i.e., within hours) and re-observed every few hours until the ultraviolent emission fades. Early-time spectra should be obtained for 10~kilonovae. The imaging sample should be large enough to subdivide the sample into $\geq 8$ redshift bins with $\gtrsim10$ events per bin, thereby permitting studies of the temporal history of elemental enrichment. 

\enabling Substantial progress would require sensitivity to objects as faint as UV magnitude = 28 to characterize a smaller sample of 25~objects within a few hours of explosion. Five of these objects should be observed with multi-epoch spectroscopy while the remainder should be imaged. The redshift distribution of the targets should permit coarse investigations of temporal evolution by subdividing the events into a few redshift bins.

\subsubsection{Flash Spectroscopy of Core-Collapse Supernovae with the Habitable Worlds Observatory (SCDD-EE-3)}
\label{sssec:flash_ccsne}
\sleads{Jen Andrews, Eric Burns}

\citet{andrews_hwo25} describes this science case, which aims to investigate the final moments of massive stars to better understand how these short-lived, extremely luminous stars form elements and enrich the interstellar medium. This study would directly address multiple questions identified by the Astro2020 panels on Compact Objects \& Energetic Phenomena; Galaxies; Interstellar Medium \& Star \& Planet Formation; and Particle Astrophysics \& Gravitation \citep{astro2020}. As discussed in Section~\ref{sec:astro2020}, these questions are focused on the physics of transients and their impact on their host galaxies. 

Massive stars explode as core-collapse supernovae (CCSNe), which are extremely bright. Prior to their eventual explosions, these stars experience mass loss, thereby creating shells of circumstellar material (CSM). When the supernova explodes, the shock wave and ejecta have to travel through the CSM and the intense X-ray/UV radiation from shock breakout can ionize the CSM. When the CSM recombines, the resulting emission features can be used to assess the composition of the CSM. Observations obtained within roughly the first week of the explosion when the primary energy source is still the explosion itself as opposed to radioactive decay and the ejecta has not yet reached the outer edge of the CSM, therefore provide a valuable opportunity to determine the state of stellar mass loss immediately prior to explosion. This technique of observing the supernovae shortly after explosion is known as flash spectroscopy \citep[e.g.,[]{khazov_et_al2016}.

Theoretical predictions for the expected rate of mass loss vary dramatically \citep[e.g.,][]{ekstrom_et_al2012, beasor_et_al2020} but some studies suggest that the apparent disagreement could be partially resolved by intensified mass loss just before explosion \citep[e.g.,][]{quataert+shiode2012,fuller2017,fuller+tsuna2024}. Past flash spectroscopy studies have been primarily conducted using ground-based telescopes observing at optical wavelengths, but UV data are essential for probing the composition and ionization state of the CSM. For instance, FUV/NUV data are needed to directly measure CSM metallicity via Fe~VI and Fe~VII lines at $120-150$~nm; measure the structure and speed of CSM winds; and detect emission from CIV, OIV, and NIV, all of which have high optical depths in the optical.

Currently, UV flash spectra have been obtained for only a handful of CCSNe. While HST has sufficient spectral resolution and wavelength coverage for UV flash spectroscopy, HST operations do not permit sufficiently fast observations. Even when executing the single ultra-rapid disruptive target-of-opportunity observation allowed per HST cycle, the delay between triggering HST and obtaining data is several days. As the name suggests, \emph{Swift} is able to respond more swiftly, but the sensitivity and resolution are too low for detailed studies of the CSM. The upcoming ULTRASAT mission \citep{shvartzvald_et_al2024} will have the necessary responsiveness, sensitivity, and resolution, but ULTRASAT will cover only $230 - 290$~nm, thereby missing crucial FUV coverage. The subsequent UVEX mission \citep{kulkarni_et_al2021} will be capable of conducting target-of-opportunity observations within 3~hours of trigger, but the wavelength range of UVEX is limited to $115 - 275$~nm, which excludes the Mg~II doublet at 279.5~nm and 280.2~nm. 

\breakthrough Breakthrough progress would require rapid-response (i.e., within 1 hour of explosion and trigger), UV ($110 - 300$~nm) spectroscopy of at least 15 CC SNe at a S/N$ \geq 5$ and spectral resolution $R \sim 10,000$. The targets are expected to have absolute magnitudes between $-16.5$ and $-21$, which translates to apparent magnitudes of $12.5 - 17$ at distances of 50~Mpc. 

\enabling Enabling progress would require observing a smaller sample of 10-15 CC~SNe on a slightly less rapid timescale (i.e., within 1~day of explosion and trigger). These observations should be obtained at a lower spectral resolution $R \sim 3000$.

\subsubsection{New Frontiers in the Study of Magnetic Massive Stars (SCDD-EE-15)}
\sleads{Alexandre David-Uraz, V\'{e}ronique Petit, Coralie Neiner, Jean-Claude Bouret, Ya\"{e}l Naz\'{e}, Christiana Erba, Miriam Garcia, Kenneth Gayley, Richard Ignace, Ji\v{r}i Krti\v{c}ka, Hugues Sana, Nicole St-Louis, Asif ud-Doula}
\label{sssec:mag_stars}

As described in \citet{david-uraz_hwo25}, the aim of this science case is to investigate how massive stars affect their surroundings. This science case focuses particularly on stellar magnetism, which differentiates from the other massive star science cases discussed in Sections \ref{sssec:massive_stars_lowZ} and \ref{sssec:vms}. The twin goals of this science case are (1) determine how and why only 10\% of massive stars have strong magnetic fields and (2) assess how massive stars and their magnetic fields co-evolve. Massive stars have a large influence on their host galaxies and are the progenitors of neutron stars and black holes, so these goals are aligned with the Astro2020 Priority Science Areas ``Unveiling the drivers of Galaxy growth'' and ``New windows on the dynamic universe'' \citep{astro2020}.

While roughly 10\% of massive stars have stable magnetic fields with strengths $\gtrsim 1000$~G, most massive stars have fields that are too weak to be detected. The weakest detected fields ($\sim100$G) are far enough above the detection limits to indicate a possible ``magnetic desert'' between stars with fields $>100$~G and those with ultra-weak fields $<1$~G \citep{auriere_et_al2007}. A possible explanation for this apparent bimodality is that strong magnetic fields are able to persist on the timescales of massive star lifetimes while magnetic fields below a threshold strength of roughly 100~G are disrupted, perhaps by the FeCZ (a thin, sub-surface convection zone). The lost fields are hypothesized to be replaced by very weak fields generated by stellar dynamos \citep{jermyn+cantiello2021}. 

Extremely weak fields ($\lesssim 1$~G) have been detected in some A stars but their existence in O and B stars has not yet been established. The non-detection of weak fields in O and B stars may be because the increased breadth of their spectral lines intensifies the challenge of detecting and measuring magnetic fields, and therefore higher sensitivity may be required to detect weak fields in OB stars compared to A stars. The FeCZ convective zone is controlled by metallicity-dependent opacity bumps, so observing massive stars with a range of metallicities could help clarify mechanisms driving the distribution of field strengths. For stars with strong fields, UV spectropolarmetry of the surrounding environment will enable investigations of the interplay between stellar magnetic fields and circumstellar material, thereby providing insight into the influence of magnetic massive stars on their host galaxies. 

Addressing these goals requires four types of observations. The first dataset should consist of spectropolarimetric observations of 60 Milky Way stars that are possible progenitors of magnetic stars. Half of the targets (30 stars) should be ``non-magnetic'' stars while the other half should be likely post-merger objects. The observations should be sensitive to magnetic fields strengths of 0.1~G for the ``non-magnetic'' stars and 1~G for the post-merger objects. Repeating the observations 2-3 times per star will increase the sensitivity of the observations and therefore their usefulness for mapping the proposed magnetic desert and evaluating different mechanisms for forming magnetic fields.

The second dataset should consist of multi-epoch snapshots of hundreds of stars beyond the Milky Way. These stars should be observed 2-3 times each at a sensitivity sufficient to detect longitudinal magnetic fields $\geq 100$~G. The sample should include both single stars and stars in multi-star systems and representatively sample massive stars at low metallicities. The requested sample size is set by the desire to recover Milky-Way-like magnetic field rates of 10\%. 

The third dataset should be very deep spectropolarimetry of evolved massive stars. The stars should be observed 2-3 times each at a sensitivity sufficient to detect fields $\sim 0.1$~G. The purpose of this dataset is to investigate if and how magnetic fields evolve with time. If evolved stars have simpler magnetic fields than equally massive main-sequence and pre-main sequence stars, then that would be consistent with Ohmic dissipation disrupting higher-order multipolar fields more quickly than lower-order fields. In addition, constraining the magnetic fields of evolved stars will test models of stellar evolution and could potentially reveal differences between magnetically active and magnetically quiet stars. 

Finally, multi-wavelength (UV/VIS/NIR) spectropolarimetric observations of 30 bright magnetic stars are required to investigate the co-evolution of massive stars and their magnetic fields. These observations will map stellar magnetospheres and reveal interactions with stellar winds. 

\breakthrough Breakthrough progress requires full Stokes (IQUV) spectropolarimetry of massive stars within the Milky Way and the Magellanic Clouds. The observations span a broad wavelength range from the FUV to the NIR ($100-1600$~nm) and achieve a polarimetric precision $< 10^{-6}$. The spectral resolution should be $\sim120,000$ in the optical (i.e., equivalent to current instrumentation) and $R>60,000$ in the UV. For ``non-magnetic'' Milky Way stars and evolved stars, the observations should be sensitive to magnetic field strengths $\geq 0.1$G. The sample should also include low-metallicity targets ($Z < 0.2 Z_\odot$) observed with sensitivity sufficient to detect fields $\geq100$~G. Targets should be observed 2-3 times each for snapshot observations or at 20 times well distributed in rotational phase for phase-revolved observations. 

\enabling Substantial progress requires obtaining partial Stokes (IV) spectropolarimetry for massive stars within the Milky Way and Molecular Clouds. The spectral resolution would be the same as for breakthrough science, but the wavelength range would be narrowed to NUV to FIR (i.e., removing the FUV). In addition, the polarimetric precision would be relaxed to $< 10^{-5}$. The target sample should include stars in the Small Magellanic Cloud ($0.2 Z_\odot$). The observations should be sensitive to fields as weak as 0.5~G for ``non-magnetic'' Milky Way stars and evolved stars and to fields $\geq 300$G for low-metallicity targets and Wolf-Rayet stars. For the survey, snapshot targets should be observed once each. For the targeted component, stars should be observed ten times with observations well-spaced in rotational phase.

\subsection{Solar Systems in Context}
\label{ssec:ssic_scdds}
The Solar Systems in Context Working Group identified \nssicall~potential science cases. Most (\nssic) of these cases were later published or posted to the SCDD Portal. 

The SSiC science cases span a wide range of topics and targets ranging from assessing the threat of impacts from near-Earth objects (Section~\ref{sssec:defense}) to characterizing the atmospheres and orbits of giant planets orbiting other host stars (Sections~\ref{sssec:giant_orbits}, \ref{sssec:reflected_giants}, \ref{sssec:transits}, and~\ref{sssec:highres_atmos}). The common bridge between these science cases is the desire to determine how our solar system compares to other planetary systems. Accordingly, several science cases aim to understand the prevalence of various planet types and system configurations (e.g., Sections~\ref{sssec:rocky_sub}, \ref{sssec:hab_system}, \ref{sssec:occ_small}, and~\ref{sssec:occ_binary}). 

SSiC science cases also investigate the formation of the solar system (Sections~\ref{sssec:solar}, \ref{sssec:venus}, \ref{sssec:mars}, and~\ref{sssec:aminos}) and other planetary systems (Sections~\ref{sssec:disk_winds}, \ref{sssec:proto}, \ref{sssec:debris}, and~\ref{sssec:exozodi}). Multiple cases also explore how the presence and composition of planetary atmospheres might evolve over time by considering atmospheric retention and loss (Sections~\ref{sssec:ozone_onset}, \ref{sssec:retention}, \ref{sssec:escape}, and \ref{sssec:exovenus}). Those studies are complemented by investigations of how the properties of solar systems change over a variety of timescales from hours to decades (Sections~\ref{sssec:mars_post}, \ref{sssec:titan}, and~\ref{sssec:solargiant}). 

Other cases explore the habitability of ocean worlds both within the solar system (Section~\ref{sssec:oceans_habitable}) and orbiting other stars (Section~\ref{sssec:id_oceans}). Additional cases focus on the longevity and detectability of surface oceans of liquid water (Sections~\ref{sssec:surface_water} and~\ref{sssec:survive_water}). Another common theme is assessing the influence of stellar and planetary magnetic fields on planetary atmospheres (Sections~\ref{sssec:bfield}, \ref{sssec:young_bfields}, and~\ref{sssec:aurorae}). Finally, SSiC cases consider the detectability of rings and moons orbiting other planets (Sections~\ref{sssec:exorings} and~\ref{sssec:exomoons}, respectively). 

\subsubsection{Atmospheres of Earth Analogs as a Function of Age (SSiC-11)}
\label{sssec:ozone_onset}
\sleads{Sarah Blunt, Eric Nielsen, Elisabeth Newton, Jessie Christiansen, Tansu Daylan, Courtney Dressing, Caleb K. Harada, Stephen R. Kane, Malena Rice, Romy Rodr\'iguez, Sabina Sagynbayeva}

This science case was developed during the START era and published in \citet{blunt_et_al2025}. The document was not reformatted for inclusion in the HWO25 Conference Proceedings, but it was posted in SCDD format on the STScI SCDD Portal.\footnote{\url{https://docs.google.com/document/d/1U0KbLAlAIImkqVbagdkM_m60AfbwTBB3OPAqMVdzTYM}} In this blurb, we adopt the definitions of breakthrough and enabling science given in the version posted on the SCDD Portal. 

The aim of this science case is to measure the time needed for Earth-like planets to develop oxygen-rich atmospheres. By measuring the amount of ozone in the atmospheres of Earth-like planets spanning a range of well-determined ages, HWO would constrain the timing and universality of the rise of oxygen. For additional discussion of the scientific motivation and approach, see \citet{blunt_et_al2025}. 

\breakthrough Breakthrough progress would require determining the onset of ozone to a precision of 1 Gyr. Achieving this goal necessitates detecting 25 Earth analogs orbiting stars with ages known to $20\%$ precision. In addition, the targets should be observed with sensitivity high enough to detect atmospheric ozone or rule out ozone levels comparable to that of modern Earth. As discussed in Section~\ref{sssec:life}, determining the abundance of ozone requires UV spectroscopy at $200-350$~nm with $SNR > 10$ at 500~nm and $R \sim 7$. 

\enabling For enabling progress, the age precision could be relaxed to 40\% and the planet sample could be reduced to 10~Earth analogs. As discussed in detail by \citet{blunt_et_al2025}, this sample size is still large enough that a non-detection of ozone would place meaningful constraints on the timescale of ozone emergence. 
 \subsubsection{Identifying Rocky Planets and Water Worlds Among Sub-Neptune-sized Exoplanets with the Habitable Worlds Observatory
 (SCDD-SSiC-3)}
\label{sssec:rocky_sub}
\sleads{Renyu Hu, Michiel Min, Max Millar-Blanchaer, Jacob Lustig-Yaeger, Tyler Robinson, Jennifer Burt, Athena Coustenis, Mario Damiano, Chuanfei Dong, Courtney Dressing, Luca Fossati, Stephen Kane, Soumil Kelkar, Tim Lichtenberg, Jean-Baptiste Ruffio, Dibyendu Sur, Armen Tokadjian, Martin Turbet}

As explained by \citet{hu_rocky_hwo25}, the purpose of this science case is to measure the frequency of various types of sub-Neptune sized planets within and beyond the habitable zone (HZ). Specifically, the program aims to determine the abundance of terrestrial planets, water worlds, and volatile-rich mini-Neptunes. Spectroscopic observations with a facility like HWO could determine the presence and composition of the atmospheres of small planets. 

Although H$_2$ lacks significant absorption features at the wavelengths likely to be covered by HWO, analysis of chemical signatures (e.g., CH$_4$/CO$_2$) or the reflectance spectrum slope can reveal whether a planet has an atmosphere dominated by H$_2$ \citep[e.g.,][]{hall_et_al2023, yang+hu2024}. Classification by chemical signatures would require NIR spectra (1.0 - 1.8 $\mu$m at $R > 40$ and $SNR > 10$) to determine the mixing ratios of CH$_4$ and CO$_2$ to a precision of at least 1.0 dex \citep{damiano+hu2022}. Classification by reflectance spectrum slope would require visible spectra with $SNR > 20$ \citep{hall_et_al2023}. Fully resolving a degeneracy between O$_2$ and N$_2$-dominated atmospheres requires $SNR > 40$ and distinguishing between atmospheres dominated by N$_2$ and CO requires observing the CO absorption feature at $1.6 \mu$m \citep{hall_et_al2023}.

Planets with atmospheres dominated by H$_2$ would be classified as mini-Neptunes if they have massive atmospheres and water worlds if they have less massive atmospheres. Planets without atmospheres dominated by H$_2$ would be further analyzed to determine whether their spectra reveal signs of volcanic outgassing (e.g., H$_2$S, SO$_2$), indications of gases that could have been dissolved in liquid water oceans (e.g., NH$_3$, HCN, SO$_2$) or surface features (e.g., ocean glint, see Section~\ref{sssec:surface_water}). Depending on those results, they will be classified as either rocky planets or water worlds. 

Fully understanding the properties of the observed planets will require determining their radii and masses. Radii could be found indirectly by using phase-resolved observations to determine the planetary albedo. Planetary masses could be determined by ground-based radial velocity observations or astrometric observations made with HWO or other facilities. 

\breakthrough At the ``breakthrough'' level, this science case would include comprehensive atmospheric characterization of roughly 100 planets within the HZs of FGK stars. The planets should have masses of $3-20 M_\oplus$ determined to a precision $>10\sigma$. The program would aim to detect and measure mixing ratios of H$_2$O, CH$_4$, CO$_2$, CO, NH$_3$, HCN, H$_2$S, and SO$_2$ to precisions $\leq 1$~dex. In addition, breakthrough progress would include assessing the dominant constituent of the atmospheres of both H$_2$-dominated and non-H$_2$-dominated planets 

Achieving this level of atmospheric characterization would require UV/O/IR reflectance spectroscopy and broadband polarimetry. The reflectance spectroscopy should be obtained at $R> 7$ and $SNR>10$ at $250-400$~nm; $R>140$ and $SNR>40$ at $400-1000$~nm; and $R> 70$ and $SNR>10$ at $1000-1800$~nm. The polarimetry should be acquired in multiple bands and multiple epochs ($5-10$ per planet) with $\pm1\%$ precision at each epoch. 

\enabling 
For enabling science, the sample size would be reduced to 50~planets. The planets should still have masses between $3M_\oplus$ and $20M_\oplus$, but the required mass precision would be decreased to $>3\sigma$. The wavelength range of the reflectance spectra would be narrowed to $400-1800$~nm. Additionally, the required SNR would be decreased to $SNR>20$ for $400-1000$~nm and the required spectral resolution would be reduced to $R>40$ for $1000-1800$~nm. 
The polarimetric capabilities would be similarly relaxed to less precise ($\pm 3\%$ per epoch) single-band polarimetry obtained at $3-5$ epochs per planet.

\subsubsection{Retention of Volatiles on Rocky Planets (SCDD-SSiC-21)} %SSSCDD
\label{sssec:retention}
\sleads{Ludmila Carone}

This science case was proposed during the START process and posted on the STScI SCDD Portal\footnote{\url{https://docs.google.com/document/d/1bPCSTe1qzhCvCo-VbBv6FpkBeNfKhxUB/}}. For more information about the technical requirements, note that elements of this concept were eventually included in other science cases including those presented in Sections~\ref{sssec:rocky_sub}, \ref{sssec:escape}, \ref{sssec:transits}.

The goal of this science case is to determine the conditions needed for Earth-mass terrestrial planets to retain significant volatiles. For this case, HWO should assess whether atmospheres exist on terrestrial planets spanning a wide range of irradiation levels orbiting stars with a range of spectral types (FGKM) and ages. HWO would differentiate between bare-rock worlds and planets with significant atmospheres by measuring the spectrally resolved geometric albedo ($A_g$) between 200~nm and 1 $\micron$. In this wavelength range, albedos below 0.2 would likely indicate that the planet does not have an atmosphere. Precisely measuring $A_g$ requires knowledge of planet radius, and therefore this science case is well-suited for application to transiting planets detected by \emph{TESS}, \emph{PLATO}, and other surveys. For planets without known radii, measurements of spectral polarization at UV and optical wavelengths could reduce the degeneracy between albedo and planet radius by constraining surface composition (for bare-rock planets), cloud coverage and composition, photochemical hazes, and surface oceans. 

\breakthrough This science case requires UV and visible spectroscopy of Earth-mass planets. At least one system younger than 1~Gyr and one system older than 1~Gyr should be observed for each combination of host star spectral type (M, K, G, or F) and planet irradiance level (1, 2, 10, 100, or 1000~$S_\odot$) for a total sample of $\geq 40$ Earth-mass planets. The data should cover $200 - 1000$~nm with spectral resolution $R\sim 100$. Ideally, the instrument would have spectropolarimetric abilities to detect features such as glint and extended red coverage ($>1500$~nm) to better constrain albedo. For this document, we adopt a red wavelength cut-off of 2000~nm to address the breakthrough request for additional NIR data. A large instantaneous field-of-regard would be beneficial for scheduling observations at particular times such as during transits for transiting planets or at angles where rainbows and halos could be visible for directly imaged planets.

\enabling The requirements for enabling progress are largely the same as for breakthrough progress, but polarimetric observations and NIR observations would not be required. Additionally, the instantaneous field-of-regard could be smaller. 

\subsubsection{Exoplanet Atmospheric Escape Observations with the Habitable Worlds Observatory (SCDD-SSiC-14)}
\label{sssec:escape}
\sleads{Leonardo A. Dos Santos, Eric D. Lopez, Luca Fossati, Antonio Garc\'{i}a Mu\~{n}oz, Shingo Kameda, Munazza K. Alam, Keighley Rockcliffe, Seth Redfield, Yuichi Ito, Joshua Lothringer, Shreyas Vissapragada, Hannah R. Wakeford, Apurva V. Oza, Girish M. Duvvuri, Raissa Estrela, Ryoya Sakata, Chuanfei Dong, and Ziyu Huang
}

Continuing the theme of Section~\ref{sssec:retention}, the science case presented by \citet{dossantos_hwo25} also considers the ability of small planets to retain their atmospheres. A facility like HWO could contribute to this investigation by probing the composition, ionization state, and extent of the exosphere around potentially Earth-like transiting exoplanets and by measuring the rate of atmospheric escape for planets spanning a range of masses and ages. If Earth analogs have exospheres similar to that of the Earth, their exospheres would consist primarily of neutral H and extend dozens of planetary radii from the surface \citep[e.g.][]{wallace_et_al1970, kameda_et_al2017, baliukin_et_al2019}. 

The hydrogen in Earth's exosphere is resupplied by photodissociation of water molecules lower in the Earth's atmosphere, so the detection of significant quantities of hydrogen in the exospheres of Earth analogs could be an indication that those planets have evaporating oceans \citep[e.g.][]{jura2004} and are undergoing rapid H escape similar to that experienced by Venus \citep{kasting+pollack1983}. In the UV, HWO could detect atmospheric escape by measuring Lyman-$\alpha$ (Ly$\alpha$) at 121~nm, C II at 133~nm, and a variety of other spectral features \citep{linssen+oklopcic2023}. Although interstellar absorption can significantly erode the Ly$\alpha$ signal for stars that are not in the immediate solar neighborhood and those that have low radial velocities, the nearby M dwarfs such as TRAPPIST-1 are attractive targets for Ly$\alpha$ observations of atmospheric mass loss \citep{dos_santos_et_al2019}. 

In addition to observing many ($\leq 20$) transits of M dwarf planets to build up the signal needed to detect exospheres similar to that of modern Earth, HWO could also construct a more general picture of the effects of atmospheric mass loss by obtaining high-resolution UV transmission spectra of roughly 50 transiting planets at a variety of temperatures, ages, and masses. Observing these bright stars will require large dynamic range and more tolerance of high count rates than HST/STIS and HST/COS.

\breakthrough For breakthrough science, HWO observations should be sensitive to exospheres with densities as low as $10^5 \, \rm cm^{-3}$, temperatures as cool as 1000~K, and radii as small as $40\, R_\oplus$ and able to measure atmospheric escape rates as small as $10^3\, \rm g/s$. The corresponding observational needs are high-resolution UV spectra ($R>45,000$, $100-300$~nm) and detectors with high dynamic range ($\geq 10,000$~cts/s/pixel). Additionally, the instrument should have an effective area $>130,000$~cm$^2$ in the UV. The high spectral resolution ($R\sim45,000$) would support additional science such as probing ISM properties (e.g., Section~\ref{sssec:dust_uv}), measuring planet-star alignment using the Rossiter-Mclaughlin effect \citep[e.g.,][]{cegla_et_al2016}, and identifying effects caused by stellar activity. 

\enabling For enabling science, the sensitivity requirements would be relaxed to exosphere radii $\gtrsim 50 \, R_\oplus$, exospheric densities $\gtrsim 10^7 \, \rm cm^{-3}$, exobase temperatures $\geq 3000$~K, and atmospheric escape rates $\geq 10^5\, \rm g/s$. The necessary spectroscopic resolution and wavelength range would be the same as for the breakthrough case, but the detector dynamic range and effective area would be reduced to $\geq 1000$~cts/s/pixel and $>73,000$~cm$^{2}$, respectively.

\subsubsection{Detecting and Characterizing Venus Analogs Orbiting Other Stars with the Habitable Worlds Observatory
 (SCDD-SSiC-18)}
\label{sssec:exovenus}
\sleads{Stephen R. Kane, Emma L. Miles, Colby M. Ostberg, Noam R. Izenberg, Sarah Blunt, Jessie Christiansen, Tansu Daylan, Courtney Dressing, Luca Fossati, Antonio Garcia Munoz, Kenneth Goodis Gordon, Caleb Harada, Renyu Hu, Ludmila Carone, Eric Nielsen, Elisabeth Newton, Malena Rice, Sabina Sagynbayeva, Christopher Stark, Peter Woitke}

As described by \citet{kane_venus_hwo25}, the goal of this science case is to better understand planetary habitability and the prevalence of runaway greenhouse states by detecting and characterizing potential Venus analogs orbiting other stars \citep{kane2014,kane2019}. Detecting such atmospheric states will provide critical insights into the evolution of habitable conditions through leveraging the solar system in-situ data to construct and diagnose terrestrial exoplanet atmospheric models \citep{kane2021,garvin2022,kane2024}. Exo-Venus candidates have previously been detected primarily via transit and radial velocity surveys, and additional candidates will be uncovered through either further precursor observations or high-contrast imaging with HWO \citep{ostberg2019,ostberg2023a}.

Directly detecting a sample of Venus analogs with HWO would require sensitivity to Earth-sized planets orbiting $0.3 - 1.5$~AU from solar-type stars. Studying Venus analogs around solar-type stars would provide the first opportunity to directly compare Venus to planets in analogous stellar environments. HWO should also be used to conduct transit and eclipse spectroscopy of Venus analogs since the majority of them orbit smaller, cooler stars. With the proposed bandpass range of HWO, it will be able to clarify our understanding of Venus analogs observed by JWST. Several Venus analogs which have been observed by JWST appear to have suffered from significant atmospheric erosion from their host star \citep[e.g. ][]{greene2023,lincowski2023,zieba2023}. Previous works have demonstrated that hazy, Venus-like atmospheres can cause planets to appear to have no atmospheres at all \citep{lustig-yaeger2019}. HWO may be able to remove this ambiguity by identifying the presence of hazes through the detection of SO$_2$ at UV wavelengths ($200-400$ nm), or by identifying scattering caused by clouds and hazes at visible wavelengths. UV observations will also reveal whether these planets may have atmospheres enriched in O$_2$ due to evaporation of H$_2$O oceans \citep[e.g.][]{kasting+pollack1983,lincowski2018,johnstone2020}. Spectropolarimetry spanning optical and near-infrared wavelengths would be useful for distinguishing between Earth analogs and Venus analogs \citep[e.g.,][]{ehrenreich2012,barstow2016,west_et_al2022,ostberg2023b}. Contemporaneous observations at visible and near-infrared wavelengths could also detect absorption, or the lack of absorption, by O$_2$ and CO$_2$ molecules that will enable further constraints of atmospheric composition and climate.

\breakthrough Breakthrough progress would require broad, contemporaneous multi-wavelength spectroscopy from 200~nm in the UV to beyond $1.5~\mu$m in the NIR. Future work is needed to determine the required spectral resolution, but it should be high enough to permit identification of molecular bandheads and polarization features. At visible and NIR wavelengths, polarimetric capabilities are requested to investigate the microphysics, composition, and possible time variability of aerosols. The ability to schedule observations at specific orbital phases is required to detect phase-dependent signals such as polarization rainbows and investigate seasonal variability. Given this timing requirement, a large instantaneous field of regard is needed to maximize the number of observation windows. The spatial resolution should be high enough to detect planets $0.3-1.5$~AU from FGK~host stars. 

\enabling For enabling science, the requirements are largely the same as for breakthrough science, but the wavelength coverage is restricted to the UV and optical ($200 \, \rm nm - 1 \, \mu m$) and polarimetry is not required. Additionally, the data need not be contemporaneous and the ability to obtain observations at particular phases would be relaxed, thereby reducing the need for a large instantaneous field of regard. 

\subsubsection{Identifying Cold Ocean Planets and Characterizing Their Geologic Activity (SCDD-SSiC-17)}
\label{sssec:id_oceans}
\sleads{Lynnae Quick, Jessica L. Noviello, Apurva V. Oza}

As discussed by \citet{quick_hwo25}, this science case would expand studies of habitability in extrasolar planetary systems by investigating the presence of subsurface oceans on ice-covered planets that could be large-scale analogs of the icy moons in our own solar system \citep[e.g.,][]{quick_et_al2020, quick_et_al2023}. The two objectives of this science case are to first determine the frequency of icy planets that could potentially contain internal oceans and then to scrutinize those planets for geyser-like activity that would serve as evidence of these subsurface oceans. The ability for evaporative transmission spectra \citep{gebek+oza2020} to probe subsurface material driven by geological activity has been demonstrated for sodium and potassium for rocky worlds \citep{oza_et_al2019} which HWO would be able to considerably advance towards cold ocean worlds. Within the wavelength range likely to be probed by HWO, the clearest indication of interior oceans would be the presence of variable absorption from water vapor or its derivatives (e.g., H at $0.12\mu$, O at $0.13\mu$m and O at $0.136\mu$m during geyser-like eruptions in UV transmission spectra. 

At longer wavelengths of 12 - 15$\mu$m, complementary secondary eclipse and phase curve measurements \citep{meyer_zu_westram_et_al2024} would be useful for constraining the dayside surface temperatures of these planets. Significant deviations from blackbody spectra and sinusoidal phase curves, alongside strong spectral signatures showing liquid or gaseous water, could indicate thermal emission from geysers. This wavelength range is likely much redder than the red cut-off of HWO, but the data could be obtained with other upcoming facilities such as LIFE \citep{quanz_et_al2022}. 

\breakthrough Breakthrough progress would require observing at least 30 exoplanets that are potential Europa analogs. These approximately Earth-sized worlds should be far enough from their host stars to have icy surfaces. For planets orbiting M or K stars, this requirement translates to distances of $0.02 - 5$~AU while planets orbiting hotter sun-like stars should be at separations of $5-30$~AU. A large instantaneous field of regard would be beneficial to maximize observing windows. Additionally, rapid response capabilities are needed to quickly observe targets if transient geysering activity is reported by other facilities. High-resolution spectroscopy is further advantageous in that the velocity profile of a geologic event can, in principle, be separated from the surface as is the case for extrasolar satellites that may be volcanically active \citep{oza_et_al2024, unni_et_al2025}.

Transiting super-Europas should be observed during transit at UV wavelengths (spectroscopy at $50-210$~nm) and during secondary eclipse at MIR wavelengths ($12 \mu$m and $15 \mu$m). The UV observations should have spectral resolution high enough to permit detection of water vapor or the products produced by the dissociation of water (i.e., H at 120~nm; O at 130~nm and 136~nm). Obtaining secondary eclipse observations at both $12 \mu$m and $15 \mu$m would constrain the dayside surface temperature and reduce potential confusion from CO$_2$, which has a prominent absorption feature at $15 \mu$m. At least five secondary eclipses should be observed at $15\mu$m to distinguish between permanent and transient hot spots suggestive of geysers. Repeated MIR observations could also constrain the libration amplitude of candidate super-Europas, providing valuable insight into which planets have subsurface oceans. These observations would advance the state-of-the-art of what the James Webb Space Telescope is able to currently probe for water vapor and cryovolcanic volatiles \citep{kleisioti_et_al2026, oza_et_al2026}. As previously observed in the solar system, icy worlds with interior oceans should have larger libration amplitudes because their liquid layers decouple their icy shells from their cores \citep[e.g.,][]{van_hoolst_et_al2008, thomas_et_al2016}. 
For non-transiting planets, high-contrast reflectance spectra would be required. Additional studies are needed to determine the required resolution, sensitivity, inner working angle, and outer working angle. 

\enabling For enabling progress, the target sample could be reduced from 30~planets to 20~planets. In addition, the wavelength range of the UV observations could be reduced to $100 - 140$~nm. At MIR wavelengths, $15\mu$m secondary eclipse observations without accompanying $12\mu$m observations would suffice as long as at least five eclipses were observed for each transiting planet. A large instantaneous field of regard would still be beneficial for scheduling observations, but rapid response capabilities would not be required. 

\subsubsection{Assessing Ocean World Habitability with HWO (SCDD-SSiC-1)}
\label{sssec:oceans_habitable}
\sleads{R. J. Cartwright, L. C. Quick, M. Neveu, T. M. Becker, J. C. Castillo-Rogez, K. France, S. Kameda, A. Roberge, U. Raut, K. L. Craft, M. W. McElwain, G. Sweetak, B. N. Sitarski, L. C. Mayorga, J. Lustig-Yaeger, G. L. Villanueva}

As discussed in \citet{cartwright_hwo25}, this science case explores the habitability of the interiors of icy bodies in the solar system. While the interior regions cannot be directly probed by HWO, spatially and spectrally resolved spectroscopy with HWO would provide valuable insights into habitability by detecting and characterizing salts, plumes, and atmospheric activity. Salts and plumes provide a window into the hidden interiors of these worlds, allowing HWO to investigate the presence of liquid water, energy sources, and bioessential elements. Lessons learned from these observations of nearby objects will also be relevant for understanding the possible habitability of much less accessible icy worlds orbiting other stars. Observations with HWO would complement and extend the baseline of in situ observations with Europa Clipper, JUICE, and possible future missions such as a Ceres lander and sample return mission, Enceladus plume sampling missions, Enceladus Orbilander, and Uranus Orbiter and Probe. 

Coverage from the EUV to the NIR is necessary for fully understanding conditions on icy worlds during the HWO era. Simultaneous observations would be ideal, but contemporaneous observations obtained within a 3-hr interval would be sufficient. The brightness of the bodies will likely necessitate saturation mitigation procedures and high dynamic range. EUV coverage at 50 nm - 150 nm is needed to detect irradiated carbonaceous residues on the surfaces of ocean worlds. Observations at slightly longer wavelengths ($150 - 200$ nm) would cover a diagnostic absorption feature of H$_2$O ice, thereby providing an opportunity to investigate the porosity and contamination of surface ice. In the near-infrared, HWO observations could detect H$_2$O vapor ($2.6 - 2.7 \mu m$) and CO$_2$ ($4.2 - 4.3 \mu m$) at wavelengths that are extremely challenging to observe from the ground due to telluric absorption. 

\breakthrough Breakthrough progress would require $50 - 5000$~nm, $SNR \gtrsim 100$ IFS spectroscopy with spectral resolution $R\sim10,000$ and pixel scale $\leq 20$~mas/spaxel over a $3'' \times 3''$ FOV. Cryocooling the IR detector to $\sim 30$~K would likely be needed to reach the desired SNR at longer wavelengths. Due to the brightness of the proposed targets and the need to observe faint plumes near bright disks, attention should be paid to saturation mitigation and dynamic range. 

\enabling For enabling science, the spectral range would be reduced to $90 - 3000$~nm with a possible gap in coverage between UV observations at $90-300$~nm and VIS/NIR observations at $400-3000$~nm. The spectral resolution requirements would be relaxed to $R\sim3000$ at $90-300$~nm and $R\sim 5000$ at $400-3000$~nm. A larger pixel scale of 30~mas/spaxel would be sufficient. Bright source mitigation strategies would still be needed but partial flux loss from bright targets would be tolerated. 

\subsubsection{Detecting Surface Liquid Water on Exoplanets (SCDD-SSiC-4)}
\label{sssec:surface_water}
\sleads{Jacob Lustig-Yaeger, Nicolas B. Cowan, Renyu Hu, L. C. Mayorga, Tyler D. Robinson}

As discussed in \citet{lustig-yaeger_liquidwater_hwo25}, the goal of this science case is to determine the prevalence of oceans on terrestrial planets orbiting within the HZ. The two avenues to ocean detection considered in this science case are spatially mapping the planetary surface and observing ocean glint. Assuming that the planet has heterogeneous surface coverage with longitudinal variations in albedo, the spin of the planet will cause various features to rotate in and out of view. While the planet will be unresolved in a single HWO image, observations obtained at multiple times (and therefore multiple rotational and orbital geometries) can be combined to construct maps of longitudinal variations in planetary albedo. The smoothness of liquid will cause significant brightness variations depending on whether the oceans or lakes are viewed at gibbous phases when they appear dark \citep[e.g.][]{cowan_et_al2009, cowan_et_al2011} or at crescent phases when they exhibit high reflectance and linear polarization \citep[e.g.,][]{robinson_et_al2010, lustig-yaeger_et_al2018, vaughan_et_al2023, roccetti_et_al2025c}. Detecting oceans via surface mapping will require photometric precision of $1\%$ over timescales of hours. Although the rotation periods of potential targets are unknown, observations with HWO could be used to both measure the rotation period and map the surface \citep[e.g.,][]{ford_et_al2001, palle_et_al2008, oakley+cash2009}.

While detecting oceans using rotational variability requires that the fraction of the surface covered by liquid water varies longitudinally, specular reflection (also known as ``glint'') can reveal oceans even on longitudinally symmetric worlds. Detecting oceans via glint requires obtaining phase-resolved reflectance measurements at crescent phase to spot the extreme increase in albedo at glancing angles. Glint observations are technically demanding because they require observing the planet at small angular separations and monitoring how the albedo varies over month-long timescales. For the baseline, the planet could either be observed continuously for months or HWO could conduct multiple visits of the planet over a series of months. The former strategy of continuous observations may impose demands on field-of-regard, scheduling, onboard data storage, data transmission, and the ease of parallel observations with other instruments. The latter strategy of repeated visits separated by observations of other targets puts less pressure on the observatory schedule, data storage, and data transmission at the expense of stricter limits on absolute instrument stability. Depending on the specific pattern of observations used for repeat visits, the second strategy may also reduce the necessary field of regard. 

Clouds will complicate assessments of liquid surface coverage because they can both shield surface water from detection and masquerade as specular reflection, but spectropolarimetry can reduce modeling degeneracies \citep{trees+stam2019, ryan+robinson2022, roccetti_et_al2025b}. In addition to reducing the degeneracy between clouds and oceans, conducting polarimetric and spectroscopic observations rather than photometric observations would increase confidence in ocean detections by allowing ocean glint to be detected using a differential measurement of polarization fraction or color rather than an absolute measurement of brightness. Polarimetric capabilities (ideally with the ability to measure Q and U simultaneously) would therefore significantly reduce the need to achieve 10\% absolute photometric stability on month-long timescales. Additionally, the wavelength dependence of the polarization signal across water vapor absorption bands can provide a further, independent diagnostic for distinguishing ocean glint from a dry surface \citep[e.g.,][]{trees+stam2022}.

\breakthrough Breakthrough studies of ocean glint would require observing 20 planets when they appear as small crescents (phase angle = $160^\circ \pm 5^\circ$). The observations should be precise enough to distinguish between the observed brightness and that expected from a Lambertian model to $\pm 0.01$ in the scattering asymmetry factor \citep[e.g.,][]{bruna_et_al2023}.

Breakthrough progress in ocean mapping would require observing and characterizing roughly a dozen Earth-like planets. The rotational variability and albedo inhomogeneity of these planets should be measured to $8\sigma$ precision and their rotational periods and broadband colors should be constrained to 10\%. 

For both ocean glint and ocean mapping, the science case demands observing planets at $400 - 1100$~nm and achieving a contrast ratio of 10$^{-11}$. For ocean glint, the data should be polarimetric with a photometric precision of $0.01 (R_p/a)^2$. The planets should be observed at ten~different epochs covering $50^\circ$ of the orbit from gibbous phase to small crescent phase, which requires an inner working angle $\leq 20$~mas \citep{vaughan_et_al2023}. The spectra should have SNR $\geq 20$ on the continuum. 

For ocean mapping, the observations should have an uncertainty $< 0.01 (R_p/a)^2$in a few hours of integration. This precision must be achieved on relatively short timescales to longitudinally resolve features on target planets. Each observation should last roughly a week and be obtained with 1-hr cadence. 

\enabling 
Enabling studies of ocean glint would require observing a smaller sample of 10~Earth-like planets when they are slightly more illuminated (phase angle = $150^\circ \pm 10^\circ$). The observations should have the precision needed to constrain scatter asymmetry factor to $\pm 0.05$, which is a factor of five reduction in precision compared to breakthrough progress. 

Enabling studies of ocean mapping would require an even smaller sample of only a few planets. The observations should be sufficient to measure rotational variability and albedo inhomogeneity to $5\sigma$. The rotation periods should still be measured to 10\%, but 20\% precision would be acceptable for measurements of broadband surface colors. 

For both investigations, enabling progress would require achieving a slightly less extreme contrast ratio of $10^{-10}$ rather than $10^{-11}$. Ocean glint should be observed over a slightly reduced wavelength range ($400 - 1000$~nm) and at fewer epochs spanning a narrower region of the orbit (6~epochs covering $40^\circ$). The corresponding inner working angle requirement would be relaxed to $\leq 50$~mas and SNR $\geq 15$ would be sufficient. 

For ocean mapping, enabling progress would require a reduced photometric precision of $0.03 (R_p/a)^2$ over a more condensed wavelength range ($500 - 1100$~nm). The observations should sample the orbit between between gibbous phase and quadrature. Each observation should include measurements taken at a 3-hour cadence over a 72-hour baseline.

\subsubsection{Birthcries and Swansongs of Habitable Worlds (SCDD-SSiC-20)}
\label{sssec:survive_water}
\sleads{Ludmila Carone}

This science case was explored during the START era and posted on the STScI SCDD Portal\footnote{\url{https://docs.google.com/document/d/1BN5MKDxp669K8DVgRn-thp8CZDFB_JedosTAxbUlUAA/}}. Rather, the core ideas of this science case were incorporated into other science cases such as those presented in Sections~\ref{sssec:rocky_sub}, \ref{sssec:transits}, and \ref{sssec:exovenus}. 

This science case investigates planetary habitability as a function of time by searching for water in the atmospheres of terrestrial planets in or near the HZs ($0.25 S_\oplus \leq S_p \leq 2 S_\oplus$) of young ($< 1$~Gyr) and old ($>4$~Gyr) systems. The primary objective is to search for planets with steam-dominated atmospheres by looking for water emission at 0.8$\mu$m and 1$\mu$m. A very young planet (age $\ll$ 1 Gyr) with a steam-dominated atmosphere might be in the process of solidifying a magma ocean into crust and building a secondary atmosphere while an older planet with a steam-based atmosphere could be experiencing a Venus-like runaway greenhouse. 

For select targets with steam-dominated atmospheres, the program would also obtain spectra between $0.3\mu$m and $1.2\mu$m to constrain the frequency of biologically relevant elements \citep[e.g., PS at $0.3-0.6\mu$m, K at $0.7\mu$m, MgO at $0.4-0.6\mu$m, SiO at $0.15-0.7\mu$m, TiO at $0.3-2\mu$m;][]{janssen_et_al2023, zilinskas_et_al2023} and probe the oxidation state of the planetary surface \citep[e.g.,][]{baumeister_et_al2023}. For planets with outgassing observations, higher spectral resolution ($R\sim1000$) is preferred. Spectroscopy at $1500 - 1700$~nm could probe HCN and potentially constrain N abundance. 

\breakthrough Breakthrough progress would require $R\sim 1000$ high-contrast spectroscopy at contrast ratios $\le 10^{-10}$ at $200 - 2000$~nm to detect possible features from H$_2$O and surface outgassing. We selected the red cut-off of 2000~nm to fully cover the $300-2000$~nm wavelength range listed for TiO in the capability table of the SCDD. If extension to $2~\mu$m is not possible, $R\sim100$ spectra over $1.5 - 1.7~\mu$m would be needed to detect HCN from surface outgassing. Observations of the host star at $100 - 200$~nm are also needed to understand the radiation environment of the planet, but those XUV data could be acquired by other facilities with smaller apertures. 

\enabling For enabling science, $R\sim100$ spectroscopy at $300 - 1200$~nm would be sufficient for detecting H$_2$O. Additionally, the $100-200$~nm observations of the host stars requested in the breakthrough case could be obtained by other facilities.

\subsubsection{Probing the Origin of Water in Planets within Habitable Zones with HWO (SCDD-SSiC-2)}
\label{sssec:hab_system}
\sleads{Yasuhiro Hasegawa, Courtney Dressing, Ludmila Carone}

The science case described by \citet{hasegawa_water_hwo25} directly addresses questions E-Q1 and E-Q3 from the Astro2020 Decadal Survey by probing how habitable environments emerge and evolve. The specific objective is to investigate the processes by which HZ planets receive water and assess possible correlations between the water abundance of inner planets and the presence of outer planets. This case requires detecting and measuring the water abundance for Earth-like planets in systems with and without giant planets. In some cases, the giant planets might be detected directly with HWO, but in most cases those planets will be found with other facilities. Rather than leading to a separate set of observational needs, this science case illustrates the complementarity of HWO and ground-based facilities for addressing the important question of volatile delivery in planetary systems. This case also illustrates the power of combining astrophysical observations of exoplanetary systems with observations of our own solar system (see Section~\ref{sssec:solar}).

\breakthrough Breakthrough science would require measurements of water abundance on 30 planets smaller than $1.75 R_\oplus$ orbiting within the HZs of Sun-like stars as well as detections or limits on giant planets ($>3.5 R_\oplus$) orbiting the same stars with semimajor axes of $\leq 30$~AU. Roughly eight epochs of imaging are expected to be required to constrain orbital properties such as semimajor axis, eccentricity, and inclination. These observations will require achieving a contrast ratio of $\sim 4 \times 10^{-11}$. For the HZ planets, water abundance would be constrained from low-resolution ($R > 40$) spectra at $700 - 1500$~nm. The giant planets are well-suited to detection via astrometric observations (by HWO or other facilities) or radial velocity observations (by other facilities). Planet masses for HZ planets and giant planets should be determined via ground-based radial velocity observations, astrometric observations with HWO (see Section~\ref{sssec:mp}), or a combination of those techniques. 

\enabling The requirements for enabling science are not included in \citet{hasegawa_water_hwo25}, but internal versions of this science case approached enabling science by decreasing the target sample substantially to 4~potentially Earth-like planets with unknown water abundance and reducing the spectral resolution to $R>17$. Because the water abundance would not be constrained, the goal would shift from studying correlations between the properties of outer planets and water abundance of temperate terrestrial planets to assessing possible correlations between the properties of outer planets and presence of temperate terrestrial planets in the same system. 

\subsubsection{Using the Habitable Worlds Observatory to survey the orbital architecture of small exoplanets (SCDD-SSiC-13)}
\label{sssec:occ_small}
\sleads{Tansu Daylan, Romy Rodriguez Martinez}

The aim of this science case is to understand the prevalence of small planets at a range of orbital separations and investigate possible correlations between the presence of an Earth-like planet and overall system architecture. This science case was explored during the START era and posted on the STScI SCDD Portal\footnote{\url{https://docs.google.com/document/d/1_xnjMUvh7jiHyqJsxBq37utkc8yVbmXR2tYU6LYAIpY/}}. We mention SSiC-13 here for reference and historical context, but the science questions and observations initially proposed for SSiC-13 were ultimately captured in the related science cases SSiC-2 \citep[][Section~\ref{sssec:hab_system}]{hasegawa_water_hwo25} and SSiC-12 \citep[][Section~\ref{sssec:giant_orbits}]{sagynbayeva_et_al2025}. 

\subsubsection{The Role of Stellar Multiplicity in the Prevalence of Small, Cool Planets (SCDD-SSiC-6)}
\label{sssec:occ_binary}
\sleads{Elisabeth Newton, Malena Rice, Kendall Sullivan, Jessie Christiansen, Sarah Blunt, Sabina Sagynbayeva}

Although our own solar system contains only one star, many stars are part of multistar systems. The science case presented by \citet{newton_hwo25} explores whether temperate small planets are preferentially located in systems with only a single star. Using high-contrast imaging, HWO could search for planets in the HZs of the primary stars in multistar systems. For detected planets, follow-up observations would be needed to determine planetary orbits and estimate planetary radii. A sample of 100 targets would be sufficient for detecting strong suppression or enhancement of the occurrence rate of super-Earths and mini-Neptunes in binary star systems rather than single star systems. Stellar binaries in which the components are separated by less than 100~AU are the most interesting for probing the effects of stellar multiplicity on habitability, but the stars in these systems will have small angular separations. Depending on the size of the desired stellar sample and the separations of the selected binaries, this science case will require advanced starlight suppression technologies to block the light from both the primary star and physically bound stellar companions. 

As discussed in Section~\ref{sssec:life} and other science cases focused on potentially habitable planets, detecting temperate terrestrial planets will require direct imaging or spectroscopy capable of achieving a contrast of $10^{-10}$ at small angular separations. For this science case, the added challenge is that the presence of a nearby stellar companion may complicate starlight suppression and wavelength control. 

Due to the distribution of stellar binaries in the Milky Way, surveying a sufficient number of binary systems will require observing systems at greater distances from Earth, which decreases the angular separation corresponding to a fixed physical separation. The ability to observe binaries separated by $\leq10''$ is particularly important for accessing physical separations $\leq 100$~AU, which is where suppression of planet formation is expected to be most pronounced. For context, the ExEP Mission Star List \citep{mamajek+stapelfeldt2024} contains 36 stars with companions within 100~AU, but only 12 of those systems have angular separations $>5''$. More work is needed to identify additional nearby binaries with separations $<100$~AU that could be suitable targets for HWO and to mature starlight suppression strategies that can be employed to search for planets in binaries with separations $\geq 3''$. For examples of ongoing work, see \citet{sirbu_hwo25_vol2} and \citet{yoneta_hwo25_vol2_binary}.

\breakthrough The internal version of this science case noted that breakthrough progress would require observing 200~FGK primary stars in binary systems with semimajor axes sampling between 30~AU and 1000~AU. The observations should be sensitive to planets with radii $1 - 1.5 R_\oplus$ within 2~AU. 

\enabling The internal version of this science case indicated that 100~FGK primary stars should be observed at the sensitivity necessary to detect planets with radii $1.3 - 2.5 R_\oplus$ within 5~AU. For the enabling science case, binary separations should be well-sampled at $100-1000$~AU and the sample should also include 15~systems with separations $<100$~AU. 

\subsubsection{Detecting and Characterizing the Magnetic Field of Exoplanets (SCDD-SSiC-26)}
\label{sssec:bfield}
\sleads{A. Strugarek, S. V. Berdyugina, V. Bourrier, J. A. Caballero, J. J. Chebly, R. Fares, A. Fludra, L. Fossati, A. Garc\'{i}a Mu\~{n}oz, L. Gkouvelis, C. Gourv\`{e}s, J. L. Grenfell, R. D. Kavanagh, K. G. Kislyakova, L. Lamy, A. F. Lanza, C. Moutou, D. Nandy, C. Neiner, A. Oklop\v{c}i\'{c}, A. Paul, V. R\'{e}ville, D. Rodgers-Lee, E. L. Shkolnik, J. D. Turner, A. A. Vidotto, F. Yang, P. Zarka}

As discussed by \citet{strugarek_et_al2025}, understanding planetary magnetic fields is essential for exploring the evolution and habitability of exoplanets. As of July 2025, exoplanetary magnetic fields have been detected only tentatively \citep[e.g.,][]{turner_et_al2021b, strugarek+shkolnik2025} and their occurrence rate and statistical properties are highly uncertain. Considering the planets of the solar system, a natural conclusion is that exoplanetary magnetospheres should be commonplace, but recent observations of brown dwarfs have demonstrated that current dynamo scaling laws may be overly simplistic \citep[e.g.,][]{kao_et_al2016, kavanagh_et_al2024}. Accordingly, while it is reasonable to assume a high occurrence rate of exoplanetary magnetic fields, the field strengths are harder to predict than their overall prevalence. 

With HWO, it would be possible to investigate both the frequency and the properties of exoplanetary magnetospheres via direct and indirect detection. Both methods would require a UV spectropolarimeter like the Pollux instrument considered for the LUVOIR and HWO concepts \citep{bouret_et_al2018, muslimov_et_al2018, muslimov_et_al2021, muslimov_et_al2024, le_gal_et_al2019, girardot_et_al2024}. Direct detection refers to observing the light that has been affected by the planetary magnetic field. In contrast, indirect detection means that the presence of a planetary field could be inferred by observing variations in the properties of the host star caused by interactions of the planetary and stellar magnetospheres \citep{cuntz_et_al2000}. 

For direct detection, a promising avenue is measuring the polarization of the He~I~1083~nm absorption triplet \citep{oklopcic_et_al2020}. These NIR lines are produced when neutral helium atoms transition from the $2^3\rm S_1$ metastable excited state to any of the three $2^3\rm P_{2,1,0}$ excited states \citep[e.g.,][and references therein]{oklopcic+hirata2018}. For the triplet to exist at all, the $2^3\rm S_1$ state must be sufficiently populated, which can be true for short-period gas giants orbiting Sun-like stars, particularly K stars \citep{oklopcic2019}. \citet{spake_et_al2018} detected He~I~1083~nm absorption in HST transmission spectra of the warm gas giant WASP-107b, and the signal has subsequently been observed for multiple close-in planets \citep[e.g.,][]{allert_et_al2018, nortmann_et_al2018, salz_et_al2018, alonso-floriano_et_al2019, zhang_et_al2022, zhang_et_al2023}. 

As the starlight travels through the planetary atmosphere, the absorption of linearly and circularly polarized light is affected by the Hanle effect and the Zeeman effect, respectively. For transiting planets with magnetic fields $\gtrsim 10$~G, spectropolarimetric observations of the He~I~1083~nm triplet should be sensitive to the amplitude and direction of the magnetic field \citep{oklopcic_et_al2020}. Polarization induced by the planetary magnetic field would be measured by comparing spectropolarimetric measurements obtained while the planet was in transit to data acquired when the planet was not in front of the host star. The dataset should contain roughly the same amount of in-transit and out-of-transit data, which corresponds to total observation times of approximately 10~hours for close-in planets with typical transit durations of roughly five hours. 

Subtracting the in-transit and out-of-transit data would remove any static polarization signal due to the intrinsic polarization of starlight \citep[e.g.,][]{berdyugina_et_al2011}. The polarization measured within the He~I~1083~nm triplet would also be compared to the polarization signal measured in the continuum to remove any polarization induced by continuum scattering \citep[e.g.,][]{carciofi+magalhaes2005, berdyugina_et_al2008}. While He~I~1083~nm triplet transmission polarimetry is unlikely to be significantly affected by stellar activity \citep{cauley_et_al2018a}, simultaneous UV or optical spectropolarimetry covering key stellar activity features would be advantageous for ensuring that the detected signal is indeed due to the planetary magnetosphere. 

While direct detection of exoplanetary magnetic fields will likely be restricted to close-in gas giants, indirect detection could be used to infer the presence of magnetic fields on low-mass planets as well as giant giants as long as the planet is sufficiently close to the star to trigger star-planet interactions \citep[i.e., the planet should orbit within the Alfv\'{e}n radius of the star, which is $10-20 R_\odot$ for the Sun;][]{cranmer_et_al2023}. Star-planet interactions would induce hot spots on the stellar surface, which could be detected via UV spectropolarimetry. Properly disentangling planetary-induced signals from intrinsic stellar variability requires also monitoring activity indicators such as H$\alpha$ and Ca II H\&K \citep{cauley_et_al2018b}. Ideally, simultaneous UV/O/IR data should be obtained on a variety of timescales (hours - years) to sufficiently sample the planetary orbital period, stellar rotation period, and stellar activity cycle. 

\breakthrough Breakthrough progress via direct detection of planetary magnetospheres requires UV/O/IR spectropolarimetry of at least 10 transiting planets within 150~pc. The data should cover $120 - 2440$~nm at a resolving power $R > 100,000$ and all four Stokes parameters (I, Q, U, V) should be observed simultaneously. The observations must be sensitive to polarization levels as low as 0.001\% (i.e., field strengths of 1~G) and have a wavelength stability $\leq 1$m/s during the transit observation. 

For indirect detection, breakthrough progress would require spectropolarimetric observations of at least five planets within 100~pc. The spectra should extend from 92~nm\footnote{Note that Table~5 of \citet{strugarek_hwo25} accidentally lists this lower limit as 920~nm rather than 92~nm. We use the correct value of 92~nm in this paper.} to 1100~nm and have resolution $R>100,000$. All four Stokes parameters should be measured simultaneously at cadences as frequent as hourly. 

\enabling For substantial progress via direct detection, the observational capabilities are the same as for breakthrough progress except that the wavelength coverage would be decreased to $270 - 2440$~nm and the resolving power would be set to $R\sim 100,000$ rather than $> 100,000$. In addition, the sample would be reduced to five transiting planets within 100~pc.

For enabling progress via indirect detection, the sample would be restricted to only five planets within 75~pc. The wavelength range would be significantly narrowed to $123-400$~nm and the spectral resolution would be decreased to $R\sim 50,000$. Additionally, the cadence of observations would be reduced to one observation every three hours.

 \subsubsection{Investigating the Role of Magnetic Fields in Shaping Young Planetary Systems with the Habitable Worlds Observatory and the POLLUX Spectropolarimeter (SCDD-SSiC-34)}
 \label{sssec:young_bfields}
 \sleads{Ana I. G\'{o}mez de Castro, Evelyne Alecian, Silvia H. P. Alencar, M. Audard, Jean-Claude Bouret, Ada Canet, Catherine Espaillat, Kevin France, Eleanora Fiorellino, Gregory Herczeg, Ignacio Mendigut\'{i}a, Christian Scheneider, Aurora Sicilia-Aguilar, Juan C Vallejo, Bonnie Zaire}
 
As described by \citet{gomez_de_castro_magnetic_hwo25}, the goal of this multidisciplinary science case is to investigate the role of magnetic fields in shaping the properties of protoplanetary disks, young stars, and young planets. Spectropolarimetric observations of young planetary systems have the potential to map magnetic field structures, thereby revealing stellar wind interactions and providing constraints on the interplanetary medium, planet accretion, and planetary mass loss. The observations will also provide insight into angular momentum transport in disks as well as the formation and impact of bipolar outflows from young stars. Accordingly, as discussed in Section~\ref{sec:astro2020}, this science case is relevant to Astro2020 science questions raised by the Panel on Exoplanets, Astrobiology, and the Solar System; the Panel on Interstellar medium and star and planet formation; and the Panel on Stars, the Sun, and Stellar Populations. 

Untangling the role of magnetic fields in star and planet formation requires obtaining multiple observations simultaneously in order to break degeneracies and expose the complicated underlying physics. Specifically, the strength and topology of the stellar magnetic field must be determined at the same time as the location and thermal structure of accretion shocks; the speed, density, and temperature of mass ejections; the mass-loss rate and speed of photo-evaporative outflows; and the distribution and properties of dust in the disk. Thus far, it has not been possible to obtain that valuable dataset because of the rarity of UV spectropolarimeters (i.e., HST/FOS, Astro-1/WUPPE, Astro-2/HUT) and the very small set of pre-main-sequence stars observed by those instruments. Observations of a much larger set of T~Tauri stars are needed to characterize the role of magnetic fields, and extending the sample to more massive Herbig stars is important for assessing possible changes in the mode and rate of accretion as a function of stellar mass and magnetic field strength. 

\breakthrough Breakthrough progress in this area requires mapping the photospheric magnetic fields of 200~young stars as well as characterizing the thermal structure of 200~accretion shocks. For 25\% of the accretion shocks, magnetic field measurements are also needed. Additionally, observations of the magnetospheric structure; studies of the connection between accretion and ejection; and measurements of winds and episodic ejections should be completed for 50~targets. Finally, the distribution of dust should be mapped in the inner regions of ten disks. 

Obtaining these measurements will require spectropolarimetric observations of roughly 60 T Tauri stars and 15 Herbig stars. The T Tauri sample will consist of 20 stars in the nearby TW Hydrae association (25-75~pc from Earth) and 40~stars in southern, ALMA-accessible associations such as Chamaelon. Each star should be observed at multiple times spanning 1.5 - 2 times the stellar rotation period (anticipated to be 2 - 9 days with the younger stars having longer rotation periods). Gaps in coverage are fine as long as the dataset includes measurements at enough rotational phases to detect stellar flares, accretion events, and other variations in stellar brightness. Observing each target 20~times for 3,600 seconds each should be sufficient. Each observation should span $100-1500$~nm at spectral resolution $R>30,000$ and achieve SNR$=10$ and $0.5\%$ polarization precision in Stokes IQUV for targets as faint as $10^{-14}$~erg~s$^{-1}$~cm$^{-2}$. Precise pointing and PSF stability are needed to ensure that each observation covers the same area of the circumstellar disk. 

\enabling In general, substantial progress requires only 50\% of the measurements needed for breakthrough progress. Photospheric magnetic field maps and the thermal structure of accretion shocks should be determined for 100~systems. Substantial progress also requires mapping the magnetic field of accretion shocks; determining the structure of the magnetosphere; probing the connection between accretion and ejection; and mapping winds and episodic ejections for 20~systems. Maps of the distribution of dust in disks are not needed.

\subsubsection{Observed Orbital Architectures of Giant Planets (SCDD-SSiC-12)}
\label{sssec:giant_orbits}
\sleads{Sabina Sagynbayeva, Asif Abbas, Stephen R. Kane, Eric L. Nielsen, William Thompson, Sarah Blunt, Malena Rice, Jessie L. Christiansen, Caleb K. Harada, Elisabeth R. Newton, Yasuhiro Hasegawa, Philip J. Armitage, Tansu Daylan}

This science case document was listed in the STScI SCDD Portal\footnote{\url{https://docs.google.com/document/d/1BsVhxQ5rempJoWAlRENdma_wGpL8tkrnzJ97QtPD9WY/edit?tab=t.0}} and led a published paper investigating the combined power of radial velocity, astrometric, and direct imaging observations to detect terrestrial habitable zone planets and more distant giant planets \citep{sagynbayeva_et_al2025}. Here, we describe the internal science case document, which aimed to understand the dynamical processes that generate the orbital architectures of giant planets. Precise measurements of the eccentricities (ideally within $\pm 0.001$), inclinations (ideally within $\pm 0.5^\circ$), and semimajor axes of giant planet orbits will test and improve models of planet formation. The dataset needed for this science case is a combination of precursor radial velocity observations from ground-based facilities \citep[e.g.,][]{howard+fulton2016, laliotis_et_al2023, harada_et_al2024a}, astrometric measurements from HWO, and high-contrast images with HWO that directly detect giant planets beyond the snow line. The goal of imaging distant, cold planets places demands on both the sensitivity (ideally $\leq 30.5$ mag in the NIR) and outer working angle (ideally 640~mas) of HWO's high contrast imager. The notional target sample includes roughly 50 stars for which RV data suggests a high likelihood of hosting a giant planet.

\breakthrough Breakthrough science requires observations of $30 - 40$ planets with masses $\leq 30 M_J$ at separations $5 - 30$~AU. The inclinations and eccentricities of their orbits should be measured to $\pm 0.5^\circ$ and $\pm0.001$, respectively. These measurements require sensitivity to planets with NIR magnitudes $\leq 30.5$ between an inner working angle of $\sim 63$~mas and an outer working angle of $\sim1440$~mas. The necessary contrast is $10^{-9} - 10^{-10}$ at separations of $0.1'' - 1''$, and the expected wavelength range for observation is $500 - 5000$~nm. 

\enabling For enabling science, HWO should observe a smaller number of less massive planets over a narrower range of separations: $25 - 30$~planets with masses $\leq 20 M_J$ at separations $5 - 30$~AU. The inclinations of the planets should be measured to $\pm2^\circ$ and their eccentricities should be measured to $\pm0.005$. Accordingly, the observations must be sensitive planets with NIR magnitudes $\leq 30$ between an inner working angle of $\sim 90$~mas and an outer working angle of $\sim270$~mas.

\subsubsection{Direct Imaging Characterization of Cool Gaseous Planets (SCDD-SSiC-15)}
\label{sssec:reflected_giants}
\sleads{Michiel Min, Jo Barstow, L. C. Mayorga, Hannah Wakeford, Jason Wang, Renyu Hu, Beth Biller, Jos\'{e} A. Caballero, Ludmila Carone, Sarah Casewell, Katy L. Chubb, Mario Damiano, Siddharth Gandhi, Antonio Garc\'{i}a Mu\~{n}oz, Christiane Helling, Finnegan Keller, Nataliea Lowson, Evert Nasedkin, Ryan MacDonald, Jean-Baptiste Ruffio, Evgenya Shkolnik, Christopher C. Stark}

As described by \citet{min_hwo25}, the goal of this science case is to characterize gas giant planets and investigate the physical processes responsible for the observed diversity of gas giant planets. The science case is particularly focused on how the composition and evolution of planetary atmospheres depends on formation location, star-planet interactions, and overall system architecture. By conducting a detailed investigation of a large sample of giant planets, this science case also aims to understand the connection, similarities, and differences between giant planets and brown dwarfs. Exploring these questions requires measurements of the composition, thermal structure, and aerosol properties for a large and diverse sample of cool and warm planets.

Phase-resolved spectroscopy over a broad wavelength range will be crucial for understanding the complexity and time variability of planetary atmospheres. While low-resolution spectra will suffice for coarse molecular analysis, higher spectral resolution ($R \gtrsim 1000$) could improve compositional constraints, increase sensitivity to trace molecules, enable measurements of planetary spin, and possibly permit the use of Doppler imaging to map surface features. Planetary mapping necessitates obtaining multiple high SNR spectra or images per rotation period, which could potentially push instrument throughput and onboard data storage. Polarimetric capabilities would allow for more sophisticated analysis of clouds and hazes, which are a key component of planetary atmospheres. 

\breakthrough Breakthrough science requires observing over 60~giant planets orbiting stars with a range of spectral types (FGKM). The planets should span a broad range of effective temperatures from as cold as 30K to as warm as 600K. Imaging and spectroscopy should be obtained over a broad wavelength range ($150 - 5000$~nm), and spectral polarimetry should be acquired at $200 - 1800$~nm. Additional work is needed to determine the necessary spectral resolution, but \citet{min_hwo25} point out that $R>1000$ has been useful for past studies of known planets. Accordingly, we adopt $R>1000$ for breakthrough-level observations in the visible and NIR. At UV wavelengths, we use the lower spectral resolution of $R=100$. For spectropolarimetry, we assume the standard spectral resolutions of $R=7$ in the UV, $R=140$ in the visible, and $R=70$ in the NIR set by LW-1 \citep[][Section~\ref{sssec:life}]{arney_hwo25} and other Living Worlds science cases. Ideally, planets should be observable at separations between $\sim 60$~mas and $1''$. The inner working angle is driven by the desire to detect scattering phenomena like rainbows that would better constrain atmospheric properties \citep[e.g.,][]{vaughan_et_al2024}, while the outer working angle is set to that needed to detect a planet with a semimajor axis of 10~AU at quadrature when orbiting a star 10~pc away from Earth. 

\enabling Enabling science requires observations of 30~giant planets orbiting FGKM stars. The planets should cover the same broad effective temperature range as for breakthrough science (30K - 600K). Compared to breakthrough observations, enabling observations would have narrower wavelength coverage ($200 - 1800$~nm) and could possibly have a lower spectral resolution ($R\sim100$). Imaging and spectroscopy should be supplemented by spectral polarimetry in some molecular bands. The inner working angle requirement would be relaxed to roughly 100~mas.

\subsubsection{Characterizing the Dynamics and Chemistry of Transiting Exoplanets with the Habitable World Observatory (SCDD-SSiC-16)} 

\label{sssec:transits}
\sleads{Hannah R. Wakeford, L. C. Mayorga, Joanna K. Barstow, Natasha E. Batalha, Ludmila Carone, Sarah L. Casewell, Theodora Karalidi, Tiffany Kataria, Erin M. May, Michiel Min}

Like the science case described in Section \ref{sssec:reflected_giants}, the science case written by \citet{wakeford_hwo25} explores the physics responsible for the observed diversity of planets. The primary scientific objectives are to measure planetary rotation rates and determine the structure, composition, circulation, and aerosol properties of planetary atmospheres. For this analysis, HWO would obtain spectroscopic phase curves for planets with orbital periods of 5-20 days and ideally even longer. Extending phase curve studies out to longer orbital periods will enable detailed investigation of atmospheric structure, composition, and circulation for planets that are much cooler than the more highly irradiated planets accessible with JWST phase curve observations (i.e., $T_{\rm eq}<500$K for HWO versus $1400 {\rm K} \lesssim T_{\rm eq} \lesssim 2600 {\rm K}$ for JWST).

\begin{figure*}
 \centering
 \includegraphics[width=1\linewidth]{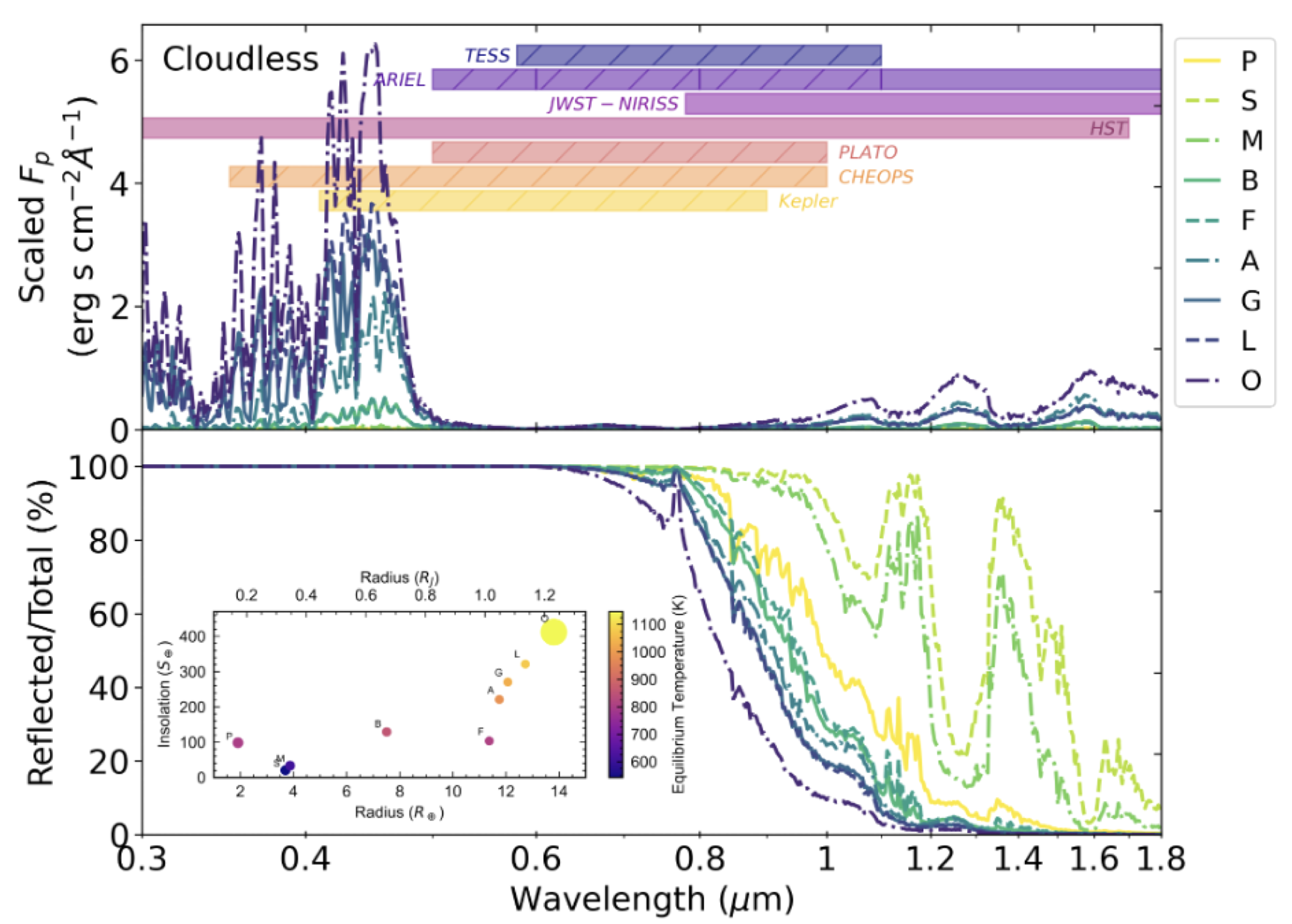}
 \caption{Predicted $R=300$ spectra of archetypal planets categorized according to the schema introduced by \citet{mayorga_et_al2019}; the insolation and radius of each planet is shown in the lower inset plot. \emph{Top:} Scaled planetary flux versus wavelength for archetypal planets (lines) versus the wavelength coverage of various facilities (shaded regions). \emph{Bottom:} Ratio of reflected flux to total flux versus wavelength for the same planets. Figure courtesy of L. C. Mayorga.}
 \label{fig:transits}
\end{figure*}

As highlighted in Figure~\ref{fig:transits}, broad wavelength coverage extending from the UV to the NIR (or even to $5\mu$m) would capture both reflected light and thermal emission, enabling HWO to conduct comprehensive characterization of planetary atmospheres. UV observations would probe high altitudes, thereby providing valuable insights into atmospheric (dis)equilibrium, aerosol properties, and the effects of photochemical processes on atmospheric composition. UV data could also reveal absorption by species such as PH$_3$ and CS. At redder wavelengths, NIR spectra from HWO would provide additional information about aerosols, atmospheric equilibrium, and heat redistribution as well as abundances for key metallicity indicators such as H$_2$O, HCN, CH$_4$, and NH$_3$ \citep{moses_et_al2011}. For aerosol studies, the combination of phase-resolved spectroscopy and polarimetric phase curves will help break degeneracies in aerosol structure, composition, and spatial distribution \citep[e.g.,][]{stam_et_al2004,karalidi_et_al2013, chubb_et_al2024}.

\breakthrough Breakthrough progress would require phase curve observations of planets with orbital periods $>20$~days and temperatures $<500$~K. There are over 1000~possible targets, but the target sample will need to be thoughtfully selected to best understand connections between observed planet properties and the underlying physics. Ideally, targets should be observable anywhere on the sky throughout the year. 

Due to the long orbital periods of the targets, the need to capture a full orbit translates into observation durations $\geq 500$~hours. Throughout the orbit, simultaneous and continuous medium-resolution spectroscopy ($R>6000$) should be obtained from 100~nm to $\geq 5000$~nm with additional polarimetry between 300~nm and $1~\mu$m. The observations should be exquisitely precise (brightness variations measured to 1~ppb), which requires both absolute pointing stability and extremely stable detectors. The need to obtain this extreme precision on targets spanning a broad magnitude range ($V=2-20$) provides motivation for single-photon-counting detectors, which have the potential to obtain efficient, high signal-to-noise observations over a broad range of flux levels \citep[e.g.,][]{crill2022, alexani_et_al2024, bryan_et_al2025}. 

\enabling For substantial progress, phase curves should be obtained for targets with slightly shorter orbital periods of $10 - 20$~days, which increases the target temperature range ($<700$~K versus $<500$~K for breakthrough science) and reduces the duration of phase curve observations to $\leq 500$~hrs per observation. The anticipated range of target magnitudes would be condensed to $V=5-18$. 

Each target would require $R\sim3000-5000$ spectra from 100~nm to 2500~nm with a precision of 100~ppb, which would require pointing stability $\leq 0.001$~pixel and active control of thermal variations in the detectors. Polarimetry is needed at 500~nm - $1~\mu$m. As for breakthrough progress, a single-photon-counting detector would be needed at UV wavelengths, but a standard IR detector would be acceptable for longer wavelengths.

\subsubsection{High-resolution Ultraviolet-to-Near-Infrared Characterization of Exoplanet Atmospheres with HWO (SCDD-SSiC-24)}
\label{sssec:highres_atmos}
\sleads{Patricio E. Cubillos, Matteo Brogi, Antonio Garc\'{i}a Mu\~{n}oz, Luca Fossati, Jasmina Blecic, Sudeshna Boro Saikia, Vincent Bourrier, Jose A. Caballero, Juan Cabrera, Andrea Chiavassa, Andrzej Fludra,  Leonardos Gkouvelis, John Lee Grenfell, Manuel Guedel,Gopal Hazra, Alvaro Labiano, Monika Lendl, Donna Rodgers-Lee, Arnaud Salvador, Ilane Schroetter, Antoine Strugarek, Benjamin Taysum, Aline Vidotto, Thomas G. Wilson 
}

\citet{cubillos_hwo25} describe the power of high-resolution spectroscopy to constrain the composition and escape of planetary atmospheres. Characterization of planetary atmospheres will advance knowledge of planet formation and help explain demographic features such as the ``Neptune desert,'' a deficit of intermediate-sized planets at short orbital periods \citep[e.g.,][]{mazeh_et_al2016}, and the ``radius valley'' marking the shortage of small planets with radii of roughly $1.8 R_\oplus$ \citep[e.g.,][]{fulton_et_al2017}. For larger planets, high-resolution spectra have the potential to break existing model degeneracies and advance the state of the art \citep[e.g.,][]{kempton+knutson2024} by precisely determining chemical abundances and clarifying the nature of aerosols.

For measuring the rate of atmospheric escape and assessing the composition of escaping gases, UV observations are essential because of their sensitivity to the low-density regions high in planetary atmospheres where escape occurs \citep{fossati_et_al2015, dos_santos_et_al2023}. FUV observations are particularly important for detecting Ly$\alpha$ absorption from escaping neutral hydrogen as well as possible C~II and O~I lines from escaping carbon and oxygen, respectively. In the NUV, observations could reveal lines from ions metals such as Mg~II, Fe~II, and Si~II, thereby refining models of star-planet interactions \citep{haswell_et_al2012} as well as the thermal structure of the atmosphere \citep[e.g.,][]{fossati_et_al2010, sing_et_al2019, sreejith_et_al2023}. At longer wavelengths, NIR observations could quantify atmospheric escape via observations of the metastable He~I triplet \citep[e.g.,][]{seager+sasselov2000, oklopcic+hirata2018}. Although studies of atmospheric escape via He~I line are usually done from ground-based observatories, space-based observations have the advantage of avoiding telluric contamination that can overpower planetary signals and could include simultaneous monitoring of UV lines to better understand how the stellar He~I lines may be varying due to potential stellar activity \citep{guilluy_et_al2020}. 

Additionally, UV spectropolarimetry has the potential to measure exoplanetary magnetic fields as discussed in Section~\ref{sssec:bfield}. Detecting multiple escaping species would provide the opportunity to investigate the impact of ongoing escape on the composition of the remaining atmosphere. Specifically, these observations could reveal which planets have secondary atmospheres generated via outgassing or impacts rather than primary atmospheres accreted from the protoplanetary disk. 

Regardless of whether a planet is experiencing ongoing atmospheric escape, high-resolution cross-correlation spectroscopy (HRCCS) can be used to detect weak signals of various molecules lower in the planetary atmosphere. As explained by \citet{snellen_et_al2010}, this method works by cross-correlating the observed spectrum with a template spectrum containing the many ro-vibrational transitions of different molecules. By comparing how the cross-correlation function changes during the orbit of the planet, astronomers can determine which gases are present in the planetary atmosphere even when the spectrum contains stellar features as well as planetary features. Although HRCCS has been traditionally performed from the ground, ground-based observations are limited by imperfect correction of telluric contamination \citep[e.g.,][]{allart_et_al2017} and by the loss of the continuum during the calibration process \citep{brogi+line2019}. Accordingly, the existing set of HRCCS observations do not typically provide tight constraints on aerosol properties because the spectral slope and absolute flux are not measurable. In contrast, space-based observations from a facility like HWO would avoid telluric contamination, thereby potentially permitting alternative approaches to data analysis that retain the continuum information needed to characterize aerosols. As demonstrated by early simulations \citep{garcia_munoz2018}, even more information could be gleaned by equipping HRCCS instruments with polarimetric capabilities and therefore distinguishing planet-induced polarization from stellar- or ISM-induced polarization.

\breakthrough Breakthrough progress in atmospheric characterization would require measuring Ly$\alpha$ and metal escape rates from 100 transiting planets ranging from Earth-like to Jupiter-like as well as He~I escape measurements with simultaneous UV stellar activity monitoring of a larger sample of 200 Earth-to-Jupiter-like planets. At higher atmospheric pressures, breakthrough progress requires constraining the full inventory of volatile and refractory elements in planets ranging from Earth-like to Jupiter-like as well as detection of the precursors of condensates (e.g., soots, MnS, MgSiO$_3$) and spatial information about their distribution (e.g., twilight/dawn limbs, patchiness). Finally, the dataset should include 50 Earth-to-Jupiter-like planets for which both the upper atmosphere and lower atmosphere have been characterized simultaneously using atmospheric escape and HRCCS observations, respectively. Assembling this dataset will require simultaneous FUV-NIR spectropolarimetry from 100~nm to 2.5$\mu$m at spectral resolution $R>100,000$ and wavelength stability $<1$~m~s$^{-1}$. Full Stokes (I, Q, U, V) spectropolarimetry should be obtained with a precision of 0.001\%. 

\enabling Substantial progress requires measuring atmospheric escape for smaller samples of slightly larger planets: 50 super-Earths to Jupiter-like planets for Ly$-\alpha$ and metal escape and 100~super-Earths to Jupiter-like planets for He~I escape. As for breakthrough progress, the NIR He~I observations should be accompanied by UV monitoring of stellar activity, but those observations could be concurrent (i.e., before, during, and/or after transit) rather than simultaneous (i.e., extending from before transit to after transit). For lower-atmosphere characterization, substantial progress requires detections or robust non-detections of multiple volatiles and refractory elements as well as clear detections of condensate precursors. A set of planets with well-characterized upper- and lower-atmospheres is still required, but the sample has been reduced to 10 Earth-to-Jupiter-like planets with near-simultaneous characterization of upper and lower atmospheres rather than 50 Earth-to-Jupiter-like planets with simultaneous observations. This dataset will require simultaneous FUV-NIR spectropolarimetry over a narrower wavelength range (120~nm to 1.7$\mu$m) at spectral resolution $R=100,000$ and wavelength stability $1$~m~s$^{-1}$. Full Stokes (I, Q, U, V) spectropolarimetry\footnote{Tables 1 and 2 of \citet{cubillos_hwo25} provide inconsistent entries about the Stokes parameters needed for substantial progress. Table~1 lists I, Q, U, and V while Table~2 lists only I and Q. We have adopted the stricter requirement of I, Q, U, and V} is still required, but the precision requirement has been relaxed by an order of magnitude to 0.01\%. 

\subsubsection{Solar System Origins (SCDD-SSiC-7)}
\label{sssec:solar}
\sleads{Kathy E. Mandt, John Noonan, Darryl Seligman, Sarah Anderson, Olga Pinto Harrington, Richard Cartwright, Maria Womack}

The aim of the science case discussed in \citet{mandt_hwo25} is to determine how the formation and migration of giant planets affected the delivery of volatiles to the inner solar system and shaped the architecture of the solar system. HWO could contribute to this science case by measuring the abundances and isotopic composition of solar system small bodies, focusing on compounds containing hydrogen, carbon, nitrogen, oxygen, and sulfur. Targets of interest include comets, Centaurs, Transneptunian Objects (most notably Pluto), asteroids, potential interstellar interlopers, and moons of the giant planets. Measurements of the surface colors of small bodies would also contribute to improving our understanding of the formation of the solar system. For transitional objects such as active asteroids, obtaining molecular and isotopic measurements before and after perihelion passage would provide new insight into interior properties. Observations at UV wavelengths ($120 - 330$ nm) are needed to measure column densities of H, C, O, S, CS, C$_2$, OH, S$_2$, H$_2$O, CO, CO$_2$. At longer wavelengths, observations between 1.5 $\mu$m and $3.3 \mu$m would be useful for detecting H$_2$O in absorption and emission. If possible, observations at even longer wavelengths ($4.2 - 4.8 \mu$m) could reveal CO$_2$ and CO. Spectral resolution better than 0.1~nm is desired to resolve atomic transitions. For newly detected comets and interstellar objects, rapid response capabilities would be advantageous. High tracking rates (120"/hr - 1800"/hr) could be necessary for observing interstellar objects. 

\breakthrough Breakthrough progress would require spectra extending from the UV (120~nm) to the NIR (4800~nm). In the UV, the spectra should have resolution $R\gtrsim2000$. JWST-like spatial resolution of $0\farcs1$/spaxel would be sufficient. Breakthrough science also requires the ability to track quickly moving objects ($\geq 500$~mas/s) and rapidly acquire observations ($\sim$day, much slower than the response times needed for the supernovae observations discussed in Sections~\ref{sssec:r_process_el} and \ref{sssec:flash_ccsne}). 

\enabling For substantial progress, the UV coverage could be restricted to $120 - 210$~nm and $250-330$~nm at lower spectral resolution of $R>600$. Similarly, the NIR coverage could be reduced to $2500 - 4800$~nm, but that change does not decrease the difficulty of the requested observations. In addition, the maximum tracking speed would be relaxed to $\geq 150$~mas/s.

\subsubsection{Venus Direct Observation with the Habitable World Observatory (SCDD-SSiC-5)}
\label{sssec:venus}
\sleads{Noam R. Izenberg, Tyler Robinson, Courtney Dressing, Michael McElwain, Giada Arney, Stephen R. Kane, Colby Ostberg, Jacob Lustig-Yaeger, Richard Cartwright, Chloe Beddingfield}

The goal of the science case presented in \citet{izenberg_hwo25} is to advance our knowledge of Venus as a member of the broader category of hot terrestrial planets within and beyond the solar system. Multi-epoch, multi-wavelength observations with HWO would constrain the evolution of the composition and morphology of Venus's clouds. HWO observations could then be compared to in situ observations to explore the connection between atmospheric features and surface features. HWO observations should include some observations of the leading and trailing hemispheres of Venus obtained in the hours or days near quadrature as well as a series of observations obtained roughly hourly over the full 4.4-day super-rotational period. Ideally, HWO observations would be obtained simultaneously at UV, optical, and NIR, but contemporaneous observations obtained within 10 minutes should still limit the motion of observable cloud features to $\lesssim 1$~pixel. Given that the goal of this program is to observe Venus as if it were an exoplanet, modest spectral resolution of $R \sim 100$ or $R \sim 200$ would be sufficient. 

UV coverage between 250~nm and 400~nm is needed to capture the region of maximum cloud opacity. The spatial resolution must be high enough that Venus spans at least 1000~pixels. Ideally, the full disk of Venus should be observable at quadrature. At that time, Venus spans 30-40'' as viewed from Earth. This science case mandates slewing quickly enough to track Venus and having a field of regard that permits viewing Venus near quadrature.

\breakthrough Overall, breakthrough science requires simultaneous observations of Venus over a broad UV/O/IR wavelength range from 200~nm to beyond $1.5~\mu$m. The dataset should include observations of different orbital phases as well as hourly observations of gibbous Venus over a full planetary super-rotational period (4.4~days). These observations will require observing targets as bright as $V = -4.8$ over a field of view of at least $40'' \times 40''$ with a pixel scale $\leq 30$ mas/pixel and the ability to track targets at Sun-angles $\gtrsim 45^\circ$ moving as quickly as $1''$/min, which is much slower than the $\geq 500$~mas/s tracking speed needed for interstellar small bodies in Section~\ref{sssec:solar}. Modest spectral resolution of $R\sim 200$ is sufficient. 

\enabling The capabilities needed for substantial progress are largely similar to those needed for breakthrough progress, but observations at different wavelengths could be obtained contemporaneously (i.e., within 10s of minutes) rather than simultaneously. In addition, the spectral resolution could also be reduced to $R\sim100$ and the wavelength range could be narrowed to 200~nm - $1.5~\mu$m.

\subsubsection{The Origin of Mars (SCDD-SSiC-22)}
\label{sssec:mars}
\sleads{Ramses Ramirez, Humberto Campins, Ramon Brasser, Richard Cartwright}

This science case was included in the STScI SCDD Portal\footnote{\url{https://docs.google.com/document/d/17PxB2qvcCXn1wgRQrm8Mgt7AxycogH47AfelpW3_Iw4/edit?tab=t.0}} but not in the HWO25 Conference Proceedings. The goal of this science case is to determine how and where Mars formed. Specifically, the science case aims to address the sources of the planetary building blocks from which Mars was assembled by measuring the abundances and spectral features of Phobos, Deimos, and a variety of asteroids within and beyond the main belt. The observations will provide new insight into the composition, size distribution, mass distribution, and spatial distribution of asteroids with a range of compositions, thereby enabling an inventory of primordial sources of CO$_2$, H$_2$O, CH$_4$, and H$_2$ that could have been used to feed young Mars. The resulting dataset will also be useful for constraining the migration history of Mars and asteroids during the epoch of planet formation.

In the UV, observations with HWO at $0.35 - 0.5\mu$m would probe the presence of hydrated minerals and organics and provide insight into space weathering. Shorter wavelength observations at $0.1-0.35\mu$m would complement NUV observations by probing space weathering, organics, and hydrated minerals. At longer wavelengths ($1.9 - 2.85 \mu$m), HWO could detect carbonates, silicas, phyllosillicates (clays), and other hydrated materials. 

\breakthrough Breakthrough progress would require spatially-resolved UV/O/IR spectroscopy ($100 - 3000$~nm) at a resolution sufficient for measuring abundances of volatiles. Saturation mitigation strategies such as neutral density filters would be required to avoid saturating on bright asteroids. Rapid response capabilities ($\sim$day) would be needed to observe main belt comets when they are actively sublimating. A large instantaneous field of regard would be advantageous for maximizing target observability and therefore increasing the odds of observing transient phenomena. 

\enabling For enabling science, the number of targets and the instantaneous field of regard could be decreased. Additionally, the maximum observation wavelength could be reduced. The science case does not state a separate wavelength range for enabling science, but it does map the spectral coverage to the minerals that could be detected. In general, reducing the red wavelength cutoff will decrease robustness of the detection of hydrated minerals such as clays but those species may still be detectable by studying proxy features at UV and visible wavelengths ($100-500$~nm).

\subsubsection{Mars, a Post-Habitable Planet?
 (SCDD-SSiC-28)}
\label{sssec:mars_post}
\sleads{Matteo Crismani, Richard Cartwright, Sara Faggi, Stephanie Milam, Geronimo Villanueva, Michael Chaffin}

This science case was not included in the HWO25 Proceedings, but it was added to the STScI SCDD Portal\footnote{\url{https://docs.google.com/document/d/1KmoZ-ywYXpzbgQsL7BnHrqdsMtBYe_TKwot2quMU97g}}. While the science case discussed in Section~\ref{sssec:mars} focuses on the formation of Mars, this science case instead proposes to use Mars as a laboratory for investigating planetary habitability. Specifically, this science case aims to use Mars as a nearby analog to more distant exoplanets that may also have been habitable earlier in their histories before they lost their liquid surface water. By observing Mars with an observatory like HWO, this science case would collect data needed to improve models of planetary climate to more accurately account for loss of planetary atmospheres and surface liquid water. In parallel, such observations would also provide more detailed information about meteorological conditions on modern Mars such as circulation patterns, cloud formation, and dust storm formation. 

Although there are many spacecraft currently operating on the surface of Mars and in Mars orbit, a powerful remote observatory such as HWO could provide unique capabilities such as high spectral resolution and high signal-to-noise. Time-resolved observations of the full disk of Mars could revolutionize modeling of the Martian atmosphere by providing data that complements the zoomed-in observations obtained by in-situ resources. Additionally, by conducting long-term monitoring of the full planet, a facility like HWO could provide advance notice of dust storms that could be detrimental to future robotic and human exploration of the Martian surface, thereby potentially advancing priorities identified by NASA's Planetary Science Directorate and Space Operations Missions Directorate, respectively. 

\breakthrough Breakthrough progress in this area would require assessing how the atmosphere and corona of Mars vary on timescales ranging from days to years. For different scientific objectives, these observations should include both isolated snapshots and long-term monitoring at various cadences. UV/VIS observations over a broad wavelength range ($80-700$~nm) are needed to detect dust, clouds, and important gases such as O$_3$ and H$_2$O, measure isotopic ratios, and quantify the escape rate of H, C, and O. Additional NIR observations at wavelengths shorter than $1.5~\mu$m are required to further explore the hydrological cycle and climate of Mars by mapping the distribution of H$_2$O, HDO, and other important molecules. High spatial resolution ($>1000$ pixels across the $3\farcs5 - 25\farcs1$ diameter disk of Mars) is needed to precisely map coronal and wind properties. The spectral resolution should be high enough to resolve the necessary molecular and isotopic lines. The science case document does not quantify the required spectral resolution but notes that Mars should be observed at a spectral resolution at least as high as that used for exoplanet observations. A large field of view ($50''$) is needed to observe the full disk and corona of Mars in a single observation. Additionally, studying Mars will require the ability to observe bright targets ($m_V = -2.94$ at closest approach, $m_V = +1.86$ at farthest approach) over a wide region of the sky. Accordingly, bright target mitigation strategies, a large instantaneous field of regard, and less restrictive solar avoidance constraints would be advantageous. Rapid response capabilities would also be beneficial for enabling observations of transient events such as dust storms and coronal responses to solar flares. 

\enabling The requirements for enabling science are similar to those for breakthrough science but the UV/VIS wavelength range could be reduced to $100-400$~nm. NIR observations at wavelengths $\leq 1.5~\mu$m would still be required. Using mosaics could reduce the required spatial resolution and field of view.

\subsubsection{Exploring Titan's Dynamic Atmosphere with the Habitable Worlds Observatory (SCDD-SSiC-27)
}
\label{sssec:titan}
\sleads{Nicholas A Lombardo,
Tommi Koskinen, 
Conor A Nixon, 
Richard J Cartwright
}

This science case was included in the STScI repository of science cases\footnote{\url{https://docs.google.com/document/d/1bJwOROlvOCkGa740a_h2h8rx9qlWMgH2}} but not in the HWO25 Proceedings. The goal of this science case is to investigate how the chemical composition and dynamics of Titan's atmosphere vary seasonally and from one Titan year to the next. Due to the combination of Titan's orbit around Saturn and Saturn's orbit around the Sun, one season on Titan lasts roughly 7.5~Earth years and the full seasonal cycle requires approximately 29.5~Earth years. Accordingly, observations every few Earth months are needed to observe variations within a single Titan season while observations over multiple decades are needed to detect changes that occur over multiple seasons. More specifically, this science case would aim quantify the rate at which atomic hydrogen escapes from Titan's atmosphere, the size distribution of aerosol particles, and the abundance of trace molecules in the upper atmosphere. Additionally, this science case would explore how Titan's atmosphere varies with time at multiple altitudes ranging from near-surface methane clouds in Titan's troposphere to the aerosol distribution in the mid to upper atmosphere. 

This science case requires observations at multiple wavelengths from the UV to the NIR. At UV wavelengths, Lyman-$\alpha$ observations of Titan's orbital environment are needed to detect escaping hydrogen. Although the Hubble Space Telescope can access that wavelength range, observations by an L2-based observatory such as HWO have the potential to enable significant advances because the instrument may be outside of the hydrogen corona of the Earth. Additionally, the combination of UV and visible observations during stellar occultations are needed to build upon existing Cassini ISS/UVIS data and refine the vertical structure and composition of Titan's atmosphere. The science case document notes that additional work is needed to determine the frequency of useful stellar occultations. 

Furthermore, mapping observations at UV, visible, and NIR wavelengths are needed to determine the size distribution of aerosol particles. Determining the integration times needed for aerosol characterization will require forward modeling that accounts for wavelength-dependent limb darkening and scattering by aerosols. In the NIR, the mapping observations should be obtained with high spatial resolution and rapid cadence to permit the detection of near-surface clouds. Finally, detecting trace gases in Titan's upper atmosphere will require disk-averaged spectra obtained at high signal-to-noise ratios. 

\breakthrough Breakthrough progress in understanding Titan's atmosphere will require low spectral resolution UV observations at 338~nm and 187~nm to map the size distribution of aerosols throughout Titan's atmosphere and at Lyman-$\alpha$ (121~nm) to detect escaping hydrogen. The NIR observations should extend to $2\mu$m and have higher spectral resolution ($R>100$) to assess the composition of Titan's stratosphere and lower spectral resolution for investigating tropospheric clouds. For science objectives requiring multiwavelength observations (e.g., aerosol properties), contemporaneous observations (i.e., within a few weeks) are needed to reduce systematic errors. Mapping observations should be obtained at a spatial resolution high enough to adequately resolve different latitudinal bins on Titan's surface; given Titan's apparent diameter of roughly $0\farcs7$ and the goal of six samples per hemisphere, the spatial resolution should be $0\farcs05$. The science case does not require a large field of view due to the sub-arcsecond size of Titan's disk, so we adopt a desired field of view of $1''\times1''$ for this science case. Rapid response capabilities and a large instantaneous field of regard would be advantageous to increase the observability of Titan and enable follow-up observations with HWO of time-sensitive phenomena reported by ground-based telescopes. 

\enabling The requirements for enabling progress are largely the same as for breakthrough science. The differences are that a large field of regard and rapid response capabilities are not required. Additionally, enabling progress may impose a less stringent requirement on spectral resolution. 

\subsubsection{Exploring Solar System Giant Planets with Habitable Worlds Observatory (SCDD-SSiC-29)}
\label{sssec:solargiant}
\sleads{Leigh N. Fletcher, Amy Simon, Michael H. Wong, Jonathan D. Nichols, Nick A. Teanby, Conor A. Nixon, Marina Galand}

As described by \citet{fletcher_hwo25}, the goal of this science case is to enrich our understanding of solar system gas giants and their satellites. In particular, the science case emphasizes the need for multiwavelength observations over a variety of timescales to understand the dynamics and variability of these fascinating worlds. Even for a single science area such as aurorae, the desired timing for time-resolved observations ranges from as short as seconds for probing the morphology of aurorae to as long as years to understand the complex connections between auroral properties, the solar wind, and the solar activity cycle. High spatial resolution would be advantageous for mapping atmospheric features and a large field of view would be preferred for observing the full disk of Jupiter. Moreover, the facility should be capable of observing bright sources moving at non-sidereal rates and have a dynamic range wide enough to allow for simultaneous study of faint moons near substantially brighter planets. 

Spectroscopic observations are needed to determine the distribution and nature of aerosols as well as atmospheric circulation and chemistry. Time-resolved spectra could also reveal the effects of impacts like Shoemaker-Levy 9 and the physics of aurorae. Precisely timed observations of the limbs of giant planets could also probe the composition and conditions at very high atmospheric altitudes by studying how FUV light from bright background stars is attenuated during occultations by the upper atmosphere. Additionally, NUV observations of giant planets will yield tighter constraints on their bond albedos and perhaps reveal why the internal heat budgets of Uranus and Neptune are so different. 

Although the paper presented a variety of photometric and spectroscopic capabilities required for this science case, it did not explicitly divide them into breakthrough and enabling progress. For this paper, we have opted to classify requirements as ``breakthrough'' unless a performance range is given. In those cases, we label the softer requirements as ``enabling'' and the more stringent requirements as ``breakthrough.'' While most of the requirements are described relative to the Jovian system, the ability to conduct high-resolution, multi-wavelength monitoring of Uranus and Neptune over a long baseline will be truly transformative. 

\breakthrough Achieving breakthrough progress would require the ability to track objects moving at non-sidereal rates as well as particular surface or atmospheric features on rotating bodies. All of the giant planets should be observable over the full wavelength range without saturation, which may require employing neutral density filters, photon-counting detectors, and/or detectors that can be read out very quickly. For monitoring the temporal evolution of atmospheric and auroral features, the observatory should be able to obtain multiple images over $20-40$~hours. 

Imaging and spectroscopy alike should be possible over a broad wavelength range including the UV ($50-400$~nm), visible ($400-900$~nm), and infrared ($900-3500$~nm). The imager should be equipped with multiple filters including a set of visible filters designed to probe NH$_3$ at $600-700$~nm, H$_2$ near 825~nm, and CH$_4$ near 890~nm. In the infrared, observations at $2.7\mu$m could probe icy clouds deep in planetary atmospheres and provide new insight into atmospheric circulation while even longer wavelength observations at $3.5\mu$m have the potential to map auroral H$_3^+$ emission. For auroral studies, simultaneous observations of the NIR and FUV would be extremely powerful to capture both outgoing (NIR) and ingoing (FUV) energy. Ideally, the FUV imager should have a field of view of at least $20''\times20''$ to fully study auroral features. At visible wavelengths, an even larger field of view of $120''$ would be useful for observing rings, moons, and plasma features at wider separations from the planetary disk (50'' for Jupiter). 

In order to produce a significant advance over current capabilities, the IFS should be capable of both wide-field observations over $>10''$ at a resolution of $0\farcs2$/spaxel and narrow field observations at ten times higher resolution ($0\farcs02$/spaxel) over a smaller $3''\times3''$ field of view. Spectral resolution of $R>3000$ in the UV, $R>1740$ at 480~nm (benchmarked to MUSE), and $R>3450$ at 930~nm (again benchmarked to MUSE) would be required to compare favorably with current instrumentation. In their Section~2.2, \citet{fletcher_hwo25} note that higher spectral resolution in the UV would enable more advanced studies of auroral properties. Specifically $R\sim12,000$ UV spectra could probe the deposition of auroral energy into planetary atmospheres and $R\sim100,000$ UV spectra could resolve the Doppler shifts of auroral features caused by atmospheric winds moving at velocities on the order of 100~m~s$^{-1}$. For this paper, we have adopted the ambitious goal of $R\sim100,000$ spectra in the UV ($50-400$~nm) and the lower goals of $R>1740$ at $400-700$~nm and $R>3450$ at $700-3500$~nm; we reserve the lower UV spectral resolution of $R\sim12,000$ as the goal for enabling science. 

For atmospheric retrievals, it would be preferable to maximize the wavelength coverage that could be obtained in single observations. A large instantaneous field of regard is preferred for accessing giant planets at a wider range of orbital phases, ideally from quadrature to quadrature. Studies of transient events like impacts and storms would benefit from the ability to repoint the telescope within 7~days (and ideally sooner).

\enabling The requirements for enabling progress are the same as for breakthrough with a few exceptions. First, the duration of atmospheric monitoring campaigns could be shortened to $10-20$~hours rather than $20-40$~hours. Second, the infrared coverage could be reduced to $\le 3\mu$m rather than $3.5\mu$m. This bluer limit would still capture the $2.7\mu$m reflectivity peak useful for assessing clouds and ices but would not permit mapping auroral H$^+_3$ emission. Third, the fields of view of the imager and IFS could be reduced to 50'' and 3'', respectively. Fourth, the spectral resolution in the UV could be reduced to $R\sim12,000$.

\subsubsection{Study of the Atmospheric Effects of Energetic Particle Precipitations on Giant Planets With the Habitable World Observatory (SCDD-SSiC-25)}
\label{sssec:aurorae}
\sleads{J.-Y. Chaufray, W. Dunn, L. Fletcher, L. Fossati, M. Galand, L. Gkouvelis, C.M. Jackman, L. Lamy, L. Roth}

As described by \citet{chaufray_hwo25}, the goal of this science case is to investigate the interactions of extra-atmospheric, high-energy particles with the atmospheres of the giant planets within the solar system. These interactions are important for providing energy to planetary atmospheres and generate UV aurorae that should be detectable by large UV telescopes like HWO. This science case is relevant to Astro2020 Question E-Q2 about the properties and diversity of planets \citep{astro2020}.

A key aim of this study is to explore how extra-atmospheric particles affect the chemistry, dynamics, and thermal structure of planetary atmospheres. The results would provide valuable insight into the magnetospheres, ionospheres, and magnetospheric-atmospheric coupling of giant planets as well as solar wind properties and the role of moons in shaping auroral structure. For instance, the Galilean moons and Enceladus influence the aurorae of Jupiter and Saturn, respectively. Neptune and Uranus are particularly compelling targets for this study because knowledge of their magnetospheres is severely limited by their extremely faint UV aurorae: HST can barely detect the UV auroral emission of Uranus while that of Neptune is currently undetectable.

\breakthrough Achieving breakthrough progress would require spectropolarimetry to constrain the atmospheric composition. The observations should cover 100-1000~nm at a spectral resolution of roughly $0.002$~nm. The data should be sensitive to features as faint as $10^{-14}$~erg~cm$^{-2}$~s$^{-1}$~{\AA}$^{-1}$. 
Observations at 110-200~nm are needed to measure the width of H$_2$ rotational lines, thereby constraining the electron energy and the atmospheric temperature at pressures of $10^{-5} - 10^{-8}$~bar. In addition, spectra at $160-200$~nm will capture absorption features from hydrocarbons such as C$_2$H$_2$ (acetylene) and C$_2$H$_6$ (ethane). Although JUNO has observed Jupiter at this $160-200$~nm wavelength range, the spectral resolution of Juno is too low to clearly distinguish those features; a higher spectral resolution of $R>20,000$ could dramatically improve our understanding of the hydrocarbons present in the atmospheres of giant planets. In addition, spectral resolution $R>50,000$ would split the D and H Lyman-$\alpha$ lines, thereby providing a measurement of the D/H ratio. Finally, observations out to 1000~nm would provide access to absorption features from other species such as CH$_4$. 

\enabling Substantial progress would require less sensitive observations obtained at a lower spectral resolution over a narrower wavelength range. Specifically, the spectra should cover 100-500~nm with spectral resolution of roughly $0.005$~nm and be sensitive to signals as faint as $10^{-13}$~erg~cm$^{-2}$~s$^{-1}$~{\AA}$^{-1}$. While polarimetric observations would not be required, the spectra should clearly detect numerous absorption lines of H$_2$ and various hydrocarbons to constrain the electron energy, atmospheric composition, and atmospheric temperature. 

\subsubsection{Detection of Amino Acids in Comets
 (SCDD-SSiC-33)}
\label{sssec:aminos}
\sleads{A. I G\'{o}mez de Castro, A. I de Isidro-G\'{o}mez}

The goal of the science case presented by \citet{gomez_de_castro_amino_hwo25} is to conduct a census of amino acids in comets passing within 1~AU of Earth. Understanding the diversity and potential enantiometric excesses of amino acids within the solar system are extremely relevant for astrobiology and may help explain why life on Earth evolved to use left-handed amino acids. To detect amino acids, this science case would exploit the polarization signature induced by the chiral structure of alanine, which is the most abundant chiral amino acid and the second-most abundant amino acid overall (after glycine, which is non-chiral.) The asymmetrical structure of alanine is advantageous for detection because an over abundance of right-handed or left-handed molecules induces a characteristic rotation signature in linearly polarized UV light at 180~nm. If equipped with a UV spectropolarimeter such as POLLUX \citep{neiner_et_al2026}, an observatory like HWO could therefore detect enantiometric excesses of alanine in nearby comets. 

\breakthrough Due to the significance and difficulty of this measurement, detecting an enantiometric excess of alanine in only a single comet would be transformative. This detection would require measuring the linear polarization of UV light between 140~nm and 220~nm. Observing roughly five comets per year for a decade (50~comets total) would provide a treasure trove of information about the distribution and chirality of prebiotic molecules in the solar system. Ideally, targets would be re-observed over time to enable studies of how space weathering affects biologically relevant materials. Robust detection of the potentially small signal will require $R\sim300$ spectropolarimetry at $140-220$~nm obtained at a signal-to-noise high enough to detect signals of 0.3\% at 180~nm for a flux of $10^{-13}$~erg~s$^{-1}$~cm$^{-2}~\text{\AA}^{-1}$. The spectrograph should have a large aperture of $5'' - 10''$ to ensure that the photon collection rate is high enough to permit detection of the polarization signal. Depending on the sensitivity of the observatory, exposure times of hours to days will be needed to reach the required accuracy of $<0.1\%$. A large instantaneous field of regard would be beneficial for maximizing target observability. Additionally, the ability to schedule target of opportunity observations would be advantageous because that would enable observations of newly discovered comets.

\enabling Enabling progress has the same observational requirements as breakthrough science ($R\sim300$ spectropolarimetry at $140-220$~nm sensitive to 0.3\% at 180~nm for flux of $10^{-13}$~erg~s$^{-1}$~cm$^{-2}~\text{\AA}^{-1}$), but an alanine detection is not required. Merely obtaining observations with the requisite sensitivity would already be substantial progress. 

\subsubsection{Disk Winds and Dispersal of Protoplanetary Disks (SCDD-SSiC-23)}
\label{sssec:disk_winds}
\sleads{Keri Hoadley, Yasuhiro Hasegawa, Thomas Haworth, J. Serena Kim, Ilaria Pascucci, Fulvia Pucci, Nicole Arulanatham, Kevin France, Neal Turner, Erika Hamden}

This science case was posted on the STScI SCDD Portal\footnote{\url{https://docs.google.com/document/d/1R5lCfu6_a2_9NgRTCtziyvpKIbeJn0Hb/}}. Although this science case takes a unique perspective, the science goals are partially addressed by the ISM-focused investigations presented in Sections~\ref{sssec:dust_extinction}, \ref{sssec:dust_uv}, and \ref{sssec:ism_uv} and the disk-focused investigations presented in Sections~\ref{sssec:proto}, \ref{sssec:debris}, and \ref{sssec:exozodi}.

The purpose of this science case is to investigate the effects of disk mass loss on planet formation by measuring the rate of mass loss at various locations within disks around protostars in the Milky Way. Observations with HWO would be designed to probe the physics of mass loss and differentiate winds driven by photoevaporation from those driven by magnetohydrodynamics. The observations would also explore how wind density and velocity are related to the launching point and driving mechanism. 

\breakthrough Breakthrough progress necessitates resolving multiple disk wind components. Accordingly, integral field spectroscopy should be obtained over $91.2 - 200$~nm with spatial resolution $\leq 50$ mas/pixel and spectral resolution $R\geq 100,000$. Longer wavelength observations would not be essential, but could potentially be useful for interpreting UV observations because IR data will provide more constraints on H$_2$ mass. At least 100~disks should be observed at distances up to 10s of kiloparsecs. In other words, the sample should include disks in the Large and Small Magellanic Clouds as well as the Milky Way galaxy. 

\enabling Enabling progress would require UV spectro-astrometry at a lower spectral resolution ($R \geq 50,000$). The wavelength range and spatial resolution capabilities would be relaxed to $100 - 200$~nm and $\leq 100$~mas/pixel, respectively. The necessary sample includes $>50$~disks at distances up to a few kiloparsecs, which is comparable to the distance currently probed by JWST. 

\subsubsection{Studying Protoplanets and Protoplanetary Disks with the Habitable Worlds Observatory (SCDD-SSiC-8)}
\label{sssec:proto}
\sleads{Bin B. Ren}

As discussed in \citet{ren_hwo25}, this science case directly addresses the discovery area identified by the Panel on the Interstellar Medium and Star and Planet Formation \citep[see Appendix~F of][]{astro2020}. Specifically, this science case aims to image planets during the epoch of formation, determine their physical properties, and investigate how planets shape the disks in which they reside. The sensitivity and spectral resolution offered by HWO would enable measurements of planetary accretion rates and variability for young planets at a range of separations as well as mapping of disk structure. Observations in polarized light will be particularly useful for revealing embedded planets \citep[e.g.,][]{wahhaj_et_al2024} and assessing the mineralogical composition \citep[e.g.,][]{debes_et_al2008} and spatial distribution \citep[e.g.,][]{chen_et_al2024} of dust that fuels growing planets. In addition to increasing the sample of directly imaged protoplanets by two orders of magnitude, high-precision astrometry with HWO would enable investigations of orbital stability and perhaps even reveal exo-Moons due to their perturbing influence on planets.

\breakthrough Breakthrough progress would require multi-epoch IFS observations of 200 protoplanets with masses $\geq 0.01 M_J$. The orbital semimajor axes of the protoplanets should be measured to a precision 0.1~AU. The data should have a spectral resolution $R\sim 10,000$ and a spatial resolution of 0.1~AU. The observations should include high-resolution polarimetric reflectance spectra with polarization fraction uncertainty $\leq 0.1\%$. Obtaining this dataset will require imaging and IFS spectroscopy over a $2' \times 2'$ field of view and spatial resolution of 5~mas at 500~nm. The instrument should be capable of observing protoplanets at separations of $0\farcs005 - 60''$ from host stars as faint as $V = 15$. Observations should cover the Lyman and Balmer lines of hydrogen, so UV and visible observations are needed. For this document, we adopt a wavelength range of $90 - 700$~nm. 

\enabling For substantial progress, the target sample would be decreased to 30~protoplanets with masses $\geq 0.1 M_J$, and the semimajor axis precision would be reduced to 0.5~AU. The targets should be observed by an IFS with spectral resolution\footnote{The resolution needed for substantial progress is noted inconsistently in \citet{ren_hwo25}. We adopt the value of $R\sim3000$ shown in their Table~2 and stated in their Section~4 rather than the higher value of $R\sim5000$ shown in their Table~1.} $R\sim3000$ and spatial resolution of 1~AU. The IFS data should be accompanied by spectropolarimetric imaging with polarization fraction uncertainty $\leq 1\%$. The needed field of view is reduced to $1' \times 1'$ for imaging and $10'' \times 10''$ for integral field spectroscopy. For both imaging and spectroscopy, the spatial resolution should be 10~mas at 500~nm, and the instrument should be capable of observing protoplanets at separations of $0\farcs02 - 5''$ from stars as faint as $V\leq 12$. 

\subsubsection{Debris Disks and their Properties with the Habitable Worlds Observatory (SCDD-SSiC-9)}
\label{sssec:debris}
\sleads{Isabel Rebollido, Yasuhiro Hasegawa, Meredith MacGregor, Bin Ren, Mark Booth, Jonathan Marshall, Courtney Dressing, Patricia Luppe}

\citet{rebollido_hwo25} presents a science case focused on advancing understanding of planet formation by measuring the structure and composition of debris disks. HWO could reveal how dust and gas are distributed within the disk, explore how composition varies throughout the disk, and find evidence of interactions of disk material with planets and minor bodies. Broad wavelength coverage from the UV to the NIR would allow HWO to detect both scattered light and thermal emission. Polarimetric capabilities would be advantageous and provide further constraints on the distribution, material properties, and formation site of dust grains within the disk. 

\breakthrough Breakthrough progress would require spatially-resolved FUV spectroscopy at resolution high enough to detect spectral features as well as NIR observations to constrain dust properties. For detecting circumstellar gas features, $R\gtrsim 80,000$ spectra at $140 -3000$~nm would be sufficient. High spatial resolution ($\lesssim 30$~mas at $1~\mu$m) would be needed to map disk features. Additionally, spectropolarimetry at resolution $R>3000$ and sensitivity high enough to detect polarization fractions $<1\%$ would improve characterization of disk solids. High-contrast imaging and spectropolarimetry should be sensitive to disk features at separations of $<1$~AU to 2000~AU from the host star. The field of view should be at least $2' \times 2'$ to increase observational efficiency by allowing full disks to be observed in a single pointing. 

\enabling The conference proceeding describing the science case does not differentiate between breakthrough and enabling progress. For this document, we adopt the approach taken in internal version of this science case, which indicated that the two progress levels would have the same requirements except for the range of disk annuli probed by the observations. Specifically, substantial progress would require observing separations of $\sim 1 - 300$~AU for substantial progress while breakthrough progress would require observations from $<1$~au out to a much larger separation of 2000~AU. Narrowing the range of disk separations would substantially decrease the minimum outer working angle and required field of view.

\subsubsection{Characterizing Exozodis in Scattered Light (SCDD-SSiC-19)}
\label{sssec:exozodi}
\sleads{John H. Debes, Steve Ertel, William C. Danchi, Yasuhiro Hasegawa, Isa Rebollido, Ramya Anche, Virginie Faramaz-Gorka, Mark Wyatt, Max Millar-Blanchaer }

The science case presented by \citet{debes_hwo25} builds on the work of ExoPAG SAG~23\footnote{\url{https://sites.google.com/view/sag23-exozodiacaldust/home}} and addresses Question E-Q1d from the Astro2020 panel on Exoplanets, Astrobiology, and the Solar System \citep[see Appendix~E][]{astro2020}. Specifically, the science case would detect and characterize exozodi dust in and near the HZs of nearby stars. Observations with HWO would enable studies of the morphology, composition, and origin of the dust and could reveal substructures pointing towards the presence of potentially habitable planets and even planets less massive than the Earth. As in Section~\ref{sssec:debris}, multi-wavelength polarimetric observations will provide valuable constraints on dust properties such as grain shape \citep[e.g., ][]{milli_et_al2024}. Determining the composition and size distribution of the dust grains is important for understanding the frequency of collisions within planetary systems and the nature of the bodies involved in the collisions. 

For instance, systems with exozodi dominated by Poynting-Robinson drag are expected to have substantially higher optical depth and thermal emission in the outer regions of the disk (e.g., $10-100$AU) than systems like the solar system for which the exozodi are primarily generated by comets \citep{rigley+wyatt2020, rigley+wyatt2022}. Given the important role of comets in delivering volatiles to the inner solar system \citep[e.g.,][]{nesvorny_et_al2010}, understanding the origin of exozodi is therefore important for investigations of habitability. Earlier work by the HOSTS survey \citep{ertel_et_al2018, ertel_et_al2020} indicates that typical FGK stars have exozodi $\lesssim 27$ times as bright as the solar system zodiacal cloud, but extrapolating from the thermal emission detected by HOSTS at mid-IR wavelengths to the scattered light that would be observed by HWO is challenging \citep[e.g.,][]{krishnanth_et_al2024, wolff_et_al2024}. Depending on the dust grain properties, a given level of thermal emission could translate to various levels of scattered light. Conclusions from the HOSTS survey are further complicated by the fact that the detected exozodis exhibit a diverse range of spatial dust distributions \citep{defrere_et_al2015, defrere_et_al2021, garreau_et_al2025}. On the other hand, resolved images of the spatial dust distribution can provide an opportunity for determining dynamical masses of detected planets \citep{bonsor_et_al2018}. 

Dust in and near the HZ is an important case of exozodiacal dust, but not the only one. Hot exozodiacal dust has been identified around $\sim20\%$ of nearby stars \citep[e.g.,][]{absil_et_al2013, absil_et_al2021, ertel_et_al2014, ertel_et_al2016, ertel_et_al2025, nunez_et_al2017, kirchschlager_et_al2020}, and is located closer to the star than the HZ \citep{kirchschlager_et_al2017}. The connection between hot exozodi and habitable-zone dust is so far poorly understood. Being able to bring significant amounts of hot dust as close to the star as it is observed and sustaining it there has major implications for the evolution and dynamics of the planetary system \citep[e.g.,][]{bonsor_et_al2012, faramaz_et_al2017, pearce_et_al2022}. Probing closer in than the HZ for a few closest systems with exozodis at the same inner working angle as needed for studying a larger sample of HZ-dust systems will allow to determine whether the hot and habitable-zone dust are spatially connected and to study the origin, dynamics, and optical properties (thus grain size, composition) of the hot dust.

If exozodi have scattering efficiencies similar to that of the Solar Zodical Cloud, observations with HWO would be sensitive to exozodi as faint as that of the Solar System. Earlier observations with Roman/CGI are anticipated to be sensitive to exozodi as faint as 20x the brightness of the Solar Zodical Cloud \citep{douglas_et_al2022}, which could help reveal the distribution of dust albedos and refine predictions for HWO. In order to thoroughly investigate the properties and origin of exozodi, at least 40 (and ideally 100) exozodi should be detected by HWO. The observations should have SNR$>5$ per resolution element and resolution high enough to enable reconstruction of disk features on scales smaller than 1~AU. 
A sample of that size and fidelity would permit measuring the occurrence rate of exozodi to better than 10\% and statistically robust statements about the distribution of disk inclinations, extent, scattering properties, and surface brightness profiles. 

\breakthrough Breakthrough progress for this science case consists of surveying $100 - 200$ stars down to a depth of 0.5 zodi and detecting exozodi in $40 - 100$ systems. These exozodi should be observed in at least two bands to characterize scattered light and studied in polarized light to determine the polarized intensity of the disk. At least 10 exozodi should display resolved asymmetries or disk substructure. In addition, $\geq 10$ exozodi should be studied spectroscopically at visible and NIR wavelengths with spectral resolution $R\sim100$. 

The portal version of this science case notes that obtaining the data needed for breakthrough progress will require direct imaging at $200 - 2000$~nm with a spatial resolution of 20~mas. The observations should be sensitive to disk features with surface brightnesses as faint as 22.5~mag/arcsec$^2$ at an inner working angle of 65~mas at 500~nm, which requires reaching a contrast level of $3 \times 10^{-11}$. The observations should extend to an outer working angle of $20 - 40~\lambda/D$ and have polarization measurement uncertainty $\leq 1\%$. 

\enabling The portal version of the science case notes that the observational needs for substantial progress are significantly relaxed. Observations should be obtained at wavelengths of $300 - 2000$~nm with spatial resolution of 40~mas. At the inner working angle, which would be increased to 150~mas at 500~nm, observations should be sensitive to features as faint as 19~mag/arcsec$^2$, which corresponds to a contrast of $1 \times 10^{-9}$. Polarization should be measured to $\leq 3\%$. The outer working angle of $20 - 40~\lambda/D$ is unchanged.

\subsubsection{Exoring Detection Characterization and Occurrence Rates
 (SCDD-SSiC-31)}
\label{sssec:exorings}
\sleads{Mary Anne Limbach, Rachel Bowens-Rubin, Sam Hopper, Elizabeth Lane, Logan Pearce}

 The goal of this science case is to establish the frequency and characteristics of planetary rings. In addition to providing valuable comparison data for improving our understanding of the rings around the giant planets in the solar system, studies of ``exorings'' are also important for understanding how the presence of rings affects the interpretation of reflected light spectra of exoplanets. This science case was posted in the STScI SCDD Portal\footnote{\url{https://docs.google.com/document/d/1Fs07W1XQvR4uy0nm5jkG4Npf8hB_PbrVdO8MuJkYo94/edit?tab=t.0}} but does not appear in the HWO25 conference proceedings. For additional details about this exoring case and the subsequent exomoon case presented in Section~\ref{sssec:exomoons}, see the peer-reviewed papers \citet{limbach_et_al2024, limbach_et_al2026}.

 A facility like HWO has the potential to determine the fraction of giant exoplanets that have ring systems as well as the physical properties of the rings such as size, orientation, albedo, composition, and thickness. HWO could also investigate how the brightness of exorings depends on geometric factors such as phase angle and inclination. Accurate treatment of rings during the data fitting process is important to ensure that ringed planets are not erroneously assumed to be ringless and therefore assumed to have radii and temperatures that are larger than their true values. 
 
\breakthrough Breakthrough progress would require directly imaging approximately 100~planets with planet-star separations of $1-10$~AU at sensitivity high enough to detect Saturn-like rings (i.e., contrast ratios $\sim 10^{-9}$). Initial ring detection will require observations at $\geq10$ distinct orbital phases to robustly identify signals caused by rings. While ring reconnaissance observations should be completed at visible wavelengths where rings are likely to be brightest, any detected rings should receive spectropolarimetric follow-up observations over a broad wavelength range (VIS - 2$\mu$m; $R\sim100$) to characterize ring particles. A large field of view ($\sim10''$) would be advantageous for detecting extended rings. Due to the need to obtain observations at multiple orbital phases, a large instantaneous field of regard would also be beneficial for this science case. 

\enabling The requirements for enabling progress are similar, but observations could be obtained at $\geq5$ orbital phases rather than the $\geq10$ orbital phases needed for breakthrough progress. Additionally, the multi-wavelength observations could have a red cutoff of $1.5\mu$m rather than $2\mu$m and a lower spectral resolution of $R\sim20$. Finally, the field of view could be reduced by a factor of ten to $\geq 1"$ and polarimetric capabilities would not be required. 

\subsubsection{Exomoon Detection with Mutual Events
 (SCDD-SSiC-32)}
\label{sssec:exomoons}
\sleads{Mary Anne Limbach}

Like Section~\ref{sssec:exorings}, which focuses on exorings, this section describes a science case document that appears on the STScI SCDD Portal\footnote{\url{https://docs.google.com/document/d/1bOa97ZAblcfkfIOmzQDZDYDxJaMN-Cue8MFVjDEDTqw}} but not in the HWO25 conference proceedings. As for the exorings case, the scientific motivation and observational goals are discussed at length in the associated peer-reviewed papers \citep{limbach_et_al2024, limbach_et_al2026}. Below, we briefly summarize the aims of this science case and the required observations. 

The focus of this science case is to determine the frequency with which terrestrial planets are orbited by moons and determine the properties of those moons. The science case specifically concentrates on moons detected via mutual events, which are moments in which one celestial body is in front or behind another \citep[e.g.,][]{cabrera+schneider2007, limbach_et_al2024}. In addition to transits and eclipses, mutual events also include occultations (i.e., when a planet passes in front of a moon) and shadows (i.e., when the shadow cast by the moon passes over the face of the planet that would otherwise be observable at that time). Due to the distances between Earth and the potential targets, exomoon detection via mutual events would rely on measuring changes in the integrated brightness of the moon-planet system rather than physically resolving the moon from its host planet. However, starlight suppression and high-contrast imaging or spectroscopy would still be needed to block out the light from the host star to permit detection of the planet+moon system. 

Depending on the properties and configuration of the planet-moon system and its orbit around its host star, the frequency of mutual events could vary by orders of magnitude from several times per day to once every few months. As lunar orbits are likely to evolve with time, the frequency of events may depend on the age of the system with younger systems more likely to exhibit frequent mutual events \citep{limbach_et_al2024}. In the case of our own Earth-Moon system, the frequency of mutual events is believed to have decreased significantly from once every 11 hours shortly after the formation of the Moon to roughly every 2.5 months (approximately 1700 hours) today. Each individual event is expected to last several hours, so observational campaigns to look for exomoons should have dwell times of a few days to increase the likelihood of catching a full event. In addition, the temporal resolution should be high enough to detect the start and end of events, which suggests that the brightness should be measured at least four times an hour. Robust detection of individual mutual events would require 10\% white light photometric precision over timescales of a few hours for systems with large moon-planet radius ratios like the Earth-moon system. Detecting moons with smaller moon-planet radius ratios or searching for mutual events using less precise data would require stacking multiple observations, which would likely be tricky given the uncertain timing of mutual events and the possibility of spacecraft systematics or astrophysical signals that could mimic moon-generated mutual events. 

\breakthrough Breakthrough progress for this science case would require monitoring $\geq 30$ terrestrial planets orbiting stars within 20~pc. Ideally, some of these planets would be in the same multiplanet system and could therefore be observed simultaneously. Theoretical studies suggest that young planets with semimajor axes of 1-5~AU are the most compelling targets because planets in closer orbits are hypothesized to be less likely to harbor moons \citep{heller_et_al2014} and more distant planets will likely be too faint to reach the required precision. The brightness of each system at optical and NIR wavelengths should be measured at 15 minute cadence over a monitoring period of several weeks. The observations should span visible and near-infrared wavelengths (out to $2\mu$m) and have a white light photometric precision of $\lesssim10\%$ over a few hours. Observations near 1.4$\mu$m and 1.9$\mu$m will be particularly important because the moon/planet flux ratio may be more favorable at those wavelengths. A large field of view ($\geq 2''$) is needed to maximize sensitivity to the desired orbital separation range ($1-5$~AU) and a large instantaneous field of regard would be beneficial for increasing the number of opportunities to observe a given system. 

\enabling The targets and capabilities needed for substantial progress are largely similar to the requirements for breakthrough science. At least 20 planets within 10~pc should be observed at 15-minute cadence for several days. The observations should still cover visible and NIR wavelengths with broadband photometric precision $\lesssim 10\%$ over several hours, but the red cut-off could be reduced to $1.5\mu$m. The field of view could be reduced to $1''$. 

\subsubsection{Utilizing Habitable Worlds Observatory for Planetary Defense
 (SCDD-SSiC-30)}
\label{sssec:defense}
\sleads{J. Dotson, Andrew S. Rivkin, Cristina Thomas, Geronimo Villanueva, Davide Farnocchia, Steven Chesley, Jessica L. Noviello}

As explained by \citet{dotson_hwo25}, the goal of this science case is to use observations by a facility like HWO to detect and characterize near-Earth objects that could potentially impact the Earth. Observations obtained from L2 by a large space-based telescope like HWO could be extremely beneficial for planetary defense because the combination of exquisite sensitivity and a non-Earth vantage point could reveal hazards that may be invisible or inaccessible from observatories on Earth or in Earth orbit. In addition to detecting potential impactors, HWO has the potential to inform risk mitigation strategies by determining the compositions, rotation rates, and shapes of near-Earth objects. Surface compositions can be constrained by multiwavelength observations and provide insight into the likely bulk compositions and densities of potential impactors. Rotation rates and shapes can be assessed by obtaining a series of observations over an extended period of time. When combined with density constraints, shape information can be used to estimate the object mass and therefore impact energy. Rotation periods can also provide information about the self-gravity and internal cohesion of a potential impactor, which are also important inputs for risk mitigation analyses. 

\breakthrough Breakthrough progress in this area would require the ability to track targets at rates of $>110$~mas/s and obtain photometry using both standard color filters (e.g., $ugriz$) for easy comparison to data obtained by ground-based telescopes and over a very wide bandpass designed to maximize SNR and increase sensitivity to faint objects. For brighter targets, the ability to obtain spectroscopic observations would also be useful. Meeting the non-sidereal tracking requirement would require both the ability to move at speeds $>110$~mas/s and a guide camera with a sufficiently large field of view. The science case would benefit significantly from the ability to look closer to the Sun than can usually be done from Earth, which may have implications for the Sun avoidance angle. Additionally, this science case would benefit from a large instantaneous field of regard as well as a field of view at least as large as those under consideration for other science cases. 

\enabling For substantial progress, the tracking requirement could be relaxed to $>75$~mas/s. Additionally, the instantaneous field of regard could be reduced.

\subsection{Living Worlds}
\label{ssec:lw_scdds}
The Living Worlds SWG approached the SCDD process differently by concentrating primarily on a single unified science case spanning the entire process of determining the frequency of Earth-like biospheres in the Milky Way. We briefly summarize this science case in Section~\ref{sssec:life}, but we refer readers to the literature for further details \citep[e.g.,][]{lyons_et_al2014, catling_et_al2018, fujii_et_al2018, krissansen-totton_et_al2018, meadows_et_al2018, schwieterman_et_al2018, walker_et_al2018, astrobio_strategy2019}. In addition to this primary science case, the Living Worlds SWG also developed ten ``auxiliary'' science cases related to the prevalence of life in the universe. The primary science case and all ten auxiliary science cases were later published or posted to the SCDD Portal. One of these cases \citep[LW-9;][Section~\ref{sssec:geochemical}]{brugman_hwo25} discusses work that should be done in advance of future observations and is therefore described in the discussion of precursor science in Section~\ref{sec:precursor}. 

The remaining nine auxiliary cases focus on observations that could be made by a facility like HWO. Section~\ref{sssec:origin} explores how to test different theories of the origin of life, and Section~\ref{sssec:prebiosignatures} considers how to constrain the frequency of \emph{pre}biotic conditions on exoplanets. The next science cases broaden the scope of the search for life by investigating the detectability of alien technologies (Section~\ref{sssec:technosignatures}) and considering how to detect life that is different from Earth life (Section~\ref{sssec:lawdki}). Other science cases concentrate on assessing the presence of vegetation or oceans on planetary surfaces (Sections~\ref{sssec:surface_bio} and~\ref{sssec:polbio}). Additionally, Section~\ref{sssec:seasonality} discusses how seasonal changes could be an indicator of life. Finally, Section~\ref{sssec:fpbio} discusses how HWO could distinguish between false positive and true positive biosignatures, and Section~\ref{sssec:mp} addresses how to determine which planets have Earth-like compositions.

\subsubsection{The Search for Life on Potentially Habitable Exoplanets (SCDD-LW-1)}
\label{sssec:life}

\sleads{Giada Arney, Niki Parenteau, Natalie Hinkel, Eric Mamajek, Joshua Krissansen-Totton, Stephanie Olson, Edward Schwieterman, Sara Walker, Kevin Fogarty, Ravi Kopparapu, Jacob Lustig-Yaeger, Mark Moussa, Sukrit Ranjan, Garima Singh, Clara Sousa-Silva, Maxwell Frissell, Samantha Gilbert-Janziek, Vincent Kofman, Natasha Latouf, Mary Anne Limbach, Rhonda Morgan, Christopher Stark, Armen Tokadjian, Anna Grace Ulses, Nicholas Wogan, Mike Wong, Amber Young}

By aiming to constrain the frequency of Earth-like global biospheres in the galaxy, the science case presented by \citet{arney_hwo25} would address the fundamental question, ``Are we alone?" This science case is directly responsive to ``The Search for Life on Exoplanets'' discovery area identified by the Astro2020 panel on Exoplanets, Astrobiology, and the Solar System \citep{astro2020} and is responsible for the first two words in the name of the observatory. Indeed, along with conducting transformative astrophysics, searching for life in the universe was one of the two primary motivations supporting Astro2020's prioritization and recommendation of the large IR/O/UV space telescope. The ability to successfully investigate this science case is therefore of utmost importance when designing HWO.

The objective of this science case is to conduct a scientifically rigorous search for biosignatures in the atmospheres of enough planets that a non-detection is statistically meaningful. Recognizing that the atmospheric composition of our own planet has changed over the eons \citep[e.g.,][]{anbar_et_al2007, lyons_et_al2014, planavsky_et_al2014, catling+zahnle2020}, and that atmospheric composition is linked the evolution of life \citep[e.g.,][]{cerling_et_al1997, chen_et_al2015, fischer_et_al2016, lenton_et_al2016, droser_et_al2017}, the Living Worlds SWG demonstrated that it would be advantageous to be able to detect life similar to that seen on ancient Earth as well as modern Earth. Specifically, the SCDD argues that HWO should be sensitive to CH$_4$ (for planets like ancient Archean Earth), O$_3$ (for planets like the modern Phanerozoic Earth and perhaps the middle-aged Proterozoic Earth), and O$_2$ (for Phanerozoic Earth). 

During the Archean eon (4 - 2.5 Gya), the atmosphere of Earth is hypothesized to have very high concentrations of CH$_4$ \citep[e.g.,][]{pavlov_et_al2000} due to the combination of a high source rate of CH$_4$ from methanogens and a low sink rate due to the low oxygen level in Earth's atmosphere. A terrestrial planet with a CH$_4$-rich atmosphere might therefore be consistent with Archean-like life, but both biotic and abiotic mechanisms can lead to high concentrations of CH$_4$. Fully understanding such a planet will require a detailed study of possible geological production pathways (e.g., volcanic outgassing) and photochemistry. Determining the abundances of other C-rich molecules like CO$_2$ and CO, measuring the UV flux from the host star will be necessary to better understand the atmospheric chemistry and assess the biological relevance of elevated CH$_4$. Planets with particularly high C/O ratios (i.e., CH$_4$/CO$_2$ $> 0.2$) may be shrouded in organic hazes that dramatically change the spectrum of the planet at bluer wavelengths as well as the temperature and habitability of the surface \citep{arney_et_al2016,arney_et_al2017, arney_et_al2018}. Observations of host stars at UV wavelengths will therefore be crucial for exploring the habitability of Archean analogs. 

In addition, NIR spectra of Archean-like planets will be needed to determine the abundances of CH$_4$, CO$_2$ and CO and assess the biological significance of CH$_4$ detections. High CO abundance would suggest an abiotic origin (e.g., volcanoes) while significant levels of CO$_2$ are unlikely to coexist with CH$_4$ in the absence of life. Assessment of planet orbits, masses, and radii will also be essential to distinguish between inhabited planets akin to Archean Earth and warm Titan-analogs \citep{krissansen-totton_et_al2022}.

A key moment in Earth's atmospheric history was the ``Great Oxidation Event,'' a dramatic increase in the abundance of atmospheric O$_2$ roughly 2.4 - 2.1~Gya at the start of the Proterozoic eon \citep[e.g.][]{lyons_et_al2014}. However, despite the large increase in O$_2$ abundance, the absolute level of O$_2$ during the mid-Proterozoic was orders of magnitude lower than that of modern Earth \citep[$\sim 0.1\%$ present atmospheric level][]{planavsky_et_al2014, lyons_et_al2021}; detection of O$_2$ spectral features in the atmosphere of a planet like Proterozoic Earth would therefore be extremely challenging. 

Detection of CH$_4$ would likely be likely nearly as daunting, but CH$_4$ abundances roughly 10~times higher than Earth's current level might be detectable via the strong CH$_4$ band at 1.65~microns. In some cases, N$_2$O could be a useful biosignature for Proterozoic planets \citep[e.g.,][]{buick2007, schwieterman_et_al2022}, but the weakness of N$_2$O spectral features blueward of 2~microns and the overlap with H$_2$O features would likely prevent robust measurements of N$_2$O for all planets without high ($\geq 1000$~ppm) abundances \citep{tokadjian_et_al2024}. 

By far, the most promising path for detecting biosignatures for Proterozoic planets appears to be combining NUV observations to search for O$_3$ via the Hartley-Huggins band ($200 - 350$~nm) and NIR observations to search for CH$_4$ (1.65~microns) and possibly N$_2$O (1.52~microns, 1.68~microns, and 1.78~microns). Without the capability to search for biosignatures in both the NUV and the NIR, HWO would likely be unable to detect life on Proterozoic planets and therefore incorrectly conclude that they are uninhabited \citep[e.g.,][]{reinhard_et_al2017}. Even if such planets harbor life that is actively generating O$_2$, the conditions on the planet could easily prevent significant quantities of O$_2$ from accumulating in the atmosphere \citep[e.g.,][]{lyons_et_al2014}. For planets like Proterozoic Earth, searching for life via disequilibrium between O$_3$ and CH$_4$ or O$_3$ and N$_2$O is a more robust approach.

In the atmosphere of modern (Phanerozoic) Earth, biologically relevant gases include O$_2$, O$_3$, CH$_4$, and CO$_2$. The combination of O$_2$ and CH$_4$ is extremely well-studied biosignature \citep[][]{hitchcock+lovelock1967}, but the rapid destruction of CH$_4$ by photochemical reactions may lead to CH$_4$ abundances that are too low to detect in the atmospheres of planets similar to modern Earth. Due to the dependence of photochemistry on the spectrum of incoming stellar radiation, biosignature detection via the simultaneous detection of O$_2$ and CH$_4$ may be more promising for planets orbiting low-mass K/M stars \citep[e.g.,][]{segura_et_al2005, arney2019}. For O$_2$ detections in the absence of detectable CH$_4$, comprehensive atmospheric investigations will be needed to determine the bulk composition and pressure of the planetary atmosphere and rule out possible false positive scenarios. One example of a false positive pathway that would lead to high O$_2$ abundance in the absence of life is photolysis of H$_2$O in the upper atmosphere and subsequent loss of H in the upper atmosphere \citep{wordsworth+pierrehumbert2014}, the likelihood of which could be estimated by determining the abundance and composition of background gases by measuring the Rayleigh scattering slope, assessing pressure broadening, or ruling out other gases \citep[e.g.,][]{hall_et_al2023, young_et_al2024}. Knowledge of the (non-O$_2$) background gas pressure is important for intepreting observations because a lack of sufficient non-condensing gases like N$_2$ could allow H$_2$O to accumulate in the upper atmosphere without being cold-trapped. 

Another false positive scenario is H escape from the atmospheres of planets whose surfaces are covered by very deep oceans \citep[$\gtrsim 50$ Earth oceans;][]{krissansen-totton_et_al2021}; this hypothesis could be rejected by detecting land features via surface mapping (see Section~\ref{sssec:surface_water}). A third scenario consists of highly-irradiated planets that have experienced runaway greenhouses and now have enhanced atmospheric abundance of O$_2$ after H has been photolyzed from H$_2$O and subsequently lost from the atmosphere \citep[e.g.,][]{luger+barnes2015, tian2015}. For this scenario, false positives could be excluded by determining the insolation received by planets throughout their orbits and measuring the abundances of CO$_2$ and H$_2$O. For planets orbiting Sun-like stars rather than M~dwarfs, additional measurements of O$_2$ abundance are needed to rule out O$_2$ partial pressures too large to be attributed to biology. One avenue is detecting O$_2-$O$_2$ CIA throughout the UV-Vis-NIR at 0.345, 0.36, 0.38, 0.445, 0.475, 0.53, 0.57, 0.63, 1.06, and 1.27 $\mu$m \citep[e.g.,][]{schwieterman_et_al2016, meadows2017, meadows_et_al2018}

For the most recent phase of Earth's history, which is sometimes called the Anthropocene, life on our planet could be detected via technosignatures (see Section~\ref{sssec:technosignatures}) as well as more conventional biosignatures. Recognizing that life in the universe may be far more varied than that seen on Earth, the ideal life detection instruments for HWO should cover as many spectral features as possible to permit the detection of both life as we know it and life as we don't know it (see Section~\ref{sssec:lawdki}). 

\breakthrough In addition to the standardized progress levels of ``breakthrough'' and ``enabling'' science, this science case also presents requirements for an advanced ``breakthrough+'' or ``super-breakthrough'' or progress level and an intermediate ``major'' progress level. Revolutionary, super-breakthrough science would require observations of 67~planets. The dataset should include low-resolution UV spectra ($250-400$~nm; $R \sim 7$), medium resolution visible spectra ($400-1100$~nm; $R \sim 140$), and medium resolution NIR spectra ($1100-1700$~nm; $R \sim 70$). 

For breakthrough science, the requested spectra should cover the same wavelength range at the same resolution, but the target sample would be reduced to 33~planets. In both cases, key spectral features should be detected at $SNR = 10$ for O$_3$ in the NUV, $SNR = 8$ for O$_2$ at 760~nm \citep{latouf_et_al2023}, $SNR = 10$ for CH$_4$ at 890~nm \citep{young_et_al2024}, $SNR = 6$ for H$_2$O at 940~nm \citep{latouf_et_al2023}, $SNR > 10$ for CO$_2$ at 1590~nm \citep{young_et_al2024}, and $SNR >20$ for CO at 1600~nm. For each planet, a decision tree like that developed by \citet{young_et_al2024} could be followed to determine the most informative order in which to accumulate additional data using successive observations each covering $10-20\%$ bandpasses.

For ``major'' progress (a tier between enabling and breakthrough), HWO should obtain visible ($400-1100$~nm; $R \sim 140$) and NIR spectra ($1100-1700$~nm; $R \sim 70$) of 26~planets. The necessary signal-to-noise ratios for H$_2$O, O$_2$, CH$_4$, CO$_2$, and CO would be unchanged, but this dataset would not have the UV coverage needed to detect O$_3$. 

\enabling Substantial progress would require visible spectra ($R \sim 140$; $450-1100$~nm) of 12~planets. The needed signal-to-noise ratios for H$_2$O, O$_2$, and CH$_4$ would be unchanged, but the data would not extend redward enough to capture the 1590~nm CO$_2$ feature or the 1600~nm CO feature. 

\subsubsection{Testing Origin-of-Life Theories with the Habitable Worlds Observatory (SCDD-LW-3)}
\label{sssec:origin}
\sleads{Sukrit Ranjan, Martin Schlecker, Nicholas Wogan, Michael Wong
}

As discussed in \citet{ranjan_origin_hwo25}, the goal of this science case is to test various theories for the origin of life by measuring the frequency of various biosignatures and assessing their correlation with various environmental conditions such as the presence of oceans. If life can originate in only a very narrow range of parameter space, then HWO would be expected to detect a lower fraction of inhabited planets compared to scenarios in which life is more flexible and can originate in a broad variety of environments. Fully exploring the diversity and habitability of environments found on temperate, terrestrial planets is likely beyond the range of HWO, especially for sub-surface and underwater features such as hydrothermal vents, but the frequency with which HWO detects life on terrestrial planets with various atmospheric and surface characteristics could provide initial insights into the flexibility or rigidity of the origin of life. In addition, comparing the frequency of biosignature detection for planets orbiting host stars with a range of spectral types and activity levels could provide valuable insight into the long-standing question of the importance of UV light for the origin of life \citep[e.g.,][]{sagan+khare1971}. 

Addressing this science case will require obtaining and analyzing UV/VIS spectra ($\geq 200$~nm) of a large sample of temperate terrestrial planets. Investigating possible correlations between biosignatures, stellar properties, and planetary surface properties will also require observations at multiple phases to search for ocean glint (see Section~\ref{sssec:surface_water} and possibly constrain the surface distribution of continents as well as an understanding of stellar flux through time, especially at shorter wavelengths (e.g., Section~\ref{sssec:survive_water}). Additional work is needed to determine the needed sample size, but initial studies suggest that over 80 planets will be needed to investigate specific theories even at the $2\sigma$ level \citep[e.g. see][on testing correlations between incident flux and atmospheric CO$_2$ partial pressure as possible evidence of a carbon-silicate weathering cycle on temperate terrestrial exoplanets]{lehmer_et_al2020}. It therefore seems highly unlikely that HWO would be able to robustly answer the question of which conditions (or sets of conditions) are needed for the evolution of life, but HWO could provide an initial estimate of the frequency with which life begins on temperate terrestrial planets and inform target selection for future missions such as LIFE \citep{quanz_et_al2022}. 

\breakthrough As described in Section~\ref{sssec:life}, breakthrough progress would require UV observations at $\geq 200$~nm to detect O$_3$ and NIR observations to detect CH$_4$ and H$_2$O. Phase-curve observations may be advantageous for assessing the land/ocean coverage as discussed in Section~\ref{sssec:surface_water}. Additionally, UV observations at higher spectral resolution may be helpful for better characterizing the host star and could support investigations of the impact of high-energy stellar radiation on the planet over the history of the system.

\enabling Substantial progress would still require UV observations to detect O$_3$, but the longer wavelength observations needed for CH$_4$ and H$_2$O could be omitted. The ``bonus'' observations such as phase-resolved observations and high-resolution UV spectra of the host stars would be omitted. 

\subsubsection{Prebiosignatures with the Habitable Worlds Observatory (SCDD-LW-4)}
\label{sssec:prebiosignatures}
\sleads{Sukrit Ranjan, Danica Adams, Michael Wong, Martin Schlecker, Nicholas Wogan, Jessica M. Weber
}

As discussed in \citet{ranjan_prebiosignatures_hwo25}, the aim of this science case is to determine the frequency with which prebiotic environments emerge in the galaxy. If the emergence of life is much rarer than the existence of habitable environments, HWO might detect prebiosignatures \citep[e.g.,][]{claringbold_et_al2023} even in the case of an unsuccessful outcome in the search for life (i.e., Section~\ref{sssec:life}). The SCDD defines ``prebiotic'' planets as those that display signs of habitability but not biosignatures. In other words, prebiotic planets are those that are habitable but not inhabited. 

There are a variety of theories for the compositions of prebiotic atmospheres, so this science case requests broadband spectra (UV - VIS needed; UV - NIR preferred) of a large sample of planets to investigate the diversity of temperate terrestrial atmospheres. Initial estimates suggest that wavelength coverage redward of 200~nm will be sufficient, but follow-up studies are underway to refine the essential wavelength range. Theoretical predictions suggest that in steady-state conditions, prebiotic worlds should have atmospheres dominated by CO$_2$ and N$_2$ with $<1\%$ H$_2$ \citep[e.g.,][]{gaillard+scaillet2014, catling+kasting2017}, but higher levels of H$_2$ may be observed in the aftermath of giant impacts \citep[e.g.,][]{zahnle_et_al2020, wogan_et_al2023}. Further work predicts low levels of HCN and CH$_4$, both of which are relevant for prebiotic chemistry. Sulfur is also important for prebiotic chemistry, but predictions are mixed on whether sulfur can accumulate in the atmospheres of prebiotic planets or whether the rate of sulfur absorption into oceans is high enough that the atmospheric concentration of sulfur remains low \citep[e.g.,][]{hu_et_al2013, ranjan_et_al2018}. Past studies have shown that molecules containing biologically accessible fixed nitrogen such as NO could be produced via lightning, impacts, and interactions with solar energetic particles \citep[e.g.,][]{mancinelli+mckay1988, airapetian_et_al2016, krasnopolsky2006}.
However, more work is needed to quantify the expected yield of nitrogen-bearing species, especially for production via lightning \citep[e.g.,][]{ardaseva_et_al2017, wong_et_al2017}.

\breakthrough Breakthrough progress requires the UV/VIS/NIR observations described in Section~\ref{sssec:life} but the UV data should extend to $200$~nm. There are no specific timing requirements for these observations because the bulk atmospheric composition is expected to be stable on the decade-long timescales over which observations would be possible with a facility like HWO. Knowledge of system ages would be beneficial, as would supplemental characterization of planets to determine that they are habitable but not inhabited. For instance, observations like those discussed in Sections~\ref{sssec:surface_water} and \ref{sssec:polbio} could establish the presence of surface liquid water. 

\enabling For substantial progress, the UV/VIS observations discussed in Section~\ref{sssec:life} would still be required but NIR data would not be needed.

\subsubsection{Technosignatures (SCDD-LW-5)}
\label{sssec:technosignatures}
\sleads{Ravikumar Kopparapu, Svetlana Berdyugina, Jacob Haqq-Misra, Thomas Beatty, Vincent Kofman, Nick Siegler, Eddie Schwieterman}

The science case discussed in \citet{kopparapu_hwo25} takes a different approach to the search for life by looking for signs of technology rather than biology. The objective of this investigation is to either detect technosignatures or conduct a deep enough search that a non-detection is statistically meaningful. There is a wide range of possible technosignatures including atmospheric features from industrial pollutants like NO$_2$ or greenhouse gases that could be used intentionally to increase planetary surface temperature; artificial structures on or orbiting planets that suggest an unusual distribution of surface albedos; and city lights that cause the nightside of the planet to emit more short-wavelength light than expected. 

The observations needed to search for certain technosignatures are very similar to those needed to search for biosignatures, and therefore HWO could conduct a serendipitous search for technosignatures in data collected for other LW and SSiC science cases. For instance, \citet{kopparapu_et_al2021} determined that 600 hours of observing time with a 15-m telescope would result in a $4\sigma$ detection of NO$_2$ for a planet with the same concentration of NO$_2$ as modern Earth. Other technosignatures are best detected at wavelengths outside the UV -- NIR range targeted by HWO and are therefore less likely to be detectable with HWO. As an example, \citet{schwieterman_et_al2024} investigated the detectability of artificial greenhouse gases (e.g., CF$_4$, C$_2$F$_6$) and concluded that while such gases could be detected at NIR wavelengths, they are significantly more detectable at longer wavelengths with JWST/MIRI than at shorter wavelengths with JWST/NIRSpec. By extension, future facilities with MIR instrumentation like LIFE \citep{quanz_et_al2022} are likely to be more capable of detecting low levels of artificial greenhouse gases. However, such gases could potentially be detectable with a facility like HWO depending on their unknown opacities at visible and NIR wavelengths. Accordingly, \citet{kopparapu_hwo25} highlight the need for precursor laboratory work to establish the line lists necessary to interpret NIR spectra of industrialized planets.

Mapping artificial structures on planetary surfaces or in orbit will require a significant investment of telescope time because of the need to obtain data in multiple bands at a range of orbital and rotational phases. Hypothetically, high-precision, multi-epoch photometry could reveal spacecraft with trajectories inconsistent with Keplerian motion and high-resolution spectra ($R > 1000$) could display emission lines from spacecraft exhaust, but these minute signals are likely beyond the reach of HWO. 

\breakthrough Breakthrough progress would entail searching for artificial structures, NO$_2$, and city lights on 30~planets. Detecting surface structures would require high-contrast imaging with photometric precision $\lesssim 1$~ppm. Detecting NO$_2$ and city lights would require high-contrast spectra at 300-600~nm and 500-900~nm, respectively. The $300-600$~nm spectra should be sensitive to NO$_2$ abundances $\geq 10^{-10}$ while the $500-900$~nm spectra should be able to detect city-induced brightness variations $\ge 1$~ppm. Polarimetric capabilities would be advantageous for detecting variations in surface albedo or laser signals.

\enabling For substantial progress, the sample size would be reduced to 10~targets. The data should permit detecting NO$_2$ abundances $\geq 10^{-9}$ and city-induced brightness variations $\sim 10$~ppm. As for breakthrough science, searching for NO$_2$ pollution and city lights would require spectra at 300-600~nm and 500-900~nm, respectively. 

\subsubsection{Searching for Life as We Don’t Know It: Detecting Signatures of Planetary-scale Complexity in Exoplanet Atmospheric Spectra (SCDD-LW-6)}
\label{sssec:lawdki}
\sleads{Sara Walker, Estelle Janin, Evgenya Shkolnik, Louie Slocombe, Leroy Cronin}

Conventionally, searches for life beyond the solar system have focused on life like that on Earth because it would presumably be easier to recognize. However, the diversity of life forms in the universe may be much vaster than the focus of current research. Maximizing HWO's life detection potential calls for a broadening of considerations and complementary strategies that expand beyond the search for Earth-like life in Earth-like environments. Accordingly, \citet{walker_hwo25} describe a science case that would explore the manifestation of extraterrestrial biospheres or technospheres -- independently of their chemical instantiation -- by proposing a unifying approach to detect signs of both Earth-like life and ``life as we don’t know it.'' 

For this science case, HWO would obtain spectra of roughly 30 planets smaller than $2 R_\oplus$ and then use a framework called Assembly Theory (a method based on measuring the chemical complexity of an ensemble of molecules to distinguish between abiotic and biotic systems; \citealt{sharma_et_al2023}, Janin et al. in prep) to assess the level of chemical selection present in the atmosphere and therefore the likelihood of the planet being inhabited. The target sample should include planets with a range of atmospheric chemistries, with a preference for older systems within which life would have had more time to emerge and evolve into planetary-scale biospheres/technospheres. Ideally, ages of targets would be known to within 2~Gyr. Broad wavelength coverage is needed to maximize detectability of a variety of atmospheric species. There is a not a specific short-wavelength cutoff, but the long-wavelength cutoff should be as high as possible (i.e., $\sim 3\mu$m). Reducing the amount of coverage at redder wavelengths would limit the confidence with which biosignatures could be detected and reduce the robustness of the approach. At shorter wavelengths, UV observations such as those considered by Section~\ref{sssec:life} would be useful for modeling planetary atmospheres and understanding photochemistry.

Precursor studies could investigate if Assembly Theory can be applied directly to reduced planetary spectra (\citealt{jirasek_et_al2024}, Janin et al. in prep) – providing a means to quantify atmospheric complexity without requiring species elucidation – or if it must be considered in the context of detected molecules and reconstructed atmospheric composition output from atmospheric retrievals. In either case, while atmospheric constituents must be detected in order to apply Assembly Theory, measurements of the abundances of various constituents are not needed.

\breakthrough Achieving breakthrough progress requires observing 30~terrestrial planets at separations of roughly 0.1~AU to 5~AU. This science case does not focus specifically on planets within the liquid water habitable zone, so the ability to access planets as close as 0.1~AU is not critical. Each target should be observed at low-to-medium spectral resolution ($100 \lesssim R \lesssim 1000$) over a broad wavelength range extending as red as possible (e.g., $\lambda \lesssim 3\mu$m). Ideally, observations should be acquired simultaneously across the full wavelength range to minimize systematic errors and increase observational efficiency by decreasing the time needed to generate a full spectrum.

\enabling For substantial progress, the long-wavelength cutoff could be reduced to $1.9\mu$m but the spectral resolution would be the same ($R\sim 100 -1000$). The required sample size (30~terrestrial planets) is the same as for breakthrough progress but the planets could be located within a narrower range of separations (0.5 - 4~AU), which would relax limits on the inner and outer working angles. Finally, observations at different bands could be acquired contemporaneously rather than simultaneously. 

\subsubsection{Living Worlds Working Group: Surface Biosignatures on Potentially Habitable Exoplanets (SCDD-LW-14)}
\label{sssec:surface_bio}
\sleads{
Niki Parenteau, Anna Grace Ulses, Connor Metz, Nancy Y. Kiang, Ligia Coelho, Edward Schwieterman, Jonathan Grone, Giulia Roccetti, Svetlana Berdyugina, Eleonora Alei, Lucas Patty, Emilie Lafleche, Taro Matsuo, Dawn Cardace, Schuyler Borges, Avi Mandell, Kenneth Goodis Gordon, Joshua Krissansen-Totton, Giada Arney, Bill Philpot, Jacob Lustig-Yeager, Stephanie Olson, Marc Neveu, Jos\'{e} A. Caballero, Yuka Fujii, William Sparks, Ludmila Carone, Mariano Battistuzzi, Daniel Whitt}

The goal of this science case is to detect or place statistically meaningful limits on the presence of surface biosignatures on exoplanets. Produced by the interaction of life with radiation, surface biosignatures are planetary-scale indications that a planet is inhabited. For instance, the color of the Earth's oceans is partially determined by the metabolic processes of microorganisms \citep{matsuo_et_al2025}. 

On Earth, one of the most notable surface biosignatures is the vegetation ``red edge,'' which refers to the dramatic increase in the reflectance of vegetation at wavelengths $\gtrsim 700$~nm. This signal is so pronounced that it can be used to map land-based vegetation in satellite data \citep[e.g., ][]{tucker_et_al1985}. For aquatic plants, the vegetation red edge is less detectable because of the low reflectivity of water at red wavelengths \citep{seager_et_al2005}, but chlorophyll abundance can be estimated through spectral ``greenness'' \citep{oreilly_et_al1998, gohin_et_al2002} and by looking for enhanced reflectivity at the blue wavelengths known as the Soret bands \citep[e.g.,][]{das_et_al2017}. Given the possibility that life on planets illuminated by hotter or colder stars might evolve to prefer different wavelengths of light \citep{kiang_et_al2007}, researchers have proposed alternative surface biosignatures such as an ``NIR edge'' for plants on M dwarf planets \citep{tinetti_et_al2006}. Pigments such as carotenoids (i.e., compounds with 40 carbon atoms) are also studied due to their prevalence across multiple kingdoms of Earth life, their ability to withstand extreme conditions, and their characteristic spectral signatures between 400~nm and 600~nm \citep[e.g.,][]{edge_et_al1997, vogl+bryant2011, schwieterman_et_al2015, vitek_et_al2017, coelho_et_al2022}. 

Additional possible surface biosignatures include bacteriorhodopsin, which produce energy by absorbing green light \citep[e.g.,][]{dassarma+schwieterman2021, sephus_et_al2022}, and bacteriochlorophylls (BChls), which collect light spanning a broad wavelength range from 400~nm to 1000~nm \citep{koblizek2015}. The anoxygenic phototrophic bacteria that use BChls likely arose during the Archean eon and are therefore interesting analogs for life detection in environments akin to that of early Earth \citep{sanroma_et_al2013}, but BChls are also used by bacteria that favor the oxic conditions of modern Earth \citep[e.g.,][]{kosmopoulos_et_al2023}. BChls could therefore be a surface biosignature for a wide range of planetary environments \citep{coelho_et_al2024}. For instance, the purple bacterium \emph{Blastochloris viridis} utilizes BChls to generate energy from light as red as $\gtrsim 1000$~nm, indicating that similar organisms might thrive on planets irradiated by M dwarfs \citep{wolstencroft+raven2002, kiang_et_al2007, coelho_et_al2024}. If present in sufficient quantities, extraterrestrial analogs to \emph{Blastochloris viridis} could be detectable by HWO-like facilities in the reflectance spectrum of Proxima Centauri b and perhaps other nearby planets \citep{metz_et_al2024}.

In addition to steady state features, surface biosignatures also include seasonal variations and transient features such as fluorescence and bioluminescence \citep{papageorgiou_et_al2007, haddock_et_al2010}. Fluorescence refers to the process by which life (including vegetation and microorganisms) absorbs UV light and re-radiates at lower energy levels, often in the red optical or near-infrared. For instance, a stellar flare might cause organisms to absorb UV radiation and then fluoresce in the red optical or near-infrared \citep{omalley-james+kaltenegger2018, omalley-james+kaltenegger2019}. These potential life signs could be distinguished from fluorescent minerals such as fluorite and calcite by their different fluorescence profiles \citep{kohler_et_al2021}. 

While fluorescence requires initial radiation from an outside source, bioluminescence occurs when organisms produce light directly. This phenomenon is observed in a variety of Earth life forms \citep{haddock_et_al2010} and is even detectable from space as enhanced emission from large patches of the ocean \citep[$\gtrsim 10,000~\rm km^2$;][]{miller_et_al2005}. Bioluminescence from large colonies of extraterrestrial marine organisms is therefore a possible surface biosignature \citep{seager_et_al2012}, but the patches of absorption seen in the Earth's oceans have low surface brightnesses that would complicate detection on distant worlds. 

In general, detecting surface biosignatures on exoplanets requires the ability to detect sharp changes in spectral properties throughout the accessible wavelength range. Although Earth organisms provide a preview of the wide range of possible pigments and spectral signatures, Earth life provides only one set of examples. Surface biosignature detection with HWO should be as flexible as possible and open to the possibility that extraterrestrial pigments may have ``edges'' at different wavelengths from Earth pigments. Accordingly, planetary spectra should have a spectral resolution of roughly $10$~nm ($R \lesssim 130$; investigations of lower limit on $R$ underway). Gaps in wavelength coverage will limit the ability to detect edges, so the spectra should not have gaps near wavelengths known to be associated with key features of Earth life (e.g., vegetation red edge). Important wavelengths include 
325~nm (BChl~c, BChl~d), 
345~nm (BChl~e), 
375~nm (BChl~a), 
400~nm (BChl~b), 
420~nm (BChl~g), 
435~nm (Chl~a), 
$442-452$~nm (Chl~c), 
450~nm (BChl~d), 
$450 - 460$~nm (BChl~c, BChl~e), 
460~nm (Chl~b), 
$500 - 650$~nm (Green Earth Hypothesis), 
575~nm (BChl~g), 
$580 - 587$~nm (Chl~c), 
590~nm (BChl~a), 
$600 - 610$~nm (BChl~b), 
$630 - 632$~nm (Chl~c), 
650~nm (Chl~b), 
670~nm (BChl~g), 
$670 - 680$~nm (Chl~a in PSII), 
700~nm (Chl~a in PSI), 
$710 - 720$~nm (Chl~d in PSII), 
$710 - 725$~nm (BChl~e), 
720~nm (Chl~f), 
$725 - 745$~nm (BChl~d), 
740~nm (Chl~d in PSI), 
$740 - 755$~nm (BChl~c), 
788~nm (BChl~g), 
$790 - 810$~nm (BChl~a), 
$830 - 920$~nm (BChl~a), 
$835 - 850$~nm (BChl~b), 
and 
$1015-1040$~nm (BChl~b). 

Part of the motivation for observing large samples of planets is to capture planets at different stages of their evolution, which provides insurance against not detecting life because the timescale for life to emerge is longer than anticipated. For instance, the choice of 11\% for the breakthrough science case reflects the fact that terrestrial vegetation and therefore surface biosignatures on Earth are believed to have emerged during the Phanerozoic Eon, which began roughly 541 Myr ago and covers the most recent 11\% of our planet's history.

\breakthrough Like the primary Living Worlds science case discussed in Section~\ref{sssec:life}, this science case considers the requirements for achieving ``super-breakthrough'' science that would be even more transformative than breakthrough science. Revolutionary, super-breakthrough science, requires obtaining spectra of at least 40~potentially inhabited planets and achieving $SNR\sim 40$ near the Vegetation Red Edge. A non-detection of the vegetation red edge in this sample would be sufficient for excluding at $2\sigma$ the hypothesis that 11\% of planets have vegetation covering $\gtrsim 7.5\%$ of their surfaces. In addition, $SNR \sim 40$ NIR spectra of 30~planets would be needed to search for the anoxygenic NIR edge. If none of those planets display an NIR edge, that would be sufficient for excluding at $2\sigma$ the hypothesis that 32\% of planets have purple anoxygenic phototrophic mat or cyanobacterial microbial mat covering $> 7.5\%$ of their surfaces.

``Regular'' breakthrough science would require obtaining spectra at the same signal-to-noise ratio ($SNR\sim 40$) but for a smaller sample of 20~planets. In that case, a non-detection of the vegetation red edge would exclude at $2\sigma$ the hypothesis that 25\% of planets have surface vegetation coverage $\gtrsim 7.5\%$. For the NIR edge, a non-detection would rule out at $2\sigma$ the hypothesis that 32\% of planets have a higher surface coverage fraction ($>15\%$) of purple anoxygenic phototrophic mat or cyanobacterial microbial mat. Overall, the abstract and conclusions of \citet{parenteau_surface_hwo25} indicate that SNR $20-40$, $R\lesssim 130$ spectra at $500-1100$~nm would be needed to robustly detect the reflectance spectra of biopigments on planets resembling Archean, Proterozoic, and Phanerozoic Earth. 

\enabling Substantial progress would require spectra of $8-9$~planets for a non-detection of vegetation red edge to exclude at $2\sigma$ the hypothesis that 50\% of planets have vegetation covering $\gtrsim 20\%$ of their surfaces.

\subsubsection{Detecting alien living worlds and photosynthetic life using imaging polarimetry with HWO (SCDD-LW-7)}
\label{sssec:polbio}
\sleads{Svetlana Berdyugina, 
Lucas Patty, 
Jonathan Grone,
Brice Demory,
Kim Bott, 
Vincent Kofman,
Giulia Roccetti, 
Kenneth Goodis Gordon, 
Frans Snik, 
Theodora Karalidi, 
Victor Trees, 
Daphne Stam, 
Mary N. Parenteau 
}

As discussed by \citet{berdyugina_imaging_hwo25}, this science case explores the power of multi-wavelength coronagraphic imaging polarimetry for robustly detecting life on terrestrial planets within the habitable zones of their host stars. Specifically, the science case aims to investigate the prevalence of photosynthetic life in the universe and how that alien photosynthetic life might differ from photosynthetic life on Earth. The study will also explore related topics such as the frequency at which atmospheres, oceans, and life occur on rocky planets. If photosynthetic life is detected, the observations could also be used to look for possible correlations between planet properties, host star properties, and the abundance and type of photosynthetic life. 

From the geological record, we know that photosynthetic life has existed on Earth for billions of years. Via laboratory investigations and remote sensing, scientists have established that terrestrial photosynthetic life has clear spectropolarimetric signatures. One of the most notable biosignatures is an extremely strong linear polarization (tens \%) associated with broad absorption bands of biological pigments (biopigments) driving photosynthesis in various organisms. Additionally, the homochirality (i.e., preference for a certain handedness of non-symmetric molecules) of 
biopigments and other complex macromolecules produces unique circular-polarization signatures. Accordingly, high-contrast low-resolution spectropolarimetry or multi-band imaging polarimetry of exoplanets using a facility like HWO could provide a novel opportunity for a robust discovery of life on exoplanets.

This science case employs a comprehensive approach consisting of two surveys and two follow-up observing programs. 
A more detailed description of this science case can be found in \citet{berdyugina_et_al2025}. Work is underway to determine the necessary observing time using the HWO Exposure Time Calculator.

Like the science case described in Section~\ref{sssec:surface_water}, Survey~1 would identify potentially habitable planets through detection of atmospheres, clouds and liquid surface water (oceans). These features would be identified by the strong linear polarization signatures ($20-40$\%) produced by Rayleigh scattering, rainbows and cloudbows, and ocean glint \citep[e.g.,][]{trees+stam2019, vaughan_et_al2023, goodis_gordon_et_al2025, roccetti_et_al2025a}. This survey would require $2-3$ visits per HZ rocky planet at $\leq 3$ selected orbital phases and any planetary axial rotation phases. 

Survey~2 would identify candidate inhabited worlds among potentially habitable planets by searching for prominent linear polarization signatures ($\le 80\%$) associated with strong and broad absorption bands reminiscent of terrestrial biopigments \citep{berdyugina_et_al2016}. 
This survey would require a single visit per potentially habitable planet, near an orbital quadrature (half-disk illumination), with measurements at $5-10$ evenly distributed planetary axial rotation phases. These observations would complement the spectroscopic searches of surface biosignatures proposed in Section~\ref{sssec:surface_bio}. 

The first follow-up observing program would obtain multi-color surface maps of living worlds, determine the distribution and abundance of alien photosynthetic organisms with exo-biopigments and correlate their properties with the atmospheric and surface compositions \citep{berdyugina+kuhn2019}. This follow-up program requires $10-20$ visits per target at selected orbital phases, each covering one planet rotation at $5-10$ evenly distributed phases. The second follow-up observing program would employ high-sensitivity circular polarization to verify the homochirality of exo-biopigments \cite[e.g.,][]{patty_et_al2021}.

\breakthrough Achieving breakthrough progress requires obtaining multi-band or low-resolution linear polarimetry at a planet-to-star flux ratio (contrast) of 10$^{-10}$ at the angular separations relevant for terrestrial planets in the habitable zones of FGKM stars. Assuming that this contrast could be achieved at angular separations $\ge$2$\lambda/D$ ($\lambda=550$\,nm, $D=8$\,m), a survey of targets within 30\,pc may result in detecting $20-156$ HZ rocky planets including $10-78$~potentially habitable planets. Depending on the frequency of photosynthetic life, the survey could result in the detection of $2-15$~inhabited worlds with biosignatures indicative of photosynthetic life. These estimates are based on assumed occurrence rates which would be verified by the proposed observations. 

For detected planets, simultaneous polarimetric observations should be obtained in at least four spectral bands with bandwidths of $10-15\%$ (or spectral resolution $\sim 50$). For Survey~1 (detection of potentially habitable planets), the bands should be centered at 350, 450, 550, and 850~nm and linear polarization should be measured with 2\% precision and 5\% accuracy. The 350~nm band is needed for Rayleigh scattering (anticipated linear polarization $\sim40\%$) while the 850~nm band would be used for ocean glint (anticipated linear polarization $\sim35\%$). Rainbows could produce linear polarization signals of roughly 20\% over the $350-850$~nm spectral range. 

For Survey~2 (detection of inhabited planets) and the first follow-up program (characterization of inhabited planets), the bands should be centered at 450, 550, 650 and 850~nm (i.e., the band at 350~nm should be replaced by one at 650~nm). Additionally, the precision and accuracy of linear polarization measurements should be improved to 1\% and 2\%, respectively. If linear spectropolarimetry is used instead of broad-band polarimetric imaging, then the three bands could cover $400-550$~nm, $550-700$~nm, and $700-850$~nm at $R\sim50-100$. 

For the second follow-up program (verifying homochirality on inhabited planets), full Stokes polarimetry would be required. $R\sim100$ spectra should be obtained at $600-750$~nm and circular polarization should be measured with 0.01\% precision and 0.1\% accuracy. The observations should be timed to occur at the orbital and axial phases at which exo-photosynthetic biosignatures are most detectable. The observations will likely need to be repeated $2-3$ times at the optimal observing geometry to build up sufficient signal-to-noise. Accordingly, this science case would strongly benefit from a large instantaneous field-of-regard to maximize observation opportunities. 

\enabling Enabling science would require observing targets within 15\,pc. Such a survey could yield detections of $5-64$ HZ rocky planets including $2-32$ potentially habitable planets. If photosynthetic life is found elsewhere in the universe, perhaps up to 6~planets would exhibit photosynthetic biosignatures. For Survey~1, simultaneous observations in three 20\% bands centered at 350, 450, and 850~nm would suffice. The absolute accuracy and precision of the linear polarimetry could be reduced to 5\% and $10\%$, respectively. For Survey~2 and the first follow-up program, the 350~nm filter should be replaced by a different 20\% filter centered at 550~nm. Additionally, the precision and accuracy of the linear polarization measurements should be improved to 2\% and 5\%, respectively. For the second follow-up program, $R\sim50$ spectra should be obtained at $600-750$~nm and circular polarization should be measured to 0.03\% precision and 0.5\% accuracy.

\subsubsection{Breathing Worlds: Leveraging Seasonality for Exo-Earth Biosphere Characterization with HWO (SCDD-LW-13)}
\label{sssec:seasonality}
\sleads{\'{E}milie A. Lafl\`{e}che, Anna Grace Ulses, Samantha Gilbert-Janizek, Jonathan Jernigan, Nicholas Wogan, Michael D. Himes, Stephanie L. Olson, Edward W. Schwieterman, Mary N. Parenteau, Joshua Krissansen-Totton, Avi Mandell}

Just as our own planetary biosphere exhibits periodic variations in atmospheric carbon dioxide concentration as vegetation grows and dies throughout the year\footnote{\url{https://climate.nasa.gov/vital-signs/carbon-dioxide/?intent=121}}, exo-biospheres might also display seasonal variations in atmospheric composition. The science case presented by \citet{lafleche_hwo25} aims to detect and measure seasonal variations that may be indicative of biological cycles. Possibilities include changes in atmospheric abundances of biologically-relevant gases \citep[e.g., small variations in O$_2$ for planets with generally low O$_2$ abundances leading to large changes in the strength of the O$_3$ Hartley-Huggins band at UV wavelengths;][]{olson_et_al2018} and variations in surface reflectance spectra due to seasonal growth and decay of vegetation (see Section~\ref{sssec:surface_bio}). In addition, periodic variations in albedo could be caused by seasonal precipitation and melting of snow and ice or changes in cloud cover, all of which could be signs of a hydrological cycle. Future work could explore the power of spectra extending redward of 1650~nm to detect seasonal changes in the abundance of CH$_4$. The targets observed for this science case will be the very best Earth-like candidates identified by the science case described in Section~\ref{sssec:life}, so their orbital properties should be well-characterized.

\breakthrough For breakthrough progress, planetary spectra should cover the UV, visible, and NIR. Coverage as blue as 200~nm will capture the important Hartley-Huggins band of O$_3$ at $200 - 350$~nm. This broad spectral range also captures the vegetation red edge near 700~nm (see Section~\ref{sssec:surface_bio}). At $\sim200-350$~nm, the observations should have SNR$\sim20$ to permit measuring O$_3$ volume mixing ratios to $10^{-10}$ precision. At $\sim700$~nm, the observations should have higher SNR ($\sim40$) to support detecting seasonally-induced differences in vegetation coverage of $\geq7.5\%$ for high-obliquity planets. 

This science case was developed using a notional observational strategy of $R\sim7$ spectra over $200-400$~nm, $R\sim140$ over $400-1000$~nm, and $R\sim 70$ over $1000-1800$~nm) but \citet{lafleche_hwo25} also explored the impact of higher spectral resolution at NUV and visible wavelengths ($R=500$ and $R=7600$, respectively). Additional work is needed to determine the optimal configuration required to address this science case. A reasonable assumption might be $R\sim 7$, $SNR\gtrsim20$ spectra at $200-350$~nm for measuring the volume mixing ratio of O$_3$ and $R\sim140$ spectra at $400-1000$~nm. The latter wavelength range is a conservative assumption because the VRE is a much narrower feature than the full 600-nm width of the $400-1000$~nm bandpass used for the simulations. For this paper, we will adopt the even more conservative assumption of using the full spectral regions considered in the simulations (i.e., $250-400$~nm and $400-1000$~nm).

Observations should be obtained simultaneously because of the increased temporal resolution and the ability to better pinpoint possible seasonal variations. Ideally, planets should be observed at $\ge 4-6$ different orbital locations. Depending on the orbital geometry, the need to capture the planet at multiple phases may place demands on the inner working angle and the instantaneous field of regard. 

\enabling For enabling progress, observations at different bands could be obtained contemporaneously rather than simultaneously. Additionally, planets should be observed at $\ge2$~different orbital phases rather than $\ge4-6$ as needed for breakthrough progress.

\subsubsection{Testing False Positive Biosignature Scenarios with HWO (SCDD-LW-10)}
\label{sssec:fpbio}
\sleads{Edward W. Schwieterman, Joshua Krissansen-Totton, Jacob Lustig-Yaeger, Sukrit Ranjan, Natalie Hinkel, Eric Mamajek, Giada Arney, Ravi Kopparapu, Victoria Meadows, Stephanie Olson, Niki Parenteau, Living Worlds SWG}

As discussed by \citet{schwieterman_hwo25}, the purpose of this science case is to investigate the reliability of potential biosignatures by looking for signs of life on planets that are unexpected to be inhabited. Specifically, this investigation would search for oxygen (as both O$_2$ and O$_3$) in the atmospheres of planets that reside inside the inner edge of the conventional liquid water HZ and measure the frequency of Venus-like planets that have oxygenated atmospheres. To determine the location of a planet relative to the HZ will require measurements of planetary parameters including semimajor axis, eccentricity, and albedo. 

To understand the significance of a detection or non-detection of O$_2$ and O$_3$, the investigation would also search for atmospheric H$_2$O and CO$_2$. The detection of high levels of water vapor could indicate that the planet is currently experiencing a runaway greenhouse while high levels of carbon dioxide, such as those seen on modern Venus, could reveal that the planet previously experienced a runaway greenhouse. In addition, detection of O$_3$ could be an indicator that detected O$_2$ is abiotic rather than biotic \citep[e.g.,][]{calder_et_al2025}. 

\breakthrough Advanced ``breakthrough+'' progress would test the hypothesis that 5\% of terrestrial planets in the Venus zone have at least as much atmospheric ozone as Proterozoic Earth. This test would require full UV/Vis/NIR ($250-1300$~nm) atmospheric characterization of 59~potentially terrestrial exoplanets. 

Standard breakthrough progress would instead test the hypothesis that 10\% of potential Venus analogs have atmospheric O$_3$ levels equal to or greater than that of Proterozoic Earth. The required data would be NUV/Vis/NIR ($250-1300$~nm) atmospheric characterization of 29~potentially terrestrial planets. Spectral resolution of $R\sim7$ in the UV and $R\sim200$ in the visible and NIR would be sufficient. To detect O$_2$ on planets with O$_2$ partial pressures at or above that of modern Earth, the spectra should cover 690~nm (O$_2$-B), 760~nm (O$_2$), 1060~nm (O$_2 - $O$_2$ collisionally-induced absorption), and 1270~nm (O$_2$ and O$_2-$O$_2$ bands). For planets with O$_2$ partial pressures significantly higher than that of modern Earth, observations in the NUV will be particularly important for capturing the O$_2-$O$_2$ collisionally-induced absorption bands at 345~nm, 360~nm, and 380~nm. Visible spectra will be needed for the additional O$_2-$O$_2$ collisionally-induced absorption bands at 445~nm, 475~nm, 530~nm, 570~nm, and 630~nm. Finally, for planets with very low O$_2$ partial pressures, NUV coverage of the O$_3$ Hartley-Huggins band\footnote{\citet{schwieterman_hwo25} state that the Hartley-Huggins bands are at wavelengths of $200-350$~nm, but they request NUV observations over a redder wavelength range from 250~nm to 400~nm.} ($250 - 400$~nm) will be essential. In all cases, the location of the target planets inside the inner edge of the HZ will demand the ability to obtain spectra at very small inner working angles. 

\enabling The hypothesis tested for enabling progress is that 20\% of potential Venus analogs have atmospheric O$_3$ at least as high as that of modern Earth. For enabling progress, the target sample would be reduced to 14~potentially terrestrial planets. Additionally, the wavelength range would be narrowed to $400-800$~nm. Finally, the spectral resolution at Vis/NIR wavelengths would be reduced to $R\sim100$.

\subsubsection{Masses of Potentially Habitable Planets Characterized by the Habitable Worlds Observatory (SCDD-LW-12)}
\label{sssec:mp}
\sleads{Kaz Gary, B. Scott Gaudi, Eduardo Bendek, Ty Robinson, Renyu Hu, Breann Sitarski, Aki Roberge, Eric Mamajek}

\citet{gary_hwo25} discusses the importance of determining planet masses for identifying potentially habitable planets and analyzing their spectra for potential habitability indicators and biosignatures. They focus specifically on the possibility of determining planet masses directly with HWO via highly precise astrometry of planet host stars. These astrometric mass constraints could supplement radial velocity mass constraints for stars amenable to high-precision radial velocity observations or supplant them entirely for stars that are poor targets for EPRV observations due to fast rotation rates or high activity levels (see Section~\ref{ssec:tss} for more details.)

Motivated by the theoretical results of \citet{damiano_et_al2025} that mass errors smaller than 10\% are needed to robustly assess planetary habitability, \citet{gary_hwo25} adopt 10\% mass precision as the target requirement for breakthrough level progress. For a typical system, realizing this mass precision necessitates measuring astrometric signals with a precision of $0.03~\mu$as per epoch. Analyzing the likely error terms and assuming notional properties for a high-resolution imager aboard an HWO-like telescope (field of view = $6' \times 6'$, Nyquist-sampled PSF, 30-min exposures, 25\% throughput, 6-m telescope), \citet{gary_hwo25} determine that astrometric observations are best made in the Gaia $G$ band and that the achievable astrometric mass precision is primarily limited by the availability of suitable reference stars. 

Accordingly, this science case is a potential driver for a larger instrumental field-of-view to increase the number of reference stars visible per observation as well as a larger instantaneous field-of-regard to maximize the orbital phase coverage of astrometric observations. The need for a suitable sample of reference stars may also be an argument for favoring non-RV biosignature targets near the galactic plane (i.e., galactic latitude $b \lesssim 40^\circ$) where the density of background stars is higher. Additionally, this science case requires a sophisticated understanding of the nature and stability of optical distortions in the telescope and instrument as well as the implementation of calibration schemes to reduce the astrometric error \citep[e.g.,][]{guyon_et_al2012, bendek_et_al2020, bendek_hwo25_vol2}. 

\breakthrough Achieving breakthrough progress requires measuring planets with masses as low as $0.1 M_\oplus$ (i.e., comparable to the mass of Mars). The corresponding observational requirement is an astrometric precision of $0.03~\mu$as per epoch. \citet{gary_hwo25} note that a plate scale of 11~mas/pixel would be required for observations in Gaia G by an instrument with a field of view of 36~arcmin$^2$ on a telescope with a diameter of 6~meters. 

\enabling For enabling science, astrometric observations should be capable of measuring planet masses as small as $1 M_\oplus$, which necessitates a per epoch astrometric precision of $0.3~\mu$as.

\subsection{Additional Science}
While the science cases developed by the working groups span a broad range of research areas, there are certainly additional topics that could be explored with HWO. For instance, the second volume of the HWO25 Conference Proceedings includes a variety of ``Broader Science Contributions'' that complement the science programs described earlier in this section. Many of these contributions focus on the search for life. For instance, \citet{haimuri_hwo25_vol2} identified a set of 20~stars they deem particularly interesting for astrobiological investigations while \citet{chavezdagostino_hwo25_vol2} analyzed archival NUV and visible data of stars to assess the potential impact of stellar flares on the atmospheric photochemistry of potentially habitable planets. By improving understanding of possible target stars, both \citet{haimuri_hwo25_vol2} and \citet{chavezdagostino_hwo25_vol2} address the goals of the Target Stars and Systems subgroup (Section~\ref{ssec:tss}). 

Continuing the theme of exploring the effects of stellar radiation on planetary atmospheres, \citet{sumida_hwo25_vol2} simulated the effects of flares on the M~dwarf planet TOI-700~d and \citet{kovacs_hwo25_vol2} modeled atmospheric escape from two planets orbiting Sun-like stars. \citet{young_hwo25_vol2} also conducting photochemical modeling and underscored the importance of UV coronagraphy like that described in Section~\ref{sssec:hc_uv_spec} for identifying potential biosignatures on planets similar to Proterozoic Earth. Emphasizing the importance of fieldwork and laboratory experiments, \citet{dincecco_hwo25_vol2} explained how geological investigations of Venus-like environments could be useful for identifying volcanic and tectonic activity on exoplanets. 

Several contributions focused specifically on how to interpret observations of potentially habitable or inhabited planets: \citet{barbosa_hwo25_vol2} presented a one-dimensional convolutional neural network; \citet{fisher_hwo25_vol2} discussed the merits of systems science; and \citet{meadows_hwo25_vol2_framework} described the statistical framework built by the Virtual Planetary Laboratory. 

More generally, \citet{fossati_hwo25_vol2} reviewed the potential benefits of the proposed Pollux high-resolution spectrograph and spectropolarimeter for exoplanet science, highlighting the previously discussed studies of planetary atmospheres \citep[][SSiC-24, Section~\ref{sssec:highres_atmos}]{cubillos_hwo25}, planetary magnetic fields \citep[][SSiC-34, Section~\ref{sssec:bfield}]{strugarek_hwo25}, giant planet aurorae \citep[][SSiC-25, Section~\ref{sssec:aurorae}]{chaufray_hwo25}, and white dwarfs \citep[][EE-7, Section~\ref{sssec:wds}]{xu_siyi_hwo25}. Similarly, \citet{malbet_hwo25_vol2} extolled the importance of extremely precise astrometry for measuring the masses of potentially habitable planets \citep[e.g.,][LW-12, Section~\ref{sssec:mp}]{gary_hwo25} and advancing knowledge of dark matter. 

At much smaller scales, \citet{holler_hwo25_vol2} described an investigation of the binary properties and ring systems of trans-Neptunian Objects, which which would extend the studies of solar system small bodies described in Section~\ref{ssec:ssic_scdds}. In addition, the simulations of habitable moon spectra presented by \citet{oza_hwo25_vol2} complement the investigation of potentially habitable icy worlds proposed in SSiC-17 \citep[][Section~\ref{sssec:id_oceans}]{quick_hwo25} and the search for exomoons described in SSiC-32 \citep[][Section~\ref{sssec:exomoons}]{limbach_et_al2024, limbach_et_al2026}. Likewise, the investigation of the composition and physical conditions of the galactic ISM proposed by \citet{ritchey_hwo25_vol2} would supplement the galactic and extragalactic ISM studies proposed by EE-2 \citep[][Section~\ref{sssec:dust_extinction}]{paladini_hwo25} and EE-11 \citep[][Section~\ref{sssec:dust_uv}]{roman-duval_hwo25}. 

At extragalactic distances, \citet{torralba_hwo25_vol2} would further studies of the distribution of dark matter on small scales, while \citet{u_hwo25_vol2} would conduct a census of intermediate-mass black holes. For more massive black holes, \citet{biedermann_hwo25_vol2} considers the potential of the proposed Pollux instrument concept to reveal the nature of the quasar Markarian~231, which is a potential binary system . The detailed analysis of \citet{biedermann_hwo25_vol2} is the focused counterpoint to the overarching study of binary black holes proposed by GG-19 \citep[][Section~\ref{sssec:smbh_pol}]{marin_et_al2025}.

\section{Key Observations \& Capabilities}
\label{sec:swg_obs}
The science cases discussed in Section~\ref{sec:scdds} were developed quasi-independently, and therefore many science cases have overlapping goals or observational needs. In this section, we synthesize the aims and goals of the science cases identified by each working group to identify the core observational datasets needed to address the majority of the science cases identified in Section~\ref{sec:scdds}. The synthesized science cases presented in this section may be useful for future investigations of the trade space for HWO and other upcoming facilities. Unless otherwise stated, the observational capabilities discussed in this section refer to those needed for breakthrough-level science. 

\subsection{Growth of Galaxies}
\label{ssec:gg_merged}
\begin{figure*}
 \centering
 \includegraphics[width=\linewidth]{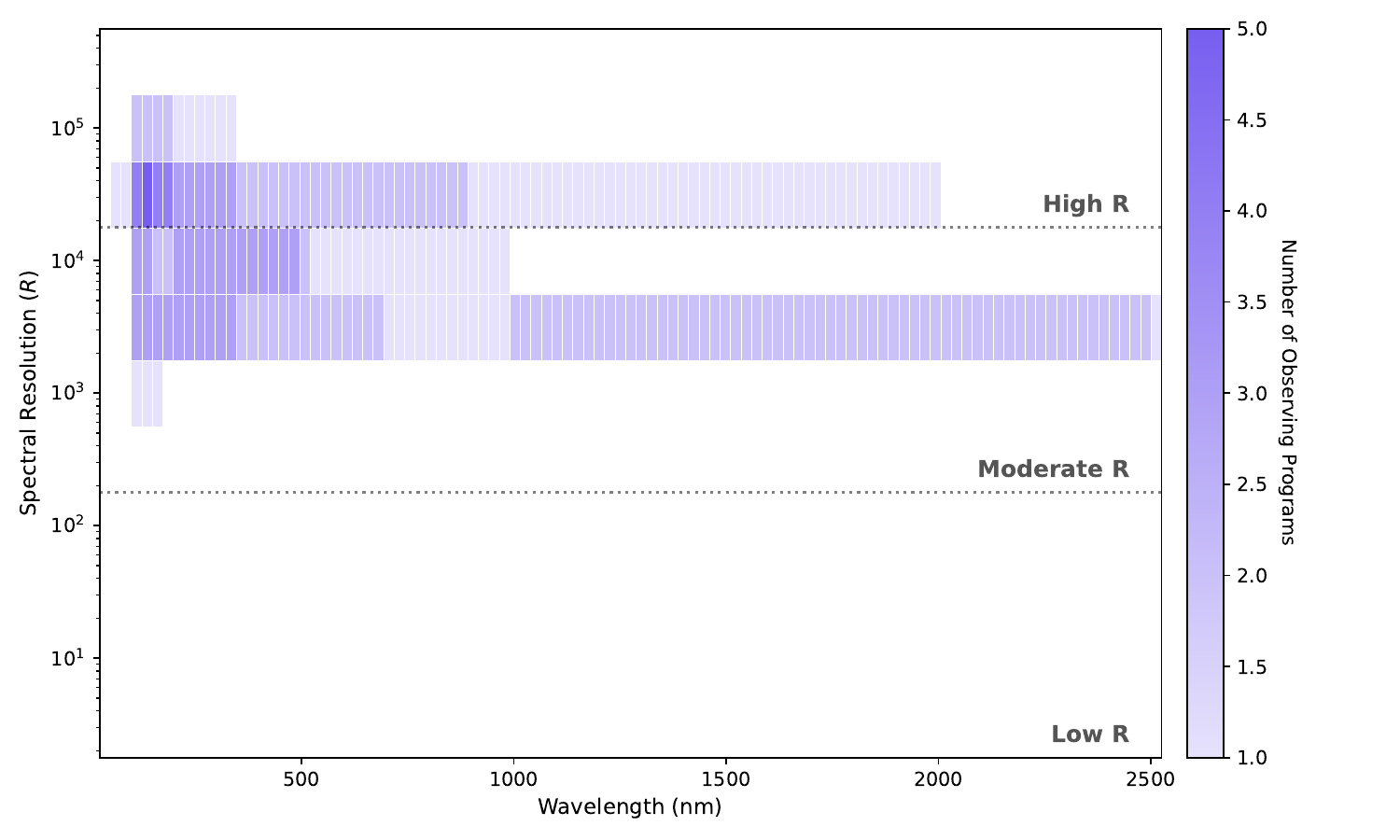}
 \caption{Demand by GG science cases for spectroscopic observations at particular combinations of spectral resolution ($R$) and wavelength ($\lambda$). As shown in the colorbar, darker colors indicate higher demand. White regions indicate combinations of $R$ and $\lambda$ that are not needed for the GG science cases presented in this paper. }
 \label{fig:gg_r_wave}
\end{figure*}

Table~\ref{tab:gg_scdds} summarizes the science questions, target samples, and observational needs for the 17~GG SCDDs. As mentioned in Section~\ref{ssec:gg_scdds}, common themes include the flow of material in and out of galaxies, reionization, the properties of black holes, and the nature of dark matter. In many cases, the desired targets (e.g., dwarf galaxies at low redshift) or observational approaches (e.g., absorption-line spectroscopy towards QSOs) are similar, thereby presenting the possibility that the datasets for various science cases could partially overlap. In Table~\ref{tab:gg_remap}, we show a remapped version of the GG science topics, target samples, and observational needs that groups observations by observational capabilities. This condensed version may be useful for future investigations of design trades. Additionally, in Figure~\ref{fig:gg_r_wave}, we show how the various observations needed for GG~science cases map onto spectral resolution versus wavelength (i.e., $R-\lambda$) space.

As shown in Table~\ref{tab:gg_remap}, the GG science cases require six types of observations: 
\begin{itemize}
 \item \textbf{GG-A} (Section~\ref{sssec:gg_a}): IFS observations at moderate spectral resolution
 \item \textbf{GG-B} (Section~\ref{sssec:gg_b}): IFS observations at high spectral resolution
 \item \textbf{GG-C} (Section~\ref{sssec:gg_c}): MOS observations at moderate spectral resolution
 \item \textbf{GG-D} (Section~\ref{sssec:gg_d}): MOS observations at high spectral resolution
 \item \textbf{GG-E} (Section~\ref{sssec:gg_e}): spectropolarimetric observations
 \item \textbf{GG-F} (Section~\ref{sssec:gg_f}): photometric observations
\end{itemize} 
When categorizing observations, we split the science cases that required multiple datasets into multiple components. For instance, GG-17 (Section~\ref{sssec:cgm_elm}) needs high spectral resolution observations with both an IFS (labeled as GG-17a and assigned to GG-B) and a MOS (GG-17b, which is part of GG-D). For GG~classification, we set $R=10^{4.25}\approx 17800$ as the boundary between moderate and high spectral resolution. In all but one case, we selected the instrument type listed in the SCDD. The exception was GG-4 (Section~\ref{sssec:bh_quiescent}), which we assigned to moderate-$R$ IFS rather than moderate-$R$ MOS due to the high spatial resolution ($\theta=10$~mas) requested by \citet{pacucci_hwo25}. We classified programs without stated instrument preferences based on their stated observational goals and similarity to other science cases. The specific spectral resolution, spatial resolution, and wavelength range needed for each investigation varies slightly. In Table~\ref{tab:gg_remap}, we display those differences by setting ``default'' observation settings and then indicating which specifications would be strengthened or softened for various observing programs. 

\subsubsection{GG-A: IFS Observations at Moderate Spectral Resolution}
\label{sssec:gg_a}
We denote the first type of observations needed for GG science cases as ``GG-A.'' Baseline GG-A observations are IFS spectra obtained at moderate spectral resolution ($R=10,000$) and high spatial resolution ($\theta \sim 20$~mas) over a $10'' \times 10''$ field of view. The default wavelength range extends from 90~nm to $2.5\mu$m. 

As shown in Table~\ref{tab:gg_remap}, baseline GG-A observations and slightly modified versions of GG-A observations could address multiple science cases. The resulting data would probe the properties of supermassive black holes (GG-2, GG-3, GG-4); the influence of AGN feedback on their host galaxies (GG-5b); the means by which ionizing radiation is produced within and escapes from galaxies (GG-9a); and how matter flows within, into, and out of galaxies (GG-17a). In addition, GG-A observations of strong lensing systems would contribute to understanding the structure of dark matter on small scales (GG-13a).

For GG-A observations of galaxies, the largest target sample is needed for GG-9a \citep{citro_hwo25}, which aims to calibrate indirect indicators of LyC escape. As discussed in Section~\ref{sssec:lyman_indirect}, that science case demands $R > 10,000$ spectroscopy for 40,000 low-redshift galaxies. That large sample could include the smaller samples needed for the other observations in GG-A. The target sample should include the following nested sub-populations:
\begin{enumerate}
 \item GG-5b: 2000~galaxies for studies of AGN feedback \citep[][Section~\ref{sssec:agn_outflow}]{zhang_hwo25}
 \begin{enumerate}
 \item GG-4: 100~dwarf galaxies with masses of $10^6 - 10^9 M_\odot$ at distances of $10-30$~Mpc to assess the relationship between SMBHs and low-mass host galaxies \citep[][Section~\ref{sssec:bh_quiescent}]{pacucci_hwo25}
 \begin{enumerate}
 \item GG-2: Low-metallicity galaxies with masses $<10^{8.5} M_\odot$ to investigate the origin and influence of SMBHs \citep[][Section~\ref{sssec:bh_mass_spin}]{cann_hwo25}
 \end{enumerate}
 \item GG-3: $>10$~Seyfert galaxies within 65~Mpc to investigate dusty tori around SMBHs \citep[][Section~\ref{sssec:bh_torus}]{gorjian_hwo25} 
 \item GG-17a: 12~galaxies spanning a broad range of masses, star formation levels, and orientations to study the exchange of matter and energy \citep[][Section~\ref{sssec:cgm_elm}]{burchett_hwo25}
 \end{enumerate}
\item GG-13a: 2000~strong lensing systems to investigate the nature of dark matter \citep[][Section~\ref{sssec:dm_lensing}]{he_hwo25} 
\end{enumerate}
In the list above, the low-metallicity, low-mass galaxy sample needed by GG-2 is envisioned as a subset of the 100-galaxy sample needed by GG-4. However, that hierarchy could flip depending on the number of galaxies needed for GG-2. Additionally, both of those samples could also contain portions of the 12-galaxy sample needed by GG-17a. 

While the galaxy sample needed by GG-9a could encompass all of the smaller galaxy samples, the data needed for that science case alone would be insufficient to address all of the goals of GG-A programs. In particular, the observations obtained for GG-9a would cover only $100-510$~nm. This blue limit is adequate for GG-5b, GG-4, and GG-13a, but the other GG-A programs would require additional data at wavelengths as short as 90~nm, as shown in Table~\ref{tab:gg_remap}. 
Furthermore, all GG-A programs except GG-17a would also require data at longer wavelengths. Depending on the program, the desired red cut-off varies from 700~nm to 5000~nm (see~Table~\ref{tab:gg_remap}). Additionally, the spatial resolution requested for GG-A observations varies from 5~mas for GG-3 to 50~mas for GG-2 and GG-17a.

\subsubsection{GG-B: IFS Observations at High Spectral Resolution}
\label{sssec:gg_b}
GG science cases also necessitate IFS observations at higher spectral resolution. For these ``GG-B'' observations, we adopt higher spectral and spatial resolution ($R=30,000$ and $\theta = 10$~mas, respectively) over a narrower wavelength range ($90-500$~nm) than GG-A. For both GG-B and GG-A, we assume a default field of view of $10''\times10''$. 

GG-B observations of galaxies would address multiple science cases. These data would provide additional insight into the flow of material within, into, and out of galaxies (GG-5a, Section~\ref{sssec:agn_outflow}) and the evolution of the production and escape of ionizing radiation over cosmic time (GG-6, Section~\ref{sssec:resolve_reionization} \& GG-7, Section~\ref{sssec:lyman_escape}). As for GG-A, the target samples could be nested to improve observational efficiency. 

For instance, the large sample of 10,000 unlensed galaxies at $z<1$ needed by GG-6 \citep[][Section~\ref{sssec:resolve_reionization}]{xu_reionization_hwo25} to probe reionization could include enough galaxies at $0\le z\le 0.3$ for the more focused study of Lyman escape at low redshift in GG-7 \citep[][Section~\ref{sssec:lyman_escape}]{carr_hwo25}. Both GG-6 and GG-7 request a red cut-off of 200~nm, but GG-6 requires a bluer UV cut-off of $50$~nm. Data obtained for GG-6 would therefore naturally include the $95-200$~nm range needed by GG-7. Additionally, while GG-7 expresses a preference for higher spectral resolution ($30,000<R<100,000$) than GG-6 ($R\sim30,000$) the $R=30,000$ spectral resolution baselined by GG-B would be formally sufficient for both programs. Accordingly, it is reasonable to assume that both GG-6 and GG-7 could be accomplished by obtaining the observations requested by \citet[][Section~\ref{sssec:resolve_reionization}]{xu_reionization_hwo25}.

The sample could also contain at least 300 AGN to investigate binary SMBHs for GG-19 \citep[][Section~\ref{sssec:smbh_pol}]{marin_et_al2025} and 50~AGN to probe AGN feedback more generally for GG-5a \citep[][Section~\ref{sssec:agn_outflow}]{zhang_hwo25}. The two AGN programs request the same blue cut-off (100~nm) and spectral resolution ($R>50,000$), but GG-19 requires spectropolarimetry while GG-5a does not. GG-19 also requires a longer red cut-off (2000~nm versus 900~nm for GG-5a). Accordingly, data obtained for GG-19 would be sufficient for GG-5a, but the reverse would not be true. Fortuitously, GG-19 requires a larger sample than GG-5a, so it may be possible for some or all of the 50 targets in the GG-5a sample to be part of the GG-19 sample of $\ge300$~AGN.

\subsubsection{GG-C: MOS Observations at Moderate Spectral Resolution}
\label{sssec:gg_c}
For other investigations, a MOS is preferable to an IFS because of the distribution of targets on the sky. We divide MOS observations into two categories based on spectral resolution. At moderate spectral resolution, we define GG-C as MOS spectroscopy at spectral resolution $R=3000$ and spatial resolution $\theta=225$~mas at \mbox{$94-350$~nm.} The notional field-of-view is $2'\times2'$, which is significantly larger than the fields of view used for the IFS observations (GG-A and GG-B).

The two science cases that could be addressed by \mbox{GG-C} observations are GG-11 \citep[][Section~\ref{sssec:ionizing_lf}]{mccandliss_hwo25} and GG-16b \citep[][Section~\ref{sssec:disk_cgm}]{borthakur_hwo25}. Compared to the GG-C baseline configuration, GG-11 demands a larger field of view (three $2'\times2'$ fields covering a total area of 12 arcmin$^2$), reduced wavelength coverage ($94-180$~nm), and lower spectral resolution ($R=1000$). GG-16b also needs a larger field of view ($6'\times 6'$) but that coverage could be obtained by tiling either the $2'\times2'$ GG-C baseline field or the $3\times(2'\times2')$ configuration requested by GG-11. Both GG-11 and GG-16b require observations of galaxies, so the samples could potentially overlap. GG-11 needs observations of 2500~galaxies with a diverse range of properties while GG-16b specifically demands observations near QSOs. Due to the broader wavelength range, higher spectral resolution, and larger field of view needed for GG-16b than for GG-11, data obtained for GG-16b would meet the guidelines of GG-11 while the inverse would not be true. 

\subsubsection{GG-D: MOS Observations at High Spectral Resolution}
\label{sssec:gg_d}
Achieving the objectives of GG science cases also demands MOS observations at higher spectral resolution. For these GG-D observations, we assume the same $94-350$~nm wavelength coverage as for \mbox{GG-C} but increase the spectral resolution substantially to $R=40,000$. We also decrease the spatial resolution to 500~mas and increase the field of view to $6'\times6'$. 

For GG-16a \citep[][Section~\ref{sssec:disk_cgm}]{borthakur_hwo25}, GG-D observations could probe the behavior of gas and metals at the disk-CGM interface. GG-D observations could also investigate the exchange of matter and energy within galaxies for GG-17b \citep[][Section~\ref{sssec:cgm_elm}]{burchett_hwo25}. Finally, for GG-18 \citep[][Section~\ref{sssec:agn_feedback}]{tillman_hwo25}, GG-D data could be used to study the relative roles of dark matter and galactic feedback in shaping structure in the universe. 

Both GG-16a and GG-18 require absorption spectra along QSO sightlines. GG-16a requires a higher spectral resolution ($R\sim100,000$) and lower spatial resolution ($\theta<1''$) while GG-18 requests a reduced wavelength range of $121.5-340.2$~nm. Accordingly, while some of the observations obtained for GG-16a could be included in the set of 750 sightlines at $z=0.7$ and $80-250$ sightlines at $z=1.6$ needed for GG-18, observations obtained for GG-18 would not satisfy the more stringent observational requirements of GG-16a without the addition of supplemental data. 

Science case GG-17 requires both IFS spectra (GG-17a, assigned to GG-A) and MOS spectra (GG-17b) of the same set of 12~galaxies. Compared to the GG-D baseline configuration, GG-17bneeds observations at lower spectral resolution ($R\sim20,000$) over a reduced wavelength range ($97.5-160$~nm and $279-290$~nm). Those observations would not be sufficient for either GG-16a or GG-18, but the higher spectral resolution data obtained for GG-16a may be useful for GG-17b. Additionally, data obtained for GG-18 could be supplemented by data at $97.5-121.5$~nm to cover the full wavelength range requested for GG-17b. However, given the need to match the GG-17b and GG-17a target samples, it may be simpler to treat GG-17b as fully independent from the other GG-D cases.

\subsubsection{GG-E: Spectropolarimetric Observations}
\label{sssec:gg_e}
One GG science case requires spectropolarimetry. Although this case is unique within GG, it is similar to other science cases within EE and SSiC. Accordingly, we define a separate category GG-E for spectropolarimetry. The default configuration is $R=120,000$ spectra at $100-1600$~nm with full Stokes polarimetry. The polarization error should be $<10^{-6}$. 

GG-E observations are needed for GG-19, which aims to study the interactions of binary supermassive black holes with their accretion disks \citep[][Section~\ref{sssec:smbh_pol}]{marin_et_al2025}. Compared to the default GG-E settings, GG-19 could use data obtained at lower spectral resolution ($R>50,000$) and relaxed polarimetric precision (polarization error $<10^{-4}$ rather than $<10^{-6}$). However, GG-19 would need data at $100-2000$~nm, which extends redder than the GG-E default of $100-1600$~nm.

\subsubsection{GG-F: Photometric Observations}
\label{sssec:gg_f}
The remaining GG investigations require photometry. We categorize these observations as GG-F. Baseline GG-F data includes UV/Vis imaging at $95-700$~nm at spatial resolution $\theta=20$~mas over a field of view of $4'\times4'$.

GG-F observations of galaxies would be useful for calibrating metrics of LyC escape as part of GG-9b \citep[][Section~\ref{sssec:lyman_indirect}]{citro_hwo25} and investigating the nature of dark matter for GG-12 \citep[][Section~\ref{sssec:dm_power}]{doppel_hwo25}. For GG-13b, GG-F observations of strong lensing systems would further contribute to studies of dark matter by constraining the dark matter halo mass function \citep[][Section~\ref{sssec:dm_lensing}]{he_hwo25}. Finally, GG-F observations of strong lenses would provide valuable insight into mergers of supermassive black holes for GG-14 (Section~\ref{sssec:smbh_mergers}). 

GG-9b requires observations of 40,000~faint galaxies at low redshift ($z<0.1$) while GG-12 needs observations of $>10$ Milky-way mass galaxies. Although the GG-9b sample is much larger than the GG-12 sample, the two investigations are focused on different mass ranges and therefore the Milky-way mass galaxies needed for GG-12 are unlikely to be suitable targets for GG-9b given that program's focus on faint, low-mass galaxies as potential drivers of reionization. Rather, the satellite galaxies detected around the Milky-way mass galaxies as part of GG-12 may be useful additions to the GG-9b sample of 40,000~faint galaxies. Science case GG-12 requests observations over a redder wavelength region than the default GG-E configuration (Vis/NIR versus UV/Vis), so satellite galaxies detected by GG-12 would need follow-up UV imaging to be added to the GG-9b sample. 

For GG-13b, the ideal dataset includes UV/Vis imaging over a narrower wavelength range of $200 - 500$~nm and higher spatial resolution of $5-10$~mas. For GG-14, the necessary dataset consists of photometric imaging at 650~nm or another wavelength between 550~nm and 900~nm. Although the spatial resolutions needed for the two science cases are similar, the wavelength ranges do not overlap and therefore the 1000~strong lensing systems needed for GG-14 cannot be a subset of the 2000~strong lensing systems needed for GG-13b unless the wavelengths of either or both of the subsamples are adjusted. The choice of 650~nm for constraining SMBH merger timescales was motivated by the desire to overlap with the $550 - 900$~nm wavelength range of Euclid's visible channel because target systems will be detected using Euclid data. Extending imaging for GG-13b slightly redder (e.g., to 600~nm) and adjusting the observation wavelength of GG-14 to match could potentially improve efficiency by allowing the resulting images to be used for both science cases. 

\subsection{Evolution of the Elements}
\begin{figure*}
 \centering
 \includegraphics[width=\linewidth]{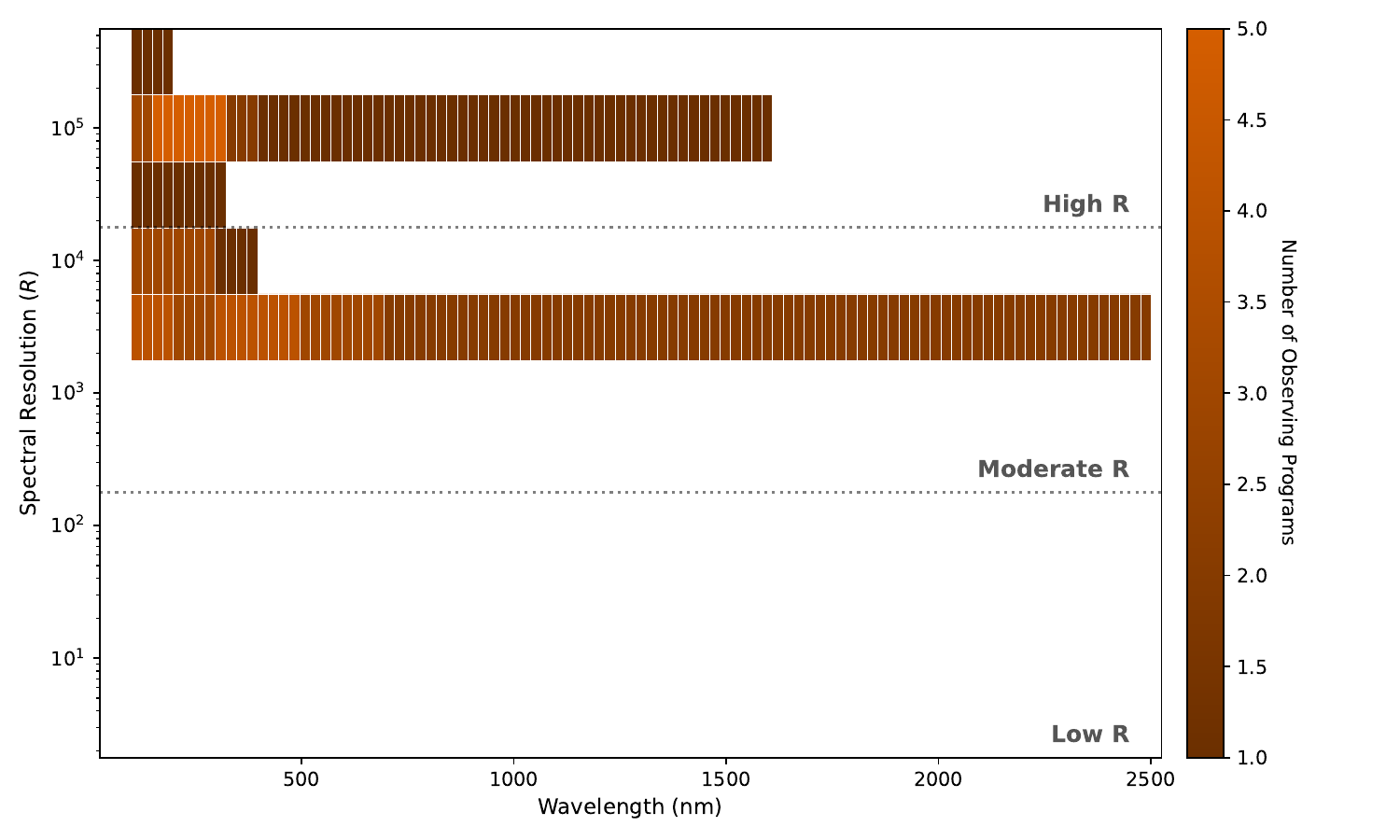}
 \caption{Same as Figure~\ref{fig:gg_r_wave} but for EE science cases.}
 \label{fig:ee_r_wave}
\end{figure*}

Table~\ref{tab:ee_scdds} summarizes the 13~EE SCDDs introduced in Section~\ref{ssec:ee_scdds}. The science cases consider the elemental compositions and properties of dust grains, exoplanets, stars, and galaxies as well as the expansion rate of the universe and fundamental physics. We highlight the range of spectral resolution and wavelength coverage needed for these science cases in Figure~\ref{fig:ee_r_wave}. As shown in Table~\ref{tab:ee_remap}, the 13~EE SCDDs can be divided into six groups based on the necessary observational capabilities: 
\begin{itemize}
 \item \textbf{EE-A} (Section~\ref{sssec:ee_a}): IFS observations
 \item \textbf{EE-B} (Section~\ref{sssec:ee_b}): MOS observations
 \item \textbf{EE-C} (Section~\ref{sssec:ee_c}): single-object spectroscopy at moderate spectral resolution
 \item \textbf{EE-D} (Section~\ref{sssec:ee_d}): single-object spectroscopy at high spectral resolution
 \item \textbf{EE-E} (Section~\ref{sssec:ee_e}): spectropolarimetric observations
 \item \textbf{EE-F} (Section~\ref{sssec:ee_f}): photometric observations
\end{itemize} 

\subsubsection{EE-A: IFS Observations}
\label{sssec:ee_a}
Three EE science cases require IFS observations of galaxies. We denote these observations as EE-A and adopt a baseline configuration of $R=10,000$ spectroscopy at $90-500$~nm. We assume that the IFS has a spatial resolution $\theta=20$~mas and covers a field of view of $10''\times10''$. These settings are identical to those used for GG-A except for the reduced long wavelength coverage ($90-500$ for EE-A versus $90-2500$ for GG-A). 

For science case EE-12 \citep[][Section~\ref{sssec:ism_uv}]{james_hwo25}, EE-A observations would be used to explore the distribution of elements within galaxies. For EE-1 \citep[][Section~\ref{sssec:massive_stars_lowZ}]{senchyna_hwo25} and EE-5 \citep[][Section~\ref{sssec:vms}]{martins_hwo25}, EE-A observations could probe the influence of massive stars on their environments. Given that EE-1, EE-5, and EE-12 request similar data, some observations could potentially be shared across those programs. For instance, EE-1 aims to study massive, low-metallicity stars in dwarf galaxies such as IZw18, which would meet the sample selection criteria for the study of nearby, low-metallicity, low-mass dwarf galaxies with high star formation rates (EE-12). In addition, EE-5 requires observations of very massive stars with a range of metallicities and distances of $0.8 - 15$~Mpc. The targets for EE-5 will therefore be within the 20-Mpc distance limit set for EE-12, meaning that the observations of low-metallicity environments obtained for EE-5 could also be used for EE-12.

However, there are slight differences between the observing strategies that would need to be addressed in order to share data between programs. In particular, EE-12 requires the baseline spectral resolution ($R>10,000$), while EE-1 and EE-5 could use lower resolution observations ($R>5000$). In addition, while all three science cases need UV observations, the blue cut-off is different for each program: 90~nm for EE-1, 95~nm for EE-12, and 100~nm for EE-5. The red cut-off varies even more significantly: 285~nm for EE-12, 500~nm for EE-1, and 700~nm for EE-5. 

Furthermore, the programs differ in their requested fields of view. For EE-12, achieving contiguous coverage over a broad $10'' \times 10''$ is critically important. The other two science cases require smaller fields of view of $3-10'' \times 3-10''$ (EE-1) and $3'' \times 3''$ (EE-5), which could be tiled to cover the full $10'' \times 10''$ needed for EE-12. Sharing data with all three programs would therefore necessitate $R>10,000$ spectra at $90-700$~nm and potentially tiling the smaller fields to build the full $10''\times10''$ field needed for EE-12. 

\subsubsection{EE-B: MOS Observations}
\label{sssec:ee_b}
The next set of EE programs are best addressed by MOS spectroscopy. For these observations, we define the EE-B baseline as $95-2500$~nm spectra at moderate spectral resolution ($R=3000$) and high spatial resolution ($\theta=10$~mas). We set the default field of view to $2'\times2'$.

This configuration is ideal for studies of galactic and extragalactic dust. For EE-2 \citep[][Section~\ref{sssec:dust_extinction}]{paladini_hwo25}, EE-B observations should be obtained for at least 22,515 diffuse sightlines towards bright OB stars and at least 126 dense sightlines towards blue AFG giants. EE-2 requires both diffuse and dense sightlines toward the Milky Way, Large Magellanic Cloud, Small Magellanic Cloud, M31, and at least one other Local Group galaxy as well as additional diffuse sightlines towards 9 more Local Group galaxies, and 5 Local Volume galaxies. For EE-11 \citep[][Section~\ref{sssec:dust_uv}]{roman-duval_hwo25}, the composition of galactic and extragalactic dust could be compared by observing 100 sightlines/galaxy for Local Group galaxies within 10~Mpc. Objective 1 of EE-11 requires $R>50,000$ spectra of Local Group galaxies as part of EE-11a while Objective~2 requires higher resolution spectra over a narrower wavelength region ($R>100,000$ at $115-315$~nm) to determine oxygen and carbon abundances for the Milky Way, LMC, and SMC as part of EE-11b. 

In theory, the smaller dataset needed for EE-11 could be part of the larger EE-2 dataset, but the spectral resolutions needed for EE-11a ($R\sim50,000$) and EE-11b ($R\sim100,000$) are much higher than the $R=3000$ baseline spectral resolution acceptable for EE-2. Conversely, using EE-11a or EE-11b observations for EE-2 would also be problematic because EE-2 requires coverage from the UV to NIR over the full $95-2500$~nm EE-B baseline while both components of EE-11 need only UV data. Using the same observations for EE-2, EE-11a, and EE-11b would require combining $R\sim50,000$ spectra at $95-115$~nm for EE-11a and EE-2, $R\sim100,000$ spectra at $115-315$~nm for all three programs, and $R=3000$ spectra at $315-2500$~nm for EE-2. 

\subsubsection{EE-C: Single-Object Spectroscopy at Moderate Spectral Resolution}
\label{sssec:ee_c}
Other EE science cases necessitate single-object spectroscopy at moderate spectral resolution. We denote this mode as EE-C and assume spectral resolution $R=10,000$ over a wavelength range of $110-500$~nm. 

This mode would be useful for two EE science cases: EE-3 \citep[][Section~\ref{sssec:flash_ccsne}]{andrews_hwo25} and EE-4a \citep[][Section~\ref{sssec:r_process_el}]{burns_hwo25}. Both of these programs aim to obtain rapid-response UV and blue optical spectroscopy of supernovae. EE-3 needs $110 - 300$~nm observations of core-collapse supernovae within one day of explosion and EE-4a requires observations of kilonovae within two days of explosion. Response times of $\leq 30 - 60$~minutes would be advantageous for both science cases. 

For these supernovae programs, the ability to obtain data very shortly after the observation is triggered is more important than the specific spectral resolution. Spectra with resolution $R\sim 10,000$ would be advantageous, but $R\sim 1000$ would be sufficient. Although we categorize these observations as EE-C for completeness, the more important constraints for these observations are slew speed, response time, and interrupt policies. 

\subsubsection{EE-D: Single-Object Spectroscopy at High Spectral Resolution}
\label{sssec:ee_d}

Additional EE science cases need higher resolution spectra of single objects. For these cases, we define EE-D as $R=100,000$ spectra at $90-310$~nm. EE-D observations of white dwarfs have the potential to constrain exoplanet compositions for EE-7a and fundamental physics for EE-7b; both programs are discussed by \citet[][Section~\ref{sssec:wds}]{xu_siyi_hwo25}. Additionally, EE-D observations of low-metallicity stars could constrain the nature of the first stars for EE-8 \citep[][Section~\ref{sssec:first_stars}]{roederer_nature_hwo25} and the production of r-process elements for EE-9 \citep[][Section~\ref{sssec:r_process_nature}]{roederer_rprocess_hwo25}.

Both EE-7a and EE-7b target white dwarfs with GALEX FUV magnitudes $\leq 20$, but they require different subsamples and spectral resolutions. EE-7a requires $R\sim60,000$ observations of 155~polluted white dwarfs for which O and Si can be measured to precision $\leq 0.1$~dex. Identifying a sufficient target sample for EE-7a will likely require observing 9500 polluted white dwarfs \citep{xu_siyi_hwo25}. In contrast, EE-7b needs $R\sim200,000$ spectra of 50~white dwarfs spanning a wide range of temperatures ($T_{\rm eff} \lesssim 60,000$K) and surface gravities ($7 \leq \log g \leq 9$). Although EE-7b needs higher spectral resolution ($R\sim200,000$) than EE-7a ($R\sim60,000$) or the EE-D default ($R=100,000$), the initial screening observations obtained to select targets for EE-7a will also be useful for identifying a statistically robust subsample of targets to reobserve at higher spectral resolution for EE-7b.

There is also potential overlap between the targets needed for EE-8 and EE-9. Both science cases focus on metal-poor stars: EE-8 requires $R\sim100,000$, $140 - 310$~nm spectra of 50~Population~III and very-low mass, second generation stars with $[\rm Fe/H]<-4$ while EE-9 needs observations of 100 metal-poor FGK stars in varied Galactic environments at the same resolution ($R \sim 100,000$) over a slightly reduced wavelength range ($150 - 310$~nm). If some of the stars observed for EE-9 are ancient and extremely metal-poor, those data could also be used for EE-8 if supplemental data were obtained at $140-150$~nm. In addition, the brighter targets in EE-8 could potentially be included in EE-9, thereby further reducing the total number of necessary spectra. 

\subsubsection{EE-E: Spectropolarimetric Observations}
\label{sssec:ee_e}
The next EE observation category is spectropolarimetry. We label this category as EE-E and adopt default settings of $R=120,000$ spectra at $100-1600$~nm with full Stokes polarimetry. The polarization error should be $<10^{-6}$. Although this category contains only a single EE science case, the EE-E configuration is identical to the defaults assumed for non-coronagraphic spectropolarimetry for all working groups. Combined, the spectropolarimetric configurations EE-E, GG-E, and SSiC-L include eight science cases. 

EE-E observations could be used for science case EE-15 \citep[][Section~\ref{sssec:mag_stars}]{david-uraz_hwo25}. EE-15 aims to investigate stellar magnetic fields by observing a variety of stars including massive stars that appear to be magnetically inactive. For EE-15, the default EE-E spectral resolution ($R=120,000$) would be needed at visible wavelengths, but UV data could be obtained at a lower spectral resolution ($R>60,000$).

\subsubsection{EE-F: Photometric Observations}
\label{sssec:ee_f}
The final type of observations needed for EE science are diffraction-limited images. We categorize these as EE-F and assume a default configuration of a $6'\times6'$ field of view observed at spatial resolution $\theta=20$~mas over a wavelength range of $95-2500$~nm.

EE-F observations of kilonovae could advance understanding of the heavy element enrichment history of the universe for EE-4b \citep[][Section~\ref{sssec:r_process_el}]{burns_hwo25}. In addition, high-resolution images of galaxies have the potential to resolve individual stars, thereby advancing studies of galaxy morphology, star formation, and chemical enrichment for EE-6 \citep[][Section~\ref{sssec:resolved_stellar_pops}]{smercina_hwo25}. Finally, for EE-10, the substantial increase in sensitivity and resolution possible with a facility like HWO could improve the cosmological distance ladder by refining distance estimates to various stellar populations in other galaxies \citep[][Section~\ref{sssec:distance_ladder}]{anand_hwo25}. 

The kilonova observations obtained for EE-4b could have a much lower spatial resolution ($\theta=1000$~mas) and would therefore not be useful for EE-6 or EE-10. For EE-4b, the primary consideration is the speed at which observations could be obtained rather than the spatial resolution. Ideally, 10 kilonovae would be observed within two days of explosion. 

The needs of EE-6 and EE-10 are more similar, and therefore it is possible that the two programs could share data. EE-6 requires a slightly smaller field of view ($5'\times 5'$) and a finer spatial resolution ($\theta=15$~mas) while EE-10 would use the default field of view and spatial resolution ($6'\times6'$, $\theta=20$~mas, respectively). Additionally, EE-10 does not need data at UV wavelengths while EE-6 requires data extending from the NUV to the NIR. Given the differences in observational strategies, it would be easier to contribute EE-6 observations to EE-10 than to use EE-10 observations for EE-6. In the former case, the $6'\times6'$ field requested for EE-10 could be obtained by tiling the smaller $5'\times5'$ field needed for EE-6. In the latter scenario, the EE-10 data would need to be supplemented by additional UV data. Additionally, the Vis/NIR data obtained for EE-10 would either need to have a higher native spatial resolution (15~mas versus 20~mas) or be acquired using a dithering scheme that permits super-resolution image reconstruction.

\subsection{Solar Systems in Context}
\begin{figure*}
 \centering
 \includegraphics[width=\linewidth]{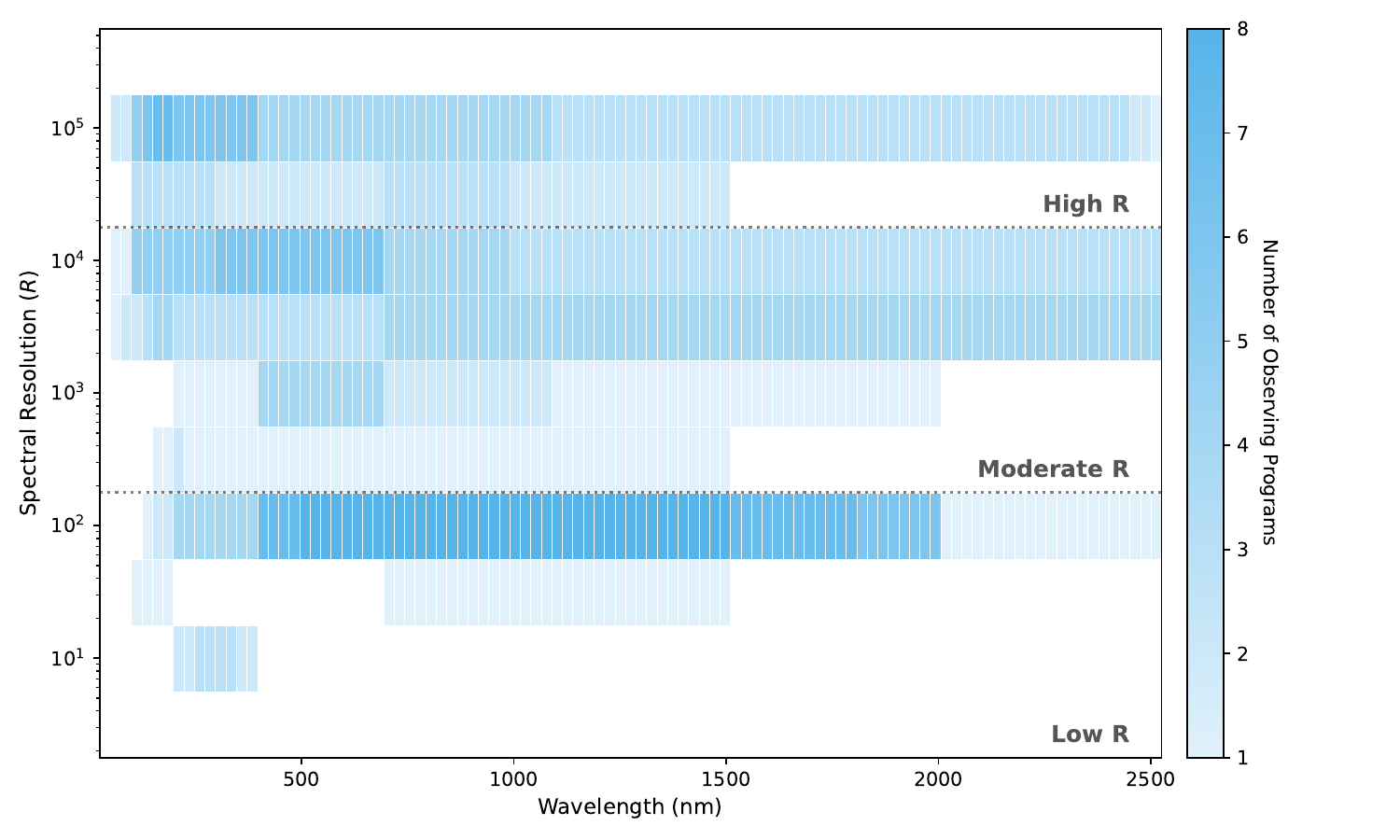}
 \caption{Same as Figure~\ref{fig:gg_r_wave} but for SSiC science cases.}
 \label{fig:ssic_r_wave}
\end{figure*}

As illustrated by the summary of Solar Systems in Context science cases in Table~\ref{tab:ssic_scdds}, placing the solar system in context requires a series of observations of planets and small bodies both within and beyond the solar system. While these investigations cover a wide parameter space and employ a variety of observational strategies, many SSiC science cases share observational capabilities or datasets. Figure~\ref{fig:ssic_r_wave} displays both the breadth of wavelengths and spectral resolutions needed to accomplish SSiC objectives as well as the concentration of programs in particular regions of $\lambda-R$ parameter space. 

In Table~\ref{tab:ssic_remap}, we reorganize the SSiC science cases into multiple categories based on the necessary capabilities. This reformatting is intended to facilitate future design trades and showcases how multiple scientific objectives could be addressed with the same instrument or data. Several of the SSiC observing programs overlap with LW programs, but we defer the discussion of cross-working group overlap until Section~\ref{sec:xswg} and focus exclusively on SSiC science cases in this section and in Table~\ref{tab:ssic_remap}. 

The objectives of SSiC science cases can be addressed with the following suite of observations:
\begin{itemize}
 \item \textbf{SSiC-A} (Section~\ref{sssec:ssic_a}): high-contrast spectroscopy in the UV
\item \textbf{SSiC-B} (Section~\ref{sssec:ssic_b}): high-contrast spectroscopy in the visible
\item \textbf{SSiC-C} (Section~\ref{sssec:ssic_c}): high-contrast spectroscopy in the NIR
\item \textbf{SSiC-D} (Section~\ref{sssec:ssic_d}): high-contrast spectropolarimetry in the UV
\item \textbf{SSiC-E} (Section~\ref{sssec:ssic_e}): high-contrast spectropolarimetry in the visible
\item \textbf{SSiC-F} (Section~\ref{sssec:ssic_f}): high-contrast spectropolarimetry in the NIR
\item \textbf{SSiC-G} (Section~\ref{sssec:ssic_g}): high-contrast polarimetric imaging
\item \textbf{SSiC-H} (Section~\ref{sssec:ssic_h}): high-contrast imaging
\item \textbf{SSiC-J} (Section~\ref{sssec:ssic_j}): spectroscopy at moderate spectral resolution
\item \textbf{SSiC-K} (Section~\ref{sssec:ssic_k}): spectroscopy at high spectral resolution
\item \textbf{SSiC-L} (Section~\ref{sssec:ssic_l}): spectropolarimetry
\item \textbf{SSiC-M} (Section~\ref{sssec:ssic_m}): IFS observations at high spatial resolution
\item \textbf{SSiC-N} (Section~\ref{sssec:ssic_n}): IFS observations at moderate spatial resolution
\item \textbf{SSiC-P} (Section~\ref{sssec:ssic_p}): photometric observations
\end{itemize} 
The observations labeled as high-contrast (i.e., SSiC-A -- SSiC-H) require achieving sensitivity to planets and disks that are much fainter than their host stars. For Earth-like planets around Sun-like stars, the required planet-star contrast ratios are $\leq10^{-10}$ at habitable zone separations. Other SSiC science cases such as detecting Venus-like planets or characterizing disk structure require sensitivity to faint objects at narrower or wider separations. For more details about the high contrast capabilities needed for SSiC science, see Section~\ref{sec:drivers}. 

\subsubsection{SSiC-A: High-Contrast Spectroscopy in the UV}
\label{sssec:ssic_a}
The first type of observations needed for SSiC science cases are UV spectra obtained at high contrast ratios. We categorize these observations as SSiC-A and adopt a baseline configuration of $R=7$ spectra at $200-400$~nm. 

SSiC-A observations are needed for seven SSiC cases. First, SSiC-11 necessitates SSiC-A data to determine the timescales required for Earth-like planets to obtain oxygenated atmospheres \citep[][Section~\ref{sssec:ozone_onset}]{blunt_et_al2025}. Similarly, SSiC-20 (Section~\ref{sssec:survive_water}) needs SSiC-A data at a much higher spectral resolution ($R\sim1000$) to assess how planetary habitability changes over time. For investigations of the retention of volatiles by terrestrial planets, SSiC-21 (Section~\ref{sssec:retention}) also needs SSiC-A data at higher-than-baseline spectral resolution of $R\sim100$. In a related science case, SSiC-18a necessitates SSiC-A data to investigate whether other stars have Venus-like planets that have experienced or are currently experiencing runaway greenhouses. These observations should use the baseline SSiC-A spectral resolution and wavelength coverage, but the observations must access close angular separations ($\ge 0.3$~AU) to observe Venus-like planets. 

Next, SSiC-3a demands SSiC-A data to investigate the nature of sub-Neptune planets \citep[][Section~\ref{sssec:rocky_sub}]{hu_rocky_hwo25}. Both SSiC-11 and SSiC-3a need data over narrower wavelength ranges than the $200-400$~nm baseline for SSiC-A: $200-350$~nm for SSiC-11 and $250-400$~nm for SSiC-3a. 

For investigations of giant planets, SSiC-15a also needs SSiC-A data \citep[][Section~\ref{sssec:reflected_giants}]{min_hwo25}. The baseline SSiC-A red cut-off would be sufficient, but the observations should extend blueward to $\lambda_\mathrm{blue} =150$~nm and be obtained at a higher spectral resolution of $R=100$. Additionally, the observations should cover a wider range of angular separations ($60-1000$~mas) to capture the wider orbits of cool gas giants. 

Finally, SSiC-8a demands SSiC-A observations at much higher spectral resolution ($R\sim10,000$) starting at a shorter wavelength blue cut-off ($\lambda_\mathrm{blue}=90$~nm) and covering a very wide field of view ($2'\times2'$). These challenging observations are designed to detect young planets and their interactions with protoplanetary disks. Planets detected by SSiC-8a may be relevant for SSiC-15a. 

Programs SSiC-11, SSiC-21a, and SSIC-20a all focus on terrestrial planets, so data obtained at high spectral resolution for program SSiC-20a could also be used for SSiC-21a and SSiC-11. Data acquired for SSiC-21a would also be sufficient for SSiC-11, but data obtained for SSiC-11 would lack the spectral resolution needed for either SSiC-20a or SSiC-21a. 

The programs SSiC-3a, SSiC-15a, and SSiC-18a focus on different types of planet, but it may be possible to obtain datasets that are useful for several of those programs as well as the terrestrial planet programs by targeting multi-planet systems that contain planets of different sizes within the region accessible to the UV coronagraph. If such systems exist, they should be observed over the broadest wavelength range possible to reduce the number of revisits needed to obtain sufficient wavelength coverage for all four programs. Additionally, the observation times and durations should be selected to obtain useful data for multiple planets. 

In theory, observations obtained for SSiC-15a could also be used for SSiC-11, SSiC-3a, SSiC-18a, and SSiC-21a if appropriate planets are detected, the achieved contrast is sufficiently high, and, for programs requiring lower spectral resolution, the noise properties of the data are compatible with spectral binning. In contrast, observations obtained for SSiC-11, SSiC-3a, and SSiC-18a will not be useful for SSiC-15a or SSiC-21a because the latter cases need significantly higher spectral resolution ($R\sim100$ versus $R=7$). Additionally, observations obtained over the narrower wavelength regions needed for SSiC-11 ($200-350$~nm) and SSiC-3a ($250-400$~nm) would need to be supplemented by additional data completing the wider wavelength range needed for SSiC-15a ($150-400$~nm) or SSiC-21a ($200-400$~nm). 

\subsubsection{SSiC-B: High-Contrast Spectroscopy in the Visible}
\label{sssec:ssic_b}
Several SSiC cases require high-contrast spectroscopy at visible wavelengths. Our baseline configuration for this mode is SSiC-B, which provides $R=140$ spectra at $400-1100$~nm. 

SSiC-B observations are needed for investigating the compositions of sub-Neptune planets in SSiC-3b \citep[][Section~\ref{sssec:rocky_sub}]{hu_rocky_hwo25} and gas giant planets for SSiC-15b \citep[][Section~\ref{sssec:reflected_giants}]{min_hwo25}. Additionally, SSiC-B data can inform models of volatile delivery to and retention on terrestrial planets for SSiC-2a \citep[][Section~\ref{sssec:hab_system}]{hasegawa_water_hwo25} and SSiC-21b (Section~\ref{sssec:retention}), respectively. SSiC-B could also provide insight into the long-term habitability of terrestrial planets for SSiC-20b (Section~\ref{sssec:survive_water}) and probe the orbital dynamics of gas giants for SSiC-12a \citep[][Section~\ref{sssec:giant_orbits}]{sagynbayeva_et_al2025}. Finally, SSiC-B observations are needed by SSiC-19a for studies of exozodi \citep[][Section~\ref{sssec:exozodi}]{debes_hwo25} and by SSiC-8b for investigating planet formation in protoplanetary disks \citep[][Section~\ref{sssec:proto}]{ren_hwo25}.

The highest spectral and spatial resolutions are requested by SSiC-8b, which demands $R\sim10,000$ spectra at a spatial resolution of $\theta=5$~mas over a shorter wavelength range of $400-700$~nm. SSiC-15b and SSiC-20b also request higher spectral resolution ($R\sim1000$) and would obtain data over the full $400-1100$~nm wavelength range of SSiC-B. For SSiC-15b, the observations should also provide access to wider angular separations of $60-1000$~mas. 

Next, SSiC-3b and SSiC-12a request observations at the default spectral resolution ($R=140$) over narrower wavelength ranges of $400-1000$~nm for SSiC-3b and $500-1100$~nm for SSiC-12a. However, as noted in Section~\ref{sssec:ssic_c}, shifting the boundary of the visible and NIR components of SSiC-3 from 1000 to the default SSiC-B red cut-off of $\lambda_\mathrm{red}=1100$~nm would simplify sharing data between SSiC-3 and other programs. The nominal $400-1100$~nm wavelength range is also acceptable for SSiC-19a and SSiC-21b, both of which request a slightly lower spectral resolution ($R\sim100$). Finally, SSiC-2a could be accomplished by obtaining spectra over the default SSiC-B wavelength range but at even lower spectral resolution of $R>40$. 

As for SSiC-A, these SSiC-B programs could potentially share data if target systems contain multiple planets that satisfy different target selection criteria and can be simultaneously observed. Program SSiC-19a is a prime example of cross-program synergy because any observations that cover the habitable zone would also provide insight into exozodi properties. For instance, programs SSiC-2a, SSiC-3b, and SSiC-21b all focus on terrestrial planets within or near the habitable zone and could contribute data to SSiC-19a in addition to sharing observations with each other. To increase observational efficiency, it would be strategic to obtain SSiC-B observations using the baseline settings of $R=140$ spectral resolution and $400-1100$~nm wavelength coverage rather than using the lower spectral resolutions requested by SSiC-2a, SSiC-19a, and SSiC-21b or the reduced wavelength coverage proposed by SSiC-3b. 

If the higher spectral resolutions requested by SSiC-8b ($R\sim10,000$), SSiC-15b ($R=1000$) and SSiC-20b ($R\sim1000$) are achievable, data obtained for those programs could be spectrally binned for use by other programs. As previously mentioned, these data would be advantageous for the exozodi studies proposed by SSiC-19a. Additionally, SSiC-20b data could be contributed to other studies of terrestrial planets by SSiC-2a, SSiC-3b, and SSiC-21b while SSiC-8b observations of gas giants forming in protoplanetary disks may be useful for the investigation of gas giant orbital dynamics proposed by SSiC-12a as well as studies of gas giant atmospheres by SSiC-15b. Due to the shorter red cut-off of SSiC-8b ($\lambda_\mathrm{red}=700$~nm), SSiC-8b observations would need to be supplemented by additional $R=1000$ data extending to the default SSiC-B red cut-off of $\lambda_\mathrm{red}=1100$~nm. 

In general, the likelihood that a given dataset would be useful for multiple programs would likely be increased by minimizing the inner working angle and maximizing the outer working angle at which high contrast could be achieved in a given observation. However, given the diversity in planet properties and occurrence rates, it would be advantageous to run simulations to determine the specific settings that would yield the highest scientific return and observational efficiency. 

\subsubsection{SSiC-C: High-Contrast Spectroscopy in the NIR}
\label{sssec:ssic_c}
Multiple programs also need high-contrast spectra at NIR wavelengths. We designate these observations as SSiC-C. The default configuration is $R=70$ spectra over $1100-1700$~nm.

The programs for which SSiC-C data are needed are the NIR components of programs that needed SSiC-B data at visible wavelengths. For instance, SSiC-3c \citep[][Section~\ref{sssec:rocky_sub}]{hu_rocky_hwo25} and SSiC-19b 
\citep[][Section~\ref{sssec:exozodi}]{debes_hwo25} require SSiC-C observations extending to longer red cut-offs of 1800~nm and 2000~nm to investigate sub-Neptune planets and exozodi, respectively. Technically, SSiC-3c also requests a shorter wavelength blue cut-off of $\lambda_\mathrm{blue}=1000$~nm, but these data could be obtained by including the $1000-1100$~nm spectral region as part of the $R>140$ visible spectra obtained for SSiC-3b rather than $R>70$ near-infrared spectra acquired for SSiC-3c. 

The observations for SSiC-19b should be obtained at a slightly higher spectral resolution of $R\sim100$ rather than the SSiC-C default of $R=70$. In contrast, investigations of volatile delivery for SSiC-2b require SSiC-C observations over a narrower wavelength range of $1100-1500$~nm at a reduced spectral resolution of $R>40$. 

Next, programs SSiC-20c and SSiC-21c necessitate NIR data at spectral resolutions of $R\sim1000$ and $R\sim100$, respectively. SSiC-20c needs observations to $\lambda_\mathrm{red}=2000$~nm to detect outgassing TiO while SSiC-21c states that data extending redward of 1500~nm would be beneficial. As discussed in Section~\ref{sssec:retention}, we adopt $\lambda_\mathrm{red}=2000$~nm for SSiC-21c. 

For investigations of giant planets, SSiC-12b \citep[][Section~\ref{sssec:giant_orbits}]{sagynbayeva_et_al2025} and SSiC-15c \citep[][Section~\ref{sssec:reflected_giants}]{min_hwo25} need SSiC-C observations extending to an even longer wavelength of 5000~nm.\footnote{We note that HWO is unlikely to obtain high-contrast spectra at wavelengths as long as the red cut-off of 5000~nm requested by SSiC-12b and SSiC-15c, but we include this value as a reminder that there is scientific merit in extending the red cut-off to the longest wavelengths that are technologically feasible.} For SSiC-15c, the observations should be obtained at a higher spectral resolution ($R=1000$) and cover wider angular separations of $60-1000$~mas to ensure access to cooler giant planets. 

Just as SSiC-A and SSiC-B data could potentially satisfy multiple programs simultaneously, SSiC-C observations also have the potential to be useful for more than one program. For instance, multiple programs focus on small planets (SSiC-2b, SSiC-3c, SSiC-20c, and SSiC-21c) or dust (SSiC-19b) within the habitable zone. Additionally, the giant planet studies proposed by SSiC-12b and SSiC-15c would consider planets at a range of orbital separations including habitable zone distances. 

To increase the likelihood of data sharing, observations should be obtained at the $R=100$ spectral resolution needed by SSiC-19b and SSiC-21c (or at least the default $R=70$ spectral resolution of SSiC-C) rather than the lower $R>40$ spectral resolution of SSiC-2b. If observations could be obtained at the higher spectral resolution of $R\sim1000$ needed for SSiC-15c and SSiC-20c, those data could be spectrally binned for use with the other programs. Additionally, the observations should cover at least the full SSiC-C wavelength range and ideally extend redward to 2000~nm or even longer wavelengths. Finally, high contrasts should be achieved over a wide range of angular separations extending from the inner edge of the habitable zone to the more distant orbits of giant planets. As we concluded for SSiC-B in Section~\ref{sssec:ssic_b}, simulations would be useful for determining the specific observational capabilities that would reduce the observation time needed to fulfill multiple SSiC-C objectives. 

\subsubsection{SSiC-D: High-Contrast Spectropolarimetry in the UV}
\label{sssec:ssic_d}
The next observations needed for SSiC science are high-contrast UV spectropolarimetry. We classify these observations as SSiC-D and adopt default values of $R=7$ spectral resolution over $200-400$~nm. 

Four SSiC observing programs require SSiC-D data. First, SSiC-15d needs SSiC-D observations to characterize clouds and hazes on gas giant planets \citep[][Section~\ref{sssec:reflected_giants}]{min_hwo25}. Second, SSiC-21d (Section~\ref{sssec:retention}) requests SSiC-D at a higher spectral resolution ($R\sim100$) to investigate the atmospheres of terrestrial planets. Third, SSiC-9a necessitates SSiC-D data with a shorter UV cut-off ($\lambda_\mathrm{blue}=140$~nm), higher spectral resolution ($R>3000$) covering a large field of view ($\geq 2' \times 2'$) at spatial resolution $\theta \lesssim 30$~mas \citep[][Section~\ref{sssec:debris}]{rebollido_hwo25}. Fourth, SSiC-8c requires SSiC-D data covering an even broader wavelength range ($90-400$~nm) at higher spectral resolution ($R=10,000$) to characterize young planets and their interactions with the surrounding protoplanetary disk \citep[][Section~\ref{sssec:proto}]{ren_et_al2025}. Additionally, SSiC-8c requests high spatial resolution ($\theta=5$~mas) over a $2' \times 2'$ field of view to study young planets and disk properties at a wide range of angular separations ($5-1000$~mas). 

Observational efficiency could be increased by using SSiC-D observations to address multiple science goals. Given that SSiC-15d aims to target gas giant planets spanning a wide range of temperatures from 30K to 600K, that program also demands access to a broad range of angular separations ($60-1000$~mas) and will therefore benefit from the wide fields of view needed by SSiC-8c and SSiC-9a. If the integration times are suitable, SSiC-8c or SSiC-9a observations could be spectrally binned to satisfy the observational goals of SSiC-15d and SSiC-21d. In contrast, SSiC-15d and SSiC-21d observations will lack the wavelength coverage, spectral resolution, and access to tight angular separations required for SSiC-8c or SSiC-9a science. Finally, SSiC-8c and SSiC-9a focus on disks of different ages (i.e., younger protoplanetary disks for SSiC-8c and older debris disks for SSiC-9a), so they cannot share data despite their similar technical requirements. 

\subsubsection{SSiC-E: High-Contrast Spectropolarimetry in the Visible}
\label{sssec:ssic_e}
Several SSiC science cases need high-contrast spectropolarimetry at visible wavelengths. We denote these observations as SSiC-E and set baseline characteristics of spectral resolution $R=140$ and wavelength coverage $400-1100$~nm. 

SSiC-E observations have the potential to investigate aerosols in the atmospheres of giant planets for SSiC-15e \citep[][Section~\ref{sssec:reflected_giants}]{min_hwo25}, runaway greenhouses on Venus-like planets for SSiC-18b \citep[][Section~\ref{sssec:exovenus}]{kane_venus_hwo25}, and atmospheric retention on terrestrial planets for SSiC-21e (Section~\ref{sssec:retention}). Additionally, SSiC-E data could constrain the frequency with which exoplanets have ring systems for SSiC-31a \citep[][Section~\ref{sssec:exorings}]{limbach_et_al2024,limbach_et_al2026}, probe interactions between young planets and disks for SSiC-8d \citep[][Section~\ref{sssec:proto}]{ren_hwo25}, and investigate the properties of debris disks for SSiC-9b \citep[][Section~\ref{sssec:debris}]{rebollido_hwo25}. The default SSiC-E wavelength range of $400-1100$~nm would be sufficient for all of these programs except for SSiC-8d, which requests a shorter wavelength red cut-off ($\lambda_\mathrm{red}=700$~nm). 

The default $R=140$ spectral resolution would also be adequate for SSiC-15e and SSiC-18b. For SSiC-31a and SSiC-21e, a lower spectral resolution of $R\sim100$ would be sufficient. In contrast, SSiC-8d and SSiC-9b demand higher spectral resolutions of $R\sim10,000$ and $R>3000$, respectively. 

In addition, SSiC-18b needs access to closer separations $\geq 0.3$~AU to detect Venus analogs while programs SSiC-9b, SSiC-15e, and SSiC-31a require coverage of wider angular separations. Program SSiC-8d requires observations spanning $5-1000$~mas to investigate disk properties at both wide and close separations. If possible, exoring detections for SSiC-31a would benefit from fields of view as large as $10'' \times 10''$ to detect ringed exoplanets at separations of up to 100~AU. If such a wide field of view is not possible, coverage out to $1''$ as requested by SSiC-15e and SSiC-8d would still enable detecting ringed planets at Saturn-like separations of 10~AU. For disk science, SSiC-8d and SSiC-9b request an even larger field of view of $\geq 2'\times 2'$.

Observational efficiency could be increased by obtaining SSiC-E observations that address multiple programs simultaneously. The most natural synergy is between program SSiC-15e to study clouds and hazes in giant planet atmospheres and program SSiC-31a to analyze exorings. The observations obtained for SSiC-31a to characterize ring properties may also provide useful insight into the atmospheres of the planets hosting the rings. Depending on the field of view and the frequency of multiplanet systems with suitable targets, it may also be possible to collect SSiC-E observations of multiple planets simultaneously. Theoretically, those observations could address the goals of all planet-focused programs: SSiC-15e, SSiC-18b, SSiC-21e, and SSiC-31a. 

However, due to the need for higher spectral resolution for disk studies, SSiC-E observations obtained for other programs would not be automatically useful for SSiC-8d and SSiC-9b. Rather, SSiC-8d and SSiC-9b observations could be spectrally binned for use by the planet-focused programs if those systems also include planets that meet the target selection criteria for the other programs. Due to the shorter wavelength range of SSiC-8d ($400-700$~nm), observations obtained for that program would need to be supplemented by additional data at $700-1100$~nm if used by other programs. 

\subsubsection{SSiC-F: High-Contrast Spectropolarimetry in the NIR}
\label{sssec:ssic_f}
A few SSiC programs also need high-contrast spectropolarimetry in the near infrared. For these SSiC-F observations, we adopt a default spectral resolution of $R=70$ and a default wavelength range of $1100-1700$~nm. 

SSiC-F observations would be useful for studying the atmospheres of gas giant planets for SSiC-15f \citep[][Section~\ref{sssec:reflected_giants}]{min_hwo25}, terrestrial planets for SSiC-21f (Section~\ref{sssec:retention}), and Venus-like planets for SSiC-18c \citep[][Section~\ref{sssec:exovenus}]{kane_venus_hwo25}. SSiC-F observations would also be helpful for characterizing ring systems for SSiC-31b \citep[][Section~\ref{sssec:exorings}]{limbach_et_al2024,limbach_et_al2026} and debris disks for SSiC-9c \citep[][Section~\ref{sssec:debris}]{rebollido_hwo25}. The default SSiC-F spectral resolution of $R=70$ would be acceptable for SSiC-15f and SSiC-18c, but SSiC-21f and SSiC-31b would require observations at a higher spectral resolution of $R\sim100$. For investigations of gas, dust, and planetary interactions in debris disks, SSiC-9c requests an even higher spectral resolution of $R>3000$ and high spatial resolution ($\theta \lesssim 30$~mas) over a wide field of view ($\geq2' \times 2'$). 

Additionally, the investigation of exorings proposed by SSiC-31b needs access to separations as large as $10''$ and the gas giant studies described in SSiC-15f necessitate observations covering $60-1000$~mas. The closest angular separations would be targeted by SSiC-18c, which requests observations at angular separations $\geq 0.3$~AU for studies of exo-Venuses. Although SSiC-18c could accommodate a relaxed red cut-off of 1500~nm, the other programs require longer wavelength red cut-offs: $\lambda_\mathrm{red}=1800$~nm for SSiC-15f; $\lambda_\mathrm{red}=2000$~nm for SSiC-21f and SSiC-31b; and $\lambda_\mathrm{red}=3000$~nm for SSiC-9c. 

To increase observational efficiency, SSiC-F observations could be used for multiple programs. Ideally, a combined program would cover a sufficiently broad range of separations (e.g., from 0.3~AU to $\geq10$~AU) over a wider wavelength range of $1100-2000$~nm at a higher spectral resolution of $R\sim100$. If $R>3000$ observations spanning $1100 - 3000$~nm can be obtained for the debris disk studies of SSiC-9c, those observations could be spectrally binned for use by all other programs needing SSiC-F observations. 

\subsubsection{SSiC-G: High-Contrast Polarimetric Imaging}
\label{sssec:ssic_g}
The next observations require both starlight suppression and polarimetry. We categorize these observations as SSiC-G and set defaults of a wavelength range of $300-1700$~nm, a field of view of $10'' \times 10''$, and full Stokes (IQUV) polarimetry with polarimetric precision $\lesssim1\%$. 

This mode is requested by four SSiC programs. First, SSiC-3d needs SSiC-G observations at multiple orbital phases to probe the atmospheres of sub-Neptune planets within the habitable zones of FGK stars \citep[][Section~\ref{sssec:rocky_sub}]{hu_rocky_hwo25}. If the wavelength range of the high-contrast UV photopolarimetry obtained for SSiC-3d matches that of the high-contrast UV spectroscopy obtained for SSiC-3a, then SSiC-3d observations would have a slightly shorter wavelength blue cut-off than the default SSiC-G observations (250~nm versus 300~nm). Likewise, matching the red cut-off used by the high-contrast NIR spectroscopic component SSiC-3c sets $\lambda_\mathrm{red}=1800$~nm for SSiC-3d. Determining the frequency of oceans on temperate terrestrial planets for SSiC-4 also necessitates SSiC-G observations at multiple orbital phases, but those data could be obtained over a narrower wavelength range ($400-1100$~nm). Next, SSiC-9d requests SSiC-G observations of a large field of view ($\geq 2' \times 2'$) over a wider wavelength range ($140-3000$~nm) to characterize debris disk properties \citep[][Section~\ref{sssec:debris}]{rebollido_hwo25}. Finally, characterization of dust and small bodies in and near the habitable zone for SSiC-19d \citep[][Section~\ref{sssec:exozodi}]{debes_hwo25} demands SSiC-G data at spatial resolution $\theta \le 20$~mas over a wavelength range that extends farther into the UV and NIR ($200-2000$~nm). 

Given that all four programs target the habitable zones of FGK stars, observations could be obtained for multiple programs simultaneously. For instance, the efficiency of SSiC-9d or SSiC-19d could be improved by supplementing existing SSiC-3d and SSiC-4 observations with additional data at UV and NIR wavelengths (i.e., $<250$~nm and $>1700$~nm for SSiC-3d targets; $<400$~nm and $>1100$~nm for SSiC-4 targets). SSiC-3d aims to observe 100~planets, which could comprise up to $50-100\%$ of the systems needed for SSiC-19d depending on planet multiplicity and the selected sample size for SSiC-19d within the goal range of $100-200$~systems. In addition, the sample of 100~temperate sub-Neptunes observed for SSiC-3d could include all $20-32$~temperate terrestrial planets needed for SSiC-4. Alternatively, data obtained for SSiC-4 could be used for SSiC-3d by obtaining supplementary data at shorter and longer wavelengths ($250-400$~nm and $1100-1700$~nm). Finally, observations that cover the wider spectral range needed by SSiC-9d ($140-3000$~nm) at the higher spatial resolution requested by SSiC-19d ($\theta \leq 20$~mas) could be used by both programs.  

\subsubsection{SSiC-H: High-Contrast Imaging}
\label{sssec:ssic_h}
Multiple SSiC observing programs demand high-contrast imaging over a broad wavelength range. We categorize these observations as SSiC-H and set the default wavelength range to $200-1100$~nm. SSiC-H observations are baselined to have a spatial resolution of $\theta=10$~mas and cover a $10''\times10''$ field of view. 

Baseline SSiC-H observations would be sufficient for detecting exorings as part of SSiC-31c \citep[][Section~\ref{sssec:exorings}]{limbach_et_al2024,limbach_et_al2026}, but other programs require variations on the baseline configuration. For instance, investigating the influence of stellar multiplicity on the occurrence rate of habitable planets for SSiC-6 demands sensitivity to small planets ($1-1.5R_\oplus$) within 2~AU of stars with binary companions separated by $\lesssim10''$ \citep[][Section~\ref{sssec:occ_binary}]{newton_hwo25}.

Furthermore, four programs would require extended NIR coverage: gas giant studies for SSiC-15g \citep[][Section~\ref{sssec:reflected_giants}]{min_hwo25} need data to a red cut-off of 1800~nm; exozodi investigations for SSiC-19c \citep[][Section~\ref{sssec:exozodi}]{debes_hwo25} and exomoon searches for SSiC-32 \citep[][Section~\ref{sssec:exomoons}]{limbach_et_al2024,limbach_et_al2026} need data to 2000~nm; and debris disk studies for SSiC-9e \citep[][Section~\ref{sssec:debris}]{rebollido_hwo25} request $\lambda_\mathrm{red}=3000$~nm. Exomoon searches also require high photometric precision ($\lesssim 10\%$ over a few hours). Compared to the baseline $10''\times10''$ configuration for SSiC-H, SSiC-15g and SSiC-32 could accommodate smaller fields of view of $1''\times1''$ and $>2''\times2''$, respectively. 

At the opposite extreme, SSiC-8e \citep[][Section~\ref{sssec:proto}]{ren_hwo25} and SSiC-9e \citep[][Section~\ref{sssec:debris}]{rebollido_hwo25} demand much larger fields of view ($\geq2'\times2'$) for studying circumstellar disks (protoplanetary disks for SSiC-8e and debris disks for SSiC-9e) and their interactions with planets. SSiC-8e also requires observations at high spatial resolution ($\theta=5$~mas) over a wavelength range that extends farther into the UV and truncates earlier in the NIR than the default SSiC-H wavelength range ($90-700$~nm versus $200-1100$~nm). For SSiC-9e, observations should cover a broader wavelength range ($140-3000$~nm) but the spatial resolution could be coarser ($\theta \lesssim 30$~mas). For exozodi studies, SSiC-19c requests an intermediate spatial resolution of $\theta = 20$~mas.

Depending on the achieved contrast at various separations and the field of view, SSiC-H observations have the potential to be useful for multiple science cases. For example, observations of gas giant planets designed for atmospheric characterization (SSiC-15g) could also be used to search for exorings (SSiC-31c) if those observations were obtained with the default SSiC-H field of view ($10''\times10''$) rather than the smaller field that would suffice for SSiC-15g ($1''\times1''$). In parallel, observations designed for SSiC-31c could also address SSiC-15g goals if the wavelength range were extended to 1800~nm. In addition, wide-field observations of disks for SSiC-8e and SSiC-9e could potentially be useful for exoring searches (SSiC-31c), exomoon detection (SSiC-32), gas giant characterization (SSiC-15g), and exozodi assessments (SSiC-19c). Due to the shorter red cut-off of SSiC-8e, observations originally obtained by SSiC-8e would need to be supplemented by additional NIR data extending to 1800~nm for use by SSiC-31c or SSiC-15g and to 2000~nm for use by SSiC-19c or SSiC-32. As SSiC-8e targets protoplanetary disks while SSiC-9e targets debris disks, those programs cannot share data with each other. Additionally, due to the difference in system ages, useful data for SSiC-31c, SSiC-32, and SSiC-19c are more likely to be shared by SSiC-9e than SSiC-8e.

Another possible synergy is that SSiC-6 observations searching for terrestrial planets in binary systems may also reveal signs of exomoons, thereby contributing to SSiC-32. However, obtaining observations that are informative for both programs would require overcoming the significant technical challenges of suppressing starlight from two stars, achieving high photometric precision ($\lesssim 10\%$ over a few hours), and extending observations to 2000~nm. 

\subsubsection{SSiC-J: Spectroscopy at Moderate Spectral Resolution}
\label{sssec:ssic_j}
Multiple SSiC programs require spectroscopy at moderate spectral resolution. For the baseline configuration for these SSiC-J observations, we adopt a wavelength range of $100-3000$~nm and a spectral resolution of $R=3000$. We note that this wavelength range is broad and would likely be addressed by multiple channels or even multiple instruments. The red limit of $3000$~nm is higher than might be expected for a passively cooled telescope, but many of the targets for the SSiC-J programs are extremely bright and will therefore be in a different noise regime than much fainter galaxies and stars observed for GG and EE programs. 

Each SSiC-J program requests a different wavelength range. Beginning in the UV, program SSiC-17 \citep[][Section~\ref{sssec:id_oceans}]{quick_hwo25} requires data at the baseline SSiC-J spectral resolution over $50-210$~nm for studies of potentially icy exoplanets. Due to this short red cut-off, SSiC-17 observations could be obtained by an instrument or channel that functions only in the UV. Similarly, SSiC-20d (Section~\ref{sssec:survive_water}) requests  data spanning $100-200$~nm.  The XUV observations that would be obtained by SSiC-20d would provide context for interpreting high-contrast UV (SSiC-20a), visible (SSiC-20b), and NIR (SSiC-20c) spectra of planetary atmospheres. A lower spectral resolution of $R\sim30$ would be sufficient for SSiC-20d, but obtaining data at higher spectral resolution may be advantageous to reduce the risk of saturation while observing bright exoplanet host stars.  

In contrast, programs SSiC-7 \citep[][Section~\ref{sssec:solar}]{mandt_hwo25} and SSiC-16a \citep[][Section~\ref{sssec:transits}]{wakeford_hwo25} require observations over a broad wavelength range extending from the UV (120~nm for SSiC-7 and 100~nm for SSiC-16a) to the NIR ($\ge4800$~nm for SSiC-7 and $\ge5000$~nm for SSiC-16a). SSiC-7 would study small bodies within the solar system and could use a spectral resolution of $R\sim2000$, which is lower than the default SSiC-J spectral resolution ($R=3000$). SSiC-16a would investigate the role of rotation in governing the dynamics, composition, and structure of planetary atmospheres. For that program, higher spectral resolution ($R>6000$) and extreme photometric precision ($\le 1$~ppb) are needed as well as saturation mitigation strategies to observe targets as bright as $V=2$. 

Program SSiC-22 also requires saturation mitigation. As described in the document posted in the STScI SCDD Portal\footnote{\url{https://docs.google.com/document/d/17PxB2qvcCXn1wgRQrm8Mgt7AxycogH47AfelpW3_Iw4/edit?tab=t.0}} and in Section~\ref{sssec:mars} of this paper, SSiC-22 would observe Phobos, Deimos, and bright asteroids to investigate the formation of Mars. For program SSiC-22, the default SSiC-J spectral resolution and wavelength range would be sufficient. 

Compared to other SSiC observing categories, SSiC-J programs are less well-suited to combining observations to increase observational efficiency because they each observe different targets. The closest possible synergy is between program SSiC-22, which would observe the moons of Mars and asteroids within and beyond the main belt, and program SSiC-7, which would observe small bodies. While many SSiC-7 targets will be icy bodies at larger separations than the predominantly rocky targets observed by SSiC-22, some asteroids may be suitable targets for both programs. In those cases, SSiC-22 data should be supplemented by observations at longer wavelengths ($3000-4800$~nm) to increase their value for SSiC-7. 

\subsubsection{SSiC-K: Spectroscopy at High Spectral Resolution}
\label{sssec:ssic_k}
One SSiC science case requires UV spectroscopy at high spectral resolution. We classify these observations as SSiC-K and match the defaults to those used for EE-D. Accordingly, we set the wavelength range to $90-310$~nm and the spectral resolution to $R=100,000$. These capabilities are stricter than needed for the SSiC programs, but the data could be spectrally binned if needed.

The SSiC program needing SSiC-K data is SSiC-14 \citep[][Section~\ref{sssec:escape}]{dossantos_hwo25}, which would observe a diverse sample of 50~transiting planets to investigate atmospheric retention. This program requests a lower spectral resolution ($R>45,000$) over a slightly reduced wavelength range ($100-300$~nm). Importantly, the detector should have a high dynamic range (10,000~cts~s$^{-1}$~px$^{-1}$) to permit observing targets that are bright in the UV. The suggested spectral resolution of $R>45,000$ is designed to mitigate the effects of stellar variability and detect small signals due to planetary motion and ISM absorption. \citet{dossantos_hwo25} note that further research is needed to determine the ideal spectral resolution for these investigations, so the baseline SSiC-K resolution of $R=100,000$ may provide important technical margin. 

\subsubsection{SSiC-L: Spectropolarimetry}
\label{sssec:ssic_l}
Multiple SSiC science cases demand spectropolarimetry. The cases vary in their wavelength and spectral resolution needs, so we set the SSiC-L baseline configuration to match that previously used for the spectropolarimetry cases EE-E and GG-E. Accordingly, SSiC-L observations cover $100-1600$~nm at spectral resolution $R=120,000$. The data include full Stokes polarimetry with polarization error $<10^{-6}$. 

The first SSiC science case requiring spectropolarimetry is SSiC-26 \citep[][Section~\ref{sssec:bfield}]{strugarek_hwo25}, which aims to detect exoplanetary magnetic fields using both direct and indirect methods. In both cases, the observations could have a spectral resolution lower than the baseline SSiC-L value ($R>100,000$ versus $R=120,000$). For direct detection (SSiC-26a), the data should extend farther into the NIR to 2440~nm and could have a redder UV cut-off of 120~nm. For indirect detection (SSiC-26b), the data should be obtained over a wavelength range extending farther into the UV but less far into the NIR (i.e., $92~\mathrm{nm}-1100~\mathrm{nm}$). Within the solar system, the investigation of the effects of solar energetic particles on giant planets proposed in SSiC-25 \citep[][Section~\ref{sssec:aurorae}]{chaufray_hwo25} needs spectropolarimetric data extending from the default SSiC-L blue cut-off ($\lambda_\mathrm{blue}=100$~nm) to a shorter wavelength red cut-off ($\lambda_\mathrm{red} = 1000$~nm) at a lower spectral resolution ($R>50,000$). 

Next, investigations of magnetic fields in young planetary systems for SSiC-34 \citep[][Section~\ref{sssec:young_bfields}]{gomez_de_castro_magnetic_hwo25} would necessitate SSiC-L data obtained at a lower spectral resolution ($R=30,000$) over a slightly narrower wavelength range ($100-1500$~nm). An even lower spectral resolution ($R>6000$) would suffice for investigating aerosols in the atmospheres of transiting planets for SSiC-16b \citep[][Section~\ref{sssec:transits}]{wakeford_hwo25}. Program SSiC-16b could also tolerate a reduced wavelength range of $300-1000$~nm. The moderate resolution phase-curve observations obtained for SSiC-16b would be complementary to the higher resolution observations of SSiC-24 \citep[][Section~\ref{sssec:highres_atmos}]{cubillos_hwo25}, which would investigate atmospheric composition and escape for a diverse set of planets. Compared to the baseline SSiC-L configuration, SSiC-24 could use a lower spectral resolution ($R>100,000$ versus $R=120,000$) but would require additional NIR coverage to 2500~nm. 

Finally, as discussed by \citet[][Section~\ref{sssec:aminos}]{gomez_de_castro_amino_hwo25}, observations of comets at a much lower spectral resolution ($R\sim300$) over a considerably narrower wavelength range ($140-220$~nm) would be useful for studying the distribution and chirality of amino acids within the solar system. The challenging aspect of SSiC-33 observations is that the polarization error would need to be reduced to $\le0.1\%$. 

SSiC-25 and SSiC-33 are distinct from the other SSiC-L programs in that they focuses on solar system targets, but it would be possible to increase observational efficiency by combining the exoplanetary and stellar observations. For instance, depending on onboard data storage, it could be possible to obtain portions of the phase curves needed for SSiC-16b using the higher spectral resolution and extended wavelength coverage needed for SSiC-24. 

\subsubsection{SSiC-M: IFS Observations at High Spatial Resolution}
\label{sssec:ssic_m}
Spatially-resolved observations at high spatial resolution are important for a wide range of solar system investigations as well as observations of the remnants of planet formation around other stars. We classify these observations as SSiC-M and adopt the same settings as GG-A. Accordingly, SSiC-M observations have a broad wavelength range of $90-2500$~nm, spectral resolution of $R=10,000$, a spatial resolution $\theta=20$~mas, and a field of view of $10''\times10''$. 

The first science case requiring SSiC-M observations is SSiC-1 \citep[][Section~\ref{sssec:oceans_habitable}]{cartwright_hwo25}, which aims to investigate the habitability of ice bodies in the solar system. Compared to baseline SSiC-M settings, SSiC-1 would require observations over a broader wavelength range ($50-5000$~nm) but could use a smaller field of view ($3''\times3''$). Investigations of the aurorae and atmospheres of giant planets for SSiC-29 \citep[][Section~\ref{sssec:solargiant}]{fletcher_hwo25} could also be accomplished using a smaller field of view ($\ge3''\times3''$). SSiC-29 can be divided into three observing programs: SSiC-29a needs $R\sim100,000$ spectra at $50-400$~nm, SSiC-29b needs $R>1740$ at $400-700$~nm, and SSiC-29c needs $R>3450$ at $700-3500$~nm. Observations of Titan for SSiC-27 (Section~\ref{sssec:titan}; posted in the STScI SCDD Portal\footnote{\url{https://docs.google.com/document/d/1bJwOROlvOCkGa740a_h2h8rx9qlWMgH2}}) could use an even smaller field of view ($1''\times1''$), a lower spatial resolution of $\theta=50$~mas, a reduced spectral resolution of $R\sim100$, and a narrower wavelength range ($121-2000$~nm).

The other SSiC-M science cases require larger fields of view. For instance, investigations of Venus for SSiC-5 \citep[][Section~\ref{sssec:venus}]{izenberg_hwo25} should cover a $40''\times40''$ region at relaxed spectral and spatial resolution ($R\sim200$ and $\theta\le30$~mas, respectively). The Venus observations could also be obtained over a narrower wavelength range extending from 200~nm to at least 1500~nm rather than the full $90-2500$~nm range baselined for SSiC-M. Observations of Mars for SSiC-28 (Section~\ref{sssec:mars_post}; posted in the STScI SCDD Portal\footnote{\url{https://docs.google.com/document/d/1KmoZ-ywYXpzbgQsL7BnHrqdsMtBYe_TKwot2quMU97g}}) need a slightly larger field of view ($\geq50''\times50''$) and could also accommodate a shorter-wavelength red cut-off ($\lambda_\mathrm{red}=1500$~nm) and a reduced spectral resolution ($R>3000$). Unlike the Venus observations, the Mars data should be obtained at a much higher spatial resolution ($\theta < 3.5$~mas) and extend significantly farther into the UV to 80~nm. 

The SSiC-M program requesting the largest field of view is SSiC-9, which would study debris disks \citep[][Section~\ref{sssec:debris}]{rebollido_hwo25}. The IFS component SSiC-9f demands an extensive field of view ($120''\times120''$), extended NIR coverage to 3000~nm, and a high spectral resolution of $R\ge80,000$. However, SSiC-9f could tolerate a relaxed blue cut-off ($\lambda_\mathrm{blue} = 140$~nm) and reduced spatial resolution ($\theta\le 30$~mas). Although the SCDD does not quantify the necessary field of view, a large field of view would also be advantageous for the closer systems observed by SSiC-23 (Section~\ref{sssec:disk_winds}). In order to investigate disk depletion and planet formation, SSiC-23 observations should have a spatial resolution $\theta \le 50$~mas and cover $91 - 200$~nm at high spectral resolution ($R\sim100,000$).

All of the science cases requesting SSiC-M observations focus on different targets, so they will each require their own datasets.

\subsubsection{SSiC-N: IFS Observations at Moderate Spatial Resolution}
\label{sssec:ssic_n}
One SSiC science case requests IFS observations at a reduced spatial resolution. Accordingly, we define SSiC-N as IFS spectroscopy obtained at wavelengths $100-2000$~nm at spectral resolution $R=3500$ and spatial resolution $\theta=200$~mas over a $10''\times10''$ field of view.

Three of the programs within SSiC-29 \citep[][Section~\ref{sssec:solargiant}]{fletcher_hwo25} require SSiC-N data to investigate the atmospheres and aurorae of solar system giant planets. First, SSiC-29d needs data at a much higher spectral resolution ($R\sim100,000$) at UV wavelengths of $50-400$~nm. Compared to the baseline SSiC-N wavelength range of $100-2000$~nm, SSiC-29d observations have a shorter blue cut-off (50~nm) and much shorter red cut-off (400~nm). Next, SSiC-29e needs SSiC-N observations over a narrower wavelength range of $400-700$~nm at a spectral resolution $R>1740$, which is lower than the baseline value of $R=3500$. Finally, SSiC-29f necessitates observations at the default SSiC-N spectral resolution of $R=3500$ beginning at a longer red cut-off (700~nm) and extending to a significantly longer red cut-off (3500~nm). The three sub-programs cover different wavelength ranges so each dataset is required for SSiC-29, but observational efficiency could be increased by maximizing the wavelength range that could be observed simultaneously. 
 
\subsubsection{SSiC-P: Photometric Observations}
\label{sssec:ssic_p}
Finally, the last type of observation needed for SSiC science is photometry. For the baseline photometric configuration SSiC-P, we set a spatial resolution of 20~mas over a field of view of $120'' \times 120''$. We list the default wavelength range as $95-2000$~nm but note that this broad coverage could be obtained by multiple instruments or different channels within a single instrument. 

SSiC-P data are needed for investigations of giant planets for SSiC-29 \citep[][Section~\ref{sssec:solargiant}]{fletcher_hwo25} as well as identification and characterization of near-Earth objects for SSiC-30 \citep[][Section~\ref{sssec:defense}]{dotson_hwo25}. SSiC-29 needs two types of imaging. SSiC-29g needs UV imaging ($50-400$~nm) over a smaller field of view ($\geq20'' \times 20''$) while SSiC-29h demands Vis/NIR multiband imaging of the default SSiC-P field of view ($120''\times120''$). Although the giant planets would span $\lesssim 50''$, the larger field of view would facilitate more efficient observations of moons and rings along with the giant planets they orbit. SSiC-29h could accommodate a relaxed blue cut-off of 400~nm but observations should extend much redder to 3500~nm. 

For SSiC-30, the most important capabilities are non-sidereal tracking at rates $>110$~mas~s$^{-1}$ and the ability to look closer to the Sun rather than the specific choice of wavelength range or field of view. \citet{dotson_hwo25} notes that current designs for the field of view ``are adequate'' and that instantaneous field of regard should be ``as large as possible.'' The instrument should have high throughput and broad Vis/NIR wavelength coverage to increase sensitivity to faint objects. In addition, brightness measurements should be directly comparable to those made by ground-based facilities, which motivates the use of standard filters. 

SSiC-29 and SSiC-30 target different objects, so the programs cannot share SSiC-P data. However, the efficiency of both programs could be increased by the ability to simultaneously obtain data at multiple wavelengths. 

\subsection{Living Worlds}
The Living Worlds science cases are summarized in Table~\ref{tab:lw_scdds}. Compared to the other SWGs, LW has the highest overlap between the objectives and datasets needed for various science cases. Although the targets and specific wavelength ranges vary slightly from one LW science case to another, each of them is connected to constraining the frequency of life in the universe. Some science cases aim to look for signs of habitability (Section~\ref{sssec:polbio}), Earth-like life (Sections~\ref{sssec:life}, \ref{sssec:surface_bio}, \& \ref{sssec:polbio}), life unlike Earth life (Section~\ref{sssec:lawdki}), and technology created by life (Section~\ref{sssec:technosignatures}) while others constrain the parameter space in which life might evolve (Sections~\ref{sssec:origin} \& \ref{sssec:prebiosignatures}) or consider the robustness of various potential biosignatures (Section~\ref{sssec:fpbio}). 

\begin{figure*}
 \centering
 \includegraphics[width=\linewidth]{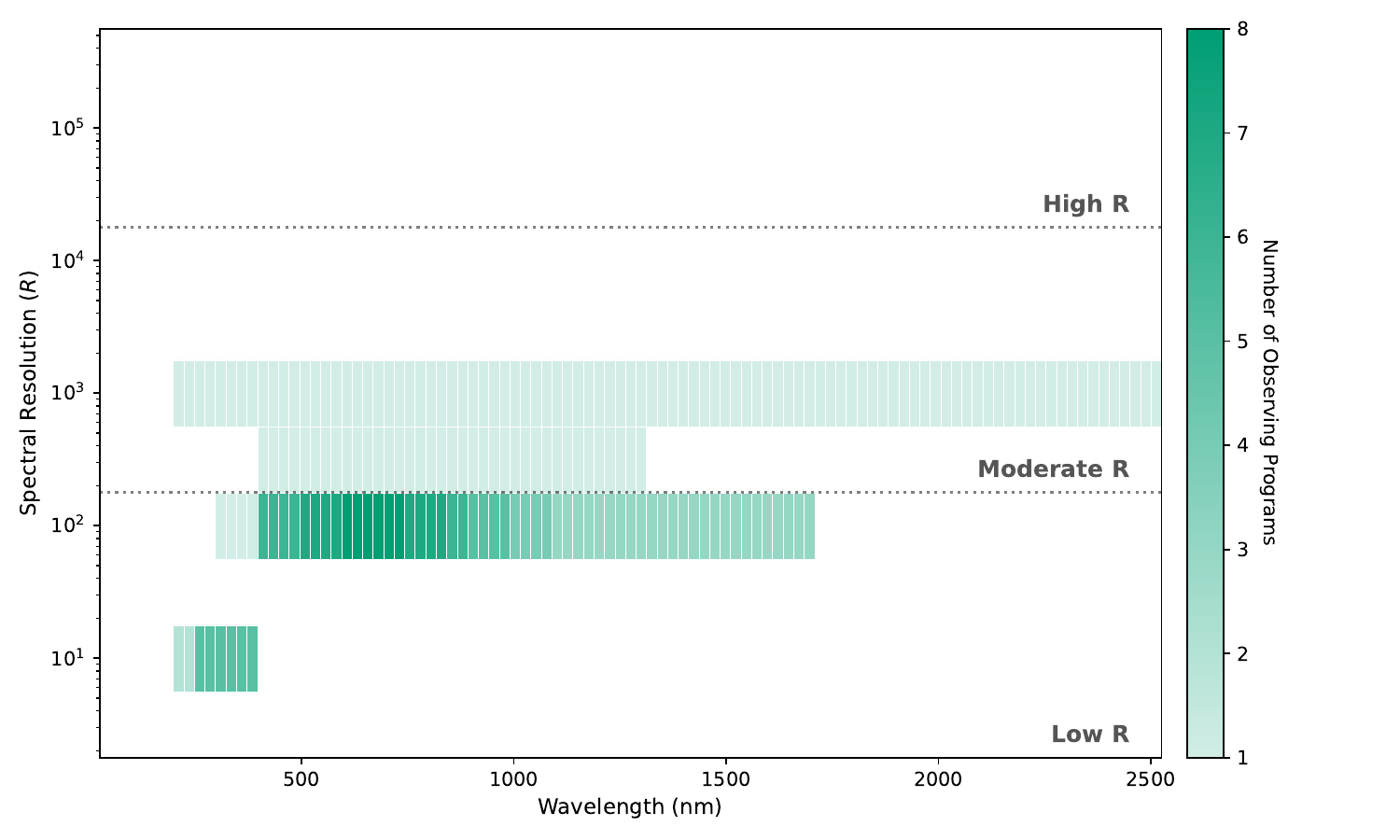}
 \caption{Same as Figure~\ref{fig:gg_r_wave} but for LW science cases.}
 \label{fig:lw_r_wave}
\end{figure*}

In Table~\ref{tab:lw_remap}, we compare the targets and capabilities needed for each of the LW science cases. We then compare the spectral resolution and wavelength coverage required for different observations in Figure~\ref{fig:lw_r_wave}. Both Table~\ref{tab:lw_remap} and Figure~\ref{fig:lw_r_wave} reveal that many science cases request identical or nearly identical observational capabilities. This agreement is a result of the high degree of coordination among LW SWG members when developing science cases and the practice of using a common set of simulations to inform multiple science cases. See LW-1 \citep[][Section~\ref{sssec:life}]{arney_hwo25} for more details about the collaborative approach used by LW and the assumptions made during the simulation process. 

Although there is an extremely high degree of similarity in the capabilities required by various LW science cases, the breadth of those capabilities is large enough to necessitate six different observation types. As shown in Table~\ref{tab:lw_remap}, the six groups are: 
\begin{itemize}
 \item \textbf{LW-A} (Section~\ref{sssec:lw_a}): high-contrast spectroscopy in the UV
 \item \textbf{LW-B} (Section~\ref{sssec:lw_b}): high-contrast spectroscopy in the visible
 \item \textbf{LW-C} (Section~\ref{sssec:lw_c}): high-contrast spectroscopy in the NIR
 \item \textbf{LW-D} (Section~\ref{sssec:lw_d}): high-contrast spectropolarimetry
 \item \textbf{LW-E} (Section~\ref{sssec:lw_e}): high-contrast polarimetric imaging
 \item \textbf{LW-G} (Section~\ref{sssec:lw_g}): astrometric observations
\end{itemize} 

The first five categories require high-contrast capabilities to block out the light from the host star while the sixth category uses the flux from the host star to measure planet masses via astrometry. For high-contrast categories, observations must be sensitive to planet-to-star flux ratios (contrasts) $\le10^{-10}$ within the habitable zone. The mapping between the instellation flux received by a planet and the planet-star angular separation depends on the stellar temperature, luminosity, and distance as well as orbital eccentricity and orientation. In general, LW science cases would benefit minimizing the inner working angle. For more details, see Section~\ref{sec:drivers}. 

\subsubsection{LW-A: High-Contrast Spectroscopy in the UV}
\label{sssec:lw_a}
The first type of observation needed to address LW objectives is high-contrast spectroscopy at UV wavelengths. We classify these observations as LW-A and set the baseline as $R=7$ spectra at $200-400$~nm. 

LW-A observations are essential for providing the UV access needed for multiple LW science cases. Notably, the LW-A wavelength range includes the $200-350$~nm Hartley-Huggins band of ozone and would therefore permit measurements of ozone abundance on potentially inhabited planets. As discussed at length in Section~\ref{ssec:lw_scdds}, constraining ozone abundance is important for assessing planetary habitability and interpreting potential biosignatures, especially when oxygen is inaccessible. 

Accordingly, LW-A observations are required to assess the frequency of Earth-like biospheres for LW-1a, investigate the origin of life for LW-3a \citep[][Section~\ref{sssec:origin}]{ranjan_origin_hwo25}, and estimate the occurrence rate of prebiotic environments for LW-4a \citep[][Section~\ref{sssec:prebiosignatures}]{ranjan_prebiosignatures_hwo25}. In addition, LW-A data are needed for LW-5a to search for technosignatures \citep[][Section~\ref{sssec:technosignatures}]{kopparapu_hwo25} and for LW-13a to search for potential seasonal variations that could be caused by life \citep[][Section~\ref{sssec:seasonality}]{lafleche_hwo25}. Finally, LW-10a requires LW-A data to determine the frequency of oxygenated atmospheres on Venus-like planets \citep[][Section~\ref{sssec:fpbio}]{schwieterman_hwo25}.

All of these science cases would use the LW-A default spectral resolution ($R=7$) and red cut-off (400~nm), but they have different blue cut-offs. Science cases LW-3a and LW-4a require stricter blue cut-offs of 200~nm to maximize sensitivity to ozone and ensure full coverage of the Hartley band. In contrast, science cases LW-13a and LW-10a follow the main LW science case LW-1a in requiring a redder UV cut-off of 250~nm. While $250-400$~nm includes the full Huggins band, it truncates the blue side of the Hartley band. When defining LW-A, we intentionally selected the stricter 200~nm limit to cover the full Huggins-Hartley system. We note that simulations by A. Young and the LW SWG indicate that a blue cut-off of 250~nm would be sufficient for ozone detection (see Figure~16 of \citealt{arney_hwo25}) but removing access to UV features $<250$~nm would reduce the ability to understand systematic effects and may limit the interpretability of low-resolution spectra, especially given the broad parameter space of atmospheric chemistry. Finally, LW-5a would use an even longer wavelength cut-off of $\lambda_\mathrm{blue}=300$~nm. 

\subsubsection{LW-B: High-Contrast Spectroscopy in the Visible}
\label{sssec:lw_b}
High-contrast spectroscopy at visible wavelengths was identified by multiple LW working groups as an essential capability. For the baseline case, we define LW-B as visible coronagraphy at spectral resolution $R=140$ at $400-1100$~nm. These parameters are identical to those requested by the primary LW science case LW-1b to determine the frequency of Earth-like global biospheres \citep[][Section~\ref{sssec:life}]{arney_hwo25}.

The default LW-B parameters are also sufficient for investigating the origin of life for LW-3b \citep[][Section~\ref{sssec:origin}]{ranjan_origin_hwo25} and measuring the frequency of prebiotic environments for LW-4b \citep[][Section~\ref{sssec:prebiosignatures}]{ranjan_prebiosignatures_hwo25}. Detecting seasonal variations for LW-13b could be addressed using LW-B observations with a relaxed red cut-off of 1000~nm \citep[][Section~\ref{sssec:seasonality}]{lafleche_hwo25} while technosignatures investigations for LW-5b could use an even shorter wavelength cut-off of $\lambda_\mathrm{red}=900$~nm. Additionally, 
detecting surface biosignatures for LW-14 could accommodate a relaxed blue cut-off of 500~nm and a lower spectral resolution of $R\lesssim130$ \citep[][Section~\ref{sssec:surface_bio}]{parenteau_surface_hwo25}. Two LW science cases would prefer higher resolution than the LW-B default of $R=140$: exploring the diversity of extraterrestrial life for LW-6 \citep[][Section~\ref{sssec:lawdki}]{walker_hwo25} requests $R\sim100-1000$ while measuring the oxygen levels of Venus analogs for LW-10b \citep[][Section~\ref{sssec:fpbio}]{schwieterman_hwo25} needs $R\sim200$. \citet{kopparapu_hwo25} do not specify the spectral resolution needed for searching for technosignatures, but higher spectral resolution would likely be advantageous for detecting emission or absorption features that could be associated with alien technology.

\subsubsection{LW-C: High-Contrast Spectroscopy in the NIR}
\label{sssec:lw_c}
Several LW science cases also require high-contrast spectroscopy in the near-infrared. We categorize these observations as LW-C. The default spectral resolution is $R=70$ and the default wavelength range is $1100-1700$~nm. 

The baseline LW-C configuration exactly matches that requested for determining the frequency of inhabited Earth-like planets for the primary LW science case LW-1c \citep[][Section~\ref{sssec:life}]{arney_hwo25}. Default LW-C (and LW-1c) observations would also be sufficient for LW-3c to investigate the origin of life \citep[][Section~\ref{sssec:origin}]{ranjan_origin_hwo25} and for LW-4c to determine the frequency of prebiotic environments \citep[][Section~\ref{sssec:prebiosignatures}]{ranjan_prebiosignatures_hwo25}. For assessing the value of oxygen as a potential biosignature, LW-10c would require LW-C observations at a higher spectral resolution of $R\sim200$ to a reduced red cut-off of 1300~nm \citep[][Section~\ref{sssec:fpbio}]{schwieterman_hwo25}.

\subsubsection{LW-D: High-Contrast Spectropolarimetric Observations}
\label{sssec:lw_d}
The next LW observations are high-contrast spectropolarimetry. We set the baseline configuration LW-D to cover $350-850$~nm at a spectral resolution of $R=100$ with full Stokes (IQUV) polarization. These default observations should have polarimetric precision $\leq1\%$.

Obtaining polarimetric observations at various orbital phases would support studies of clouds, surface compositions, surface biosignatures, and potentially even technosignatures. Specifically, two LW-7 observing programs (LW-7a \& LW-7b) require spectropolarimetry to characterize inhabited planets and search for evidence of homochirality \citep[][Section~\ref{sssec:polbio}]{berdyugina_imaging_hwo25}. For initial characterization of potentially inhabited planets (LW-7a), a relaxed blue cut-off of 400~nm, lower spectral resolution ($R\sim50-100$), and reduced polarimetric capabilities (Stokes IQU only) would be sufficient. 

For detailed follow-up studies of potentially inhabited planets to investigate the chirality of putative biosignatures (LW-7b), enhanced polarimetric precision ($\leq 0.01\%$) over a narrower wavelength range ($600-750$~nm) would be needed. Achieving the goals of LW-7b also requires measuring circular polarization, thereby justifying the choice of full Stokes (IQUV) for the baseline LW-D configuration and driving the need for exquisite polarimetric precision. 

\subsubsection{LW-E: High-Contrast Polarimetric Imaging}
\label{sssec:lw_e}
The next set of LW observations is high-contrast polarimetric imaging. We classify these observations as LW-E. As for the spectropolarimetric baseline LW-D, the default polarimetric imaging configuration LW-E should include full Stokes (IQUV) polarimetry at polarimetric precision $\leq1\%$. For LW-E, we set the baseline wavelength coverage to $300-1700$~nm. Although the default red limit of 1700~nm is much longer than needed for LW science cases, we select this value to match the wavelength range of the SSiC polarimetric imaging configuration SSiC-G (see Section~\ref{sssec:ssic_g}). Given the broad wavelength range, these observations would likely be obtained using multiple channels. If the instrument includes both an NIR channel and a visible channel, LW science cases would not impact the design of the NIR channel. 

There are two LW science cases requiring LW-E observations. First, searching for technosignatures for LW-5c demands LW-E observations at $300 - 900$~nm but not NIR wavelengths \citep[][Section~\ref{sssec:technosignatures}]{kopparapu_hwo25}. Second, searching for surface liquid water and potential bio-pigments for LW-7c would require observations at multiple wavelengths between 350~nm and 850~nm.\footnote{Note that the science case SSiC-4 \citep[][Section~\ref{sssec:surface_water}]{lustig-yaeger_liquidwater_hwo25}, which also aims to look for surface liquid water, requests a redder wavelength range of $400-1100$~nm. Like the other SSiC polarimetric imaging cases, SSiC-4 is classified as part of SSiC-G, which shares default settings with LW-E. Our default configuration for polarimetric imaging therefore satisfies the needs of both surface water science cases (SSiC-4 \& LW-7).} As for LW-5c, extension into the NIR would not be needed for LW-7c. Additionally, LW-7c does not require detecting circular polarization, so Stokes IQU would suffice. 

\subsubsection{LW-F: High-Contrast Imaging}
\label{sssec:lw_f}
The only LW science case that explicitly requests high-contrast imaging is LW-5d \citep[][Section~\ref{sssec:technosignatures}]{kopparapu_hwo25}, but imaging is likely a prerequisite for the spectroscopic observations needed for other LW science cases. For LW-5d, time-series imaging at visible wavelengths is needed to obtain phase curves that could reveal signatures of artificial structures on planetary surfaces. 

\subsubsection{LW-G: Astrometric Observations}
\label{sssec:lw_g}
The final category of LW observations is wide-field imaging with high astrometric precision, which we categorize as LW-G. These observations should consist of $400-700$~nm imaging at spatial resolution $\theta=11$~mas over a field of view of $6'\times6'$. The observations must have astrometric precision no worse than $0.03 \mu$as per epoch.

The demand for astrometric imaging is driven by LW-12, which aims to determine the masses of potentially habitable planets by measuring the astrometric wobble induced in the host star positions by their planets \citep[][Section~\ref{sssec:mp}]{gary_hwo25}. LW-12 is the only astrometric case in the LW portfolio, so we set the defaults for LW-G astrometry to match the needs of LW-12. As discussed in Section~\ref{sssec:mp}, the most important factors for LW-12 are access to a wide field of view, high spatial resolution, and high astrometric precision. 

\section{Interdisciplinary Connections Across Science Working Groups}
\label{sec:xswg}
While practical, the delineations between SWGs are somewhat artificial. In this section, we consider the full portfolio of science presented in this paper. As might be expected, the strongest overlaps are between GG and EE and between SSiC and LW. We highlight a few common themes below. Additionally, in Table~\ref{tab:planetgrid}, we highlight the synergies between SSiC and LW high-contrast science cases by categorizing the necessary observations by target category and insolation flux. As the table reveals, observations of circumstellar material and planets of all sizes could be potentially useful for multiple science cases. A prime example of the observational overlap between science cases is the search for technosignatures described in LW-5 \citep[][Section~\ref{sssec:technosignatures}]{kopparapu_hwo25}: LW-5 appears in every grid cell of Table~\ref{tab:planetgrid} because nearly all high-contrast observations of planetary systems have the potential to reveal evidence of technosignatures.

\input{exoplanet_grid}

\subsection{Star formation in low-metallicity environments}
\label{ssec:xswg_lowz}

Multiple GG and EE science cases require spatially resolved spectroscopy of galaxies, and the same data could be used for multiple science cases. The low-mass, metal-poor galaxy IZw18 \citep{aloisi_et_al2007, annibali_et_al2013} at 18.9~Mpc is an attractive target for GG science cases GG-4 \citep[][Section~\ref{sssec:bh_quiescent}]{pacucci_hwo25}, GG-11 \citep[][Section~\ref{sssec:ionizing_lf}]{mccandliss_hwo25}, and GG-17 \citep[][Section~\ref{sssec:cgm_elm}]{burchett_hwo25} as well as EE~science cases EE-1 \citep[][Section~\ref{sssec:massive_stars_lowZ}]{senchyna_hwo25}, EE-5 \citep[][Section~\ref{sssec:vms}]{martins_hwo25}, and EE-12 \citep[][Section~\ref{sssec:ism_uv}]{james_hwo25}.

In addition, EE-12 \citep[][Section~\ref{sssec:ism_uv}]{james_hwo25} requires UV IFS observations ($R>10,000$, $95-285$~nm, $\leq 0.1 \times 0.1''$ spaxels, $10'' \times 10''$ FOV) of local analogs to high redshift galaxies. These targets should have high star formation rates, low masses and low metallicities, which means they will also be compelling targets for GG-2 \citep[][Section~\ref{sssec:bh_mass_spin}]{cann_hwo25}, which focuses on low-metallicity dwarf galaxies with masses $< 10^{8.5} M_\odot$, and GG-17 \citep[][Section~\ref{sssec:cgm_elm}]{burchett_hwo25}, which considers a small set of 12~galaxies with diverse properties. Some of the galaxies observed for EE-12 could also be included in the sample of 100~galaxies needed for GG-5 \citep[][Section~\ref{sssec:agn_outflow}]{zhang_hwo25}. If observations for EE-12 are observed at higher spectral resolution over a broader wavelength range, they could also be included in studies of the ionization of the universe by galaxies and their constituents, especially LyC clusters. For instance, $R \geq 20,000$ at $100 - 510$~nm and photometry at $\lambda <100$~nm would be suitable for GG-9 \citep[][Section~\ref{sssec:lyman_indirect}]{citro_hwo25}
; $R\geq 30,000$ at $\lambda \geq 90$~nm for GG-6 \citep[][Section~\ref{sssec:resolve_reionization}]{xu_reionization_hwo25}; and $R\geq 30,000$ at $\lambda \geq 60$~nm for GG-7 \citep[][Section~\ref{sssec:lyman_escape}]{carr_hwo25}.

\subsection{The lives and deaths of massive stars}
\label{ssec:xswg_stars}
Similarly, GG IFS observations of galaxies including massive stars will be useful for the EE science cases discussed in EE-1 \citep[][Section~\ref{sssec:massive_stars_lowZ}]{senchyna_hwo25}
and EE-5 \citep[][Section~\ref{sssec:vms}]{martins_hwo25}. The observational needs of the EE science cases ($R\gtrsim5000$ at $90 - 500$~nm at $\leq 20$~mas over a $>3'' \times 3''$ FOV for EE-1; $R>5000$ at $100 - 700$~nm at 5~mas resolution over a $3'' \times 3''$ FOV for EE-5) suggest that those observations would pair well with those needed for GG studies of AGN feedback for GG-5 \citep[][Section~\ref{sssec:agn_outflow}; $R>5000$ at $1-10$~pc resolution over $100 - 5000$~nm]{zhang_hwo25}; the origin and influence of SMBHs for GG-2 \citep[][Section~\ref{sssec:bh_mass_spin}; $R=3000$ at 1~mas resolution over $90 - 2500$~nm and a $1' \times 1'$ FOV]{cann_hwo25}; and the exchange of matter and energy within galaxies for GG-17 \citep[][Section~\ref{sssec:cgm_elm}; $R=10,000$ at $97.5\,{\rm nm} \leq \lambda \leq 160$~nm and 5~mas resolution over a $10'' \times 10''$~FOV]{burchett_hwo25}.

\subsection{Dust \& extinction}
\label{ssec:xswg_dust}
Furthermore, GG observations towards local group galaxies and local volume galaxies may also be useful for the EE study of extragalactic dust grains in EE-2 \citep[][Section~\ref{sssec:dust_extinction}]{paladini_hwo25} for observations at low spectral resolution ($R\sim 3000$) or EE-11 \citep[][Section~\ref{sssec:dust_uv}]{roman-duval_hwo25} for observations at high spectral resolution ($R\sim50,000$). Complementary GG science cases at low spectral resolution include GG-5 \citep[][Section~\ref{sssec:agn_outflow}]{zhang_hwo25}, GG-2 \citep[][Section~\ref{sssec:bh_mass_spin}]{cann_hwo25}, and GG-17 \citep[][Section~\ref{sssec:cgm_elm}]{burchett_hwo25} while those at high spectral resolution include GG-6 \citep[][Section~\ref{sssec:resolve_reionization}]{xu_reionization_hwo25}, GG-7 \citep[][Section~\ref{sssec:lyman_escape}]{carr_hwo25}, GG-9 \citep[][Section~\ref{sssec:lyman_indirect}]{citro_hwo25}, and GG-16 \citep[][Section~\ref{sssec:disk_cgm}]{borthakur_hwo25}.

\subsection{Terrestrial planets}
\label{ssec:xswg_terrestrial}
Switching from galaxies to exoplanets, some of the potentially habitable and inhabited planets needed for the LW science cases will be detected in binary star systems by SSiC-6 \citep[][Section~\ref{sssec:occ_binary}]{newton_hwo25}. In addition, the SSiC investigations of the frequency of small planets with various compositions in SSiC-3 \citep[][Section~\ref{sssec:rocky_sub}]{hu_rocky_hwo25} and the conditions for terrestrial planets to retain volatiles in SSiC-14 \citep[][Section~\ref{sssec:escape}]{dossantos_hwo25} and SSiC-21 (Section~\ref{sssec:retention}) will be useful for identifying terrestrial planets with atmospheres that could be searched for biosignatures as part of LW-1 \citep[][Section~\ref{sssec:life}]{arney_hwo25}, LW-14 \citep[][Section~\ref{sssec:surface_bio}]{parenteau_surface_hwo25}, \& LW-7 \citep[][Section~\ref{sssec:polbio}]{berdyugina_imaging_hwo25} or signals that appear to be biosignatures but are actually false positives for LW-10 \citep[][Section~\ref{sssec:fpbio}]{schwieterman_hwo25}. Furthermore, the planet mass measurements resulting from LW-12 \citep[][Section~\ref{sssec:mp}]{gary_hwo25} would be important for interpreting the spectra of terrestrial planets obtained for SSiC-1 \citep[][Section~\ref{sssec:ozone_onset}]{blunt_et_al2025}, SSiC-2 \citep[][Section~\ref{sssec:hab_system}]{hasegawa_water_hwo25}
SSiC-3 \citep[][Section~\ref{sssec:rocky_sub}]{hu_rocky_hwo25}, SSiC-18 \citep[][Section~\ref{sssec:exovenus}]{kane_venus_hwo25}, SSiC-20 (Section~\ref{sssec:survive_water}), and SSiC-21 (Section~\ref{sssec:retention}).

\subsection{Habitability}
\label{ssec:xswg_habitable}
Searches for interior oceans on icy exoplanets for SSiC-17 \citep[][Section~\ref{sssec:id_oceans}]{quick_hwo25} and improved characterization of icy worlds within the solar system for SSiC-1 \citep[][Section~\ref{sssec:oceans_habitable}]{cartwright_hwo25} will also advance understanding of the range of habitable locations in the universe for LW-4 \citep[][Section~\ref{sssec:prebiosignatures}]{ranjan_prebiosignatures_hwo25} and the diversity of environments in which life could evolve for LW-3 \citep[][Section~\ref{sssec:origin}]{ranjan_origin_hwo25}. In addition, the SSiC investigations of correlations between the water abundances of inner planets and the presence of outer planets in exoplanetary systems for SSiC-2 \citep[][Section~\ref{sssec:hab_system}]{hasegawa_water_hwo25}, the presence of surface oceans on terrestrial exoplanets for SSiC-4 \citep[][Section~\ref{sssec:surface_water}]{lustig-yaeger_liquidwater_hwo25}, and studies of how the giant planets in the solar system affected the delivery of volatiles to the inner solar system for SSiC-7 \citep[][Section~\ref{sssec:solar}]{mandt_hwo25} will provide further insight into possible connections between the architectures of planetary systems and the emergence of life for LW-3 \citep[][Section~\ref{sssec:origin}]{ranjan_origin_hwo25}.

\subsection{Biosignatures}
\label{ssec:xswg_biosig}
All planetary systems studied for SSiC science cases will be useful for the investigations of the prevalence of technosignatures proposed by LW-5 \citep[][Section~\ref{sssec:technosignatures}]{kopparapu_hwo25} and planets with a wide range of properties will be intriguing targets for life unlike Earth life through LW-6 \citep[][ Section~\ref{sssec:lawdki}]{walker_hwo25}. 

Additionally, SSiC-1 \citep[][Section~\ref{sssec:ozone_onset}]{blunt_et_al2025} and SSiC-18 \citep[][Section~\ref{sssec:exovenus}]{kane_venus_hwo25} probe the timescales upon which habitable environments and life arise and die, which in turn affects the number and fraction of planets that are habitable or inhabited at any given instant. Those investigations are therefore relevant for LW-1 \citep[][Section~\ref{sssec:life}]{arney_hwo25}, LW-14 \citep[][Section~\ref{sssec:surface_bio}]{parenteau_surface_hwo25}, LW-7 \citep[][Section~\ref{sssec:polbio}]{berdyugina_imaging_hwo25}, \& LW-13 \citep[][Section~\ref{sssec:seasonality}]{lafleche_hwo25}. Understanding the conditions on Venus-like exoplanets through SSiC-18 \citep[][Section~\ref{sssec:exovenus}]{kane_venus_hwo25} and viewing our own planet Venus as an exoplanet for SSiC-5 \citep[][Section~\ref{sssec:venus}]{izenberg_hwo25} will also contribute to assessing the reliability of various biosignatures such as oxygen for LW-10 \citep[][Section~\ref{sssec:fpbio}] {schwieterman_hwo25}. 

\section{Scientific Drivers for Tackling Technical Challenges}
\label{sec:drivers}
\begin{figure*}
 \centering
 \includegraphics[width=1\linewidth]{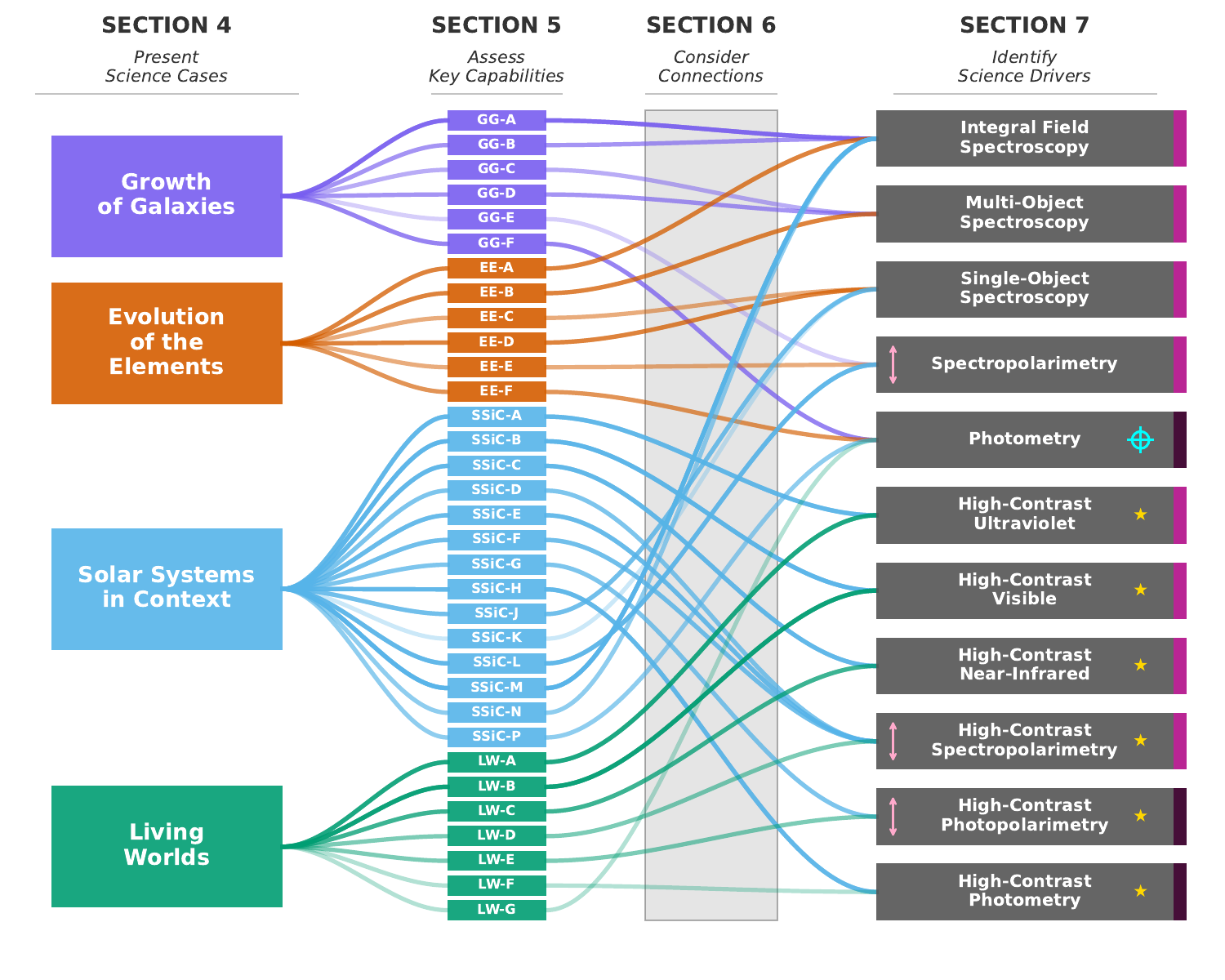}
 \caption{Visualization of the procedure used to connect scientific objectives to technical capabilities. The leftmost column shows the four Science Working Groups and represents the science cases presented in Section~\ref{sec:scdds}. The second column from the left shows how the individual science cases were grouped by observational needs in Section~\ref{sec:swg_obs}. The opacities of the lines connecting the working group boxes to the grouped observations indicate the number of science cases mapped to each observation, with darker lines indicating a higher number of science cases. Next, the shaded gray box emphasizes the overlap between science cases identified by different working groups; we discuss those interdisciplinary connections in Section~\ref{sec:xswg}. Finally, the rightmost column lists the observation types needed to accomplish the investigations proposed by the SCDDs. In Section~\ref{sec:drivers}, we consider how the various science cases could potentially drive instrument and telescope performance. The vertical stripes on the right sides of the observation boxes differentiate spectroscopy (magenta) from photometry (maroon) while the symbols within the box identify polarimetry (pink double-headed arrows), high-contrast observations (yellow stars), and astrometry (cyan crosshair).}
 \label{fig:core}
\end{figure*}

In Section~\ref{sec:scdds}, we described the science cases identified by the START and Science Working Groups. We then identified common observational needs within and across science working groups in Sections~\ref{sec:swg_obs} and~\ref{sec:xswg}, respectively. As part of those discussions, we explored which investigations could share datasets or observing modes. We summarize this process graphically in Figure~\ref{fig:core}. 

\begin{figure*}[tb]
 \centering
 \includegraphics[width=\linewidth]{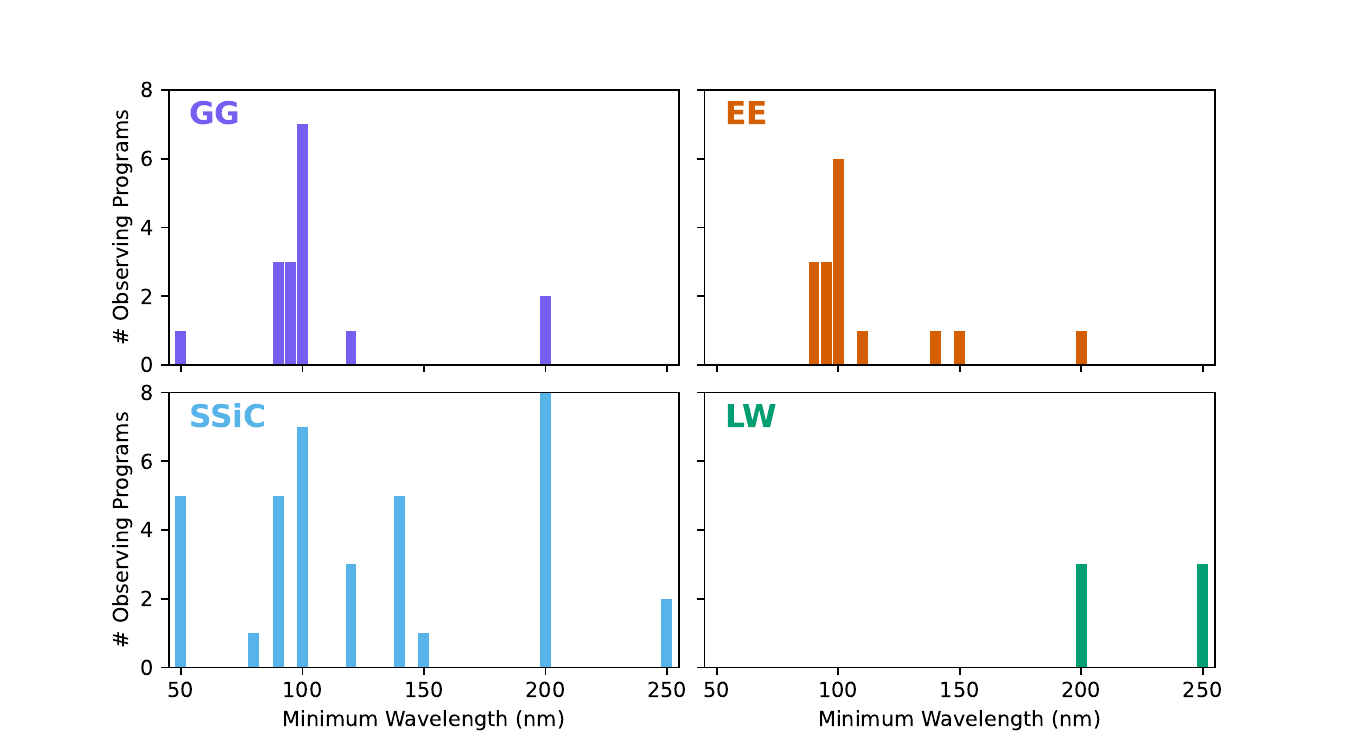}
 \includegraphics[width=\linewidth]{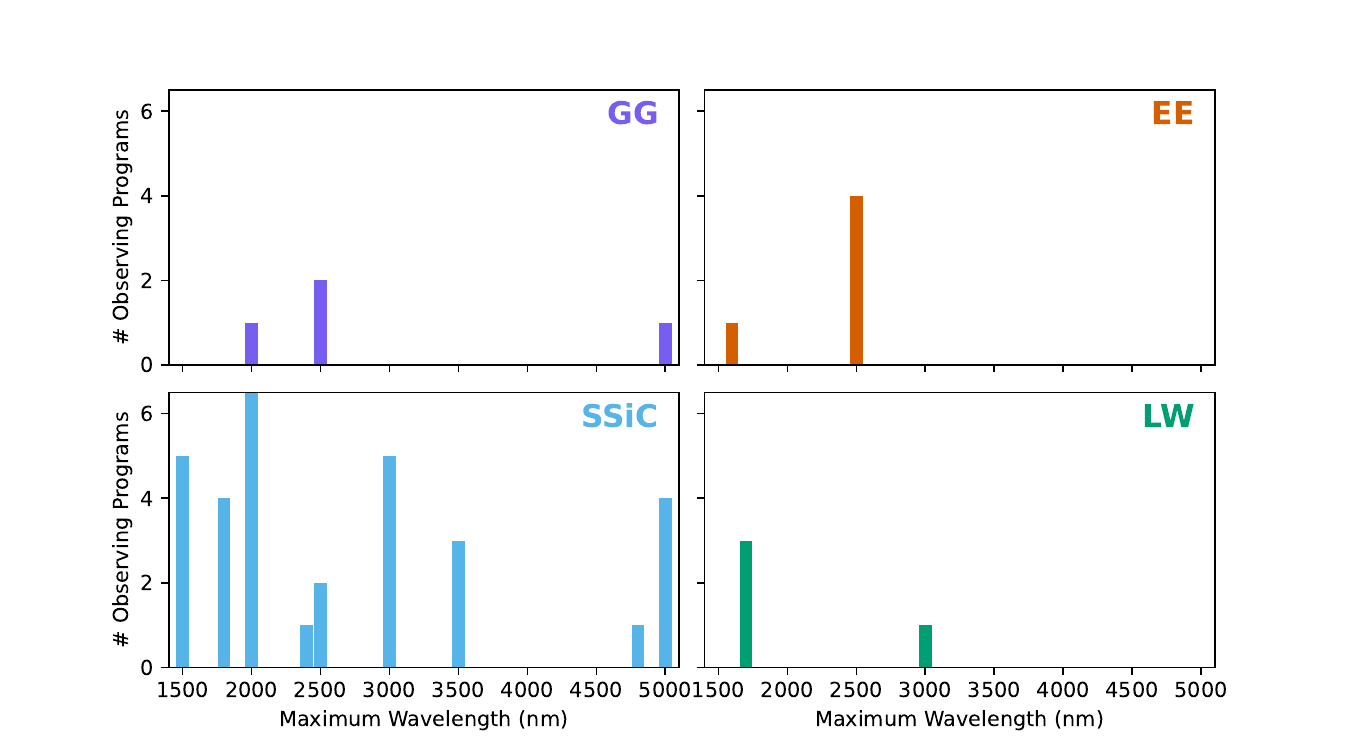}
 \caption{Distribution of requested short-wavelength cut-offs $\le 250$~nm (top) and long-wavelength cut-offs $\ge 1500$~nm (bottom) for observing programs associated with science cases proposed by the Growth of Galaxies (GG, top left, purple), Evolution of the Elements (EE, top right, orange), Solar Systems in Context (SSiC, bottom left, light blue), and Living Worlds (LW, bottom right, green) Science Working Groups (SWGs).}
 \label{fig:wave}
\end{figure*}

In this section, we shift the focus to particular engineering approaches and discuss the scientific motivation for undertaking various technical challenges. While this paper is motivated by HWO, this section is a general discussion rather than a prescription for HWO development or an official description of the HWO project. We begin by presenting wavelength coverage, spatial resolution, and spectral resolution needed to accomplish the investigations described in Section~\ref{sec:scdds}. In Figure~\ref{fig:wave}, we show the short-wavelength (top) and long-wavelength (bottom) cut-offs requested by programs within each SWG. Demand for observations $\le 100$~nm is driven by GG, EE, and SSiC. In the NIR, all four SWGs include programs with long-wavelength cut-offs $\ge 2000$~nm. The majority of long wavelength requests come from SSiC, which is unsurprising because that SWG contributed almost as many science cases (\nssic) as GG (\ngg), EE (\nee), and LW (\nlw) combined. 

\begin{figure*}[tb]
 \centering
 \includegraphics[width=\linewidth]{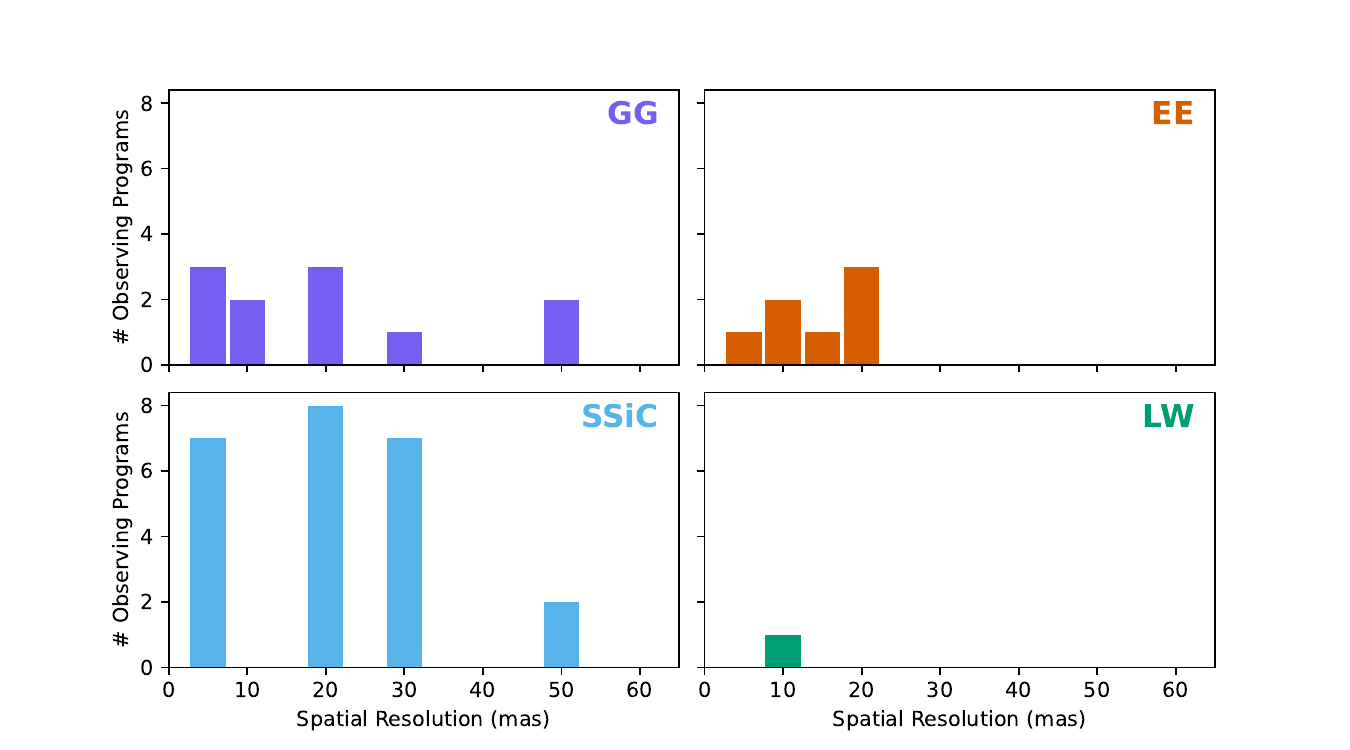}
 \includegraphics[width=\linewidth]{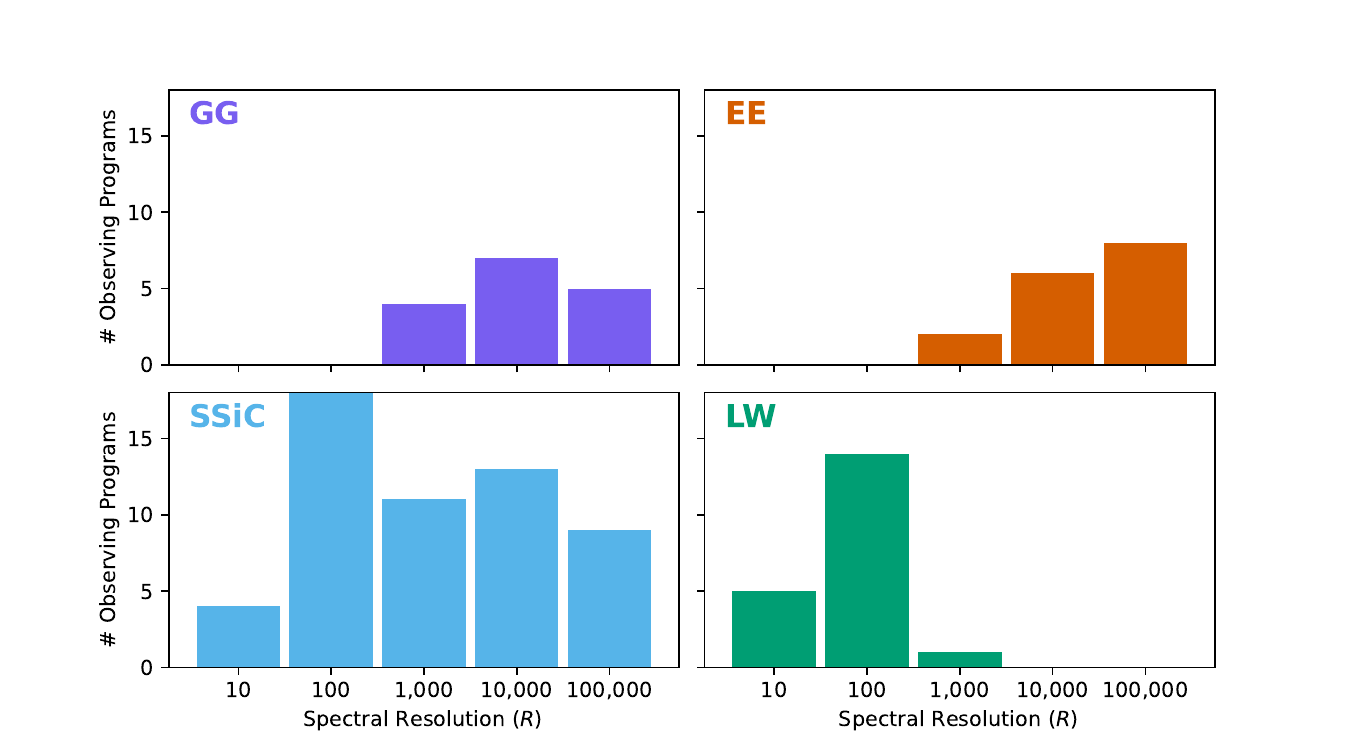}
 \caption{Same as Figure~\ref{fig:wave} but for the coarsest acceptable spatial resolution (top) and the lowest acceptable spectral resolution (bottom).}
 \label{fig:resolution}
\end{figure*}

Next, in Figure~\ref{fig:resolution}, we display the coarsest spatial resolution (top) and lowest spectral resolution (bottom) needed for breakthrough science. For spatial resolution, the axis range is set to highlight the smallest spatial scales, so the only programs shown are those that need fine spatial resolution. All four SWGs include programs requesting spatial resolution $\le 10$~mas, demonstrating the demand for diffraction-limited imaging.

As shown in the bottom panel of Figure~\ref{fig:resolution}, the minimum spectral resolution varies across multiple orders of magnitude from $R<10$ to $R>100,000$. The programs proposed by SSiC span the full range, while GG and EE programs are concentrated at the high-resolution end. In contrast, LW programs are clustered at lower spectral resolutions.

\begin{figure*}[tb]
 \centering
 \includegraphics[width=\linewidth]{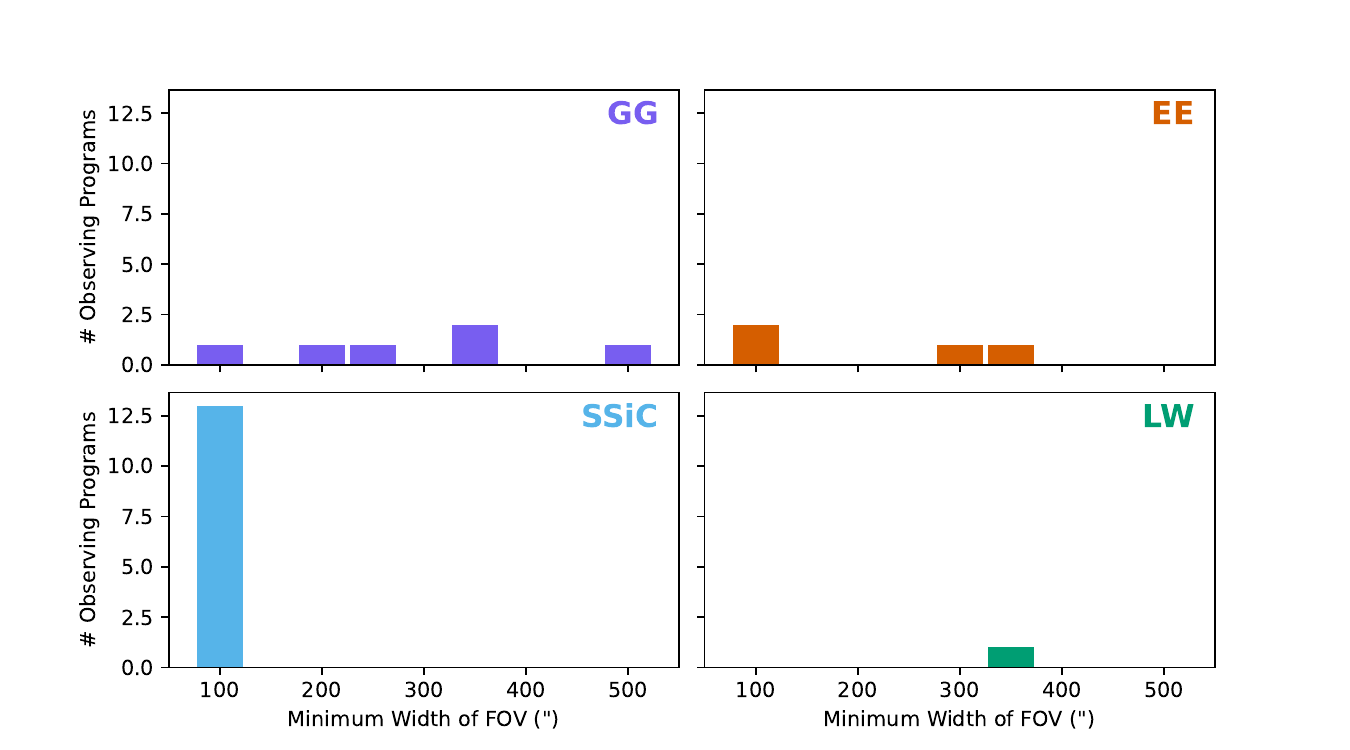}
 \caption{Same as Figure~\ref{fig:wave} but for minimum field of view width.}
 \label{fig:fov}
\end{figure*}

Finally, Figure~\ref{fig:fov} displays the minimum field of view requested for programs proposed by each SWG. The dimensions shown do not map one-to-one with target regions because many programs would tile smaller instrument footprints to cover a larger target region. More work is needed to determine ideal instrumental fields of view. We recommend that such analyses consider the angular sizes, geometry, and distribution of potential target sources as well as observational efficiency, multi-wavelength capabilities, and the importance of obtaining broad wavelength coverage of entire regions simultaneously rather than building up wavelength or spatial coverage using multiple sequential observations. 

\begin{figure*}
 \centering
\includegraphics[width=1\linewidth]{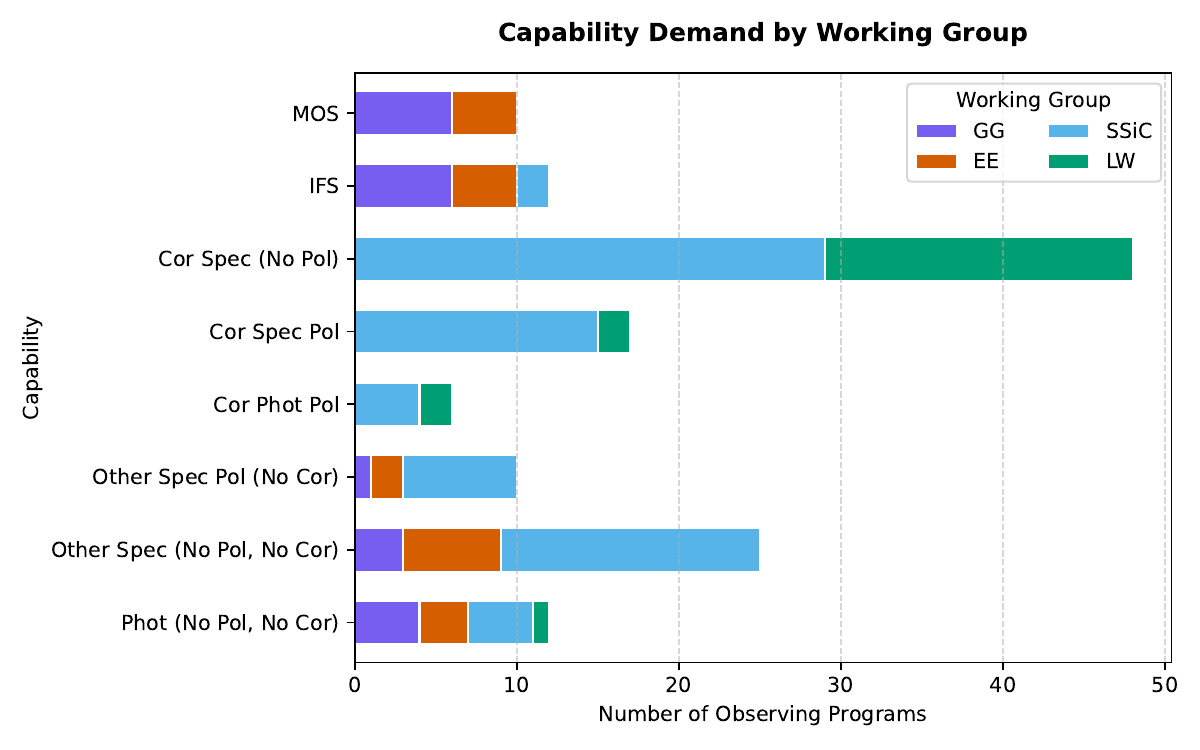}
 \caption{Number of science cases requiring particular capabilities. As shown by the legend, the different colors indicate observing programs from different working groups. }
 \label{fig:modes}
\end{figure*}

As discussed in Section~\ref{sec:swg_obs}, the observational capabilities needed for the science cases outlined in Section~\ref{sec:scdds} can be grouped into categories that could correspond to different instruments or settings. Table~\ref{tab:modes} displays the full list of the consolidated observing modes identified in Section~\ref{sec:swg_obs}. The number of observing programs requesting each mode is depicted in Figure~\ref{fig:modes}. In the following subsections, we consider the scientific drivers that push the limits of each configuration. We discuss the configurations in the order shown in Table~\ref{tab:modes} beginning with IFS observations in Section~\ref{sssec:ifs} and ending with photometric observations in Section~\ref{sssec:phot}. For clarity, each subsection is structured in the same way, skipping any categories that are not relevant for an individual configuration:
\begin{itemize}
 \item Blue wavelength limit ($\lambda_{\mathrm{blue}}$)
 \item Red wavelength limit ($\lambda_{\mathrm{red}}$)
 \item Spectral resolution ($R$)
 \item Spatial resolution ($\theta$)
 \item Field of view (FOV)
\end{itemize}

In addition to placing demands on instrumental capabilities, the ability to accomplish various science cases also depends on operational capabilities such as slew speeds, Sun avoidance zones, rapid response times, and onboard data storage. We discuss these attributes and potential science drivers at the end of this section. 

\subsection{Spectroscopy (without starlight suppression)}
\subsubsection{IFS Observations}
\label{sssec:ifs}
This category encompasses science cases grouped into GG-A (Section~\ref{sssec:gg_a}), GG-B (Section~\ref{sssec:gg_b}), EE-A (Section~\ref{sssec:ee_a}), SSiC-M (Section~\ref{sssec:ssic_m}), and SSiC-N (Section~\ref{sssec:ssic_n}). The targets range from distant galaxies to small bodies in the solar system. 

As shown in Table~\ref{tab:modes}, we have classified GG-A as the baseline configuration. GG-A consists of $90-2500$~nm spectroscopy with spectral resolution $R=10,000$ and spatial resolution $\theta=20$~mas over a $10''\times10''$ field of view. Mode SSiC-M is identical to GG-A, while EE-A is nearly identical and differs only in the choice of a shorter wavelength of 500~nm for the red cut-off. Mode GG-B has the same 500~nm red cut-off as EE-A and a higher spectral resolution ($R=30,000$). Finally, mode SSiC-N has lower spectral resolution ($R=3500$), reduced spatial resolution ($\theta=200$~mas), and a slightly reduced red cut-off ($\lambda_\mathrm{red}=2000$~nm).

For bright solar system targets, saturation mitigation strategies will be important. The ability to observe bright targets is most critical for observations of Venus by SSiC-5 \citep[][Section~\ref{sssec:venus}]{izenberg_hwo25}, Titan by SSiC-27 (Section~\ref{sssec:titan}), and Mars by SSiC-28 (Section~\ref{sssec:mars_post}), the giant planets by SSiC-29 \citep[][Section~\ref{sssec:solargiant}]{fletcher_hwo25}. 

\paragraph{Blue Wavelength Limit} For multiple GG, EE, and SSiC science cases, observing Lyman-$\alpha$ (Ly$\alpha$) and Lyman continuum (LyC) emission is critical. The specific choice of lower wavelength limit depends on the redshift of the target objects, but it would be strategic to ensure a cut-off below rest-frame Lyman limit ($91.2$~nm) to avoid eliminating the possibility of exploring hydrogen ionization in nearby systems. Increasing the UV cut-off to 95~nm or 100~nm would limit LyC studies to $z \ge 0.04$ and $z \ge 0.1$, respectively. A higher UV cut-off of 120~nm would further restrict studies to $z \ge 0.3$, thereby completely eliminating access to the sample of $0 \le z \le 0.3$ galaxies needed by GG-7 for studies of LyC escape \citep[][Section~\ref{sssec:lyman_escape}]{carr_hwo25}. 

The most extreme blue limits are requested by GG-6 \citep[][Section~\ref{sssec:resolve_reionization}]{xu_reionization_hwo25}, SSiC-1 \citep[][Section~\ref{sssec:oceans_habitable}]{cartwright_hwo25} and SSiC-29 \citep[][Section~\ref{sssec:solargiant}]{fletcher_hwo25}. GG-6 aims to study the role of LyC clusters in ionizing the universe. While a blue cut-off of 90~nm would be sufficient for enabling level science, \citet{xu_reionization_hwo25} note that a lower cut-off of 50~nm would enable measurements of the helium ionizing edge, thereby opening new parameter space and providing valuable new constraints on the spectral energy distribution of the photons responsible for reionization. 

Within the solar system, SSiC-1 and SSiC-29 would investigate icy worlds and giant planet aurorae, respectively. For SSiC-1, EUV data at $50-90$~nm would provide access to spectral features from irradiated carbon on the surfaces of solar system bodies. Understanding the distribution and processing of surface carbon is important for assessing habitability and the role of radiation in shaping these environments. For SSiC-29, EUV data are needed to detect the H$_2$ Lyman series ($80-180$~nm) and He~I emission lines ($50-60$~nm), both of which are important for understanding auroral chemistry and energy deposition into planetary atmospheres.

\paragraph{Red Wavelength Limit}
The longest wavelength red cut-off requested is 5000~nm for studies of AGN feedback by GG-5 (Section~\ref{sssec:agn_feedback}) and solar system icy worlds by SSiC-1 (Section~\ref{sssec:id_oceans}). The next reddest are 3500~nm for investigations of solar system giant planets by SSiC-29 (Section~\ref{sssec:solargiant}), 3000~nm for observations of debris disks for SSiC-9 (Section~\ref{sssec:debris}), and 2500~nm for probing the formation and evolution of supermassive black holes for GG-2 (Section~\ref{sssec:bh_mass_spin}). All other SCDDs in this category request $\lambda_\mathrm{red}\le2~\mu$m.

For GG-5 \citep[Section~\ref{sssec:agn_feedback}][]{zhang_hwo25}, NIR data are needed to peer through dusty regions to test models of AGN tori and then pinpoint the origins of AGN-driven gas outflows. Additionally, NIR data provide insight to the kinematics and mass of molecular gas in outflows (e.g., via H$_2$ at $2.12\mu$m and CO at $4.67\mu$m) and the possible relationship between molecular gas outflows and star formation (via polycyclic aromatic hydrocarbons at $3.3~\mu$m). 

For SSiC-1 \citep[Section~\ref{sssec:id_oceans}][]{cartwright_hwo25}, the motivation for NIR data is to monitor the rate and composition of gasses escaping from the interiors of solar system worlds with suspected or potential interior oceans of liquid water. NIR spectroscopy would provide valuable access to important features such as H$_2$O ($2.6-2.7~\mu$m) and CO$_2$ ($4.2-4.3~\mu$m) that are difficult to study with ground-based observatories due to significant telluric absorption. Long-term monitoring such as that possible with a facility like HWO is needed to understand the temporal behavior of geyser activity and the extent of surface-interior interactions on icy worlds. In general, NIR data would improve understanding of volatile cycling, which is an important factor in habitability assessments. 

The choice of $\lambda_\mathrm{red} = 3.5~\mu$m
for SSiC-29 \citep[Section~\ref{sssec:solargiant}][]{fletcher_hwo25} was driven by the desire to observe auroral H$^+_3$ emission, which would provide valuable insight into giant planet magnetospheres and how their atmospheres are modified by interactions with the solar wind. Although UV data alone could reveal auroral energy deposition, the combination of UV and NIR data enables a much more comprehensive view of planetary atmospheres and magnetospheres by capturing both energy deposition (in the UV) and release (in the NIR). NIR data out to $\leq3~\mu$m are also important for mapping deep clouds via the reflectivity peak at $2.7~\mu$m and probing atmospheric circulation by mapping the distribution of volatiles such as NH$_3$, H$_2$S, and CH$_4$. A red cut-off $\geq1.6~\mu$m is needed to detect CH$_4$, H$_2$S, and H$_2$ to support studies of atmospheric composition. 

SSiC-9 \citep[Section~\ref{sssec:debris}][]{rebollido_hwo25} requests $\lambda_\mathrm{red} = 3~\mu$m to capture possible H$_2$O features at $2.8\mu$~m. However, the longer red cut-off $\lambda_\mathrm{red}$ proposed by GG-5 or SSiC-1 would be advantageous because that redward extension would open access to CO ($4.5~\mu$m).

Next, GG-2 (Section~\ref{sssec:bh_mass_spin}) requests $\lambda_\mathrm{red}=2.5~\mu$m for studies of local intermediate mass black holes. 
As explained by \citet[Section~\ref{sssec:bh_mass_spin}][]{cann_hwo25}, access to [Si~VI] at 1963~nm is valuable because it is $>1$~dex brighter than the brightest optical lines ([Ne~V] at 330 and 340~nm). The proposed cut-off ($\lambda_\mathrm{red}=2.5~\mu$m) ensures that [Si~VI] would be detectable for galaxies at redshifts $z\le0.1$, which is necessary to build a sufficiently large target sample. Additionally, for the closest targets, the suggested cut-off could also enable observations of redder coronal lines such as [Al~IX] at 2044~nm and [Ca~VIII] at 2321~nm.

\paragraph{Spectral Resolution} 
The spectral resolution requested for these science cases ranges from $R>100$ for studies of Titan by SSiC-27 (Section~\ref{sssec:titan}) to $R=100,000$ for investigating planetary aurorae through SSiC-29 \citep[][Section~\ref{sssec:solargiant}]{fletcher_hwo25} and protostellar disks through SSiC-23 (Section~\ref{sssec:disk_winds}). Numerous science cases also request intermediate spectral resolutions. For instance, $R>5000$ spectra are needed for studies of extremely metal-poor massive stars by EE-1 (Section~\ref{sssec:massive_stars_lowZ}) and very massive stars by EE-5 (Section~\ref{sssec:vms}). 

As explained by \citet[][Section~\ref{sssec:solargiant}]{fletcher_hwo25}, the motivation for obtaining $R=100,000$ for SSiC-29 is to resolve the kinematics of neutral and ion auroral winds. A lower spectral resolution of $R\sim12,000$ would be sufficient for investigating auroral energy deposition and circulation patterns in the upper atmosphere. For SSiC-23 (Section~\ref{sssec:disk_winds}), $R>100,000$ spectral resolution is needed to measure kinematics of low-velocity components and resolve emission line profiles in protostellar disks.

The next highest spectral resolution ($R>80,000$) is requested by SSiC-9 (Section~\ref{sssec:debris}) for studies of debris disks. \citet{rebollido_hwo25} justifies that resolution request by pointing out that the gas velocities in debris disks are typically confined within a narrow velocity range (i.e., $<50-100$~km/s) and that high spectral resolution is needed to resolve gas motions in debris disks and avoid confusion with ISM and stellar features, which could be blended with disk features at lower spectral resolution. 

The third highest spectral resolution is requested by GG-5. As discussed in Section~\ref{sssec:agn_feedback}, \citet{zhang_hwo25} requests $R>50,000$ UV spectra (and \mbox{$R>5000$~NIR} spectra) to probe AGN feedback. However, the specific choice of $R>50,000$ is not motivated in the SCDD and may have been selected to match the upper resolution limit proposed for the LUMOS instrument during the LUVOIR mission concept study. Rather, \citet[][their Section~4]{zhang_hwo25} notes that $R>30,000$ spectra are needed to resolve and characterize outflows and inflows in galaxies. Presumably, the higher value of $R>50,000$ shown for breakthrough science in their Table~2 would lead to tighter constraints on the effects of AGN outflows on their host galaxies. 

Spectral resolution of at least $R=30,000$ is also needed for studies of the production and escape of ionizing radiation by GG-6 (Section~\ref{sssec:resolve_reionization}) and GG-7 (Section~\ref{sssec:lyman_escape}). As explained by \citet[][see their Table~2 and their Section~4.0.5]{carr_hwo25}, spectral resolution $R=30,000 - 100,000$ would provide valuable insight into gas structures as fine as the turbulent scale of cold clouds. 

The remaining science cases require spectral resolutions $R\leq10,000$. In addition to SSiC-27 (Section~\ref{sssec:titan}) which as previously noted needs $R>100$ spectra for studies of Titan, the other low spectral resolution case is SSiC-5 (Section~\ref{sssec:venus}) which requests $R=200$ spectra to investigate Venus. The next lowest spectral resolutions are $R=1740 - 3500$ spectra of giant planets for SSiC-29 (Section~\ref{sssec:solargiant}), which as mentioned earlier also requires complementary data at higher spectral resolution. 

\paragraph{Spatial Resolution}
While the majority of these science cases request spatial resolutions of $20-30$~mas, there are outliers at both extremes. At the smallest scales, GG-7 \citep[][Section~\ref{sssec:lyman_escape}]{carr_hwo25} requests spatial resolution of $0.3-30$~mas to investigate the escape of ionizing photons from galaxies. This fine spatial scale is motivated by the desire to resolve individual super star clusters and can be interpreted as a request for the highest spatial resolution possible. Likewise, GG-5 \citep[][Section~\ref{sssec:agn_outflow}]{zhang_hwo25} demands the highest spatial resolution possible (ideally $\theta=3$~mas) to resolve AGN tori down to physical scales of $\sim1$~pc. GG-3 \citep[][Section~\ref{sssec:bh_torus}]{gorjian_hwo25} also aims to study black hole tori and requests a similarly fine spatial resolution ($\theta=5$~mas). In addition, SSiC-28 (Section~\ref{sssec:mars_post}) requests $\theta = 3.5$~mas for observations of Mars, and EE-5 \citep[][Section~\ref{sssec:vms}]{martins_hwo25} requests $\theta = 5$~mas for studies of extremely massive stars. Those choices are motivated by the desire to identify small features on the surface and in the atmosphere of Mars (e.g., the corona) and spatially resolve massive stars in crowded regions, respectively. Finally, GG-4 \citep[][Section~\ref{sssec:bh_quiescent}]{pacucci_hwo25} requests $\theta=10$~mas to spatially resolve the gravitational spheres of influence of quiescent black holes and measure their effects on stellar kinematics. 

At the opposite extreme, a spatial resolution \mbox{$\theta=50$~mas} would be sufficient for studies of Titan for SSiC-27 (Section~\ref{sssec:titan}). That value is set by the desire to obtain six measurements per hemisphere to Nyquist sample the $0\farcs7$ disk of Titan. A spatial resolution of \mbox{$\theta=50$~mas} would also be appropriate for GG-17a \citep[][Section~\ref{sssec:cgm_elm}]{burchett_hwo25}, which aims to resolve features on physical scales of roughly 10~pc. This resolution would be fine enough to investigate how clouds of cool gas ``shatter'' to form smaller cloudlets within the circumgalactic medium. Likewise, spatial resolution $\theta \lesssim 50$~mas would suffice for the protostellar disk observations of SSiC-23. In addition, SSiC-29 \citep[][Section~\ref{sssec:solargiant}]{fletcher_hwo25} requests a combination of low and high spatial resolution observations ($\theta=200$~mas and $\theta=20$~mas) of giant planets to provide both broader context and a zoomed-in view of small-scale features. 

\paragraph{Field of view} 
The default $10''\times10''$ field of view would be sufficient for most science cases in this category, but a few planetary cases specifically request larger fields of view. Additionally, some GG, EE, and SSiC science cases could be accomplished with a smaller field of view. The largest field of view is requested by SSiC-9 \citep[][Section~\ref{sssec:debris}]{rebollido_hwo25} to capture the full extent of debris disks in a single observation to maximize efficiency. The data needed for SSiC-9 could also be obtained by tiling a smaller field of view across the targeted disks. SSiC-23 (Section~\ref{sssec:disk_winds}) does not quantify the necessary field of view, but like the debris disk observations of SSiC-9, the protostellar disk observations of SSiC-23 would likely benefit from an increased instrument footprint.

The next largest requests are $>50''\times50''$ for observations of Mars by SSiC-28 (Section~\ref{sssec:mars_post}) and $40''\times40''$ for observations of Venus by SSiC-5 \citep[][Section~\ref{sssec:venus}]{izenberg_hwo25}. In both cases, the requested field of view is driven by the desire to capture the full area of interest (i.e., the disk and corona of Mars and the disk of Venus) in a single observation. Given the brightness of these targets, the high anticipated overhead-to-integration ratio, and the desire to study temporal variability, it may be more advantageous to re-image the pupil to observe a larger field at reduced spatial resolution rather than tile smaller fields of view.

For the studies of giant planets described in \mbox{SSiC-29}, the combination of lower spatial resolution observations over a wide field ($\theta=200$~mas; $>10''\times10''$) and higher spatial resolution observations over a smaller field ($\theta=20$~mas; $>3'' \times 3''$) would be advantageous. That dual strategy would enable resolving small-scale features while also providing valuable context and increased observational efficiency by capturing the full auroral region of Jupiter or the full ring system of Uranus in a single observation.

A smaller $3''\times3''$ field of view would also be sufficient for investigations of massive stars for EE-5 \citep[][Section~\ref{sssec:vms}]{martins_hwo25} and solar system icy worlds by SSiC-1 \citep[][Section~\ref{sssec:id_oceans}]{cartwright_hwo25}. An even smaller $1''\times1''$ field would suffice for the observations of Titan described in SSiC-27 (Section~\ref{sssec:titan}).

\subsubsection{MOS Observations}
\label{sssec:mos}
This category consists of GG-C, GG-D, and EE-B. Modes GG-C and GG-D cover an identical wavelength range of $94-350$~nm while mode EE-B adds visible and NIR coverage for a broader wavelength range of $95-2500$~nm. Modes GG-C and EE-B offer $R=3000$ spectra over a $2'\times2'$ field of view while GG-D features a higher spectral resolution ($R=40,000$) over a larger $6'\times6'$ field of view. The three modes differ significantly in spatial resolution with EE-B having fine spatial resolution of $\theta=10$~mas and GG-C and GG-D having coarse resolutions of \mbox{$\theta=225$~mas} and \mbox{$\theta=500$~mas}, respectively.

\paragraph{Blue Wavelength Limit} 
All of the science cases in this category require UV data. The shortest wavelengths requested for the blue cut-off are 94~nm for studies of the interface between galactic disks and the circumgalactic medium for GG-16b \citep[][Section~\ref{sssec:disk_cgm}]{borthakur_hwo25} and 97.5~nm for investigations of structure within the circumgalactic medium by GG-17b \citep[][Section~\ref{sssec:cgm_elm}]{burchett_hwo25}. GG-11 \citep[][Section~\ref{sssec:ionizing_lf}]{mccandliss_hwo25} and EE-2 \citep[][Section~\ref{sssec:dust_extinction}]{paladini_hwo25} request a slightly longer wavelength blue cut-off of 100~nm for studies of ionizing radiation and dust, respectively. 

For GG-16b \citep[][Section~\ref{sssec:disk_cgm}]{borthakur_hwo25}, a blue cut-off of 94~nm would enable observations of over 300~lines including H~I Ly$\delta$ (95.0~nm), H~I Ly$\gamma$ (97.2~nm), C~III (97.7~nm), and the Werner and Lyman bands of H$_2$. Access to the higher order Lyman lines of atomic hydrogen would support determination of H~I column densities for individual clouds. Additionally, C~III (97.7~nm) would map moderately ionized gas and could be combined with measurements of highly ionized gas (via O~VI at 103.1~nm) and neutral gas (via O~I at 103.9~nm) to build a holistic picture of the behavior of gas at the disk-CGM interface. 

Access to C~III at 97.7~nm is also the justification for choice of $\lambda_\mathrm{blue} = 97.5$~nm for GG-17b \citep[][Section~\ref{sssec:cgm_elm}]{burchett_hwo25}. Measurements of C~III as well as other gases in various ionization states are needed to reveal the multiphase nature of the CGM and probe the cycling of material through galaxies. In addition to C~III, other lines of interest include O~VI at 103.8~nm, Si~III at 120.6~nm, and H~I Ly$\alpha$ at 121.6~nm. 

GG-11 aims to probe the evolution of the spectrum of ionizing radiation over the history of the universe \citep[][Section~\ref{sssec:ionizing_lf}]{mccandliss_hwo25}. For this science case, a redder cut-off of 100~nm would be sufficient because the redshifts of the selected targets ($0.2 \lesssim z\lesssim 1.2$) would cause their ionizing radiation to be Doppler shifted into the sensitivity range. Specifically, \citet{mccandliss_hwo25} notes $\lambda_\mathrm{blue}=100$~nm would be sufficient to constrain to $\le 0.3\%$ the fraction of ionizing radiation that escapes from L$^*$~galaxies. 

Finally, $\lambda_\mathrm{blue}=100$~nm is also well-matched to the studies of the formation, evolution, and properties of galactic and extragalactic dust proposed by EE-2 \citep[][Section~\ref{sssec:dust_extinction}]{paladini_hwo25}. Past observations have shown that dust extinction does not depend linearly on reddening, especially at UV wavelengths. Rather, dust extinction curves reveal a rise at FUV wavelengths. For EE-2, \citet{paladini_hwo25} therefore request $\lambda_\mathrm{blue}=100$~nm to capture the FUV rise and therefore probe the distribution of the small dust grains that are likely responsible for that feature. 

\paragraph{Red Wavelength Limit} 
For nearly all science cases in this category, UV data alone are sufficient. The exception is EE-2 \citep[][Section~\ref{sssec:dust_extinction}]{paladini_hwo25}, which requests broad wavelength coverage spanning from the UV to $\lambda_\mathrm{red}=2500$~nm. The addition of NIR data is essential for measuring how the slope of the extinction curve changes at longer wavelengths. The NIR spectral slope is a key diagnostic of grain growth and therefore provides a wealth of data about the size distribution of galactic and extragalactic dust. 

The next longest red cut-offs are $\lambda_\mathrm{red}=350$~nm for studies of the disk-CGM interface by GG-16 \citep[][Section~\ref{sssec:disk_cgm}]{borthakur_hwo25}, $\lambda_\mathrm{red}=350$~nm for investigations of AGN feedback by GG-18 \citep[][Section~\ref{sssec:agn_feedback}]{tillman_hwo25}, and $\lambda_\mathrm{red}=315$~nm for additional studies of dust by EE-11 \citep[][Section~\ref{sssec:dust_uv}]{roman-duval_hwo25}. The rationale for why EE-11 selected a significantly shorter red cut-off than EE-2 even though both studies focus on dust is that EE-11 primarily concentrates on dust composition while EE-2 also considers the dust grain size distribution. For the composition studies proposed by EE-11, high-resolution UV spectra ($R\sim50,000$ and $R\sim100,000$ at $95-315$~nm) are needed to detect spectral features caused by electronic transitions in neutral gas. In contrast, EE-2 requests lower spectral resolution data over a broader wavelength range ($R\sim3000$ at $100-2500$~nm) to measure the overall slope of the dust extinction curve. 

\paragraph{Spectral Resolution} 
The spectral resolutions requested for these science cases vary by two orders of magnitude from $R\sim 1000$ to $R\sim100,000$. This wide range is also seen within individual science cases. For instance, GG-16 needs both $R\sim100,000$ spectra and $R\gtrsim 5000$ spectra \citep[][Section~\ref{sssec:disk_cgm}]{borthakur_hwo25}. Additionally, EE-11 requests spectra at 
$R\sim100,000$ and $R\sim50,000$ \citep[][Section~\ref{sssec:dust_uv}]{roman-duval_hwo25}. A slightly lower spectra resolution of $R\sim40,000$ would be sufficient for GG-18. Finally, GG-17b demands $R\sim20,000$ spectra and GG-11 needs $R\sim1000$ data. 

At the high resolution end, $R\sim100,000$ spectra are needed to measure outflow rates and kinematics for individual clouds for GG-16 \citep[][Section~\ref{sssec:disk_cgm}]{borthakur_hwo25} and detect very weak transitions to quantify the amount of oxygen and carbon present in dust for Objective~2 of EE-11 \citep[][Section~\ref{sssec:dust_uv}]{roman-duval_hwo25}. For Objective~1 of EE-11, a lower resolution of $R\sim50,000$ would be sufficient to measure the abundances of other important dust ingredients such as Fe, Si, Mg, S, and Zn. GG-16 also requests an even lower spectral resolution of $R\gtrsim5000$ for emission spectroscopy sensitive to strong emission features such as Ly$\alpha$, C~III, and Mg~II to complement the absorption spectra obtained at higher spectral resolution. The combination of absorption and emission data will constrain the density of the CGM. 

For GG-18 \citep[][Section~\ref{sssec:agn_feedback}]{tillman_hwo25} and GG-17b \citep[][Section~\ref{sssec:cgm_elm}]{burchett_hwo25}, the requested spectral resolutions of $R\sim40,000$ and $R\sim20,000$, respectively, are driven by the need to understand the kinematics of small regions. In both cases, a higher spectral resolution would be advantageous. \citet[][their Section~2.2]{burchett_hwo25} specifically highlight the possible advantages of $R\sim100,000$ for resolving the kinematics of individual clouds, but they question the feasibility of $R\sim100,000$ spectral resolution. Further studies are needed to assess the relative impacts of high spectral resolution and high spatial resolution on resolving cloud kinematics, as well as the feasibility of increased resolution in both spectral and spatial dimensions. 

At the opposite extreme, GG-11 requests much lower spectral resolution ($R\sim1000$) for observations of ionizing flux because the targets are expected to be faint. However, \citet[][their Section~2]{mccandliss_hwo25} also notes that higher spectral resolution at rest wavelengths $\geq91.2$~nm would be useful for constraining the properties of galactic outflows. The investigations proposed in GG-11 could then be enhanced by connecting those results to the insights into ionizing radiation provided by the $R\sim1000$ data they request from $\lambda_\mathrm{blue} = 100$~nm to $\lambda_\mathrm{red} = 180$~nm. 

\paragraph{Spatial Resolution} 
The spatial resolutions proposed for these observations extend from 10~mas to $<1000$~mas. The finest spatial resolution ($\theta = 10$~mas) is requested by EE-2 while the coarsest ($\theta < 1000$~mas) is requested by GG-16. The rationale for the selection of $\theta=10$~mas for EE-2 is that fine spatial resolution is required to resolve and measure absorption towards individual stars in galaxies at distances up to 10~Mpc. In turn, access to $\leq 10$~Mpc is needed to fulfill the science goal of investigating how the properties of dust in other galaxies might differ from those of Milky Way dust.

For GG-11, \citet{mccandliss_hwo25} note that while high spatial resolution is beneficial for studies of escape, targeted observations of small fields at high spatial resolution would be better made with an IFS rather than a MOS. Essentially, they argue that coarse spatial resolution ($\theta \sim 225$~mas) is sufficient for the science case described in GG-11 but concur with the authors of science cases such as GG-6 \citep[][Section~\ref{sssec:resolve_reionization}]{xu_reionization_hwo25}, GG-7 \citep[][Section~\ref{sssec:lyman_escape}]{carr_hwo25}, and GG-9 \citep[][Section~\ref{sssec:lyman_indirect}]{citro_hwo25} that high spatial resolution is ideal for IFS-based observational approaches. This same approach is taken by \citet[][Section~\ref{sssec:cgm_elm}]{burchett_hwo25} for studies of galactic winds and sub-structure proposed in GG-17: they request $\theta=500$~mas spatial resolution for MOS observations for GG-17b, which is in this observation category and finer spatial resolution ($\theta=50$~mas) for the IFS component GG-17a, which is described in Section~\ref{sssec:ifs}.

For GG-16 \citep[][Section~\ref{sssec:disk_cgm}]{borthakur_hwo25}, the requested spatial resolution ($\theta < 1000$~mas) would be sufficient for resolving physical scales of roughly 50~pc and linking gas flows to specific star clusters or associations within galaxies. 

Other science cases do not explicitly quantify the necessary spatial resolution. For instance, for GG-18 \citep[][Section~\ref{sssec:agn_feedback}]{tillman_hwo25}, the spatial resolution and field of view should be matched to the expected distribution of QSO sightlines needed for the proposed investigations of AGN feedback. 

\paragraph{Field of view}
Most of these science cases request a large field of view. The largest proposed fields of view are $8'\times8'$ for GG-17b \citet[][Section~\ref{sssec:cgm_elm}]{burchett_hwo25}, $6'\times6'$ for GG-16b \citep[][Section~\ref{sssec:disk_cgm}]{borthakur_hwo25}, and $3 \times [2' \times 2']$ for GG-11 \citep[][Section~\ref{sssec:ionizing_lf}]{mccandliss_hwo25}. For GG-11, the notation indicates that the field would be comprised of three modules, each of which would cover a $2' \times 2'$ field. The justifications for these large fields of view are the desire to map full targets in a single pointing \citep[e.g., see Section 2.1 of][]{carr_hwo25} and simultaneously observe background sources along with target galaxies \citep[e.g., see Section~5.6 of][]{borthakur_hwo25}. In general, for MOS (and IFS) observations, additional work is needed to consider the density of possible sources on the sky and understand how various combinations of field-of-view, spatial resolution, and spectral resolution would affect the resulting observational efficiency.

\subsubsection{Single-Object Spectroscopy}
\label{sssec:sos}
This mode includes EE-C, EE-D, SSiC-J, and SSiC-K. We designate EE-D as the baseline configuration for single-object spectroscopy. EE-D and the identical mode SSiC-K feature $R=100,000$ spectra from 90~nm to 310~nm. EE-C offers lower spectral resolution ($R=10,000$) over a wavelength range that is shifted slightly redward ($110-500$~nm) while SSiC-J provides even lower spectral resolution ($R=3000$) over a broad wavelength range of $100-3000$~nm. Observations of bright targets such as main-belt asteroids and nearby exoplanet host stars will require saturation mitigation strategies. Some of the science cases in this mode request long-duration observations and could therefore amass large data sets that could place demands on onboard storage and data downlink rates. The clearest example is SSiC-16, which aims to acquire spectroscopic phase curves of exoplanets spanning orbital periods $>20$~days. 

\paragraph{Blue Wavelength Limit}
All of the science cases in this category require UV observations. The shortest wavelength blue-cut-offs requested are 50~nm for searching for potentially habitable icy worlds orbiting other stars for SSiC-17 \citep[][Section~\ref{sssec:id_oceans}]{quick_hwo25}, 90~nm for investigations of exoplanet compositions and fundamental physics via white dwarf spectroscopy for EE-7 \citep[][Section~\ref{sssec:wds}]{xu_siyi_hwo25}, and 100~nm for probing atmospheric escape from transiting exoplanets for SSiC-14 \citep[][Section~\ref{sssec:escape}]{dossantos_hwo25}.

For SSiC-17, the proposed wavelength range of $50-210$~nm was inspired by the $55-206$~nm wavelength range of Europa-UVS, a UV spectrograph on the Europa Clipper spacecraft that will search for water and its dissociation products on the surface and in the atmosphere of Europa \citep{retherford_et_al2024}. Specific features of interest include atomic hydrogen at 121.6~nm and atomic oxygen at 130.4~nm and 135.6~nm. 

For EE-7 \citep[][Section~\ref{sssec:wds}]{xu_siyi_hwo25}, the rationale for $\lambda_\mathrm{blue} = 90$~nm is the need to access a variety of spectral lines that are important for planetary composition (e.g., S, C, O, Mg, Fe, and Si) and measurements of fundamental constants (e.g., H$_2$ and highly ionized Fe and Ni). As shown in Table~4 of \citet{xu_siyi_hwo25}, the bluest lines of interest for planetary compositions include C~II at 90.4~nm, S~II at 90.7~nm, and N~II at 91.6~nm. For fundamental physics, H$_2$ has numerous transitions between 90~nm and 160~nm while Fe~IV/V/VI and Ni~IV/V/VI exhibit thousands of transitions between 100~nm and 160~nm. 

As explained by \citet[][Section~\ref{sssec:escape}]{dossantos_hwo25}, a redder cut-off of $\lambda_\mathrm{blue} = 100$~nm would permit measuring the transit depth at 121~nm (Ly$\alpha$) and 133~nm (C~II) to detect escaping hydrogen and carbon, respectively. An even redder cut-off of $\lambda_\mathrm{blue} = 120$~nm would be sufficient for investigating the origin of the solar system via the observation of small bodies proposed in SSiC-7 \citep[][Section~\ref{sssec:solar}]{mandt_hwo25}. Finally, blue cut-offs of 140~nm and 150~nm would be adequate for investigating the natures of the first stars and r-process elements for EE-8 \citep[][Section~\ref{sssec:first_stars}]{roederer_nature_hwo25} and EE-9 \citep[][Section~\ref{sssec:r_process_nature}]{roederer_rprocess_hwo25}, respectively. In both cases, obtaining spectra at wavelengths $\ge140$~nm or $\ge150$~nm would open up new exploration space and provide the opportunity to detect previously undetected lines in the spectra of metal-poor stars. 

\paragraph{Red Wavelength Limit}
Three of the science cases in this category request data extending to the NIR ($\lambda_\mathrm{red} = 3000-5000$~nm). The remaining science cases exclusively request UV data. However, multiple science cases such as the investigations of r-process elements described in EE-4 \citep[][Section~\ref{sssec:r_process_el}]{burns_hwo25} would benefit from combining space-based data obtained at wavelengths inaccessible from the ground (e.g., $\lambda \lesssim 400$~nm) with ground-based observations at other wavelengths. 
The science case with the longest wavelength red-cut off is SSiC-16a \citep[][Section~\ref{sssec:transits}]{wakeford_hwo25}, which requests spectroscopic phase curves of exoplanets extending to 5000~nm. This extended IR coverage would provide access to features of CO$_2$ and CO to better constrain formation location as well as features caused by sulfur-bearing molecules that would enable further insight planet formation, atmospheric evolution, and photochemistry. A much more conservative cut-off of $\lambda_\mathrm{red}=1600$~nm would encompass features from H$_2$O, HCN, CH$_4$, and NH$_3$. These features could constrain metallicity and may provide some constraints on planet formation. 
The next longest cut-off ($\lambda_\mathrm{red} = 4800$~nm) is requested by SSiC-7 for investigations of the origin of the solar sytem via observations of small bodies \citep[][Section~\ref{sssec:solar}]{mandt_hwo25}. As for SSiC-16a, the red cut-off for SSiC-7 is set by the need to capture CO$_2$ and CO features at $4200-4800$~nm. Access to $1500-3300$~nm is also important for detecting H$_2$O in emission and absorption. 

The third science case requesting IR data is SSiC-22 (Section~\ref{sssec:mars}), which proposes $\lambda_\mathrm{red} = 3000$~nm to explore the origin of Mars. Like SSiC-7, SSiC-22 would involve observations of small bodies. The selected red cut-off ($\lambda_\mathrm{red} = 3000$~nm) would provide access to features at 1900~nm, 2700~nm, and $2200-2500$~nm caused by carbonates and other hydrated minerals. 

\paragraph{Spectral Resolution}
The requested spectral resolution varies by two orders of magnitude from $R\sim30$ to $R=200,000$. The highest spectral resolution is essential for the fundamental physics aspects of EE-7 \citep[][Section~\ref{sssec:wds}]{xu_siyi_hwo25} because evaluating possible variations in $\alpha$ and $\mu$ necessitates measuring subtle differences in line positions. For the component of EE-7 focused on determining the composition of accreted planets (i.e., EE-7a), a lower spectral resolution of $R>60,000$ would be sufficient to disentangle signatures of planet accretion from photospheric, interstellar, and circumstellar lines. 

An intermediate spectral resolution of $R=100,000$ would be ideal for the studies of the nature of the first stars and r-process elements described in EE-8 \citep[][Section~\ref{sssec:first_stars}]{roederer_nature_hwo25} and EE-9 \citep[][Section~\ref{sssec:r_process_nature}]{roederer_rprocess_hwo25}, respectively. That resolution would support the detection, de-blending, and elemental abundance analyses of crucial lines, especially faint, poorly studied lines of $r$-process elements. 

For the studies of atmospheric escape proposed in SSiC-14 \citep[][Section~\ref{sssec:escape}]{dossantos_hwo25}, a slightly lower spectral resolution ($R\sim45,000$) would be advantageous for studies of atmospheric escape because it would enable more sophisticated treatment of stellar activity. High spectral resolution would also provide the opportunity to measure the spin-orbit alignment of the planet as well as ISM properties along the line of sight to the system. However, more research is needed to determine the optimal spectral resolution for various investigations of atmospheric escape \citet[][see their Section~4.2]{dossantos_hwo25}. 

Significantly lower spectral resolutions of $R=2000 - 6000$ are requested for other science cases in this category. For instance, $R>6000$ spectra are needed to probe the effects of rotation on the atmospheres of transiting planets for SSiC-16 \citep[][Section~\ref{sssec:transits}]{wakeford_hwo25} and $R\sim2000$ spectra would be sufficient for observing small bodies to investigate the origin of the Solar System for SSiC-7 \citep[][Section~\ref{sssec:solar}]{mandt_hwo25}. Finally, $R\sim30$ would suffice for characterizing the XUV spectra of planet host stars for SSiC-20d (Section~\ref{sssec:survive_water}). For the transient-focused science cases EE-3 on core-collapse supernovae \citep[][Section~\ref{sssec:flash_ccsne}]{andrews_hwo25} and EE-4 on kilonovae \citep[][Section~\ref{sssec:r_process_el}]{burns_hwo25}, the specific spectral resolution is less important than the ability to obtain observations within 1 or 2 days of explosion, respectively.

\subsubsection{Spectropolarimetry}
\label{sssec:specpol}
By design, the spectropolarimetric configurations \mbox{GG-E}, EE-E, and SSIC-L are identical. This mode would provide $R=120,000$ spectra from $100-1600$~nm with full Stokes polarimetry. For completeness, we designate GG-E as the baseline configuration.

\paragraph{Blue Wavelength Limit} 
All of the science cases in this category request UV data. The shortest wavelength blue cut-off is requested by SSiC-26 \citep[][Section~\ref{sssec:bfield}]{strugarek_hwo25} for indirect measurements of the magnetic fields of exoplanets (i.e., observing program SSiC-26b).
Although the strict requirement is access to Ly$\alpha$ (121.6~nm), \citet{strugarek_hwo25} request a bluer limit of $\lambda_\mathrm{blue}=92$~nm to provide additional UV coverage. 

The second bluest cut-off is $\lambda_\mathrm{blue}=100$~nm, which is needed for investigations of binary black hole candidates by GG-19 \citep[][Section~\ref{sssec:smbh_pol}]{marin_et_al2025}, the magnetic fields of massive stars by EE-15 \citep[][Section~\ref{sssec:mag_stars}]{david-uraz_hwo25}, the effects of solar energetic particles on solar system giant planets for SSiC-25 \citep[][Section~\ref{sssec:aurorae}]{chaufray_hwo25}, the composition and escape of planetary atmospheres for SSiC-24 \citep[][Section~\ref{sssec:highres_atmos}]{cubillos_hwo25}, and the role of magnetic fields in shaping young planetary systems by SSiC-34 \citep[][Section~\ref{sssec:young_bfields}]{gomez_de_castro_magnetic_hwo25}. In the case of GG-19, FUV data are needed because the wavelengths and intensities of peaks in the FUV spectral energy distribution are a powerful probe of accretion disk properties. For EE-15, the choice of $\lambda_\mathrm{blue}=100$~nm is necessary to measure stellar wind properties and detect wind-sensitive spectral lines such as the P~V doublet at 111.8~nm and 112.8~nm. The same $\lambda_\mathrm{blue}=100$~nm cu-off is appropriate for SSiC-24 because it captures Ly$\alpha$ and a variety of other lines that important for determining the compositions and mass loss rates of planetary atmospheres. For instance, FUV observations could reveal the loss of heavy elements from planetary atmospheres by measuring transit depths of C~II and O~I. For SSiC-34, the selection of $\lambda_\mathrm{blue}=100$~nm is important for probing magnetospheric properties via observations of the young stars and brown dwarfs as well as their inner disks. 

Next, a longer wavelength cut-off of $\lambda_\mathrm{blue}=120$~nm would be sufficient for SSiC-26a, which is the direct detection component of SSiC-26 \citep[][Section~\ref{sssec:bfield}]{strugarek_hwo25}. For direct detection, the primary dataset would be measurements of the He~I triplet at 1083~nm. However, observations of UV lines indicative of the conditions of the stellar atmosphere (e.g., Si~III at 120.7~nm, O~V at 121.8~nm, N~V at 123.9~nm, and Fe~XII at 124.2~nm) are needed to identify stellar contamination that may affect the interpretation of the NIR data. 

In comparison, SSiC-33 needs a slightly longer blue cut-off of $\lambda_\mathrm{blue}=140$~nm for studies of amino acids in comets \citep[][Section~\ref{sssec:aminos}]{gomez_de_castro_amino_hwo25}. Lastly, the investigations of the atmospheric dynamics of exoplanets proposed by SSiC-16 could accommodate a much longer wavelength cut-off of $\lambda_\mathrm{blue}=300$~nm \citep[][Section~\ref{sssec:transits}]{wakeford_hwo25}.

\paragraph{Red Wavelength Limit} 
The requested red cut-offs vary by over an order magnitude from 220~nm to 2500~nm. The longest proposed red wavelength cut-offs are $\lambda_\mathrm{red} = 2500$~nm for studies of planetary atmospheres by SSiC-24 \citep[][Section~\ref{sssec:highres_atmos}]{cubillos_hwo25} and $\lambda_\mathrm{red} = 2440$~nm for direct observations of planetary magnetic fields by SSiC-26a \citep[][Section~\ref{sssec:bfield}]{strugarek_hwo25}. The next longest cut-off is $\lambda_\mathrm{red} = 2000$~nm for identification of black hole binaries by GG-19 \citep[][Section~\ref{sssec:smbh_pol}]{marin_et_al2025}. The remaining science cases propose red cut-offs between 220~nm and 1600~nm. 

For SSiC-24, a long wavelength cut-off of $\lambda_\mathrm{red} = 2500$~nm would increase the number of species that could be observed, thereby refining assessments of atmospheric composition and escape. Broad wavelength coverage would also provide more insight into the overall shape of the spectrum, which could constrain aerosol properties such as composition and the particle size distribution. Additionally, an extended wavelength range is beneficial for building a holistic picture of atmospheric composition and circulation because UV observations are most sensitive to the upper atmosphere while NIR data access higher pressures closer to the planetary surface. 

Broad wavelength coverage would also be beneficial for SSiC-26 to disentangle signals of planetary magnetic fields from stellar activity or other systematics. Finally, for GG-19, observations of polarized NIR light would be essential to determine the structure and conditions of the accretion disks around black holes. 

\paragraph{Spectral Resolution}
The highest requested spectral resolution is $R=120,000$ for visible-wavelength studies of stellar magnetism for EE-15 \citep[][Section~\ref{sssec:mag_stars}]{david-uraz_hwo25}. EE-15 also requests $R=60,000$ spectroscopy in the UV. The combination of high-resolution UV and visible spectropolarimetry is necessary to complete the scientific objectives of measuring mass loss rates, assessing stellar abundances, and mapping the magnetospheres of massive stars. 

High spectral resolution ($R=100,000$) is also needed for studies of the atmospheres and magnetic fields of exoplanets by SSiC-24 \citep[][Section~\ref{sssec:highres_atmos}]{cubillos_et_al2025} and SSiC-26 \citep[][Section~\ref{sssec:bfield}]{strugarek_hwo25}, respectively. For SSiC-24, high spectral resolution is essential for isolating signatures of planetary magnetic fields and atmospheres from those of their host stars. Similarly, SSiC-26a (direct field measurements) mandates high spectral resolution to assess the orientations of planetary magnetic fields via measurements of the Hanle and Zeeman effects. For SSiC-26b (indirect field measurements), high spectral resolution would facilitate detailed mapping of the thermal and magnetic structures within stellar atmospheres, thereby providing essential inputs needed to infer the properties of planetary magnetic fields by modeling the hot spots they may induce on their host stars. 

Next, $R=50,000$ spectral resolution would be sufficient for the investigations of binary black holes proposed by GG-19 \citep[][Section~\ref{sssec:smbh_pol}]{marin_et_al2025} and the effects of solar energetic particles on giant planets proposed by SSiC-25 \citep[][Section\ref{sssec:aurorae}]{chaufray_hwo25}, while a lower value of $R=30,000$ would suffice for probing the influence of magnetic fields on young stars and planetary systems for SSiC-34 \citep[][Section~\ref{sssec:young_bfields}]{gomez_de_castro_magnetic_hwo25}. Finally, the lowest spectral resolution ($R=300$) is requested for observations of amino acids in comets by SSiC-33 \citep[][Section~\ref{sssec:aminos}]{gomez_de_castro_amino_hwo25} because a coarse resolution is appropriate due to the characteristic morphology of the polarization signature induced by each enantiomer of alanine. 

\paragraph{Polarization}
The most precise polarimetry is needed for the investigations of stellar magnetism proposed by EE-15 \citep[][Section~\ref{sssec:mag_stars}]{david-uraz_hwo25}. For GG-19, polarimetric errors $<0.01\%$ would be sufficient for investigating binary black holes \citep[][Section~\ref{sssec:smbh_pol}]{martins_hwo25}. Finally, due to the strong signal expected from chiral amino acids, polarimetric errors $<0.1\%$ would be appropriate for the studies of alanine proposed in SSiC-33 \citep[][Section~\ref{sssec:aminos}]{gomez_de_castro_amino_hwo25}.

\subsection{Photometry (without starlight suppression)}
\label{sssec:phot}

The photometric modes are GG-F, EE-F, SSiC-P, and LW-G. The baseline configuration EE-F features $\theta=20$~mas imaging at $95-2500$~nm over a $6'\times6'$ field of view. LW-G would image the same $6'\times6'$ field of view but at a higher spatial resolution ($\theta=11$~mas) and only in the visible ($400-700$~nm). Mode GG-F spans both the UV and visible ($95-700$~nm) at the same $\theta=20$~mas spatial resolution as EE-F but over a smaller field of view ($4'\times4'$). Finally, SSiC-P has a slightly shorter red cut-off than EE-F (2000~nm versus 2500~nm) and a much smaller field of view ($2'\times2'$).

Due to the large wavelength range, this configuration likely represents multiple imaging channels or perhaps even multiple instruments. In that case, SSiC-P and the baseline mode EE-F would use the UV, visible, and NIR channels. In contrast, mode GG-F would use only the UV and visible channels. Finally, mode LW-G would exclusively use the visible channel. We discuss these capabilities in Section~\ref{sssec:genphot}, which covers general photometric needs, and in Section~\ref{sssec:astrometry}, which focuses on particular concerns for astrometric imaging. 

\subsubsection{General Imaging}
\label{sssec:genphot}

\paragraph{Blue Wavelength Limit} 
UV imaging is needed for multiple studies. The shortest blue cut-off requested is 50~nm for investigations of solar system giant planets for SSiC-29 \citep[][Section~\ref{sssec:solargiant}]{fletcher_hwo25}. This cut-off is specifically selected to provide sensitivity to helium emission. Additionally, time-resolved imaging at FUV wavelengths is needed for auroral studies. \citet{fletcher_hwo25} does not quote an exact wavelength range for the time-resolved photometry, but they mention that studies with HST/STIS typically focus on $\lambda \ge 127.5$~nm. 

Next, the studies of Lyman continuum escape from galaxies proposed in GG-9 demand observations at wavelengths $<100$~nm \citep[][Section~\ref{sssec:lyman_indirect}]{citro_hwo25}. A longer wavelength cut-off of $\lambda_\mathrm{blue}=200$~nm is requested by GG-13 for imaging of strong lensing systems to investigate the dark matter halo mass function \citep[][Section~\ref{sssec:dm_lensing}]{he_hwo25}. Programs EE-4 \citep[][Section~\ref{sssec:r_process_el}]{burns_hwo25} and EE-6 \citep[][Section~\ref{sssec:resolved_stellar_pops}]{smercina_hwo25} both request UV observations for studies of heavy element enrichment and galaxy evolution, respectively, but neither specifies a particular UV cut-off. For insight into filter selection for extragalactic studies and spectral energy distribution fitting of galactic properties, see \citet{cook_hwo25_vol2}. In brief, \citet{cook_hwo25_vol2} recommend combining wide UV coverage akin to that provided by GALEX with finer sampling in the NUV (e.g., UVOT filters) to better constrain dust parameters. The remaining science cases in this category do not request UV observations. 

\paragraph{Red Wavelength Limit} 
Roughly half of the science cases in this category request observations at NIR wavelengths. As for the blue limit, the most aggressive red limit is requested by SSiC-29 \citep[][Section~\ref{sssec:solargiant}]{fletcher_hwo25} for investigations of giant planets. The specific choice of $\lambda_\mathrm{red} = 3500$~nm would advance studies of giant planet magnetospheres by capturing auroral H$^+_3$ emission. That red limit would also provide access to deep clouds via the reflectivity peak at 2700~nm.

The other science cases needing NIR imaging are the studies of resolved stellar populations in EE-6 \citep[][Section~\ref{sssec:resolved_stellar_pops}]{smercina_hwo25}, the cosmic distance ladder in EE-10 \citep[][Section~\ref{sssec:distance_ladder}]{anand_hwo25}, the warm dark matter particle mass in GG-12 \citep[][Section~\ref{sssec:dm_power}]{doppel_hwo25}, and near-Earth objects in SSiC-30 \citep[][Section~\ref{sssec:defense}]{dotson_hwo25}. None of those science cases include a specific red limit, so we conservatively set $\lambda_\mathrm{red} = 2500$~nm when including those science cases in figures or tables. The remaining science cases in this category do not request NIR data.

\paragraph{Spatial Resolution} 
The necessary spatial resolution varies by over two orders of magnitude from 5~mas to 1000~mas. The finest resolution (5~mas) is requested for the dark matter investigations proposed by GG-13 \citep[][Section~\ref{sssec:dm_lensing}]{he_hwo25}. The goal of GG-13 is to detect very low-mass dark matter halos, and high spatial resolution is essential to push to the low halo masses ($\lesssim 10^7 M_\odot$) where a cut-off in the dark matter halo function might occur. 

The second finest resolution (11~mas) is needed for astrometric imaging by LW-12 \citep[][Section~\ref{sssec:mp}]{gary_hwo25}, which is discussed below in Section~\ref{sssec:astrometry}.

The third finest resolution (15~mas) is needed to sufficiently resolve individual stars for the study of galaxy evolution described in EE-6 \citep[][Section~\ref{sssec:resolved_stellar_pops}]{smercina_hwo25}. The stars that would be targeted by EE-6 are in crowded fields, and thus high spatial resolution, well-characterized PSFs, and exquisite sensitivity are crucial for obtaining photometry with the precision and accuracy needed to probe the morphological, chemical, and star formation evolution of galaxies. 

Next, slightly coarser spatial resolution of $\theta=20$~mas would be appropriate for the studies of the cosmic distance ladder, warm dark matter, and mergers of supermassive black holes presented in EE-10 \citep[][Section~\ref{sssec:distance_ladder}]{anand_et_al2022}, GG-12 \citep[][Section~\ref{sssec:dm_power}]{doppel_hwo25}, and GG-14 (Section~\ref{sssec:smbh_mergers}), respectively. For the kilonova observations needed by EE-4 to probe the history of r-process enrichment, a significantly coarser resolution of $1''$ would be acceptable for source localization. For the remaining science cases in this category, the spatial resolution needed for imaging is not specifically stated. In the case of the giant planet observations proposed by SSiC-29 \citep[][Section~\ref{sssec:solargiant}]{fletcher_hwo25}, a spatial resolution of 20~mas would match the finer resolution requested for IFS observations (i.e., observing programs SSiC-29a,b,c).

\paragraph{Field of view} 
The largest field of view requested by programs in this category is $6'\times6'$. For EE-10 \citep[][Section~\ref{sssec:distance_ladder}]{anand_hwo25}, this wide field is needed to observe large regions of target galaxies in a single pointing and therefore efficiently build up the necessary sample of intermediate distance indicators such as Cepheid variable, tip of the red giant branch, and J-region asymptotic giant branch stars. As explained below in Section~\ref{sssec:astrometry}, a $6'\times6'$ field is also needed for the astrometric observations proposed by LW-12 \citep[][Section~\ref{sssec:mp}]{gary_hwo25}.

Next, a slightly smaller field of view ($5'\times5'$) is needed for the extragalactic studies of resolved stellar populations proposed by EE-6 \citep[][Section~\ref{sssec:resolved_stellar_pops}]{smercina_hwo25}. As for EE-10, this large field is motivated by the goal of increasing observational efficiency by decreasing the number of pointings needed to fully cover a target galaxy. Similarly, a field of view large enough to cover regions containing multiple lenses would increase the efficiency of the dark matter halo mass studies proposed by GG-13 \citep[][Section~\ref{sssec:dm_lensing}]{he_hwo25}. However, \citet[][their Table~2]{he_hwo25} specifically note that tiled observations would be acceptable. Indeed, for investigating the nature of warm dark matter for GG-12, \citet[][]{doppel_hwo25} proposed searching for dwarf galaxies by tiling a series of fields each spanning at least $4'\times 3'$. 

A smaller field of $2' \times 2'$ would be acceptable for the studies of solar system giant planets proposed by SSiC-29 \citep[][Section~\ref{sssec:solargiant}]{fletcher_hwo25}. Although the $50''$ disk of Jupiter spans less than half that width, a larger $2'\times2'$ field would encompass other important attributes of the Jovian system such as rings, moons, and plasma tori. 

\subsubsection{Astrometric Imaging}
\label{sssec:astrometry}
The astrometric capabilities are driven by LW-12 \citep[][Section~\ref{sssec:mp}]{gary_hwo25}, which aims to measure planet masses. As discussed in Section~\ref{sec:scdds}, planet masses are a key ingredient for selecting targets and interpreting observations for multiple SSiC and LW science cases. While the extremely precise radial velocity observations described in Table~\ref{tab:tss} may be sufficient for some systems, other targets, especially massive stars that are likely to have faster rotation rates, will not be amenable to RVs. Even for ``quiet'' stars with low stellar activity levels and slower rotation rates, achieving the $< 10$~cm~s$^{-1}$ precision needed to robustly measure the masses of potentially habitable planets may not be viable. Accordingly, the ability to measure planet masses via astrometry is potentially mission-critical for ensuring the successful completion of LW and SSiC science goals enumerated in Sections~\ref{ssec:lw_scdds} and~\ref{ssec:ssic_scdds}, respectively. 

\paragraph{Blue Wavelength Limit} 
LW-12 requests a blue cut-off of $\lambda_\mathrm{blue}=400$~nm, which is significantly redder than the bluest values requested in Section~\ref{sssec:genphot}.

\paragraph{Red Wavelength Limit} 
As for the blue cut-off, LW-12 does not drive the red cut-off. The requested value ($\lambda_\mathrm{red}=700$~nm) is bluer than that needed for other imaging cases discussed in Section~\ref{sssec:genphot}.

\paragraph{Spatial Resolution \& Astrometric Precision} 
For LW-12 \citep[][Section~\ref{sssec:mp}]{gary_hwo25}, high spatial resolution ($\theta = 11$~mas) is essential for astrometric mass measurement of planet host stars. The specific spatial resolution of 11~mas was selected to correspond to a per epoch astrometric precision of $\sim0.3\,\mu\mathrm{as}$, which is the value needed to ensure robust mass measurements for habitable planets with masses $\geq0.1 M_\oplus$ (i.e., planets at least as massive as Mars). 

\paragraph{Field of view} 
As discussed in Section~\ref{sssec:genphot}, EE-10 \citep[][Section~\ref{sssec:distance_ladder}]{anand_hwo25} requests a $6' \times 6'$ field of view. LW-12 \citep[][Section~\ref{sssec:mp}]{gary_hwo25} shares this request. For LW-12, the large field is even more crucial because the resulting astrometric precision depends sensitively on the number of comparison stars, which in turn is set by the field of view. The ability to interpret planetary habitability and potential biosignatures depends on planet mass, and therefore reducing the field of view could be detrimental to nearly all of the LW objectives presented in Section~\ref{ssec:lw_scdds}. Planet masses are also essential for many of the SSiC programs described in Section~\ref{ssec:ssic_scdds}.

\subsection{High-Contrast Observations}
\label{ssec:hc}
In this section, we discuss the observations that require suppressing starlight to detect significantly fainter objects such as planets, disks, and moons. The level to which the starlight must be suppressed and the angular separations of interest are important for instrument design, but these parameters depend on the goals of each individual science case and vary by orders of magnitude. In Section~\ref{ssec:hc_sim}, we briefly discuss how the physical and orbital characteristics of planets would potentially affect their detectability. In the subsequent sections, we document the physical separations and contrasts requested in individual SCDDs. However, to enable future studies to build upon past trade studies \citep[e.g.,][]{belikov_hwo25_vol2_coronagraph, morgan_hwo25_vol2, tokadjian_hwo25_vol2} to further explore the relationships between mission architecture and science yield, we also include physical characteristics such as semimajor axis, orbital eccentricity, planet radius, and albedo when provided. As previously discussed in Section~\ref{sec:xswg} and shown in Table~\ref{tab:planetgrid}, multiple SSiC and LW science cases are associated with high-contrast observations of terrestrial planets, sub-Neptunes, gas giants, and circumstellar material at various insolation flux levels. 

In order to highlight the technical similarities between these science cases, we summarize the types of observations needed for each one in Table~\ref{tab:hc_obs}. As indicated in Table~\ref{tab:hc_obs}, when discussing high-contrast observations we follow the LW and SSiC SCDDs in classifying observations as UV for $\lambda < 400~\mathrm{nm}$, visible for $400~\mathrm{nm} \le \lambda \le 1100~\mathrm{nm}$, and NIR for $\lambda>1100~\mathrm{nm}$. The three most requested observation types are UV, visible, and NIR spectroscopy, which are needed by 15, 16, and 13 science cases, respectively. After those modes, the next-most popular configurations are visible polarimetric imaging, visible photometry, visible spectropolarimetry, NIR polarimetric imaging, and UV polarimetric imaging. Each of those five modes is requested by $5-8$ science cases with the exact count depending on the specific polarimetric modes needed for the investigation of technosignatures and photosynthetic life proposed by LW-5 \citep[][Section~\ref{sssec:technosignatures}]{kopparapu_hwo25} and LW-7 \citep[][Section~\ref{sssec:polbio}]{berdyugina_imaging_hwo25}, respectively. Next, photometry and spectropolarimetry at UV and NIR wavelengths are each needed for three science cases. 

\input{hc_obsgrid}
 
\subsubsection{Connecting Physical Properties to Observables}
\label{ssec:hc_sim}
Quoting specific needs for inner working angles (IWAs) and outer working angles (OWAs) is intrinsically challenging because those values depend on the distances to the targeted planetary systems as well as the orientations of planetary orbits and the locations of planets within their orbits during observations. In many cases, setting numerical limits on the inner and outer working angles is further complicated by the need to detect planets prior to characterization and the corresponding uncertainty regarding which particular stars will need to be observed. The frequencies with which certain types of planets occur and any possible correlations between stellar properties and planet occurrence will establish the range of angular separations over which certain planets could possibly be observed. Accordingly, simulations incorporating various occurrence paradigms \citep[e.g., ][]{stark_et_al2014, savransky+garrett2016, bixel_et_al2020, delacroix_et_al2016, stark_et_al2019, stark_et_al2020, stark2022} are needed to translate demands for planetary samples with particular physical characteristics (e.g., mass, semimajor axis) into the necessary sensitivity as a function of angular separation and contrast ratio. 

Furthermore, as various science cases may demand different sensitivities at a given separation, listing IWA, OWA, or contrast as a single, fixed number does not fully capture the complexity of the relationship between coronagraphic performance and science yield. Rather, the aforementioned end-to-end simulations of planet detectability are needed to assess the impact of various design decision and establish benchmarks for necessary IWA, OWA, and achieved starlight suppression as a function of angular separation. Those simulations could also explore more nuanced topics such as the possible benefits and disadvantages of achieving deeper contrast over a narrower region. Such an approach would likely result in higher SNR spectra for planets within the high-priority zone, but it would necessitate localizing planets prior to characterization and would potentially reduce the ability to boost efficiency by simultaneously obtaining spectra for multiple planets within a system or to search for additional fainter planets using the data obtained while characterizing a previously detected planet. For more details, see Dressing et al. (\emph{in prep}). 

\subsubsection{High-Contrast UV Spectroscopy}
\label{sssec:hc_uv_spec}
This category consists of SSiC-A and LW-A. The two modes are identical, but we designate SSiC-A as the baseline configuration. Both modes nominally consist of low-resolution ($R=7$) spectroscopy at $200-400$~nm obtained at a contrast of $10^{-10}$.

\paragraph{Blue Wavelength Limit}
The shortest wavelength observations are needed for observations of young planets and protoplanetary disks by SSiC-8a \citep[][Section~\ref{sssec:proto}]{ren_hwo25}. The SCDD does not quantify the wavelength range, but it does mention the need to observe Lyman and Balmer lines of hydrogen. Accordingly, we have adopted $\lambda_\mathrm{blue}=90$~nm for that science case. The specific value should be determined if that science case becomes a driver.   

The next bluest cut-off is $\lambda_\mathrm{blue}=150$~nm for breakthrough observations of gas giants in reflected light proposed by SSiC-15a \citep[][Section~\ref{sssec:reflected_giants}]{min_hwo25}.
The SCDD does not clearly differentiate between the science possible with this more aggressive blue limit compared to the 200~nm limit requested for enabling science; \citet{min_hwo25} mention that further work is needed to assess the amount of UV data needed to determine the abundances of various species and constrain aerosol properties. 

For other science cases, the short wavelength limits are divided between $\lambda_\mathrm{blue}=200$~nm and $\lambda_\mathrm{blue}=250$~nm. The bluer cut-off is needed for SSiC-20a (Section~\ref{sssec:survive_water}), SSiC-21a (Section~\ref{sssec:retention}), LW-3a \citep[][Section~\ref{sssec:origin}]{ranjan_et_al2025origin}, LW-4a \citep[][Section~\ref{sssec:prebiosignatures}]{ranjan_et_al2025prebio}, SSiC-11 \citep[][Section~\ref{sssec:ozone_onset}]{blunt_et_al2025}, and SSiC-18a \citep[][Section~\ref{sssec:exovenus}]{kane_venus_hwo25}. In contrast, a redder limit of $\lambda_\mathrm{blue}=250$~nm is requested for LW-1a \citep[][Section~\ref{sssec:life}]{arney_hwo25}, LW-10a \citep[][Section~\ref{sssec:fpbio}]{schwieterman_hwo25}, LW-13a \citep[][Section~\ref{sssec:seasonality}]{lafleche_hwo25}, and SSiC-3a \citep[][Section~\ref{sssec:rocky_sub}]{hu_rocky_hwo25}. 

The choice of $\lambda_\mathrm{blue}=200$~nm for six science cases was motivated by the need to determine ozone abundances. As discussed in Section~\ref{ssec:lw_scdds} and~\ref{sssec:lw_a}, ozone abundance is an important factor when interpreting potential biosignatures and assessing planetary habitability. A blue cut-off of 200~nm would permit observing the entire $200-350$~nm Hartley-Huggins band. In contrast, the redder cut-off of $250$~nm would remove access to the bluer ozone features, thereby decreasing sensitivity to ozone and complicating the interpretation of oxygen detections or upper limits. 

For the observations of Venus analogs proposed in SSiC-18a \citep[][Section~\ref{sssec:exovenus}]{kane_venus_hwo25}, $\lambda_\mathrm{blue}=200$~nm is also needed to detect and measure the abundance of SO$_2$. Assessing both O$_3$ and SO$_2$ is important for better constraining the overall redox state of the planet and therefore potential energy sources for life as well as for resolving degeneracies between atmospheric composition and the presence of aerosols. Interestingly, \citet[][see their Section~4]{hu_rocky_hwo25} also express an interest in measuring O$_3$ and SO$_2$ (as well as H$_2$S and photochemical products of both H$_2$ and SO$_2$), but they instead select a UV wavelength range of $250-400$~nm. 

In Figure~16 of their proposal for LW-1, \citet{arney_hwo25} display the results of a sensitivity analysis exploring the effects of different UV cut-offs on the detectability of O$_3$. The analysis, which was conducted by A.~Young, assumes the fixed low spectral resolution selected for LW UV observations ($R=7$) and accounts for possible contamination of O$_3$ signatures by SO$_2$. The figure shows results for 72~different models using each combination of six short wavelength cut-offs (200~nm, 240~nm, 250~nm, 264~nm, 288~nm, and 346~nm), three spectroscopic signal-to-noise ratios (10, 20, and 40), and four atmospheric models designed to be variations of Proterozoic Earth. Two of the atmospheric models have O$_2$ abundances set to 1\% of the present atmospheric level on Earth while the other two have only 0.1\% as much O$_2$ as modern Earth. At each O$_2$ level, one atmospheric model also includes high SO$_2$ (inferred to be a surface flux of $10^{12}$ molecules~cm$^{-2}$~s$^{-1}$ given the discussion by \citet{arney_hwo25} of their preceding figure). For context, the SO$_2$ flux due to volcanoes on present-day Earth is estimated to be $1-3.5 \times 10^9$~ molecules~cm$^{-2}$~s$^{-1}$, which is three orders of magnitude lower than the ``high SO$_2$'' case considered by Young. 

Based on this analysis, \citet{arney_hwo25} conclude that $\lambda_\mathrm{blue}=250$~nm is sufficient for detecting O$_3$ and mitigating confusion from SO$_2$ in this particular test of four atmospheric compositions, but the conclusion may not hold for more diverse ensembles of planetary atmospheres or when incorporating more realistic noise models. Figure~16 of \citet{arney_hwo25} does not include errors on the retrieved O$_3$ and SO$_2$ abundances, so it is not possible to ascertain how the improved SNR alters the posterior distribution. The values included in their Figure~16 reveal that the inferred O$_3$ abundance depends on SNR and can vary by up to 0.5~dex based on the choice of SNR, but \citet{arney_hwo25} do not compare the amplitude of that shift to the width of the posterior or to the precision with which O$_3$ should be measured to achieve their science goals. Further simulations including non-white noise, a wider variety of atmospheric compositions, and a range of spectral resolutions, cut-offs, and SNRs are needed to determine the instrumental capabilities needed to constrain atmospheric abundances. Moreover, the performance of those simulations should be evaluated with respect to the accuracy and precision needed for biosignature detection.

Prior to the completion of those next-generation simulations, we strongly advocate for the more ambitious $\lambda_\mathrm{blue}=200$~nm proposed by LW-3a \citep[][Section~\ref{sssec:origin}]{ranjan_et_al2025origin} and other cases focused on ozone. If viable, the even shorter cut-off of $\lambda_\mathrm{blue}=150$~nm requested by SSiC-15a \citep[][Section~\ref{sssec:reflected_giants}]{min_hwo25} would provide more extensive UV coverage and additional technical margin to disentangle true planetary signals from stellar or spacecraft systematics. 

\paragraph{Red Wavelength Limit} 
In nearly all cases, these programs request a red cut-off of $\lambda_\mathrm{red} = 400$~nm, which is the UV/visible channel boundary assumed in LW coronagraphic simulations. The exception is SSiC-11 \citep[][Section~\ref{sssec:ozone_onset}]{blunt_et_al2025}, which requests a bluer cut-off of $\lambda_\mathrm{red} = 350$~nm. This bluer limit still covers the Hartley-Huggins ozone band as needed for the science goal of SSiC-11 to determine the timescales over which terrestrial planets acquire oxygen-rich atmospheres. 

\paragraph{Spectral Resolution} 
The highest spectral resolution ($R\sim 10,000$) is needed for the investigations of protoplanetary disks proposed by SSiC-8a \citep[][Section~\ref{sssec:proto}]{ren_hwo25}. This high spectral resolution would reveal the shapes of accretion lines, thereby providing insight into the growth of protoplanets.  

Next, SSiC-20a (Section~\ref{sssec:survive_water}) requests $R\sim1000$ spectra to investigate how habitability depends on system age. The stated value of $R\sim1000$ would provide enhanced sensitivity to surface outgassing of molecules such as PS, K, MgO, SiO, and TiO. However, a more modest spectral resolution ($R\sim100$) would be sufficient for detecting water emission features and constraining outgassing more generally. 

A spectral resolution of $R\sim100$ is requested by SSiC-15a and SSiC-21a to assess the atmospheres of giant planets and terrestrial planets, respectively. The other programs request the default $R=7$ spectral resolution used in LW simulations of UV spectroscopy \citep{arney_hwo25}. However, as noted previously, additional studies are needed to determine the ideal observational settings for each science case. 

\paragraph{Small Physical Separations} 
The tightest physical separations would be probed by the science cases focused on highly irradiated planets. For instance, SSiC-18a \citep[][Section~\ref{sssec:exovenus}]{kane_venus_hwo25} aims to detect Venus-like planets orbiting FGK stars and therefore demands sensitivity to planets with semimajor axes as small as 0.3~AU. Similarly, the analysis of false positive biosignatures proposed by LW-10 necessitates measuring oxygen levels in the atmospheres of terrestrial planets orbiting interior to the inner edge of the liquid water habitable zone.

\paragraph{Large Physical Separations} 
The largest physical separations would be targeted by SSiC-8a \citep[][Section~\ref{sssec:proto}]{ren_hwo25}, which aims to study protoplanetary disks and protoplanets up to 1000~AU from their host stars. The next most distant separations would be observed by SSiC-15a \citep[][Section~\ref{sssec:reflected_giants}]{min_hwo25}, which is the UV component of a UV/Vis/NIR science case that would investigate gas giant planets with temperatures as cool as 30~K. The relationship between planetary irradiation temperature and planet-star angular separation depends on both stellar luminosity and the distance to the system. SSiC-15 would target planets orbiting FGKM stars, so targets at the same irradiation temperature could have quite different semimajor axes. For instance, a 30~K planet would need to orbit much more closely to an M dwarf than to an F star. \citet{min_hwo25} also point out that the accessibility of a given planet would depend on the wavelength of observation, with UV observations generally more favorable for close-in planets due to the decrease in IWA with decreasing wavelength at a fixed telescope aperture. Conversely, if a coronagraph has a fixed OWA, it would be easier to observe widely separated planets at NIR wavelengths than at UV wavelengths. UV observations are therefore unlikely to drive the OWA needed for SSiC-15. If UV observations of cool planets are needed, \citet{min_hwo25} suggest that more distant systems could be observed to decrease the angular separation corresponding to a fixed physical separation. However, additional work is needed to assess the landscape of potential targets and understand how the information content of the proposed planetary spectra would depend on design decisions such as IWA, OWA, contrast, and target star brightness limits. 

\subsubsection{High-Contrast Visible Spectroscopy}
\label{sssec:hc_vis_spec}
The two modes for high-contrast visible spectroscopy are SSiC-B and LW-B. As for high-contrast UV spectroscopy, we refer to SSiC-B as the baseline configuration but the two modes are identical. The nominal configuration is $R=140$ spectroscopy at $400-1100$~nm obtained at a contrast of $10^{-10}$.

\paragraph{Blue Wavelength Limit} 
All but two science cases use the default blue limit of 400~nm. The exceptions are SSiC-12a \citep[][Section~\ref{sssec:giant_orbits}]{sagynbayeva_et_al2025} and LW-14 \citep[][Section~\ref{sssec:surface_bio}]{parenteau_surface_hwo25}, both of which request $\lambda_\mathrm{blue} = 500$~nm. SSiC-12a is focused on orbit determination for giant planets, so the particular choice of $\lambda_\mathrm{blue}$ is not significant as long as the planet reflects sufficient light at observable wavelengths. In contrast, LW-14 is focused on photosynthetic life and requests $\lambda_\mathrm{blue}=500$~nm because a redder cut-off would reduce sensitivity to ``green oceans'' that could be an indication of photosynthetic life.

\paragraph{Red Wavelength Limit} 
Nearly all science cases request the default red limit of $1100$~nm. However, SSiC-3b \citep[][Section~\ref{sssec:rocky_sub}]{hu_rocky_hwo25}, SSiC-8b \citep[][Section~\ref{sssec:proto}]{ren_hwo25}, and LW-13b \citep[][Section~\ref{sssec:seasonality}]{lafleche_hwo25} propose shorter wavelength red limits of $\lambda_\mathrm{red}=700-1000$~nm. SSiC-3b, SSiC-8b, and LW-13b are the visible components of the multi-wavelength science cases SSiC-3, SSiC-8, and LW-13. SSiC-3 is focused on the densities of sub-Neptunes, SSiC-8 studies protoplanetary disks, and  LW-13 would investigate possible seasonal changes in the atmospheric and surface properties of potentially inhabited planets. All three science cases also include spectroscopy in the UV (SSiC-3a, SSiC-8a, and LW-13a). Programs SSiC-3 and LW-13 also request lower spectral resolution data in the NIR (SSiC-3c and LW-13c), so using the default $\lambda_\mathrm{red}=1100$~nm for those programs would still provide data requested by the science cases but at a higher spectral resolution. If needed, observations redward of 1000~nm obtained at the baseline SSiC-B resolution of $R=140$ could be binned to the $R=70$ spectral resolution requested for the NIR components of SSiC-3 and LW-13.

\paragraph{Spectral Resolution} 
The default spectral resolution ($R=140$) is appropriate for the majority of observing programs in this category, but the range of requested spectral resolutions extends from $R>40$ to $R\sim10,000$. As for high-contrast UV spectroscopy (Section~\ref{sssec:hc_uv_spec}), the visible components of SSiC-8 and SSiC-20 request spectral resolutions of $R\sim10,000$ and $R\sim1000$, respectively. For SSiC-8b, the high spectral resolution would facilitate studies of protoplanetary accretion by resolving line profiles. For SSiC-20b, $R\sim1000$ data would enable more nuanced studies of planetary outgassing but lower spectral resolutions of $R\sim100$ at $400-1000$~nm and $R\sim30$ at $1000-1200$ would still permit detection of steam-dominated atmospheres or high rates of XUV photolysis of water indicative of a runaway greenhouse.  

SSiC-15b proposes $R>1000$ spectra for characterization of gas giants in reflected light \citep[][Section~\ref{sssec:reflected_giants}]{min_hwo25}. Although lower spectral resolution ($R=100$) may be sufficient for rough determinations of atmospheric chemistry, $R>1000$ spectra have the potential to improve the accuracy of measurements of atmospheric composition and may provide sensitivity to isotopologues and other trace molecules. Additionally, higher spectral resolution will permit studies of atmospheric and orbital dynamics via measurements of radial velocities, spin-rotation alignment, and possibly even Doppler imaging of two-dimensional atmospheric features. Although \citet{min_hwo25} note that additional simulation work is needed to explore the impact of spectral resolution on science return, they also mention that conducting observations at higher spectral resolution could potentially facilitate better discrimination between planetary features and residual speckle noise. 

For LW-6 \citep[][Section~\ref{sssec:lawdki}]{walker_hwo25}, the SCDD requests a wide range of spectral resolutions from as low as $R\sim100$ to as high as $R\sim1000$. The range includes the default spectral resolution of $R=140$ whereas the higher limit would increase the ability to interpret overlapping spectral features. Next, LW-10b \citep[][Section~\ref{sssec:fpbio}]{schwieterman_hwo25} proposes a spectral resolution of $R\sim100-200$. The upper limit of that range is about 40\% higher than the baseline value of $R=140$. Although \citet{schwieterman_hwo25} note agreement with the $R=140$ spectral resolution requested for other LW science cases, they do not elaborate on the specific scientific motivation for requesting a broader range of $R\sim100-200$. 

Sub-baseline spectral resolutions of $R\sim130$, $R\sim100$, and $R\lesssim100$ are needed for investigations of surface biosignatures by LW-14 \citep[][Section~\ref{sssec:surface_bio}]{parenteau_surface_hwo25}, atmospheric retention on terrestrial planets by SSiC-21b (Section~\ref{sssec:retention}), and exozodi by SSiC-19a \citep[][Section~\ref{sssec:exozodi}]{debes_hwo25}, respectively. For LW-14, \citet{parenteau_surface_hwo25} note that past work has identified $R<130$ as the upper limit of the spectral resolution needed for the detection of the vegetation red edge but that further work is needed to determine the lowest resolution at which vegetation could be robustly distinguished from minerals with similar spectra. In the case of SSiC-21b, a spectral resolution of $R\sim100$ would be sufficient for determining the wavelength-dependent albedos and probing the atmospheric and surface properties of terrestrial planets. The spectral resolution for SSiC-19a is driven by the need to determine the composition and size distribution of the dust grains comprising the exozodi. Finally, the lowest spectral resolution ($R>40$) is needed by SSiC-2a for the detection of broad water features in the spectra of potentially habitable planets \citep[][Section~\ref{sssec:hab_system}]{hasegawa_water_hwo25}. 

\paragraph{Spatial Resolution}
For the exozodi study described in SSiC-19a \citep[][Section~\ref{sssec:exozodi}]{debes_hwo25}, key goals are to map the spatial distribution of dust and determine the frequency at which planets smaller than Earth reside within the habitable zone. Accordingly, observations must be sensitive to features significantly smaller than 1~AU. For reference, \citet{debes_hwo25} request a spatial resolution of $\theta=20$~mas. For the observations of protoplanetary disks and protoplanets discussed in SSiC-8b, \citet[][Section~\ref{sssec:proto}]{ren_hwo25} request an even finer spatial resolution of $\theta = 5$~mas. If 5~mas resolution cannot be achieved, conducting observations at the highest feasible spatial resolution would increase sensitivity to small-scale features within protoplanetary disks (ideally $\lesssim 0.1$~AU) and could potentially resolve circumplanetary accretion disks. 

\paragraph{Small Physical Separations}
The smallest physical separations would be probed by the study of terrestrial-like planets in the habitable zone proposed by LW-10b \citep[][Section~\ref{sssec:fpbio}]{schwieterman_hwo25} and the exploration of exozodi in and near the habitable zone described in SSiC-19a \citep[][Section~\ref{sssec:exozodi}]{debes_hwo25}. 

\paragraph{Large Physical Separations}
As for UV high-contrast spectroscopy, sensitivity to the largest spatial separations would be needed for the investigations of protoplanetary disks and gas giants proposed by SSiC-8b \citep[][Section~\ref{sssec:proto}]{ren_hwo25} and SSiC-15b \citep[][Section~\ref{sssec:reflected_giants}]{min_hwo25}, respectively.

\subsubsection{High-Contrast NIR Spectroscopy}
\label{sssec:hc_nir_spec}
This category consists of the identical modes SSiC-C and LW-C. The observations consist of $R=70$ spectra from 1100~nm to 1700~nm obtained at a contrast ratio of $10^{-10}$. As for high-contrast spectroscopy at UV and visible wavelengths, we designate the SSiC mode as the baseline configuration. 

\paragraph{Blue Wavelength Limit} 
All of the science cases in this category share a blue wavelength limit of $\lambda_\mathrm{blue} = 1100$~nm. This value does not have a particular scientific significance, but was selected because it is the notional transition point between the visible and NIR channels in the default configuration used for Living Worlds science cases. 

\paragraph{Red Wavelength Limit} 
The requested red wavelength limit ranges from $\lambda_\mathrm{red}=1300$~nm to $\lambda_\mathrm{red}=5000$~nm. The longest wavelength cut-off is requested for studies of gas giant exoplanets by SSiC-15b \citep[][Section~\ref{sssec:reflected_giants}]{min_hwo25} and SSiC-12b \citep[][Section~\ref{sssec:giant_orbits}]{sagynbayeva_et_al2025}. For the characterization of cool gas giants proposed by SSiC-15b, the choice of $\lambda_\mathrm{red}=5000$~nm is motivated by the possibility of observing auroral emission in H$_3^+$. A red limit of $\lambda_\mathrm{red}=5000$~nm would also improve constraints on molecular abundance by increasing the number of observable features for NH$_3$, PH$_3$, and other important molecules. If such a red limit is not feasible, \citet{min_hwo25} state that $\lambda_\mathrm{red}\ge 1700$~nm is needed to provide sensitivity to H$_2$S and therefore some insight into sulfur chemistry.

For SSiC-12b, the goal is to determine the orbits of giant planets. \citet{sagynbayeva_et_al2025} does not discuss the necessary wavelength range, but $\lambda_\mathrm{red}=5000$~nm is listed in the version of SSiC-12b submitted to the SCDD Portal\footnote{\url{https://docs.google.com/document/d/1BsVhxQ5rempJoWAlRENdma_wGpL8tkrnzJ97QtPD9WY/}}. However, the posted SCDD does not discuss the rationale for the selected wavelength range. Additional justification is needed if SSiC-12b is considered as a potential wavelength driver. 

The next reddest cut-off requested is $\lambda_\mathrm{red}=2000$~nm for SSiC-19b \citep[][Section~\ref{sssec:exozodi}]{debes_hwo25}. As for SSiC-12b, the wavelength limit comes from the version of the science case posted in the SCDD Portal\footnote{\url{https://docs.google.com/document/d/18rw8LQERZOZ9BbzSvzPTegzaYICmeXDq/}} rather than final publication. \citet{debes_hwo25} does mention the importance of NIR data for constraining the composition and size distribution of dust grains, but neither version of the SCDD includes a detailed justification supporting the choice of $\lambda_\mathrm{red}=2000$~nm. As shown in Table~\ref{tab:ssic_remap}, $\lambda_\mathrm{red}=2000$~nm is also listed for the study of habitability over time described in SSiC-20c (Section~\ref{sssec:survive_water}) and the investigation of volatile retention on terrestrial planets proposed by SSiC-21c (Section~\ref{sssec:retention}). For SSiC-20c, long-wavelength data would provide more insight into surface outgassing, while for SSiC-21c, additional NIR data would better constrain planetary albedo. However, as mentioned when defining SSiC-C observations in Section~\ref{sssec:ssic_c}, the indicated value of $\lambda_\mathrm{red}=2000$~nm is an approximation rather than a firm limit. For SSiC-20c, coverage to at least $1700$~nm is needed to capture outgassing HCN while for SSiC-21c, data should extend to at least $1000$~nm.

The third reddest cut-off of $\lambda_\mathrm{red}=1800$~nm is proposed by SSiC-3c \citep[][Section~\ref{sssec:rocky_sub}]{hu_rocky_hwo25}. This science case aims to investigate the nature and compositional diversity of sub-Neptune planets. For those investigations, spectroscopic observations covering $1400-1800$~nm could reveal features from a variety of important molecules such as CH$_4$, CO, CO$_2$, H$_2$O, and NH$_3$. 

The remaining science cases in this category request red cut-offs at or below the default configuration of $\lambda_\mathrm{red}=1700$~nm. For example, the search for life on terrestrial exoplanets proposed in LW-1c \citep[][Section~\ref{sssec:life}]{arney_hwo25}, and the study of prebiosignatures described in LW-4c \citep[][Section~\ref{sssec:prebiosignatures}]{ranjan_prebiosignatures_hwo25} request $\lambda_\mathrm{red}=1700$~nm, and we assume the same wavelength range for the investigation of origin-of-life theories outlined in LW-3c \citep[][Section~\ref{sssec:origin}]{ranjan_origin_hwo25}. In contrast, a shorter wavelength cut-off of $\lambda_\mathrm{red}=1500$~nm would suffice for the investigations of volatile delivery to HZ planets described in SSiC-2c \citep[][Section~\ref{sssec:hab_system}]{hasegawa_water_hwo25}. Finally, an even shorter cut-off of $\lambda_\mathrm{red}=1300$~nm would be adequate for the study of false positive biosignatures proposed in LW-10c \citep[][Section~\ref{sssec:fpbio}]{schwieterman_hwo25}.

\paragraph{Spectral Resolution} 
The spectral resolution requested by observations in this category ranges from $R=40$ to $R=1000$. The highest spectral resolution is requested by SSiC-15c for characterization of gas giants \citep[][Section~\ref{sssec:reflected_giants}]{min_hwo25} and SSiC-20c for assessing the time-dependence of planetary habitability (Section~\ref{sssec:survive_water}). For SSiC-15c, conducting observations at $R=1000$ could potentially refine assessments of atmospheric composition and orbital dynamics but $R=100$ spectra would still provide valuable insight into the dominant constituents of planetary atmospheres. Similarly, while $R=1000$ spectra could provide more insight into surface outgassing, a lower spectral resolution of $R\sim100$ would be sufficient for identifying steam-dominated atmospheres and detecting high rates of water loss consistent with ongoing runaway greenhouses.  For both programs, additional work is needed to determine whether the improved scientific insight possible from higher spectral resolution data warrants any associated increase in technical complexity. 

The next highest spectral resolution ($R\sim100$) is requested by SSiC-19b for studies of exozodi \citep[][Section~\ref{sssec:exozodi}]{debes_hwo25} and by SSiC-21c for probing volatile retention on terrestrial planets (Section~\ref{sssec:retention}). For SSiC-19b, the rationale for selecting this specific number is not discussed in the version of the SCDD posted in the SCDD Portal, and the necessary spectral resolution is not addressed in the published version of the SCDD \citep{debes_hwo25}. Given that an aim of the SCDD is to measure the visible - NIR spectral slope to determine the dust scattering efficiency, it seems likely that the baseline spectral resolution of $R=70$ would be considered to satisfy the request for $R\sim100$. Similarly, $R=70$ would likely be sufficient for measuring the wavelength dependence of planetary albedo and investigating the composition of planetary surfaces and atmospheres for SSiC-21c.

As explained by \citet{arney_hwo25}, $R=70$ spectral resolution at NIR wavelengths was deemed sufficient for the search for life proposed in LW-1c (Section~\ref{sssec:life}). Accordingly, LW-4c \citep[][Section~\ref{sssec:prebiosignatures}]{ranjan_prebiosignatures_hwo25} and SSiC-3c \citep[][Section~\ref{sssec:rocky_sub}]{hu_rocky_hwo25} adopted the same spectral resolution for their studies of prebiosignatures and the nature of sub-Neptunes, respectively. We also assume this default value for the investigations of origin-of-life theories presented in LW-3c \citep[][Section~\ref{sssec:origin}]{ranjan_origin_hwo25} and giant planet orbits proposed in SSiC-12b \citep[][Section~\ref{sssec:giant_orbits}]{sagynbayeva_et_al2025}. As the goal of SSiC-12b is orbit determination, the choice of spectral resolution is likely unimportant. Indeed, the SCDD does not discuss the motivation for requesting spectroscopic observations rather than photometric observations; presumably, the spectra could be used to assess atmospheric composition. 

Finally, SSiC-2 \citep[][Section~\ref{sssec:hab_system}]{hasegawa_water_hwo25} requests a lower spectral resolution of $R=40$ to measure water abundances on terrestrial planets. The total wavelength range for the combined visible program (SSiC-2a; $700-1100$~nm) and NIR program (SSiC-2b; $1100-1500$~nm) includes multiple water absorption features (i.e., at 0.97, 1.19, and $1.45~\mu$m) and the spectral resolution matches the ``baseline projected performance'' of $R=40$ and $\mathrm{SNR}=10$ spectra from $0.975 - 1.8\mu$m listed in the Science Traceability Matrix developed for the HabEx mission concept \citep[][see their Table~5.1-2]{gaudi_et_al2020}. 

\paragraph{Small Physical Separations}
The majority of the science cases in this category focus on planets within the habitable zones of their host stars, but LW-10c \citep[][Section~\ref{sssec:fpbio}]{schwieterman_hwo25} proposes to observe more highly irradiated planets in the ``Venus-zone'' (i.e., interior to the inner edge of the habitable zone). Accordingly, LW-10c would drive the need for sensitivity at close physical separations. 

\paragraph{Large Physical Separations}
At the opposite extreme, access to the widest physical separations is requested for the investigations of giant planets proposed by SSiC-12c \citep[][Section~\ref{sssec:giant_orbits}]{sagynbayeva_et_al2025} and SSiC-15c \citep[][Section~\ref{sssec:reflected_giants}]{min_hwo25}. For instance, SSiC-15c aims to study planets as cool as 30~K, which requires observing planets at separations of $10-100$~AU depending on system properties. In contrast, SSiC-12c took a different approach to science case exploration by exploring the population of giant planets that could be detected assuming notional instrument capabilities (i.e., IWA = 60~mas, OWA = 360 -- 2520~mas) rather than determining the range of physical separations needed to address their underlying science questions.

\subsubsection{High-Contrast Spectropolarimetry}
\label{sssec:hc_specpol}
This broad category encompasses all high-contrast spectropolarimetry. The baseline configuration SSiC-E consists of full Stokes spectropolarimetry obtained at the default spectral resolution ($R=140$) and wavelength range ($400-1100$~nm) for high-contrast visible spectroscopy (Section~\ref{sssec:hc_vis_spec}). Essentially, SSiC-E can be thought of as SSiC-B (or the equivalent mode LW-B) with the addition of polarimetry. The configuration LW-D is very similar to the baseline configuration SSiC-E but is shifted towards bluer wavelengths ($\lambda_\mathrm{blue}=350$, $\lambda_\mathrm{red}=850$~nm) and has a lower spectral resolution ($R=100$). 

Finally, modes SSiC-D and SSiC-F are the UV and NIR analogs to SSiC-E, respectively. The UV mode SSiC-D covers $200-400$~nm at $R=7$ and is therefore the polarimetric version of SSiC-A (and its equivalent LW-A). In the NIR, mode SSiC-F spans $1100-1700$~nm at $R=70$. Mode SSiC-F is akin to SSiC-C (or the identical mode LW-C) with added polarimetry. 

\paragraph{Blue Wavelength Limit} 
The bluest data are requested for the study of protoplanetary disks and young planets proposed by SSiC-8a \citep[][Section~\ref{sssec:proto}]{ren_hwo25}. For that program, access to the Lyman and Balmer lines of hydrogen is needed to assess the rate, variability, and geometry of planetary accretion. Although Figure~3 of \citet{ren_hwo25} is shaded to indicate a notional HWO wavelength range extending from roughly $90$~nm to $>1~\mu$m, the SCDD does not list a specific numerical wavelength range; we conservatively adopt $\lambda_\mathrm{blue}=90$~nm to fully capture the Lyman series.

Next, SSiC-9a requests $\lambda_\mathrm{blue}=140$~nm for observations of debris disks \citep[][Section~\ref{sssec:debris}]{rebollido_hwo25}. The motivation for this specific number is not explained in detail in the SCDD, but the wavelength range was selected to provide sensitivity to important sublimation, evaporation, and dissociation products as well as organics. 

The third shortest cut-off ($\lambda_\mathrm{blue}=200$~nm) is requested by SSiC-15d for characterization of gas giants in reflected light \citep[][Section~\ref{sssec:reflected_giants}]{min_hwo25}. Polarimetric data will provide valuable insight into clouds and hazes, but the specific choice of $\lambda_\mathrm{blue}=200$~nm for UV spectropolarimetry is not explained in detail in the SCDD. Indeed, in their Section~4, \citet{min_hwo25} state that additional work is needed to determine the relationship between science return and the selected $\lambda_\mathrm{blue}$ for UV spectroscopy, which would include observations with and without polarimetry.

The fourth shortest blue cut-off ($\lambda_\mathrm{blue}=300$~nm) is proposed by LW-5a \citep[][Section~\ref{sssec:technosignatures}]{kopparapu_hwo25}. The aim of this program is to search for technosignatures, which motivates collecting as much data as possible to search for signatures of intelligent life such as reflections from artificial structures. In their Figure~3, \citet{kopparapu_hwo25} show that polarimetric signatures from artificial structures on planetary surfaces could be detectable within the same $300-600$~nm wavelength range needed for spectroscopic studies of potential industrialization by searching for nitrogen dioxide pollution.

\paragraph{Red Wavelength Limit} 
The longest wavelength red cut-off ($\lambda_\mathrm{red}=3000$~nm) is proposed by SSiC-9c \citep[][Section~\ref{sssec:debris}]{rebollido_hwo25} for studies of debris disks. This wavelength range would capture the water feature at $2.8~\mu$m and provide insight into dust properties. Next, $\lambda_\mathrm{red}=2000$~nm is needed for the studies of atmospheric retention and exorings by SSiC-21f (Section~\ref{sssec:retention}) and SSiC-31b \citep[][Section~\ref{sssec:exorings}]{limbach_et_al2024, limbach_et_al2026}, respectively. For SSiC-21f, extended NIR coverage provides more insight into cloud and haze properties, but the specific choice of $\lambda_\mathrm{red}=2000$~nm is just an approximation as discussed in Section~\ref{sssec:retention}. The aim of SSiC-31b is to search for rings around exoplanets and then characterize detected ``exorings.'' NIR spectropolarimetry would constrain the geometry of ring systems as well as the compositions and scattering properties of ring particles. 

The next reddest data are needed for studies of gas giants by SSiC-15f \citep[][Section~\ref{sssec:reflected_giants}]{min_hwo25}. For that science case, the proposed red cut-off of $\lambda_\mathrm{red}=1800$~nm was selected to improve characterization of hazes, clouds, and gases in planetary atmospheres. As for the blue cut-off, \citet{min_hwo25} do not provide a detailed discussion of the specific choice of $\lambda_\mathrm{red}=1800$~nm rather than other NIR wavelengths. 

The fourth longest wavelength red cut-off ($\lambda_\mathrm{red}\gtrsim1500$~nm) was proposed by SSiC-18 \citep[][Section~\ref{sssec:exovenus}]{kane_venus_hwo25}. This project aims to detect and characterize Venus-like exoplanets orbiting other stars. NIR spectropolarimetry would provide important leverage for distinguishing between Earth-like atmospheres and Venus-like atmospheres by constraining the refractive index, sizes, and variability of particles within planetary atmospheres.

\paragraph{Spectral Resolution} 
The requested spectral resolution varies by two orders of magnitude. The highest spectral resolution ($R=10,000$) is requested for studies of protoplanetary disks and accreting planets in the UV (SSiC-8a) and visible (SSiC-8b) As explained in \citet[][Section~\ref{sssec:proto}]{ren_hwo25}, this high spectral resolution would significantly advance studies of planet formation by resolving the shapes of accretion lines. The next highest spectral resolution ($R>3000$) is needed for the investigations of debris disks proposed by SSiC-9 \citep[][Section~\ref{sssec:debris}]{rebollido_hwo25}. The observations would cover UV (SSiC-9a), visible (SSiC-9b), and NIR (SSiC-9c) wavelengths.

The technosignatures investigation proposed in LW-5a does not quantify the necessary spectral resolution, so we assume a moderately low value of $R=500$. The actual spectral resolution needed for technosignature detection likely depends on the nature of potential signals and the wavelength of observation. 

Finally, the remaining programs in this category either request $R=100$ or were assigned standard resolutions corresponding to those of the equivalent non-polarimetric high-contrast spectroscopy modes. Specifically, $R=100$ was proposed for the investigation of alien photosynthetic life by LW-7a and LW-7b \citep[][Section~\ref{sssec:polbio}]{berdyugina_imaging_hwo25} as well as the characterization of exorings by SSiC-31a and SSiC-31b \citep[][Section~\ref{sssec:exorings}]{limbach_et_al2024, limbach_et_al2026}. SSiC-15 \citep[][Section~\ref{sssec:reflected_giants}]{min_hwo25} and SSiC-18 \citep[][Section~\ref{sssec:exovenus}]{kane_venus_hwo25} did not state specific spectral resolutions for spectropolarimetric observations, so we assume $R=7$ for the UV component SSiC-15d; $R=140$ for the visible components SSiC-15e and SSiC-18b; and $R=70$ for the NIR components SSiC-15f and SSiC-18c.

\paragraph{Small Physical Separations}
The program requiring sensitivity at the smallest physical separation is SSiC-18 \citep[][Section~\ref{sssec:exovenus}]{kane_venus_hwo25}. SSiC-18 aims to study Venus analogs and therefore needs to be able to observe planets at separations as small as $0.3$~AU at visible (SSiC-18b) and NIR (SSiC-18c) wavelengths.

\paragraph{Large Physical Separations}
Several of the science cases propose observing planets or disks at wide separations. At the widest extreme, SSiC-8 \citep[][Section~\ref{sssec:proto}]{ren_hwo25} aims to image protoplanets at separations up to 1000~AU from their host stars in the UV (SSiC-8a) and visible (SSiC-8b). Despite that ambitious goal, Figure~2 of \citet{ren_hwo25} indicates that the majority of imaged protoplanets would likely be within 100~AU. 

The next widest observations are needed for the studies of debris disks and gas giants proposed by SSiC-9 \citep[][Section~\ref{sssec:debris}]{rebollido_hwo25} and SSiC-15 \citep[][Section~\ref{sssec:reflected_giants}]{min_hwo25}, respectively. For SSiC-9, observations should cover a large field of view ($\geq 2' \times 2'$). SSiC-15 would study planets as cold as 30~K in the UV (SSiC-15d), visible (SSiC-15e), and NIR (SSiC-15f). Accordingly, SSiC-15 would need sensitivity to planets at separations of 10--100~AU depending on system properties. Finally, for the characterization of exorings, program SSiC-31 \citep[][Section~\ref{sssec:exorings}]{limbach_et_al2024, limbach_et_al2026} would need to image planets at separations up to 10~AU in the visible (SSiC-31a) and NIR (SSiC-31b). 

\subsubsection{High-Contrast Polarimetric Imaging}
\label{sssec:hc_imaging_pol}
High-contrast polarimetric imaging is categorized as SSiC-G and LW-E. The two modes are identical, and we opt to designate SSiC-G as the baseline configuration. SSiC-G (and LW-E) observations consist of high-contrast imaging spanning 300 -- 1700~nm with full Stokes polarimetry and polarimetric precision $\leq 1\%$. Due to the broad wavelength range, these observations may necessitate using multiple channels or multiple instruments. 

\paragraph{Blue Wavelength Limit} 
The blue cut-off needed for these observations ranges from 140~nm to 400~nm. The shortest wavelength observations are needed for the debris disk observations proposed by SSiC-9 \citep[][Section~\ref{sssec:debris}]{rebollido_hwo25}. As the diffraction limit will be smaller at UV wavelengths, these observations could potentially reveal small-scale features (i.e., $\lesssim 1$~AU) such as gaps shaped by orbital resonances with exoplanets. 

The next bluest data ($\lambda_\mathrm{blue}=200$~nm) are requested by SSiC-19d for studies of exozodi (Section~\ref{sssec:exozodi}). The published version of the science case \citep{debes_hwo25} does not quantify the necessary wavelength range, but $\lambda_\mathrm{blue}=200$~nm is specified in the version posted to the SCDD Portal.\footnote{\url{https://docs.google.com/document/d/18rw8LQERZOZ9BbzSvzPTegzaYICmeXDq/}}. Both documents discuss the importance of polarimetric observations at multiple wavelengths for determining the shapes, sizes, and compositions of dust grains. Determining those dust properties is essential for achieving the stated objective of connecting the current architectures of mature planetary systems to the conditions present at the time of planet formation. 

SSiC-3 \citep[][Section~\ref{sssec:rocky_sub}]{hu_rocky_hwo25} does not specify the exact wavelength range over which polarimetric imaging is needed for investigating the nature of sub-Neptunes, but it seems reasonable to adopt the same wavelength range as for spectroscopy. By that logic, SSiC-3d would have the third bluest short-wavelength cut-off ($\lambda_\mathrm{blue}=250$~nm). Following the same line of reasoning and also matching the polarimetric imaging wavelength range to the spectroscopic wavelength range for the technosignatures investigation proposed in LW-5 \citep[][Section~\ref{sssec:technosignatures}]{kopparapu_hwo25}, LW-5c could accommodate a slightly more relaxed cut-off of $\lambda_\mathrm{blue}=300$~nm.

\paragraph{Red Wavelength Limit} 
As for the blue wavelength limit, the broadest red cut-off ($\lambda=3000$~nm) is requested by SSiC-9 for observations of debris disks \citep[][Section~\ref{sssec:debris}]{rebollido_hwo25}. The selected wavelength range would span the $2.8~\mu$m water feature and enable constraining dust properties. 

The next longest wavelength red cut-off ($\lambda_\mathrm{red}=2000$~nm) is requested by the portal version of SSiC-19 for studies of exozodi \citep[][Section~\ref{sssec:exozodi}]{debes_hwo25}. The general rationale for NIR polarimetry is the need to measure dust scattering properties to constrain the sizes, shapes, and compositions of dust grains, but SSiC-19 does not provide a detailed discussion of the specific choice of 2000~nm as opposed to other NIR wavelengths.

Continuing to assume that the wavelength range for polarimetric imaging should match the spectroscopic wavelength range for the sub-Neptune study described in SSiC-3 \citep[][Section~\ref{sssec:rocky_sub}]{hu_rocky_hwo25}, the second reddest cut-off ($\lambda_\mathrm{red} = 1800$~nm) would be needed for SSiC-3d. The remaining science cases in this category propose much shorter wavelength cut-offs of 850 --1100~nm. 

\paragraph{Physical Separations}
While SSiC-9d requests coverage of a wide field of view ($\geq 2'\times2'$) to better capture full disks, the other science cases in this category are primarily concerned with observing planets or disk structures in the habitable zones of their host stars. Accordingly, these programs demand sensitivity to planets within the habitable zone, ideally at multiple orbital phases. For the search for surface oceans described in SSiC-4 \citep[][Section~\ref{sssec:surface_water}]{lustig-yaeger_liquidwater_hwo25}, access to planets at small separations is crucial because detecting ocean glint mandates observing changes in planet reflectivity as the phase angle decreases from gibbous to crescent. Observations at different orbital phases are also important for assessing atmospheric properties and potential biopigments as proposed by LW-7 \citep[][Section~\ref{sssec:polbio}]{berdyugina_imaging_hwo25}. 

However, given that technological civilizations may have ventured beyond their home planets and therefore placed detectable signs on worlds beyond the habitable zone, the technosignatures investigation proposed in LW-5 \citep[][Section~\ref{sssec:technosignatures}]{kopparapu_hwo25} is more agnostic regarding the specific choice of target planet. Additionally, small planets with a range of equilibrium temperatures would be appropriate for the study of sub-Neptunes described in SSiC-3 \citep[][Section~\ref{sssec:rocky_sub}]{hu_rocky_hwo25}. Both LW-5 and SSiC-3 could therefore target planets interior to the inner edge or exterior to the outer edge of the habitable zone. Coverage of a broader annulus and therefore a wider range of disk temperatures would also be useful for the investigation of exozodi proposed by SSiC-19d \citep[][Section~\ref{sssec:exozodi}]{debes_hwo25}. 

\subsubsection{High-Contrast Imaging (without polarimetry)}
\label{sssec:hc_imaging}
The final category of high-contrast observations is photometry without polarimetry. These observations are classified as SSiC-H and LW-F. The modes are nearly identical, but the baseline configuration SSiC-H has broader wavelength coverage ($200-1100$~nm versus $300-900$~nm for LW-F). Both modes consist of imaging a $10''\times10''$ field of view at a contrast ratio of $10^{-10}$. For SSiC-6 \citep[][Section~\ref{sssec:occ_binary}]{newton_hwo25}, these observations should be sensitive to planets orbiting FGK primaries in stellar binaries with orbital separations of $30-1000$~AU as well as planets orbiting single stars.

\paragraph{Blue Wavelength Limit} 
The short wavelength cut-off needed for programs in this category ranges from 90~nm to 400~nm. The science case with the most ambitious blue cut-off is SSiC-8c \citep[][Section~\ref{sssec:proto}]{ren_hwo25}, which aims to study young planets and protoplanetary disks. For that program, a short blue cut-off at $\lambda_\mathrm{blue} = 90$~nm would capture accretion signatures from growing protoplanets. The next bluest cut-off is $\lambda_\mathrm{blue} =140$~nm for observations of debris disks by SSiC-9e \citep[][Section~\ref{sssec:debris}]{rebollido_hwo25}. Table~4.1 of \citet{rebollido_hwo25} indicates that the  selected wavelength range would provide sensitivity to organics and sublimation products, but those signatures would most likely be detected spectroscopically. Although we assume the same wavelength range for photometry as for spectroscopy, the science case does not elaborate on the specific wavelength range needed for photometry. 

The third bluest cut-off ($\lambda_\mathrm{blue} =200$~nm) is requested for studies of gas giant planets and exozodi by SSiC-15g \citep[][Section~\ref{sssec:reflected_giants}]{min_hwo25} and SSiC-19c \citep[][Section~\ref{sssec:exozodi}]{debes_hwo25}, respectively. For SSiC-15g, precise, time-resolved photometry could be used to construct longitudinal maps of planetary albedo, which varies with wavelength and therefore benefits from broad wavelength coverage. For \mbox{SSiC-19c}, multi-band photometry is needed to constrain dust properties. We adopt the short wavelength cut-off listed in the portal version of SSiC-19 ($\lambda_\mathrm{blue} =200$~nm), but the published version of the science case \citep{debes_hwo25} emphasizes the need for Vis/NIR imaging in at least three bands. Accordingly, the scientific return from adding UV imaging and the motivation for the specific choice of $\lambda_\mathrm{blue} = 200$~nm are somewhat unclear.

\paragraph{Red Wavelength Limit}
The proposed red cut-offs extend from 700~nm to 3000~nm. The longest wavelength data are needed for studies of debris disks by SSiC-9e \citep[][Section~\ref{sssec:debris}]{rebollido_hwo25}, exozodi by SSiC-19c \citep[][Section~\ref{sssec:exozodi}]{debes_hwo25}, and exomoons by SSiC-32 \citep[][Section~\ref{sssec:exomoons}]{limbach_et_al2024, limbach_et_al2026}. For SSiC-9e ($\lambda_\mathrm{red}=3000$~nm) and SSiC-19c ($\lambda_\mathrm{red}=2000$~nm), NIR data would be valuable for constraining the composition and size of dust particles. For SSiC-32, extending observations to 2000~nm would likely improve moon detectability because the moon-to-planet contrast ratio is expected to be more favorable at longer wavelengths.\footnote{Portal version of SSiC-32: \url{https://docs.google.com/document/d/1bOa97ZAblcfkfIOmzQDZDYDxJaMN-Cue8MFVjDEDTqw/edit?tab=t.0}} The next longest wavelength cut-off is $\lambda_\mathrm{red} = 1800$~nm, which is proposed by SSiC-15g for the albedo mapping of gas giants \citep[][Section~\ref{sssec:reflected_giants}]{min_hwo25}. 

\paragraph{Small Physical Separations}
Several science cases in this category focus on planets within the habitable zone. Accordingly, high-contrast photometry should be capable of observing planets and disks at separations as close as the inner edge of the habitable zone. Observations of planets interior to the inner edge of the habitable zone (i.e., as close as 0.5~AU) could also provide useful data for the studies of giant planets proposed in SSiC-15g \citep[][Section~\ref{sssec:reflected_giants}]{min_hwo25}, young planets and protoplanetary disks discussed in SSiC-8 \citep[][Section~\ref{sssec:proto}]{ren_hwo25}, and debris disks presented in SSiC-9 \citep[][Section~\ref{sssec:debris}]{rebollido_hwo25}. 

\paragraph{Large Physical Separations}
Sensitivity to planets and disks at wide orbital separations is also important for these science cases. The widest coverage is requested by SSiC-8c \citep[][Section~\ref{sssec:proto}]{ren_hwo25}, which aims to detect disk structures and protoplanets at distances of up to 1000~AU.\footnote{In practice, such observations of planets and disk structures 1000~AU from their host stars will likely be possible only for more distant systems and may be better accomplished using other facilities.} Access to a wide field of view ($\geq 2' \times 2'$) would also be useful for the studies of debris disks addressed in SSiC-9 \citep[][Section~\ref{sssec:debris}]{rebollido_hwo25}. 

For SSiC-15g \citep[][Section~\ref{sssec:reflected_giants}]{min_hwo25}, observations of planets as cool as 30~K are needed. Such planets would be $\sim10-100$~AU from their host stars with the exact distance depending on stellar, planetary, and orbital properties. Additionally, the studies of exorings and exomoons described by \citet{limbach_et_al2024, limbach_et_al2026} necessitate observing planets at distances of $1-10$~AU for exorings (SSiC-31; Section~\ref{sssec:exorings}) and $1-5$~AU for exomoons (SSiC-32; Section~\ref{sssec:exomoons}). 

\subsection{Mission Operations} 
In this section, we consider the effects of mission operations on the ability of a facility like HWO to carry out the science programs proposed in Section~\ref{sec:scdds}. Rather than list each science case repeatedly, we instead highlight specific examples of which science cases might be affected by different choices and explain why those science cases demand certain telescope or operational capabilities. For instance, we could have listed nearly every astrobiological investigation in the following categories but we instead opt to focus on the science cases that are most strongly affected. 

\subsubsection{Simultaneous \& Contemporaneous Observations}
\label{sssec:simultaneous}
For observations of transient or time-variable sources such as supernovae, exoplanets, and solar system objects, it would be scientifically advantageous if data could be obtained contemporaneously or even simultaneously at multiple wavelengths. For instance, observations obtained within 10~minutes would be useful for studying Venus for SSiC-5 \citep[][Section~\ref{sssec:venus}]{izenberg_hwo25} while studies of debris disks for SSiC-9 \citep[][Section~\ref{sssec:debris}]{rebollido_hwo25} would benefit from simultaneous observations in different bandpasses to characterize exocomets as well as contemporaneous (timescales of days - years) observations for studies of collisions. 

For extragalactic science, GG-2 \citep[][Section~\ref{sssec:bh_mass_spin}]{cann_hwo25} demands simultaneous observations extending from the UV to the NIR to measure black hole spins via continuum-fitting of AGN disk emission. The disk emission varies rapidly and therefore piecing together spectra of different wavelength chunks obtained at multiple epochs would not be effective. In addition, GG-5 \citep[][Section~\ref{sssec:agn_outflow}]{zhang_hwo25} requests simultaneous UV and NIR observations to better correct for dust extinction when determining AGN properties and investigating the impact of AGN on their host galaxies. Due to the time variability of AGN, the study of supermassive black hole binaries proposed by GG-19 \citep[][Section~\ref{sssec:smbh_pol}]{martins_hwo25} also necessitates simultaneous, broad spectral coverage: spectropolarimetry spanning 100~nm to 2~$\mu$m.

Regardless of potential target variability, increasing the wavelength range that could be covered in a given observation would increase the efficiency of programs requiring broad spectroscopic coverage such as investigating the origin of $r$-process elements via stellar spectroscopy for EE-9 \citep[][Section~\ref{sssec:r_process_nature}]{roederer_nature_hwo25} and improving the cosmic distance ladder via observations of fainter Cepheid variables and tip-of-the-red-giant-branch stars for EE-10 \citep[][Section~\ref{sssec:distance_ladder}]{anand_hwo25}. Many other science cases would also benefit from the improved efficiency of multi-channel observations of the same target as well as parallel observations of other targets.

The studies that would likely benefit the most from the ability to simultaneously obtain data covering a wide spectral bandpass are high-contrast observations for LW and SSiC investigations of planetary habitability (e.g., Section~\ref{ssec:xswg_habitable}) and potential biosignatures (e.g., Section~\ref{ssec:xswg_biosig}). As discussed more in Section~\ref{sssec:duration}, these studies demand long observation times to accumulate the signal-to-noise ratio needed to detect or rule out subtle features. An additional complication is that many abiotic and biotic signals are time variable and therefore stitching together multiwavelength data sets obtained at different epochs may significantly complicate interpretation. Given the necessity of broad wavelength coverage for robust assessments of the existence of habitable environments and life on exoplanets as well as the significant time savings associated with obtaining data in multiple bandpasses simultaneously rather than sequentially, the astrobiological science cases strongly push towards a design capable of obtaining high-contrast observations simultaneously at multiple wavelengths. 

When considering simultaneous observations, we use the same language as HST\footnote{\url{https://hst-docs.stsci.edu/acsihb/chapter-7-observing-techniques/7-9-parallel-observations}} and JWST.\footnote{\url{https://jwst-docs.stsci.edu/methods-and-roadmaps/jwst-parallel-observations}} Specifically, we refer to the observation with the highest priority as the ``primary'' observation and the other observation(s) as ``parallel'' observation(s). If those supplementary observations are associated with the primary observation, then they are deemed ``coordinated parallels.'' If they are unrelated observations from a different observing program, then they are classified as ``pure parallels.'' 

While implementation of a ``pure parallels'' mode could increase overall operational efficiency by enabling serendipitous observations of unrelated targets with different instruments during long-duration exoplanet observations (e.g., multiple deep fields), the scientific value of such observations may be reduced by operational factors such as restrictions on instrument filter changes to reduce vibrations that would compromise the exquisite stability needed to reach high contrast levels and the need to prioritize limited onboard data storage for the primary observation. Accordingly, the theoretical efficiency gain may not be realized unless the total per-filter integration times are similar for both programs. 

Moreover, a ``pure parallel'' operations mode would still result in observing various spectral regions of exoplanet spectra non-simultaneously (or perhaps even non-contemporaneously, given viewing constraints; see Section~\ref{sssec:field}). Accordingly, it would be strategic to prioritize multi-channel high-contrast observations of single targets (i.e., ``coordinated parallels'') over multi-instrument observations of different targets (i.e., ``pure parallels''). If coordinated parallel high-contrast observations are not possible, then the true viability of pure parallel observations must be considered when estimating the operational efficiency gain from parallel observations. 

\subsubsection{Field of Regard}
\label{sssec:field}
While a broad field of regard is beneficial for nearly all observations and would simplify scheduling, transient and (exo)planetary science cases are the strongest drivers towards large field of regard. For instance, increasing the instantaneous field of regard would increase the fraction of supernovae that could be observed for EE-4 \citep[][Section~\ref{sssec:r_process_el}]{burns_hwo25} and EE-3 \citep[][Section~\ref{sssec:flash_ccsne}]{andrews_hwo25}. 

For biosignature investigations, observations at multiple phases are needed to map planetary surfaces and potentially detect glint and continents for SSiC-4 \citep[][Section~\ref{sssec:surface_water}]{lustig-yaeger_liquidwater_hwo25} and LW-7 \citep[][Section~\ref{sssec:polbio}]{berdyugina_imaging_hwo25}. Phase-resolved observations are also necessary for investigating seasonal variations for LW-13 \citep[][Section~\ref{sssec:seasonality}]{lafleche_hwo25}. For all directly imaged planets (Sections~\ref{ssec:ssic_scdds} \& \ref{ssec:lw_scdds}), observations at multiple phases are also needed to determine orbital parameters and reduce the degeneracy between planet size and albedo. In addition, technosignature searches for artificial illumination on planetary nightsides require observations precisely timed to occur when the planet is at high ($>90^\circ$) phase angle \citep[][Section~\ref{sssec:technosignatures}]{kopparapu_hwo25}.

For transiting planets, the need to time observations to coincide with transit or eclipse favors a broader field of regard because that would increase the number of scheduling opportunities. The investigations focused on transiting planets include SSiC-14 \citep[][Section~\ref{sssec:escape}]{dossantos_hwo25}, SSiC-16 \citep[][Section~\ref{sssec:transits}]{wakeford_hwo25}, SSiC-17 \citep[][Section~\ref{sssec:id_oceans}]{quick_hwo25}, SSiC-18 \citep[][Section~\ref{sssec:exovenus}]{kane_venus_hwo25}, SSiC-24 \citep[][Section~\ref{sssec:highres_atmos}]{cubillos_hwo25}, and SSiC-26 \citep[][Section~\ref{sssec:bfield}]{strugarek_hwo25}.

Within the solar system, obtaining observations of Venus near quadrature for SSiC-5 motivate the ability to look closer to the Sun (i.e., reduce the ``solar avoidance angle'') and observe Venus near quadrature \citep[][Section~\ref{sssec:venus}]{izenberg_hwo25}. A reduced solar avoidance angle would also increase the observability of near-Earth objects for SSiC-30 \citep[][Section~\ref{sssec:defense}]{dotson_hwo25}, thereby improving the ability to detect objects that could be potentially destructive to life on Earth. Decreasing the solar avoidance angle would also provide more opportunities to observe Titan for SSiC-27 (Section~\ref{sssec:titan}), Mars for SSiC-28 (Section~\ref{sssec:mars_post}), solar system giant planets for SSiC-29 \citep[][Section~\ref{sssec:solargiant}]{fletcher_hwo25}, and comets for SSiC-33 \citep[][Section~\ref{sssec:aminos}]{gomez_de_castro_amino_hwo25}. 

\subsubsection{Observation Duration}
\label{sssec:duration}
For most science cases, faint features would likely be detected by combining multiple integrations that can be interrupted as needed for spacecraft operations. However, a few programs request either uninterrupted observations or extreme absolute stability so that observations before and after interruptions can be successfully combined. 

One of the science cases requesting the longest duration observations is SSiC-16 \citep[][Section~\ref{sssec:transits}]{wakeford_hwo25}, which would study the atmospheric dynamics and chemistry of transiting planets via phase curves. The ability of a phase curve to constrain surface and atmospheric features is highly dependent on the measurement precision, and therefore the ideal data set for SSIC-16 consists of continuous, uninterrupted phase curves spanning at least a full orbital period. SSiC-16 aims to observe planets with periods longer than 20~days, which would result in phase curve observations exceeding 500~hours per observation. The need for extended duration observations could potentially be relaxed if the observations have exquisite absolute precision, but stitching together partial phase curves generally results in lower precision and more persistent systematics compared to obtaining a single continuous phase curve \citep[e.g.,][]{lewis_et_al2013}. 

The other program proposing extremely long-duration observations is SSiC-32 \citep[][Section~\ref{sssec:exomoons}]{limbach_et_al2024, limbach_et_al2026}, which aims to detect exomoons by monitoring the brightness of star-planet systems. For that program, the rarity of the mutual events needed for exomoon detection means that brightnesses of target systems should be monitored during observing campaigns as long as several weeks. As for SSiC-16, it may be possible to reduce the duration of individual observations if the photometry has sufficient absolute precision, but frequent breaks to downlink data or perform other spacecraft operations would reduce the duty cycle and therefore increase the risk that the mutual events that could have revealed the presence of exomoons fall in gaps between observations. 
When the ephemeris of a transiting planet is known in advance and a full phase curve is not needed, the observation duration can be significantly reduced. For instance, observation durations of $\sim30$~hours would be needed to study transiting planets in the habitable zones of Sun-like stars for SSiC-14 \citep[][Section~\ref{sssec:escape}]{dossantos_hwo25} and SSiC-24 \citep[][Section~\ref{sssec:highres_atmos}]{cubillos_hwo25}. An even shorter observation duration ($<10$~hours) would be sufficient for SSiC-26 \citep[][Section~\ref{sssec:bfield}]{strugarek_hwo25}, which aims to directly measure planetary magnetic fields by observing transits of short-period planets and determining the polarization of escaping He~I gas. 

For the studies of cool gas giants in reflected light proposed by SSiC-15 \citep[][Section~\ref{sssec:reflected_giants}]{min_hwo25}, $\sim 10 -24$~hours of continuous data would be beneficial for measuring planetary rotation periods. Detecting liquid surface water on potentially habitable planets by observing ocean glint as proposed by SSiC-4 \citep[][Section~\ref{sssec:surface_water}]{lustig-yaeger_liquidwater_hwo25} would require observations over several months, but obtaining such a long sequence of data will almost certainly necessitate stitching together different observations. Accordingly, SSiC-4 and other investigations of the potential habitability indicators and biosignatures are likely drivers for both the duration of individual observations and absolute instrument stability. Key examples include the search for life on exoplanets presented by LW-1 \citep[][Section~\ref{sssec:life}]{arney_hwo25}, the search for surface biosignatures described by LW-14 \citep[][Section~\ref{sssec:surface_bio}]{parenteau_surface_hwo25}, and the investigation of the relationship between ozone abundances and planet age proposed by SSiC-11 \citep[][Section~\ref{sssec:ozone_onset}]{blunt_et_al2025}.

Investigations of weather and surface properties in the solar system would also benefit from the ability to obtain observations over timescales of many hours. For example, for SSiC-5 \citep[][Section~\ref{sssec:venus}]{izenberg_hwo25}, a series of multi-hour observations of Venus obtained over several days are needed to study how Venus's clouds change over time. Multi-hour observations are also necessary for studying the variability of clouds and aurorae on solar system giant planets for SSiC-29 \citep[][Section~\ref{sssec:solargiant}]{fletcher_hwo25}. 

\subsubsection{Rapid Response}
\label{sssec:rapid}
Unsurprisingly, rapid response capabilities are most valuable for transient phenomena. In all cases, onboard decision making would be helpful for adapting to potentially unknown target properties (e.g., brightness in HWO bandpass) and obtaining multi-wavelength, multi-epoch datasets. The need to review, accept, and transmit observing requests quickly could also inform decisions about ground support and operational policies. 

The most rapid response time requested is 30 minutes for breakthrough-quality observations of neutron star mergers to advance studies of r-process elements for EE-4 \citep[][Section~\ref{sssec:r_process_el}]{burns_hwo25}. If a 30-min response time is not achievable, observations within a few hours would yield enabling science. The second fastest request is set by supernova studies: observations within one~hour of trigger and within one~day of explosion would enable breakthrough-level flash spectroscopy to better understand the state of massive stars prior to their deaths as core collapse supernovae for EE-3 \citep[][Section~\ref{sssec:flash_ccsne}]{andrews_hwo25}. 

Target-of-opportunity and rapid response capabilities will also be useful for solar system science, but the resulting demands on response time are less strict. Obtaining observations within a few days would be beneficial for characterizing the orbits and properties of newly discovered comets and interstellar objects for SSiC-7 \citep[][Section~\ref{sssec:solar}]{mandt_hwo25} as well as capturing observations of main belt comets when they are observed by other facilities to be actively sublimating for SSiC-22 (Section~\ref{sssec:mars}). Additionally, fast response times would facilitate observations of transient phenomena such as active geysers on cold ocean planets for SSiC-17 \citep[][Section~\ref{sssec:id_oceans}]{quick_hwo25} and impacts or storms on giant planets for SSiC-29 \citep[][Section~\ref{sssec:solargiant}; response times $<7$~days preferred]{fletcher_hwo25}. 

\subsubsection{Target Tracking}
\label{sssec:tracking}
Tracking needs would be set by the need to observe solar system objects such as Venus for SSiC-5 \citep[][Section~\ref{sssec:venus}]{izenberg_hwo25} and interstellar interlopers like 'Oumuamua (500~mas/s at closest approach) for SSiC-7 \citep[][Section~\ref{sssec:solar}]{mandt_hwo25}. The ability to guide and track at non-sidereal rates is also essential for studying solar system ocean worlds for SSiC-1 \citep[][Section~\ref{sssec:oceans_habitable}]{cartwright_hwo25}, Titan for SSiC-27 (Section~\ref{sssec:titan}), and Mars for SSiC-28 (Section~\ref{sssec:mars_post}). Furthermore, non-sidereal tracking is needed for observations of gas giant systems for SSiC-25 \citep[][Section~\ref{sssec:aurorae}]{chaufray_hwo25} and SSiC-29 \citep[][Section~\ref{sssec:solargiant}]{fletcher_hwo25} as well as observations of small bodies by SSiC-22 (Section~\ref{sssec:mars}) to probe the origin of Mars and by SSiC-33 \citep[][Section~\ref{sssec:aminos}]{gomez_de_castro_amino_hwo25} to investigate the chirality of amino acids. 

Other targets, especially extragalactic targets, will move at slower rates. While the targets needing the highest tracking rates are within the solar system, understanding their properties is critical to advancing exoplanetary science by placing the solar system in context of other planetary systems. 

\subsubsection{Bright Targets and Dynamic Range}
\label{sssec:saturation}
Saturation is primarily a concern for (exo)planetary observations. Within the solar system, HWO would ideally be capable of observing Venus for SSiC-5 \citep[][Section~\ref{sssec:venus}]{izenberg_hwo25}; bright asteroids like Ceres and Vesta to probe the origin of Mars for SSiC-22 (Section~\ref{sssec:mars}); and the ocean worlds of the outer solar system for SSiC-1 \citep[][Section~\ref{sssec:oceans_habitable}]{cartwright_hwo25}. High dynamic range will be particularly important for detecting faint plumes near the bright disks of Enceladus and Europa as well as studying other moons and rings of solar system giant planets for SSiC-29 \citep[][Section~\ref{sssec:solargiant}]{fletcher_hwo25}. 

For exoplanetary science, the observatory should be capable of conducting transmission spectroscopy of bright stars for SSiC-14 \citep[][Section~\ref{sssec:escape}]{dossantos_hwo25}, SSiC-16 \citep[][Section~\ref{sssec:transits}; $m_V \ge 2$ ]{wakeford_hwo25}, and SSiC-26 \citep[][Section~\ref{sssec:bfield}]{strugarek_hwo25}. Additionally, SSiC-6 \citep[][Section~\ref{sssec:occ_binary}]{newton_hwo25} necessitates the ability to obtain high-contrast observations of planets orbiting stars in bright, close binaries with separations $\ge3''$. 

For extragalactic science, improving the cosmic distance ladder would demand observing bright Cepheids in nearby galaxies (Section~\ref{sssec:distance_ladder}). The brightest Cepheids in the Large and Small Magellanic Clouds are expected to have magnitudes $m_V \gtrsim 14$. The Cepheids are therefore considerably fainter than the bright stars that will be targeted for transmission spectroscopy, but the Cepheids would ideally be observed via a wide-field imager while transmission and emission spectra are single-object spectra obtained at moderate spectral resolution ($R > 3000$). In both cases, dispersing light, spatial scanning, and reading out subarrays may be useful. Finally, astrometric determination of masses and orbits of potentially habitable and inhabited planets for LW-12 \citep[][Section~\ref{sssec:mp}]{gary_hwo25} would necessitate accurate and precise measurements of stellar positions for both bright LW target stars and fainter background stars. 

\subsubsection{Data Volume and Downlink}
\label{sssec:downlink}
Many of the science cases presented in Section~\ref{sec:scdds} would collect large volumes of data. Even accounting for onboard compression and analysis, these programs would also necessitate transmitting large data volumes back to Earth. Demands for onboard data storage would likely be driven by long-duration SSiC and LW programs such as those described in Section~\ref{sssec:duration}, while the cadence of communications with Earth may be driven by the time-sensitive programs presented in Section~\ref{sssec:rapid}. Fast downlink speeds and high downlink bandwidth would be advantageous for all programs, but especially those collecting massive spectroscopic data sets such as the IFS and MOS programs described in Sections~\ref{sssec:ifs} and \ref{sssec:mos}. 

The high data generation rate from the individual programs described in Section~\ref{sec:scdds} could be further increased by the implementation of parallel observations to improve observational efficiency (e.g., pure parallel observations of extragalactic deep fields during long LW and SSiC high-contrast observations; see Section~\ref{sssec:simultaneous}). Accordingly, it would be strategic to investigate the ability to transmit data during long-duration observations to reduce the demand on onboard storage and decrease the lag between collection and analysis of data. Without this capability, data storage may limit the duration of individual observations and the implementation of parallel observations, thereby reducing the scientific return of the observatory. Ensuring high downlink rates and cadences potentially places demands on the spacecraft itself as well as the availability and capacity of ground stations or intermediate, space-based relay stations. 

\section{Connections to Astro2020}
\label{sec:astro2020}
As mentioned in Section~\ref{sec:intro}, Astro2020 \citep{astro2020} declared that HWO (then known as IR/O/UV) ``would directly address two-thirds of the key science questions identified [by Astro2020] and will contribute to addressing many of the others.'' We now support that assertion by mapping the science cases described in Section~\ref{sec:scdds} to the 24~questions and six~discovery areas identified by Astro2020. In Table~\ref{tab:astro2020}, we list the HWO science cases related to each Astro2020 question. For reference, we also include a column showing the capability of LUVOIR-B to address those science questions according to the Astro2020 Electromagnetic Observations from Space~1 (EOS1) panel \citep[Appendix I][]{astro2020}. 

\begin{figure*}
 \centering
 \includegraphics[width=1\linewidth]{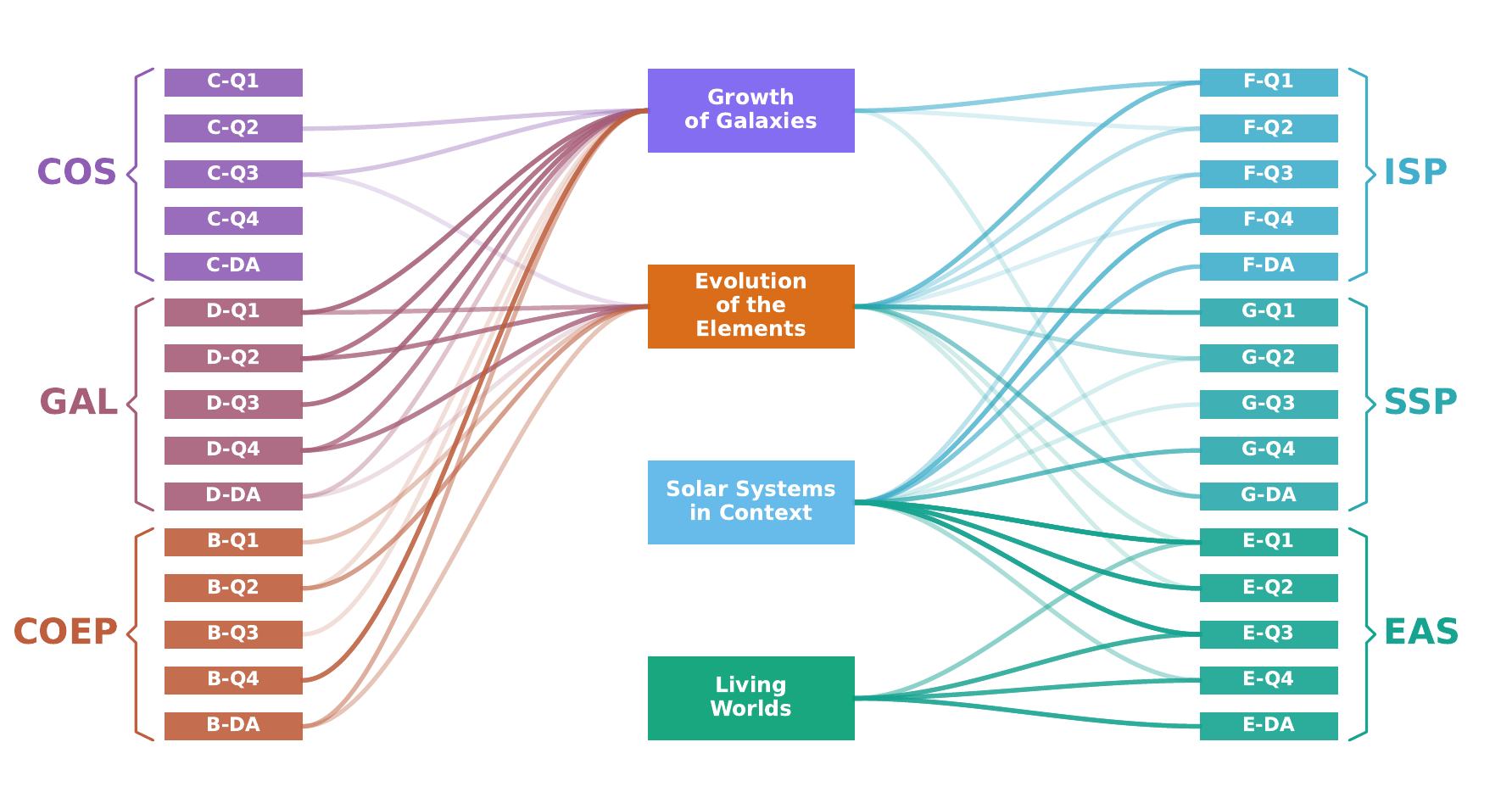}
 \caption{Mapping of the science cases proposed by the Science Working Groups (center) to the questions and discovery areas identified by Astro2020 (left and right). The transparencies of the lines indicate the number of connections with darker lines indicating stronger links between the SWG and the Astro2020 topic. The connections are listed explicitly in Table~\ref{tab:astro2020}.}
 \label{fig:astro2020}
\end{figure*}

Overall, the EOS1 panel concluded that LUVOIR-B could fully address nine and partially address 13 of the 30~science questions or discovery areas identified by the science panels.\footnote{Interestingly, the EOS1 panel also counted 9 fully addressed and 13 partially addressed science questions and discovery areas for HabEx-4H and HabEx-3.2S despite their smaller apertures. Presumably, the coarse classification of science questions as either ``fully'' or ``partially'' addressed masks differences in the degree to which each possible observatory could address the selected issue.} In contrast, Table~\ref{tab:astro2020} and the visual summary in Figure~\ref{fig:astro2020} reveal that the \ncases~science cases presented in this paper at least partially address all of the science questions and discovery areas except C-Q1 (``What set the hot Big Bang in motion?''), C-Q4 (``How will measurements of gravitational waves reshape our cosmological view?'') and C-DA (``The dark ages as a cosmological probe''). Accordingly, undertaking all \ncases~science cases discussed in Section~\ref{sec:scdds} would address 90\% of the science questions and discovery areas identified by Astro2020.

\subsection{Cosmology (COS)}
The panel on Cosmology (COS) focused on the origin of the universe, cosmic expansion, structure formation, and the properties of dark matter. The panel also considered how gravitational wave science could advance knowledge of the universe. While most of the investigations identified by the COS panel require observations at radio or microwave wavelengths \citep[see their Table C.2 in][]{astro2020}, two of the four questions raised by the COS panel overlap with the science cases presented in Section~\ref{sec:scdds}. Specifically, GG-12 (Section~\ref{sssec:dm_power}) and GG-13 (Section~\ref{sssec:dm_lensing}) both address ``...the properties of dark matter...'' (C-Q2). In addition, GG-13 (Section~\ref{sssec:dm_lensing}), GG-18 (Section~\ref{sssec:agn_feedback}), and EE-10 (Section~\ref{sssec:distance_ladder}) consider cosmic expansion (C-Q3). 

\subsection{Galaxies (GAL)}
As shown by Table~\ref{tab:astro2020}, the questions and discovery area identified by the panel on Galaxies (GAL) are extremely well-suited to HWO: all GAL topics are fully or partially addressed by GG or EE science cases. For example, GAL question D-Q1 asks, ``How did the intergalactic medium and the first sources of radiation evolve from cosmic dawn through the epoch of reionization,'' which is considered by no fewer than nine~GG science cases and four~EE science cases. Questions~D-Q2 (``How do gas, metals, and dust flow into, through, and out of galaxies'') and D-Q4 (``How do the histories of galaxies and their dark matter halos shape their observable properties?'') are similarly covered by a mixture of GG and EE science cases while D-Q3 (growth of supermassive black holes) is addressed by nine~GG cases. 

The investigation of AGN outflows and their impact on their host galaxies proposed by GG-5 \citep[Section~\ref{sssec:agn_outflow}][]{zhang_hwo25} addresses all four GAL questions. In addition, three GAL questions are addressed by the studies of black hole spin in GG-2 (Section~\ref{sssec:bh_mass_spin}), reionization in GG-6 (Section~\ref{sssec:resolve_reionization}), Lyman continuum escape in GG-7 (Section~\ref{sssec:lyman_escape}), merging of supermassive black holes in GG-14 (Section~\ref{sssec:smbh_mergers}), AGN feedback in GG-18 (Section~\ref{sssec:agn_feedback}), the first stars in EE-8 (Section~\ref{sssec:first_stars}), and the distribution of elements within galaxies in EE-12 (Section~\ref{sssec:ism_uv}). Finally, D-DA (``Mapping the circumgalactic medium and intergalactic medium in emission'') is covered by EE-12 (Section~\ref{sssec:ism_uv}) along with GG-16 (Section~\ref{sssec:disk_cgm}), which probes the behavior of metals and gases at the interfaces between galactic disks and the circumgalactic medium, and GG-17 (Section~\ref{sssec:cgm_elm}), which studies the exchange of matter and energy within galaxies. 

\subsection{Compact Objects and Energetic Phenomena (COEP)}
The panel on Compact Objects and Energetic Phenomena (COEP) questioned the nature of neutron stars, black holes, jets from compact objects, and explosions. For their discovery area, they selected combined studies of electromagnetic radiation, particles, and gravitational waves. While a facility like HWO would not directly detect gravitational waves, it could support those studies by detecting the electromagnetic counterparts to gravitational waves. 

The four COEP questions and discovery area are well-covered by GG and EE science cases. As shown in Figure~\ref{fig:astro2020}, COEP goals are addressed by nine GG science cases and five EE science cases. The number of related science cases per question ranges from one for B-Q3 (jets of compact objects) to nine for B-Q4 (seeding and growth of supermassive black holes). Questions B-Q1 (``What are the mass and spin distributions of neutron stars and stellar mass black holes?'') and B-Q2 (``What powers the diversity of explosive phenomena across the electromagnetic spectrum?'') have two and five matches, respectively. 

The studies of r-process elements in kilonovae described in EE-4 (Section~\ref{sssec:r_process_el}) and the magnetic properties of massive stars described in EE-15 (Section~\ref{sssec:mag_stars}) are relevant to the stellar mass black holes that are the focus of B-Q1. Along with the investigations of quiescent black holes in GG-4 (Section~\ref{sssec:bh_quiescent}), core-collapse supernovae in EE-3 (Section~\ref{sssec:flash_ccsne}), and the physics of the formation of r-process elements in EE-9 (Section~\ref{sssec:r_process_nature}), science cases EE-4 and EE-15 also address B-Q2. The sole science case that is related to B-Q3 is GG-5 (Section~\ref{sssec:agn_outflow}), which considers the launching of jets in AGN and the resulting impact on star formation. In contrast, there are many connections for B-Q4 because multiple GG science cases are focused on massive black holes and AGN: GG-2 (Section~\ref{sssec:bh_mass_spin}), GG-3 (Section~\ref{sssec:bh_torus}), GG-4 (Section~\ref{sssec:bh_quiescent}), GG-5 (Section~\ref{sssec:agn_outflow}), GG-14 (Section~\ref{sssec:smbh_mergers}), GG-16 (Section~\ref{sssec:disk_cgm}), GG-17 (Section~\ref{sssec:cgm_elm}), GG-18 (Section~\ref{sssec:agn_feedback}), and GG-19 (Section~\ref{sssec:smbh_pol}). 

Finally, as shown in Table~\ref{tab:astro2020}, the COEP discovery area is well addressed by a combination of three GG and two EE science cases. Four of the cases (GG-5, GG-14, GG-19, and EE-4) also address COEP questions and are discussed above. The additional science case is EE-7 (Section~\ref{sssec:wds}), which has the potential to advance knowledge of fundamental physics by looking for deviations from general relativity in high-resolution spectra of white dwarfs. 

\subsection{Interstellar Medium and Star and Planet Formation (ISP)}
The panel on Interstellar Medium and Star and Planet Formation (ISP) identified observational needs spanning the electromagnetic spectrum from radio wavelengths to the UV \citep[Appendix~F][]{astro2020}. The panel proposed a diverse mix of questions that are addressed by a combination of GG, EE, and SSiC science cases. The questions with the strongest overlap with the SCDDs are F-Q1 (``How do star-forming structures arise from, and interact with, the diffuse ISM?'') with ten matches and F-Q4 (``Is planet formation fast or slow?'') with eight~matches. Questions F-Q2 (``What regulates the structures and motions within molecular clouds?'') and F-Q3 (``How does gas flow from parsec scales down to protostars and disks?'') have three and four matches, respectively. The discovery area F-DA (``Detecting and Characterizing Forming Planets'') has five matches. 

The ISP questions are ordered from large to small spatial scales, and that pattern is reflected in Table~\ref{tab:astro2020} and Figure~\ref{fig:astro2020}, which show that GG science cases map only to F-Q1 and F-Q2. In contrast, SSiC science cases map only to F-Q3, F-Q4, and F-DA. EE science cases map to all four questions but not to the discovery area. 

\subsection{Stars, the Sun, and Stellar Populations (SSP)}
Next, the panel on Stars, the Sun, and Stellar Populations (SSP) raised questions about the most extreme stars (G-Q1), the effects of stellar multiplicity on stellar evolution (G-Q2), how other stars would appear if they could be studied as well as the Sun (G-Q3), and the formation of space weather (G-Q4). The SSP discovery area G-DA of ``industrial scale spectroscopy'' refers to measuring the spectra of $10^9$ stars in and beyond the Milky Way. The SSP questions and discovery area are addressed by a combination of GG, EE, and SSiC science cases. Unlike ISP, which showed a smooth transition from GG through EE to SSiC, three of the four SSP questions map exclusively to one SWG: G-Q1 is covered by EE while G-Q3 and G-Q4 are addressed by SSiC. The intermediate question G-Q2 is relevant to the studies of low-metallicity massive stars, stellar magnetism, and the role of magnetic fields on young planetary systems proposed in EE-1 (Section~\ref{sssec:massive_stars_lowZ}), EE-15 (Section~\ref{sssec:mag_stars}), and SSiC-34 (Section~\ref{sssec:young_bfields}), respectively. SSiC-34 is also the only science case that addresses G-Q3 regarding obtaining solar-like data for other stars. 

For extreme stars and stellar populations (G-Q1), several EE science cases will contribute to advancing knowledge of specific populations such as extremely massive stars through EE-1 (Section~\ref{sssec:massive_stars_lowZ}) and EE-5 (Section~\ref{sssec:vms}); polluted white dwarfs through EE-7 (Section~\ref{sssec:wds}); the earliest stars through EE-8 (Section~\ref{sssec:first_stars}) and EE-9 (Section~\ref{sssec:r_process_nature}); distant Cepheid variable stars through EE-10 (Section~\ref{sssec:distance_ladder}); and magnetic stars through EE-15 (Section~\ref{sssec:mag_stars}). By examining a large sample of stars in dwarf galaxy IZw18 (Section~\ref{sssec:massive_stars_lowZ}), EE science will also enable comparative studies of single stars, binary stars, and those in higher multiplicity systems (G-Q2). 

Questions G-Q3 and G-Q4 also require time-domain observations of the Sun, but a facility like HWO could contribute to answering those questions by obtaining spectra, time-resolved photometry, and even time-series spectroscopy of other stars through observations like those proposed in SSiC-14 (Section~\ref{sssec:escape}) and SSiC-24 (Section~\ref{sssec:highres_atmos}), SSiC-26 (Section~\ref{sssec:bfield}), and SSiC-34 (Section~\ref{sssec:young_bfields}). SSiC-25 (Section~\ref{sssec:aurorae}) and SSiC-29 (Section~\ref{sssec:solargiant}) would also help address G-Q4 by observing aurorae on solar system giant planets. Furthermore, understanding the atmospheric properties and history of planets for various SSiC and LW science goals will require considering how the irradiation from the host star has affected the planet, thereby indirectly supporting SSP investigations of space weather (G-Q4). Lastly, for G-DA, the most relevant science cases are the large spectroscopic programs described in GG-11 (Section~\ref{sssec:ionizing_lf}), EE-1 (Section~\ref{sssec:massive_stars_lowZ}),EE-2 (Section~\ref{sssec:dust_extinction}), EE-7 (Section~\ref{sssec:wds}), and EE-8 (Section~\ref{sssec:first_stars}), but stellar spectra obtained during other spatially-resolved GG observations of galaxies or in parallel mode during SSiC and LW observations could also be considered. 

\subsection{Exoplanets, Astrobiology, and the Solar System (EAS)}
The science topics raised by the Exoplanets, Astrobiology, and the Solar System (EAS) area also overlap fully with the HWO science cases in this document. In general, questions focused on demographics, system architectures, and atmospheric physics are addressed by SSiC science cases while those focused on habitability and biosignatures are addressed by LW science cases. Table~\ref{tab:astro2020} and Figure~\ref{fig:astro2020} reveal a shift in the relative mix of SSiC and LW science cases through the EAS category. 

E-Q1 is focused on demographics and is addressed nearly entirely by SSiC (20 science cases) with a few additions from LW (three science cases) and even EE (one science case). E-Q2 zooms in on the properties of individual planets, which nearly triples the number of LW matches compared to E-Q1. In total, one~EE science case, 20~SSiC science cases, and eight~LW science cases are relevant to the questions about planet properties and diversity posed by E-Q2. For both E-Q1 and E-Q2, the EE match is EE-7 (Section~\ref{sssec:wds}), which would obtain spectra of polluted white dwarfs to investigate the compositions of accreted planets.

Next, E-Q3 focuses on planetary habitability and is addressed by LW (eight science cases) and SSiC (14 science cases). E-Q4 focuses on detecting and interpreting biosignatures, and is therefore dominated by LW science cases (eight matches) rather than SSiC (two matches). Finally, the EAS discovery area E-DA (``The Search for Life on Exoplanets'') matches perfectly with the ten science cases identified by LW.

\section{The Need for HWO}
\label{sec:need_hwo}
In this section, we compare the observational capabilities needed to address the science cases presented in Section~\ref{sec:scdds} to the capabilities of facilities that are anticipated to be operational in the 2030s and 2040s to determine which science questions are best addressed with HWO. The classification of ``best addressed'' includes those science questions that would be uniquely addressed by HWO as well as those for which HWO would have a significant advantage over other facilities in terms of sensitivity, sample size, efficiency, or other observational parameters. 

\begin{figure*}
 \centering
 \includegraphics[width=1\linewidth]{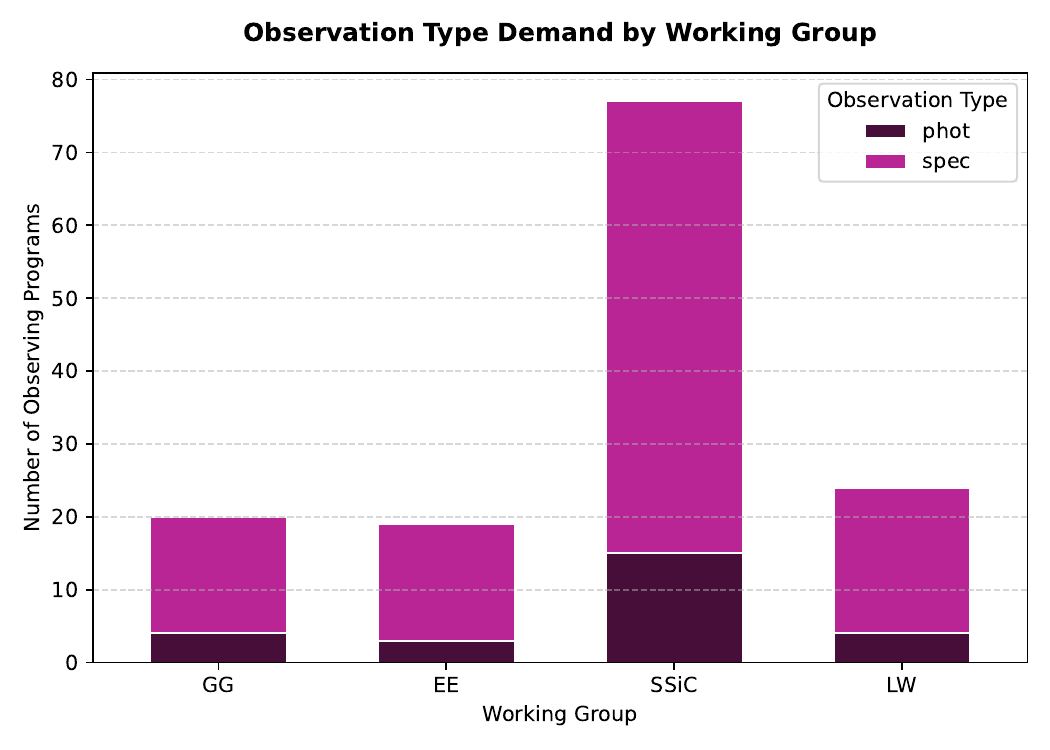}
 \caption{Stacked histogram showing the demand for photometric observations (plum) and spectroscopic observations (magenta) for science cases developed by each Science Working Group (GG = Growth of Galaxies, EE = Evolution of the Elements, SSiC = Solar Systems in Context, LW = Living Worlds).}
 \label{fig:obstype}
\end{figure*}

As depicted in Figure~\ref{fig:obstype}, spectroscopic observations significantly outnumber photometric observations for all SWGs. Evolution of the Elements demands the largest percentage of spectroscopic observations (84\%), while Growth of Galaxies needs the lowest (80\%). These high values are nearly identical, illustrating the necessity of spectroscopic observations for addressing the science questions posed in Section~\ref{sec:scdds}.

\begin{figure*}
 \centering
 \includegraphics[width=1\linewidth]{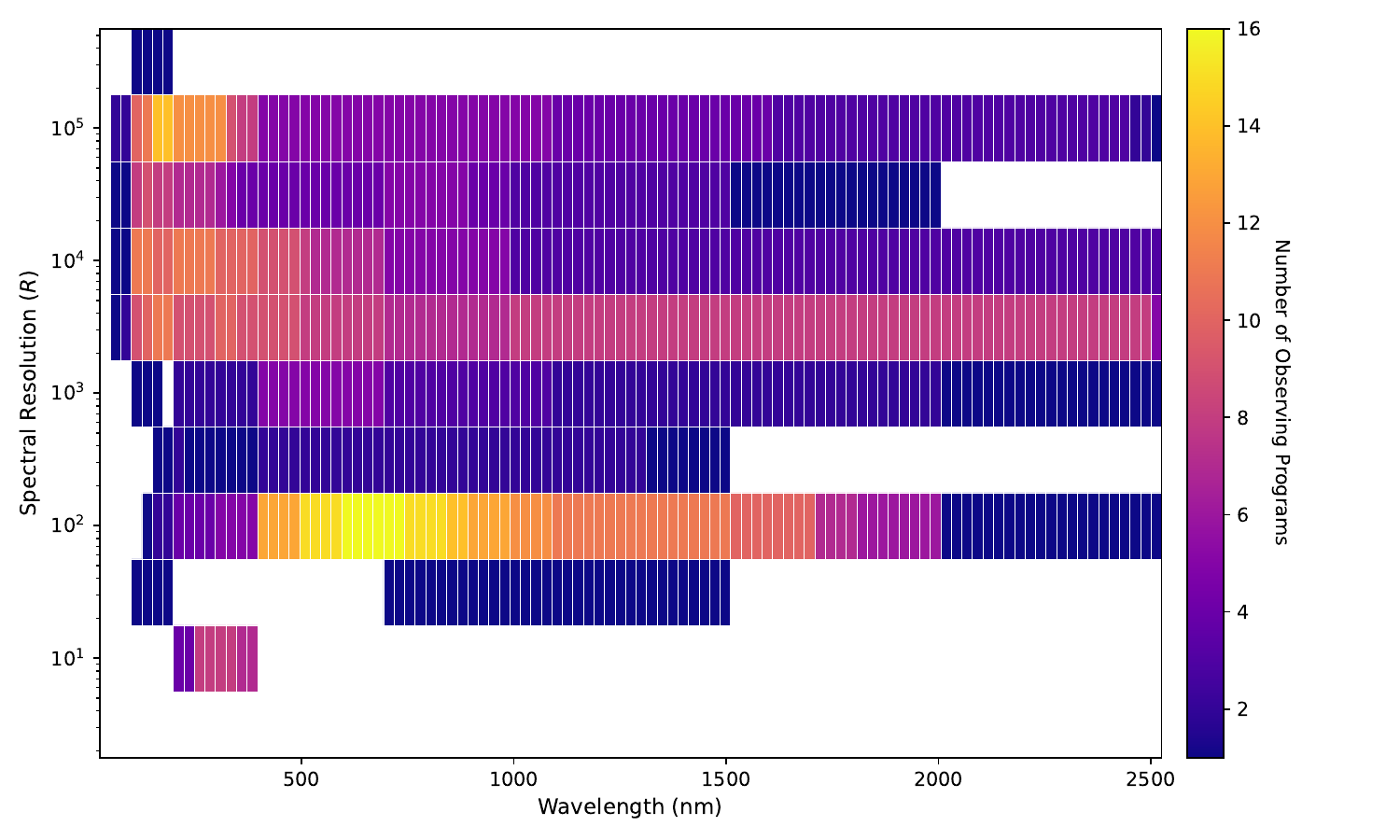}
 \caption{Demand for spectroscopic observations at particular spectral resolutions and wavelengths. As shown by the colorbar, the cell shading indicates the number of observing programs that require each combination of parameters. White regions indicate regions of parameter space that are not required for the science cases presented in this paper.}
 \label{fig:all_r_wave}
\end{figure*}

In Figure~\ref{fig:all_r_wave}, we display the wavelength ranges and spectral resolutions requested for all spectroscopic observations. This figure is effectively the sum of the individual plots for each SWG shown in Figures~\ref{fig:gg_r_wave}, \ref{fig:ee_r_wave}, \ref{fig:ssic_r_wave}, and~\ref{fig:lw_r_wave}. Each observing program has the same weight so the combined figure more heavily reflects the demands of the Solar Systems in Context SWG (62~spectroscopic observing programs) than those of Growth of Galaxies (16~programs), Evolution of the Elements (16~programs), or Living Worlds (20~programs). The pile-up of requests at $R=100$ is due to the extensive use of Vis/NIR high-contrast spectroscopy ($400-1100$~nm at $R=140$; $1100-1700$~nm at $R=70$) by science cases developed by the Solar Systems in Context and Living Worlds SWGs. 

In the UV, the small cluster at low spectral resolution ($R=10$) is caused by the overlap of SSiC and LW observing programs requesting UV high-contrast spectroscopy at $R=7$ at either $200-400$~nm or $250-400$~nm. In contrast, the UV spike at moderate spectral resolution ($R=2000-20,000$) is due to a combination of science cases proposed by GG, EE, and SSiC. At even higher spectral resolution ($R\gtrsim100,000$), the majority of requests for UV spectroscopy are from EE but there are also contributions from GG and SSiC. Those three SWGs (GG, EE, and SSiC) are also responsible for the stripe of increased demand for moderate spectral resolution ($R\sim5000$) NIR spectroscopy.

Fundamentally, the main capabilities needed to address the science cases presented in Section~\ref{sec:scdds} are high sensitivity, broad wavelength coverage from the UV to the NIR, high spatial resolution, and extreme starlight suppression. High sensitivity could be achieved via selecting large mirrors to maximize collecting area and strategically designing optical paths and coatings to maintain throughput. While next-generation, extremely-large ground based facilities will have substantially larger collecting areas than a space-based observatory like HWO, those facilities will be forced to contend with the Earth's atmosphere. Observations with ELTs will be restricted to wavelength ranges without significant telluric absorption, thereby preventing observations at the UV wavelengths needed for numerous science cases, especially those related to reionization. 

In addition, although the theoretical diffraction limit of a telescope scales with the inverse of the aperture, ground-based ELTs will require sophisticated adaptive and active optics to mitigate the effects of atmospheric turbulence and enable observations at spatial resolution comparable to the diffraction limit. Single conjugate adaptive optics systems are generally limited to small fields of view, which means that an ELT equipped with a ``standard'' AO system would not be able to obtain diffraction-limited imaging or spatially resolved spectra of extended sources such as nearby galaxies \citep{rigaut+neichel2018}. Work is ongoing to design AO systems capable of correcting wavefront distortions over larger angular scales \citep{chun_et_al2016, rigaut_et_al2014}, but the imagers, IFSs, and MOSs under design for ELTs have fields of view much smaller than that expected for a facility like HWO, implying that many mosaicked observations would be needed to map important sources such as nearby, low-redshift galaxies and the disks of solar system objects. 

A large, space-based observatory capable of accomplishing the investigations outlined in Section~\ref{sec:scdds} would have significantly higher angular resolution and sensitivity than HST, UVEx \citep{kulkarni_et_al2021}, or CASTOR \citep{cote_et_al2019}. For reference, the 2.4-m diameter of HST corresponds to a diffraction-limit of 10.5~mas at 100~nm while telescopes with apertures of $6 - 10$~m would have diffraction limits of $4.2 - 2.5$~mas at the same wavelength. Considering only the increase in collecting area and neglecting improvements in optical path and coatings, telescopes with diameters of $6 - 10$~m would also have $6-17$ times the collecting power of HST, thereby opening the door to studying more distant, fainter galaxies at the same fidelity as the closest galaxies.

For SSiC and LW science, the enhanced stability of a next-generation space-based observatory is as important as the larger collecting area and smaller diffraction limit. Searching for life on exoplanets (e.g., Section~\ref{ssec:xswg_biosig}) requires suppressing the light of the host star by many orders of magnitude to reveal planets up to $10^{11}$ times fainter than their host stars. Furthermore, detecting planets in the HZ necessitates a telescope aperture large enough that the detectors receives sufficient light from planets receiving Earth-like levels of irradiation and that the planets are spatially resolved from their host star. 

While ground-based ELTs will have tiny diffraction limits due to their large apertures, the additional disturbances present in a ground-based observing environment will almost certainly prevent reaching the extreme $10^{-11}$ contrast limit necessary to detect and characterize Earth-like planets in the HZs of Sun-like stars. A space-based, large-aperture, extremely stable facility is needed for direct imaging and spectroscopy of potential Earth analogs orbiting Sun-like stars. In contrast, the very large apertures and correspondingly small diffraction limits of the ELTs will be better suited to directly imaging potentially Earth-like planets orbiting M~dwarfs due to the low luminosities of M dwarfs and therefore the small angular separations between potentially habitable M~dwarf planets and their host stars \citep[e.g.,][]{guyon_et_al2012}. 

Rather than reiterate the many science cases that would benefit from a facility like HWO, it is more straightforward to highlight a few science cases that could more easily be investigated with other facilities. First, GG-14 (Section~\ref{sssec:smbh_mergers}) demands imaging of strong lenses to constrain the merger timescales of supermassive black holes. Ideally, imaging would be conducted during the $550 - 900$~nm wavelength range of Euclid VIS's imager, so UV data are not needed. The necessary spatial resolution ($20$~mas) is within reach of ELTs, but advanced adaptive optics would be needed to ensure that deviations introduced by atmospheric turbulence are not mistaken for astrophysical perturbations generated by black holes. 

Second, EE-10 \citep[][Section~\ref{sssec:distance_ladder}]{anand_hwo25} needs visible and NIR photometry of Cepheid variable stars, tip of the red giant branch stars, and J-region asymptotic giant branch stars in Cepheid host galaxies. The data should have a pixel scale of 20~mas/pixel, an absolute flux calibration uncertainty of $0.5\%$, and cover a broad $6' \times 6'$ field of view. While such observations could potentially be obtained by an ELT using an idealized adaptive optics system, extensive mosaicking would be needed to cover the full $6' \times 6'$ target region using the narrower field-of-view of an AO-equipped ELT imager. 

In addition, EE-10 demands the same extreme image stability as GG-14 as well as the added difficulty that the observations are best obtained at visible wavelengths while current AO systems work best in the NIR. Conducting the observations needed for EE-10 using an ELT would therefore necessitate a highly stable, visible-light AO system and a larger investment of telescope time than would be needed using a space-based imager with a larger field of view. Third, if EE-10 can be conducted using ELTs, the same would likely be true for GG-12 \citep[][Section~\ref{sssec:dm_power}]{doppel_hwo25}, which aims to detect ultra-faint dwarf galaxies by obtaining $4' \times 3'$ Vis/NIR imaging of Milky-way-mass galaxies at a spatial resolution of 20~mas. 

Fourth, SSiC-12 \citep[][Section~\ref{sssec:giant_orbits}]{sagynbayeva_et_al2025} necessitates precise characterization of the orbits of giant planets. Ground-based radial velocity observations alone will not break the degeneracy between planet mass and orbital inclination, so additional astrometric or imaging data will be needed to fully describe planetary orbits. Giant planets at the desired semimajor axis range of $5 - 30$~AU will be at large angular separations, especially for nearby stars, so ground-based facilities may be capable of reaching the necessary contrast of $10^{-9} - 10^{-10}$ at separations of $0.1'' - 1''$. Indeed, depending on the selected outer working angle of the coronagraphic imager, giant planets at $5 - 30$~AU may be too far from their host stars to be detected by space-based observatories at most orbital phases and system orientations. For the same reasons, the giant planet observations needed for SSiC-2 (Section~\ref{sssec:hab_system}) may also be better obtained by ELTs than by space-based observatories. 

Other science cases such as GG-5 \citep[][Section~\ref{sssec:agn_feedback}]{zhang_hwo25}, EE-2 \citep[][Section~\ref{sssec:dust_extinction}]{paladini_hwo25}, and EE-6 \citep[][Section~\ref{sssec:resolved_stellar_pops}]{smercina_hwo25} are potentially well-suited to space-ground collaborations in which UV data are obtained from space and Vis/NIR observations are obtained from the ground. For instance, for the study of AGN feedback proposed by GG-5, space-based UV/Vis IFS observations of 2000~galaxies could be supplemented by ground-based NIR IFS data of the same targets. Likewise, the broad wavelength range ($100-2500$~nm) needed for the investigation of galactic and extragalactic dust described in EE-2 could be completed by obtaining UV spectroscopy from space and Vis/NIR spectra from the ground. In addition, the multiwavelength photometric dataset needed to investigate the morphologies of galaxies for EE-6 could consist of space-based UV images and ground-based Vis/NIR images. 

Similarly, the solar system observations of icy bodies \citep[][SSiC-1; Section~\ref{sssec:oceans_habitable}]{cartwright_hwo25}, asteroids (SSiC-22; Section~\ref{sssec:mars}), Venus \citep[][SSiC-5; Section~\ref{sssec:venus}]{izenberg_hwo25}, Mars (SSiC-28; Section~\ref{sssec:mars_post}), Titan (SSiC-27; Section~\ref{sssec:titan}), and giant planets \citep[][SSiC-29; Section~\ref{sssec:solargiant}]{fletcher_hwo25} could be accomplished by supplementing space-based UV observations with Vis/NIR data obtained by ground-based facilities. Depending on instrument performance, ELTs may also be an advantageous choice for obtaining the Vis/NIR portions of the multiwavelength datasets needed for studies of protoplanetary disks, debris disks, and exozodi as part of SSiC-8 \citep[][Section~\ref{sssec:proto}]{ren_hwo25}, SSiC-9 \citep[][Section~\ref{sssec:debris}]{rebollido_hwo25}, and SSiC-19 \citep[][Section~\ref{sssec:exozodi}]{debes_hwo25}, respectively. Finally, the planet mass measurements needed for many SSiC and LW science cases could be accomplished by combining the astrometric observations described in LW-12 \citep[][Section~\ref{sssec:mp}]{gary_hwo25} with ground-based radial velocity observations for some targets. 

The vast majority (93\%) of the science cases presented in Section~\ref{sec:scdds} require observations at UV wavelengths, starlight suppression at contrast levels $\leq 10^{-10}$, or both. While many of those science cases could be enhanced by precursor observations to inform target selection, complementary multi-wavelength observations by other facilities \citep[e.g.,][]{andersen_hwo25_vol2, saracco_hwo25_vol2}, and/or detailed follow-up observations over small regions, those 62~science cases could not be sufficiently addressed at the breakthrough scientific return level without a facility like HWO. The science cases that need neither UV observations nor extreme starlight suppression are GG-12, GG-14, EE-10, SSiC-30, and LW-12. For the planetary defense observations proposed by SSiC-30 \citep[][Section~\ref{sssec:defense}]{dotson_hwo25}, the likely location of a future flagship at L2 could provide the crucial capability to detect potential impactors that could not be detected from Earth, but realizing this advantage may place demands on the instantaneous field of regard and the Sun avoidance angle as discussed in Section~\ref{sssec:field}. In the case of the astrometric planet mass and orbit determinations described in LW-12 \citep[][Section~\ref{sssec:mp}]{gary_hwo25}, robustly identifying and classifying terrestrial planets necessitates an extremely stable platform and exquisite astrometric precision. For the study of the merger timescales of supermassive black holes described in GG-14 (Section~\ref{sssec:smbh_mergers}), the motivation for requesting space-based observations is that ground-based facilities may not be able to achieve the PSF stability needed to robustly measure merger rates, especially if the timescale is fast. However, the portal version of the SCDD\footnote{\url{https://docs.google.com/document/d/14ZGS_XTkN8vtiaMKtJylhcOS-2EsEL0q6AuBMRA-o_A/edit?tab=t.0}} notes that additional work is needed to quantify the necessary PSF stability and therefore more fully justify the need for space-based observations with a facility like HWO. In the case of the survey of ultra-faint dwarf galaxies proposed by GG-12 (Section~\ref{sssec:dm_power}), \citet{doppel_hwo25} argues that space-based observations are needed due to the faintness of the targets and the need to resolve individual stars at visible wavelengths. However, further work is needed to assess the viability of completing GG-12 using a wide-field imager on an ELT equipped with a visible-light AO system. Finally, EE-10 \citep[][Section~\ref{sssec:distance_ladder}]{anand_hwo25} requests observations from an HWO-like facility because refining the distance ladder demands resolving and precisely measuring the brightnesses of individual stars in galaxies at distances up to 100~Mpc. \citet{anand_hwo25} states that current AO systems are not sufficient for the proposed observations due to their reduced precision at bluer wavelengths but the justification for space-based observations for EE-10 would be strengthened by additional exploration of the potential capabilities of the next generation of visible-light AO systems for ELTs.

\section{Preparing for Future Missions}
\label{sec:precursor}
The scientific return of a given facility depends not just on the raw capabilities of the facility but also on how the facility is used. Theoretical analyses, computational simulations, laboratory experiments, and observational programs with other facilities can all enrich the scientific return of future facilities by informing design trades and operational strategies. Work that must be done early in a project life-cycle in order to determine the facility architecture is deemed ``precursor science'' while pre-observation projects that inform operations rather than design are designated as ``preparatory science.'' Although by definition, preparatory science does not influence mission architecture, it can include observations that must be collected over long timescales and must therefore begin many years (or even decades) before a new facility begins operations. Science return can also be enhanced
by conducting coordinated observations by other facilities at exactly the same time (i.e., ``simultaneous science'') or nearly the same time (i.e., ``contemporaneous science''). 

Identifying necessary precursor and preparatory observations is vital to mission success because that early work establishes the foundation on which future science will be based. Maximizing scientific return also demands early assessment of the urgency of precursor and preparatory science needs so that crucial information is available in time to support critical design and operational decisions. Advance planning is particularly essential for precursor and preparatory investigations requiring observations over a long baseline because it is not possible to compensate for initial delays by devoting more resources later in the project; rather, the completion date would be extended by the same duration as the initial delay. 

In this section, we present two examples of investigations that would enrich the scientific return of an observatory like HWO. Both of these examples are related to exoplanet science, but precursor and preparatory studies would be useful for the full range of potential HWO observing programs presented in Section~\ref{sec:scdds}. First, in Section~\ref{ssec:tss}, we present an analysis by the Target Stars and Systems subgroup of precursor science needed to support the search for life on other planets. 
Second, in Section~\ref{sssec:geochemical}, we motivate preparatory laboratory investigations that would provide insight into which stars might host geochemically habitable planets. 

\subsection{Precursor and Preparatory Observations of Planetary Systems and Planet Host Stars}
\label{ssec:tss}

As mentioned in Section~\ref{sec:wg}, the Living Worlds SWG included the Target Stars \& Systems (TSS) subgroup. TSS was tasked with considering current knowledge of stars that might be potentially targeted for biosignature searches (i.e., ``target stars'') and with assessing what additional information should be learned about those stars in order to maximize the scientific return of future observations by a facility like HWO. As reflected by the final word in the subgroup name (i.e., ``systems''), TSS was also responsible for investigating any stellar companions, substellar companions, and circumstellar matter (i.e. dust disks) that might be associated with the target stars. The purpose of TSS was unique within the SWGs because the onus was not to develop observing programs that could be conducted with HWO but rather to support other SSiC and LW SCDDs by identifying and compiling information about target stars and systems.

Specifically, the objectives of TSS were to:
\begin{itemize}
\itemsep0em
\item Assemble and maintain a table of potential target stars and properties to inform science, architecture, and yield studies;
\item Assess which stars and systems should be observed by other facilities as part of preparatory, precursor, and/or contemporaneous studies to support SSiC and LW science;
\item Determine which target star properties are most important to target selection and the interpretation of exoplanet spectra, and when those properties should be measured; and
\item Interface with other START/TAG subgroups that might need information related to target stars and systems.
\end{itemize}

In order to accomplish these objectives, the TSS leads established task groups and solicited feedback from other HWO WG members as well as the broader scientific community (see Appendix~\ref{app:tss} for additional details). A key outcome from the work of the TSS subgroup was the development of a more extensive list of potential target stars. The astrobiological science cases presented in Sections~\ref{ssec:ssic_scdds} and~\ref{ssec:lw_scdds} demand observing certain numbers of potentially inhabited planets, but those planets must first be detected. Any work that can be done to accelerate the process of identifying potentially habitable and potentially inhabited planets would improve mission efficiency by reducing the amount of time needed to discover planets prior to characterization. The magnitude of this efficiency gain depends on the full science portfolio of the observatory, the ability to use exo-Earth search observations for other SSiC investigations such as the studies of giant planets proposed by SSiC-12 \citep[][Section~\ref{sssec:giant_orbits}]{sagynbayeva_et_al2025} and SSiC-15 \citep[Section~\ref{sssec:reflected_giants}][]{min_hwo25}, and the viability of conducting parallel observations using other instruments (e.g., long stares for GG science cases) while looking for exo-Earths using the high-contrast instrument. 

That caveat aside, exo-Earth science cases would benefit significantly from the ability to strategically select potential targets. Ideally, all HWO exo-Earth survey targets should be characterized as thoroughly as possible prior to mission start. To that end, \citet{mamajek+stapelfeldt2024} presented an initial catalog of 164 potential targets that became known as the ``ExEP Mission Star List'' (EMSL). \citet{harada_et_al2024a} expanded upon the EMSL by collecting and incorporating additional stellar parameters. These supplemental parameters include stellar abundances because, as discussed in Section~\ref{sssec:geochemical}, the chemical compositions of host stars are related to compositions of the exoplanets that orbit them. For example, Mg, Si, and Fe in the star are directly correlated with the make-up of the planet \citep{hinkel_et_al2018, schulze_et_al2021, unterborn_et_al2023, schulze_et_al2026}, which in turn impacts the determination of mass, age, and interior planetary processes that could impact interpretation of possible biosignatures. Thus, measuring the host star's specific elemental abundances (e.g., more than bulk [M/H] or [Fe/H]) is of critical importance to the characterization of the planet \citep{hinkel_et_al2024}.

While the addition of more stellar properties enables more comprehensive target selection, the number of stars included in the EMSL is too small to fully investigate the impact of various observatory architectures on the potential exo-Earth yield. To expand the sample of potential targets, \citet{tuchow_et_al2024} published the Habitable Worlds Observatory Preliminary Input Catalog (HPIC), which contains roughly 80~times as many targets as the EMSL. The large, 13,000-star size of the HPIC is advantageous for HWO yield simulations, but the HPIC is much less detailed than the supplemented EMSL \citep{harada_et_al2024a}. Thoroughly assessing the impact of potential mission design choices demands a catalog that combines the breadth of HPIC \citep{tuchow_et_al2024} with the depth of the EMSL \citep{mamajek+stapelfeldt2024, harada_et_al2024a}. 

Accordingly, TSS developed and released the HWO Target Stars and Systems 2025 (TSS25) list which classifies targets into three tiers based on the likelihood that they would be observed by a facility like HWO \citep{tuchow_et_al2025}. The 164~targets classified as most compelling were assigned to Tier 1 while 12,285 significantly less attractive targets were placed in Tier~3. In the middle, 495~Tier~2 targets might be suitable targets depending on architecture decisions. We summarize the information that should be collected about TSS25 targets in Table~\ref{tab:tss}. 

\input{properties_tss}

The TSS subgroup recognizes that obtaining the information shown in Table~\ref{tab:tss} would necessitate a large amount of effort by the astronomical community. Additionally, they acknowledge that substantial work will be done to investigate stars that do not end up being selected as exo-Earth target stars for a facility like HWO. However, given that several of the items in Table~\ref{tab:tss} necessitate obtaining data over decades, it would be beneficial to the success of future missions like HWO to observe as many viable targets as possible. Moreover, all of the observations collected as part of this effort would contribute to advancing the overall understanding of stellar and planetary systems. 

\subsection{Preparatory Investigations of Geochemical Habitability (SCDD-LW-9)}
\label{sssec:geochemical}
\sleads{Kara Brugman, Christy B. Till}

As discussed in Sections~\ref{ssec:ssic_scdds} and~\ref{ssec:lw_scdds}, multiple SSiC and LW science cases are dedicated to investigating habitable planets. Accordingly, the scientific return of an HWO-like mission would be enhanced if it were possible to increase the rate at which such systems are observed. A possible shortcut is to ascertain in advance which stars are most likely to host habitable planets. In LW-9, \citet{brugman_hwo25} describes how insight into the likelihood of geochemically habitable planets could be obtained via a preparatory program that includes laboratory experiments. Currently, the effect of rocky, silicate exoplanet compositional variability on exoplanet habitability, planetary evolution model outcomes, and observation target selection is largely unknown and unexplored. 

However, by conducting geochemical experiments, \citet{brugman_et_al2021} found that planet compositions that deviate only slightly from Earth-like ratios of Mg/Si or Ca/Al could lead to conditions that are detrimental to long-term habitability. For instance, the cycling of bioessential elements could be suppressed and the surface of the planet could be covered by a persistent magma ocean. Obtaining these results required empirical investigations because the findings cannot be predicted by existing geochemical/thermodynamic models. Additional empirical studies are necessary to understand the extent to which rocky silicate planet compositions can vary from that of Earth while maintaining both geological characteristics necessary for long-term habitability and geochemical relationships that we are able to interpret. 

The necessary empirical studies are complex and demand a significant investment of time and resources. Accordingly, a coordinated, multi-lab effort would be needed to return and synthesize results in a timely manner. The anticipated experimental results would provide critical information to assess which stars are likely to be the best targets for observations of habitable worlds. Additionally, the laboratory studies would enable extending existing geochemical and planet-atmosphere evolution models beyond solar system-like rocky silicate planet compositions. Finally, the results would help identify future experiments and astronomical observations to further improve our understanding of exoplanet geochemistry and habitability.

\section{Conclusions and Future Work}
\label{sec:conc}
The Science, Technology, Architecture Review Team for the Habitable Worlds Observatory mission concept engaged the global scientific community in forming working groups and developing science cases. The science working groups were divided into Growth of Galaxies, Evolution of the Elements, Solar Systems in Context, and Living Worlds, each of which had multiple subgroups or task forces devoted to different topics. In total, working group members developed over \ncases~science cases covering a broad portfolio of astrophysics and planetary science. In Section~\ref{ssec:scdd_review}, we briefly review the observational needs resulting from those science cases. We then summarize our mapping from science cases to potential observational capabilities in Section~\ref{ssec:bridge} before discussing next steps in Section~\ref{ssec:next}.

\subsection{Science Case Summary}
\label{ssec:scdd_review}
Access to wavelengths shorter than the Lyman limit ($<91.2$~nm) is essential for many science cases, particularly those related to reionization and Lyman continuum escape such as GG-6 \citep[][Section~\ref{sssec:resolve_reionization}]{xu_reionization_hwo25} or the properties and dusty tori of black holes like GG-2 \citep[][Section~\ref{sssec:bh_mass_spin}]{cann_hwo25} and GG-3 \citep[][Section~\ref{sssec:bh_torus}]{gorjian_hwo25}. Additionally, NIR coverage to 2~microns (or even beyond) would be advantageous for a broad swath of targets from the properties of dust in the Milky Way and nearby galaxies for EE-2 \citep[][Section~\ref{sssec:dust_extinction}]{paladini_hwo25} to the influence of AGN feedback on their host galaxies \citep[][Section~\ref{sssec:agn_outflow}]{zhang_hwo25}.

Completing the investigations described in Section~\ref{sec:scdds} would require photometry, single-object spectroscopy, and multi-object spectroscopy across a broad UV/Vis/NIR wavelength range. Furthermore, spectropolarimetry would be needed for several programs including investigations of the magnetism of massive stars \citep[][EE-15; Section~\ref{sssec:mag_stars}]{david-uraz_hwo25}, the mergers of black hole binaries \citep[][GG-19; Section~\ref{sssec:smbh_pol}]{marin_et_al2025}, and the magnetic fields of exoplanets \citep[][SSiC-26; Section~\ref{sssec:bfield}]{strugarek_hwo25}. Other observations would be obtained at high contrast like the investigations of the prevalence of prebiotic environments and surface biosignatures described by LW-4 \citep[][Section~\ref{sssec:prebiosignatures}]{ranjan_prebiosignatures_hwo25} and LW-14 \citep[][Section~\ref{sssec:surface_bio}]{parenteau_surface_hwo25}, respectively. As summarized by \citet{karalidi_hwo25_vol2}, other programs necessitate high-contrast spectropolarimetry and photopolarimetry such as constraining the presence of photosynthetic life on terrestrial planets for LW-7 \citep[][Section~\ref{sssec:polbio}]{berdyugina_imaging_hwo25}. 

While many observations of spatially-resolved sources will be best-served by an IFS like that described in Section~\ref{sssec:ifs}, other observations may be more efficiently conducted using a spectrograph with multiple shutters or slits as discussed in Section~\ref{sssec:mos}. For instance, the studies of quiescent black holes proposed in GG-4 \citep[][Section~\ref{sssec:bh_quiescent}]{pacucci_hwo25} and very massive stars proposed in EE-5 \citep[][Section~\ref{sssec:vms}]{martins_et_al2025} call for IFS observations while those of the evolution of the ionizing photon luminosity function in GG-11 \citep[][Section~\ref{sssec:ionizing_lf}]{mccandliss_hwo25} and the composition of extragalactic dust for EE-11 \citep[][Section~\ref{sssec:dust_uv}]{roman-duval_hwo25} favor MOS observations. 

Still other science cases such as the exploring nature of the first stars for EE-8 \citep[][Section~\ref{sssec:first_stars}]{roederer_nature_hwo25} and probing fundamental physics and exoplanet compositions via white dwarf observations for EE-7 \citep[][Section~\ref{sssec:wds}]{xu_siyi_hwo25} necessitate single object spectroscopy as discussed in Section~\ref{sssec:sos}. The variety of spectroscopic programs described in Section~\ref{sec:scdds} motivates providing multiple ways to obtain spectra (and spectropolarimetry) for resolved and unresolved sources. 

Direct imaging and spectroscopy of exoplanets (i.e., most SSiC and LW science cases) necessitates advanced starlight suppression technology to observe planets $10^{-11}$ times fainter than their host stars at close angular separations. In addition, studying the effects of stellar multiplicity on habitability for SSiC-6 \citep[][Section~\ref{sssec:occ_binary}]{newton_hwo25} needs the ability to directly observe planets orbiting stars with stellar companions as close as $3''$. Direct imaging and spectroscopy of exoplanets also demands high photometric stability on timescales ranging from hours to months to search for variations in planetary surface and atmospheric properties and detect potentially subtle indications of life. Access to planets at multiple orbital phases is essential for certain science cases such as detecting oceans via glint for SSiC-4 \citep[][Section~\ref{sssec:surface_water}]{lustig-yaeger_liquidwater_hwo25} and quantifying seasonal variations for LW-13 \citep[][Section~\ref{sssec:seasonality}]{lafleche_hwo25}, which consequently favors a large instantaneous field of regard.

For nearly all science cases, the number of feasible targets and the information content of the resulting data scale with increasing light collection rate and improved angular resolution, which generates a strong push towards larger apertures. A large aperture would improve observational efficiency by reducing the time needed to reach a given survey depth and increase access to the faintest galaxies and stars that are essential targets for investigations such as investigating the nature of dark matter via GG-12 \citep[][Section~\ref{sssec:dm_power}]{doppel_hwo25} and GG-13 \citep[][Section~\ref{sssec:dm_lensing}]{he_hwo25}. For biosignature and technosignature searches like those proposed in LW-1 and LW-5, respectively, an increased photon collection rate would be extremely beneficial for reducing the total integration time needed to detect subtle features. An enhanced photon rate would also enable higher temporal sampling, which would be essential for increasing sensitivity to features that vary with time or longitude such as storm systems for SSiC-15 \citep[][Section~\ref{sssec:reflected_giants}]{min_hwo25}, artificial surface structures for LW-5 \citep[][Section~\ref{sssec:technosignatures}]{kopparapu_hwo25}, and potential biologically-induced seasonal changes in atmospheric composition for LW-13 \citep[][Section~\ref{sssec:seasonality}]{lafleche_hwo25}. 

Simultaneously, conducting the suite of investigations described in Section~\ref{sec:scdds} also necessitates the ability to observe bright sources such as transiting planet host stars ($2 < V < 20$; for SSiC-16 \citep[][Section~\ref{sssec:transits}]{wakeford_hwo25}, Venus for SSiC-5 \citep[][Section~\ref{sssec:venus}]{izenberg_hwo25}, and solar system icy worlds for SSiC-1 \citep[][Section~\ref{sssec:oceans_habitable}]{cartwright_hwo25}, which motivates saturation mitigation strategies, high dynamic range, and clever detector modes as highlighted in Section~\ref{sssec:saturation}. Furthermore, exquisitely precise astrometry for both bright exoplanet host stars and fainter reference stars is essential for measuring the masses and determining the orbits of potentially habitable and inhabited planets for LW-12 \citep[][Section~\ref{sssec:mp}]{gary_hwo25}.

Finally, carrying out this scientific portfolio demands the ability to observe quickly moving objects for investigations of solar system interlopers and small bodies (tracking rate $\geq 500$~mas/s) for SSiC-7 \citep[][Section~\ref{sssec:solar}]{mandt_hwo25}. Additionally, the ability to quickly obtain target-of-opportunity observations is needed for studies of transient phenomena such as supernovae (response time $\leq 30$~min) for EE-3 \citep[][Section~\ref{sssec:flash_ccsne}]{andrews_hwo25} and EE-4 \citep[][Section~\ref{sssec:r_process_el}]{burns_hwo25}; newly discovered comets (response time $\lesssim$ days) for SSiC-7 \citep[][Section~\ref{sssec:solar}]{mandt_hwo25}; and sublimating comets for SSiC-22 (Section~\ref{sssec:mars}). A large instantaneous field of regard would increase the observability of potential targets and thereby support studies of transient objects as well as phase-dependent observations of exoplanets, which are essential for studying planetary atmospheres, surface properties, orbits, habitability, and possible biosignatures. Solar system studies in general and observations of Venus for SSiC-5 \citep[][Section~\ref{sssec:venus}]{izenberg_hwo25} and near-Earth objects for SSiC-30 \citep[][Section~\ref{sssec:defense}]{dotson_hwo25} in particular would also benefit from a large instantaneous field of regard and a reduced Sun avoidance angle. 

In nearly all cases, as discussed in Section~\ref{sssec:simultaneous}, observations of potentially variable sources such as exoplanets, supernovae, and AGN disks would benefit significantly from obtaining simultaneous observations over a broad wavelength range. In some cases such as the investigation of Titan proposed by SSiC-27 (Section~\ref{sssec:titan}), observing different wavelengths contemporaneously within a few weeks would be sufficient. 

Moreover, the ability to collect observations using multiple instruments or modes simultaneously would improve overall observational efficiency by enabling parallel observations such as the acquisition of extragalactic deep fields during long direct spectroscopic searches for biosignatures. The large resulting data volume may increase demand for onboard data storage, processing, and data transmission as discussed in Section~\ref{sssec:downlink}. Additionally, target-of-opportunity observations and observations of targets with unknown or variable brightnesses may benefit from adaptive observational schemes such as autonomous changes to exposure times or filters \citep[e.g.,][]{thompson_et_al2018, lawson_et_al2025}; these ideas were considered by the Artificial Intelligence (AI) and Machine Learning (ML) Working Group. 

As discussed by \citet{shporer_hwo25_vol2}, AI may also be able to assist in biosignature identification and dynamic observation scheduling that optimizes the timing of future observations of particular targets based on past data. Additionally, AI/ML could potentially be integrated into mission development, mission operations, observation planning, and data interpretation more generally. For instance, AI/ML has been incorporated into the Vera C. Rubin Observatory's Legacy Survey of Space and Time \citep{lsst2026} and engineers are working to optimize the incorporation of convolutional neural networks and artificial intelligence classifiers in satellite operations \citep[e.g.,][]{diana+dini2024, furano_et_al2020, giuffrida_et_al2022}. Additionally, \citet{jimenez_et_al2023} discussed how language models could be used to address software engineering challenges, \citet{gottweis_et_al2025} illustrated how AI could be used to generate and assess hypotheses, and \citet{lu_et_al2024} demonstrated that autonomous AI agents could potentially manage the full research cycle. 

\subsection{Connecting Science to Engineering}
\label{ssec:bridge}
The goal of this paper was to distill the numerous SCDDs into a set of observing programs and identify the observational capabilities needed to accomplish the goals of those programs. We first reviewed the science cases developed by each Science Working Group in Section~\ref{sec:scdds}. In Section~\ref{sec:swg_obs}, we then grouped the science cases within each SWG by observational capabilities and assigned those groups alphabetical indicators. For instance, we defined ``EE-A'' as IFS observations that would support EE science cases and ``SSiC-H'' as high-contrast polarimetric imaging for SSiC science cases. When a science case required several different types of observations, we tracked the various observing programs within a science case using lowercase letters. As an example, we split the reionization science case GG-9 \citep[][Section~\ref{sssec:lyman_indirect}]{citro_hwo25} into GG-9a for spectroscopy and GG-9b for photometry. Program GG-9a was subsequently classified as part of GG-A for moderate resolution IFS spectroscopy in support of GG science while GG-9b was assigned to GG-F with the other photometric observations supporting GG science.

Next, in Section~\ref{sec:xswg} we identified scientific synergies among the science cases proposed by different SWGs. In Section~\ref{sec:drivers}, we then compared the observation types across WGs and identified the science cases that could potentially drive the design of various observing modes. For instance, in Section~\ref{sssec:ifs} we assessed the needs for IFS data across GG, EE, and SSiC science cases by comparing modes GG-A (moderate spectral resolution for GG science), GG-B (high $R$) to EE-A (moderate $R$), SSiC-J (moderate $R$), and SSiC-K (low $R$). Across the relevant science cases, we determined that the blue wavelength limit was set by the need to observe UV features of H, H$_2$, and He while studies of AGN feedback and solar system icy worlds drove the red wavelength limit. 

After exploring the mapping between scientific goals and observational capabilities, we returned to science in Section~\ref{sec:astro2020} to address how the science cases connected to the questions and discovery areas identified by the Astro2020 Decadal Survey. In total, we found that the science cases in Section~\ref{sec:scdds} would address 90\% of the 30 science questions and discovery areas. As shown in Figure~\ref{fig:astro2020}, the panel on Cosmology is the only science panel that is not fully addressed by the science program described in this paper. For the other five panels (Galaxies; Compact Objects and Energetic Phenomena; Interstellar Medium and Star and Planet Formation; Stars, the Sun, and Stellar Populations; and Exoplanets, Astrobiology, and the Solar System) all questions and discovery areas are addressed by at least one of the science cases described in Section~\ref{sec:scdds}. 

Having established the relevance of these science cases to Astro2020, we discussed the need for a new facility like HWO in Section~\ref{sec:need_hwo}. We then described important work needed to prepare for future observations in Section~\ref{sec:precursor} before concluding in this section. 

In summary, the science working groups launched by START have developed an inaugural set of \ncases~science cases that could be conducted with a facility like HWO. The design of HWO and its specific science portfolio are not yet determined, but the work of the START, TAG, and working groups has helped inform early investment in technology maturation and scientific studies. START, and the TSS subgroup in particular, also identified precursor and preparatory investigations that should be obtained to support the further maturation of HWO and other concepts. Many of those investigations, especially the precursor studies that could potentially inform architecture, should commence as soon as possible. 

\subsection{Next Steps}
\label{ssec:next}
In June 2025, NASA announced the HWO Community Science \& Instrument Team (CSIT). Along with the HWO Technology Maturation Project Office, this group is now guiding the next phase of HWO development. As stated on the NASA HWO website\footnote{\url{https://science.nasa.gov/astrophysics/programs/habitable-worlds-observatory/news/}}, the CSIT was created to ensure that the scientific community can provide input into the early-stage development of HWO. Additional objectives of the CSIT are to conduct scientific studies to assist in setting ``primary science objectives and requirements,'' provide scientific input on plans for developing the technology necessary to achieve those goals, and consider notional instrument concepts presented by the HWO Technology Maturation Project Office.

The community remains heavily invested in maturing science cases for HWO and other future facilities. Much of the community-led work is now organized by the HWO Science Interest Group (SIG), which organizes a virtual seminar series and maintains a list of HWO-related activities\footnote{\url{https://science.nasa.gov/astrophysics/programs/cosmic-origins/community/hwo-sig}}. The activities include other SIGs and SAGs (Science Analysis Groups) such as the HWO Red Wavelength Limit SAG,\footnote{\url{https://science.nasa.gov/astrophysics/programs/cosmic-origins/community/red-limit-science-analysis-group-red-limit-sag/}} which is exploring the need for observations at NIR wavelengths. In parallel, the Spatially Resolved UV Spectroscopy Science Analysis Group in Support of Habitable Worlds Observatory\footnote{\url{https://science.nasa.gov/astrophysics/programs/cosmic-origins/community/sruvs-sag/}} is exploring the need for short-wavelength observations and the status of UV technology development. Moving beyond HWO, NASA Astrophysics has begun the Astrophysics Strategic Technology and Research Accelerator (ASTRA) Initiative,\footnote{\url{https://science.nasa.gov/astrophysics/programs/cosmic-origins/studies/astra-initiative/}} which is an opportunity for scientists, engineers, and international partners to work to develop the next generation of potential mission concepts. 

Interest in HWO extends beyond the boundaries of the United States: the co-authors of this paper are affiliated with institutions in 31~different countries, and the European community is organizing a large meeting on HWO near Paris, France in Fall 2026. As highlighted in Figure~\ref{fig:map}, the combination of the HWO Working Groups and the author list of this paper includes individuals with primary affiliations in 41~US states, the District of Columbia, and Puerto Rico as well as 37~other countries. Additionally, the START membership included ex officio international representatives (see Table~\ref{tab:start}); the CSIT has continued this practice by including ex officio representatives from CSA, ESA, JAXA, and the European HWO Program Coordination Group.\footnote{\url{https://science.nasa.gov/astrophysics/programs/habitable-worlds-observatory/team/}} 

Moreover, potential international partners are pursuing significant work on instrument concepts such as the Pollux spectrograph and spectropolarimeter \citep{neiner_et_al2026, neiner_hwo25_vol2, girardot_hwo25_vol2} and a possible UK-led high-resolution imager covering a broad NUV -- NIR wavelength range and designed for extremely precise astrometry \citep[e.g.,][]{van_eylen_et_al2025}. Additionally, French researchers are exploring a potential detector \citep{lizzana_hwo25_vol2} and an instrument concept \citep{amiaux_hwo25_vol2} that could obtain the extremely precise astrometry needed for exoplanet mass measurement \citep[][LW-12, Section~\ref{sssec:mp}]{gary_hwo25} and other astrometric science cases \citep[e.g.,][]{malbet_hwo25_vol2}.

As discussed by \citet{gomez_de_castro_hwo25_vol2}, the Observatorio Ultravioleta Lunar (Lunar Ultraviolet Observatory; OUL) was recently selected as a scientific payload by the Agencia Espacial Espa\~{n}ola (Spanish Space Agency). OUL would conduct observations of Earth as an exoplanet that could provide useful context for interpreting future spectra of potentially habitable or inhabited exoplanets. 

In addition, JAXA is conducting independent studies of a far-ultraviolet integral field spectrograph and high-resolution spectrograph \citep{kameda_hwo25_vol2} as well as exploration of coronagraph designs \citep{enya_et_al2026}. JAXA is also maturing UV technology via the upcoming Life-environmentology, Astronomy, and PlanetarY Ultraviolet Telescope Assembly (LAPYUTA) mission \citep{tsuchiya_et_al2025}. Finally, CSA issued a Request for Information regarding possible Canadian participation in HWO\footnote{\url{https://canadabuys.canada.ca/en/tender-opportunities/tender-notice/cb-887-68106765}}, and is maturing UV technology for the upcoming Cosmological Advanced Survey Telescope for Optical and ultraviolet Research \citep[CASTOR;][]{cote_et_al2012, cote_et_al2019, cote_et_al2025} mission.

\include{scdd_refs}

\include{gg_scdds}
\include{gg_remap}

\include{ee_scdds}
\include{ee_remap}

\include{ssic_scdds}
\include{ssic_remap}

\include{lw_scdds}
\include{lw_remap}

\include{all_modes}

\include{scdd_astro2020_mapping}

\begin{contribution}
C. D. Dressing wrote the manuscript and developed the classification scheme used to transform observational capabilities into potential observing modes. She analyzed, described, and synthesized all science cases, assessing connections between science cases and alignment with the priorities of the Astro2020 Decadal Survey. C.~D. Dressing produced all tables and all figures except for the small subset in Section~\ref{sec:scdds} that were reproduced with permission from SCDDs (Figures~\ref{fig:quiescent}--\ref{fig:transits}). During the START era, she led the effort to involve community members in the development of science cases and coordinated science case maturation along with other members of the HWO leadership. She also coordinated the author list for this paper and managed the opt-in authorship process. M.~W. McElwain coordinated the internal review of the manuscript by the HWO Project Office and, in tandem with E. Mamajek and P. Chen, managed the NASA approval process. 

The science case leads listed at the start of the blurbs in Section~\ref{sec:scdds} developed the science cases presented in this paper. C.~D. Dressing wrote the accompanying science case summaries as well as the vast majority of the paper. The exceptions are the description of the work of the Target Stars \& Systems Working Group (draft version contributed by N. Hinkel and the Target Stars \& Systems Working Group; revised into Section~\ref{ssec:tss} and Appendix~\ref{app:tss} by C. D. Dressing); the discussion of LW-9 in Section~\ref{sssec:geochemical} (draft text contributed by K. Brugman and revised by C. D. Dressing); the first two paragraphs describing SSiC-18 in Section~\ref{sssec:exovenus} (initial text by C. D. Dressing; revised version by S. R. Kane, E. L. Miles, and C. M. Ostberg); and the fourth paragraph of the description of GG-7 and the captions for the associated visuals (Figure~\ref{fig:gg7_diagnostics} and Figure~\ref{fig:gg7_clusters}) in Section~\ref{sssec:lyman_escape} (draft text contributed by C. Carr and revised by C. D. Dressing).

The author list is divided into three alphabetical groups. In addition to the individuals mentioned above, the co-authors listed in the first alphabetical group led one or more science cases or the Target Stars \& Systems Working Group. The co-authors listed in the second alphabetical group actively participated in science case development in at least one of the following ways:
\begin{itemize}
 \item Contributed to a science case
 \item Discussed a science case
 \item Developed figures for a science case
 \item Provided feedback on or edited a science case
 \item Gave a presentation about a science case
 \item Led a working group, subgroup, or task group
\end{itemize}
Finally, the third group includes authors who contributed in other ways, such as helping to establish the working group structure; participating as a member of the START, TAG, HWO Technology Maturation Project Office, and/or working groups; and reading and commenting on this manuscript.
\end{contribution}

\begin{acknowledgments}
This work would not have been possible without decades of contributions from the global scientific and engineering community. The authors gratefully acknowledge the work of everyone who contributed to the studies that laid the essential groundwork for a future mission like the Habitable Worlds Observatory. We thank NASA Research and Education Support Services (NRESS) for providing logistical support for meetings of the HWO START, TAG, and Working Groups. Additionally, we recognize Yesenia Arroyo for her service to HWO and contributions to the website. 

We also thank Janice C. Lee, Jessica Noviello, Stephanie LaMassa, and Marc Postman for their work in editing and publishing the conference proceedings from HWO25 \citep{HWO25_proceedings, HWO25_proceedings_part2}. Furthermore, we acknowledge Megan Ansdell, Julie Crooke, Shawn Domagal-Goldman, Lee Feinberg, Joshua Pepper, and Aki Roberge for their leadership and feedback on the manuscript at various stages. 

This research has made use of the Astrophysics Data System, funded by NASA under Cooperative Agreement 80NSSC25M7105. We acknowledges use of the Hypatia Catalog Database, an online compilation of stellar abundance data as described in \citet{hinkel_et_al2014}. 

C.~D.~Dressing acknowledges support from the David and Lucile Packard Foundation (2019-69648), the NASA Precursor Science Program (80NSSC23K1476), the University of California Berkeley, and the Watson and Marilyn Alberts Chair in the Search for Extraterrestrial Intelligence. C.~D.~Dressing is grateful to NASA for the invitation to co-chair the START and lead this paper. Additionally, C.~D.~Dressing thanks Catherine Bonanno,  David Charbonneau, Scott Gaudi, Andrea Leistra, and Jessica Lu for their support and insight during this multi-year writing process. 

\input{latex_acknowledgements_list_10august2026}
\end{acknowledgments}

\vspace{5mm}
%\facilities{HST (COS)}

\software{
\texttt{NumPy} \citep{harris2020_numpy},
          \texttt{pandas} \citep{mckinney2010_pandas},
          \texttt{Matplotlib} \citep{hunter2007_matplotlib},
          \texttt{Plotly} \citep{plotly2015},
          \texttt{Kaleido} \citep{Kaleido2020}
          }
\appendix

\section{START Timeline, Membership, and Deliverables}
\label{app:start}
In April 2023, NASA issued a Dear Colleague letter\footnote{\url{https://smd-cms.nasa.gov/wp-content/uploads/2023/07/dcl-gomap-hwo-start-fy23-signed.pdf}} announcing the formation of the Science, Technology, Architecture Review Team (START). After reviewing applications, a NASA-led panel selected the individuals listed in Table~\ref{tab:start}. START included ex officio representatives from NASA Program Offices, industry, the Canadian Space Agency (CSA), the European Space Agency (ESA), and the Japan Aerospace Exploration Agency (JAXA). The ex officio representatives from international space agencies were selected by those agencies and joined START in October 2023 (JAXA), January 2024 (ESA), and February 2024 (CSA).

The START Terms of Reference state that START should deliver a final report including ``assessments, evaluations, and discussions of activities described above as well as:
\begin{enumerate}
\item{analysis results from science and technical activities;}
\item{a complete list of open assessments or analyses that the START was not able to
perform, either because of a lack of time, resources, or other limitations;}
\item{a roadmap to enable future quantification of those unaddressed relationships;}
\item{open-source publication of tools used to produce these relationships, including
documentation for those tools; and}
\item{assessment of the level of fidelity of models needed in the future to execute trades.''}
\end{enumerate}

In Section~\ref{sec:scdds}, we summarize the results of the analyses conducted by START (Item 1 above). When discussing each science case, we highlight open scientific assessments (Item 2) and identify potential areas for future research (Item 3). The tool documentation requested for Item~4 can be found in the Github repositories\footnote{\url{https://github.com/HWO-GOMAP-Working-Groups}} established during the START era as repositories for notes and open-source software. Finally, we address Item~5 when discussing potential scientific drivers (Section~\ref{sec:drivers}) and the status of the initial exploration of the scientific discovery space for a facility like HWO (Section~\ref{sec:conc}).

According to the Terms of Reference, START activities were anticipated to ``last roughly between 18 months up to 3 years'' and would continue ``until HWO receives formulation authorization and a Pre-Phase A Project office is created, or until otherwise terminated by the NASA Astrophysics Division Director.'' At the direction of Congress, NASA established the HWO Technology Maturation Project Office at NASA Goddard on 1 August 2024. Opening this office necessitated disbanding START/TAG and transitioning to a model in which working group activities were managed directly by the new HWO Technology Maturation Project Office. TAG activities were largely absorbed by the Technology Maturation Project Office while many former START members continued participating in the science working groups. 

In total, START lasted just under a year from early September 2023 until 31 July 2024. During that time, START and TAG held three in-person meetings in October 2023, March 2024, and June 2024. The latter two meetings included the working group co-chairs, who also met in October 2024 under the direction of the newly established HWO Technology Maturation Project Office. In keeping with the community engagement goal of START and the working groups, all four meetings\footnote{\url{https://science.nasa.gov/astrophysics/programs/habitable-worlds-observatory/meetings/}} were open to online participation by community members. 

\input{start_members}

\section{The Structure of the Target Stars and Systems Subgroup}
\label{app:tss}
To accomplish the TSS objectives, the TSS leads made a concerted effort to engage with a variety of the other WG/SWG/JWGs in order to identify those properties that were most fundamental to the science cases (as described in the SCDDs) anticipated by HWO. The science needs of the larger HWO community were compiled during in-person meetings as well as online forms, which helped inform the direction of the TSS operations. As a result, the TSS was separated into 6 task groups: 
\begin{itemize}
\itemsep0em
\item \textit{Catalogs \& Databases Task Group}: identify a stellar population that could be directly imaged by a range of potential HWO architectures to detect Earth-sized planets; define a tiered target list (i.e., Tier 1/2/3) that coarsely estimates the likelihood that particular stars would be observed by HWO, recognizing that the mission architecture is not currently defined;
\item \textit{Fundamental Properties Task Group}: compile inherent, chemical, and kinematic stellar properties (with a focus on observables needed to derive these properties) and their associated uncertainties that have a direct influence on the formation and evolution of its planetary system;
\item \textit{Activity \& Rotation Task Group}: collate stellar activity measurements within the literature and rank them based on the data quality and resolution, as well as identify the number/types of stars observed since the stellar activity can impact the measured size and chemical composition of exoplanet atmospheres;
\item \textit{High Energy Emission Task Group}: explore 15+ current, retired, and upcoming instruments ranging from the near-UV to the X-ray (e.g., 1--3200\,\AA) and categorize their photometric and spectroscopic observations, due to the impact of high energy radiation on exoplanet atmospheres and planetary life; 
\item \textit{Multiple Stars Task Group}: determine the multiplicity rate of HWO target stars by cross-referencing several large repositories (e.g., \textit{Gaia} DR3, the Washington Double Star catalog) and by employing a variety of techniques (e.g., speckle imaging, adaptive optics) so that the architecture of a planetary system may be accurately characterized; and 
\item \textit{Extreme Precision Radial Velocities (EPRV) Task Group}: ascertain the state of the field for EPRV monitoring of potential HWO targets, for which Earth-like planets would induce signals with semiamplitudes of roughly 10\,cm\,s$^{-1}$. 
\end{itemize}
\noindent
The results from these task groups are to be summarized in a number of published papers as well as an upcoming white paper report, which will include the community-driven property spreadsheet along with associated science cases. The TSS task groups have already published a summary of high-energy observations \citep{peacock_et_al2025}, an Activity and Rotation Catalog \citep{fetherolf_et_al2026}, and a catalog of speckle imaging observations to assess stellar multiplicity \citep{hartman_et_al2026}.

\bibliography{hwo_start}
\bibliographystyle{aasjournalv7}

%% the collaboration and author truncation commands are used. It has to
%% Workaround for AASTeX v7.0.1 bug when using \suppressAffiliations (recommended by vulpicastor at https://github.com/AASJournals/AASTeX7/issues/22)
\global\suppressAffiliationsfalse
\allauthors
\end{document}

%% file: latex_author_list_10august2026.tex
\author[0000-0001-9897-9680]{Danica Adams}
\email{dadams@fas.harvard.edu}
\affiliation{Department of Earth \& Planetary Sciences, Harvard University, Cambridge, MA}

\author[0000-0001-5260-7179]{Evelyne Alecian}
\email{evelyne.alecian@univ-grenoble-alpes.fr}
\affiliation{Univ. Grenoble Alpes, CNRS, IPAG, 38000 Grenoble, France}

\author[0000-0002-5259-2314]{Gagandeep Anand}
\email{ganand@stsci.edu}
\affiliation{Space Telescope Science Institute, 3700 San Martin Drive, Baltimore, MD, 21218, USA}

\author[0000-0001-6285-267X]{Giada Arney}
\email{giada.n.arney@nasa.gov}
\affiliation{NASA Goddard Space Flight Center, Greenbelt, MD, USA}

\author[0000-0003-3267-5398]{Sarah Gomes Aroucha Barbosa}
\email{sarahg.aroucha@gmail.com}
\affiliation{Universidade Federal do Cear\'a}

\author[0000-0002-7116-3259]{Martin Barstow}
\email{mab@le.ac.uk}
\affiliation{School of Physics and Astronomy, University of Leicester, University Road, Leicester, LE1 7RH, UK}

\author[0000-0003-3726-5419]{Joanna K. Barstow}
\email{jo.barstow@open.ac.uk}
\affiliation{School of Physical Sciences, The Open University, Milton Keynes, UK}

\author[0000-0002-1691-8217]{Rachael L. Beaton}
\email{rbeaton@stsci.edu}
\affiliation{Space Telescope Science Institute, 3700 San Martin Drive, Baltimore, MD, 21218, USA}

\author[0000-0002-9408-8925]{Eduardo Bendek}
\email{eduardo.bendek@nasa.gov}
\affiliation{NASA Ames Research Center}
\affiliation{SETI Institute, Mountain View, CA}

\author[0000-0002-2238-7416]{Svetlana Berdyugina}
\email{svetlana.berdyugina@irsol.usi.ch}
\affiliation{Istituto ricerche solari Aldo e Celle Dacc\'o (IRSOL), Universit\'a Svizzera italiana, Switzerland}

% ...

%% file: roadmap.tex
\begin{table*}[htb]
\centering
\caption{Overview of Paper Structure and Content}
\label{tab:roadmap}
\vspace{0.5em} 
\begin{tabular}{| >{\centering\arraybackslash}p{0.5in}  >{\raggedright\arraybackslash}p{2in}  >{\raggedright\arraybackslash}p{4in} |}
\hline
\textbf{Section} & \textbf{Title} & \textbf{Description} \\ \hline
\ref{sec:intro} & Introduction & Explains the context, role, scope, and structure of the paper\\ \hline
\ref{sec:start} & The Science, Technology, Architecture Review Team (START) & Describes the goals and work of START \\ \hline
\ref{sec:wg} & Working Groups & Introduces the Science Working Groups and outlines the science case development process\\ \hline
\ref{sec:scdds} & Potential Science Cases & Presents the science cases developed by each working group  \\ \hline
\ref{sec:swg_obs} & Key Observations \& Capabilities & Assesses the key capabilities and datasets needed to accomplish the objectives of the science cases presented in Section~\ref{sec:scdds} \\ \hline
\ref{sec:xswg} & Interdisciplinary Connections Across Science Working Groups & Considers the overlap between the observations needed for science cases developed by different working groups\\ \hline
\ref{sec:drivers} & Scientific Drivers for Tackling Technical Challenges & Identifies the science cases that present the most significant technical challenges \\ \hline
\ref{sec:astro2020} & Connections to Astro2020 & Reveals how the science cases developed by the working groups are responsive to the questions and discovery areas identified by Astro2020 \\ \hline
\ref{sec:need_hwo} & The Need for HWO & Demonstrates how completing the investigations proposed in Section~\ref{sec:scdds} demands new capabilities and instruments \\ \hline
\ref{sec:precursor} & Necessary Precursor Science & Explores observations that should be done in advance to maximize the scientific productivity of a facility like HWO \\ \hline
\ref{sec:conc} & Conclusions and Next Steps & Reviews core findings and discusses possible future work  \\ \hline
\hline
App.~\ref{app:start} & START Timeline, Membership, and Deliverables & Provides additional details regarding the formation, composition, and charge of START as well as a summary of deliverables \\
\hline
App.~\ref{app:tss} & The Structure of the Target Stars and Systems Subgroup & Describes the objectives and outcomes of the six task groups within the Target Stars and Systems Subgroup\\
\hline
\hline
\end{tabular}
\end{table*}

%% file: exoplanet_grid.tex
\begin{table*}[t]
\centering
\caption{Comparison of High-Contrast Target Samples for SSiC \& LW Science Cases}
\label{tab:planetgrid}
\renewcommand{\arraystretch}{1.5} % Adjusts row height padding cleanly
\begin{tabular}{|>{\centering}p{1in}|>{\raggedright}p{1.5in}|>{\raggedright}p{1.875in}|>{\raggedright\arraybackslash}p{1.5in}|}
\hline 
\textbf{Category} & \textbf{Hot} & \textbf{Temperate} & \textbf{Cold} \\ 
\hline \hline

\textbf{Terrestrial Planets}    & 
\mbox{SSiC-6\tablenotemark{a} (\ref{sssec:occ_binary})}, \mbox{SSiC-18 (\ref{sssec:exovenus})}, \mbox{SSiC-21 (\ref{sssec:retention})}, \mbox{LW-5 (\ref{sssec:technosignatures})}, \mbox{LW-6 (\ref{sssec:lawdki})}, \mbox{LW-10 (\ref{sssec:fpbio})} & 
\mbox{SSiC-2 (\ref{sssec:hab_system})}, \mbox{SSiC-3 (\ref{sssec:rocky_sub})}, \mbox{SSiC-4 (\ref{sssec:surface_water})}, \mbox{SSiC-6\tablenotemark{a} (\ref{sssec:occ_binary})}, \mbox{SSiC-11 (\ref{sssec:ozone_onset})}, \mbox{SSiC-18 (\ref{sssec:exovenus})}, \mbox{SSiC-20 (\ref{sssec:survive_water})}, \mbox{SSiC-21 (\ref{sssec:retention})}, \mbox{SSiC-32 (\ref{sssec:exomoons})}, \mbox{LW-1 (\ref{sssec:life})}, \mbox{LW-3 (\ref{sssec:origin})}, \mbox{LW-4 (\ref{sssec:prebiosignatures})}, \mbox{LW-5 (\ref{sssec:technosignatures})}, \mbox{LW-6 (\ref{sssec:lawdki})}, \mbox{LW-7 (\ref{sssec:polbio})}, \mbox{LW-13 (\ref{sssec:seasonality})}, \mbox{LW-14 (\ref{sssec:surface_bio})} & 
\mbox{SSiC-32 (\ref{sssec:exomoons})}, \mbox{LW-5 (\ref{sssec:technosignatures})}, \mbox{LW-6 (\ref{sssec:lawdki})}  \\ \hline

\textbf{Sub-Neptunes}           & 
\mbox{SSiC-8 (\ref{sssec:proto})}, \mbox{LW-5 (\ref{sssec:technosignatures})} & 
\mbox{SSiC-3 (\ref{sssec:rocky_sub})}, \mbox{SSiC-8 (\ref{sssec:proto})}, \mbox{SSiC-31 (\ref{sssec:exorings})}, \mbox{LW-5 (\ref{sssec:technosignatures})} & 
\mbox{SSiC-8 (\ref{sssec:proto})}, \mbox{SSiC-31 (\ref{sssec:exorings})}, \mbox{LW-5 (\ref{sssec:technosignatures})} \\ \hline

\textbf{Gas Giants}             & 
\mbox{SSiC-8 (\ref{sssec:proto})}, \mbox{SSiC-15 (\ref{sssec:reflected_giants})}, \mbox{LW-5 (\ref{sssec:technosignatures})} & 
\mbox{SSiC-8 (\ref{sssec:proto})}, \mbox{SSiC-15 (\ref{sssec:reflected_giants})}, \mbox{SSiC-31 (\ref{sssec:exorings})}, \mbox{LW-5 (\ref{sssec:technosignatures})} & 
\mbox{SSiC-8 (\ref{sssec:proto})}, \mbox{SSiC-2\tablenotemark{b} (\ref{sssec:hab_system})}, \mbox{SSiC-12 (\ref{sssec:giant_orbits})}, \mbox{SSiC-15 (\ref{sssec:reflected_giants})}, \mbox{SSiC-31 (\ref{sssec:exorings})}, \mbox{LW-5 (\ref{sssec:technosignatures})} \\ \hline

\textbf{Circumstellar Material} & 
\mbox{SSiC-8 (\ref{sssec:proto})}, \mbox{SSiC-9 (\ref{sssec:debris})}, \mbox{SSiC-23 (\ref{sssec:disk_winds})}, \mbox{LW-5 (\ref{sssec:technosignatures})} & 
\mbox{SSiC-8 (\ref{sssec:proto})}, \mbox{SSiC-9 (\ref{sssec:debris})}, \mbox{SSiC-19 (\ref{sssec:exozodi})}, \mbox{SSiC-23 (\ref{sssec:disk_winds})}, \mbox{LW-5 (\ref{sssec:technosignatures})} & 
\mbox{SSiC-8 (\ref{sssec:proto})}, \mbox{SSiC-9 (\ref{sssec:debris})}, \mbox{SSiC-23 (\ref{sssec:disk_winds})}, \mbox{LW-5 (\ref{sssec:technosignatures})} \\ \hline

\end{tabular}
\tablenotetext{a}{For SSiC-6 \citep[][Section~\ref{sssec:occ_binary}]{newton_hwo25}, sensitivity to planets in binary systems is necessary.}
\tablenotetext{b}{For SSiC-2 \citep[][Section~\ref{sssec:hab_system}]{hasegawa_water_hwo25}, the giant planets could be detected using other facilities.}
\end{table*}

%% file: hc_obsgrid.tex
\begin{table*}[t]
\centering
\caption{Comparison of High-Contrast Capabilities Needed for SSiC \& LW Science Cases}
\label{tab:hc_obs}
\renewcommand{\arraystretch}{1.5} % Adjusts row height padding cleanly

\begin{tabular}{
|>{\centering\bfseries}p{1.1in}
|>{\raggedright\arraybackslash}p{1.1in}
|>{\raggedright\arraybackslash}p{2.1in}
|>{\raggedright\arraybackslash}p{1.1in}
|>{\raggedright\arraybackslash}p{1.1in}|
}
\hline 
\textbf{Wavelength} & \centering\textbf{Photometry} & \centering\textbf{Spectroscopy} & \centering\textbf{Spectro-polarimetry} & \centering\arraybackslash\textbf{Polarimetric Imaging}\\ 
\hline \hline

UV\linebreak ($< 400$~nm) &  
\mbox{SSiC-8 (\ref{sssec:proto})}, \mbox{SSiC-15 (\ref{sssec:reflected_giants})}, \mbox{SSiC-19 (\ref{sssec:exozodi})}  &  

\mbox{SSiC-3 (\ref{sssec:rocky_sub})}, \mbox{SSiC-8 (\ref{sssec:proto})}, \mbox{SSiC-11 (\ref{sssec:ozone_onset})}, \mbox{SSiC-12 (\ref{sssec:giant_orbits})}, \mbox{SSiC-15 (\ref{sssec:reflected_giants})}, \mbox{SSiC-18 (\ref{sssec:exovenus})}, \mbox{SSiC-20 (\ref{sssec:survive_water})}, \mbox{SSiC-21 (\ref{sssec:retention})}, \mbox{LW-1 (\ref{sssec:life})}, \mbox{LW-3 (\ref{sssec:origin})}, \mbox{LW-4 (\ref{sssec:prebiosignatures})}, \mbox{LW-5 (\ref{sssec:technosignatures})}, \mbox{LW-6 (\ref{sssec:lawdki})}, \mbox{LW-10 (\ref{sssec:fpbio})}, \mbox{LW-13 (\ref{sssec:seasonality})} &  

\mbox{SSiC-8 (\ref{sssec:proto})}, \mbox{SSiC-9 (\ref{sssec:disk_winds})}, \mbox{SSiC-21 (\ref{sssec:retention})} &  

\mbox{SSiC-3 (\ref{sssec:rocky_sub})}, \mbox{SSiC-9 (\ref{sssec:disk_winds})}, \mbox{SSiC-15 (\ref{sssec:reflected_giants})}, \mbox{SSiC-19 (\ref{sssec:exozodi})}, \mbox{LW-5\tablenotemark{a} (\ref{sssec:technosignatures})} \\ \hline

Visible &  

\mbox{SSiC-8 (\ref{sssec:proto})}, \mbox{SSiC-15 (\ref{sssec:reflected_giants})}, \mbox{SSiC-19 (\ref{sssec:exozodi})}, \mbox{SSiC-31 (\ref{sssec:exorings})}, \mbox{SSiC-32 (\ref{sssec:exomoons})}, \mbox{LW-5 (\ref{sssec:technosignatures})} &  

\mbox{SSiC-3 (\ref{sssec:rocky_sub})}, \mbox{SSiC-6 (\ref{sssec:occ_binary})}, \mbox{SSiC-8 (\ref{sssec:proto})}, \mbox{SSiC-15 (\ref{sssec:reflected_giants})}, \mbox{SSiC-18 (\ref{sssec:exovenus})}, \mbox{SSiC-19 (\ref{sssec:exozodi})}, \mbox{SSiC-20 (\ref{sssec:survive_water})}, \mbox{SSiC-21 (\ref{sssec:retention})}, \mbox{LW-1 (\ref{sssec:life})}, \mbox{LW-3 (\ref{sssec:origin})}, \mbox{LW-4 (\ref{sssec:prebiosignatures})}, \mbox{LW-5 (\ref{sssec:technosignatures})}, \mbox{LW-6 (\ref{sssec:lawdki})}, \mbox{LW-10 (\ref{sssec:fpbio})}, \mbox{LW-13 (\ref{sssec:seasonality})}, \mbox{LW-14 (\ref{sssec:surface_bio})} &  

\mbox{SSiC-4 (\ref{sssec:surface_water})}, \mbox{SSiC-8 (\ref{sssec:proto})}, \mbox{SSiC-9 (\ref{sssec:disk_winds})}, \mbox{SSiC-12 (\ref{sssec:giant_orbits})}, \mbox{SSiC-18 (\ref{sssec:exovenus})}, \mbox{SSiC-21 (\ref{sssec:retention})}, \mbox{LW-7\tablenotemark{b} (\ref{sssec:polbio})} &  

\mbox{SSiC-3 (\ref{sssec:rocky_sub})}, \mbox{SSiC-4 (\ref{sssec:surface_water})}, \mbox{SSiC-9 (\ref{sssec:disk_winds})}, \mbox{SSiC-15 (\ref{sssec:reflected_giants})}, \mbox{SSiC-19 (\ref{sssec:exozodi})}, \mbox{SSiC-31 (\ref{sssec:exorings})}, \mbox{LW-5\tablenotemark{a} (\ref{sssec:technosignatures})}, \mbox{LW-7\tablenotemark{b} (\ref{sssec:polbio})} \\ \hline

NIR\linebreak ($> 1100$~nm) &  

\mbox{SSiC-15 (\ref{sssec:reflected_giants})}, \mbox{SSiC-19 (\ref{sssec:exozodi})}, \mbox{SSiC-32 (\ref{sssec:exomoons})} &  
\mbox{SSiC-2 (\ref{sssec:hab_system})}, \mbox{SSiC-3 (\ref{sssec:rocky_sub})}, \mbox{SSiC-15 (\ref{sssec:reflected_giants})}, \mbox{SSiC-18 (\ref{sssec:exovenus})}, \mbox{SSiC-19 (\ref{sssec:exozodi})}, \mbox{SSiC-20 (\ref{sssec:survive_water})}, \mbox{SSiC-21 (\ref{sssec:retention})}, \mbox{LW-1 (\ref{sssec:life})}, \mbox{LW-3 (\ref{sssec:origin})}, \mbox{LW-4 (\ref{sssec:prebiosignatures})}, \mbox{LW-6 (\ref{sssec:lawdki})}, \mbox{LW-10 (\ref{sssec:fpbio})}, \mbox{LW-14 (\ref{sssec:surface_bio})} &  

\mbox{SSiC-9 (\ref{sssec:disk_winds})}, \mbox{SSiC-18 (\ref{sssec:exovenus})}, \mbox{SSiC-21 (\ref{sssec:retention})} &  

\mbox{SSiC-3 (\ref{sssec:rocky_sub})}, \mbox{SSiC-9 (\ref{sssec:disk_winds})}, \mbox{SSiC-15 (\ref{sssec:reflected_giants})}, \mbox{SSiC-19 (\ref{sssec:exozodi})}, \mbox{SSiC-31 (\ref{sssec:exorings})}\\ \hline
\end{tabular}
\tablenotetext{a}{LW-5 requests polarimetry but the specific nature of the polarimetry is uncertain. We have listed LW-5 under UV and visible polarimetric imaging to cover the wavelength range of $300-900$~nm, but other modes may also be needed. }
\tablenotetext{b}{The first few objectives of LW-7 could be accomplished with either high-contrast polarimetric imaging or spectropolarimetry, but one objective demands high-contrast spectropolarimetry at $600-750$~nm. }
\end{table*}

%% file: properties_tss.tex
\begin{center}
\begin{longtable}{
| >{\centering}p{0.75in} %property
>{\raggedright\arraybackslash}p{2in} %motivation
>{\raggedright\arraybackslash}p{2.75in} %status
>{\centering\arraybackslash}p{1in} | %SCDDs
} %Thanks, https://tex.stackexchange.com/questions/200284/centering-defined-width-columns-with-long-table for tips for centering columns!
\caption{Stellar and System Properties Needed for Exo-Earth Science \label{tab:tss}}\\

\hline \multicolumn{1}{|c}{\textbf{Property}} & \multicolumn{1}{c}{\textbf{Scientific Motivation}}  & \multicolumn{1}{c}{\textbf{Current Status \& Path Forward}}  & \multicolumn{1}{c}{\textbf{SCDDs}}\\ \hline 
\endfirsthead

\multicolumn{4}{c}%
{{\bfseries \tablename\ \thetable{} -- continued from previous page}} \\

\hline \multicolumn{1}{|c}{\textbf{Property}} & \multicolumn{1}{c}{\textbf{Scientific Motivation}}  & \multicolumn{1}{c}{\textbf{Current Status \& Path Forward}}  & \multicolumn{1}{c}{\textbf{SCDDs}}\\   \hline \hline
\endhead

\hline \multicolumn{4}{|r|}{{\emph{Continued on next page}}} \\ \hline
\endfoot

\hline \hline
\endlastfoot
%PO ID & Section & Title & Wavelength (nm) & spatial_res (mas) & R & Type  & $\cdots$ & 
%PO ID (SCDD-) & Section & Title & Wavelength (nm) & spatial_res (mas) & R & Type  & $\cdots$ & 
\multicolumn{4}{|l|}{\textbf{\emph{Stellar Properties}}} \\ \hline
Stellar elemental abundances & Planet compositions are connected to stellar properties, so knowledge of stellar abundances might help identify stars that are more likely to host habitable or inhabited planets \citep{hinkel_et_al2024} & Some potential targets have measured abundances \citep{harada_et_al2024a}, but most stars in the longer TSS25 list do not. Additionally, stars with measured abundances often lack constraints for key elements \citep{hinkel_et_al2014}. & SSiC-15 (\ref{sssec:reflected_giants}), SSiC-16 (\ref{sssec:transits}), SSiC-24 (\ref{sssec:highres_atmos}), LW-3 (\ref{sssec:origin}), LW-4 (\ref{sssec:prebiosignatures}), LW-6 (\ref{sssec:lawdki}), LW-9 (\ref{sssec:geochemical})  \\ \hline
Stellar age & Planetary ages are assumed to be the same as stellar ages. Age information enables studies of the timing of crucial biological and geological developments. & Considering 659 Tier~1 and Tier~2 targets, \citet{ware_et_al2026} found that roughly 20\% have gyrochronal ages and about 5\% have asteroseismic ages. Focusing on Tier~1 targets, the TSS WG determined that 10\% have ages measured to the desired precision of $\le20$\%. Achieving $\le20\%$ age precision is theoretically possible for all Tier~1 targets via a coordinated effort. Important measurements include continued TESS 20-second cadence observations and EPRV campaigns to support asteroseismic age determinations for FGK~stars \citep[e.g.][]{huber_et_al2024,campante_et_al2024} and improved calibration of age-rotation-activity relations for M-dwarfs \citep[e.g.][]{bouma_et_al2023}. &  SSiC-11 (\ref{sssec:ozone_onset}), SSiC-14 (\ref{sssec:escape}), SSiC-32 (\ref{sssec:exomoons}), LW-3 (\ref{sssec:origin}), LW-4 (\ref{sssec:prebiosignatures}), LW-5 (\ref{sssec:technosignatures}), LW-6 (\ref{sssec:lawdki}) \\ \hline
Stellar mass & Planet masses are determined relative to stellar masses, and  $\le10\%$ mass precision is needed to interpret atmospheric spectra \citep{damiano_et_al2025}. Stellar masses are important for planetary orbits and dynamics. The effectiveness of astrometric and radial velocity observations for measuring planet masses depends on stellar mass.  & The precision and accuracy of mass measurements are highly heterogeneous. More observations and analyses are needed to determine robust masses for potential targets. & SSiC-2 (\ref{sssec:hab_system}), SSiC-12 (\ref{sssec:giant_orbits}), SSiC-15 (\ref{sssec:reflected_giants}), SSiC-16 (\ref{sssec:transits}), SSiC-17 (\ref{sssec:id_oceans}), SSiC-24 (\ref{sssec:highres_atmos}), LW-1 (\ref{sssec:life}), LW-12 (\ref{sssec:mp}) \\ \hline
Stellar radius or angular diameter & Stellar radii are needed for fundamental measurements of planetary radii and stellar effective temperatures, which directly affect the irradiation flux experienced by planets. Stellar and planetary radii need to be calculated to high accuracy ($\lesssim$20\%) for precise characterization of the planet's effective temperature and interior structure. & Only $\sim$60\% of Tier 1 targets currently have measured angular sizes. Such measurements are within the capabilities of current optical long-baseline interferometers such as the CHARA Array \citep[e.g.][]{boyajian_et_al2012} and VLTI \citep[e.g.][]{rains_et_al2020}. &  SSiC-16 (\ref{sssec:transits}), SSiC-17 (\ref{sssec:id_oceans}), SSiC-24 (\ref{sssec:highres_atmos}), SSiC-31 (\ref{sssec:exorings}), SSiC-32 (\ref{sssec:exomoons}), LW-1 (\ref{sssec:life}), LW-10 (\ref{sssec:fpbio}) \\ \hline
Stellar luminosity & The insolation flux experienced by planets and the resulting planetary temperatures depend on stellar luminosities. Luminosity uncertainties $\lesssim10$\,\% are needed to obtain $\lesssim5$\,\% uncertainty in the habitable zone distance boundaries for a given insolation level. & Improving luminosity estimates will require new observational efforts to measure accurate bolometric fluxes. & SSiC-14 (\ref{sssec:escape}), SSiC-16 (\ref{sssec:transits}), SSiC-24 (\ref{sssec:highres_atmos}), SSiC-18 (\ref{sssec:exovenus}), LW-1 (\ref{sssec:life}),  LW-5 (\ref{sssec:technosignatures}), LW-10 (\ref{sssec:fpbio})  \\ \hline
X-ray and EUV flux & Flux at X-ray and EUV wavelengths strongly contributes to planetary atmospheric escape and therefore affects planetary habitability. X-ray and EUV flux levels should therefore be considered in yield simulations to increase the likelihood of detecting a sufficient sample of habitable planets. X-ray and EUV fluxes can also provide information about stellar ages, which are needed to interpret planetary spectra. & Due to the difficulty of observing EUV fluxes, few measurements have been determined for G-type stars $<1$~Gyr or FKM-type stars of any age. In an analysis of archival data of 98~Tier~1 targets, \citet{peacock_et_al2025} found that while
71\% of stars have usable X-ray photometry, only 26\% have usable X-ray spectra. In the EUV, the numbers are even lower: 25\% of targets have usable EUV photometry and only 2\% have usable EUV spectra. A concerted effort is needed to determine X-ray and EUV fluxes for potential target stars.  & SSiC-14 (\ref{sssec:escape}), SSiC-16 (\ref{sssec:transits}), SSiC-18 (\ref{sssec:exovenus}), SSiC-24 (\ref{sssec:highres_atmos}), LW-1 (\ref{sssec:life}), LW-3 (\ref{sssec:origin}), LW-4 (\ref{sssec:prebiosignatures}), LW-10 (\ref{sssec:fpbio}), LW-14 (\ref{sssec:surface_bio}) \\ \hline
FUV and NUV flux & FUV and NUV spectra are essential inputs for photochemical modeling of planetary atmospheres including interpretation of potential biosignatures and identification of false positive biosignatures. Additionally, NUV flux affects the exposure times that would be needed to detect planets at NUV wavelengths and the resulting planet yield. & Analyzing archival observations of 98 Tier~1 targets, \citet{peacock_et_al2025} found that stars were more likely to have usable FUV spectra (40\%) or NUV spectra (62\%) than FUV photometry (35\%) or NUV photometry (8\%). Narrow-band NUV photometry should be obtained for potential targets.  

& SSiC-8 (\ref{sssec:proto}), SSiC-9 (\ref{sssec:debris}), SSiC-14 (\ref{sssec:escape}), SSiC-15 (\ref{sssec:reflected_giants}), SSiC-16 (\ref{sssec:transits}), SSiC-17 (\ref{sssec:id_oceans}), SSiC-18 (\ref{sssec:exovenus}), SSiC-24 (\ref{sssec:highres_atmos}), SSiC-26 (\ref{sssec:bfield}), SSiC-34 (\ref{sssec:young_bfields}), LW-1 (\ref{sssec:life}), LW-3 (\ref{sssec:origin}), LW-4 (\ref{sssec:prebiosignatures}), LW-10 (\ref{sssec:fpbio}), LW-14 (\ref{sssec:surface_bio})  \\ \hline 
Stellar variability and activity cycles & Stellar variability induces radial velocity signals with amplitudes and timescales that can be misinterpreted as planetary signals or complicate the measurement of genuine planetary signals. A detailed understanding of stellar variability over multiple timescales is needed to accurately determine the presence and masses of planets. & \citet{fetherolf_et_al2026} inspected archival observations of potential targets and found that $<20\%$ had measured stellar activity cycles.  In order to optimize HWO target selection towards stars that are quiescent at the time of HWO observations, the activity cycles of all targets need to be well understood. Accordingly, all potential targets should have repeated measurements of stellar magnetic activity across a long baseline such that the typical activity variations over many-year cycles are well documented \citep[e.g.,][]{isaacson_et_al2024, fetherolf_et_al2026}.   & SSiC-3 (\ref{sssec:rocky_sub}), SSiC-16 (\ref{sssec:transits}), SSiC-24 (\ref{sssec:highres_atmos}), SSiC-26 (\ref{sssec:bfield}), SSiC-34 (\ref{sssec:young_bfields}), LW-1 (\ref{sssec:life}), LW-13 (\ref{sssec:seasonality}), LW-14 (\ref{sssec:surface_bio}) \\ \hline
Ca H\&K emission & Measuring Ca H\&K emission can constrain ages, overall stellar activity level, and high-energy emission, to varying degrees of precision. Ages are necessary for understanding the timing of various planetary events while the stellar activity and high-energy emission are important for interpreting planetary spectra. & When cataloging archival observations of potential targets in their Activity and Rotation Catalog, \citet{fetherolf_et_al2026} found that $75\%$ of Tier~1, 63\% of Tier~2, and 10\% of Tier~3 targets had reported values of $\log R'_\mathrm{HK}$ while 90\%, 79\%, and 17\% had reported $S$-indices. Ca H\&K emission should be measured for all targets, and stars with existing measurements should be monitored to understand how their Ca H\&K emission varies over time. & SSiC-11 (\ref{sssec:ozone_onset}), SSiC-16 (\ref{sssec:transits}), SSiC-24 (\ref{sssec:highres_atmos}), SSiC-26 (\ref{sssec:bfield}), SSiC-34 (\ref{sssec:young_bfields}), LW-1 (\ref{sssec:life}), LW-3 (\ref{sssec:origin}), LW-4 (\ref{sssec:prebiosignatures}), LW-6 (\ref{sssec:lawdki}), LW-10 (\ref{sssec:fpbio}), LW-13 (\ref{sssec:seasonality}), LW-14 (\ref{sssec:surface_bio}) \\ \hline 

\pagebreak
\multicolumn{4}{|l|}{\emph{\textbf{System Properties}}} \\ \hline
Stellar multiplicity & Bound stellar companions can influence the location and stability of habitable orbits. Depending on the luminosity ratio and angular separation, the presence of a stellar companion within the field of view could also limit the effectiveness of starlight suppression and therefore reduce the sensitivity to potentially inhabited planets. & The multiplicities and proper motions of the target stars and their companions must be assessed in order to determine their expected locations and habitable zone accessibility at the time of future observations. The multiplicity sub-group has obtained and published high-resolution imaging of potential targets for 80 Tier~1 targets \citep{hartman_et_al2026}. & SSiC-6 (\ref{sssec:occ_binary}), LW-1 (\ref{sssec:life}) \\ \hline
Presence of other planets & Massive planets have the potential to render habitable zone orbits unstable to Earth-like planets. Additionally, other planets generate astrometric and radial velocity signals that can complicate the detection and mass measurement of potentially habitable planets. For habitable planets, mass measurements are needed to interpret atmospheric spectra and analyze potential biosignatures.  & For over 50\% of Tier~1 targets, archival RV data are insufficient to detect Neptune-mass planets within the habitable zone \citep[e.g.,][]{harada_et_al2024b}, which limits stability analyses of these systems \citep[e.g.,][]{kane_et_al2024b}. Continued long-duration RV monitoring is needed to complete the census of planets orbiting potential target stars.  Even with the upcoming release of Gaia DR4, not all potential HWO targets will have Gaia solutions. Additionally, published astrometric measurements of the full HPIC sample (Tiers 1-3) should be compiled and analyzed in order to plan future astrometry campaigns to maximize sensitivity to potential stellar and substellar companions. & SSiC-2 (\ref{sssec:hab_system}), SSiC-3 (\ref{sssec:rocky_sub}), SSiC-12 (\ref{sssec:giant_orbits}), SSiC-15 (\ref{sssec:reflected_giants}), SSiC-16 (\ref{sssec:transits}), SSiC-24 (\ref{sssec:highres_atmos}), SSiC-31 (\ref{sssec:exorings}), SSiC-32 (\ref{sssec:exomoons}), LW-1 (\ref{sssec:life}), LW-5 (\ref{sssec:technosignatures}) \\ \hline
Properties of potentially inhabited planets & Knowledge of the orbits and masses of temperate terrestrial planets is needed to plan observations for times when planets are at accessible angular separations and when certain features are most detectable (e.g., ocean glint). & Long-baseline, extremely precise radial velocity observations are needed to ensure the detection and characterization of potentially inhabited planets \citep{morgan_et_al2021}.  Depending on the stellar and planet properties, the RV errors must be $\le5-50$~cm s$^{-1}$.  & SSiC-2 (\ref{sssec:hab_system}), SSiC-3 (\ref{sssec:rocky_sub}), SSiC-4 (\ref{sssec:surface_water}), SSiC-32 (\ref{sssec:exomoons}), LW-1 (\ref{sssec:life}), LW-5 (\ref{sssec:technosignatures}), LW-7 (\ref{sssec:polbio}), LW-12 (\ref{sssec:mp}), LW-13 (\ref{sssec:seasonality}), LW-14 (\ref{sssec:surface_bio}) \\ \hline 
Presence of circumstellar material & The amount and distribution of circumstellar material affects exoplanet detectability. Some systems could be too dusty to detect exo-Earths. & Only $\sim2-3\%$ of Tier~1 targets have been observed by the most sensitive studies \citep[e.g.,][]{ertel_et_al2020}. Current state-of-the-art methods can probe only large amounts of dust. Upcoming facilities such as Roman will be more sensitive to lower amounts of dust, but those observations will generally be in regions of the disk exterior to the habitable zone. Improved sensitivity is needed to assess dust levels within the habitable zone or worst-case scenarios should be assumed for yield modeling. & SSiC-8 (\ref{sssec:proto}), SSiC-9 (\ref{sssec:debris}), SSiC-19 (\ref{sssec:exozodi}), LW-1 (\ref{sssec:life}), LW-5 (\ref{sssec:technosignatures}) \\ \hline
\hline \hline
\pagebreak
\multicolumn{4}{|l|}{\emph{\textbf{Line-of-Sight Properties}}} \\ \hline
ISM column density, especially for H~I & Interactions with the interstellar medium markedly impact UV spectral features (e.g., Lyman-$\alpha$ absorption, Mg II h\&k absorption, attenuation and flux reduction of EUV flux near $40-91.2$~nm). The complicating presence of the ISM directly impacts measurements of the high energy stellar flux and can complicate studies of how stellar radiation affects planetary atmospheres. & Preparatory ISM measurements are needed to interpret observations of potential exo-Earths and to determine the intrinsic line flux from HWO observations. & EE-2 (\ref{sssec:dust_extinction}), EE-8 (\ref{sssec:dust_uv}), SSiC-14 (\ref{sssec:escape}), SSiC-16 (\ref{sssec:transits}), SSiC-18 (\ref{sssec:exovenus}), SSiC-24 (\ref{sssec:highres_atmos}), LW-1 (\ref{sssec:life}), LW-3 (\ref{sssec:origin}), LW-4 (\ref{sssec:prebiosignatures}), LW-10 (\ref{sssec:fpbio}), LW-14 (\ref{sssec:surface_bio})  \\ \hline
Presence of background or foreground stars & Stars at small projected separations from target stars can decrease the effectiveness of starlight suppression, thereby reducing the sensitivity to potentially inhabited planets. & The proper motions of target stars and the population of anticipated background objects must be determined to verify that their habitable zones will be accessible to direct imaging during the era in which observations could be obtained. The multiplicity sub-group has obtained and published high-resolution imaging of potential targets for 80 Tier~1 targets \citep{hartman_et_al2026}. & SSiC-6 (\ref{sssec:occ_binary}), LW-1 (\ref{sssec:life}) \\ \hline
\end{longtable}
\end{center}

%% file: scdd_refs.tex
\begin{center}
\begin{longtable}{
| >{\centering}p{0.55in} 
>{\centering}p{0.35in} 
>{\centering}p{2.6in} 
>{\centering}p{1in} 
>{\centering}p{1in} 
>{\centering\arraybackslash}p{1in} | 
} 
\caption{References for HWO Science Cases. The PO IDs given in the first column are the science case identifiers (IDs) assigned by the HWO Project Office (PO) without the ``SCDD-'' prefix. IDs marked with an asterisk (*) were considered during the START era but not listed in the STScI SCDD Portal, included in the HWO25 Conference Proceedings, or published in the literature. The numbers in the Paper Section column refer to the sections in this paper. Some science cases used different titles for HWO25 Conference Proceedings, peer-reviewed articles, and internal submissions. \label{tab:refs}} \\

\hline \multicolumn{1}{|c}{\textbf{PO}} & \multicolumn{1}{c}{\textbf{Paper}}  & \multicolumn{1}{c}{\textbf{SCDD}} &  \multicolumn{1}{c}{\textbf{SCDD}} & \multicolumn{1}{c}{\textbf{Peer-Reviewed}} & \multicolumn{1}{c|}{\textbf{HWO25}}\\ 
 \multicolumn{1}{|c}{\textbf{ID}} & \multicolumn{1}{c}{\textbf{Section}} & \multicolumn{1}{c}{\textbf{Title}} & \multicolumn{1}{c}{\textbf{Leads}} & \multicolumn{1}{c}{\textbf{Articles}} & \multicolumn{1}{c|}{\textbf{Proceedings}}\\ \hline 
\endfirsthead

\multicolumn{6}{c}%
{{\bfseries \tablename\ \thetable{} -- continued from previous page}} \\
%\hline \multicolumn{1}{|c}{\textbf{PO ID}} & \multicolumn{1}{c}{\textbf{Section}} & \multicolumn{1}{c}{\textbf{Title}} & \multicolumn{1}{c}{\textbf{Leads}} & \multicolumn{1}{c}{\textbf{Peer-Reviewed Articles}} & \multicolumn{1}{c}{\textbf{HWO25 Proceedings}}\\ \hline \hline 
\hline \multicolumn{1}{|c}{\textbf{PO}} & \multicolumn{1}{c}{\textbf{Paper}}  & \multicolumn{1}{c}{\textbf{SCDD}} &  \multicolumn{1}{c}{\textbf{SCDD}} & \multicolumn{1}{c}{\textbf{Peer-Reviewed}} & \multicolumn{1}{c}{\textbf{HWO25}}\\ 
 \multicolumn{1}{|c}{\textbf{ID}} & \multicolumn{1}{c}{\textbf{Section}} & \multicolumn{1}{c}{\textbf{Title}} & \multicolumn{1}{c}{\textbf{Leads}} & \multicolumn{1}{c}{\textbf{Articles}} & \multicolumn{1}{c}{\textbf{Proceedings}}\\ \hline \hline
\endhead

\hline \multicolumn{6}{|r|}{{\emph{Continued on next page}}} \\ \hline
\endfoot

\hline \hline
\endlastfoot
%PO ID & Section & Title & Wavelength (nm) & spatial_res (mas) & R & Type  & $\cdots$ & 
%PO ID (SCDD-) & Section & Title & Wavelength (nm) & spatial_res (mas) & R & Type  & $\cdots$ & 
GG-5 & \ref{sssec:agn_outflow} & Deciphering The Launching of Multi-phase AGN-driven Outflows and Their (Spatially Resolved) Multi-scale Impact & L. Zhang, G. Kaur, T. Gao et al.  & $\cdots$ & \citep{zhang_hwo25} \\ \hline
GG-4 & \ref{sssec:bh_quiescent} & Exploring the Quiescent Black Hole Population of Nearby Dwarf Galaxies with the Habitable World Observatory & F. Pacucci  & $\cdots$ & \citep{pacucci_hwo25} \\ \hline
GG-2 & \ref{sssec:bh_mass_spin} & The Formation and Evolution of Supermassive Black Holes: IMBH Mass and Spin Functions & J. Cann, K. L. Smith, F. Civano et al.  & $\cdots$ & \citep{cann_hwo25} \\ \hline
GG-1* & \ref{sssec:bh_energy} & Probing Energy Extraction from Black Holes with HWO & M. Singha, P. Senchyna  & $\cdots$  & $\cdots$ \\ \hline
GG-3 & \ref{sssec:bh_torus} & Imaging the Dusty Torus Around Supermassive Black Holes & V. Gorjian, C. Packham, E. Hicks, \& S. La Massa  & $\cdots$ & \citep{gorjian_hwo25} \\ \hline
GG-6 & \ref{sssec:resolve_reionization} & Spatially Resolving the Fundamental Elements of Reionization in Galaxies & X. Xu, S. McCandliss, A. Strom et al.  & $\cdots$ & \citep{xu_reionization_hwo25} \\ \hline
GG-7 & \ref{sssec:lyman_escape} & How Do Ionizing Photons Escape from Star-Forming Galaxies? & C. Carr, R. Cen, S. Flury et al.  & $\cdots$ & \citep{carr_hwo25} \\ \hline
GG-8* & \ref{sssec:green_peas} & Tracking Cosmic Reionization via Green Pea Galaxies with HWO & M. Singha  & $\cdots$  & $\cdots$ \\ \hline
GG-9 & \ref{sssec:lyman_indirect} & Identifying the Contributors to Cosmic Reionization with the Habitable World Observatory: LyC Escape Calibration in Faint Galaxies & A. Citro, C. M. Scarlata, C. A. Carr et al.  & $\cdots$ & \citep{citro_hwo25} \\ \hline
GG-11 & \ref{sssec:ionizing_lf} & Evolution of the Ionizing Photon Luminosity Function & S. McCandliss, S. Ravindranath, S. Malhotra et al.  & $\cdots$ & \citep{mccandliss_hwo25} \\ \hline
GG-12 & \ref{sssec:dm_power} & Dark Matter through Dwarf Satellite Galaxies in Milky-Way Analogue Galaxies in the Local Universe & J. Doppel, E. O. Nadler, L. Moustakas et al.  & $\cdots$ & \citep{doppel_hwo25}\\ \hline
GG-13 & \ref{sssec:dm_lensing} & Habitable Worlds Observatory: Strong Lensing Constraints on Dark Matter Halo Mass Function down to $10^7 {\rm M}_\odot$ & Q. He, D. Lagattuta, M. Jauzac et al.  & $\cdots$ & \citep{he_hwo25} \\ \hline
GG-14 & \ref{sssec:smbh_mergers} & Measuring SMBH Merger Timescales with HWO & J. Nightingale  & $\cdots$ & SCDD portal\tablenotemark{a} \\ \hline
GG-15* & \ref{sssec:cgm_map} & A High Spatial and Spectral Resolution Absorption Map of the Inner CGM Enabled by HWO & J. Burchett  & $\cdots$ & $\cdots$\\ \hline
GG-16 & \ref{sssec:disk_cgm} & Characterizing Gas Flows Through Observations of the Disk-Circumgalactic Medium Interface with the Habitable Worlds Observatory & S. Borthakur, J. N. Burchett, F. Cashman et al. & \citep{borthakur_et_al2025} & \citep{borthakur_hwo25}  \\ \hline
GG-17 & \ref{sssec:cgm_elm} & Mapping Galactic Winds and Small-scale Structure in the Circumgalactic Medium with Habitable Worlds Observatory & J. N. Burchett, D. M. Lokhorst, Y. Faerman et al. & \citep{burchett_et_al2025} & \citep{burchett_hwo25} \\ \hline
GG-18 & \ref{sssec:agn_feedback} & Active Galactic Nuclei Feedback Effects on the Intergalactic Medium & M. T. Tillman, J. N. Burchett, B. Burkhart et al. & \citep{tillman_et_al2025b} & \citep{tillman_hwo25}  \\ \hline %paper title: ``Reconstructing galactic feedback history via the Lyman-$\alpha$ forest with Habitable Worlds Observatory''
GG-19 & \ref{sssec:smbh_pol} & Unveiling Supermassive Black Hole Binaries with FUV-to-NIR Spectropolarimetry & F. Marin, J. Biedermann, \& T. Barnouin & \citep{marin_et_al2025} & SCDD Portal\tablenotemark{b} \\ 
\hline

EE-1 & \ref{sssec:massive_stars_lowZ} & Echoes of the First Stars: Massive Star Evolution in Extremely Metal-Poor Environments with the Habitable Worlds Observatory & P. Senchyna, C. Hawcroft, M. Garcia et al.  & $\cdots$ & \citep{senchyna_hwo25} \\ \hline
EE-5  & \ref{sssec:vms} & Very Massive Stars with the Habitable Worlds Observatory & F. Martins, A. Wofford, M. Garcia et al.  & $\cdots$ & \citep{martins_hwo25} \\ \hline
EE-6  & \ref{sssec:resolved_stellar_pops} & Resolving Individual Stars in Nearby Large Galaxies with the Habitable Worlds Observatory & A. Smercina, T. Fetherolf, E. W. Koch et al.  & $\cdots$ & \citep{smercina_hwo25}  \\ \hline
EE-7 & \ref{sssec:wds} & White Dwarfs as Probes of Extrasolar Planet Compositions and Fundamental Astrophysics & S. Xu, M. Barstow, A. Buchan et al.  & $\cdots$ & \citep{xu_siyi_hwo25} \\ \hline
EE-8 & \ref{sssec:first_stars} & Habitable Worlds Observatory: The Nature of the First Stars & I. Roederer, R. Ezzeddine, \& J. Sobeck  & $\cdots$ & \citep{roederer_nature_hwo25} \\ \hline
EE-9 & \ref{sssec:r_process_nature} & Habitable Worlds Observatory: The Nature of the Astrophysical r-process & I. Roederer \& R. Ezzeddine  & $\cdots$ & \citep{roederer_rprocess_hwo25} \\ \hline
EE-10 & \ref{sssec:distance_ladder} & Envisioning the Distance Ladder in the Era of the Habitable Worlds Observatory & G. Anand, M. Durbin, R. Beaton et al.  & $\cdots$ & \citep{anand_hwo25} \\ \hline
EE-2 & \ref{sssec:dust_extinction} & Dust Extinction Beyond the Milky Way & R. Paladini, S. Salim, M. Boquien et al.  & $\cdots$ & \citep{paladini_hwo25} \\ \hline
EE-11 & \ref{sssec:dust_uv} & Probing the Variations of Interstellar Dust Abundance and Properties Within and Between Galaxies with HWO UV Spectroscopy in the Local Volume & J. Roman-Duval, M. Boquien, \& Y. Choi  & $\cdots$ & \citep{roman-duval_hwo25} \\ \hline
EE-12 & \ref{sssec:ism_uv} & Probing the Full Depth of ISM Properties with a UV-IFU & B. James \& D. Berg  & $\cdots$ & \citep{james_hwo25} \\ \hline
EE-4 & \ref{sssec:r_process_el} & The Heavy Element Enrichment History of the Universe & E. Burns \& J. Andrews  & $\cdots$ & \citep{burns_hwo25} \\ \hline
EE-3 & \ref{sssec:flash_ccsne} & Flash Spectroscopy of Core-Collapse Supernovae with the Habitable Worlds Observatory & J. Andrews \& E. Burns  & $\cdots$ & \citep{andrews_hwo25} \\ \hline
EE-15 & \ref{sssec:mag_stars} & New Frontiers in the Study of Magnetic Massive Stars with the Habitable Worlds Observatory & A. David-Uraz  & $\cdots$ & \citep{david-uraz_hwo25} \\ \hline
SSiC-11 & \ref{sssec:ozone_onset} & Atmospheres of Earth Analogs as a Function of Age & S. Blunt, E. L. Nielsen, E. R. Newton et al. & \citep{blunt_et_al2025} & SCDD portal\tablenotemark{c} \\ \hline
SSiC-3 & \ref{sssec:rocky_sub} & Identifying Rocky Planets and Water Worlds Among Sub-Neptune-sized Exoplanets with the Habitable Worlds Observatory & R. Hu, M. Min, M. Millar-Blanchaer et al.  & $\cdots$ & \citep{hu_rocky_hwo25} \\ \hline
SSiC-21 & \ref{sssec:retention} & Retention of Volatiles on Rocky Planets & L. Carone  & $\cdots$  & SCDD Portal\tablenotemark{d} \\ \hline
SSiC-14 & \ref{sssec:escape} & Exoplanet Atmospheric Escape Observations with the Habitable Worlds Observatory & L. A. Dos Santos, E. D. Lopez, L. Fossati et al.  & $\cdots$ & \citep{dossantos_hwo25} \\ \hline
SSiC-18 & \ref{sssec:exovenus} & Detecting and Characterizing Venus Analogs Orbiting Other Stars with the Habitable Worlds Observatory & S. Kane, E. L. Miles, C. M. Ostberg et al.  & $\cdots$ & \citep{kane_venus_hwo25} \\ \hline
SSiC-17 & \ref{sssec:id_oceans} & Identifying Cold Ocean Planets and Characterizing Their Geologic Activity & L. Quick, J. L. Noviello, \& A. V. Oza  & $\cdots$ & \citep{quick_hwo25} \\ \hline
SSiC-1 & \ref{sssec:oceans_habitable} & Assessing Ocean World Habitability with HWO & R. Cartwright, L. C. Quick, M. Neveu et al.  & $\cdots$ & \citep{cartwright_hwo25} \\ \hline
SSiC-4 & \ref{sssec:surface_water} & Detecting Surface Liquid Water on Exoplanets & J. Lustig-Yaeger, N. Cowan, R. Hu et al. & $\cdots$ & \citep{lustig-yaeger_liquidwater_hwo25} \\ \hline
SSiC-20 & \ref{sssec:survive_water} & Birthcries and Swansongs of Habitable Worlds & L. Carone  & $\cdots$  & SCDD Portal\tablenotemark{e} \\ \hline
SSiC-2 & \ref{sssec:hab_system} & Probing the Origin of Water in Planets within Habitable Zones with HWO & Y. Hasegawa, C. Dressing, \& L. Carone  & $\cdots$ & \citep{hasegawa_water_hwo25} \\ \hline
SSiC-13* & \ref{sssec:occ_small} & Using the Habitable Worlds Observatory to Survey the Orbital Architecture of Small Exoplanets & T. Daylan \& R. Rodriguez Martinez & $\cdots$  & SCDD Portal\tablenotemark{f} \\ \hline
SSiC-6 & \ref{sssec:occ_binary} & The Role of Stellar Multiplicity in the Prevalence of Small, Cool Planets & E. Newton, M. Rice, K. Sullivan et al. & Newton et al. (in prep) & \citep{newton_hwo25} \\ \hline
SSiC-26 & \ref{sssec:bfield} & Detecting and Characterising the Magnetic Field of Exoplanets & A. Strugarek, S. V. Berdyugina, V. Bourrier et al. & \citep{strugarek_et_al2025} & \citep{strugarek_hwo25} \\ \hline
SSiC-34 & \ref{sssec:young_bfields} & Investigating the Role of Magnetic Fields in Shaping Young Planetary Systems with the Habitable Worlds Observatory and the POLLUX Spectropolarimeter & A. I. G\'{o}mez de Castro, E. Alecian, S. H. P. Alencar et al.  & $\cdots$ & \citep{gomez_de_castro_magnetic_hwo25} \\ \hline 
SSiC-12 & \ref{sssec:giant_orbits} & Observed Orbital Architectures of Giant Planets & S. Sagynbayeva, A. Abbas, S. Kane et al. & \citep{sagynbayeva_et_al2025} & SCDD portal\tablenotemark{g} \\ \hline
SSiC-15 & \ref{sssec:reflected_giants} & Direct Imaging Characterization of Cool Gaseous Planets & M. Min, J. Barstow, L. C. Mayorga et al. & $\cdots$ & \citep{min_hwo25} \\ \hline
SSiC-16 & \ref{sssec:transits} & Characterizing the Dynamics and Chemistry of Transiting Exoplanets with the Habitable World Observatory & H. R. Wakeford, L. C. Mayorga, J. K. Barstow et al.  & $\cdots$ & \citep{wakeford_hwo25} \\ \hline
SSiC-24 & \ref{sssec:highres_atmos} & High-resolution Ultraviolet-to-Near-Infrared Characterization of Exoplanet Atmospheres with HWO & P. Cubillos, M. Brogi, A. Garc\'ia Mu\~noz  & $\cdots$ & \citep{cubillos_hwo25} \\ \hline
SSiC-7 & \ref{sssec:solar} & Solar System Origins  & K. Mandt, J. Noonan, D. Seligman et al.  & $\cdots$ & \citep{mandt_hwo25} \\ \hline
SSiC-5 & \ref{sssec:venus} & Venus Direct Observation with the Habitable World Observatory & N. Izenberg, T. Robinson, C. Dressing et al.  & $\cdots$ & \citep{izenberg_hwo25}  \\ \hline
SSiC-22 & \ref{sssec:mars} & The Origin of Mars  & R. Ramirez, H. Campins, R. Brasser, \& R. Cartwright  & $\cdots$ & SCDD Portal\tablenotemark{h} \\ \hline
SSiC-28 & \ref{sssec:mars_post} & Investigating Mars with HWO & M. Crismani, R. Cartwright, S. Faggi et al.  & $\cdots$ & 
SCDD Portal\tablenotemark{i} \\ \hline 
SSiC-27 & \ref{sssec:titan} & Exploring Titan's Dynamic Atmosphere with the Habitable Worlds Observatory & N. Lombardo  & $\cdots$ & SCDD Portal\tablenotemark{j} \\ \hline 
SSiC-29 & \ref{sssec:solargiant} & Exploring Solar System Giant Planets with Habitable Worlds Observatory & L. N. Fletcher, A. Simon, M. H. Wong et al.  & $\cdots$ & \citep{fletcher_hwo25}
\\ \hline 
SSiC-25 & \ref{sssec:aurorae} & Study of the Atmospheric Effects of Energetic Particle Precipitations on Giant Planets With the Habitable World Observatory & J.-Y. Chaufray, W. Dunn, L. N. Fletcher et al.  & $\cdots$ & \citep{chaufray_hwo25} \\ \hline
SSiC-33 & \ref{sssec:aminos} & Detection of Amino Acids in Comets & A. I. G\'{o}mez de Castro \& A. I. de Isidro-G\'{o}mez  & $\cdots$ & \citep{gomez_de_castro_amino_hwo25}
\\ \hline 
SSiC-23 & \ref{sssec:disk_winds} & Disk Winds and Dispersal of Protoplanetary Disks & K. Hoadley, Y. Hasegawa, T. Haworth et al.  & $\cdots$  &  SCDD portal\tablenotemark{k} \\ \hline %formal title: "UV H2 fluorescent lines as a probe of dispersal mechanisms of protoplanetary disks"
SSiC-8 & \ref{sssec:proto} & Studying Protoplanets and Protoplanetary Disks with the Habitable Worlds Observatory & B. Ren  & $\cdots$ & \citep{ren_hwo25} \\ \hline
SSiC-9 & \ref{sssec:debris} & Debris Disks and their Properties with the Habitable Worlds Observatory & I. Rebollido, Y. Hasegawa, M. MacGregor et al.  & $\cdots$ & \citep{rebollido_hwo25} \\ \hline
SSiC-19 & \ref{sssec:exozodi} & Characterizing Exozodis in Scattered Light & J. H. Debes, S. Ertel, W. C. Danchi et al.  & $\cdots$ & \citep{debes_hwo25} \\ \hline
SSiC-31 & \ref{sssec:exorings} & Exoring Detection, Characterization and Occurrence Rates & M. A. Limbach, R. Bowens-Rubin, S. Hopper et al. & \citep{limbach_et_al2024} & SCDD Portal\tablenotemark{l}
\\ \hline %https://docs.google.com/document/d/1Fs07W1XQvR4uy0nm5jkG4Npf8hB_PbrVdO8MuJkYo94/edit?tab=t.0
SSiC-32 & \ref{sssec:exomoons} & Exomoon Detection with Mutual Events & M. A. Limbach & \citep{limbach_et_al2024, limbach_et_al2026} & SCDD Portal\tablenotemark{m}
\\ \hline %https://docs.google.com/document/d/1Fs07W1XQvR4uy0nm5jkG4Npf8hB_PbrVdO8MuJkYo94/edit?tab=t.0#heading=h.ek6wyshtwzwp
SSiC-30 & \ref{sssec:defense} & Utilizing Habitable Worlds Observatory for Planetary Defense & J. Dotson, A. S. Rivkin, C. Thomas et al.  & $\cdots$ & \citep{dotson_hwo25} \\ \hline
LW-1 & \ref{sssec:life}  & The Search for Life on Potentially Habitable Exoplanets & G. Arney, M. N. Parenteau, N. Hinkel et al.  & $\cdots$ & \citep{arney_hwo25} \\ \hline
LW-3 & \ref{sssec:origin} & Testing Origin-of-Life Theories with the Habitable Worlds Observatory & S. Ranjan, M. Schlecker, N. Wogan, \& M. Wong  & $\cdots$ & \citep{ranjan_origin_hwo25} \\ \hline
LW-4 & \ref{sssec:prebiosignatures} & Prebiosignatures with the Habitable Worlds Observatory & S. Ranjan, D. Adams, M. Wong et al.  & $\cdots$ & \citep{ranjan_prebiosignatures_hwo25} \\ \hline
LW-5 & \ref{sssec:technosignatures} & Technosignatures & R. Kopparapu, S. Berdyugina, J. Haqq-Misra et al. & $\cdots$ & \citep{kopparapu_hwo25} \\ \hline
LW-6 & \ref{sssec:lawdki} & Searching for Life as We Don’t Know It: Detecting Signatures of Planetary-scale Complexity in Exoplanet Atmospheric Spectra & S. Walker, E. Janin, E. Shkolnik et al. & $\cdots$ & \citep{walker_hwo25}\\ \hline
LW-14 & \ref{sssec:surface_bio} & Living Worlds Working Group: Surface Biosignatures on Potentially Habitable Exoplanets & M. N. Parenteau, T. Matsuo & Matsuo, Parenteau et al. (in prep) & \citep{parenteau_surface_hwo25} \\ \hline
LW-7 & \ref{sssec:polbio} & Detecting Alien Living Worlds and Photosynthetic Life using Imaging Polarimetry with the HWO Coronagraph & S. Berdyugina, L. Patty, J. Grone et al.  & $\cdots$ & \citep{berdyugina_imaging_hwo25}\\ \hline
LW-13 & \ref{sssec:seasonality} & Breathing Worlds: Leveraging Seasonality for Exo-Earth Biosphere Characterization with HWO & \'{E}. Lafl\`{e}che,  A. G. Ulses, S. Gilbert-Janizek et al. & $\cdots$ & \citep{lafleche_hwo25} \\ \hline
LW-9\tablenotemark{n} & \ref{sssec:geochemical} & Geochemical Habitability (precursor science) & K. Brugman \& C. Till  & $\cdots$ & \citep{brugman_hwo25} \\ \hline
LW-10 & \ref{sssec:fpbio} & Testing False Positive Biosignature Scenarios with the Habitable Worlds Observatory & E. Schwieterman, J. Krissansen-Totton, J. Lustig-Yaeger et al.  & $\cdots$ & \citep{schwieterman_hwo25} \\ \hline
\hline
LW-12 & \ref{sssec:mp} & Masses of Potentially Habitable Planets Characterized by the Habitable Worlds Observatory & K. Gary, B. S. Gaudi, E. Bendek et al.  & $\cdots$ & \citep{gary_hwo25}\\
\hline \hline
\end{longtable}

\tablenotetext{a}{GG-14 was posted to the SCDD Portal. The link is \url{https://docs.google.com/document/d/14ZGS_XTkN8vtiaMKtJylhcOS-2EsEL0q6AuBMRA-o_A}}

\tablenotetext{b}{GG-19 was posted to the SCDD Portal. The link is \url{https://docs.google.com/document/d/1WJ9Cw5DVZk33IP3kr_q2bGwm02nbnv--9fXcdVA-o2I}}

\tablenotetext{c}{SSiC-11 was posted to the SCDD Portal. The link is \url{https://docs.google.com/document/d/1U0KbLAlAIImkqVbagdkM_m60AfbwTBB3OPAqMVdzTYM}}

\tablenotetext{d}{SSiC-21 was posted to the SCDD Portal. The link is \url{https://docs.google.com/document/d/1bPCSTe1qzhCvCo-VbBv6FpkBeNfKhxUB/}}

\tablenotetext{e}{SSiC-20 was posted to the SCDD Portal. The link is \url{https://docs.google.com/document/d/1BN5MKDxp669K8DVgRn-thp8CZDFB_JedosTAxbUlUAA/}}

\tablenotetext{f}{SSiC-13 was posted to the SCDD Portal. The link is \url{https://docs.google.com/document/d/1_xnjMUvh7jiHyqJsxBq37utkc8yVbmXR2tYU6LYAIpY/}}

\tablenotetext{g}{SSiC-12 was posted to the SCDD Portal. The link is \url{https://docs.google.com/document/d/1BsVhxQ5rempJoWAlRENdma_wGpL8tkrnzJ97QtPD9WY}}

\tablenotetext{h}{SSiC-22 was posted to the SCDD Portal. The link is \url{https://docs.google.com/document/d/17PxB2qvcCXn1wgRQrm8Mgt7AxycogH47AfelpW3_Iw4}}

\tablenotetext{i}{SSiC-28 was posted to the SCDD Portal. The link is \url{https://docs.google.com/document/d/1KmoZ-ywYXpzbgQsL7BnHrqdsMtBYe_TKwot2quMU97g}}

\tablenotetext{j}{SSiC-27 was posted to the SCDD Portal. The link is \url{https://docs.google.com/document/d/1bJwOROlvOCkGa740a_h2h8rx9qlWMgH2}}

\tablenotetext{k}{SSiC-23 was posted to the SCDD Portal. The link is \url{https://docs.google.com/document/d/1R5lCfu6_a2_9NgRTCtziyvpKIbeJn0Hb/}}

%%%%%%

\tablenotetext{l}{SSiC-31 was posted to the SCDD Portal. The link is \url{https://docs.google.com/document/d/1Fs07W1XQvR4uy0nm5jkG4Npf8hB_PbrVdO8MuJkYo94}}

\tablenotetext{m}{SSiC-32 was posted to the SCDD Portal. The link is \url{https://docs.google.com/document/d/1bOa97ZAblcfkfIOmzQDZDYDxJaMN-Cue8MFVjDEDTqw}}

\tablenotetext{n}{LW-9 describes precursor science and is therefore discussed in Section~\ref{sssec:geochemical} rather than Section~\ref{ssec:lw_scdds}.}

\end{center}

%% file: gg_scdds.tex
\begin{center}
\begin{longtable}{
| >{\centering}p{0.25in} 
>{\centering}p{0.25in} 
>{\centering}p{1.5in} 
>{\centering}p{1.5in} 
>{\centering}p{1.25in} 
>{\centering\arraybackslash}p{1.25in} | 
} 
\caption{Science questions, target samples, and observational capabilities for growth of galaxies science cases. The numbers in the ID column refer to the identification number assigned by the project office. All of these science cases are for the growth of galaxies, so the ``SCDD-GG-'' prefix has been omitted.} \label{tab:gg_scdds} \\

\hline \multicolumn{1}{|c}{\textbf{ID}} & \multicolumn{1}{c}{\textbf{Section}} & \multicolumn{1}{c}{\textbf{Question}} & 
\multicolumn{1}{c}{\textbf{Targets}} &
\multicolumn{1}{c}{\textbf{Capabilities}} &
\multicolumn{1}{c|}{\textbf{Astro2020}}\\ \hline 
\endfirsthead

\multicolumn{6}{c}%
{{\bfseries \tablename\ \thetable{} -- continued from previous page}} \\
\hline \multicolumn{1}{|c}{\textbf{ID}} & \multicolumn{1}{c}{\textbf{Section}} & \multicolumn{1}{c}{\textbf{Question}} & 
\multicolumn{1}{c}{\textbf{Targets}} &
\multicolumn{1}{c}{\textbf{Capabilities}} &
\multicolumn{1}{c|}{\textbf{Astro2020}}\\ \hline 
\endhead

\hline \multicolumn{6}{|r|}{{\emph{Continued on next page}}} \\ \hline
\endfoot

\hline \hline
\endlastfoot
5 & \ref{sssec:agn_outflow} & How does AGN feedback drive the evolution of galaxies? & 2000 galaxies (1950 with NIR only; 50 with UV/Vis/NIR) & UV/Vis IFS ($100-900$~nm, $R>50,000$, $\theta=3$~mas), NIR IFS ($1-5\mu$m, $R > 5000$, $\theta=30$~mas, FOV$=10''\times 10''$) & B-DA, B-Q3, B-Q4, D-Q1, D-Q2, D-Q3, D-Q4  \\
\hline
4 & \ref{sssec:bh_quiescent} & What is the relationship between SMBHs and their host galaxies for low-mass galaxies? &  100 dwarf galaxies ($10-30$ Mpc; $10^6-10^9 M_\odot$) & UV/Vis MOS ($100-1000$~nm, $R\sim8000$, $\theta=10$~mas) & B-Q2, B-Q4, D-Q1, D-Q3 \\
\hline
2 & \ref{sssec:bh_mass_spin} & What is the origin of SMBHs and how do they affect their host galaxies? & low-metallicity galaxies with masses ($<10^{8.5}M_\odot$) &  IFS ($90-2500$~nm, $R>3000$, $\theta=50$~mas, FOV diam. $=1'$ & B-Q4, D-Q1, D-Q3, D-Q4 \\
\hline

%NOT INCLUDED %1 & \ref{sssec:bh_energy} & What is the relationship between SMBHs and their hosts, and how do SMBHs affect their surroundings? &  active galaxies ($z < 0.06$; radiatively efficient \& inefficient systems) & IFU, $R>5000$, 12~mas res, UV $\geq 100$~nm, NIR, polarization & B-Q2, B-Q3, B-Q3a, B-Q3b, B-DA, D-Q3, D-Q3d \\ %wavelength mentioned in text of SCDD

\hline
3 & \ref{sssec:bh_torus} & What are the properties of the dusty tori surrounding SMBHs? & $>10$ nearby Seyfert galaxies ($<65$ Mpc) & spectroscopy ($90-700$~nm, $R > 2700$, $\theta=5$~mas) & B-Q4, D-Q2, D-Q3 \\
\hline 

6 & \ref{sssec:resolve_reionization} & How did galaxies and their LyC clusters ionize the universe? & 10,000~unlensed galaxies at $z<1$ & spectroscopy ($50-200$~nm, $R=30,000$, $\theta=10$~mas, FOV$=10''\times10''$) & D-Q1, D-Q2, D-Q4, F-Q2 \\
\hline 

7 & \ref{sssec:lyman_escape} & How does ionizing radiation escape from galaxies and what sources are responsible for ionizing photons? & low-redshift galaxies ($0\le z \le 0.3$) & IFS ($95 - 200$~nm, $30,000 < R < 100,000$, $\theta = 0.3-30$~mas, FOV$\ge 3''\times3''$) & D-Q1, D-Q2, D-Q4, F-Q1   \\
\hline

%NOT INCLUDED 8 & \ref{sssec:green_peas} & How does the efficiency of LyC escape vary for Green Pea galaxies, and what are the drivers of reionization? & 500 - 1000 ``Green Pea'' galaxies at $z < 1$ & IFU, $8,000 < R < 10,000$, 12~mas resolution, $2'' \times 2''$ FOV, $\geq 90$~nm & D-Q1b, D-Q1c, D-Q2, D-Q2b, D-Q2c, D-Q4c, D-Q4d, F-Q1c \\
\hline

9 & \ref{sssec:lyman_indirect} & What fraction of LyC photons can escape galaxies, and what are the main contributors to reionization? & 40,000 faint, low-redshift galaxies ($M_{\rm UV} \le -13$, $m_{\rm UV}\le 25$, $z<0.1$)& spectroscopy ($100 - 510$~nm, $R > 10,000$), photometry ($< 100$~nm) &  D-Q1, D-Q2, F-Q1  \\
\hline

11 & \ref{sssec:ionizing_lf} & How has the number and ionizing luminosity of galaxies varied over cosmic time? & 2500 galaxies with a range of sizes \& environments & MOS ($100-180$~nm, $R=1000$, $\theta=225$~mas, FOV$=3\times[2'\times2']$)   &  D-Q1, G-DA  \\
\hline

12 & \ref{sssec:dm_power} & What is the nature of dark matter? & $>10$ Milky Way-mass host galaxies & photometry ($\ge 2$ VIS/NIR filters, $\theta=20$~mas, FOV$=4'\times3'$) & C-Q2, D-Q4   \\
\hline

13 & \ref{sssec:dm_lensing} & What is the nature of dark matter, and how does dark matter affect the evolution of the universe? & 2000 strong lensing systems & IFS ($200-500$~nm, $R\sim10,000$, $\theta=20$~mas, FOV=$20''\times 20''$); photometry ($200-500$~nm; $\theta=5-15$~mas, FOV$=4'\times4'$) & C-Q2, C-Q3  \\
\hline

14 & \ref{sssec:smbh_mergers} & How quickly do SMBHs merge? & 1000 strong lenses & photometry (650~nm or another $\lambda$ in $550-900$~nm, $\theta=20$~mas) & B-Q4, B-DA, D-Q1, D-Q3, D-Q4 \\
\hline 

%NOT INCLUDED 15 & \ref{sssec:cgm_map} & How does the inflow and outflow of matter affect the growth of galaxies? & $\geq6$ QSO sightlines towards $>100$ low-z galaxies & $R=100,000$, $6' \times 6'$ FOV, $\geq 90$~nm & D-Q2, D-Q2a, D-Q2b, D-Q2d, F-Q1c  \\
\hline

16 & \ref{sssec:disk_cgm} & How do metals and gases behave at the interface between galactic disks and the circumgalactic medium? & $4-6$ QSO absorption sightlines towards 25~low-z  galaxies; absorption sightlines towards $2-3$ background galaxies per target galaxy; emission \& absorption spectra of 100s of QSOs across $30-40$\% of the sky to trace MW CGM and probe 10s of galaxies & MOS ($94-350$~nm, $R\sim 100,000$ for absorption \& $R\gtrsim5000$ for emission, $\theta<1''$, FOV$=6'\times6'$) & B-Q4,  D-Q2, D-Q3, D-DA, F-Q1 \\
\hline

17 & \ref{sssec:cgm_elm} & How do galaxies exchange matter and energy? & 12 galaxies (range of masses; quiescent \& star-forming; edge-on \& face-on) & IFS ($97.5 -160$~nm \& $279-290$~nm, $R\sim10,000$, $\theta=50$~mas, FOV$=10''\times10''$); MOS ($97.5 - 160$~nm \& $279-290$~nm, $R\sim20,000$, $\theta=500$~mas resolution, FOV$=8'\times8'$)  & B-Q4, D-Q2, D-Q3, D-DA, F-Q1 \\
\hline

18 & \ref{sssec:agn_feedback} & What are the relative effects of dark matter and galactic feedback on the growth of structure in the universe? & 750 QSO sightlines at $z=0.7$ \& 80-250 QSO sightlines at $z=1.6$ & spectroscopy ($121.5-340.2$~nm, $R>40,000$) & B-Q4, C-Q3, D-Q1, D-Q2, D-Q3 \\
\hline

19 & \ref{sssec:smbh_pol} & How do binary supermassive black holes interact with their accretion disks? & $\ge 300$~AGN at a redshift high enough that the full AGN core is smaller than the instrument aperture & spectropolarimetry ($100-2000$~nm, $R>50,000$, polarization error $<0.01\%$) &  B-Q4, B-DA, D-Q3\\

\hline \hline
\end{longtable}
\end{center}

%% file: gg_remap.tex
%Reformatted 1 June, 2026
\setlength{\tabcolsep}{0.05in} %minimal horizontal padding between columns

\begin{center}
\begin{longtable}{
| >{\centering\arraybackslash}p{0.75in}                   % ID (Centered)
  >{\centering\arraybackslash}p{0.5in}                    % Section (Centered)
  >{\raggedright\arraybackslash}p{1.5in}                  % Focus (Automated Ragged-Right)
  >{\raggedright\arraybackslash}p{1.5in}                  % Targets (Automated Ragged-Right)
  >{\raggedright\arraybackslash}p{\cw} |                  % Capabilities (Automated Ragged-Right)
} 
\caption{Summarized Observations for Growth of Galaxies Science Cases} \label{tab:gg_remap} \\

\hline \multicolumn{1}{|c}{\textbf{ID}} & \multicolumn{1}{c}{\textbf{Section}} & \multicolumn{1}{c}{\textbf{Focus}} & 
\multicolumn{1}{c}{\textbf{Targets}} &
\multicolumn{1}{c|}{\textbf{Capabilities}} \\ \hline 
\endfirsthead

\multicolumn{5}{c}%
{{\bfseries \tablename\ \thetable{} -- continued from previous page}} \\
\hline \multicolumn{1}{|c}{\textbf{ID}} & \multicolumn{1}{c}{\textbf{Section}} & \multicolumn{1}{c}{\textbf{Focus}} & 
\multicolumn{1}{c}{\textbf{Targets}} &
\multicolumn{1}{c|}{\textbf{Capabilities}} \\ \hline 
\endhead

\hline \multicolumn{5}{|r|}{{\emph{Continued on next page}}} \\ \hline
\endfoot
\hline \hline
\endlastfoot
\rowcolor{colorGG}
GG-A & \ref{sssec:gg_a} & \emph{IFS Observations At Moderate Spectral Resolution} & galaxies (including strong lensing systems) &
UV/VIS/NIR IFS ($90-2500$~nm, $R=10,000$, $\theta=20$~mas, FOV$=10''\times10''$) \\
\hline
5b & \ref{sssec:agn_outflow} & AGN feedback  & 2000 galaxies (1950 with NIR only; 50 with UV/Vis/NIR) & \harder NIR cut-off ($5\mu$m) \easier blue cut-off (1000~nm), spectral resolution ($R>5000$), spatial resolution ($\theta=30$~mas) \\
\hline
4 & \ref{sssec:bh_quiescent} & Relationship between SMBHs and low-mass host galaxies\tablenotemark{a} &  100 dwarf galaxies ($10-30$ Mpc; $10^6-10^9 M_\odot$) & \harder spatial resolution ($\theta=10$~mas) \easier narrower wavelength range ($100-1000$~nm), spectral resolution ($R\sim8000$)  \\
\hline
2 & \ref{sssec:bh_mass_spin} & Origin \& influence of SMBHs & low-metallicity galaxies with masses ($<10^{8.5}M_\odot$) & \harder N/A \easier spectral resolution ($R\sim8000$), spatial resolution ($\theta=50$~mas) \\ 
\hline
3 & \ref{sssec:bh_torus} & Dusty tori surrounding SMBHs & $>10$ nearby Seyfert galaxies ($<65$ Mpc) & \harder spatial resolution ($\theta=5$~mas)  \easier spectral resolution ($R> 3000$), wavelength range ($90-700$~nm) \\
\hline
9a & \ref{sssec:lyman_indirect} & Reionization \& LyC escape & 40,000 faint, low-redshift galaxies ($M_{\rm UV} \le -13$, $m_{\rm UV}\le 25$, $z<0.1$)& \harder N/A \easier narrower wavelength range ($100-510$~nm), spatial resolution ($\theta=30$~mas) \\
\hline
13a & \ref{sssec:dm_lensing} & Dark matter & 2000 strong lensing systems & 
\harder N/A
\easier narrower wavelength range ($200-500$~nm) \\
\hline
17a & \ref{sssec:cgm_elm} & Exchange of matter \& energy in galaxies & 12 galaxies (range of masses; quiescent \& star-forming; edge-on \& face-on) & \harder N/A \easier narrower wavelength range ($97.5 -160$~nm \& $279-290$~nm), spatial resolution ($\theta=50$~mas)\\
\hline \hline
\rowcolor{colorGG}
GG-B & \ref{sssec:gg_b} & \emph{IFS Observations At High Spectral Resolution} & galaxies (including AGN) & UV/VIS IFS ($95-500$~nm, $R=30,000$, $\theta=10$~mas, FOV$=10''\times10''$) \\
\hline
5a & \ref{sssec:agn_outflow} & AGN feedback & 50 galaxies & \harder spectral resolution ($R>50,000$), spatial resolution ($\theta=3$~mas), red cut-off (900~nm) \easier UV cut-off (100~nm) \\
\hline
6 & \ref{sssec:resolve_reionization} & Role of galaxies and their LyC clusters in ionizing the universe? & 10,000~unlensed galaxies at $z<1$ & \harder UV cut-off (50~nm) \easier red cut-off (200~nm) \\
\hline
7 & \ref{sssec:lyman_escape} & Source \& escape of ionizing radiation & low-redshift galaxies ($0\le z \le 0.3$) & 
\harder spectral resolution ($30,000 < R < 100,000$), spatial resolution ($\theta = 0.3-30$~mas)
\easier red cut-off (200~nm), field-of-view ($3-10'' \times 3-10''$) \\
\hline \hline
\rowcolor{colorGG}
GG-C & \ref{sssec:gg_c} & \emph{MOS Observations At Moderate Spectral Resolution} & galaxies & UV MOS ($94-350$~nm, $R=3000$, $\theta=225$~mas, FOV$=2'\times2'$)\\
\hline
11 & \ref{sssec:ionizing_lf} & Ionizing luminosity of galaxies & 2500 galaxies with a range of sizes \& environments & 
\harder FOV ($3 \times [2' \times 2']$) \easier spectral resolution ($R=1000$), narrower wavelength range ($100-180$~nm) \\
\hline
16b & \ref{sssec:disk_cgm} & IGM/CGM interface & $4-6$ emission spectra of galaxies near 100s of QSOs across $30-40$\% of the sky \& 10s of extragalactic targets & \harder spectral resolution ($R\gtrsim 5000$), FOV ($6'\times6'$) \easier spatial resolution ($\theta < 1000$~mas)\\
\hline \hline
\rowcolor{colorGG}
GG-D & \ref{sssec:gg_d} & \emph{MOS Observations At High Spectral Resolution} & galaxies  & UV MOS ($94-350$~nm, $R=40,000$, $\theta=500$~mas, FOV$=6'\times6'$)\\
\hline
16a & \ref{sssec:disk_cgm} & IGM/CGM interface & $4-6$ QSO absorption sightlines towards 25~low-z  galaxies; absorption sightlines towards $2-3$ background galaxies per target galaxy; absorption spectra of 100s of QSOs across $30-40$\% of the sky & \harder spectral resolution ($R\sim 100,000$) \easier spatial resolution ($\theta<1''$)\\
\hline
17b & \ref{sssec:cgm_elm} & Exchange of matter \& energy in galaxies & 12 galaxies (range of masses; quiescent \& star-forming; edge-on \& face-on) &  
\harder FOV ($8'\times8'$) \easier narrower wavelength range ($97.5 - 160$~nm \& $279-290$~nm), spectral resolution ($R\sim20,000$) \\
\hline
18 & \ref{sssec:agn_feedback} & Dark matter & 750 QSO sightlines at $z=0.7$ \& 80-250 QSO sightlines at $z=1.6$ & \harder N/A \easier narrower wavelength range ($121.5-340.2$~nm)\\
\hline \hline
\rowcolor{colorGG}
GG-E & \ref{sssec:gg_e} & \emph{Spectropolarimetric Observations} & targets & UV/VIS IFS ($100-1600$~nm, $R=120,000$, full Stokes, polarization error $<10^{-6}$) \\
\hline 
19 & \ref{sssec:smbh_pol} & Interaction of binary SMBHs with their accretion disks & $\ge 300$~AGN & \harder red cut-off (2000~nm) \easier spectral resolution ($R>50,000$), polarization error ($<0.01\%$) \\
\hline \hline
\rowcolor{colorGG}
GG-F & \ref{sssec:gg_f} & \emph{Photometric Observations} & galaxies (including strong lenses) & UV/Vis Imager ($95-700$~nm, $\theta = 20$~mas, FOV=$4'\times 4'$) \\
\hline
9b & \ref{sssec:lyman_indirect} & Reionization \& LyC escape & 40,000 faint, low-redshift galaxies ($M_{\rm UV} \le -13$, $m_{\rm UV}\le 25$, $z<0.1$)& \harder N/A \easier blue cut-off ($< 100$~nm) \\
\hline 
12 & \ref{sssec:dm_power} & Dark matter & $>10$ Milky Way-mass host galaxies & \harder red cut-off (includes NIR) \easier blue cut-off (Vis/NIR only; no UV), FOV ($4'\times3'$)  \\
\hline
13b & \ref{sssec:dm_lensing} & Dark matter & 2000 strong lensing systems & 
\harder spatial resolution ($\theta=5-15$~mas) \easier narrower wavelength range ($200-500$~nm)\\
\hline
14 & \ref{sssec:smbh_mergers} & SMBH mergers & 1000 strong lenses & \harder red cut-off (observations would ideally be at 650~nm but could be anywhere in the range $550-900$~nm) \easier blue cut-off (Vis/NIR only; no UV) \\
\hline \hline
\end{longtable}
\tablenotetext{a}{For the science case described in Section~\ref{sssec:bh_quiescent}, \citep{pacucci_hwo25} requested a MOS, but the high spatial resolution causes the science case to fit better in this IFS category.}
\end{center}

%% file: ee_scdds.tex
\begin{center}
\begin{longtable}{
| >{\centering}p{0.25in} 
>{\centering}p{0.25in} 
>{\centering}p{1.5in} 
>{\centering}p{1.5in} 
>{\centering}p{1.25in} 
>{\centering\arraybackslash}p{1.25in} | 
} 
\caption{Science questions, target samples, and observational capabilities for evolution of the elements science cases. The numbers in the ID column refer to the identification number assigned by the project office. All of these science cases are for the evolution of the elements, so the ``SCDD-EE-'' prefix has been omitted. Science cases addressing multiple distinct science goals (e.g., Section~\ref{sssec:wds}) are listed once for each goal. The acronyms MW, LMC, and SMC refer to the Milky Way, Large Magellanic Cloud, and Small Magellanic Cloud, respectively.} \label{tab:ee_scdds} \\

\hline \multicolumn{1}{|c}{\textbf{ID}} & \multicolumn{1}{c}{\textbf{Section}} & \multicolumn{1}{c}{\textbf{Question}} & 
\multicolumn{1}{c}{\textbf{Targets}} &
\multicolumn{1}{c}{\textbf{Capabilities}} &
\multicolumn{1}{c|}{\textbf{Astro2020}}\\ \hline 
\endfirsthead

\multicolumn{6}{c}%
{{\bfseries \tablename\ \thetable{} -- continued from previous page}} \\
\hline \multicolumn{1}{|c}{\textbf{ID}} & \multicolumn{1}{c}{\textbf{Section}} & \multicolumn{1}{c}{\textbf{Question}} & 
\multicolumn{1}{c}{\textbf{Targets}} &
\multicolumn{1}{c}{\textbf{Capabilities}} &
\multicolumn{1}{c|}{\textbf{Astro2020}}\\ \hline 
\endhead

\hline \multicolumn{6}{|r|}{{\emph{Continued on next page}}} \\ \hline
\endfoot

\hline \hline
\endlastfoot
1 & \ref{sssec:massive_stars_lowZ} & How did ancient, very metal-poor massive stars irradiate and enrich the universe? & $\gg 10$ massive, low-metallicity stars ($\gtrsim 30 M_\odot$, $\geq 0.03 Z_\odot$) in $1-10$ low-metallicity ($Z < 0.1Z_\odot$) dwarf galaxies  such as IZw18 & IFS ($90-200$~nm \& $300-500$~nm, $R\gtrsim5000$, $\theta \lesssim 20$~mas, FOV=$3-10''$) & F-Q3, G-Q1, G-Q2, G-DA\\
\hline
5 & \ref{sssec:vms} & How do very massive stars influence their environments, and how do their properties depend on metallicity? & $\geq 500$ very massive stars at distances of $0.8 - 15$~Mpc with a range of metallicities ($0.01 - 2 Z_\odot$) &  IFS ($100-700$~nm, $R> 5000$, $\theta=5$~mas, FOV=$3''\times3''$) & D-Q1, D-Q2, G-Q1 \\
\hline
6 & \ref{sssec:resolved_stellar_pops} & How is the observed diversity of galaxy morphologies connected to the interaction, assembly, star formation, and enrichment history of galaxies? &  Depth 1: 100 galaxies within 5~Mpc with masses $10^8 - 10^{11} M_\odot$ and a range of metallicities \& star formation rates;  Depth 2: hundreds of galaxies within 20~Mpc; Depth 3: thousands of galaxies within 50~Mpc &  photometry (UV/optical/NIR, $\theta=15$~mas, FOV=0.007 deg$^2$, depths of $UV < 32.5$, $B < 34$, \& $I < 33$) & D-Q2, D-Q4, F-Q1, F-Q3, G-Q1 \\
\hline

7a & \ref{sssec:wds} & What are the compositions of exoplanets accreted onto white dwarfs?  & $\sim9500$~polluted white dwarfs with GALEX FUV magnitudes $\leq20$ & spectroscopy ($>90$~nm, $R\sim 60,000$) & E-Q1, E-Q2, G-DA, G-Q1\\
\hline
7b & \ref{sssec:wds} & Does general relativity hold in regions with strong gravitational fields? How do the values of fundamental constants depend on the strength of gravity? & 50 white dwarfs with GALEX FUV magnitudes $\leq20$ and a range of surface gravities ($7 \leq \log g \leq 9$) \& temperatures & spectroscopy ($>90$~nm, $R\sim 200,000$) & B-DA, G-DA, G-Q1\\
\hline
8 & \ref{sssec:first_stars} & What was the nature of the first stars, and how did they enrich the universe? & 50 stars with [Fe/H]$<-4$ \& $V< 18$  &  spectroscopy ($140 - 310$~nm, $R\sim 100,000$,  $\theta \leq 100$~mas) & D-Q1, D-Q2, D-Q4, G-DA, G-Q1 \\
\hline
9 & \ref{sssec:r_process_nature} & How are rapid neutron capture process (r-process) elements produced in stars? & 100 metal-poor FGK stars with $V \leq 17$  in varied Galactic environments & spectroscopy ($150 - 310$~nm, $R\sim 100,000$) & B-Q2, D-Q2, D-Q4, G-Q1\\
\hline

10 & \ref{sssec:distance_ladder} & What are the distances to distant galaxies? How quickly is the universe expanding? & Cepheids, Tip of the Red Giant Branch stars, and J-region Asymptotic Giant Branch stars in $\sim100$ Cepheid host galaxies at distances $\leq 100$~Mpc & photometry (Vis/NIR, $\theta = 20$~mas, FOV=$6' \times 6'$, depths of $V < 33$ \& $I < 32$)
& C-Q3, D-Q4, G-Q1\\
\hline

2 & \ref{sssec:dust_extinction} & What are the properties of extragalactic dust grains, and how do they differ from galactic dust grains?  & Objective~1:  $\ge 22,515$ diffuse sightlines to bright OB stars (20,000 towards the MW; 2000 LMC; 400 SMC; 100 M31; 10 local group galaxies; 5 local volume galaxies); Objective~2: $\ge 126$~dense sightlines to blue A/F/G giants (100 MW; 10 LMC; 10 SMC; 5 M31; 1 local group galaxy)  & 
MOS ($100-2500$~nm, $R\sim 3000$, $\theta = 10$~mas)
& D-Q4, F-Q1, F-Q2, F-Q4, G-DA \\
\hline
11 & \ref{sssec:dust_uv} & How do the composition and properties of interstellar dust vary within and among galaxies in the Local Volume? & 100~sightlines/galaxy towards the MW, LMC, SMC, and other galaxies within 10~Mpc. Sightlines should span a range of column densities ($20\rm~cm^{-2} < \log N(H) < 22 \rm~cm^{-2}$) and metallicities ($1 - 300\%$ solar) & MOS ($95 - 315$~nm, $R\sim50,000$ for Objective~1 \& $R>100,000$ for Objective~2, FOV $\geq 2' \times 2'$) & D-Q2, D-Q4, F-Q1\\
\hline
12 & \ref{sssec:ism_uv} & How are elements distributed within galaxies, and how does the distribution of elements depend on galactic properties and cosmic time? & $\sim 100$ nearby galaxies ($z < 0.025$, $D < 100$~Mpc) with low metallicities ($12 + \log (\rm O/H) < 8.5$), high star-formation rates ($> 1 M_\odot/\rm {yr}$), \& low masses ($M < 10^{10} M_\odot$) & IFS ($95 - 285$~nm, $R > 10,000$, $\theta \leq 100$~mas, FOV$>10'' \times 10''$)  & D-DA, D-Q1, D-Q2, D-Q4, F-Q1, F-Q2 \\ 
\hline

4 & \ref{sssec:r_process_el} & What is the nature of kilonova explosions, and how do remnant objects, jets, and ejecta contribute to the resulting emission? & 80 kilonovae (10~each in 8~redshift bins) including 10~kilonovae within two~days of explosion & imaging (UV, $\theta = 1''$, depths of UV mag $\le30$), spectroscopy (UV, low $R$), rapid-response ($\leq 30$~minutes), high-cadence monitoring (every few hours) & B-Q1, B-Q2, B-DA, D-Q1, G-Q1 \\
\hline

3 & \ref{sssec:flash_ccsne} & How do short-lived, extremely luminous, massive stars form elements and enrich the interstellar medium? & 15~core-collapse supernovae within one~day of explosion & spectroscopy ($110-300$~nm, $R\sim 10,000$), rapid-response ($\leq 1$~hour after trigger) & B-Q2, D-Q2, F-Q1 \\
\hline

15 & \ref{sssec:mag_stars} & Why do only 10\% of massive stars have strong magnetic fields, and how do massive stars co-evolve with their magnetic fields? & 220 MW targets (60 ``non-magnetic'' stars [30~MS \& 30~evolved], 30 post-merger objects, 100~evolved stars, 30 bright magnetic stars), 100s of extragalactic massive stars  & spectropolarimetry ($100-1600$~nm, $R>60,000$ in UV \& $R \sim120,000$ in VIS/NIR, full Stokes (IQUV), polarization error $<10^{-6}$) & B-Q1, B-Q2, F-Q1, G-Q1, G-Q2 \\
\hline

\hline \hline
\end{longtable}
\end{center}

%% file: ee_remap.tex
%Reformatted 1 June, 2026
\setlength{\tabcolsep}{0.05in} %minimal horizontal padding between columns

\begin{center}
\begin{longtable}{
| >{\centering\arraybackslash}p{0.75in}                   % ID (Centered)
  >{\centering\arraybackslash}p{0.5in}                    % Section (Centered)
  >{\raggedright\arraybackslash}p{1.5in}                  % Focus (Automated Ragged-Right)
  >{\raggedright\arraybackslash}p{1.5in}                  % Targets (Automated Ragged-Right)
  >{\raggedright\arraybackslash}p{\cw} |                  % Capabilities (Automated Ragged-Right)
} 
\caption{Summarized Observations for Evolution of the Elements Science Cases} \label{tab:ee_remap} \\

\hline \multicolumn{1}{|c}{\textbf{ID}} & \multicolumn{1}{c}{\textbf{Section}} & \multicolumn{1}{c}{\textbf{Focus}} & 
\multicolumn{1}{c}{\textbf{Targets}} &
\multicolumn{1}{c|}{\textbf{Capabilities}} \\ \hline 
\endfirsthead

\multicolumn{5}{c}%
{{\bfseries \tablename\ \thetable{} -- continued from previous page}} \\
\hline \multicolumn{1}{|c}{\textbf{ID}} & \multicolumn{1}{c}{\textbf{Section}} & \multicolumn{1}{c}{\textbf{Focus}} & 
\multicolumn{1}{c}{\textbf{Targets}} &
\multicolumn{1}{c|}{\textbf{Capabilities}} \\ \hline 
\endhead

\hline \multicolumn{5}{|r|}{{\emph{Continued on next page}}} \\ \hline
\endfoot
\hline \hline
\endlastfoot
\rowcolor{colorEE}
EE-A & \ref{sssec:ee_a} & \emph{IFS Observations} & very massive stars \& low-metallicity galaxies & UV/VIS IFS ($90-500$~nm, $R=10,000$, $\theta=20$~mas, FOV$=10''\times10''$) \\
\hline 
1a, 1b & \ref{sssec:massive_stars_lowZ} & Very metal-poor massive stars & $\gg 10$ massive, low-metallicity stars ($\gtrsim 30 M_\odot$, $\geq 0.03 Z_\odot$) in $1-10$ low-metallicity ($Z < 0.1Z_\odot$) dwarf galaxies such as IZw18 & \harder N/A \easier discontinuous spectral coverage (need data at $90-200$~nm \& $300-500$~nm; gap at $200-300$~nm), spectral resolution ($R\gtrsim5000$), FOV ($3-10'' \times 3-10''$) \\
\hline
5 & \ref{sssec:vms} & Very massive stars & $\geq 500$ very massive stars at distances of $0.8 - 15$~Mpc with a range of metallicities ($0.01 - 2 Z_\odot$) &  \harder spatial resolution ($\theta=5$~mas), red cut-off (700~nm) \easier UV cut-off (100~nm), spectral resolution ($R> 5000$), FOV ($3''\times3''$)\\
\hline
12 & \ref{sssec:ism_uv} & Chemical enrichment of galaxies & $\sim 100$ nearby galaxies ($z < 0.025$, $D < 100$~Mpc) with low metallicities ($12 + \log (\rm O/H) < 8.5$), high star-formation rates ($> 1 M_\odot/\rm {yr}$), \& low masses ($M < 10^{10} M_\odot$) & \harder N/A \easier narrower wavelength range ($95-285$~nm), spatial resolution ($\theta \leq 100$~mas) \\
\hline \hline
\rowcolor{colorEE}
EE-B & \ref{sssec:ee_b} & \emph{MOS Observations} & MW, LMC, SMC, \& other nearby galaxies  & UV/VIS/NIR MOS ($95-2500$~nm, $R=3000$, $\theta=10$~mas, FOV$=2'\times2'$) \\
\hline 
2a, 2b & \ref{sssec:dust_extinction} & Dust extinction in the Milky Way \& other galaxies & Objective~1:  $\ge 22,515$ diffuse sightlines to bright OB stars (20,000 towards the MW; 2000 LMC; 400 SMC; 100 M31; 10 local group galaxies; 5 local volume galaxies); Objective~2: $\ge 126$~dense sightlines to blue A/F/G giants (100 MW; 10 LMC; 10 SMC; 5 M31; 1 local group galaxy)  & \harder N/A, \easier UV cut-off (100~nm) \\
\hline
11a & \ref{sssec:dust_uv} & Composition of galactic \& extragalactic dust & 20-60~sightlines/galaxy towards OB stars within the MW, LMC, \& SMC. Sightlines should span a range of column densities ($20\rm~cm^{-2} < \log N(H) < 23 \rm~cm^{-2}$) and metallicities ($1 - 300\%$ solar) & \harder spectral resolution ($R\sim50,000$) \easier red cut-off (315~nm) \\
\hline
11b & \ref{sssec:dust_uv} & Composition of galactic \& extragalactic dust & 100~sightlines/galaxy local volume galaxies within 10~Mpc. Sightlines should span a range of column densities ($20\rm~cm^{-2} < \log N(H) < 22 \rm~cm^{-2}$) and metallicities ($1 - 300\%$ solar) & \harder spectral resolution ($R>100,000$) \easier narrower wavelength range ($115-315$~nm)\\
\hline \hline
\rowcolor{colorEE}
EE-C & \ref{sssec:ee_c} & \emph{Single-Object Spectroscopy At Moderate Spectral Resolution} & transient events & UV/VIS ($110-400$~nm, $R=10,000$) \\
\hline 
3 & \ref{sssec:flash_ccsne} & Core-collapse supernovae & 15~core-collapse supernovae within one~day of explosion & \harder N/A \easier red cut-off (300~nm) \\
\hline
4a & \ref{sssec:r_process_el} & Kilonovae & 80 kilonovae (10~each in 8~redshift bins) including 10~kilonovae within two~days of explosion & \harder N/A \easier N/A \\
\hline \hline
\rowcolor{colorEE}
EE-D & \ref{sssec:ee_d} & \emph{Single-Object Spectroscopy At High Spectral Resolution} & white dwarfs \& low-metallicity stars & UV ($90-310$~nm, $R=100,000$) \\
\hline 
7a & \ref{sssec:wds} & Exoplanet composition  & $\sim9500$~polluted white dwarfs with GALEX FUV magnitudes $\leq20$ & \harder red cut-off? (value not specified in SCDD) \easier spectral resolution ($R\sim 60,000$) \\
\hline
7b & \ref{sssec:wds} & Fundamental physics & 50 white dwarfs with GALEX FUV magnitudes $\leq20$ and a range of surface gravities ($7 \leq \log g \leq 9$) \& temperatures & \harder spectral resolution ($R\sim 200,000$), red cut-off? (value not specified in SCDD) \easier N/A\\
\hline
8 & \ref{sssec:first_stars} & Nature of the first stars & 50 stars with [Fe/H]$<-4$ \& $V< 18$  & \harder spatial resolution ($\theta \leq 100$~mas; for resolving targets) \easier UV cut-off (140~nm)  \\
\hline
9 & \ref{sssec:r_process_nature} & Production of r-process elements & 100 metal-poor FGK stars with $V \leq 17$  in varied Galactic environments & \harder N/A \easier UV cut-off (150~nm) \\
\hline \hline
\rowcolor{colorEE}
EE-E & \ref{sssec:ee_e} & \emph{Spectropolarimetric Observations} & targets & UV/VIS IFS ($100-1600$~nm, $R=120,000$, full Stokes, polarization error $<10^{-6}$) \\
\hline 
15a, 15b & \ref{sssec:mag_stars} & Stellar magnetic fields & 220 MW targets (60 ``non-magnetic'' stars [30~MS \& 30~evolved], 30 post-merger objects, 100~evolved stars, 30 bright magnetic stars), 100s of extragalactic massive stars  & \harder N/A \easier spectral resolution ($R>60,000$ in UV, $R\sim120,000$ in Vis) \\ 
\hline \hline
\rowcolor{colorEE}
EE-F & \ref{sssec:ee_f} & \emph{Photometric Observations} & Kilonovae \& galaxies & UV/VIS/NIR Imager ($95-2500$~nm, $\theta=20$~mas, FOV$=6'\times6'$) \\
\hline 
4b & \ref{sssec:r_process_el} & Kilonovae & 80 kilonovae (10~each in 8~redshift bins) including 10~kilonovae within two~days of explosion & \harder N/A \easier red cut-off (UV data only), spatial resolution (1000~mas)\\
\hline
6 & \ref{sssec:resolved_stellar_pops} & Morphology of galaxies &  Depth 1: 100 galaxies within 5~Mpc with masses $10^8 - 10^{11} M_\odot$ and a range of metallicities \& star formation rates;  Depth 2: hundreds of galaxies within 20~Mpc; Depth 3: thousands of galaxies within 50~Mpc & \harder spatial resolution ($\theta=15$~mas) \easier FOV ($5' \times 5'$) \\
\hline
10 & \ref{sssec:distance_ladder} & Distance ladder & Cepheids, Tip of the Red Giant Branch stars, and J-region Asymptotic Giant Branch stars in $\sim100$ Cepheid host galaxies at distances $\leq 100$~Mpc & \harder N/A \easier blue cut-off (data at Vis/NIR only) \\
\end{longtable}
\end{center}

%% file: ssic_scdds.tex
\begin{center}
\begin{longtable}{
| >{\centering}p{0.25in} 
>{\centering}p{0.25in} 
>{\centering}p{1.5in} 
>{\centering}p{1.5in} 
>{\centering}p{1.25in} 
>{\centering\arraybackslash}p{1.25in} | 
} 
\caption{Science questions, target samples, and observational capabilities for solar systems in context science cases. The numbers in the ID column refer to the identification number assigned by the project office. All of these science cases are for solar systems in context, so the ``SCDD-SSiC-'' prefix has been omitted.} \label{tab:ssic_scdds} \\

\hline \multicolumn{1}{|c}{\textbf{ID}} & \multicolumn{1}{c}{\textbf{Section}} & \multicolumn{1}{c}{\textbf{Question}} & 
\multicolumn{1}{c}{\textbf{Targets}} &
\multicolumn{1}{c}{\textbf{Capabilities}} &
\multicolumn{1}{c|}{\textbf{Astro2020}}\\ \hline 
\endfirsthead

\multicolumn{6}{c}%
{{\bfseries \tablename\ \thetable{} -- continued from previous page}} \\
\hline \multicolumn{1}{|c}{\textbf{ID}} & \multicolumn{1}{c}{\textbf{Section}} & \multicolumn{1}{c}{\textbf{Question}} & 
\multicolumn{1}{c}{\textbf{Targets}} &
\multicolumn{1}{c}{\textbf{Capabilities}} &
\multicolumn{1}{c|}{\textbf{Astro2020}}\\ \hline 
\endhead

\hline \multicolumn{6}{|r|}{{\emph{Continued on next page}}} \\ \hline
\endfoot

\hline \hline
\endlastfoot
11 & \ref{sssec:ozone_onset} & What are the timescales required for Earth-like planets to develop oxygen-rich atmospheres? & 25 Earth-like planets & high-contrast spectroscopy ($200-350$~nm, $R \sim 7$) & E-Q2, E-Q3 \\
\hline
3 & \ref{sssec:rocky_sub} & What are the frequencies of sub-Neptune sized planets with various compositions? & 100 planets within the habitable zones of FGK stars (planet masses $3 - 20M_\oplus$; measured to $>10\sigma$ precision) & high-contrast spectroscopy ($250-400$~nm at $R>7$,  $400 - 1000$~nm at $R>140$, \& $1000 - 1800$~nm at $R>70$); polarimetry (multiple broad bands, error $=1\%$) & E-Q1, E-Q2, E-Q3\\
\hline
21 & \ref{sssec:retention} & What are the conditions required for Earth-mass terrestrial planets to retain significant volatiles? & $\geq 40$ terrestrial planets spanning a wide range of irradiation levels orbiting stars with a range of spectral types (FGKM) and ages & high-contrast spectroscopy \& spectropolarimetry ($200-2000$~nm, $R\sim100$) & E-Q1, E-Q2, E-Q3 \\
\hline
14 & \ref{sssec:escape} & How well do small planets retain their atmospheres? & Roughly 50 transiting planets at a variety of temperatures, ages, and masses & spectroscopy ($100-300$~nm, $R > 45,000$) & E-Q2, E-Q3, G-Q4 \\
\hline
18 & \ref{sssec:exovenus} & Do other stars host Venus-like planets that have experienced or are experiencing runaway greenhouses? & Earth-sized planets orbiting $0.3 - 1.5$~AU from solar-type stars & high-contrast spectroscopy ($200 - >1500$~nm; includes polarimetry in Vis/NIR) & E-Q1, E-Q2, E-Q3\\
\hline
17 & \ref{sssec:id_oceans} & Do planets orbiting other stars have interior oceans? How common are such planets? & $\ge 30$ transiting, potentially icy exoplanets & spectroscopy ($50-210$~nm) & E-Q1, E-Q2, E-Q3 \\
\hline
1 & \ref{sssec:oceans_habitable} & How habitable are icy bodies in the solar system? & solar system icy bodies &  IFS ($50-5000$~nm,  $R \gtrsim 10,000$, $\theta \le 20$~mas, FOV$=3''\times3''$) & E-Q2, E-Q3, E-Q4\\
\hline
4 & \ref{sssec:surface_water} & How common are oceans on temperate terrestrial planets? & terrestrial planets orbiting within the habitable zone (20 for ocean glint; $\sim 12$ for ocean mapping) & high-contrast, polarimetric multi-band photometry or spectroscopy ($400-1100$~nm)&  E-Q2, E-Q3 \\
\hline
20 & \ref{sssec:survive_water} & How does planetary habitability vary over time? What is the frequency of surface oceans on planets of different ages? & terrestrial planets in or near the habitable zones ($0.25 S_\oplus \leq S_p \leq 2 S_\oplus$) of young ($< 1$~Gyr) and old ($>4$~Gyr) systems & high-contrast spectroscopy ($200 - 2000$~nm, $R\sim 1000$);  spectroscopy ($100-200$~nm, $R\sim30$) & E-Q1, E-Q2, E-Q3\\
\hline
2 & \ref{sssec:hab_system} & How do habitable zone planets acquire their water? Are there correlations between the water abundance of inner planets and the presence of outer planets? & 30~planetary systems containing Sun-like stars and habitable zone planets smaller than $1.75 R_\oplus$ &  high-contrast spectroscopy ($700 - 1500$~nm, $R > 40$) & E-Q1, E-Q3 \\
\hline
%NOT INCLUDED 13 & \ref{sssec:occ_small} & How prevalent are small planets, and is the presence of Earth-like planets correlated with  overall system architecture? & 100~nearby stars & $600 - 1500$~nm direct imaging sensitive to $0.8 - 1.2 R_\oplus$ planets receiving $0.5 - 1.3 S_\oplus$; contrast $10^{-11}$ & E-Q1, E-Q1a, E-Q1e \\
\hline
6 & \ref{sssec:occ_binary} & How does the presence of a nearby stellar companion affect the formation of habitable planets? & 200 FGK primary stars in binaries with semimajor axes $30 - 1000$~AU & direct imaging sensitive to planets with radii $1 - 1.5 R_\oplus$ within 2~AU & E-Q1, F-Q4 \\
\hline
26 & \ref{sssec:bfield} & How commonly do exoplanets have magnetic fields, and what are the properties of their fields? & direct detection: $\ge 10$ transiting planets within 150~pc; indirect detection: $\ge5$ planets within 100~pc & direct detection: spectropolarimetry ($120-2440$~nm, $R > 100,000$, full Stokes IQUV); indirect detection: spectropolarimetry ($92-1100$~nm, $R > 100,000$, full Stokes IQUV) & E-Q2, G-Q4 \\
\hline
34 & \ref{sssec:young_bfields} & What is the role of magnetic fields in shaping young planetary systems? & 200~young stars including 60~T~Tauri stars \& 15~Herbig stars & spectropolarimetry ($100-1500$~nm, $R>30,000$, full Stokes IQUV) & E-Q1, F-DA, F-Q3, F-Q4, G-Q2, G-Q3, G-Q4 \\
\hline
12 & \ref{sssec:giant_orbits} & How do dynamical processes shape the orbits of giant planets? & $30 - 40$ planets ($M_p \leq 30 M_J$; $5~\rm AU < a < 30~\rm AU$) & high-contrast spectroscopy ($500 - 5000$~nm; sensitive to NIR magnitudes $\leq 30.5$ at $\sim 63-1440$~mas) & E-Q1  \\
\hline
15 & \ref{sssec:reflected_giants} & What are the properties of individual gas giant planets, and which physical processes lead to the observed diversity of giant planets? & $>60$ giant planets orbiting a wide range of host stars & high-contrast imaging \& spectroscopy ($150 - 5000$~nm, separations $\sim 60-1000$~mas); high-contrast spectropolarimetry ($200 - 1800$~nm, separations $\sim 60-1000$~mas) & E-Q1, E-Q2 \\
\hline
16 & \ref{sssec:transits} & How does planetary rotation affect the dynamics, radiative properties, and chemistry of planetary atmospheres? & transiting planets with orbital periods $>20$~days and temperatures $<500$~K orbiting stars with $2 < V < 20$ & spectroscopy ($100 - \ge 5000$~nm; $R>6000$; 1~ppb precision; duration $\ge 500$~hr); spectropolarimetry ($300 - 1000$~nm) & E-Q2 \\
\hline
24 & \ref{sssec:highres_atmos} & What are the compositions of exoplanet atmospheres and how are they modified by escape? & 200 transiting planets ranging from Earth-like to Jupiter-like & spectropolarimetry ($100-2500$~nm, $R>100,000$, full Stokes IQUV) & E-Q1, E-Q2, G-Q4 \\ \hline
7 & \ref{sssec:solar} &  How did the formation and migration of giant planets affect the delivery of volatiles to the inner solar system and shape the architecture of the solar system? & small bodies in the solar system &  spectroscopy ($120-4800$~nm; $R\sim2000$); fast tracking ($\geq 500$~mas/s); rapid response ($\sim1$~day) & E-Q1, F-Q4 \\
\hline
5 & \ref{sssec:venus} & How do the morphology and composition of Venus's clouds evolve? & Venus & spectroscopy ($200->1500$~nm; $R\sim200$; $\theta \leq 30$~mas/pixel) & E-Q2, E-Q3\\
\hline
22 & \ref{sssec:mars} & How and where did Mars form? & Phobos, Deimos, asteroids within and beyond the main belt & spectroscopy ($100 - 3000$~nm); saturation mitigation; rapid response ($\sim1$~day) & E-Q1, F-Q4\\
\hline
28 & \ref{sssec:mars_post} &  How do the atmosphere and corona of Mars vary with time? & Mars & spectroscopy ($80-700$~nm \& NIR $\le1500$~nm; $\theta< 3.5$~mas; FOV$\ge50''$) & E-Q2, E-Q3 \\ \hline
27 & \ref{sssec:titan} & How do the chemical composition and dynamics of Titan's atmosphere vary with time? & Titan & spectroscopy ($121-2000$~nm; $R>100$, $\theta=50$~mas) &  E-Q2, E-Q3 \\ \hline

29 & \ref{sssec:solargiant} & How do planetary aurorae vary with time, and what are the connections between auroral properties, the solar wind, and the solar activity cycle? & solar system giant planets & photometry ($50-3500$~nm w/ filters at $600-700$~nm, 825~nm, \& 890~nm; UV FOV$\ge20''\times20''$; visible FOV$\sim120''\times120''$); spectroscopy ($50-3500$~nm; $R\sim100,000$ in UV, $R>1740$ at 480~nm, \& $R>3450$ at 930~nm;  $\theta=200$~mas over FOV$\ge10''\times 10''$ \& $\theta=20$~mas over FOV$=3''\times 3''$); rapid response ($\le 7$ days) & E-Q2, G-Q4  \\ 
\hline
25 & \ref{sssec:aurorae} & How do solar energetic particles affect the atmospheres of solar system giant planets? & solar system giant planets & spectropolarimetry ($100-1000$~nm; $R>50,000$) & E-Q2, G-Q4  \\ 
\hline
33 & \ref{sssec:aminos} & What is the distribution of amino acids within the solar system, and why does Earth life use left-handed amino acids? & comets & spectropolarimetry ($140-220$~nm; $R\sim300$; polarization error $\le0.1\%$) & E-Q1, E-Q3, E-Q4\\ 
\hline
23 & \ref{sssec:disk_winds} & How are disks depleted by winds, and how does depletion affect planet formation? & $\ge 100$ protostellar disks in the Milky Way, LMC, and SMC &  spectroscopy ($91 - 200$~nm, $R\geq 100,000$; $\theta \leq 50$~mas) & F-Q3, F-Q4, F-DA \\

\hline
8 & \ref{sssec:proto} & What are the properties of young planets, and how do they shape the protoplanetary disks in which they reside? & 200 protoplanets with masses $\geq 0.01 M_J$ orbiting stars with $V \leq 15$ & high-contrast photometry \& spectroscopy ($90-700$~nm; $R\sim 10,000$; $\theta=5~\rm mas-1'$; FOV$=2' \times 2'$; separations of $0\farcs005 - 60''$) & E-Q1, F-Q4, F-DA \\
\hline
9 & \ref{sssec:debris} & How are dust and gas distributed within disks, and do disks show signs of interactions with planets and planetesimals? & debris disks & spectroscopy ($140 - 3000$~nm; $R\gtrsim 80,000$; $\theta \lesssim 30$~mas; FOV$\ge 2' \times 2'$); high-contrast imaging ($1000-3000$~nm; $\theta \lesssim 30$~mas; FOV$\ge 2' \times 2'$); high-contrast spectropolarimetry ($1000-3000$~nm; $R>3000$; $\theta \lesssim 30$~mas; FOV$\ge 2' \times 2'$; polarimetric precision $<1\%$) & E-Q1, F-Q4, F-DA\\
\hline
19 & \ref{sssec:exozodi} & How are the distributions of dust and small bodies in and near the habitable zone of a mature planetary system connected to the dynamical state and history of the system? & $100 - 200$ FGK stars  & high-contrast imaging ($200-2000$~nm; $R\sim100$; $\theta=20$~mas); high-contrast spectroscopy ($400-2000$~nm; $R\sim100$; $\theta=20$~mas); high-contrast polarimetric imaging ($200-2000$~nm; $\theta=20$~mas;  polarization error $\leq 3\%$) & E-Q1, F-DA\\
\hline
31 & \ref{sssec:exorings} & What are the properties and prevalence of planetary rings? & $\sim100$~planets with planet-star separations of $1-10$~AU & detection: high-contrast imaging (Vis; FOV$\sim10''\times 10''$); characterization: high-contrast spectropolarimetry (VIS - 2$\mu$m; $R\sim100$; FOV$\sim10''\times 10''$) & E-Q1, E-Q2 \\
\hline
32 & \ref{sssec:exomoons} & What are the properties and prevalence of moons orbiting terrestrial planets? & $\geq 30$ terrestrial planets with semimajor axes $1-5$~AU orbiting stars within 20~pc & high-contrast imaging (Vis$-2000$~nm; photometric precision $\lesssim10\%$ over a few hours; FOV$>2''\times 2''$) & E-Q1, E-Q2 \\
\hline
30 & \ref{sssec:defense} & What are the properties and prevalence of near-Earth objects that could potentially impact the Earth? & Near-Earth objects & photometry (wide bandpass \& standard filters); non-sidereal tracking ($\ge 110$~mas/s) & E-Q1  \\

\hline \hline
\end{longtable}
\end{center}

%% file: ssic_remap.tex
%Reformatted 1 June, 2026
\setlength{\tabcolsep}{0.05in} %minimal horizontal padding between columns

\begin{center}
\begin{longtable}{
| >{\centering\arraybackslash}p{0.75in}                   % ID (Centered)
  >{\centering\arraybackslash}p{0.5in}                    % Section (Centered)
  >{\raggedright\arraybackslash}p{1.5in}                  % Focus (Automated Ragged-Right)
  >{\raggedright\arraybackslash}p{1.5in}                  % Targets (Automated Ragged-Right)
  >{\raggedright\arraybackslash}p{\cw} |                  % Capabilities (Automated Ragged-Right)
} 
\caption{Summarized Observations for Solar Systems in Context Science Cases} \label{tab:ssic_remap} \\

\hline \multicolumn{1}{|c}{\textbf{ID}} & \multicolumn{1}{c}{\textbf{Section}} & \multicolumn{1}{c}{\textbf{Focus}} & 
\multicolumn{1}{c}{\textbf{Targets}} &
\multicolumn{1}{c|}{\textbf{Capabilities}} \\ \hline 
\endfirsthead

\multicolumn{5}{c}%
{{\bfseries \tablename\ \thetable{} -- continued from previous page}} \\
\hline \multicolumn{1}{|c}{\textbf{ID}} & \multicolumn{1}{c}{\textbf{Section}} & \multicolumn{1}{c}{\textbf{Focus}} & 
\multicolumn{1}{c}{\textbf{Targets}} &
\multicolumn{1}{c|}{\textbf{Capabilities}} \\ \hline 
\endhead

\hline \multicolumn{5}{|r|}{{\emph{Continued on next page}}} \\ \hline
\endfoot
\hline \hline
\endlastfoot
\rowcolor{colorSSiC}
SSiC-A & \ref{sssec:ssic_a} & \emph{High-Contrast Spectroscopy in the UV} & planets \& FGK stars & Coronagraph ($200-400$~nm, $R=7$) \\
\hline 
11 & \ref{sssec:ozone_onset} & Atmospheres of terrestrial planets & 25 Earth-like planets & \harder N/A \easier red cut-off (350~nm) \\
\hline 
3a & \ref{sssec:rocky_sub} & Sub-Neptune sized planets & 100 planets within the habitable zones of FGK stars (planet masses $3 - 20M_\oplus$; measured to $>10\sigma$ precision) & \harder N/A \easier UV cut-off (250~nm) \\
\hline
21a & \ref{sssec:retention} & Atmospheric retention by terrestrial planets & $\geq 40$ terrestrial planets spanning a wide range of irradiation levels, ages, and host star spectral types (FGKM) & \harder spectral resolution ($R\sim100$) \easier N/A \\
\hline
18a & \ref{sssec:exovenus} & Do other stars host Venus-like planets that have experienced or are experiencing runaway greenhouses? & Earth-sized planets orbiting $0.3 - 1.5$~AU from nearby FGK stars & \harder closer angular separations ($\ge0.3$~AU) \easier N/A \\
\hline 
20a & \ref{sssec:survive_water} & Planetary habitability over time & terrestrial planets in or near the habitable zones ($0.25 S_\oplus \leq S_p \leq 2 S_\oplus$) of young ($< 1$~Gyr) and old ($>4$~Gyr) systems  & \harder  spectral resolution ($R\sim1000$) \easier N/A \\
\hline 
15a & \ref{sssec:reflected_giants} & Gas giants & $>60$ giant planets orbiting a wide range of host stars & \harder blue cut-off (150~nm), spectral resolution ($R=100$), wider angular separations ($\sim 60-1000$~mas) \easier N/A \\
\hline 
8a & \ref{sssec:proto} & Young planets \& interactions with protoplanetary disks & 200 protoplanets with masses $\geq 0.01 M_J$ orbiting stars with $V \leq 15$ & 
\harder UV cut-off (90~nm), spectral resolution ($R\sim 10,000$), spatial resolution ($\theta=5$~mas), FOV ($2' \times 2'$), wider angular separations ($\theta=5~\rm mas-1'$) \easier N/A \\ \hline \hline

\rowcolor{colorSSiC}
SSiC-B & \ref{sssec:ssic_b} & \emph{High-Contrast Spectroscopy in the Visible} & planets, disks, \& FGK stars & Coronagraph ($400-1100$~nm, $R=140$) \\
\hline 
3b & \ref{sssec:rocky_sub} & Sub-Neptune sized planets & 100 planets within the habitable zones of FGK stars (planet masses $3 - 20M_\oplus$; measured to $>10\sigma$ precision) & \harder N/A \easier red cut-off (1000~nm) \\
\hline 
21b & \ref{sssec:retention} & Atmospheric retention by terrestrial planets & $\geq 40$ terrestrial planets spanning a wide range of irradiation levels, ages, and host star spectral types (FGKM) & \harder N/A \easier spectral resolution ($R\sim100$) \\
\hline
20b & \ref{sssec:survive_water} & Planetary habitability over time & terrestrial planets in or near the habitable zones ($0.25 S_\oplus \leq S_p \leq 2 S_\oplus$) of young ($< 1$~Gyr) and old ($>4$~Gyr) systems  & \harder  spectral resolution ($R\sim1000$) \easier N/A \\
\hline 
2a & \ref{sssec:hab_system} & Volatile delivery to terrestrial planets & 30~planetary systems containing Sun-like stars and habitable zone planets smaller than $1.75 R_\oplus$ & \harder \easier spectral resolution ($R > 40$) \\
\hline
12a & \ref{sssec:giant_orbits} & Orbital dynamics of giant planets & $30 - 40$ planets ($M_p \leq 30 M_J$; $5~\rm AU < a < 30~\rm AU$) & \harder N/A \easier blue cut-off (500~nm) \\
\hline
15b & \ref{sssec:reflected_giants} & Gas giants & $>60$ giant planets orbiting a wide range of host stars & \harder spectral resolution ($R=1000$), wider angular separations ($\sim 60-1000$~mas) \easier N/A \\
\hline 
8b & \ref{sssec:proto} & Young planets \& interactions with protoplanetary disks & 200 protoplanets with masses $\geq 0.01 M_J$ orbiting stars with $V \leq 15$ & \harder spectral resolution ($R\sim 10,000$), spatial resolution ($\theta=5$~mas), FOV ($2' \times 2'$), wider angular separations ($\theta=5~\rm mas-1'$) \easier red cut-off (700~nm) \\
\hline
19a & \ref{sssec:exozodi} & Distributions of dust and small bodies in and near the habitable zone & $100 - 200$ FGK stars  & \harder spatial resolution ($\theta=20$~mas) \easier spectral resolution ($R\sim100$) \\
\hline \hline

\rowcolor{colorSSiC}
SSiC-C & \ref{sssec:ssic_c} & \emph{High-Contrast Spectroscopy in the NIR} & terrestrial planets & Coronagraph ($1100-1700$~nm, $R=70$) \\
\hline
3c & \ref{sssec:rocky_sub} & Sub-Neptune sized planets & 100 planets within the habitable zones of FGK stars (planet masses $3 - 20M_\oplus$; measured to $>10\sigma$ precision) & \harder wider wavelength range ($1000-1800$~nm) \easier N/A \\
\hline
21c & \ref{sssec:retention} & Atmospheric retention by terrestrial planets & $\geq 40$ terrestrial planets spanning a wide range of irradiation levels, ages, and host star spectral types (FGKM) & \harder red cut-off (2000~nm), spectral resolution ($R\sim100$) \easier N/A \\
\hline
20c & \ref{sssec:survive_water} & Planetary habitability over time & terrestrial planets in or near the habitable zones ($0.25 S_\oplus \leq S_p \leq 2 S_\oplus$) of young ($< 1$~Gyr) and old ($>4$~Gyr) systems  & \harder  red cut-off (2000~nm), spectral resolution ($R\sim1000$) \easier N/A \\
\hline 
2b & \ref{sssec:hab_system} & Volatile delivery to terrestrial planets & 30~planetary systems containing Sun-like stars and habitable zone planets smaller than $1.75 R_\oplus$ & \harder \easier red cut-off (1500~nm), spectral resolution ($R > 40$) \\
\hline
12b & \ref{sssec:giant_orbits} & Orbital dynamics of giant planets & $30 - 40$ planets ($M_p \leq 30 M_J$; $5~\rm AU < a < 30~\rm AU$) & \harder red cut-off (5000~nm) \easier N/A \\
\hline
15c & \ref{sssec:reflected_giants} & Gas giants & $>60$ giant planets orbiting a wide range of host stars & \harder red cut-off (5000~nm), spectral resolution ($R=1000$), wider angular separations ($\sim 60-1000$~mas) \easier N/A \\
\hline 
19b & \ref{sssec:exozodi} & Distributions of dust and small bodies in and near the habitable zone & $100 - 200$ FGK stars  & \harder red cut-off (2000~nm), spectral resolution ($R\sim100$), spatial resolution ($\theta=20$~mas) \easier N/A \\

\hline \hline
\rowcolor{colorSSiC}
SSiC-D & \ref{sssec:ssic_d} & \emph{High-Contrast Spectropolarimetry in the UV} & planets, protoplanetary disks, \& FGK stars & Coronagraph ($200-400$~nm, $R=7$, polarimetry) \\
\hline 
21d & \ref{sssec:retention} & Atmospheric retention by terrestrial planets & $\geq 40$ terrestrial planets spanning a wide range of irradiation levels, ages, and host star spectral types (FGKM) & \harder spectral resolution ($R\sim100$) \easier N/A \\
\hline
15d & \ref{sssec:reflected_giants} & Gas giants & $>60$ giant planets orbiting a wide range of host stars & \harder wider angular separations ($\sim 60-1000$~mas) \easier N/A \\ \hline 
8c & \ref{sssec:proto} & Young planets \& interactions with protoplanetary disks & 200 protoplanets with masses $\geq 0.01 M_J$ orbiting stars with $V \leq 15$ & 
\harder UV cut-off (90~nm), spectral resolution ($R\sim 10,000$), spatial resolution ($\theta=5$~mas), FOV ($2' \times 2'$), wider angular separations ($\theta=5~\rm mas-1'$) \easier N/A \\ \hline 
9a & \ref{sssec:debris} & Distribution of dust \& gas in debris disks \& interactions with planets & debris disks & 
\harder UV cut-off (140~nm), spectral resolution ($R > 3000$), spatial resolution ($\theta\lesssim 30$~mas), FOV ($\geq 2' \times 2'$) \easier N/A \\
\hline \hline

\rowcolor{colorSSiC}
SSiC-E & \ref{sssec:ssic_e} & \emph{High-Contrast Spectropolarimetry in the Visible} & planets, protoplanetary disks, \& FGK stars & Coronagraph ($400-1100$~nm, $R=140$, polarimetry) \\
\hline 
21e & \ref{sssec:retention} & Atmospheric retention by terrestrial planets & $\geq 40$ terrestrial planets spanning a wide range of irradiation levels, ages, and host star spectral types (FGKM) & \harder N/A \easier  spectral resolution ($R\sim100$)\\
\hline
18b & \ref{sssec:exovenus} & Do other stars host Venus-like planets that have experienced or are experiencing runaway greenhouses? & Earth-sized planets orbiting $0.3 - 1.5$~AU from nearby FGK stars & \harder closer angular separations ($\ge0.3$~AU) \easier N/A \\
\hline 
15e & \ref{sssec:reflected_giants} & Gas giants & $>60$ giant planets orbiting a wide range of host stars & \harder wider angular separations ($\sim 60-1000$~mas) \easier N/A \\
\hline
8d & \ref{sssec:proto} & Young planets \& interactions with protoplanetary disks & 200 protoplanets with masses $\geq 0.01 M_J$ orbiting stars with $V \leq 15$ & \harder spectral resolution ($R\sim 10,000$), spatial resolution ($\theta=5$~mas), FOV ($2' \times 2'$), wider angular separations ($\theta=5~\rm mas-1'$) \easier red cut-off (700~nm) \\
\hline
9b & \ref{sssec:debris} & Distribution of dust \& gas in debris disks \& interactions with planets & debris disks & 
\harder spectral resolution ($R > 3000$), spatial resolution ($\theta\lesssim 30$~mas), FOV ($\geq 2' \times 2'$) \easier N/A \\
\hline
31a & \ref{sssec:exorings} & Exorings & $\sim100$~planets with planet-star separations of $1-10$~AU and evidence of rings & \harder FOV ($\sim10''\times 10''$) \easier spectral resolution ($R\sim100$) \\

\hline \hline
\rowcolor{colorSSiC}
SSiC-F & \ref{sssec:ssic_f} & \emph{High-Contrast Spectropolarimetry in the NIR} & planets & Coronagraph ($1100-1700$~nm, $R=70$, polarimetry) \\
\hline
21f & \ref{sssec:retention} & Atmospheric retention by terrestrial planets & $\geq 40$ terrestrial planets spanning a wide range of irradiation levels, ages, and host star spectral types (FGKM) & \harder red cut-off (2000~nm), spectral resolution ($R\sim100$) \easier N/A \\
\hline
18c & \ref{sssec:exovenus} & Do other stars host Venus-like planets that have experienced or are experiencing runaway greenhouses? & Earth-sized planets orbiting $0.3 - 1.5$~AU from nearby FGK stars & \harder closer angular separations ($\ge0.3$~AU) \easier red cut-off (1500~nm) \\
\hline 
15f & \ref{sssec:reflected_giants} & Gas giants & $>60$ giant planets orbiting a wide range of host stars & \harder red cut-off (1800~nm), wider angular separations ($\sim 60-1000$~mas) \easier N/A \\
\hline
9c & \ref{sssec:debris} & Distribution of dust \& gas in debris disks \& interactions with planets & debris disks & 
\harder red cut-off (3000~nm), spectral resolution ($R > 3000$), spatial resolution ($\theta\lesssim 30$~mas), FOV ($\geq 2' \times 2'$) \easier N/A \\
\hline
31b & \ref{sssec:exorings} & Exorings & $\sim100$~planets with planet-star separations of $1-10$~AU and evidence of rings & \harder red cut-off (2000~nm), spectral resolution ($R\sim100$), FOV ($\sim10''\times 10''$) \easier N/A \\
\hline \hline
\pagebreak
\rowcolor{colorSSiC}
SSiC-G & \ref{sssec:ssic_g} & \emph{High-Contrast Polarimetric Imaging} & planets \& FGK stars & Coronagraph ($300-1700$~nm, full Stokes (IQUV), polarimetric precision $\le 1\%$) \\
\hline 
3d & \ref{sssec:rocky_sub} & Sub-Neptune sized planets & 100 planets within the habitable zones of FGK stars (planet masses $3 - 20M_\oplus$; measured to $>10\sigma$ precision) & \harder broader wavelength range ($250-1800$~nm) \easier N/A \\
\hline
4 & \ref{sssec:surface_water} & How common are oceans on temperate terrestrial planets? & terrestrial planets orbiting within the habitable zone (20 for ocean glint; $\sim 12$ for ocean mapping) & \harder N/A \easier narrower wavelength range ($400-1100$~nm)\\
\hline
9d & \ref{sssec:debris} & Distribution of dust \& gas in debris disks \& interactions with planets & debris disks & 
\harder wider wavelength range ($140-3000$~nm), spatial resolution ($\theta\lesssim 30$~mas), FOV ($\geq 2' \times 2'$) \easier N/A \\
\hline
19d & \ref{sssec:exozodi} & Distributions of dust and small bodies in and near the habitable zone & $100 - 200$ FGK stars  & \harder wider wavelength range ($200-2000$~nm), spatial resolution ($\theta\le 20$~mas) \easier N/A \\
\hline \hline

\rowcolor{colorSSiC}
SSiC-H & \ref{sssec:ssic_h} & \emph{High-Contrast Imaging} & planets \& binary stars & Coronagraph ($200-1100$~nm, $\theta=10$~mas, FOV$=10''\times10''$) \\
\hline 
6 & \ref{sssec:occ_binary} & Habitable planets in binary systems & 200 FGK primary stars in binaries with semimajor axes $30 - 1000$~AU & \harder ability to observe binaries separated by $\leq10''$; sensitivity to planets with radii $1 - 1.5 R_\oplus$ within 2~AU \easier N/A \\
\hline
15g & \ref{sssec:reflected_giants} & Gas giants & $>60$ giant planets orbiting a wide range of host stars & \harder red cut-off (1800~nm) \easier FOV ($1''\times1''$) \\
\hline
8e & \ref{sssec:proto} & Young planets \& interactions with protoplanetary disks & 200 protoplanets with masses $\geq 0.01 M_J$ orbiting stars with $V \leq 15$ & 
\harder UV cut-off (90~nm), spatial resolution ($\theta=5$~mas), FOV ($2'\times2'$) \easier red cut-off (700~nm) \\
\hline 
9e & \ref{sssec:debris} & Distribution of dust \& gas in debris disks \& interactions with planets & debris disks & 
\harder wider wavelength range ($140-3000$~nm), FOV ($\geq 2' \times 2'$) \easier spatial resolution ($\theta\lesssim 30$~mas) \\
\hline 
19c & \ref{sssec:exozodi} & Distributions of dust and small bodies in and near the habitable zone & $100 - 200$ FGK stars  & \harder red cut-off (2000~nm) \easier spatial resolution ($\theta=20$~mas)\\
\hline 
31c & \ref{sssec:exorings} & Exorings & $\sim100$~planets with planet-star separations of $1-10$~AU & \harder N/A  \easier N/A \\
\hline 
32 & \ref{sssec:exomoons} & Exomoons & $\geq 30$ terrestrial planets with semimajor axes $1-5$~AU orbiting stars within 20~pc & \harder red cut-off (2000~nm), photometric precision $\lesssim10\%$ over a few hours  \easier FOV ($>2''\times 2''$) \\
\hline \hline

\rowcolor{colorSSiC}
SSiC-J & \ref{sssec:ssic_j} & \emph{Spectroscopy At Moderate Spectral Resolution} & transiting planets \& small bodies in the solar system  & Spectrograph ($100-3000$~nm, $R=3,000$) \\
\hline 
17 & \ref{sssec:id_oceans} & Exoplanets with interior oceans  & $\ge 30$ transiting, potentially icy exoplanets & \harder UV cut-off (50~nm) \easier red cut-off (210~nm) \\
\hline
20d & \ref{sssec:survive_water} & Planetary habitability over time & young ($< 1$~Gyr) and old ($>4$~Gyr) stars with terrestrial planets in or near the habitable zones ($0.25 S_\oplus \leq S_p \leq 2 S_\oplus$)  & \harder  N/A \easier red cut-off (200~nm), spectral resolution ($R\sim30$) \\
\hline 
16a & \ref{sssec:transits} & Role of rotation in planetary atmospheres & transiting planets with orbital periods $>20$~days and temperatures $<500$~K orbiting stars with $2 < V < 20$ & \harder red cut-off ($\ge 5000$~nm), spectral resolution ($R>6000$), photometric precision (1~ppb) \easier N/A \\
\hline
7 & \ref{sssec:solar} &  Influence of giant planets on volatile delivery in the solar system & small bodies in the solar system & \harder red cut-off (4800~nm)  \easier blue cut-off (120~nm), spectral resolution ($R\sim2000$) \\
\hline
22 & \ref{sssec:mars} & Formation of Mars & Phobos, Deimos, asteroids within and beyond the main belt & \harder red cut-off (3000~nm), saturation mitigation \easier N/A \\

\hline \hline
\rowcolor{colorSSiC}
SSiC-K & \ref{sssec:ssic_k} & \emph{UV Spectroscopy At High Spectral Resolution} & planets \& protostars & Spectrograph ($90-310$~nm, $R=100,000$)  \\
\hline 
14 & \ref{sssec:escape} & Atmospheric retention for small planets & $\sim50$ transiting planets at a variety of temperatures, ages, and masses & \harder N/A \easier wavelength range ($100-300$~nm), spectral resolution ($R > 45,000$) \\
\hline \hline
\rowcolor{colorSSiC}
SSiC-L & \ref{sssec:ssic_l} & \emph{Spectropolarimetry} & planets (transiting \& nearby), young stars, \& comets  & Spectropolarimeter ($100-1600$~nm, $R=120,000$, polarization (error $<10^{-6}$) \\
\hline
26a & \ref{sssec:bfield} & Exoplanetary magnetic fields & direct detection: $\ge 10$ transiting planets within 150~pc & \harder red cut-off (2440~nm) \easier UV cut-off (120~nm), spectral resolution ($R > 100,000$) \\
\hline
26b & \ref{sssec:bfield} & Exoplanetary magnetic fields & $\ge5$ planets within 100~pc & \harder UV cut-off (92~nm)  \easier red cut-off (1100~nm), spectral resolution ($R > 100,000$) \\
\hline
34 & \ref{sssec:young_bfields} & Role of magnetic fields in young planetary systems & 200~young stars including 60~T~Tauri stars \& 15~Herbig stars & \harder N/A \easier red cut-off (1500~nm), spectral resolution ($R > 30,000$) \\
\hline 
16b & \ref{sssec:transits} & Role of rotation in planetary atmospheres & transiting planets with orbital periods $>20$~days and temperatures $<500$~K orbiting stars with $2 < V < 20$ & \harder N/A \easier narrower wavelength range ($300 - 1000$~nm), spectral resolution ($R>6000$) \\
\hline 
24 & \ref{sssec:highres_atmos} & Composition \& escape of exoplanet atmospheres & 200 transiting planets ranging from Earth-like to Jupiter-like & \harder red cut-off (2500~nm) \easier spectral resolution ($R>100,000$) \\
\hline
25 & \ref{sssec:aurorae} & Planetary aurorae & solar system giant planets & \harder N/A \easier red cut-off (1000~nm), spectral resolution ($R>50,000$)   \\ 
\hline
33 & \ref{sssec:aminos} & Distribution of amino acids within the solar system & comets & \harder polarization error ($\le0.1\%$)  \easier narrower wavelength range ($140-220$~nm), spectral resolution ($R\sim300$) \\
\hline \hline
\rowcolor{colorSSiC}
SSiC-M & \ref{sssec:ssic_m} & \emph{IFS Observations at High Spatial Resolution} & debris disks, Mars, Titan, \& icy bodies in the solar system & IFS ($90-2500$~nm, $R=10,000$, $\theta=20$~mas, FOV=$10''\times10''$) \\
\hline 
1 & \ref{sssec:oceans_habitable} & Habitability of icy bodies in the solar system & solar system icy bodies & \harder wider wavelength range ($50-5000$~nm), spectral resolution ($R \gtrsim 10,000$) \easier FOV ($3''\times3''$) \\
\hline
5 & \ref{sssec:venus} & Morphology \& composition of Venus's clouds & Venus & \harder FOV ($\ge40''\times40''$) \easier narrower wavelength range ($200->1500$~nm), spectral resolution ($R\sim200$), spatial resolution ($\theta \leq 30$~mas) \\
\hline
28a, 28b & \ref{sssec:mars_post} & Atmosphere \& corona of Mars & Mars & \harder UV cut-off (80~nm), spatial resolution ($\theta< 3.5$~mas), FOV ($\ge50'' \times 50''$)  \easier red cut-off ($\le1500$~nm), spectral resolution ($R>3000$)  \\ 
\hline
27 & \ref{sssec:titan} & Composition \& dynamics of Titan's atmosphere & Titan & \harder N/A \easier narrower wavelength range ($121-2000$~nm), spectral resolution ($R>100$), spatial resolution ($\theta=50$~mas), FOV ($1''\times1''$) \\ 
\hline
29a & \ref{sssec:solargiant} & Planetary aurorae & solar system giant planets & \harder UV cut-off (50~nm), spectral resolution ($R\sim100,000$) \easier red cut-off (400~nm), FOV ($\ge3''\times 3''$) \\
\hline
29b & \ref{sssec:solargiant} & Planetary aurorae & solar system giant planets & \harder N/A \easier narrower wavelength range ($400-700$~nm), spectral resolution ($R>1740$), FOV ($\ge3''\times 3''$) \\
\hline
29c & \ref{sssec:solargiant} & Planetary aurorae & solar system giant planets & \harder spectral resolution ($R>3450$), red cut-off (3500~nm) \easier blue cut-off (700~nm), FOV ($\ge3''\times 3''$) \\
\hline 
23 & \ref{sssec:disk_winds} & Disk depletion and planet formation & $\ge 100$ protostellar disks in the Milky Way, LMC, and SMC & \harder spectral resolution ($R\sim100,000$) \easier narrower wavelength range ($91-200$~nm), spatial resolution ($\theta \leq 50$~mas) \\
\hline
9f & \ref{sssec:debris} & Distribution of dust \& gas in debris disks \& interactions with planets & debris disks & \harder red cut-off (3000~nm), spectral resolution ($R\gtrsim 80,000$), FOV ($120''\times120''$) \easier blue cut-off (140~nm), spatial resolution ($\theta \lesssim 30$~mas) \\
\hline \hline

\rowcolor{colorSSiC}
SSiC-N & \ref{sssec:ssic_n} & \emph{IFS Observations at Moderate Spatial Resolution} & targets & IFS ($100-2000$~nm, $R=3500$, $\theta=200$~mas, FOV$=10''\times10''$) \\
\hline
29d & \ref{sssec:solargiant} & Planetary aurorae & solar system giant planets & \harder UV cut-off (50~nm), spectral resolution ($R\sim100,000$) \easier red cut-off (400~nm) \\
\hline
29e & \ref{sssec:solargiant} & Planetary aurorae & solar system giant planets & \harder N/A \easier narrower wavelength range ($400-700$~nm), spectral resolution ($R>1740$)\\
\hline
29f & \ref{sssec:solargiant} & Planetary aurorae & solar system giant planets & \harder red cut-off (3500~nm) \easier blue cut-off (700~nm) \\
\hline \hline

\rowcolor{colorSSiC}
SSiC-P & \ref{sssec:ssic_p} & \emph{Photometric Observations} & targets & Imager ($95-2000$~nm, $\theta=20$~mas, FOV$=120''\times120''$) \\
\hline 
29g & \ref{sssec:solargiant} & Planetary aurorae & solar system giant planets & \harder UV cut-off (50~nm) \easier red cut-off (400~nm), FOV ($\ge20''\times20''$) \\
\hline
29h & \ref{sssec:solargiant} & Planetary aurorae & solar system giant planets & \harder red cut-off (3500~nm), multiple filters ($600-700$~nm, 825~nm, \& 890~nm) \easier blue cut-off (400~nm), FOV ($\sim120''\times120''$)  \\ 
\hline
30 & \ref{sssec:defense} & Near-Earth objects & near-Earth objects that could potentially impact the Earth & \harder wide bandpass \& standard filters \easier narrower wavelength range (Vis/NIR only) \\
\hline \hline
\end{longtable}
\end{center}

%% file: lw_scdds.tex
\begin{center}
\begin{longtable}{
| >{\centering}p{0.25in} 
>{\centering}p{0.25in} 
>{\centering}p{1.5in} 
>{\centering}p{1.5in} 
>{\centering}p{1.25in} 
>{\centering\arraybackslash}p{1.25in} | 
} 
\caption{Science questions, target samples, and observational capabilities for living worlds science cases. The numbers in the ID column refer to the identification number assigned by the project office. All of these science cases are for living worlds, so the ``SCDD-LW-'' prefix has been omitted. When ranges of numbers are given, the easier number corresponds to the capability for breakthrough science and the harder number is for super-breakthrough science.} \label{tab:lw_scdds} \\

\hline \multicolumn{1}{|c}{\textbf{ID}} & \multicolumn{1}{c}{\textbf{Section}} & \multicolumn{1}{c}{\textbf{Question}} & 
\multicolumn{1}{c}{\textbf{Targets}} &
\multicolumn{1}{c}{\textbf{Capabilities}} &
\multicolumn{1}{c|}{\textbf{Astro2020}}\\ \hline 
\endfirsthead

\multicolumn{6}{c}%
{{\bfseries \tablename\ \thetable{} -- continued from previous page}} \\
\hline \multicolumn{1}{|c}{\textbf{ID}} & \multicolumn{1}{c}{\textbf{Section}} & \multicolumn{1}{c}{\textbf{Question}} & 
\multicolumn{1}{c}{\textbf{Targets}} &
\multicolumn{1}{c}{\textbf{Capabilities}} &
\multicolumn{1}{c|}{\textbf{Astro2020}}\\ \hline 
\endhead

\hline \multicolumn{6}{|r|}{{\emph{Continued on next page}}} \\ \hline
\endfoot

\hline \hline
\endlastfoot 
1 & \ref{sssec:life} & What is the frequency of Earth-like global biospheres in the galaxy? & $33-67$ temperate terrestrial planets & high-contrast spectroscopy ($250-400$~nm at $R \sim 7$; $400-1100$~nm at $R \sim 140$; $1100-1700$~nm at $R \sim 70$) & E-Q1, E-Q2, E-Q3, E-Q4, E-DA \\
\hline
3 & \ref{sssec:origin} & What is the range of conditions within which life can originate? & temperate terrestrial planets &  high-contrast spectroscopy (UV $\geq 200$~nm, Vis, \& NIR) & E-Q2, E-Q3, E-DA \\
\hline
4 & \ref{sssec:prebiosignatures} & At what frequency do prebiotic environments emerge in the galaxy? & temperate terrestrial planets &  high-contrast spectroscopy ($200-400$~nm at $R \sim 7$; $400-1100$~nm at $R \sim 140$; $1100-1700$~nm at $R \sim 70$) & E-Q2, E-Q3, E-DA \\
\hline
5 & \ref{sssec:technosignatures} & Are there signs of technology elsewhere in the galaxy? & planets &  high-contrast spectroscopy ($300 - 900$~nm), photometry (Vis), \& polarimetry (Vis) & E-Q4, E-DA \\
\hline
6 & \ref{sssec:lawdki} & How diverse is extraterrestrial life, and does life \emph{unlike} Earth life exist elsewhere in the universe? & 30 terrestrial planets with separations of $0.1 - 5$~AU & high-contrast spectroscopy (UV $-3000$~nm, $R \sim 100 - 1000$) & E-Q4, E-DA \\
\hline
14 & \ref{sssec:surface_bio} & Do planets orbiting other stars have detectable global biospheres? & $20-40$~potentially inhabited planets & high-contrast spectroscopy ($500-1100$~nm at $R \lesssim 130$)  & E-Q4, E-DA \\
\hline
7 & \ref{sssec:polbio} & Do planets orbiting other stars have surface liquid water or detectable global biospheres? & temperate terrestrial planets & high-contrast polarimetric imaging or spectroscopy ($10\%$ bandpasses at 350, 450, 550, \& 850~nm or $400-850$~nm spectra at $R\sim50-100$; linear polarization); high-contrast spectropolarimetry ($600-750$~nm, $R\sim100$, circular polarization)
& E-Q1, E-Q2, E-Q3, E-Q4, E-DA \\
\hline
13 & \ref{sssec:seasonality} & Do other planets display seasonal variations that might indicate biological cycles? & temperate terrestrial planets & high-contrast spectroscopy ($250 - 400$~nm at $R\sim7$; $400 - 1000$~nm at $R\sim140$) at $6-8$ different orbital locations & E-Q2, E-Q3, E-Q4, E-DA \\
\hline
10 & \ref{sssec:fpbio} & How reliable are various biosignatures? How frequently is oxygen found in the atmospheres of planets that are not expected to be inhabited? & terrestrial planets in the Venus zone & high-contrast spectroscopy ($250 - 400$~nm at $R\sim7$;  $400-1300$~nm at $R\sim200$) & E-Q2, E-Q3, E-Q4, E-DA\\ 
\hline
12 & \ref{sssec:mp} & What are the masses of potentially habitable planets? & planet host stars & photometry ($\theta = 11$~mas, FOV=$6'\times 6'$, astrometric precision $=0.03~\mu$as per epoch) & E-Q1, E-Q2, E-Q3, E-Q4, E-DA \\
\hline
\end{longtable}
\end{center}

%% file: lw_remap.tex
%Reformatted 1 June, 2026
\setlength{\tabcolsep}{0.05in} %minimal horizontal padding between columns

\begin{center}
\begin{longtable}{
| >{\centering\arraybackslash}p{0.75in}                   % ID (Centered)
  >{\centering\arraybackslash}p{0.5in}                    % Section (Centered)
  >{\raggedright\arraybackslash}p{1.5in}                  % Focus (Automated Ragged-Right)
  >{\raggedright\arraybackslash}p{1.5in}                  % Targets (Automated Ragged-Right)
  >{\raggedright\arraybackslash}p{\cw} |                  % Capabilities (Automated Ragged-Right)
} 
\caption{Summarized Observations for Living Worlds Science Cases} \label{tab:lw_remap} \\

\hline \multicolumn{1}{|c}{\textbf{ID}} & \multicolumn{1}{c}{\textbf{Section}} & \multicolumn{1}{c}{\textbf{Focus}} & 
\multicolumn{1}{c}{\textbf{Targets}} &
\multicolumn{1}{c|}{\textbf{Capabilities}} \\ \hline 
\endfirsthead

\multicolumn{5}{c}%
{{\bfseries \tablename\ \thetable{} -- continued from previous page}} \\
\hline \multicolumn{1}{|c}{\textbf{ID}} & \multicolumn{1}{c}{\textbf{Section}} & \multicolumn{1}{c}{\textbf{Focus}} & 
\multicolumn{1}{c}{\textbf{Targets}} &
\multicolumn{1}{c|}{\textbf{Capabilities}} \\ \hline 
\endhead

\hline \multicolumn{5}{|r|}{{\emph{Continued on next page}}} \\ \hline
\endfoot
\hline \hline
\endlastfoot
\rowcolor{colorLW}
LW-A & \ref{sssec:lw_a} & \emph{High-Contrast Spectroscopy in the UV} & terrestrial planets & Coronagraph ($200-400$~nm, $R=7$) \\
\hline 
1a & \ref{sssec:life} & Frequency of Earth-like global biospheres & $33-67$ temperate terrestrial planets & \harder N/A \easier UV cut-off (250~nm) \\
\hline
3a & \ref{sssec:origin} & Origin of life & temperate terrestrial planets &  \harder N/A \easier N/A \\
\hline
4a & \ref{sssec:prebiosignatures} & Frequency of prebiotic environments & temperate terrestrial planets & \harder N/A \easier N/A \\
\hline
5a & \ref{sssec:technosignatures} & Technosignatures & planets & \harder N/A \easier UV cut-off (300~nm) \\
\hline
13a & \ref{sssec:seasonality} & Seasonal variations & temperate terrestrial planets &  \harder N/A \easier UV cut-off (250~nm) \\
\hline
10a & \ref{sssec:fpbio} & Oxygen levels of Venus analogs & terrestrial planets in the Venus zone & \harder N/A \easier UV cut-off (250~nm) \\
\hline \hline

\rowcolor{colorLW}
LW-B & \ref{sssec:lw_b} & \emph{High-Contrast Spectroscopy in the Visible} & terrestrial planets & Coronagraph ($400-1100$~nm, $R=140$) \\
\hline 
1b & \ref{sssec:life} & Frequency of Earth-like global biospheres & $33-67$ temperate terrestrial planets & \harder N/A \easier N/A \\
\hline
3b & \ref{sssec:origin} & Origin of life & temperate terrestrial planets &  \harder N/A \easier N/A \\
\hline
4b & \ref{sssec:prebiosignatures} & Frequency of prebiotic environments & temperate terrestrial planets &   \harder N/A \easier N/A \\
\hline
5b & \ref{sssec:technosignatures} & Technosignatures & planets & \harder N/A \easier red cut-off (900~nm) \\
\hline
6 & \ref{sssec:lawdki} & Diversity of extraterrestrial life & 30 terrestrial planets with separations of $0.1 - 5$~AU & \harder spectral resolution ($R\sim 100 - 1000$) \easier N/A \\
\hline
13b & \ref{sssec:seasonality} & Seasonal variations & temperate terrestrial planets &  \harder N/A \easier red cut-off (1000~nm) \\
\hline
14 & \ref{sssec:surface_bio} & Detection of global biospheres & $20-40$~potentially inhabited planets & \harder N/A \easier blue cut-off (500~nm), spectral resolution ($R \lesssim 130$) \\
\hline
10b & \ref{sssec:fpbio} & Oxygen levels of Venus analogs & terrestrial planets in the Venus zone & \harder spectral resolution ($R\sim200$) \easier N/A \\
\hline \hline

\rowcolor{colorLW}
LW-C & \ref{sssec:lw_c} & \emph{High-Contrast Spectroscopy in the NIR} & terrestrial planets & Coronagraph ($1100-1700$~nm, $R=70$) \\
\hline 
1c & \ref{sssec:life} & Frequency of Earth-like global biospheres & $33-67$ temperate terrestrial planets & \harder N/A \easier N/A \\
\hline
3c & \ref{sssec:origin} & Origin of life & temperate terrestrial planets &  \harder N/A \easier N/A \\
\hline
4c & \ref{sssec:prebiosignatures} & Frequency of prebiotic environments & temperate terrestrial planets & \harder N/A \easier N/A \\
\hline
10c & \ref{sssec:fpbio} & Oxygen levels of Venus analogs & terrestrial planets in the Venus zone & \harder spectral resolution ($R\sim200$) \easier red cut-off (1300~nm) \\
\hline \hline
\rowcolor{colorLW}
LW-D & \ref{sssec:lw_d} & \emph{High-Contrast Spectropolarimetry} & planets & Coronagraph ($350-850$~nm, $R=100$, full Stokes (IQUV), polarimetric precision $\le 1\%$)  \\
\hline 
7a & \ref{sssec:polbio} & Surface distribution of potential exo-biopigments  & temperate terrestrial planets & \harder N/A  \easier blue cut-off (400~nm), spectral resolution ($R\sim50-100$), linear polarization (Stokes IQU) \\
\hline
7b & \ref{sssec:polbio} & Chirality of potential exo-biopigments & temperate terrestrial planets & \harder polarimetric precision $\le 0.01\%$  \easier narrower wavelength range ($600-750$~nm) \\
\hline \hline
\rowcolor{colorLW}
LW-E & \ref{sssec:lw_e} & \emph{High-Contrast Polarimetric Imaging} & planets & Coronagraph ($300-1700$~nm, full Stokes (IQUV), polarimetric precision $\le 1\%$) \\
\hline 
5c & \ref{sssec:technosignatures} & Technosignatures & planets & \harder N/A  \easier red cut-off (900~nm) \\
\hline
7c & \ref{sssec:polbio} & Surface liquid water \& global biospheres & temperate terrestrial planets & \harder multiple filters ($10\%$ bandpasses at 350, 450, 550, \& 850~nm)  \easier narrower wavelength range ($350-850$~nm), linear polarization (Stokes IQU) \\
\hline \hline
\rowcolor{colorLW}
LW-F & \ref{sssec:lw_f} & \emph{High-Contrast Imaging} & planets & Coronagraph ($300-900$~nm) \\
\hline 
5d & \ref{sssec:technosignatures} & Technosignatures & planets & \harder N/A  \easier N/A \\
\hline \hline
\rowcolor{colorLW}
LW-G & \ref{sssec:lw_g} & \emph{Astrometric Observations} & planet host stars & Imager ($400-700$~nm, $\theta=11$~mas, FOV$=6'\times6'$, astrometric precision $=0.03~\mu$as per epoch) \\
\hline 
12 & \ref{sssec:mp} & Planet masses & planet host stars & \harder N/A \easier N/A \\
\hline \hline
\end{longtable}
\end{center}

%% file: all_modes.tex
\setlength{\tabcolsep}{0.05in} %minimal horizontal padding between columns

\begin{center}
\begin{longtable}{
| >{\centering\arraybackslash}p{0.5in}                    % ID 
  >{\centering\arraybackslash}p{0.5in}                    % Section 
  >{\raggedleft\arraybackslash}p{.4in}                   % Red 
  >{\raggedleft\arraybackslash}p{.4in}                   % Blue 
  >{\raggedleft\arraybackslash}p{0.5in}                  % R 
    >{\raggedleft\arraybackslash}p{.35in}                 % theta 
  >{\centering\arraybackslash}p{.5in}                   %  FOV
  >{\centering\arraybackslash}p{1.4in}                    % notes 
   >{\raggedleft\arraybackslash}p{1.45in} | % science cases 
} 
\caption{Summarized Observing Modes for All Science Cases} \label{tab:modes} \\

\hline \multicolumn{1}{|c}{\textbf{}} & \multicolumn{1}{c}{\textbf{}} & \multicolumn{1}{r}{\textbf{$\lambda_{\rm blue}$}} & 
\multicolumn{1}{r}{\textbf{$\lambda_{\rm red}$}} &
\multicolumn{1}{r}{} & 
\multicolumn{1}{r}{\textbf{$\theta$}} &
\multicolumn{1}{c}{} & 
\multicolumn{1}{c}{} &
\multicolumn{1}{c|}{}\\
\multicolumn{1}{|c}{\textbf{ID}} & \multicolumn{1}{c}{\textbf{Section}} & \multicolumn{1}{r}{(nm)} & 
\multicolumn{1}{r}{(nm)} &
\multicolumn{1}{r}{\textbf{$R$}} & 
\multicolumn{1}{r}{(mas)} &
\multicolumn{1}{c}{\textbf{FOV}} & 
\multicolumn{1}{c}{\textbf{Notes}} &
\multicolumn{1}{r|}{\textbf{SCDDs}}\\
\hline \hline
\endfirsthead

\multicolumn{9}{c}%
{{\bfseries \tablename\ \thetable{} -- continued from previous page}} \\
\hline \multicolumn{1}{|c}{\textbf{}} & \multicolumn{1}{c}{\textbf{}} & \multicolumn{1}{r}{\textbf{$\lambda_{\rm blue}$}} & 
\multicolumn{1}{r}{\textbf{$\lambda_{\rm red}$}} &
\multicolumn{1}{r}{} & 
\multicolumn{1}{r}{\textbf{$\theta$}} &
\multicolumn{1}{c}{} & 
\multicolumn{1}{c}{} &
\multicolumn{1}{c|}{}\\
\multicolumn{1}{|c}{\textbf{ID}} & \multicolumn{1}{c}{\textbf{Section}} & \multicolumn{1}{r}{(nm)} & 
\multicolumn{1}{r}{(nm)} &
\multicolumn{1}{r}{\textbf{$R$}} & 
\multicolumn{1}{r}{(mas)} &
\multicolumn{1}{c}{\textbf{FOV}} & 
\multicolumn{1}{c}{\textbf{Notes}} &
\multicolumn{1}{r|}{\textbf{SCDDs}}\\ \hline
\endhead

\hline \multicolumn{9}{|r|}{{\emph{Continued on next page}}} \\ \hline
\endfoot
\hline \hline
\endlastfoot
\multicolumn{9}{|c|}{IFS Observations (Section~\ref{sssec:ifs})}\\
\hline 

\rowcolor{colorGG} % 
GG-A & \ref{sssec:gg_a} & 90 & 2500 & 10,000  & 20 & $10''\times10''$ & \textbf{BASELINE} & 
  \mbox{5b~(\ref{sssec:agn_outflow})}, \mbox{4~(\ref{sssec:bh_quiescent})}, 
  \mbox{2~(\ref{sssec:bh_mass_spin})}, \mbox{3~(\ref{sssec:bh_torus})}, 
  \mbox{9a~(\ref{sssec:lyman_indirect})}, \mbox{13a~(\ref{sssec:dm_lensing})}, 
  \mbox{17a~(\ref{sssec:cgm_elm})}
\\
\hline
\rowcolor{colorGG}
GG-B & \ref{sssec:gg_b} & 95 & 500 &  30,000 & 10 & $10''\times10''$ &  UV/blue, High  $R$ & 
  \mbox{5a~(\ref{sssec:agn_outflow})}, \mbox{6~(\ref{sssec:resolve_reionization})}, 
  \mbox{7~(\ref{sssec:lyman_escape})}
\\
\hline
\rowcolor{colorEE}
EE-A & \ref{sssec:ee_a} & 90 & 500 & 10,000 & 20 &  $10''\times10''$ & $\approx$GG-A ($\downarrow \lambda_{\rm{red}}$) & 
  \mbox{1~(\ref{sssec:massive_stars_lowZ})}, \mbox{12~(\ref{sssec:ism_uv})}, 
  \mbox{5~(\ref{sssec:vms})}
\\
\hline 
\rowcolor{colorSSiC}
SSiC-M & \ref{sssec:ssic_m} & 90 & 2500 & 10,000 & 20 &  $10''\times10''$ & $\equiv$ GG-A & 
  \mbox{1~(\ref{sssec:oceans_habitable})}, \mbox{5~(\ref{sssec:venus})}, 
  \mbox{28~(\ref{sssec:mars_post})}, \mbox{27~(\ref{sssec:titan})}, 
  \mbox{29a~(\ref{sssec:solargiant})}, \mbox{29b~(\ref{sssec:solargiant})}, 
  \mbox{29c~(\ref{sssec:solargiant})}, \mbox{9a~(\ref{sssec:debris})}, 
  \mbox{9b~(\ref{sssec:debris})}
\\
\hline 
\rowcolor{colorSSiC}
SSiC-N & \ref{sssec:ssic_n} & 100 & 2000 & 3500 & 200 &  $10''\times10''$ & NIR, Low $R$  & 
  \mbox{29d~(\ref{sssec:solargiant})}, \mbox{29e~(\ref{sssec:solargiant})}, 
  \mbox{29f~(\ref{sssec:solargiant})}
\\
\hline \hline
\multicolumn{9}{c|}{MOS Observations (Section~\ref{sssec:mos})}\\
\hline
\rowcolor{colorGG}
GG-C & \ref{sssec:gg_c} & 94 & 350 & 3000 & 225 & $2'\times2'$ & UV Only, Low $R$, \mbox{Coarse $\theta$} & 
  \mbox{11~(\ref{sssec:ionizing_lf})}, \mbox{16b~(\ref{sssec:disk_cgm})}
\\
\hline
\rowcolor{colorGG}
GG-D & \ref{sssec:gg_d} & 94 & 350 & 40,000 &  500 & $6'\times6'$ & UV Only, High  $R$, \mbox{Coarse $\theta$}  & 
  \mbox{16a~(\ref{sssec:disk_cgm})}, \mbox{17b~(\ref{sssec:cgm_elm})}, 
  \mbox{18~(\ref{sssec:agn_feedback})}
\\
\hline
\rowcolor{colorEE}
EE-B & \ref{sssec:ee_b} & 95 & 2500 & 3000 & 10 & $2'\times2'$ & UV/Vis/NIR, \mbox{Low $R$}, Fine $\theta$  & 
  \mbox{2~(\ref{sssec:dust_extinction})}, \mbox{11a~(\ref{sssec:dust_uv})}, 
  \mbox{11b~(\ref{sssec:dust_uv})}
\\
\hline \hline
\multicolumn{9}{|c|}{Single-Object Spectroscopy  (Section~\ref{sssec:sos})}\\
\hline
\rowcolor{colorEE}
EE-C & \ref{sssec:ee_c} &110 & 500 & 10,000 & & & UV/blue, Medium $R$ & 
  \mbox{3~(\ref{sssec:flash_ccsne})}, \mbox{4a~(\ref{sssec:r_process_el})}
\\
\hline 
\rowcolor{colorEE}
EE-D & \ref{sssec:ee_d} &  90 & 310 & 100,000 & & & \textbf{BASELINE} & 
  \mbox{7a~(\ref{sssec:wds})}, \mbox{7b~(\ref{sssec:wds})}, 
  \mbox{8~(\ref{sssec:first_stars})}, \mbox{9~(\ref{sssec:r_process_nature})}
\\
\hline 
\rowcolor{colorSSiC}
SSiC-J & \ref{sssec:ssic_j} & 100 & 3000 & 3,000 & & & UV/Vis/NIR, Low $R$ & 
  \mbox{17~(\ref{sssec:id_oceans})}, \mbox{16a~(\ref{sssec:transits})}, 
  \mbox{7~(\ref{sssec:solar})}, \mbox{22~(\ref{sssec:mars})}
\\
\hline 
\rowcolor{colorSSiC}
SSiC-K & \ref{sssec:ssic_k} & 90 & 310 & 100,000 & & & $\equiv$~EE-D& 
  \mbox{14~(\ref{sssec:escape})}
\\
\hline \hline
\multicolumn{9}{|c|}{Spectropolarimetry (Section~\ref{sssec:specpol})}\\
\hline
\rowcolor{colorGG}
GG-E & \ref{sssec:gg_e} & 100 & 1600 & 120,000 & & & \textbf{BASELINE}  & 
  \mbox{19~(\ref{sssec:smbh_pol})}
\\
\hline
\rowcolor{colorEE}
EE-E & \ref{sssec:ee_e} &  100 & 1600 & 120,000 & & & $\equiv$ GG-E & 
  \mbox{15~(\ref{sssec:mag_stars})}
\\
\hline 
\rowcolor{colorSSiC}
SSiC-L & \ref{sssec:ssic_l} & 100 & 1600 & 120,000 & & & $\equiv$ GG-E & 
  \mbox{26a~(\ref{sssec:bfield})}, \mbox{26b~(\ref{sssec:bfield})}, 
  \mbox{34~(\ref{sssec:young_bfields})}, \mbox{16b~(\ref{sssec:transits})}, 
  \mbox{24~(\ref{sssec:highres_atmos})}, \mbox{33~(\ref{sssec:aminos})}
\\
\hline \hline
\multicolumn{9}{|c|}{Photometric Observations (Section~\ref{sssec:phot})}\\
\hline
\rowcolor{colorGG}
GG-F & \ref{sssec:gg_f} & 95 & 700 & & 20 & $4'\times 4'$ & $\approx$ EE-F ($\downarrow \lambda_{\mathrm{red}},\mathrm{FOV}$) & 
  \mbox{9b~(\ref{sssec:lyman_indirect})}, \mbox{12~(\ref{sssec:dm_power})}, 
  \mbox{13b~(\ref{sssec:dm_lensing})}, \mbox{14~(\ref{sssec:smbh_mergers})}
\\
\hline
\rowcolor{colorEE}
EE-F & \ref{sssec:ee_f} & 95 & 2500 & & 20 & $6'\times6'$ & \textbf{BASELINE} & 
  \mbox{4b~(\ref{sssec:r_process_el})}, \mbox{6~(\ref{sssec:resolved_stellar_pops})}, 
  \mbox{10~(\ref{sssec:distance_ladder})}
\\
\hline 
\rowcolor{colorSSiC}
SSiC-P & \ref{sssec:ssic_p} & 95 & 2000 & & 20 & $2'\times2'$ & $\approx$ EE-F ($\downarrow \lambda_{\mathrm{red}},\mathrm{FOV}$) & 
  \mbox{29g~(\ref{sssec:solargiant})}, \mbox{29h~(\ref{sssec:solargiant})}, 
  \mbox{30~(\ref{sssec:defense})}
\\
\hline 
\rowcolor{colorLW}
LW-G & \ref{sssec:lw_g} & 400 & 700 & & 11 & $6'\times6'$ & $\approx$ EE-F ($\uparrow \lambda_{\mathrm{blue}},\downarrow \lambda_{\mathrm{red}},\downarrow\theta$), astrometric precision per epoch $=0.03~\mu$as  & 
  \mbox{12~(\ref{sssec:mp})}
\\
\hline \hline
\pagebreak
\multicolumn{9}{|c|}{High-Contrast UV Spectroscopy  (Section~\ref{sssec:hc_uv_spec})}\\
\hline
\rowcolor{colorSSiC}
SSiC-A & \ref{sssec:ssic_a} & 200 & 400 & 7 & & & \textbf{BASELINE} & 
  \mbox{11~(\ref{sssec:ozone_onset})}, \mbox{3a~(\ref{sssec:rocky_sub})}, 
  \mbox{18a~(\ref{sssec:exovenus})}, \mbox{15a~(\ref{sssec:reflected_giants})}, 
  \mbox{19a~(\ref{sssec:exozodi})}
\\
\hline
\rowcolor{colorLW}
LW-A & \ref{sssec:lw_a} & 200 & 400 & 7 & & & $\equiv$ SSiC-A & 
  \mbox{1a~(\ref{sssec:life})}, \mbox{3a~(\ref{sssec:origin})}, 
  \mbox{4a~(\ref{sssec:prebiosignatures})}, \mbox{5a~(\ref{sssec:technosignatures})}, \mbox{8a~(\ref{sssec:seasonality})}, 
  \mbox{10a~(\ref{sssec:fpbio})}\\
\hline \hline
%\pagebreak
\multicolumn{9}{|c|}{High-Contrast Visible Spectroscopy (Section~\ref{sssec:hc_vis_spec})}\\
\hline
\rowcolor{colorSSiC}
SSiC-B & \ref{sssec:ssic_b} &  400 & 1100 & 140 & & & \textbf{BASELINE} & 
  \mbox{3b~(\ref{sssec:rocky_sub})}, \mbox{2a~(\ref{sssec:hab_system})}, 
  \mbox{12a~(\ref{sssec:giant_orbits})}, \mbox{15b~(\ref{sssec:reflected_giants})}, 
  \mbox{19b~(\ref{sssec:exozodi})}
\\
\hline
\rowcolor{colorLW}
LW-B & \ref{sssec:lw_b} &  400 & 1100 & 140 & & & $\equiv$ SSiC-B & 
  \mbox{1b~(\ref{sssec:life})}, \mbox{3b~(\ref{sssec:origin})}, 
  \mbox{4b~(\ref{sssec:prebiosignatures})}, \mbox{5b~(\ref{sssec:technosignatures})}, \mbox{6~(\ref{sssec:lawdki})}, 
  \mbox{8b~(\ref{sssec:seasonality})}, \mbox{14~(\ref{sssec:surface_bio})}, 
  \mbox{10b~(\ref{sssec:fpbio})}\\
\hline \hline
\multicolumn{9}{|c|}{High-Contrast NIR Spectroscopy (Section~\ref{sssec:hc_nir_spec})}\\
\hline
\rowcolor{colorSSiC}
SSiC-C & \ref{sssec:ssic_c} & 1100 & 1700 & 70 & & & \textbf{BASELINE} & 
  \mbox{3c~(\ref{sssec:rocky_sub})}, \mbox{2b~(\ref{sssec:hab_system})}, 
  \mbox{12b~(\ref{sssec:giant_orbits})}, \mbox{15c~(\ref{sssec:reflected_giants})}, 
  \mbox{19c~(\ref{sssec:exozodi})} \\
\hline 
\rowcolor{colorLW}
LW-C & \ref{sssec:lw_c} & 1100 & 1700 & 70 & & & $\equiv$ SSiC-C & 
  \mbox{1c~(\ref{sssec:life})}, \mbox{3c~(\ref{sssec:origin})}, 
  \mbox{4c~(\ref{sssec:prebiosignatures})}, \mbox{10c~(\ref{sssec:fpbio})}
\\
\hline \hline
\multicolumn{9}{|c|}{High-Contrast Spectropolarimetry (Section~\ref{sssec:hc_specpol})}\\
\hline
\rowcolor{colorSSiC}
SSiC-D & \ref{sssec:ssic_d} & 200 & 400 & 7 & & & UV & 
  \mbox{15d~(\ref{sssec:reflected_giants})}, \mbox{8a~(\ref{sssec:proto})}
\\
\hline
\rowcolor{colorSSiC}
SSiC-E & \ref{sssec:ssic_e} & 400 & 1100 & 140 & & & \textbf{ BASELINE} & 
  \mbox{18b~(\ref{sssec:exovenus})}, \mbox{15e~(\ref{sssec:reflected_giants})}, 
  \mbox{8b~(\ref{sssec:proto})}, \mbox{31a~(\ref{sssec:exorings})}
\\
\hline
\rowcolor{colorSSiC}
SSiC-F & \ref{sssec:ssic_f} & 1100 & 1700 & 70 & & & NIR & 
  \mbox{18c~(\ref{sssec:exovenus})}, \mbox{15f~(\ref{sssec:reflected_giants})}, 
  \mbox{31b~(\ref{sssec:exorings})}
\\
\hline
\rowcolor{colorLW}
LW-D & \ref{sssec:lw_d} & 350 & 850 & 100 & & & $\approx$ SSiC-E ($\downarrow \lambda_{\mathrm{blue}}, \lambda_{\mathrm{red}},R$)  & 
  \mbox{7a~(\ref{sssec:polbio})}, 
  \mbox{7b~(\ref{sssec:polbio})}
\\
%Long version: ($\downarrow \lambda_{\mathrm{blue}}$, $\downarrow \lambda_{\mathrm{red}}$, $\downarrow R$) 
\hline \hline
\multicolumn{9}{|c|}{High-Contrast Polarimetric Imaging (Section~\ref{sssec:hc_imaging_pol})}\\
\hline 
\rowcolor{colorSSiC}
SSiC-G & \ref{sssec:ssic_g} & 300 & 1700 & & & & \textbf{BASELINE} & 
  \mbox{3c~(\ref{sssec:rocky_sub})}, \mbox{4~(\ref{sssec:surface_water})}, 
  \mbox{19c~(\ref{sssec:exozodi})}
\\
\hline
\rowcolor{colorLW}
LW-E & \ref{sssec:lw_e} & 300 & 1700 & & &  & $\equiv$ SSiC-G & 
  \mbox{5c~(\ref{sssec:technosignatures})}, \mbox{7c~(\ref{sssec:polbio})}
\\
\hline \hline

\multicolumn{9}{|c|}{High-Contrast Imaging (Section~\ref{sssec:hc_imaging})}\\
\hline
\rowcolor{colorSSiC}
SSiC-H & \ref{sssec:ssic_h} & 200 & 1100 &  & & $10''\times10''$ & \textbf{BASELINE}  & 
  \mbox{6~(\ref{sssec:occ_binary})}, \mbox{15g~(\ref{sssec:reflected_giants})}, 
  \mbox{8c~(\ref{sssec:proto})}, \mbox{31c~(\ref{sssec:exorings})}, 
  \mbox{32~(\ref{sssec:exomoons})}
\\
\hline
\rowcolor{colorLW}
LW-F & \ref{sssec:lw_f} & 300 & 900 &  & & $10''\times10''$ & $\approx$ SSiC-H ($\uparrow \lambda_{\rm{blue}}$, $\downarrow \lambda_{\rm{red}}$) & 
  \mbox{5d~(\ref{sssec:technosignatures})}
\\
\hline \hline

\end{longtable}
\end{center}

%% file: scdd_astro2020_mapping.tex
\begin{center}
\begin{longtable}{
| >{\centering\arraybackslash}m{0.5in} % Column 1 (Panel): Centered
  >{\centering\arraybackslash}m{0.5in} % Column 2 (Ref): Centered
  p{2.9in}                             % Column 3 (Question): Left-aligned top
  >{\raggedright\arraybackslash}p{2.1in} % Column 4 (Science Cases): Left-aligned ragged
  >{\centering\arraybackslash}m{0.5in} | % Column 5 (LUVOIR-B): Centered
}
\caption{Mapping of science questions and discovery areas from Astro2020 \citep{astro2020} to science cases identified for HWO. The LUVOIR-B assessments in the rightmost column are from Table~I.1 of \citep{astro2020}. The notation is \textbf{Y} = Yes, \textbf{N} = No, and \textbf{P} = Partially. Science case LW-9 is marked by an asterisk because it describes foundational laboratory science rather than a space-based observing program.} \label{tab:astro2020} \\

\hline 
% Header row with explicit centering for all columns
\multicolumn{1}{|c}{\textbf{Panel}} & \multicolumn{1}{c}{\textbf{Ref}} & \multicolumn{1}{c}{\textbf{Question}} & \multicolumn{1}{c}{\textbf{Science Cases (Links)}} & \multicolumn{1}{c|}{\textbf{LUVOIR-B}}\\ \hline 
\endfirsthead

\multicolumn{5}{c}%
{{\bfseries \tablename\ \thetable{} -- continued from previous page}} \\
\hline 
\multicolumn{1}{|c}{\textbf{Panel}} & \multicolumn{1}{c}{\textbf{Ref}} & \multicolumn{1}{c}{\textbf{Question}} & \multicolumn{1}{c}{\textbf{Science Cases (Links)}} & \multicolumn{1}{c|}{\textbf{LUVOIR-B}}\\ \hline 
\endhead

\hline \multicolumn{5}{|r|}{{\emph{Continued on next page}}} \\ \hline
\endfoot

\hline \hline
\endlastfoot

% --- COS PANEL ---
\cellcolor{panelCOS}{\color{white}\textbf{COS}} & C-Q1 & What set the hot Big Bang in motion? & & N \\ \hline
\cellcolor{panelCOS}{\color{white}\textbf{COS}} & C-Q2 & What are the properties of dark matter and the dark sector? & \pref[-12]{GG}{sssec:dm_power}, \pref[-13]{GG}{sssec:dm_lensing} & Y \\ \hline
\cellcolor{panelCOS}{\color{white}\textbf{COS}} & C-Q3 & What physics drives the cosmic expansion and the large-scale evolution of the universe? & \pref[-13]{GG}{sssec:dm_lensing}, \pref[-18]{GG}{sssec:agn_feedback}, \pref[-10]{EE}{sssec:distance_ladder} & N \\ \hline
\cellcolor{panelCOS}{\color{white}\textbf{COS}} & C-Q4 & How will measurements of gravitational waves reshape our cosmological view? & & N \\ \hline
\cellcolor{panelCOS}{\color{white}\textbf{COS}} & C-DA & \emph{The Dark Ages as a Cosmological Probe} & & N \\ \hline
\hline

% --- GAL PANEL ---
\cellcolor{panelGAL}{\color{white}\textbf{GAL}} & D-Q1 & How did the intergalactic medium and the first sources of radiation evolve from cosmic dawn through the epoch of reionization? & \pref[-2]{GG}{sssec:bh_mass_spin}, \pref[-4]{GG}{sssec:bh_quiescent}, \pref[-5]{GG}{sssec:agn_outflow}, \pref[-6]{GG}{sssec:resolve_reionization}, \pref[-7]{GG}{sssec:lyman_escape}, \pref[-9]{GG}{sssec:lyman_indirect}, \pref[-11]{GG}{sssec:ionizing_lf}, \pref[-14]{GG}{sssec:smbh_mergers},\pref[-18]{GG}{sssec:agn_feedback}, \pref[-4]{EE}{sssec:r_process_el}, \pref[-5]{EE}{sssec:vms}, \pref[-8]{EE}{sssec:first_stars}, \pref[-12]{EE}{sssec:ism_uv} & P \\ \hline
\cellcolor{panelGAL}{\color{white}\textbf{GAL}} & D-Q2 & How do gas, metals, and dust flow into, through, and out of galaxies? & \pref[-3]{GG}{sssec:bh_torus}, \pref[-5]{GG}{sssec:agn_outflow}, \pref[-6]{GG}{sssec:resolve_reionization}, \pref[-7]{GG}{sssec:lyman_escape}, \pref[-9]{GG}{sssec:lyman_indirect},  \pref[-16]{GG}{sssec:disk_cgm}, \pref[-17]{GG}{sssec:cgm_elm}, \pref[-18]{GG}{sssec:agn_feedback}, \pref[-3]{EE}{sssec:flash_ccsne}, \pref[-5]{EE}{sssec:vms}, \pref[-6]{EE}{sssec:resolved_stellar_pops}, \pref[-8]{EE}{sssec:first_stars}, \pref[-9]{EE}{sssec:r_process_nature}, \pref[-11]{EE}{sssec:dust_uv}, \pref[-12]{EE}{sssec:ism_uv} & P \\ \hline
\cellcolor{panelGAL}{\color{white}\textbf{GAL}} & D-Q3 & How do supermassive black holes form and how is their growth coupled to the evolution of their host galaxies? & \pref[-2]{GG}{sssec:bh_mass_spin},  \pref[-3]{GG}{sssec:bh_torus}, \pref[-4]{GG}{sssec:bh_quiescent}, \pref[-5]{GG}{sssec:agn_outflow}, \pref[-14]{GG}{sssec:smbh_mergers}, \pref[-16]{GG}{sssec:disk_cgm}, \pref[-17]{GG}{sssec:cgm_elm}, \pref[-18]{GG}{sssec:agn_feedback},  \pref[-19]{GG}{sssec:smbh_pol} & P \\ \hline
\cellcolor{panelGAL}{\color{white}\textbf{GAL}} & D-Q4 & How do the histories of galaxies and their dark matter halos shape their observable properties? & \pref[-2]{GG}{sssec:bh_mass_spin}, \pref[-5]{GG}{sssec:agn_outflow}, \pref[-6]{GG}{sssec:resolve_reionization}, \pref[-7]{GG}{sssec:lyman_escape}, \pref[-12]{GG}{sssec:dm_power}, \pref[-14]{GG}{sssec:smbh_mergers},  \pref[-2]{EE}{sssec:dust_extinction}, \pref[-6]{EE}{sssec:resolved_stellar_pops}, \pref[-8]{EE}{sssec:first_stars}, \pref[-9]{EE}{sssec:r_process_nature}, \pref[-10]{EE}{sssec:distance_ladder}, \pref[-11]{EE}{sssec:dust_uv}, \pref[-12]{EE}{sssec:ism_uv} & P \\ \hline
\cellcolor{panelGAL}{\color{white}\textbf{GAL}} & D-DA & \emph{Mapping the Circumgalactic Medium and Intergalactic Medium in Emission} & \pref[-16]{GG}{sssec:disk_cgm}, \pref[-17]{GG}{sssec:cgm_elm}, \pref[-12]{EE}{sssec:ism_uv} & P \\ \hline
\hline

% --- COEP PANEL ---
\cellcolor{panelCOEP}{\color{white}\textbf{COEP}} & B-Q1 & What are the mass and spin distributions of neutron stars and stellar-mass black holes? &  \pref[-4]{EE}{sssec:r_process_el}, \pref[-15]{EE}{sssec:mag_stars}  & N \\ \hline
\cellcolor{panelCOEP}{\color{white}\textbf{COEP}} & B-Q2 & What powers the diversity of explosive phenomena across the electromagnetic spectrum? & \pref[-4]{GG}{sssec:bh_quiescent}, \pref[-3]{EE}{sssec:flash_ccsne}, \pref[-4]{EE}{sssec:r_process_el}, \pref[-9]{EE}{sssec:r_process_nature}, \pref[-15]{EE}{sssec:mag_stars}  & Y \\ \hline
\cellcolor{panelCOEP}{\color{white}\textbf{COEP}} & B-Q3 & Why do some compact objects eject material at nearly-light-speed jets, and what is that material made of? & \pref[-5]{GG}{sssec:agn_outflow} & N \\ \hline
\cellcolor{panelCOEP}{\color{white}\textbf{COEP}} & B-Q4 & What seeds supermassive black holes and how do they grow? & \pref[-2]{GG}{sssec:bh_mass_spin}, \pref[-3]{GG}{sssec:bh_torus}, \pref[-5]{GG}{sssec:agn_outflow}, \pref[-4]{GG}{sssec:bh_quiescent}, \pref[-14]{GG}{sssec:smbh_mergers},  \pref[-16]{GG}{sssec:disk_cgm}, \pref[-17]{GG}{sssec:cgm_elm}, \pref[-18]{GG}{sssec:agn_feedback}, \pref[-19]{GG}{sssec:smbh_pol} & N \\ \hline
\cellcolor{panelCOEP} {\color{white}\textbf{COEP}} & B-DA & \emph{Transforming our View of the Universe by Combining New Information from Light, Particles, and Gravitational Waves} & \pref[-5]{GG}{sssec:agn_outflow}, \pref[-14]{GG}{sssec:smbh_mergers}, \pref[-19]{GG}{sssec:smbh_pol},  \pref[-4]{EE}{sssec:r_process_el}, \pref[-7]{EE}{sssec:wds}  & Y \\ \hline
\hline

%\pagebreak

% --- ISP PANEL ---
\cellcolor{panelISP}{\color{white}\textbf{ISP}} & F-Q1 & How do star-forming structures arise from, and interact with, the diffuse ISM? & \pref[-7]{GG}{sssec:lyman_escape}, \pref[-9]{GG}{sssec:lyman_indirect}, \pref[-16]{GG}{sssec:disk_cgm}, \pref[-17]{GG}{sssec:cgm_elm}, \pref[-2]{EE}{sssec:dust_extinction}, \pref[-3]{EE}{sssec:flash_ccsne}, \pref[-6]{EE}{sssec:resolved_stellar_pops}, \pref[-11]{EE}{sssec:dust_uv}, \pref[-12]{EE}{sssec:ism_uv}, \pref[-15]{EE}{sssec:mag_stars} & P \\ \hline
\cellcolor{panelISP}{\color{white}\textbf{ISP}} & F-Q2 & What regulates the structures and motions within molecular clouds? & \pref[-6]{GG}{sssec:resolve_reionization},  \pref[-2]{EE}{sssec:dust_extinction}, \pref[-12]{EE}{sssec:ism_uv} & N \\ \hline
\cellcolor{panelISP}{\color{white}\textbf{ISP}} & F-Q3 & How does gas flow from parsec scales down to protostars and disks? & \pref[-1]{EE}{sssec:massive_stars_lowZ}, \pref[-6]{EE}{sssec:resolved_stellar_pops},  \pref[-23]{SSiC}{sssec:disk_winds}, \pref[-34]{SSiC}{sssec:young_bfields} & Y \\ \hline
\cellcolor{panelISP}{\color{white}\textbf{ISP}} & F-Q4 & Is Planet Formation Fast or Slow? & \pref[-2]{EE}{sssec:dust_extinction}, \pref[-6]{SSiC}{sssec:occ_binary}, \pref[-7]{SSiC}{sssec:solar}, \pref[-8]{SSiC}{sssec:proto}, \pref[-9]{SSiC}{sssec:debris}, \pref[-22]{SSiC}{sssec:mars}, \pref[-23]{SSiC}{sssec:disk_winds},  \pref[-34]{SSiC}{sssec:young_bfields} & P \\ \hline
\cellcolor{panelISP}{\color{white}\textbf{ISP}} & F-DA & \emph{Detecting and Characterizing Forming Planets} & \pref[-8]{SSiC}{sssec:proto}, \pref[-9]{SSiC}{sssec:debris}, \pref[-19]{SSiC}{sssec:exozodi}, \pref[-23]{SSiC}{sssec:disk_winds}, \pref[-34]{SSiC}{sssec:young_bfields} & P \\ \hline
\hline

% --- SSP PANEL ---
\cellcolor{panelSSP}{\color{white}\textbf{SSP}} & G-Q1 & What are the most extreme stars and stellar populations? & \pref[-1]{EE}{sssec:massive_stars_lowZ}, \pref[-4]{EE}{sssec:r_process_el}, \pref[-5]{EE}{sssec:vms}, \pref[-6]{EE}{sssec:resolved_stellar_pops}, \pref[-7]{EE}{sssec:wds}, \pref[-8]{EE}{sssec:first_stars}, \pref[-9]{EE}{sssec:r_process_nature}, \pref[-10]{EE}{sssec:distance_ladder}, \pref[-15]{EE}{sssec:mag_stars} & P \\ \hline
\cellcolor{panelSSP}{\color{white}\textbf{SSP}} & G-Q2 & How does multiplicity affect the way a star lives and dies? & \pref[-1]{EE}{sssec:massive_stars_lowZ}, \pref[-15]{EE}{sssec:mag_stars}, \pref[-34]{SSiC}{sssec:young_bfields} & P \\ \hline
\cellcolor{panelSSP}{\color{white}\textbf{SSP}} & G-Q3 & What would stars look like if we view them like we do the Sun? & \pref[-34]{SSiC}{sssec:young_bfields} & P \\ \hline
\cellcolor{panelSSP}{\color{white}\textbf{SSP}} & G-Q4 & How do the Sun and other stars create space weather? &  \pref[-14]{SSiC}{sssec:escape},  \pref[-24]{SSiC}{sssec:highres_atmos}, \pref[-25]{SSiC}{sssec:aurorae}, \pref[-26]{SSiC}{sssec:bfield}, \pref[-29]{SSiC}{sssec:solargiant}, \pref[-34]{SSiC}{sssec:young_bfields} & P \\ \hline
\cellcolor{panelSSP}{\color{white}\textbf{SSP}} & G-DA & \emph{“Industrial Scale” Spectroscopy} & \pref[-11]{GG}{sssec:ionizing_lf}, \pref[-1]{EE}{sssec:massive_stars_lowZ}, \pref[-2]{EE}{sssec:dust_extinction}, \pref[-7]{EE}{sssec:wds}, \pref[-8]{EE}{sssec:first_stars} & P \\ \hline

% --- EAS PANEL ---
\cellcolor{panelEAS}{\color{white}\textbf{EAS}} & E-Q1 & What is the range of planetary system architectures, and is the configuration of the solar system common? & \pref[-7]{EE}{sssec:wds},  \pref[-2]{SSiC}{sssec:hab_system},  \pref[-3]{SSiC}{sssec:rocky_sub}, \pref[-6]{SSiC}{sssec:occ_binary}, \pref[-7]{SSiC}{sssec:solar}, \pref[-8]{SSiC}{sssec:proto}, \pref[-9]{SSiC}{sssec:debris}, \pref[-12]{SSiC}{sssec:giant_orbits}, \pref[-15]{SSiC}{sssec:reflected_giants}, \pref[-17]{SSiC}{sssec:id_oceans}, \pref[-18]{SSiC}{sssec:exovenus}, \pref[-19]{SSiC}{sssec:exozodi}, \pref[-20]{SSiC}{sssec:survive_water}, \pref[-21]{SSiC}{sssec:retention}, \pref[-22]{SSiC}{sssec:mars}, \pref[-24]{SSiC}{sssec:highres_atmos}, \pref[-30]{SSiC}{sssec:defense}, \pref[-31]{SSiC}{sssec:exorings}, \pref[-32]{SSiC}{sssec:exomoons}, \pref[-33]{SSiC}{sssec:aminos},   \pref[-34]{SSiC}{sssec:young_bfields}, \pref[-1]{LW}{sssec:life}, \pref[-7]{LW}{sssec:polbio}, \pref[-12]{LW}{sssec:mp}  & Y \\ 
\hline
\cellcolor{panelEAS}{\color{white}\textbf{EAS}} & E-Q2 & What are the properties of individual planets, and which processes lead to planetary diversity? & \pref[-7]{EE}{sssec:wds}, \pref[-1]{SSiC}{sssec:oceans_habitable}, \pref[-3]{SSiC}{sssec:rocky_sub}, \pref[-4]{SSiC}{sssec:surface_water}, \pref[-5]{SSiC}{sssec:venus}, \pref[-11]{SSiC}{sssec:ozone_onset}, \pref[-14]{SSiC}{sssec:escape}, \pref[-15]{SSiC}{sssec:reflected_giants}, \pref[-16]{SSiC}{sssec:transits}, \pref[-17]{SSiC}{sssec:id_oceans}, \pref[-18]{SSiC}{sssec:exovenus}, \pref[-20]{SSiC}{sssec:survive_water}, \pref[-21]{SSiC}{sssec:retention},  \pref[-24]{SSiC}{sssec:highres_atmos},  
\pref[-25]{SSiC}{sssec:aurorae}, \pref[-26]{SSiC}{sssec:bfield}, \pref[-27]{SSiC}{sssec:titan}, \pref[-28]{SSiC}{sssec:mars_post}, \pref[-29]{SSiC}{sssec:solargiant}, \pref[-31]{SSiC}{sssec:exorings}, \pref[-32]{SSiC}{sssec:exomoons},   \pref[-1]{LW}{sssec:life}, \pref[-3]{LW}{sssec:origin}, \pref[-4]{LW}{sssec:prebiosignatures}, \pref[-7]{LW}{sssec:polbio}, \pref[-9*]{LW}{sssec:geochemical}, \pref[-10]{LW}{sssec:fpbio}, \pref[-12]{LW}{sssec:mp}, \pref[-13]{LW}{sssec:seasonality}  & Y \\ \hline
\cellcolor{panelEAS}{\color{white}\textbf{EAS}} & E-Q3 & How do habitable environments arise and evolve within the context of their planetary systems? &  \pref[-1]{SSiC}{sssec:oceans_habitable}, \pref[-2]{SSiC}{sssec:hab_system},  \pref[-3]{SSiC}{sssec:rocky_sub}, \pref[-4]{SSiC}{sssec:surface_water}, \pref[-5]{SSiC}{sssec:venus}, \pref[-11]{SSiC}{sssec:ozone_onset}, \pref[-14]{SSiC}{sssec:escape}, \pref[-17]{SSiC}{sssec:id_oceans}, \pref[-18]{SSiC}{sssec:exovenus}, \pref[-20]{SSiC}{sssec:survive_water}, \pref[-21]{SSiC}{sssec:retention}, \pref[-27]{SSiC}{sssec:titan},  \pref[-28]{SSiC}{sssec:mars_post}, \pref[-33]{SSiC}{sssec:aminos}, \pref[-1]{LW}{sssec:life}, \pref[-3]{LW}{sssec:origin}, \pref[-4]{LW}{sssec:prebiosignatures}, \pref[-7]{LW}{sssec:polbio}, \pref[-9*]{LW}{sssec:geochemical}, \pref[-10]{LW}{sssec:fpbio}, \pref[-12]{LW}{sssec:mp}, \pref[-13]{LW}{sssec:seasonality}  & Y \\ 
\hline
\cellcolor{panelEAS}{\color{white}\textbf{EAS}} & E-Q4 & How can signs of life be identified and interpreted in the context of their planetary environments? & \pref[-1]{SSiC}{sssec:oceans_habitable}, \pref[-33]{SSiC}{sssec:aminos}, \pref[-1]{LW}{sssec:life},  \pref[-5]{LW}{sssec:technosignatures}, \pref[-6]{LW}{sssec:lawdki}, \pref[-7]{LW}{sssec:polbio}, \pref[-10]{LW}{sssec:fpbio}, \pref[-12]{LW}{sssec:mp}, \pref[-13]{LW}{sssec:seasonality}, \pref[-14]{LW}{sssec:surface_bio}& Y \\ \hline
\cellcolor{panelEAS}{\color{white}\textbf{EAS}} & E-DA & \emph{The Search for Life on Exoplanets} & \pref[-1]{LW}{sssec:life}, \pref[-3]{LW}{sssec:origin}, \pref[-4]{LW}{sssec:prebiosignatures}, \pref[-5]{LW}{sssec:technosignatures}, \pref[-6]{LW}{sssec:lawdki},  \pref[-14]{LW}{sssec:surface_bio}, \pref[-7]{LW}{sssec:polbio},  \pref[-13]{LW}{sssec:seasonality}, \pref[-10]{LW}{sssec:fpbio}, \pref[-12]{LW}{sssec:mp}
& Y \\ \hline
\hline

\end{longtable}
\end{center}

%% file: latex_acknowledgements_list_10august2026.tex
D. J. Adams is funded by NASA through the NASA Hubble Fellowship Program Grant HST-HF2-51523.001-A awarded by the Space Telescope Science Institute, which is operated by the Association of Universities for Research in Astronomy, Inc. for NASA, under contract NAS5-26555.
E. Alecian received funding from the French Agence Nationale de la Recherche (ANR) through the project PROMETHEE (ANR-22-CE31-0020).
J. Biedermann would like to acknowledge the support of the CNES, the CNRS, the University of Strasbourg, the PNHE and the PNCG.
J. N. Burchett is grateful for funding support from the National Science Foundation under Grant Number 2327438.
J. M. Cann is supported by NASA under award number 80GSFC24M0006.
C. Carr is supported by NSFC grant W2433001 and the NSFC Talent-Introduction Program.
T. Daylan was partially supported by the McDonnell Center for the Space Sciences at Washington University in St. Louis.
J. E. Doppel is supported by the United Kingdom Research and Innovation (UKRI) Future Leaders Fellowship ``Using Cosmic Beasts to uncover the Nature of Dark Matter'' (grant number MR/X006069/1).
R. Ezzeddine acknowledges support from NSF grant AST 2206263, HST GO15657 and GO-15951, and NASA ARP grant 80NSSC24K0899.
T. Fetherolf acknowledges support from an appointment through the NASA Postdoctoral Program at the NASA Astrobiology Center, administered by Oak Ridge Associated Universities under contract with NASA.
A. I. G\'omez de Castro has acted as a scientific representative of the European Space Agency (ESA) in the NASA HWO START and acknowledges the support of ESA and the Ministry of Science \& Innovation of Spain (Ref:PID2020-116726RB-I00,PID2023-147740OB-I00).
Y. Hasegawa acknowledges that this research was carried out in part at the Jet Propulsion Laboratory, California Institute of Technology, under a contract with the National Aeronautics and Space Administration (80NM0018D0004). Y. Hasegawa is supported by JPL/Caltech.
D. J. Lagattuta is supported by STFC grants ST/T000244/1 and ST/W002612/1. D.J.L. is also partially supported by the United Kingdom Research and Innovation (UKRI) Future Leaders Fellowship ``Using Cosmic Beasts to uncover the Nature of Dark Matter'' (grant number MR/S017216/1).
J. Lustig-Yaeger acknowledges internal support from Johns Hopkins APL.
E. Mamajek acknowledges that part of this research was carried out at the Jet Propulsion Laboratory, California Institute of Technology, under a contract with the National Aeronautics and Space Administration (80NM0018D0004).
L. A. Moustakas acknowledges that part of this research was carried out at the Jet Propulsion Laboratory, California Institute of Technology, under contract with the National Aeronautics and Space Administration (80NM0018D0004).
J. W. Nightingale is supported by an STFC/UKRI Ernest Rutherford Fellowship, Project Reference: ST/X003086/1.
A. Smercina is supported by NASA through the NASA Hubble Fellowship grant HST-HF2-51567 awarded by the Space Telescope Science Institute, which is operated by the Association of Universities for Research in Astronomy, Inc., for NASA, under contract NAS5-26555.
A. Strugarek acknowledges funding from the European Union's Horizon 2020 research and innovation programme (grant agreement no. 776403 ExoplANETS-A), the PLATO/CNES grant at CEA/IRFU/DAp, and the European Research Council project ExoMagnets (grant agreement no. 101125367).
V. U gratefully acknowledges partial funding support from NASA ADSPS grant \#80NSSC25K0169, National Science Foundation (NSF) Astronomy and Astrophysics Research Grant \#2536603, as well as STScI grant \#JWST-GO-08391.001-A, which was provided by NASA through a grant from the Space Telescope Science Institute, which is operated by the Association of Universities for Research in Astronomy, Inc., under NASA contract.
S. Xu is supported by the international Gemini Observatory, a program of NSF NOIRLab, which is managed by the Association of Universities for Research in Astronomy (AURA) under a cooperative agreement with the U.S. National Science Foundation, on behalf of the Gemini partnership of Argentina, Brazil, Canada, Chile, the Republic of Korea, and the United States of America.
L. Zhang acknowledges grant support from the Space Telescope Science Institute (ID: JWST-GO-01670; JWST-GO-03535; JWST-GO-04972).
E. Bellocchi acknowledges support from the Spanish grants PID2022-138621NB-I00 and PID2021-123417OB- I00, funded by MCIN/AEI/10.13039/501100011033/FEDER, EU.
S. E. I. Bosman is supported by the Deutsche Forschungsgemeinschaft (DFG) under Emmy Noether grant number BO 5771/1-1.
J. A. Burt acknowledges that part of this research was carried out at the Jet Propulsion Laboratory, California Institute of Technology, under contract with the National Aeronautics and Space Administration (80NM0018D0004).
J. A. Caballero acknowledges financial support from the Agencia Estatal de Investigaci\'on (AEI/10.13039/501100011033) of the Ministerio de Ciencia e Innovaci\'on and the ERDF ``A way of making Europe'' through project PID2022-137241NB-C42.
S. Ertel is supported by NASA through grants 80NSSC23K1473 (PI: Ertel), 80NSSC21K0394 (PI: Ertel), and 80NSSC23K0288 (PI: Faramaz-Gorka).
R. Estrela acknowledges that part of this research was carried out at the Jet Propulsion Laboratory, California Institute of Technology, under contract with the National Aeronautics and Space Administration (80NM0018D0004).
T. Gao acknowledges support from ARC Discovery Project DP210101945.
M. Garcia gratefully acknowledges support by grants PID2022-137779OB-C41 and PID2022-140483NB-C22, funded by the Spanish Ministry of Science, Innovation and Universities/State Agency of Research MICIU/AEI/10.13039/501100011033 and by ``ERDF A way of making Europe'', and grant MAD4SPACE, TEC-2024/TEC-182 from Comunidad de Madrid (Spain).
L. Gkouvelis acknowledges financial support from the Severo Ochoa grant CEX2021-001131-S funded by MCIN/AEI/10.13039/501100011033 and Ministerio de Ciencia e Innovación through the project PID2022-137241NB-C43.
C. K. Harada acknowledges support from the National Science Foundation (NSF) Graduate Research Fellowship Program (GRFP) under Grant No. DGE 2146752 and the NASA Astrophysics Decadal Survey Precursor Science (ADSPS) program under Grant No. 80NSSC23K1476.
S. Hernandez is thankful for support from the European Space Agency (ESA).
C. M. Jackman is supported by Taighde \'eireann ‐ Research Ireland awards: Laureate Consolidator award SOLMEX, and President of Ireland award 18/FRL/6199.
A. Kiessling is supported by JPL, which is run under contract by the California Institute of Technology for NASA.
K. G. Kislyakova acknowledges funding from the European Research Council project EASE (grant agreement no. 101123041).
J. Krissansen-Totton was supported by NASA Astrophysics Decadal Survey Precursor Science grant 80NSSC23K1471.
N. Liu was supported by the NASA grants 80NSSC23K1034 and 80NSSC24K0132.
B. Mennesson acknowledges that part of this research was carried out at the Jet Propulsion Laboratory, California Institute of Technology, under contract with the National Aeronautics and Space Administration (80NM0018D0004).
M. Neveu was supported by NASA under award number 80GSFC24M0006.
S. Peacock acknowledges support from NASA under award number 80GSFC24M0006.
S. Ravindranath acknowledges that this work is partially based upon work supported by NASA under award number 80GSFC24M0006.
D. Rodgers-Lee would like to acknowledge that this publication has emanated in part from research conducted with the financial support of Taighde {\'E}ireann = Research Ireland under Grant number 21/PATH-S/9339.
R. J. Smith was supported by the STFC through the Durham Astronomy Consolidated Grant (ST/X001075/1).
M. von Wietersheim-Kramsta gratefully acknowledges the support from UKRI STFC consolidated grant ST/X001075/1, from Innovate UK, and from the UK Space Agency.
D. J. Wilson was supported by NASA Astrophysics Decadal Survey Precursor Science grant 80NSSC23K1474.
T. G. Wilson acknowledges support from the University of Warwick and UKSA.
N. F. Wogan was supported by the NASA Postdoctoral Program.
M. L. Wong is funded by NASA through the NASA Hubble Fellowship Program Grant HST-HF2-51521.001-A awarded by the Space Telescope Science Institute, which is operated by the Association of Universities for Research in Astronomy, Inc., for NASA, under contract NAS5-26555.
T. Zingales acknowledges support from CHEOPS ASI-INAF agreement n. 2019-29-HH.0, NVIDIA Academic Hardware Grant Program for the use of the Titan V GPU card and the Italian MUR Departments of Excellence grant 2023-2027 ``Quantum Frontiers''.
D. Huber acknowledges support from the Alfred P. Sloan Foundation and the National Aeronautics and Space Administration (80NSSC23K0434, 80NSSC23K0435, 80NSSC21K0652).
B. Lewis notes that this material is partially based upon work supported by the National Science Foundation Astronomy \& Astrophysics Postdoctoral Fellowship Award No. 2401654. Any opinions, findings, and conclusions or recommendations expressed in this material are those of the authors(s) and do not necessarily reflect the views of the National Science Foundation.
L. Mancini acknowledges financial contribution from PRIN MUR 2022 project 2022J4H55R.
F. J. Pozuelos acknowledges financial support from the Severo Ochoa grant CEX2021-001131-S funded by MCIN/AEI/10.13039/501100011033 and Ministerio de Ciencia e Innovaci\'on through the project PID2022-137241NB-C43.
L. D. Vega acknowledges support from the Heising-Simons Postdoctoral Launch Program and from NASA under award No. 80GSFC21M0002 and 80GSFC24M0006.

%% file: start_members.tex
\begin{table*}
    \centering
    \begin{tabular}{|l|l|}
    \hline
    \hline
        \textbf{Person} & \textbf{Affiliation} \\
        \hline
Giada Arney & NASA Goddard Space Flight Center\\
\hline
Natasha Batalha & NASA Ames Research Center\\
\hline
Eric Burns & Louisiana State University\\
\hline
Jessie Christiansen & NASA Exoplanet Science Institute\\
\hline
\emph{Courtney Dressing (Co-Chair)} & \emph{University of California, Berkeley}\\
\hline
Kevin France & University of Colorado, Boulder\\
\hline
Scott Gaudi & Ohio State University\\
\hline
Renyu Hu & NASA Jet Propulsion Laboratory\\
\hline
Alina Kiessling & NASA Jet Propulsion Laboratory\\
\hline
Janice Lee & Space Telescope Science Institute\\
\hline
Bruce Macintosh & University of California Observatories\\
\hline
\emph{John O’Meara (Co-Chair)} & \emph{W. M. Keck Observatory}\\
\hline
Jim Oschmann & Marinus Consulting\\
\hline
Rachel Osten & Space Telescope Science Institute\\
\hline
Chris Packham & University of Texas, San Antonio\\
\hline
Lynnae Quick & NASA Goddard Space Flight Center\\
\hline
Jason Rhodes & NASA Jet Propulsion Laboratory\\
\hline
Jane Rigby & NASA Goddard Space Flight Center\\
\hline
Tyler Robinson & University of Arizona\\
\hline
Dmitry Savransky & Cornell University\\
\hline
Evan Scannapieco & Arizona State University\\
\hline
Evgenya Shkolnik & Arizona State University\\
\hline
\hline
Charlie Atkinson (ex officio) & Northrop Grumman\\
\hline
Matthew East (ex officio)& L3Harris\\
\hline
Alison Nordt (ex officio)& Lockheed Martin\\
\hline
Erik Wilkinson (ex officio)& BAE Systems (formerly Ball Aerospace)\\
\hline
Eric Mamajek (ex officio)& NASA Exoplanet Exploration Program Office\\
\hline
Swara Ravindranath (ex officio)& NASA Cosmic Origins Program Office\\
\hline
Miyazaki Satoshi (ex officio)& JAXA / NAOJ\\
\hline
Takahiro Sumi (ex officio)& JAXA / Osaka\\
\hline
Ana Gomez de Castro (ex officio)& ESA / Madrid\\
\hline
Michiel Min (ex officio)& ESA / Amsterdam\\
\hline
David Mouillet (ex officio)& ESA / Grenoble\\
\hline
Christian Marois (ex officio)& CSA / NRC-Herzberg\\
    \hline
    \hline
    \end{tabular}
    \caption{START members and affiliations}
    \label{tab:start}
\end{table*}

%% file: hwo_start.bib
@ARTICLE{abbott_et_al2016,
       author = {{Abbott}, B.~P. and {Abbott}, R. and {Abbott}, T.~D. and {Abernathy}, M.~R. and {Acernese}, F. and {Ackley}, K. and {Adams}, C. and {Adams}, T. and {Addesso}, P. and {Adhikari}, R.~X. and {Adya}, V.~B. and {Affeldt}, C. and {Agathos}, M. and {Agatsuma}, K. and {Aggarwal}, N. and {Aguiar}, O.~D. and {Aiello}, L. and {Ain}, A. and {Ajith}, P. and {Allen}, B. and {Allocca}, A. and {Altin}, P.~A. and {Anderson}, S.~B. and {Anderson}, W.~G. and {Arai}, K. and {Arain}, M.~A. and {Araya}, M.~C. and {Arceneaux}, C.~C. and {Areeda}, J.~S. and {Arnaud}, N. and {Arun}, K.~G. and {Ascenzi}, S. and {Ashton}, G. and {Ast}, M. and {Aston}, S.~M. and {Astone}, P. and {Aufmuth}, P. and {Aulbert}, C. and {Babak}, S. and {Bacon}, P. and {Bader}, M.~K.~M. and {Baker}, P.~T. and {Baldaccini}, F. and {Ballardin}, G. and {Ballmer}, S.~W. and {Barayoga}, J.~C. and {Barclay}, S.~E. and {Barish}, B.~C. and {Barker}, D. and {Barone}, F. and {Barr}, B. and {Barsotti}, L. and {Barsuglia}, M. and {Barta}, D. and {Bartlett}, J. and {Barton}, M.~A. and {Bartos}, I. and {Bassiri}, R. and {Basti}, A. and {Batch}, J.~C. and {Baune}, C. and {Bavigadda}, V. and {Bazzan}, M. and {Behnke}, B. and {Bejger}, M. and {Belczynski}, C. and {Bell}, A.~S. and {Bell}, C.~J. and {Berger}, B.~K. and {Bergman}, J. and {Bergmann}, G. and {Berry}, C.~P.~L. and {Bersanetti}, D. and {Bertolini}, A. and {Betzwieser}, J. and {Bhagwat}, S. and {Bhandare}, R. and {Bilenko}, I.~A. and {Billingsley}, G. and {Birch}, J. and {Birney}, I.~A. and {Birnholtz}, O. and {Biscans}, S. and {Bisht}, A. and {Bitossi}, M. and {Biwer}, C. and {Bizouard}, M.~A. and {Blackburn}, J.~K. and {Blair}, C.~D. and {Blair}, D.~G. and {Blair}, R.~M. and {Bloemen}, S. and {Bock}, O. and {Bodiya}, T.~P. and {Boer}, M. and {Bogaert}, G. and {Bogan}, C. and {Bohe}, A. and {Bojtos}, P. and {Bond}, C. and {Bondu}, F. and {Bonnand}, R. and {Boom}, B.~A. and {Bork}, R. and {Boschi}, V. and {Bose}, S. and {Bouffanais}, Y. and {Bozzi}, A. and {Bradaschia}, C. and {Brady}, P.~R. and {Braginsky}, V.~B. and {Branchesi}, M. and {Brau}, J.~E. and {Briant}, T. and {Brillet}, A. and {Brinkmann}, M. and {Brisson}, V. and {Brockill}, P. and {Brooks}, A.~F. and {Brown}, D.~A. and {Brown}, D.~D. and {Brown}, N.~M. and {Buchanan}, C.~C. and {Buikema}, A. and {Bulik}, T. and {Bulten}, H.~J. and {Buonanno}, A. and {Buskulic}, D. and {Buy}, C. and {Byer}, R.~L. and {Cabero}, M. and {Cadonati}, L. and {Cagnoli}, G. and {Cahillane}, C. and {Bustillo}, J. Calder{\'o}n and {Callister}, T. and {Calloni}, E. and {Camp}, J.~B. and {Cannon}, K.~C. and {Cao}, J. and {Capano}, C.~D. and {Capocasa}, E. and {Carbognani}, F. and {Caride}, S. and {Diaz}, J. Casanueva and {Casentini}, C. and {Caudill}, S. and {Cavagli{\`a}}, M. and {Cavalier}, F. and {Cavalieri}, R. and {Cella}, G. and {Cepeda}, C.~B. and {Baiardi}, L. Cerboni and {Cerretani}, G. and {Cesarini}, E. and {Chakraborty}, R. and {Chalermsongsak}, T. and {Chamberlin}, S.~J. and {Chan}, M. and {Chao}, S. and {Charlton}, P. and {Chassande-Mottin}, E. and {Chen}, H.~Y. and {Chen}, Y. and {Cheng}, C. and {Chincarini}, A. and {Chiummo}, A. and {Cho}, H.~S. and {Cho}, M. and {Chow}, J.~H. and {Christensen}, N. and {Chu}, Q. and {Chua}, S. and {Chung}, S. and {Ciani}, G. and {Clara}, F. and {Clark}, J.~A. and {Cleva}, F. and {Coccia}, E. and {Cohadon}, P.-F. and {Colla}, A. and {Collette}, C.~G. and {Cominsky}, L. and {Constancio}, M. and {Conte}, A. and {Conti}, L. and {Cook}, D. and {Corbitt}, T.~R. and {Cornish}, N. and {Corsi}, A. and {Cortese}, S. and {Costa}, C.~A. and {Coughlin}, M.~W. and {Coughlin}, S.~B. and {Coulon}, J.-P. and {Countryman}, S.~T. and {Couvares}, P. and {Cowan}, E.~E. and {Coward}, D.~M. and {Cowart}, M.~J.},
        title = "{Observation of Gravitational Waves from a Binary Black Hole Merger}",
      journal = {\prl},
         year = 2016,
        month = feb,
       volume = {116},
       number = {6},
          eid = {061102},
        pages = {061102},
          doi = {10.1103/PhysRevLett.116.061102},
archivePrefix = {arXiv},
       eprint = {1602.03837},
 primaryClass = {gr-qc},
       adsurl = {https://ui.adsabs.harvard.edu/abs/2016PhRvL.116f1102A}
}

@ARTICLE{abbott_et_al2019,
       author = {{Abbott}, B.~P. and {Abbott}, R. and {Abbott}, T.~D. and {Abraham}, S. and {Acernese}, F. and {Ackley}, K. and {Adams}, C. and {Adhikari}, R.~X. and {Adya}, V.~B. and {Affeldt}, C. and {Agathos}, M. and {Agatsuma}, K. and {Aggarwal}, N. and {Aguiar}, O.~D. and {Aiello}, L. and {Ain}, A. and {Ajith}, P. and {Allen}, G. and {Allocca}, A. and {Aloy}, M.~A. and {Altin}, P.~A. and {Amato}, A. and {Ananyeva}, A. and {Anderson}, S.~B. and {Anderson}, W.~G. and {Angelova}, S.~V. and {Antier}, S. and {Appert}, S. and {Arai}, K. and {Araya}, M.~C. and {Areeda}, J.~S. and {Ar{\`e}ne}, M. and {Arnaud}, N. and {Arun}, K.~G. and {Ascenzi}, S. and {Ashton}, G. and {Aston}, S.~M. and {Astone}, P. and {Aubin}, F. and {Aufmuth}, P. and {AultONeal}, K. and {Austin}, C. and {Avendano}, V. and {Avila-Alvarez}, A. and {Babak}, S. and {Bacon}, P. and {Badaracco}, F. and {Bader}, M.~K.~M. and {Bae}, S. and {Baker}, P.~T. and {Baldaccini}, F. and {Ballardin}, G. and {Ballmer}, S.~W. and {Banagiri}, S. and {Barayoga}, J.~C. and {Barclay}, S.~E. and {Barish}, B.~C. and {Barker}, D. and {Barkett}, K. and {Barnum}, S. and {Barone}, F. and {Barr}, B. and {Barsotti}, L. and {Barsuglia}, M. and {Barta}, D. and {Bartlett}, J. and {Bartos}, I. and {Bassiri}, R. and {Basti}, A. and {Bawaj}, M. and {Bayley}, J.~C. and {Bazzan}, M. and {B{\'e}csy}, B. and {Bejger}, M. and {Belahcene}, I. and {Bell}, A.~S. and {Beniwal}, D. and {Berger}, B.~K. and {Bergmann}, G. and {Bernuzzi}, S. and {Bero}, J.~J. and {Berry}, C.~P.~L. and {Bersanetti}, D. and {Bertolini}, A. and {Betzwieser}, J. and {Bhandare}, R. and {Bidler}, J. and {Bilenko}, I.~A. and {Bilgili}, S.~A. and {Billingsley}, G. and {Birch}, J. and {Birney}, R. and {Birnholtz}, O. and {Biscans}, S. and {Biscoveanu}, S. and {Bisht}, A. and {Bitossi}, M. and {Bizouard}, M.~A. and {Blackburn}, J.~K. and {Blackman}, J. and {Blair}, C.~D. and {Blair}, D.~G. and {Blair}, R.~M. and {Bloemen}, S. and {Bode}, N. and {Boer}, M. and {Boetzel}, Y. and {Bogaert}, G. and {Bondu}, F. and {Bonilla}, E. and {Bonnand}, R. and {Booker}, P. and {Boom}, B.~A. and {Booth}, C.~D. and {Bork}, R. and {Boschi}, V. and {Bose}, S. and {Bossie}, K. and {Bossilkov}, V. and {Bosveld}, J. and {Bouffanais}, Y. and {Bozzi}, A. and {Bradaschia}, C. and {Brady}, P.~R. and {Bramley}, A. and {Branchesi}, M. and {Brau}, J.~E. and {Briant}, T. and {Briggs}, J.~H. and {Brighenti}, F. and {Brillet}, A. and {Brinkmann}, M. and {Brisson}, V. and {Brockill}, P. and {Brooks}, A.~F. and {Brown}, D.~D. and {Brunett}, S. and {Buikema}, A. and {Bulik}, T. and {Bulten}, H.~J. and {Buonanno}, A. and {Buskulic}, D. and {Bustamante Rosell}, M.~J. and {Buy}, C. and {Byer}, R.~L. and {Cabero}, M. and {Cadonati}, L. and {Cagnoli}, G. and {Cahillane}, C. and {Calder{\'o}n Bustillo}, J. and {Callister}, T.~A. and {Calloni}, E. and {Camp}, J.~B. and {Campbell}, W.~A. and {Canepa}, M. and {Cannon}, K.~C. and {Cao}, H. and {Cao}, J. and {Capocasa}, E. and {Carbognani}, F. and {Caride}, S. and {Carney}, M.~F. and {Carullo}, G. and {Casanueva Diaz}, J. and {Casentini}, C. and {Caudill}, S. and {Cavagli{\`a}}, M. and {Cavalier}, F. and {Cavalieri}, R. and {Cella}, G. and {Cerd{\'a}-Dur{\'a}n}, P. and {Cerretani}, G. and {Cesarini}, E. and {Chaibi}, O. and {Chakravarti}, K. and {Chamberlin}, S.~J. and {Chan}, M. and {Chao}, S. and {Charlton}, P. and {Chase}, E.~A. and {Chassande-Mottin}, E. and {Chatterjee}, D. and {Chaturvedi}, M. and {Chatziioannou}, K. and {Cheeseboro}, B.~D. and {Chen}, H.~Y. and {Chen}, X. and {Chen}, Y. and {Cheng}, H.-P. and {Cheong}, C.~K. and {Chia}, H.~Y. and {Chincarini}, A. and {Chiummo}, A. and {Cho}, G. and {Cho}, H.~S. and {Cho}, M. and {Christensen}, N. and {Chu}, Q. and {Chua}, S. and {Chung}, K.~W.},
        title = "{GWTC-1: A Gravitational-Wave Transient Catalog of Compact Binary Mergers Observed by LIGO and Virgo during the First and Second Observing Runs}",
      journal = {Physical Review X},
         year = 2019,
        month = jul,
       volume = {9},
       number = {3},
          eid = {031040},
        pages = {031040},
          doi = {10.1103/PhysRevX.9.031040},
archivePrefix = {arXiv},
       eprint = {1811.12907},
 primaryClass = {astro-ph.HE},
       adsurl = {https://ui.adsabs.harvard.edu/abs/2019PhRvX...9c1040A}
}

@ARTICLE{airapetian_et_al2016,
       author = {{Airapetian}, V.~S. and {Glocer}, A. and {Gronoff}, G. and {H{\'e}brard}, E. and {Danchi}, W.},
        title = "{Prebiotic chemistry and atmospheric warming of early Earth by an active young Sun}",
      journal = {Nature Geoscience},
         year = 2016,
        month = jun,
       volume = {9},
       number = {6},
        pages = {452-455},
          doi = {10.1038/ngeo2719},
       adsurl = {https://ui.adsabs.harvard.edu/abs/2016NatGe...9..452A}
}

@ARTICLE{ardaseva_et_al2017,
       author = {{Ardaseva}, Aleksandra and {Rimmer}, Paul B. and {Waldmann}, Ingo and {Rocchetto}, Marco and {Yurchenko}, Sergey N. and {Helling}, Christiane and {Tennyson}, Jonathan},
        title = "{Lightning chemistry on Earth-like exoplanets}",
      journal = {\mnras},
         year = 2017,
        month = sep,
       volume = {470},
       number = {1},
        pages = {187-196},
          doi = {10.1093/mnras/stx1012},
archivePrefix = {arXiv},
       eprint = {1704.07917},
 primaryClass = {astro-ph.EP},
       adsurl = {https://ui.adsabs.harvard.edu/abs/2017MNRAS.470..187A}
}

@ARTICLE{absil_et_al2013,
       author = {{Absil}, O. and {Defr{\`e}re}, D. and {Coud{\'e} du Foresto}, V. and {Di Folco}, E. and {M{\'e}rand}, A. and {Augereau}, J. -C. and {Ertel}, S. and {Hanot}, C. and {Kervella}, P. and {Mollier}, B. and {Scott}, N. and {Che}, X. and {Monnier}, J.~D. and {Thureau}, N. and {Tuthill}, P.~G. and {ten Brummelaar}, T.~A. and {McAlister}, H.~A. and {Sturmann}, J. and {Sturmann}, L. and {Turner}, N.},
        title = "{A near-infrared interferometric survey of debris-disc stars. III. First statistics based on 42 stars observed with CHARA/FLUOR}",
      journal = {\aap},
         year = 2013,
        month = jul,
       volume = {555},
          eid = {A104},
        pages = {A104},
          doi = {10.1051/0004-6361/201321673},
archivePrefix = {arXiv},
       eprint = {1307.2488},
 primaryClass = {astro-ph.SR},
       adsurl = {https://ui.adsabs.harvard.edu/abs/2013A&A...555A.104A}
}

@ARTICLE{absil_et_al2021,
       author = {{Absil}, O. and {Marion}, L. and {Ertel}, S. and {Defr{\`e}re}, D. and {Kennedy}, G.~M. and {Romagnolo}, A. and {Le Bouquin}, J. -B. and {Christiaens}, V. and {Milli}, J. and {Bonsor}, A. and {Olofsson}, J. and {Su}, K.~Y.~L. and {Augereau}, J. -C.},
        title = "{A near-infrared interferometric survey of debris-disk stars. VII. The hot-to-warm dust connection}",
      journal = {\aap},
         year = 2021,
        month = jul,
       volume = {651},
          eid = {A45},
        pages = {A45},
          doi = {10.1051/0004-6361/202140561},
archivePrefix = {arXiv},
       eprint = {2104.14216},
 primaryClass = {astro-ph.EP},
       adsurl = {https://ui.adsabs.harvard.edu/abs/2021A&A...651A..45A}
}

@ARTICLE{alexandroff_et_al2015,
       author = {{Alexandroff}, Rachael M. and {Heckman}, Timothy M. and {Borthakur}, Sanchayeeta and {Overzier}, Roderik and {Leitherer}, Claus},
        title = "{Indirect Evidence for Escaping Ionizing Photons in Local Lyman Break Galaxy Analogs}",
      journal = {\apj},
         year = 2015,
        month = sep,
       volume = {810},
       number = {2},
          eid = {104},
        pages = {104},
          doi = {10.1088/0004-637X/810/2/104},
archivePrefix = {arXiv},
       eprint = {1504.02446},
 primaryClass = {astro-ph.GA},
       adsurl = {https://ui.adsabs.harvard.edu/abs/2015ApJ...810..104A}
}

@INPROCEEDINGS{alexani_et_al2024,
       author = {{Alexani}, Edwin and {Figer}, Donald F. and {Gatkine}, Pradip},
        title = "{Simulation of a single photon counting photonic spectrograph for exoplanet atmospheric characterization}",
    booktitle = {Space Telescopes and Instrumentation 2024: Optical, Infrared, and Millimeter Wave},
         year = 2024,
       editor = {{Coyle}, Laura E. and {Matsuura}, Shuji and {Perrin}, Marshall D.},
       series = {Society of Photo-Optical Instrumentation Engineers (SPIE) Conference Series},
       volume = {13092},
        month = aug,
          eid = {130925T},
        pages = {130925T},
          doi = {10.1117/12.3016761},
       adsurl = {https://ui.adsabs.harvard.edu/abs/2024SPIE13092E..5TA}
}

@ARTICLE{amorin_et_al2024,
       author = {{Amor{\'\i}n}, R.~O. and {Rodr{\'\i}guez-Henr{\'\i}quez}, M. and {Fern{\'a}ndez}, V. and {V{\'\i}lchez}, J.~M. and {Marques-Chaves}, R. and {Schaerer}, D. and {Izotov}, Y.~I. and {Firpo}, V. and {Guseva}, N. and {Jaskot}, A.~E. and {Komarova}, L. and {Mu{\~n}oz-Vergara}, D. and {Oey}, M.~S. and {Bait}, O. and {Carr}, C. and {Chisholm}, J. and {Ferguson}, H. and {Flury}, S.~R. and {Giavalisco}, M. and {Hayes}, M.~J. and {Henry}, A. and {Ji}, Z. and {King}, W. and {Leclercq}, F. and {{\"O}stlin}, G. and {Pentericci}, L. and {Saldana-Lopez}, A. and {Thuan}, T.~X. and {Trebitsch}, M. and {Wang}, B. and {Worseck}, G. and {Xu}, X.},
        title = "{Ubiquitous broad-line emission and the relation between ionized gas outflows and Lyman continuum escape in Green Pea galaxies}",
      journal = {\aap},
         year = 2024,
        month = feb,
       volume = {682},
          eid = {L25},
        pages = {L25},
          doi = {10.1051/0004-6361/202449175},
archivePrefix = {arXiv},
       eprint = {2401.04278},
 primaryClass = {astro-ph.GA},
       adsurl = {https://ui.adsabs.harvard.edu/abs/2024A&A...682L..25A}
}

@ARTICLE{aller+greenstein1960,
       author = {{Aller}, Lawrence H. and {Greenstein}, Jesse L.},
        title = "{The Abundances of the Elements in G-Type Subdwarfs}",
      journal = {\apjs},
         year = 1960,
        month = nov,
       volume = {5},
        pages = {139},
          doi = {10.1086/190054},
       adsurl = {https://ui.adsabs.harvard.edu/abs/1960ApJS....5..139A}
}

@ARTICLE{allart_et_al2017,
       author = {{Allart}, R. and {Lovis}, C. and {Pino}, L. and {Wyttenbach}, A. and {Ehrenreich}, D. and {Pepe}, F.},
        title = "{Search for water vapor in the high-resolution transmission spectrum of HD 189733b in the visible}",
      journal = {\aap},
         year = 2017,
        month = oct,
       volume = {606},
          eid = {A144},
        pages = {A144},
          doi = {10.1051/0004-6361/201730814},
archivePrefix = {arXiv},
       eprint = {1706.00027},
 primaryClass = {astro-ph.EP},
       adsurl = {https://ui.adsabs.harvard.edu/abs/2017A&A...606A.144A}
}

@ARTICLE{allert_et_al2018,
       author = {{Allart}, R. and {Bourrier}, V. and {Lovis}, C. and {Ehrenreich}, D. and {Spake}, J.~J. and {Wyttenbach}, A. and {Pino}, L. and {Pepe}, F. and {Sing}, D.~K. and {Lecavelier des Etangs}, A.},
        title = "{Spectrally resolved helium absorption from the extended atmosphere of a warm Neptune-mass exoplanet}",
      journal = {Science},
         year = 2018,
        month = dec,
       volume = {362},
       number = {6421},
        pages = {1384-1387},
          doi = {10.1126/science.aat5879},
archivePrefix = {arXiv},
       eprint = {1812.02189},
 primaryClass = {astro-ph.EP},
       adsurl = {https://ui.adsabs.harvard.edu/abs/2018Sci...362.1384A}
}

@ARTICLE{alonso-floriano_et_al2019,
       author = {{Alonso-Floriano}, F.~J. and {Snellen}, I.~A.~G. and {Czesla}, S. and {Bauer}, F.~F. and {Salz}, M. and {Lamp{\'o}n}, M. and {Lara}, L.~M. and {Nagel}, E. and {L{\'o}pez-Puertas}, M. and {Nortmann}, L. and {S{\'a}nchez-L{\'o}pez}, A. and {Sanz-Forcada}, J. and {Caballero}, J.~A. and {Reiners}, A. and {Ribas}, I. and {Quirrenbach}, A. and {Amado}, P.~J. and {Aceituno}, J. and {Anglada-Escud{\'e}}, G. and {B{\'e}jar}, V.~J.~S. and {Brinkm{\"o}ller}, M. and {Hatzes}, A.~P. and {Henning}, Th. and {Kaminski}, A. and {K{\"u}rster}, M. and {Labarga}, F. and {Montes}, D. and {Pall{\'e}}, E. and {Schmitt}, J.~H.~M.~M. and {Zapatero Osorio}, M.~R.},
        title = "{He I {\ensuremath{\lambda}} 10 830 {\r{A}} in the transmission spectrum of HD209458 b}",
      journal = {\aap},
         year = 2019,
        month = sep,
       volume = {629},
          eid = {A110},
        pages = {A110},
          doi = {10.1051/0004-6361/201935979},
archivePrefix = {arXiv},
       eprint = {1907.13425},
 primaryClass = {astro-ph.EP},
       adsurl = {https://ui.adsabs.harvard.edu/abs/2019A&A...629A.110A}
}

@ARTICLE{aloisi_et_al2007,
       author = {{Aloisi}, A. and {Clementini}, G. and {Tosi}, M. and {Annibali}, F. and {Contreras}, R. and {Fiorentino}, G. and {Mack}, J. and {Marconi}, M. and {Musella}, I. and {Saha}, A. and {Sirianni}, M. and {van der Marel}, R.~P.},
        title = "{I Zw 18 Revisited with HST ACS and Cepheids: New Distance and Age}",
      journal = {\apjl},
         year = 2007,
        month = oct,
       volume = {667},
       number = {2},
        pages = {L151-L154},
          doi = {10.1086/522368},
archivePrefix = {arXiv},
       eprint = {0707.2371},
 primaryClass = {astro-ph},
       adsurl = {https://ui.adsabs.harvard.edu/abs/2007ApJ...667L.151A}
}

@ARTICLE{anand_et_al2022,
       author = {{Anand}, Gagandeep S. and {Tully}, R. Brent and {Rizzi}, Luca and {Riess}, Adam G. and {Yuan}, Wenlong},
        title = "{Comparing Tip of the Red Giant Branch Distance Scales: An Independent Reduction of the Carnegie-Chicago Hubble Program and the Value of the Hubble Constant}",
      journal = {\apj},
         year = 2022,
        month = jun,
       volume = {932},
       number = {1},
          eid = {15},
        pages = {15},
          doi = {10.3847/1538-4357/ac68df},
archivePrefix = {arXiv},
       eprint = {2108.00007},
 primaryClass = {astro-ph.CO},
       adsurl = {https://ui.adsabs.harvard.edu/abs/2022ApJ...932...15A}
}

@ARTICLE{anbar_et_al2007,
       author = {{Anbar}, Ariel D. and {Duan}, Yun and {Lyons}, Timothy W. and {Arnold}, Gail L. and {Kendall}, Brian and {Creaser}, Robert A. and {Kaufman}, Alan J. and {Gordon}, Gwyneth W. and {Scott}, Clinton and {Garvin}, Jessica and {Buick}, Roger},
        title = "{A Whiff of Oxygen Before the Great Oxidation Event?}",
      journal = {Science},
         year = 2007,
        month = sep,
       volume = {317},
       number = {5846},
        pages = {1903},
          doi = {10.1126/science.1140325},
       adsurl = {https://ui.adsabs.harvard.edu/abs/2007Sci...317.1903A}
}

@ARTICLE{angel+woolf1997,
       author = {{Angel}, J.~R.~P. and {Woolf}, N.~J.},
        title = "{An Imaging Nulling Interferometer to Study Extrasolar Planets}",
      journal = {\apj},
         year = 1997,
        month = jan,
       volume = {475},
       number = {1},
        pages = {373-379},
          doi = {10.1086/303529},
       adsurl = {https://ui.adsabs.harvard.edu/abs/1997ApJ...475..373A}
}

@ARTICLE{annibali_et_al2013,
       author = {{Annibali}, F. and {Cignoni}, M. and {Tosi}, M. and {van der Marel}, R.~P. and {Aloisi}, A. and {Clementini}, G. and {Contreras Ramos}, R. and {Fiorentino}, G. and {Marconi}, M. and {Musella}, I.},
        title = "{The Star Formation History of the Very Metal-poor Blue Compact Dwarf I Zw 18 from HST/ACS Data}",
      journal = {\aj},
         year = 2013,
        month = dec,
       volume = {146},
       number = {6},
          eid = {144},
        pages = {144},
          doi = {10.1088/0004-6256/146/6/144},
archivePrefix = {arXiv},
       eprint = {1303.3909},
 primaryClass = {astro-ph.CO},
       adsurl = {https://ui.adsabs.harvard.edu/abs/2013AJ....146..144A}
}

@ARTICLE{armus_et_al2021,
       author = {{Armus}, L. and {Megeath}, S.~T. and {Corrales}, L. and {Marengo}, M. and {Kirkpatrick}, A. and {Smith}, J.~D. and {Meyer}, M. and {Gezari}, S. and {Kraft}, R.~P. and {McCandliss}, S. and {Tuttle}, S. and {Elvis}, M. and {Bentz}, M. and {Binder}, B. and {Civano}, F. and {Dragomir}, D. and {Espaillat}, C. and {Finkelstein}, S. and {Fox}, D.~B. and {Greenhouse}, M. and {Hamden}, E. and {Kauffmann}, J. and {Khullar}, G. and {Lazio}, J. and {Lee}, J. and {Lillie}, C. and {Lightsey}, P. and {Mushotzky}, R. and {Scarlata}, C. and {Scowen}, P. and {Tremblay}, G.~R. and {Wang}, Q.~D. and {Wolk}, S.},
        title = "{Great Observatories: The Past and Future of Panchromatic Astrophysics}",
      journal = {arXiv e-prints},
         year = 2021,
        month = mar,
          eid = {arXiv:2104.00023},
        pages = {arXiv:2104.00023},
          doi = {10.48550/arXiv.2104.00023},
archivePrefix = {arXiv},
       eprint = {2104.00023},
 primaryClass = {astro-ph.IM},
       adsurl = {https://ui.adsabs.harvard.edu/abs/2021arXiv210400023A}
}

@ARTICLE{arrigoni_battaia_et_al2016,
       author = {{Arrigoni Battaia}, Fabrizio and {Hennawi}, Joseph F. and {Cantalupo}, Sebastiano and {Prochaska}, J. Xavier},
        title = "{The Stacked LY{\ensuremath{\alpha}} Emission Profile from the Circum-Galactic Medium of z {\ensuremath{\sim}} 2 Quasars}",
      journal = {\apj},
         year = 2016,
        month = sep,
       volume = {829},
       number = {1},
          eid = {3},
        pages = {3},
          doi = {10.3847/0004-637X/829/1/3},
archivePrefix = {arXiv},
       eprint = {1604.02942},
 primaryClass = {astro-ph.CO},
       adsurl = {https://ui.adsabs.harvard.edu/abs/2016ApJ...829....3A}
}

@ARTICLE{antonucci1993,
       author = {{Antonucci}, Robert},
        title = "{Unified models for active galactic nuclei and quasars.}",
      journal = {\araa},
         year = 1993,
        month = jan,
       volume = {31},
        pages = {473-521},
          doi = {10.1146/annurev.aa.31.090193.002353},
       adsurl = {https://ui.adsabs.harvard.edu/abs/1993ARA&A..31..473A}
}

@BOOK{astro2020,
  author    = {{National Academies of Sciences, Engineering, and Medicine}},
  title     = {Pathways to Discovery in Astronomy and Astrophysics for the 2020s: Highlights of a Decadal Survey},
  doi       = {10.17226/26933},
  url       = {https://nap.nationalacademies.org/catalog/26933/pathways-to-discovery-in-astronomy-and-astrophysics-for-the-2020s},
  year      = {2023},
  publisher = {The National Academies Press},
  address   = {Washington, DC}
}

@ARTICLE{atek_et_al2015,
       author = {{Atek}, Hakim and {Richard}, Johan and {Kneib}, Jean-Paul and {Jauzac}, Mathilde and {Schaerer}, Daniel and {Clement}, Benjamin and {Limousin}, Marceau and {Jullo}, Eric and {Natarajan}, Priyamvada and {Egami}, Eiichi and {Ebeling}, Harald},
        title = "{New Constraints on the Faint End of the UV Luminosity Function at z \raisebox{-0.5ex}\textasciitilde 7-8 Using the Gravitational Lensing of the Hubble Frontier Fields Cluster A2744}",
      journal = {\apj},
         year = 2015,
        month = feb,
       volume = {800},
       number = {1},
          eid = {18},
        pages = {18},
          doi = {10.1088/0004-637X/800/1/18},
archivePrefix = {arXiv},
       eprint = {1409.0512},
 primaryClass = {astro-ph.GA},
       adsurl = {https://ui.adsabs.harvard.edu/abs/2015ApJ...800...18A}
}

@ARTICLE{arcavi2018,
       author = {{Arcavi}, Iair},
        title = "{The First Hours of the GW170817 Kilonova and the Importance of Early Optical and Ultraviolet Observations for Constraining Emission Models}",
      journal = {\apjl},
         year = 2018,
        month = mar,
       volume = {855},
       number = {2},
          eid = {L23},
        pages = {L23},
          doi = {10.3847/2041-8213/aab267},
archivePrefix = {arXiv},
       eprint = {1802.02164},
 primaryClass = {astro-ph.HE},
       adsurl = {https://ui.adsabs.harvard.edu/abs/2018ApJ...855L..23A}
}

@ARTICLE{arney_et_al2016,
       author = {{Arney}, Giada and {Domagal-Goldman}, Shawn D. and {Meadows}, Victoria S. and {Wolf}, Eric T. and {Schwieterman}, Edward and {Charnay}, Benjamin and {Claire}, Mark and {H{\'e}brard}, Eric and {Trainer}, Melissa G.},
        title = "{The Pale Orange Dot: The Spectrum and Habitability of Hazy Archean Earth}",
      journal = {Astrobiology},
         year = 2016,
        month = nov,
       volume = {16},
       number = {11},
        pages = {873-899},
          doi = {10.1089/ast.2015.1422},
archivePrefix = {arXiv},
       eprint = {1610.04515},
 primaryClass = {astro-ph.EP},
       adsurl = {https://ui.adsabs.harvard.edu/abs/2016AsBio..16..873A}
}

@ARTICLE{arney_et_al2017,
       author = {{Arney}, Giada N. and {Meadows}, Victoria S. and {Domagal-Goldman}, Shawn D. and {Deming}, Drake and {Robinson}, Tyler D. and {Tovar}, Guadalupe and {Wolf}, Eric T. and {Schwieterman}, Edward},
        title = "{Pale Orange Dots: The Impact of Organic Haze on the Habitability and Detectability of Earthlike Exoplanets}",
      journal = {\apj},
         year = 2017,
        month = feb,
       volume = {836},
       number = {1},
          eid = {49},
        pages = {49},
          doi = {10.3847/1538-4357/836/1/49},
archivePrefix = {arXiv},
       eprint = {1702.02994},
 primaryClass = {astro-ph.EP},
       adsurl = {https://ui.adsabs.harvard.edu/abs/2017ApJ...836...49A}
}

@ARTICLE{arney_et_al2018,
       author = {{Arney}, Giada and {Domagal-Goldman}, Shawn D. and {Meadows}, Victoria S.},
        title = "{Organic Haze as a Biosignature in Anoxic Earth-like Atmospheres}",
      journal = {Astrobiology},
         year = 2018,
        month = mar,
       volume = {18},
       number = {3},
        pages = {311-329},
          doi = {10.1089/ast.2017.1666},
archivePrefix = {arXiv},
       eprint = {1711.01675},
 primaryClass = {astro-ph.EP},
       adsurl = {https://ui.adsabs.harvard.edu/abs/2018AsBio..18..311A}
}

@ARTICLE{arney2019,
       author = {{Arney}, Giada N.},
        title = "{The K Dwarf Advantage for Biosignatures on Directly Imaged Exoplanets}",
      journal = {\apjl},
         year = 2019,
        month = mar,
       volume = {873},
       number = {1},
          eid = {L7},
        pages = {L7},
          doi = {10.3847/2041-8213/ab0651},
archivePrefix = {arXiv},
       eprint = {2001.10458},
 primaryClass = {astro-ph.EP},
       adsurl = {https://ui.adsabs.harvard.edu/abs/2019ApJ...873L...7A}
}

@BOOK{astrobio_strategy2019,
  author    = "{National Academies of Sciences, Engineering, and Medicine}",
  title     = "An Astrobiology Strategy for the Search for Life in the Universe",
  isbn      = "978-0-309-48416-9",
  doi       = "10.17226/25252",
  url       = "https://nap.nationalacademies.org/catalog/25252/an-astrobiology-strategy-for-the-search-for-life-in-the-universe",
  year      = 2019,
  publisher = "The National Academies Press",
  address   = "Washington, DC"
}

@ARTICLE{auriere_et_al2007,
       author = {{Auri{\`e}re}, M. and {Wade}, G.~A. and {Silvester}, J. and {Ligni{\`e}res}, F. and {Bagnulo}, S. and {Bale}, K. and {Dintrans}, B. and {Donati}, J.~F. and {Folsom}, C.~P. and {Gruberbauer}, M. and {Hui Bon Hoa}, A. and {Jeffers}, S. and {Johnson}, N. and {Landstreet}, J.~D. and {L{\`e}bre}, A. and {Lueftinger}, T. and {Marsden}, S. and {Mouillet}, D. and {Naseri}, S. and {Paletou}, F. and {Petit}, P. and {Power}, J. and {Rincon}, F. and {Strasser}, S. and {Toqu{\'e}}, N.},
        title = "{Weak magnetic fields in Ap/Bp stars. Evidence for a dipole field lower limit and a tentative interpretation of the magnetic dichotomy}",
      journal = {\aap},
         year = 2007,
        month = dec,
       volume = {475},
       number = {3},
        pages = {1053-1065},
          doi = {10.1051/0004-6361:20078189},
archivePrefix = {arXiv},
       eprint = {0710.1554},
 primaryClass = {astro-ph},
       adsurl = {https://ui.adsabs.harvard.edu/abs/2007A&A...475.1053A}
}

@ARTICLE{bagdonaite_et_al2014,
       author = {{Bagdonaite}, J. and {Ubachs}, W. and {Murphy}, M.~T. and {Whitmore}, J.~B.},
        title = "{Analysis of Molecular Hydrogen Absorption toward QSO B0642-5038 for a Varying Proton-to-electron Mass Ratio}",
      journal = {\apj},
         year = 2014,
        month = feb,
       volume = {782},
       number = {1},
          eid = {10},
        pages = {10},
          doi = {10.1088/0004-637X/782/1/10},
archivePrefix = {arXiv},
       eprint = {1308.1330},
 primaryClass = {astro-ph.CO},
       adsurl = {https://ui.adsabs.harvard.edu/abs/2014ApJ...782...10B}
}

@MISC{bahcall+odell1979,
       author = {{Bahcall}, J.~N. and {Odell}, C.~R.},
        title = "{The Space Telescope Observatory}",
 howpublished = {NASA Technical Memorandum, NASA/STI Accession number: 19790018872},
         year = 1979,
        month = jun,
        pages = {18872},
       adsurl = {https://ui.adsabs.harvard.edu/abs/1979ntrs.rept18872B}
}

@ARTICLE{baliukin_et_al2019,
       author = {{Baliukin}, I.~I. and {Bertaux}, J. -L. and {Qu{\'e}merais}, E. and {Izmodenov}, V.~V. and {Schmidt}, W.},
        title = "{SWAN/SOHO Lyman-{\ensuremath{\alpha}} Mapping: The Hydrogen Geocorona Extends Well Beyond the Moon}",
      journal = {Journal of Geophysical Research (Space Physics)},
         year = 2019,
        month = feb,
       volume = {124},
       number = {2},
        pages = {861-885},
          doi = {10.1029/2018JA026136},
       adsurl = {https://ui.adsabs.harvard.edu/abs/2019JGRA..124..861B}
}

@ARTICLE{barrow_et_al2020,
       author = {{Barrow}, Kirk S.~S. and {Robertson}, Brant E. and {Ellis}, Richard S. and {Nakajima}, Kimihiko and {Saxena}, Aayush and {Stark}, Daniel P. and {Tang}, Mengtao},
        title = "{The Lyman Continuum Escape Survey: Connecting Time-dependent [O III] and [O II] Line Emission with Lyman Continuum Escape Fraction in Simulations of Galaxy Formation}",
      journal = {\apjl},
         year = 2020,
        month = oct,
       volume = {902},
       number = {2},
          eid = {L39},
        pages = {L39},
          doi = {10.3847/2041-8213/abbd8e},
archivePrefix = {arXiv},
       eprint = {2010.00592},
 primaryClass = {astro-ph.GA},
       adsurl = {https://ui.adsabs.harvard.edu/abs/2020ApJ...902L..39B}
}

@ARTICLE{barstow2016,
       author = {{Barstow}, J.~K. and {Aigrain}, S. and {Irwin}, P.~G.~J. and {Kendrew}, S. and {Fletcher}, L.~N.},
        title = "{Telling twins apart: exo-Earths and Venuses with transit spectroscopy}",
      journal = {\mnras},
         year = 2016,
        month = may,
       volume = {458},
       number = {3},
        pages = {2657-2666},
          doi = {10.1093/mnras/stw489},
archivePrefix = {arXiv},
       eprint = {1602.08277},
 primaryClass = {astro-ph.EP},
       adsurl = {https://ui.adsabs.harvard.edu/abs/2016MNRAS.458.2657B}
}

@ARTICLE{baumeister_et_al2023,
       author = {{Baumeister}, Philipp and {Tosi}, Nicola and {Brachmann}, Caroline and {Grenfell}, John Lee and {Noack}, Lena},
        title = "{Redox state and interior structure control on the long-term habitability of stagnant-lid planets}",
      journal = {\aap},
         year = 2023,
        month = jul,
       volume = {675},
          eid = {A122},
        pages = {A122},
          doi = {10.1051/0004-6361/202245791},
archivePrefix = {arXiv},
       eprint = {2301.03466},
 primaryClass = {astro-ph.EP},
       adsurl = {https://ui.adsabs.harvard.edu/abs/2023A&A...675A.122B}
}

@ARTICLE{beasor_et_al2020,
       author = {{Beasor}, Emma R. and {Davies}, Ben and {Smith}, Nathan and {van Loon}, Jacco Th and {Gehrz}, Robert D. and {Figer}, Donald F.},
        title = "{A new mass-loss rate prescription for red supergiants}",
      journal = {\mnras},
         year = 2020,
        month = mar,
       volume = {492},
       number = {4},
        pages = {5994-6006},
          doi = {10.1093/mnras/staa255},
archivePrefix = {arXiv},
       eprint = {2001.07222},
 primaryClass = {astro-ph.SR},
       adsurl = {https://ui.adsabs.harvard.edu/abs/2020MNRAS.492.5994B}
}

@ARTICLE{beaton_et_al2016,
       author = {{Beaton}, Rachael L. and {Freedman}, Wendy L. and {Madore}, Barry F. and {Bono}, Giuseppe and {Carlson}, Erika K. and {Clementini}, Gisella and {Durbin}, Meredith J. and {Garofalo}, Alessia and {Hatt}, Dylan and {Jang}, In Sung and {Kollmeier}, Juna A. and {Lee}, Myung Gyoon and {Monson}, Andrew J. and {Rich}, Jeffrey A. and {Scowcroft}, Victoria and {Seibert}, Mark and {Sturch}, Laura and {Yang}, Soung-Chul},
        title = "{The Carnegie-Chicago Hubble Program. I. An Independent Approach to the Extragalactic Distance Scale Using Only Population II Distance Indicators}",
      journal = {\apj},
         year = 2016,
        month = dec,
       volume = {832},
       number = {2},
          eid = {210},
        pages = {210},
          doi = {10.3847/0004-637X/832/2/210},
archivePrefix = {arXiv},
       eprint = {1604.01788},
 primaryClass = {astro-ph.CO},
       adsurl = {https://ui.adsabs.harvard.edu/abs/2016ApJ...832..210B}
}

@ARTICLE{beaton_et_al2018,
       author = {{Beaton}, Rachael L. and {Bono}, Giuseppe and {Braga}, Vittorio Francesco and {Dall'Ora}, Massimo and {Fiorentino}, Giuliana and {Jang}, In Sung and {Mart{\'\i}nez-V{\'a}zquez}, Clara E. and {Matsunaga}, Noriyuki and {Monelli}, Matteo and {Neeley}, Jillian R. and {Salaris}, Maurizio},
        title = "{Old-Aged Primary Distance Indicators}",
      journal = {\ssr},
         year = 2018,
        month = dec,
       volume = {214},
       number = {8},
          eid = {113},
        pages = {113},
          doi = {10.1007/s11214-018-0542-1},
archivePrefix = {arXiv},
       eprint = {1808.09191},
 primaryClass = {astro-ph.GA},
       adsurl = {https://ui.adsabs.harvard.edu/abs/2018SSRv..214..113B}
}

@ARTICLE{beaton_et_al2019,
       author = {{Beaton}, Rachael L. and {Seibert}, Mark and {Hatt}, Dylan and {Freedman}, Wendy L. and {Hoyt}, Taylor J. and {Jang}, In Sung and {Lee}, Myung Gyoon and {Madore}, Barry F. and {Monson}, Andrew J. and {Neeley}, Jillian R. and {Rich}, Jeffrey A. and {Scowcroft}, Victoria},
        title = "{The Carnegie-Chicago Hubble Program. VII. The Distance to M101 via the Optical Tip of the Red Giant Branch Method}",
      journal = {\apj},
         year = 2019,
        month = nov,
       volume = {885},
       number = {2},
          eid = {141},
        pages = {141},
          doi = {10.3847/1538-4357/ab4263},
archivePrefix = {arXiv},
       eprint = {1908.06120},
 primaryClass = {astro-ph.GA},
       adsurl = {https://ui.adsabs.harvard.edu/abs/2019ApJ...885..141B}
}

@ARTICLE{beers+christlieb2005,
       author = {{Beers}, Timothy C. and {Christlieb}, Norbert},
        title = "{The Discovery and Analysis of Very Metal-Poor Stars in the Galaxy}",
      journal = {\araa},
         year = 2005,
        month = sep,
       volume = {43},
       number = {1},
        pages = {531-580},
          doi = {10.1146/annurev.astro.42.053102.134057},
       adsurl = {https://ui.adsabs.harvard.edu/abs/2005ARA&A..43..531B}
}

@ARTICLE{bellovary_et_al2019,
       author = {{Bellovary}, Jillian M. and {Cleary}, Colleen E. and {Munshi}, Ferah and {Tremmel}, Michael and {Christensen}, Charlotte R. and {Brooks}, Alyson and {Quinn}, Thomas R.},
        title = "{Multimessenger signatures of massive black holes in dwarf galaxies}",
      journal = {\mnras},
         year = 2019,
        month = jan,
       volume = {482},
       number = {3},
        pages = {2913-2923},
          doi = {10.1093/mnras/sty2842},
archivePrefix = {arXiv},
       eprint = {1806.00471},
 primaryClass = {astro-ph.GA},
       adsurl = {https://ui.adsabs.harvard.edu/abs/2019MNRAS.482.2913B}
}

@INPROCEEDINGS{bendek_et_al2020,
       author = {{Bendek}, Eduardo and {Noyes}, Matthew and {Flores}, Catalina and {Belikov}, Ruslan and {Sirbu}, Dan and {Mejia Prada}, Camilo and {Tuthill}, Peter and {Guyon}, Olivier},
        title = "{Status of NASA's stellar astrometry testbeds for exoplanet detection: Science and technology overview}",
    booktitle = {Space Telescopes and Instrumentation 2020: Optical, Infrared, and Millimeter Wave},
         year = 2020,
       editor = {{Lystrup}, Makenzie and {Perrin}, Marshall D.},
       series = {Society of Photo-Optical Instrumentation Engineers (SPIE) Conference Series},
       volume = {11443},
        month = dec,
          eid = {114433V},
        pages = {114433V},
          doi = {10.1117/12.2562994},
       adsurl = {https://ui.adsabs.harvard.edu/abs/2020SPIE11443E..3VB}
}

@ARTICLE{bennert_et_al2021,
       author = {{Bennert}, Vardha N. and {Treu}, Tommaso and {Ding}, Xuheng and {Stomberg}, Isak and {Birrer}, Simon and {Snyder}, Tomas and {Malkan}, Matthew A. and {Stephens}, Andrew W. and {Auger}, Matthew W.},
        title = "{A Local Baseline of the Black Hole Mass Scaling Relations for Active Galaxies. IV. Correlations Between M$_{BH}$ and Host Galaxy {\ensuremath{\sigma}}, Stellar Mass, and Luminosity}",
      journal = {\apj},
         year = 2021,
        month = nov,
       volume = {921},
       number = {1},
          eid = {36},
        pages = {36},
          doi = {10.3847/1538-4357/ac151a},
archivePrefix = {arXiv},
       eprint = {2101.10355},
 primaryClass = {astro-ph.GA},
       adsurl = {https://ui.adsabs.harvard.edu/abs/2021ApJ...921...36B}
}

@ARTICLE{berdyugina_et_al2008,
       author = {{Berdyugina}, S.~V. and {Berdyugin}, A.~V. and {Fluri}, D.~M. and {Piirola}, V.},
        title = "{First Detection of Polarized Scattered Light from an Exoplanetary Atmosphere}",
      journal = {\apjl},
         year = 2008,
        month = jan,
       volume = {673},
       number = {1},
        pages = {L83},
          doi = {10.1086/527320},
archivePrefix = {arXiv},
       eprint = {0712.0193},
 primaryClass = {astro-ph},
       adsurl = {https://ui.adsabs.harvard.edu/abs/2008ApJ...673L..83B}
}

@ARTICLE{berdyugina_et_al2011,
       author = {{Berdyugina}, S.~V. and {Berdyugin}, A.~V. and {Fluri}, D.~M. and {Piirola}, V.},
        title = "{Polarized Reflected Light from the Exoplanet HD189733b: First Multicolor Observations and Confirmation of Detection}",
      journal = {\apjl},
         year = 2011,
        month = feb,
       volume = {728},
       number = {1},
          eid = {L6},
        pages = {L6},
          doi = {10.1088/2041-8205/728/1/L6},
archivePrefix = {arXiv},
       eprint = {1101.0059},
 primaryClass = {astro-ph.EP},
       adsurl = {https://ui.adsabs.harvard.edu/abs/2011ApJ...728L...6B}
}

@ARTICLE{berdyugina_et_al2016,
       author = {{Berdyugina}, Svetlana V. and {Kuhn}, Jeff R. and {Harrington}, David M. and {{\v{S}}antl-Temkiv}, Tina and {Messersmith}, E. John},
        title = "{Remote sensing of life: polarimetric signatures of photosynthetic pigments as sensitive biomarkers}",
      journal = {International Journal of Astrobiology},
         year = 2016,
        month = jan,
       volume = {15},
       number = {1},
        pages = {45-56},
          doi = {10.1017/S1473550415000129},
       adsurl = {https://ui.adsabs.harvard.edu/abs/2016IJAsB..15...45B}
}

@article{berdyugina+kuhn2019,
  title = {Surface {{Imaging}} of {{Proxima}} b and {{Other Exoplanets}}: {{Albedo Maps}}, {{Biosignatures}}, and {{Technosignatures}}},
  shorttitle = {Surface {{Imaging}} of {{Proxima}} b and {{Other Exoplanets}}},
  author = {Berdyugina, S. V. and Kuhn, J. R.},
  year = {2019},
  month = dec,
  journal = {The Astronomical Journal},
  volume = {158},
  pages = {246},
  publisher = {IOP},
  issn = {0004-6256},
  doi = {10.3847/1538-3881/ab2df3},
  urldate = {2024-04-17}
}

@ARTICLE{berdyugina_et_al2025,
       author = {{Berdyugina}, Svetlana and {Patty}, Lucas and {Grone}, Jonathan and {Demory}, Brice and {Bott}, Kim and {Kofman}, Vincent and {Roccetti}, Giulia and {Goodis Gordon}, Kenneth and {Snik}, Frans and {Karalidi}, Theodora and {Trees}, Victor and {Stam}, Daphne and {Parenteau}, Mary N.},
        title = "{Detecting alien living worlds and photosynthetic life using imaging polarimetry with the HWO coronagraph}",
      journal = {arXiv e-prints},
         year = 2025,
        month = jul,
          eid = {arXiv:2507.03819},
        pages = {arXiv:2507.03819},
          doi = {10.48550/arXiv.2507.03819},
archivePrefix = {arXiv},
       eprint = {2507.03819},
 primaryClass = {astro-ph.EP},
       adsurl = {https://ui.adsabs.harvard.edu/abs/2025arXiv250703819B}
}

@ARTICLE{berengut_et_al2013,
       author = {{Berengut}, J.~C. and {Flambaum}, V.~V. and {Ong}, A. and {Webb}, J.~K. and {Barrow}, John D. and {Barstow}, M.~A. and {Preval}, S.~P. and {Holberg}, J.~B.},
        title = "{Limits on the Dependence of the Fine-Structure Constant on Gravitational Potential from White-Dwarf Spectra}",
      journal = {\prl},
         year = 2013,
        month = jul,
       volume = {111},
       number = {1},
          eid = {010801},
        pages = {010801},
          doi = {10.1103/PhysRevLett.111.010801},
archivePrefix = {arXiv},
       eprint = {1305.1337},
 primaryClass = {astro-ph.CO},
       adsurl = {https://ui.adsabs.harvard.edu/abs/2013PhRvL.111a0801B}
}

@ARTICLE{berg_et_al2019,
       author = {{Berg}, Danielle A. and {Erb}, Dawn K. and {Henry}, Richard B.~C. and {Skillman}, Evan D. and {McQuinn}, Kristen B.~W.},
        title = "{The Chemical Evolution of Carbon, Nitrogen, and Oxygen in Metal-poor Dwarf Galaxies}",
      journal = {\apj},
         year = 2019,
        month = mar,
       volume = {874},
       number = {1},
          eid = {93},
        pages = {93},
          doi = {10.3847/1538-4357/ab020a},
archivePrefix = {arXiv},
       eprint = {1901.08160},
 primaryClass = {astro-ph.GA},
       adsurl = {https://ui.adsabs.harvard.edu/abs/2019ApJ...874...93B}
}

@ARTICLE{berg_et_al2024,
       author = {{Berg}, Danielle A. and {Skillman}, Evan D. and {Chisholm}, John and {Pogge}, Richard W. and {Gazagnes}, Simon and {Rogers}, Noah S.~J. and {Erb}, Dawn K. and {Arellano-C{\'o}rdova}, Karla Z. and {Leitherer}, Claus and {Appel}, Jackie and {Moustakas}, John},
        title = "{CHAOS. VIII. Far-ultraviolet Spectra of M101 and the Impact of Wolf{\textendash}Rayet Stars}",
      journal = {\apj},
         year = 2024,
        month = aug,
       volume = {971},
       number = {1},
          eid = {87},
        pages = {87},
          doi = {10.3847/1538-4357/ad5292},
archivePrefix = {arXiv},
       eprint = {2405.19477},
 primaryClass = {astro-ph.GA},
       adsurl = {https://ui.adsabs.harvard.edu/abs/2024ApJ...971...87B}
}

@ARTICLE{bestenlehner_et_al2014,
       author = {{Bestenlehner}, J.~M. and {Gr{\"a}fener}, G. and {Vink}, J.~S. and {Najarro}, F. and {de Koter}, A. and {Sana}, H. and {Evans}, C.~J. and {Crowther}, P.~A. and {H{\'e}nault-Brunet}, V. and {Herrero}, A. and {Langer}, N. and {Schneider}, F.~R.~N. and {Sim{\'o}n-D{\'\i}az}, S. and {Taylor}, W.~D. and {Walborn}, N.~R.},
        title = "{The VLT-FLAMES Tarantula Survey. XVII. Physical and wind properties of massive stars at the top of the main sequence}",
      journal = {\aap},
         year = 2014,
        month = oct,
       volume = {570},
          eid = {A38},
        pages = {A38},
          doi = {10.1051/0004-6361/201423643},
archivePrefix = {arXiv},
       eprint = {1407.1837},
 primaryClass = {astro-ph.SR},
       adsurl = {https://ui.adsabs.harvard.edu/abs/2014A&A...570A..38B}
}

@ARTICLE{bestenlehner_et_al2020,
       author = {{Bestenlehner}, Joachim M. and {Crowther}, Paul A. and {Caballero-Nieves}, Saida M. and {Schneider}, Fabian R.~N. and {Sim{\'o}n-D{\'\i}az}, Sergio and {Brands}, Sarah A. and {de Koter}, Alex and {Gr{\"a}fener}, G{\"o}tz and {Herrero}, Artemio and {Langer}, Norbert and {Lennon}, Daniel J. and {Ma{\'\i}z Apell{\'a}niz}, Jesus and {Puls}, Joachim and {Vink}, Jorick S.},
        title = "{The R136 star cluster dissected with Hubble Space Telescope/STIS - II. Physical properties of the most massive stars in R136}",
      journal = {\mnras},
         year = 2020,
        month = dec,
       volume = {499},
       number = {2},
        pages = {1918-1936},
          doi = {10.1093/mnras/staa2801},
archivePrefix = {arXiv},
       eprint = {2009.05136},
 primaryClass = {astro-ph.SR},
       adsurl = {https://ui.adsabs.harvard.edu/abs/2020MNRAS.499.1918B}
}

@INPROCEEDINGS{bitten_et_al2019,
  author={Bitten, Robert E. and Shinn, Stephen A. and Emmons, Debra L.},
  booktitle={2019 IEEE Aerospace Conference}, 
  title={Challenges and Potential Solutions to Develop and Fund NASA Flagship Missions}, 
  year={2019},
  volume={},
  number={},
  pages={1-13},
  doi={10.1109/AERO.2019.8741920}}

@ARTICLE{bixel_et_al2020,
       author = {{Bixel}, Alex and {Apai}, D{\'a}niel},
        title = "{Identifying Exo-Earth Candidates in Direct Imaging Data through Bayesian Classification}",
      journal = {\aj},
         year = 2020,
        month = jan,
       volume = {159},
       number = {1},
          eid = {3},
        pages = {3},
          doi = {10.3847/1538-3881/ab5222},
archivePrefix = {arXiv},
       eprint = {1910.13440},
 primaryClass = {astro-ph.EP},
       adsurl = {https://ui.adsabs.harvard.edu/abs/2020AJ....159....3B}
}

@ARTICLE{bluhm+de_boer2001,
       author = {{Bluhm}, H. and {de Boer}, K.~S.},
        title = "{H$_{2}$, HD, and CO at the edge of 30 Dor in the LMC: The line of sight to Sk-69 246}",
      journal = {\aap},
         year = 2001,
        month = nov,
       volume = {379},
        pages = {82-89},
          doi = {10.1051/0004-6361:20011302},
archivePrefix = {arXiv},
       eprint = {astro-ph/0109297},
 primaryClass = {astro-ph},
       adsurl = {https://ui.adsabs.harvard.edu/abs/2001A&A...379...82B}
}

@ARTICLE{blumenthal_et_al1984,
       author = {{Blumenthal}, G.~R. and {Faber}, S.~M. and {Primack}, J.~R. and {Rees}, M.~J.},
        title = "{Formation of galaxies and large-scale structure with cold dark matter.}",
      journal = {\nat},
         year = 1984,
        month = oct,
       volume = {311},
        pages = {517-525},
          doi = {10.1038/311517a0},
       adsurl = {https://ui.adsabs.harvard.edu/abs/1984Natur.311..517B}
}

@ARTICLE{blunt_et_al2025,
       author = {{Blunt}, Sarah and {Nielsen}, Eric L. and {Newton}, Elisabeth R. and {Christiansen}, Jessie and {Daylan}, Tansu and {Dressing}, Courtney and {Harada}, Caleb K. and {Kane}, Stephen R. and {Rice}, Malena and {Mart{\'\i}nez}, Romy Rodr{\'\i}guez and {Sagynbayeva}, Sabina},
        title = "{Statistical method for constraining the capability of the Habitable Worlds Observatory to understand ozone onset time in Earth analogs}",
      journal = {Journal of Astronomical Telescopes, Instruments, and Systems},
         year = 2025,
        month = oct,
       volume = {11},
          eid = {042214},
        pages = {042214},
          doi = {10.1117/1.JATIS.11.4.042214},
archivePrefix = {arXiv},
       eprint = {2507.06188},
 primaryClass = {astro-ph.EP},
       adsurl = {https://ui.adsabs.harvard.edu/abs/2025JATIS..11d2214B}
}

@ARTICLE{bonsor_et_al2018,
       author = {{Bonsor}, Amy and {Wyatt}, Mark C. and {Kral}, Quentin and {Kennedy}, Grant and {Shannon}, Andrew and {Ertel}, Steve},
        title = "{Using warm dust to constrain unseen planets}",
      journal = {\mnras},
         year = 2018,
        month = nov,
       volume = {480},
       number = {4},
        pages = {5560-5579},
          doi = {10.1093/mnras/sty2200},
archivePrefix = {arXiv},
       eprint = {1807.11536},
 primaryClass = {astro-ph.EP},
       adsurl = {https://ui.adsabs.harvard.edu/abs/2018MNRAS.480.5560B}
}

@ARTICLE{bonsor_et_al2012,
       author = {{Bonsor}, A. and {Augereau}, J. -C. and {Th{\'e}bault}, P.},
        title = "{Scattering of small bodies by planets: a potential origin for exozodiacal dust?}",
      journal = {\aap},
         year = 2012,
        month = dec,
       volume = {548},
          eid = {A104},
        pages = {A104},
          doi = {10.1051/0004-6361/201220005},
archivePrefix = {arXiv},
       eprint = {1209.6033},
 primaryClass = {astro-ph.EP},
       adsurl = {https://ui.adsabs.harvard.edu/abs/2012A&A...548A.104B}
}

@ARTICLE{borthakur_et_al2025,
       author = {{Borthakur}, Sanchayeeta and {Burchett}, Joseph N. and {Cashman}, Frances and {Fox}, Andrew J. and {Zheng}, Yong and {French}, David M. and {Bordoloi}, Rongmon and {Koplitz}, Brad},
        title = "{Characterizing gas flows through observations of the disk-circumgalactic medium interface with the Habitable Worlds Observatory}",
      journal = {Journal of Astronomical Telescopes, Instruments, and Systems},
         year = 2025,
        month = oct,
       volume = {11},
          eid = {042207},
        pages = {042207},
          doi = {10.1117/1.JATIS.11.4.042207},
archivePrefix = {arXiv},
       eprint = {2506.10517},
 primaryClass = {astro-ph.GA},
       adsurl = {https://ui.adsabs.harvard.edu/abs/2025JATIS..11d2207B}
}

@ARTICLE{bouma_et_al2023,
       author = {{Bouma}, Luke G. and {Palumbo}, Elsa K. and {Hillenbrand}, Lynne A.},
        title = "{The Empirical Limits of Gyrochronology}",
      journal = {\apjl},
         year = 2023,
        month = apr,
       volume = {947},
       number = {1},
          eid = {L3},
        pages = {L3},
          doi = {10.3847/2041-8213/acc589},
archivePrefix = {arXiv},
       eprint = {2303.08830},
 primaryClass = {astro-ph.SR},
       adsurl = {https://ui.adsabs.harvard.edu/abs/2023ApJ...947L...3B}
}

@ARTICLE{bouret_et_al2003,
       author = {{Bouret}, J. -C. and {Lanz}, T. and {Hillier}, D.~J. and {Heap}, S.~R. and {Hubeny}, I. and {Lennon}, D.~J. and {Smith}, L.~J. and {Evans}, C.~J.},
        title = "{Quantitative Spectroscopy of O Stars at Low Metallicity: O Dwarfs in NGC 346}",
      journal = {\apj},
         year = 2003,
        month = oct,
       volume = {595},
       number = {2},
        pages = {1182-1205},
          doi = {10.1086/377368},
archivePrefix = {arXiv},
       eprint = {astro-ph/0301454},
 primaryClass = {astro-ph},
       adsurl = {https://ui.adsabs.harvard.edu/abs/2003ApJ...595.1182B}
}

@ARTICLE{bouret_et_al2013,
       author = {{Bouret}, J. -C. and {Lanz}, T. and {Martins}, F. and {Marcolino}, W.~L.~F. and {Hillier}, D.~J. and {Depagne}, E. and {Hubeny}, I.},
        title = "{Massive stars at low metallicity. Evolution and surface abundances of O dwarfs in the SMC}",
      journal = {\aap},
         year = 2013,
        month = jul,
       volume = {555},
          eid = {A1},
        pages = {A1},
          doi = {10.1051/0004-6361/201220798},
archivePrefix = {arXiv},
       eprint = {1304.6923},
 primaryClass = {astro-ph.SR},
       adsurl = {https://ui.adsabs.harvard.edu/abs/2013A&A...555A...1B}
}

@ARTICLE{bouret_et_al2015,
       author = {{Bouret}, J. -C. and {Lanz}, T. and {Hillier}, D.~J. and {Martins}, F. and {Marcolino}, W.~L.~F. and {Depagne}, E.},
        title = "{No breakdown of the radiatively driven wind theory in low-metallicity environments}",
      journal = {\mnras},
         year = 2015,
        month = may,
       volume = {449},
       number = {2},
        pages = {1545-1569},
          doi = {10.1093/mnras/stv379},
archivePrefix = {arXiv},
       eprint = {1502.05641},
 primaryClass = {astro-ph.SR},
       adsurl = {https://ui.adsabs.harvard.edu/abs/2015MNRAS.449.1545B}
}

@INPROCEEDINGS{bouret_et_al2018,
       author = {{Bouret}, Jean-Claude and {Neiner}, Coralie and {G{\'o}mez de Castro}, Ana I. and {Evans}, Chris and {Gaensicke}, Boris and {Shore}, Steve and {Fossati}, Luca and {Gry}, C{\'e}cile and {Charlot}, St{\'e}phane and {Marin}, Fr{\'e}d{\'e}ric and {Noterdaeme}, Pasquier and {Chaufray}, Jean-Yves},
        title = "{The science case for POLLUX: a high-resolution UV spectropolarimeter onboard LUVOIR}",
    booktitle = {Space Telescopes and Instrumentation 2018: Ultraviolet to Gamma Ray},
         year = 2018,
       editor = {{den Herder}, Jan-Willem A. and {Nikzad}, Shouleh and {Nakazawa}, Kazuhiro},
       series = {Society of Photo-Optical Instrumentation Engineers (SPIE) Conference Series},
       volume = {10699},
        month = jul,
          eid = {106993B},
        pages = {106993B},
          doi = {10.1117/12.2312621},
archivePrefix = {arXiv},
       eprint = {1805.10021},
 primaryClass = {astro-ph.IM},
       adsurl = {https://ui.adsabs.harvard.edu/abs/2018SPIE10699E..3BB}
}

@ARTICLE{boyajian_et_al2012,
       author = {{Boyajian}, Tabetha S. and {von Braun}, Kaspar and {van Belle}, Gerard and {McAlister}, Harold A. and {ten Brummelaar}, Theo A. and {Kane}, Stephen R. and {Muirhead}, Philip S. and {Jones}, Jeremy and {White}, Russel and {Schaefer}, Gail and {Ciardi}, David and {Henry}, Todd and {L{\'o}pez-Morales}, Mercedes and {Ridgway}, Stephen and {Gies}, Douglas and {Jao}, Wei-Chun and {Rojas-Ayala}, B{\'a}rbara and {Parks}, J. Robert and {Sturmann}, Laszlo and {Sturmann}, Judit and {Turner}, Nils H. and {Farrington}, Chris and {Goldfinger}, P.~J. and {Berger}, David H.},
        title = "{Stellar Diameters and Temperatures. II. Main-sequence K- and M-stars}",
      journal = {\apj},
         year = 2012,
        month = oct,
       volume = {757},
       number = {2},
          eid = {112},
        pages = {112},
          doi = {10.1088/0004-637X/757/2/112},
archivePrefix = {arXiv},
       eprint = {1208.2431},
 primaryClass = {astro-ph.SR},
       adsurl = {https://ui.adsabs.harvard.edu/abs/2012ApJ...757..112B}
}

@ARTICLE{boyarsky_et_al2019,
       author = {{Boyarsky}, A. and {Drewes}, M. and {Lasserre}, T. and {Mertens}, S. and {Ruchayskiy}, O.},
        title = "{Sterile neutrino Dark Matter}",
      journal = {Progress in Particle and Nuclear Physics},
         year = 2019,
        month = jan,
       volume = {104},
        pages = {1-45},
          doi = {10.1016/j.ppnp.2018.07.004},
archivePrefix = {arXiv},
       eprint = {1807.07938},
 primaryClass = {hep-ph},
       adsurl = {https://ui.adsabs.harvard.edu/abs/2019PrPNP.104....1B}
}

@ARTICLE{brands_et_al2022,
       author = {{Brands}, Sarah A. and {de Koter}, Alex and {Bestenlehner}, Joachim M. and {Crowther}, Paul A. and {Sundqvist}, Jon O. and {Puls}, Joachim and {Caballero-Nieves}, Saida M. and {Abdul-Masih}, Michael and {Driessen}, Florian A. and {Garc{\'\i}a}, Miriam and {Geen}, Sam and {Gr{\"a}fener}, G{\"o}tz and {Hawcroft}, Calum and {Kaper}, Lex and {Keszthelyi}, Zsolt and {Langer}, Norbert and {Sana}, Hugues and {Schneider}, Fabian R.~N. and {Shenar}, Tomer and {Vink}, Jorick S.},
        title = "{The R136 star cluster dissected with Hubble Space Telescope/STIS. III. The most massive stars and their clumped winds}",
      journal = {\aap},
         year = 2022,
        month = jul,
       volume = {663},
          eid = {A36},
        pages = {A36},
          doi = {10.1051/0004-6361/202142742},
archivePrefix = {arXiv},
       eprint = {2202.11080},
 primaryClass = {astro-ph.SR},
       adsurl = {https://ui.adsabs.harvard.edu/abs/2022A&A...663A..36B}
}

@ARTICLE{brogi+line2019,
       author = {{Brogi}, Matteo and {Line}, Michael R.},
        title = "{Retrieving Temperatures and Abundances of Exoplanet Atmospheres with High-resolution Cross-correlation Spectroscopy}",
      journal = {\aj},
         year = 2019,
        month = mar,
       volume = {157},
       number = {3},
          eid = {114},
        pages = {114},
          doi = {10.3847/1538-3881/aaffd3},
archivePrefix = {arXiv},
       eprint = {1811.01681},
 primaryClass = {astro-ph.EP},
       adsurl = {https://ui.adsabs.harvard.edu/abs/2019AJ....157..114B}
}

@ARTICLE{bromm+larson2004,
       author = {{Bromm}, Volker and {Larson}, Richard B.},
        title = "{The First Stars}",
      journal = {\araa},
         year = 2004,
        month = sep,
       volume = {42},
       number = {1},
        pages = {79-118},
          doi = {10.1146/annurev.astro.42.053102.134034},
archivePrefix = {arXiv},
       eprint = {astro-ph/0311019},
 primaryClass = {astro-ph},
       adsurl = {https://ui.adsabs.harvard.edu/abs/2004ARA&A..42...79B}
}

@ARTICLE{bromm2013,
       author = {{Bromm}, Volker},
        title = "{Formation of the first stars}",
      journal = {Reports on Progress in Physics},
         year = 2013,
        month = nov,
       volume = {76},
       number = {11},
          eid = {112901},
        pages = {112901},
          doi = {10.1088/0034-4885/76/11/112901},
archivePrefix = {arXiv},
       eprint = {1305.5178},
 primaryClass = {astro-ph.CO},
       adsurl = {https://ui.adsabs.harvard.edu/abs/2013RPPh...76k2901B}
}

@ARTICLE{brown_et_al2002,
       author = {{Brown}, Thomas M. and {Heap}, Sara R. and {Hubeny}, Ivan and {Lanz}, Thierry and {Lindler}, Don},
        title = "{Isolating Clusters with Wolf-Rayet Stars in I Zw 18}",
      journal = {\apjl},
         year = 2002,
        month = nov,
       volume = {579},
       number = {2},
        pages = {L75-L78},
          doi = {10.1086/345336},
archivePrefix = {arXiv},
       eprint = {astro-ph/0210089},
 primaryClass = {astro-ph},
       adsurl = {https://ui.adsabs.harvard.edu/abs/2002ApJ...579L..75B}
}

@ARTICLE{breuval_et_al2024,
       author = {{Breuval}, Louise and {Riess}, Adam G. and {Casertano}, Stefano and {Yuan}, Wenlong and {Macri}, Lucas M. and {Romaniello}, Martino and {Murakami}, Yukei S. and {Scolnic}, Daniel and {Anand}, Gagandeep S. and {Soszy{\'n}ski}, Igor},
        title = "{Small Magellanic Cloud Cepheids Observed with the Hubble Space Telescope Provide a New Anchor for the SH0ES Distance Ladder}",
      journal = {\apj},
         year = 2024,
        month = sep,
       volume = {973},
       number = {1},
          eid = {30},
        pages = {30},
          doi = {10.3847/1538-4357/ad630e},
archivePrefix = {arXiv},
       eprint = {2404.08038},
 primaryClass = {astro-ph.CO},
       adsurl = {https://ui.adsabs.harvard.edu/abs/2024ApJ...973...30B}
}

@ARTICLE{brugman_et_al2021,
       author = {{Brugman}, K. and {Phillips}, M.~G. and {Till}, C.~B.},
        title = "{Experimental Determination of Mantle Solidi and Melt Compositions for Two Likely Rocky Exoplanet Compositions}",
      journal = {Journal of Geophysical Research (Planets)},
         year = 2021,
        month = jul,
       volume = {126},
       number = {7},
          eid = {e06731},
        pages = {e06731},
          doi = {10.1029/2020JE00673110.31223/x5wg7z},
       adsurl = {https://ui.adsabs.harvard.edu/abs/2021JGRE..12606731B}
}

@ARTICLE{bruna_et_al2023,
       author = {{Bruna}, Margaret and {Cowan}, Nicolas B. and {Sheffler}, Julia and {Haggard}, Hal M. and {Bourdon}, Audrey and {M{\^a}lin}, Mathilde},
        title = "{Combining photometry and astrometry to improve orbit retrieval of directly imaged exoplanets}",
      journal = {\mnras},
         year = 2023,
        month = feb,
       volume = {519},
       number = {1},
        pages = {460-470},
          doi = {10.1093/mnras/stac3521},
archivePrefix = {arXiv},
       eprint = {2208.08447},
 primaryClass = {astro-ph.EP},
       adsurl = {https://ui.adsabs.harvard.edu/abs/2023MNRAS.519..460B}
}

@ARTICLE{bryan_et_al2025,
       author = {{Bryan}, Sean and {Barnaby}, Hugh and {Basha}, Oketa and {Bradford}, C. Matt and {Chamberlin}, Kathryn and {Cothard}, Nicholas and {Dahal}, Sumit and {Essinger-Hileman}, Thomas and {Geist}, Alessandro and {Glenn}, Jason and {Jamison-Hooks}, Tracee and {Karthikeyan}, Abarna and {Mauskopf}, Philip and {Miles}, Lynn and {Bailey Newman}, Sanetra and {Roberson}, Cody and {Rostem}, Karwan and {Sinclair}, Adrian},
        title = "{KID Detector Readout Electronics Development for Habitable Worlds Observatory}",
      journal = {arXiv e-prints},
         year = 2025,
        month = sep,
          eid = {arXiv:2509.14363},
        pages = {arXiv:2509.14363},
          doi = {10.48550/arXiv.2509.14363},
archivePrefix = {arXiv},
       eprint = {2509.14363},
 primaryClass = {astro-ph.IM},
       adsurl = {https://ui.adsabs.harvard.edu/abs/2025arXiv250914363B}
}

@article{buick2007,
author = {BUICK, R.},
title = {Did the Proterozoic ‘Canfield Ocean’ cause a laughing gas greenhouse?},
journal = {Geobiology},
volume = {5},
number = {2},
pages = {97-100},
doi = {https://doi.org/10.1111/j.1472-4669.2007.00110.x},
url = {https://onlinelibrary.wiley.com/doi/abs/10.1111/j.1472-4669.2007.00110.x},
eprint = {https://onlinelibrary.wiley.com/doi/pdf/10.1111/j.1472-4669.2007.00110.x},
year = {2007}
}

@ARTICLE{bunker_et_al2023,
       author = {{Bunker}, Andrew J. and {Saxena}, Aayush and {Cameron}, Alex J. and {Willott}, Chris J. and {Curtis-Lake}, Emma and {Jakobsen}, Peter and {Carniani}, Stefano and {Smit}, Renske and {Maiolino}, Roberto and {Witstok}, Joris and {Curti}, Mirko and {D'Eugenio}, Francesco and {Jones}, Gareth C. and {Ferruit}, Pierre and {Arribas}, Santiago and {Charlot}, Stephane and {Chevallard}, Jacopo and {Giardino}, Giovanna and {de Graaff}, Anna and {Looser}, Tobias J. and {L{\"u}tzgendorf}, Nora and {Maseda}, Michael V. and {Rawle}, Tim and {Rix}, Hans-Walter and {Del Pino}, Bruno Rodr{\'\i}guez and {Alberts}, Stacey and {Egami}, Eiichi and {Eisenstein}, Daniel J. and {Endsley}, Ryan and {Hainline}, Kevin and {Hausen}, Ryan and {Johnson}, Benjamin D. and {Rieke}, George and {Rieke}, Marcia and {Robertson}, Brant E. and {Shivaei}, Irene and {Stark}, Daniel P. and {Sun}, Fengwu and {Tacchella}, Sandro and {Tang}, Mengtao and {Williams}, Christina C. and {Willmer}, Christopher N.~A. and {Baker}, William M. and {Baum}, Stefi and {Bhatawdekar}, Rachana and {Bowler}, Rebecca and {Boyett}, Kristan and {Chen}, Zuyi and {Circosta}, Chiara and {Helton}, Jakob M. and {Ji}, Zhiyuan and {Kumari}, Nimisha and {Lyu}, Jianwei and {Nelson}, Erica and {Parlanti}, Eleonora and {Perna}, Michele and {Sandles}, Lester and {Scholtz}, Jan and {Suess}, Katherine A. and {Topping}, Michael W. and {{\"U}bler}, Hannah and {Wallace}, Imaan E.~B. and {Whitler}, Lily},
        title = "{JADES NIRSpec Spectroscopy of GN-z11: Lyman-{\ensuremath{\alpha}} emission and possible enhanced nitrogen abundance in a z = 10.60 luminous galaxy}",
      journal = {\aap},
         year = 2023,
        month = sep,
       volume = {677},
          eid = {A88},
        pages = {A88},
          doi = {10.1051/0004-6361/202346159},
archivePrefix = {arXiv},
       eprint = {2302.07256},
 primaryClass = {astro-ph.GA},
       adsurl = {https://ui.adsabs.harvard.edu/abs/2023A&A...677A..88B}
}

@ARTICLE{burchett_et_al2021,
       author = {{Burchett}, Joseph N. and {Rubin}, Kate H.~R. and {Prochaska}, J. Xavier and {Coil}, Alison L. and {Vaught}, Ryan Rickards and {Hennawi}, Joseph F.},
        title = "{Circumgalactic Mg II Emission from an Isotropic Starburst Galaxy Outflow Mapped by KCWI}",
      journal = {\apj},
         year = 2021,
        month = mar,
       volume = {909},
       number = {2},
          eid = {151},
        pages = {151},
          doi = {10.3847/1538-4357/abd4e0},
archivePrefix = {arXiv},
       eprint = {2005.03017},
 primaryClass = {astro-ph.GA},
       adsurl = {https://ui.adsabs.harvard.edu/abs/2021ApJ...909..151B}
}

@ARTICLE{burns2020,
       author = {{Burns}, Eric},
        title = "{Neutron star mergers and how to study them}",
      journal = {Living Reviews in Relativity},
         year = 2020,
        month = dec,
       volume = {23},
       number = {1},
          eid = {4},
        pages = {4},
          doi = {10.1007/s41114-020-00028-7},
archivePrefix = {arXiv},
       eprint = {1909.06085},
 primaryClass = {astro-ph.HE},
       adsurl = {https://ui.adsabs.harvard.edu/abs/2020LRR....23....4B}
}

@article{burchett_et_al2025,
author = {Joseph N. Burchett and Deborah M. Lokhorst and Yakov Faerman and Kevin France and Kate H. R. Rubin and David S. N. Rupke and Sanchayeeta Borthakur},
title = {{Mapping galactic winds and small-scale structure in the circumgalactic medium with Habitable Worlds Observatory}},
volume = {11},
journal = {Journal of Astronomical Telescopes, Instruments, and Systems},
number = {4},
publisher = {SPIE},
pages = {042213},
year = {2025},
doi = {10.1117/1.JATIS.11.4.042213},
URL = {https://doi.org/10.1117/1.JATIS.11.4.042213}
}

@INPROCEEDINGS{cabrera+schneider2007,
       author = {{Cabrera}, J. and {Schneider}, J.},
        title = "{Detecting Exomoons}",
    booktitle = {Transiting Extrapolar Planets Workshop},
         year = 2007,
       editor = {{Afonso}, C. and {Weldrake}, D. and {Henning}, Th.},
       series = {Astronomical Society of the Pacific Conference Series},
       volume = {366},
        month = jul,
        pages = {242},
       adsurl = {https://ui.adsabs.harvard.edu/abs/2007ASPC..366..242C}
}

@ARTICLE{calder_et_al2025,
       author = {{Calder}, Robb and {Shorttle}, Oliver and {Jordan}, Sean and {Rimmer}, Paul and {Constantinou}, Tereza},
        title = "{Abiotic ozone in the observable atmospheres of Venus and Venus-like exoplanets}",
      journal = {\mnras},
         year = 2025,
        month = jul,
       volume = {540},
       number = {3},
        pages = {2432-2450},
          doi = {10.1093/mnras/staf851},
       adsurl = {https://ui.adsabs.harvard.edu/abs/2025MNRAS.540.2432C}
}

@ARTICLE{cameron_et_al2023,
       author = {{Cameron}, Alex J. and {Saxena}, Aayush and {Bunker}, Andrew J. and {D'Eugenio}, Francesco and {Carniani}, Stefano and {Maiolino}, Roberto and {Curtis-Lake}, Emma and {Ferruit}, Pierre and {Jakobsen}, Peter and {Arribas}, Santiago and {Bonaventura}, Nina and {Charlot}, Stephane and {Chevallard}, Jacopo and {Curti}, Mirko and {Looser}, Tobias J. and {Maseda}, Michael V. and {Rawle}, Tim and {Rodr{\'\i}guez Del Pino}, Bruno and {Smit}, Renske and {{\"U}bler}, Hannah and {Willott}, Chris and {Witstok}, Joris and {Egami}, Eiichi and {Eisenstein}, Daniel J. and {Johnson}, Benjamin D. and {Hainline}, Kevin and {Rieke}, Marcia and {Robertson}, Brant E. and {Stark}, Daniel P. and {Tacchella}, Sandro and {Williams}, Christina C. and {Willmer}, Christopher N.~A. and {Bhatawdekar}, Rachana and {Bowler}, Rebecca and {Boyett}, Kristan and {Circosta}, Chiara and {Helton}, Jakob M. and {Jones}, Gareth C. and {Kumari}, Nimisha and {Ji}, Zhiyuan and {Nelson}, Erica and {Parlanti}, Eleonora and {Sandles}, Lester and {Scholtz}, Jan and {Sun}, Fengwu},
        title = "{JADES: Probing interstellar medium conditions at z {\ensuremath{\sim}} 5.5-9.5 with ultra-deep JWST/NIRSpec spectroscopy}",
      journal = {\aap},
         year = 2023,
        month = sep,
       volume = {677},
          eid = {A115},
        pages = {A115},
          doi = {10.1051/0004-6361/202346107},
archivePrefix = {arXiv},
       eprint = {2302.04298},
 primaryClass = {astro-ph.GA},
       adsurl = {https://ui.adsabs.harvard.edu/abs/2023A&A...677A.115C}
}

@ARTICLE{campante_et_al2024,
       author = {{Campante}, T.~L. and {Kjeldsen}, H. and {Li}, Y. and {Lund}, M.~N. and {Silva}, A.~M. and {Corsaro}, E. and {Gomes da Silva}, J. and {Martins}, J.~H.~C. and {Adibekyan}, V. and {Azevedo Silva}, T. and {Bedding}, T.~R. and {Bossini}, D. and {Buzasi}, D.~L. and {Chaplin}, W.~J. and {Costa}, R.~R. and {Cunha}, M.~S. and {Cristo}, E. and {Faria}, J.~P. and {Garc{\'\i}a}, R.~A. and {Huber}, D. and {Lundkvist}, M.~S. and {Metcalfe}, T.~S. and {Monteiro}, M.~J.~P.~F.~G. and {Neitzel}, A.~W. and {Nielsen}, M.~B. and {Poretti}, E. and {Santos}, N.~C. and {Sousa}, S.~G.},
        title = "{Expanding the frontiers of cool-dwarf asteroseismology with ESPRESSO. Detection of solar-like oscillations in the K5 dwarf ϵ Indi}",
      journal = {\aap},
         year = 2024,
        month = mar,
       volume = {683},
          eid = {L16},
        pages = {L16},
          doi = {10.1051/0004-6361/202449197},
archivePrefix = {arXiv},
       eprint = {2403.16333},
 primaryClass = {astro-ph.SR},
       adsurl = {https://ui.adsabs.harvard.edu/abs/2024A&A...683L..16C}
}

@ARTICLE{carciofi+magalhaes2005,
       author = {{Carciofi}, A.~C. and {Magalh{\~a}es}, A.~M.},
        title = "{The Polarization Signature of Extrasolar Planet Transiting Cool Dwarfs}",
      journal = {\apj},
         year = 2005,
        month = dec,
       volume = {635},
       number = {1},
        pages = {570-577},
          doi = {10.1086/497064},
archivePrefix = {arXiv},
       eprint = {astro-ph/0508343},
 primaryClass = {astro-ph},
       adsurl = {https://ui.adsabs.harvard.edu/abs/2005ApJ...635..570C}
}

@ARTICLE{cardamone_et_al2009,
       author = {{Cardamone}, Carolin and {Schawinski}, Kevin and {Sarzi}, Marc and {Bamford}, Steven P. and {Bennert}, Nicola and {Urry}, C.~M. and {Lintott}, Chris and {Keel}, William C. and {Parejko}, John and {Nichol}, Robert C. and {Thomas}, Daniel and {Andreescu}, Dan and {Murray}, Phil and {Raddick}, M. Jordan and {Slosar}, An{\v{z}}e and {Szalay}, Alex and {Vandenberg}, Jan},
        title = "{Galaxy Zoo Green Peas: discovery of a class of compact extremely star-forming galaxies}",
      journal = {\mnras},
         year = 2009,
        month = nov,
       volume = {399},
       number = {3},
        pages = {1191-1205},
          doi = {10.1111/j.1365-2966.2009.15383.x},
archivePrefix = {arXiv},
       eprint = {0907.4155},
 primaryClass = {astro-ph.CO},
       adsurl = {https://ui.adsabs.harvard.edu/abs/2009MNRAS.399.1191C}
}

@ARTICLE{cardelli_et_al1989,
       author = {{Cardelli}, Jason A. and {Clayton}, Geoffrey C. and {Mathis}, John S.},
        title = "{The Relationship between Infrared, Optical, and Ultraviolet Extinction}",
      journal = {\apj},
         year = 1989,
        month = oct,
       volume = {345},
        pages = {245},
          doi = {10.1086/167900},
       adsurl = {https://ui.adsabs.harvard.edu/abs/1989ApJ...345..245C}
}

@ARTICLE{cai_et_al2019,
       author = {{Cai}, Zheng and {Cantalupo}, Sebastiano and {Prochaska}, J. Xavier and {Arrigoni Battaia}, Fabrizio and {Burchett}, Joe and {Li}, Qiong and {Chisholm}, John and {Bundy}, Kevin and {Hennawi}, Joseph F.},
        title = "{Evolution of the Cool Gas in the Circumgalactic Medium of Massive Halos: A Keck Cosmic Web Imager Survey of Ly{\ensuremath{\alpha}} Emission around QSOs at z {\ensuremath{\approx}} 2}",
      journal = {\apjs},
         year = 2019,
        month = dec,
       volume = {245},
       number = {2},
          eid = {23},
        pages = {23},
          doi = {10.3847/1538-4365/ab4796},
archivePrefix = {arXiv},
       eprint = {1909.11098},
 primaryClass = {astro-ph.GA},
       adsurl = {https://ui.adsabs.harvard.edu/abs/2019ApJS..245...23C}
}

@ARTICLE{carr_et_al2025a,
       author = {{Carr}, Cody A. and {Cen}, Renyue and {Scarlata}, Claudia and {Xu}, Xinfeng and {Henry}, Alaina and {Marques-Chaves}, Rui and {Schaerer}, Daniel and {Amor{\'\i}n}, Ricardo O. and {Oey}, M.~S. and {Komarova}, Lena and {Flury}, Sophia and {Jaskot}, Anne and {Saldana-Lopez}, Alberto and {Ji}, Zhiyuan and {Huberty}, Mason and {Heckman}, Timothy and {{\"O}stlin}, G{\"o}ran and {Bait}, Omkar and {Hayes}, Matthew James and {Thuan}, Trinh and {Ravindranath}, Swara and {Berg}, Danielle A. and {Giavalisco}, Mauro and {Rutkowski}, Michael and {Borthakur}, Sanchayeeta and {Chisholm}, John and {Ferguson}, Harry C. and {Michel-Dansac}, Leo and {Verhamme}, Anne and {Worseck}, G{\'a}bor},
        title = "{The Effect of Radiation and Supernovae Feedback on LyC Escape in Local Star-forming Galaxies}",
      journal = {\apj},
         year = 2025,
        month = apr,
       volume = {982},
       number = {2},
          eid = {137},
        pages = {137},
          doi = {10.3847/1538-4357/adb72f},
archivePrefix = {arXiv},
       eprint = {2409.05180},
 primaryClass = {astro-ph.GA},
       adsurl = {https://ui.adsabs.harvard.edu/abs/2025ApJ...982..137C}
}

@article{carr_jatis_25,
  title={Resolving the origins and pathways of ionizing radiation
escape with UV integral field spectroscopy},
  author={Carr, Cody A and Cen, Renyue and Fleming, Brian and Flury,
Sophia and McCandliss, Stephan and Oey, MS and Scarlata, Claudia and
Strom, Allison},
  journal={Journal of Astronomical Telescopes, Instruments, and Systems},
  volume={12},
  number={4},
  pages={041019--041019},
  year={2026},
  publisher={Society of Photo-Optical Instrumentation Engineers}
}

@ARTICLE{cantalupo_et_al2014,
       author = {{Cantalupo}, Sebastiano and {Arrigoni-Battaia}, Fabrizio and {Prochaska}, J. Xavier and {Hennawi}, Joseph F. and {Madau}, Piero},
        title = "{A cosmic web filament revealed in Lyman-{\ensuremath{\alpha}} emission around a luminous high-redshift quasar}",
      journal = {\nat},
         year = 2014,
        month = feb,
       volume = {506},
       number = {7486},
        pages = {63-66},
          doi = {10.1038/nature12898},
archivePrefix = {arXiv},
       eprint = {1401.4469},
 primaryClass = {astro-ph.CO},
       adsurl = {https://ui.adsabs.harvard.edu/abs/2014Natur.506...63C}
}

@ARTICLE{castellano_et_al2024,
       author = {{Castellano}, Marco and {Napolitano}, Lorenzo and {Fontana}, Adriano and {Roberts-Borsani}, Guido and {Treu}, Tommaso and {Vanzella}, Eros and {Zavala}, Jorge A. and {Arrabal Haro}, Pablo and {Calabr{\`o}}, Antonello and {Llerena}, Mario and {Mascia}, Sara and {Merlin}, Emiliano and {Paris}, Diego and {Pentericci}, Laura and {Santini}, Paola and {Bakx}, Tom J.~L.~C. and {Bergamini}, Pietro and {Cupani}, Guido and {Dickinson}, Mark and {Filippenko}, Alexei V. and {Glazebrook}, Karl and {Grillo}, Claudio and {Kelly}, Patrick L. and {Malkan}, Matthew A. and {Mason}, Charlotte A. and {Morishita}, Takahiro and {Nanayakkara}, Themiya and {Rosati}, Piero and {Sani}, Eleonora and {Wang}, Xin and {Yoon}, Ilsang},
        title = "{JWST NIRSpec Spectroscopy of the Remarkable Bright Galaxy GHZ2/GLASS-z12 at Redshift 12.34}",
      journal = {\apj},
         year = 2024,
        month = sep,
       volume = {972},
       number = {2},
          eid = {143},
        pages = {143},
          doi = {10.3847/1538-4357/ad5f88},
archivePrefix = {arXiv},
       eprint = {2403.10238},
 primaryClass = {astro-ph.GA},
       adsurl = {https://ui.adsabs.harvard.edu/abs/2024ApJ...972..143C}
}

@ARTICLE{catling+zahnle2020,
       author = {{Catling}, David C. and {Zahnle}, Kevin J.},
        title = "{The Archean atmosphere}",
      journal = {Science Advances},
         year = 2020,
        month = feb,
       volume = {6},
       number = {9},
          eid = {eaax1420},
        pages = {eaax1420},
          doi = {10.1126/sciadv.aax1420},
       adsurl = {https://ui.adsabs.harvard.edu/abs/2020SciA....6.1420C}
}

@ARTICLE{catelan2009,
       author = {{Catelan}, M.},
        title = "{Horizontal branch stars: the interplay between observations and theory, and insights into the formation of the Galaxy}",
      journal = {\apss},
         year = 2009,
        month = apr,
       volume = {320},
       number = {4},
        pages = {261-309},
          doi = {10.1007/s10509-009-9987-8},
archivePrefix = {arXiv},
       eprint = {astro-ph/0507464},
 primaryClass = {astro-ph},
       adsurl = {https://ui.adsabs.harvard.edu/abs/2009Ap&SS.320..261C}
}

@ARTICLE{catling_et_al2018,
       author = {{Catling}, David C. and {Krissansen-Totton}, Joshua and {Kiang}, Nancy Y. and {Crisp}, David and {Robinson}, Tyler D. and {DasSarma}, Shiladitya and {Rushby}, Andrew J. and {Del Genio}, Anthony and {Bains}, William and {Domagal-Goldman}, Shawn},
        title = "{Exoplanet Biosignatures: A Framework for Their Assessment}",
      journal = {Astrobiology},
         year = 2018,
        month = jun,
       volume = {18},
       number = {6},
        pages = {709-738},
          doi = {10.1089/ast.2017.1737},
archivePrefix = {arXiv},
       eprint = {1705.06381},
 primaryClass = {astro-ph.EP},
       adsurl = {https://ui.adsabs.harvard.edu/abs/2018AsBio..18..709C}
}

@BOOK{catling+kasting2017,
       author = {{Catling}, David C. and {Kasting}, James F.},
        title = "{Atmospheric Evolution on Inhabited and Lifeless Worlds}",
         year = 2017,
        publisher = "{Cambridge University Press}",
       adsurl = {https://ui.adsabs.harvard.edu/abs/2017aeil.book.....C}
}

@ARTICLE{charbonnel_et_al2023,
       author = {{Charbonnel}, C. and {Schaerer}, D. and {Prantzos}, N. and {Ram{\'\i}rez-Galeano}, L. and {Fragos}, T. and {Kuruvanthodi}, A. and {Marques-Chaves}, R. and {Gieles}, M.},
        title = "{N-enhancement in GN-z11: First evidence for supermassive stars nucleosynthesis in proto-globular clusters-like conditions at high redshift?}",
      journal = {\aap},
         year = 2023,
        month = may,
       volume = {673},
          eid = {L7},
        pages = {L7},
          doi = {10.1051/0004-6361/202346410},
archivePrefix = {arXiv},
       eprint = {2303.07955},
 primaryClass = {astro-ph.GA},
       adsurl = {https://ui.adsabs.harvard.edu/abs/2023A&A...673L...7C}
}

@ARTICLE{cauley_et_al2018a,
       author = {{Cauley}, P. Wilson and {Kuckein}, Christoph and {Redfield}, Seth and {Shkolnik}, Evgenya L. and {Denker}, Carsten and {Llama}, Joe and {Verma}, Meetu},
        title = "{The Effects of Stellar Activity on Optical High-resolution Exoplanet Transmission Spectra}",
      journal = {\aj},
         year = 2018,
        month = nov,
       volume = {156},
       number = {5},
          eid = {189},
        pages = {189},
          doi = {10.3847/1538-3881/aaddf9},
archivePrefix = {arXiv},
       eprint = {1808.09558},
 primaryClass = {astro-ph.EP},
       adsurl = {https://ui.adsabs.harvard.edu/abs/2018AJ....156..189C}
}

@ARTICLE{cauley_et_al2018b,
       author = {{Cauley}, P. Wilson and {Shkolnik}, Evgenya L. and {Llama}, Joe and {Bourrier}, Vincent and {Moutou}, Claire},
        title = "{Evidence of Magnetic Star-Planet Interactions in the HD 189733 System from Orbitally Phased Ca II K Variations}",
      journal = {\aj},
         year = 2018,
        month = dec,
       volume = {156},
       number = {6},
          eid = {262},
        pages = {262},
          doi = {10.3847/1538-3881/aae841},
archivePrefix = {arXiv},
       eprint = {1810.05253},
 primaryClass = {astro-ph.EP},
       adsurl = {https://ui.adsabs.harvard.edu/abs/2018AJ....156..262C}
}

@ARTICLE{chen_et_al2023,
       author = {{Chen}, Hsiao-Wen and {Qu}, Zhijie and {Rauch}, Michael and {Chen}, Mandy C. and {Zahedy}, Fakhri S. and {Johnson}, Sean D. and {Schaye}, Joop and {Rudie}, Gwen C. and {Boettcher}, Erin and {Cantalupo}, Sebastiano and {Faucher-Gigu{\`e}re}, Claude-Andr{\'e} and {Greene}, Jenny E. and {Lopez}, Sebastian and {Simcoe}, Robert A.},
        title = "{The Cosmic Ultraviolet Baryon Survey: Empirical Characterization of Turbulence in the Cool Circumgalactic Medium}",
      journal = {\apjl},
         year = 2023,
        month = sep,
       volume = {955},
       number = {1},
          eid = {L25},
        pages = {L25},
          doi = {10.3847/2041-8213/acf85b},
archivePrefix = {arXiv},
       eprint = {2309.05699},
 primaryClass = {astro-ph.GA},
       adsurl = {https://ui.adsabs.harvard.edu/abs/2023ApJ...955L..25C}
}

@ARTICLE{chen_et_al2024,
       author = {{Chen}, Minghan and {Lawson}, Kellen and {Brandt}, Timothy D. and {Lewis}, Briley L. and {Uyama}, Taichi and {Millar-Blanchaer}, Max and {Tazaki}, Ryo and {Currie}, Thayne},
        title = "{Multiband polarimetric imaging of HD 34700 with SCExAO/CHARIS}",
      journal = {\mnras},
         year = 2024,
        month = sep,
       volume = {533},
       number = {2},
        pages = {2473-2487},
          doi = {10.1093/mnras/stae1957},
archivePrefix = {arXiv},
       eprint = {2408.09038},
 primaryClass = {astro-ph.SR},
       adsurl = {https://ui.adsabs.harvard.edu/abs/2024MNRAS.533.2473C}
}

@ARTICLE{chisholm_et_al2016,
       author = {{Chisholm}, John and {Tremonti}, Christy A. and {Leitherer}, Claus and {Chen}, Yanmei and {Wofford}, Aida},
        title = "{Shining a light on galactic outflows: photoionized outflows}",
      journal = {\mnras},
         year = 2016,
        month = apr,
       volume = {457},
       number = {3},
        pages = {3133-3161},
          doi = {10.1093/mnras/stw178},
archivePrefix = {arXiv},
       eprint = {1601.05090},
 primaryClass = {astro-ph.GA},
       adsurl = {https://ui.adsabs.harvard.edu/abs/2016MNRAS.457.3133C}
}

@ARTICLE{chisholm_et_al2022,
       author = {{Chisholm}, J. and {Saldana-Lopez}, A. and {Flury}, S. and {Schaerer}, D. and {Jaskot}, A. and {Amor{\'\i}n}, R. and {Atek}, H. and {Finkelstein}, S.~L. and {Fleming}, B. and {Ferguson}, H. and {Fern{\'a}ndez}, V. and {Giavalisco}, M. and {Hayes}, M. and {Heckman}, T. and {Henry}, A. and {Ji}, Z. and {Marques-Chaves}, R. and {Mauerhofer}, V. and {McCandliss}, S. and {Oey}, M.~S. and {{\"O}stlin}, G. and {Rutkowski}, M. and {Scarlata}, C. and {Thuan}, T. and {Trebitsch}, M. and {Wang}, B. and {Worseck}, G. and {Xu}, X.},
        title = "{The far-ultraviolet continuum slope as a Lyman Continuum escape estimator at high redshift}",
      journal = {\mnras},
         year = 2022,
        month = dec,
       volume = {517},
       number = {4},
        pages = {5104-5120},
          doi = {10.1093/mnras/stac2874},
archivePrefix = {arXiv},
       eprint = {2207.05771},
 primaryClass = {astro-ph.GA},
       adsurl = {https://ui.adsabs.harvard.edu/abs/2022MNRAS.517.5104C}
}

@ARTICLE{chen_et_al2015,
       author = {{Chen}, Xi and {Ling}, Hong-Fei and {Vance}, Derek and {Shields-Zhou}, Graham A. and {Zhu}, Maoyan and {Poulton}, Simon W. and {Och}, Lawrence M. and {Jiang}, Shao-Yong and {Li}, Da and {Cremonese}, Lorenzo and {Archer}, Corey},
        title = "{Rise to modern levels of ocean oxygenation coincided with the Cambrian radiation of animals}",
      journal = {Nature Communications},
         year = 2015,
        month = may,
       volume = {6},
          eid = {7142},
        pages = {7142},
          doi = {10.1038/ncomms8142},
       adsurl = {https://ui.adsabs.harvard.edu/abs/2015NatCo...6.7142C}
}

@ARTICLE{chisholm_et_al2017,
       author = {{Chisholm}, J. and {Orlitov{\'a}}, I. and {Schaerer}, D. and {Verhamme}, A. and {Worseck}, G. and {Izotov}, Y.~I. and {Thuan}, T.~X. and {Guseva}, N.~G.},
        title = "{Do galaxies that leak ionizing photons have extreme outflows?}",
      journal = {\aap},
         year = 2017,
        month = sep,
       volume = {605},
          eid = {A67},
        pages = {A67},
          doi = {10.1051/0004-6361/201730610},
archivePrefix = {arXiv},
       eprint = {1707.01913},
 primaryClass = {astro-ph.GA},
       adsurl = {https://ui.adsabs.harvard.edu/abs/2017A&A...605A..67C}
}

@ARTICLE{chisholm_et_al2019,
       author = {{Chisholm}, J. and {Rigby}, J.~R. and {Bayliss}, M. and {Berg}, D.~A. and {Dahle}, H. and {Gladders}, M. and {Sharon}, K.},
        title = "{Constraining the Metallicities, Ages, Star Formation Histories, and Ionizing Continua of Extragalactic Massive Star Populations}",
      journal = {\apj},
         year = 2019,
        month = sep,
       volume = {882},
       number = {2},
          eid = {182},
        pages = {182},
          doi = {10.3847/1538-4357/ab3104},
archivePrefix = {arXiv},
       eprint = {1905.04314},
 primaryClass = {astro-ph.GA},
       adsurl = {https://ui.adsabs.harvard.edu/abs/2019ApJ...882..182C}
}

@ARTICLE{choustikov_et_al2024a,
       author = {{Choustikov}, Nicholas and {Katz}, Harley and {Saxena}, Aayush and {Cameron}, Alex J. and {Devriendt}, Julien and {Slyz}, Adrianne and {Rosdahl}, Joki and {Blaizot}, Jeremy and {Michel-Dansac}, Leo},
        title = "{The Physics of Indirect Estimators of Lyman Continuum Escape and their Application to High-Redshift JWST Galaxies}",
      journal = {\mnras},
         year = 2024,
        month = apr,
       volume = {529},
       number = {4},
        pages = {3751-3767},
          doi = {10.1093/mnras/stae776},
archivePrefix = {arXiv},
       eprint = {2304.08526},
 primaryClass = {astro-ph.GA},
       adsurl = {https://ui.adsabs.harvard.edu/abs/2024MNRAS.529.3751C}
}

@ARTICLE{choustikov_et_al2024b,
       author = {{Choustikov}, Nicholas and {Katz}, Harley and {Saxena}, Aayush and {Garel}, Thibault and {Devriendt}, Julien and {Slyz}, Adrianne and {Kimm}, Taysun and {Blaizot}, Jeremy and {Rosdahl}, Joki},
        title = "{The great escape: understanding the connection between Ly {\ensuremath{\alpha}} emission and LyC escape in simulated JWST analogues}",
      journal = {\mnras},
         year = 2024,
        month = aug,
       volume = {532},
       number = {2},
        pages = {2463-2484},
          doi = {10.1093/mnras/stae1586},
archivePrefix = {arXiv},
       eprint = {2401.09557},
 primaryClass = {astro-ph.GA},
       adsurl = {https://ui.adsabs.harvard.edu/abs/2024MNRAS.532.2463C}
}

@ARTICLE{chubb_et_al2024,
       author = {{Chubb}, Katy L. and {Stam}, Daphne M. and {Helling}, Christiane and {Samra}, Dominic and {Carone}, Ludmila},
        title = "{Modelling reflected polarized light from close-in giant exoplanet WASP-96b using PolHEx (Polarization of hot exoplanets)}",
      journal = {\mnras},
         year = 2024,
        month = jan,
       volume = {527},
       number = {3},
        pages = {4955-4982},
          doi = {10.1093/mnras/stad3413},
archivePrefix = {arXiv},
       eprint = {2307.09601},
 primaryClass = {astro-ph.EP},
       adsurl = {https://ui.adsabs.harvard.edu/abs/2024MNRAS.527.4955C}
}

@INPROCEEDINGS{chun_et_al2016,
       author = {{Chun}, Mark R. and {Lai}, Olivier and {Toomey}, Douglas and {Lu}, Jessica R. and {Service}, Max and {Baranec}, Christoph and {Thibault}, Simon and {Brousseau}, Denis and {Hayano}, Yutaka and {Oya}, Shin and {Santi}, Shane and {Kingery}, Christopher and {Loss}, Keith and {Gardiner}, John and {Steele}, Brad},
        title = "{Imaka: a ground-layer adaptive optics system on Maunakea}",
    booktitle = {Adaptive Optics Systems V},
         year = 2016,
       editor = {{Marchetti}, Enrico and {Close}, Laird M. and {V{\'e}ran}, Jean-Pierre},
       series = {Society of Photo-Optical Instrumentation Engineers (SPIE) Conference Series},
       volume = {9909},
        month = jul,
          eid = {990902},
        pages = {990902},
          doi = {10.1117/12.2233208},
archivePrefix = {arXiv},
       eprint = {1608.01804},
 primaryClass = {astro-ph.IM},
       adsurl = {https://ui.adsabs.harvard.edu/abs/2016SPIE.9909E..02C}
}

@ARTICLE{cegla_et_al2016,
       author = {{Cegla}, H.~M. and {Lovis}, C. and {Bourrier}, V. and {Beeck}, B. and {Watson}, C.~A. and {Pepe}, F.},
        title = "{The Rossiter-McLaughlin effect reloaded: Probing the 3D spin-orbit geometry, differential stellar rotation, and the spatially-resolved stellar spectrum of star-planet systems}",
      journal = {\aap},
         year = 2016,
        month = apr,
       volume = {588},
          eid = {A127},
        pages = {A127},
          doi = {10.1051/0004-6361/201527794},
archivePrefix = {arXiv},
       eprint = {1602.00322},
 primaryClass = {astro-ph.EP},
       adsurl = {https://ui.adsabs.harvard.edu/abs/2016A&A...588A.127C}
}

@ARTICLE{cen2020,
       author = {{Cen}, Renyue},
        title = "{Physics of Prodigious Lyman Continuum Leakers}",
      journal = {\apjl},
         year = 2020,
        month = jan,
       volume = {889},
       number = {1},
          eid = {L22},
        pages = {L22},
          doi = {10.3847/2041-8213/ab6560},
archivePrefix = {arXiv},
       eprint = {2001.11083},
 primaryClass = {astro-ph.GA},
       adsurl = {https://ui.adsabs.harvard.edu/abs/2020ApJ...889L..22C}
}

@ARTICLE{cerling_et_al1997,
       author = {{Cerling}, Thure E. and {Harris}, John M. and {MacFadden}, Bruce J. and {Leakey}, Meave G. and {Quade}, Jay and {Eisenmann}, Vera and {Ehleringer}, James R.},
        title = "{Global vegetation change through the Miocene/Pliocene boundary}",
      journal = {\nat},
         year = 1997,
        month = sep,
       volume = {389},
       number = {6647},
        pages = {153-158},
          doi = {10.1038/38229},
       adsurl = {https://ui.adsabs.harvard.edu/abs/1997Natur.389..153C}
}

@ARTICLE{cicone_et_al2014,
       author = {{Cicone}, C. and {Maiolino}, R. and {Sturm}, E. and {Graci{\'a}-Carpio}, J. and {Feruglio}, C. and {Neri}, R. and {Aalto}, S. and {Davies}, R. and {Fiore}, F. and {Fischer}, J. and {Garc{\'\i}a-Burillo}, S. and {Gonz{\'a}lez-Alfonso}, E. and {Hailey-Dunsheath}, S. and {Piconcelli}, E. and {Veilleux}, S.},
        title = "{Massive molecular outflows and evidence for AGN feedback from CO observations}",
      journal = {\aap},
         year = 2014,
        month = feb,
       volume = {562},
          eid = {A21},
        pages = {A21},
          doi = {10.1051/0004-6361/201322464},
archivePrefix = {arXiv},
       eprint = {1311.2595},
 primaryClass = {astro-ph.CO},
       adsurl = {https://ui.adsabs.harvard.edu/abs/2014A&A...562A..21C}
}

@ARTICLE{cicone_et_al2017,
       author = {{Cicone}, C. and {Bothwell}, M. and {Wagg}, J. and {M{\o}ller}, P. and {De Breuck}, C. and {Zhang}, Z. and {Mart{\'\i}n}, S. and {Maiolino}, R. and {Severgnini}, P. and {Aravena}, M. and {Belfiore}, F. and {Espada}, D. and {Fl{\"u}tsch}, A. and {Impellizzeri}, V. and {Peng}, Y. and {Raj}, M.~A. and {Ram{\'\i}rez-Olivencia}, N. and {Riechers}, D. and {Schawinski}, K.},
        title = "{The final data release of ALLSMOG: a survey of CO in typical local low-M$_{{\ensuremath{*}}}$ star-forming galaxies}",
      journal = {\aap},
         year = 2017,
        month = aug,
       volume = {604},
          eid = {A53},
        pages = {A53},
          doi = {10.1051/0004-6361/201730605},
archivePrefix = {arXiv},
       eprint = {1705.05851},
 primaryClass = {astro-ph.GA},
       adsurl = {https://ui.adsabs.harvard.edu/abs/2017A&A...604A..53C}
}

@ARTICLE{claringbold_et_al2023,
       author = {{Claringbold}, A.~B. and {Rimmer}, P.~B. and {Rugheimer}, S. and {Shorttle}, O.},
        title = "{Prebiosignature Molecules Can Be Detected in Temperate Exoplanet Atmospheres with JWST}",
      journal = {\aj},
         year = 2023,
        month = aug,
       volume = {166},
       number = {2},
          eid = {39},
        pages = {39},
          doi = {10.3847/1538-3881/acdacc},
archivePrefix = {arXiv},
       eprint = {2306.02897},
 primaryClass = {astro-ph.EP},
       adsurl = {https://ui.adsabs.harvard.edu/abs/2023AJ....166...39C}
}

@ARTICLE{clayton_et_al2003,
       author = {{Clayton}, Geoffrey C. and {Wolff}, Michael J. and {Sofia}, Ulysses J. and {Gordon}, K.~D. and {Misselt}, K.~A.},
        title = "{Dust Grain Size Distributions from MRN to MEM}",
      journal = {\apj},
         year = 2003,
        month = may,
       volume = {588},
       number = {2},
        pages = {871-880},
          doi = {10.1086/374316},
archivePrefix = {arXiv},
       eprint = {astro-ph/0301488},
 primaryClass = {astro-ph},
       adsurl = {https://ui.adsabs.harvard.edu/abs/2003ApJ...588..871C}
}

@article{coelho_et_al2022,
author = {Coelho, L\'{\i}gia F. and Madden, Jack and Kaltenegger, Lisa and Zinder, Stephen and Philpot, William and Esqu\'{\i}vel, M. Gl\'{o}ria and Can\'{a}rio, Jo\~{a}o and Costa, Rodrigo and Vincent, Warwick F. and Martins, Zita},
title = {Color Catalogue of Life in Ice: Surface Biosignatures on Icy Worlds},
journal = {Astrobiology},
volume = {22},
number = {3},
pages = {313-321},
year = {2022},
doi = {10.1089/ast.2021.0008},
    note ={PMID: 34964651},

URL = { 
    
        https://doi.org/10.1089/ast.2021.0008
    
    

},
eprint = { 
    
        https://doi.org/10.1089/ast.2021.0008
    
    

}
}

@ARTICLE{coelho_et_al2024,
       author = {{Coelho}, L{\'\i}gia Fonseca and {Kaltenegger}, Lisa and {Zinder}, Stephen and {Philpot}, William and {Price}, Taylor L. and {Hamilton}, Trinity L.},
        title = "{Purple is the new green: biopigments and spectra of Earth-like purple worlds}",
      journal = {\mnras},
         year = 2024,
        month = may,
       volume = {530},
       number = {2},
        pages = {1363-1368},
          doi = {10.1093/mnras/stae601},
archivePrefix = {arXiv},
       eprint = {2404.10105},
 primaryClass = {astro-ph.EP},
       adsurl = {https://ui.adsabs.harvard.edu/abs/2024MNRAS.530.1363C}
}

@INPROCEEDINGS{cote_et_al2012,
       author = {{C{\^o}te}, Patrick and {Scott}, Alan and {Balogh}, Michael and {Buckingham}, Ron and {Aldridge}, David and {Carlberg}, Ray and {Chen}, Weiguo and {Dupuis}, Jean and {Evans}, Clinton and {Drissen}, Laurent and {Fraser}, Wes and {Grandmont}, Frederic and {Harrison}, Paul and {Hutchings}, John and {Kavelaars}, JJ and {Landry}, John-Thomas and {Lange}, Christian and {Laurin}, Denis and {Patel}, Tarun and {Pillay}, Venka and {Piche}, Louis and {Rader}, Andres and {Robert}, Carmelle and {Sawicki}, Marchin and {Sorba}, Robert and {Theriault}, Guillaume and {Van Waerbeke}, Ludovic},
        title = "{CASTOR: the Cosmological Advanced Survey Telescope for Optical and Ultraviolet Research}",
    booktitle = {Space Telescopes and Instrumentation 2012: Optical, Infrared, and Millimeter Wave},
         year = 2012,
       editor = {{Clampin}, Mark C. and {Fazio}, Giovanni G. and {MacEwen}, Howard A. and {Oschmann}, Jr., Jacobus M.},
       series = {Society of Photo-Optical Instrumentation Engineers (SPIE) Conference Series},
       volume = {8442},
        month = sep,
          eid = {844215},
        pages = {844215},
          doi = {10.1117/12.926198},
       adsurl = {https://ui.adsabs.harvard.edu/abs/2012SPIE.8442E..15C}
}

@INPROCEEDINGS{cote_et_al2019,
       author = {{Cote}, Patrick and {Abraham}, Bob and {Balogh}, Michael and {Capak}, Peter and {Carlberg}, Ray and {Cowan}, Nick and {Djazovski}, Oleg and {Drissen}, Laurent and {Drout}, Maria and {Dupuis}, Jean and {Evans}, Chris and {Fantin}, Nicholas and {Ferrarese}, Laura and {Fraser}, Wes and {Gallagher}, Sarah and {Girard}, Terry and {Gleisinger}, Robert and {Grandmont}, Frederic and {Hall}, Patrick and {Hellmich}, Martin and {Hardy}, Tim and {Harrison}, Paul and {Hlozek}, Renee and {Haggard}, Daryl and {Henault-Brunet}, Vincent and {Hutchings}, John and {Khatu}, Viraja and {Kavelaars}, JJ and {Laurin}, Denis and {Lavigne}, Jean-Francois and {Lisman}, Doug and {Marois}, Christian and {McCabe}, David and {Metchev}, Stanimir and {Moutard}, Thibaud and {Netterfield}, Barth and {Nikzad}, Shouleh and {Ouellette}, Nathalie and {Pass}, Emily and {Parker}, Laura and {Pazder}, John and {Percival}, Will and {Rhodes}, Jason and {Robert}, Carmelle and {Rowe}, Jason and {Sanchez-Janssen}, Ruben and {Sivakoff}, Greg and {Shapiro}, Charles and {Sawicki}, Marcin and {Scott}, Alan and {Van Waerbeke}, Ludovic and {Venn}, Kim},
        title = "{CASTOR: A Flagship Canadian Space Telescope}",
    booktitle = {Canadian Long Range Plan for Astronomy and Astrophysics White Papers},
         year = 2019,
       volume = {2020},
        month = oct,
          eid = {18},
        pages = {18},
          doi = {10.5281/zenodo.3758463},
       adsurl = {https://ui.adsabs.harvard.edu/abs/2019clrp.2020...18C}
}

@ARTICLE{cote_et_al2025,
       author = {{C{\^o}t{\'e}}, Patrick and {Woods}, Tyrone E. and {Hutchings}, John B. and {Rhodes}, Jason D. and {S{\'a}nchez-Janssen}, Rub{\'e}n. and {Scott}, Alan D. and {Pazder}, John and {Amenouche}, Melissa and {Balogh}, Michael and {Blouin}, Simon and {Cournoyer}, Alain and {Drout}, Maria R. and {Kuzmin}, Nick and {Mack}, Katherine J. and {Ferrarese}, Laura and {Fraser}, Wesley C. and {Gallagher}, Sarah C. and {Grandmont}, Fr{\'e}d{\'e}ric and {Haggard}, Daryl and {Harrison}, Paul and {H{\'e}nault-Brunet}, Vincent and {Kavelaars}, J.~J. and {Khatu}, Viraja and {Roediger}, Joel C. and {Rowe}, Jason and {Sawicki}, Marcin and {Skottfelt}, Jesper and {Taylor}, Matt and {van Waerbeke}, Ludo and {Amen}, Laurie and {Bansal}, Dhananjhay and {Bergeron}, Martin and {Brown}, Toby and {Burley}, Greg and {Chand}, Hum and {Cheng}, Isaac and {Cloutier}, Ryan and {Dickson}, Nolan and {Djazovski}, Oleg and {Damjanov}, Ivana and {Doherty}, James and {Finner}, Kyle and {Del Valle Espinosa}, Macarena Garc{\'\i}a. and {Glover}, Jennifer and {G{\'o}mez de Castro}, Ana I. and {Graur}, Or and {Hardy}, Tim and {Kao}, Michelle and {Leahy}, Denis and {Lokhorst}, Deborah and {Malz}, Alex I. and {Man}, Allison and {Marshall}, Madeline A. and {McGee}, Sean and {McKenzie}, Ryan and {Michaud}, Kai and {More}, Surhud S. and {Morris}, David and {Morris}, Patrick W. and {Moutard}, Thibaud and {Naqvi}, Wasi and {Nicholl}, Matt and {Noirot}, Ga{\"e}l. and {Oey}, M.~S. and {Opitom}, Cyrielle and {Salim}, Samir and {Scott}, Bryan R. and {Shapiro}, Charles A. and {Stern}, Daniel and {Subramaniam}, Annapurni and {Thilke}, David and {Wevers}, Ivan and {Vorobiev}, Dmitri and {Yung}, L.~Y. Aaron and {Zamkotsian}, Fr{\'e}d{\'e}ric and {Aigrain}, Suzanne and {Alavi}, Anahita and {Barstow}, Martin and {Bartosik}, Peter and {Bluhm}, Hadleigh and {Bovy}, Jo and {Cameron}, Peter and {Carlberg}, Raymond G. and {Christiansen}, Jessie L. and {Chen}, Yuyang and {Crowther}, Paul and {Dage}, Kristen and {Dotter}, Aaron L. and {Dufour}, Patrick and {Dupuis}, Jean and {Dryer}, Ben and {Duara}, Angaraj and {Eadie}, Gwendolyn M. and {Eduardo}, Marielle R. and {Estrada-Carpenter}, Vincente and {Fabbro}, S{\'e}bastien and {Faisst}, Andreas and {Ford}, Nicole M. and {Fraser}, Morgan and {Gaensicke}, Boris T. and {Ganesh}, Shashkiran and {Gandhi}, Poshak and {Graham}, Melissa L. and {Hamel}, Rebecca and {Hellmich}, Martin and {Hennessy}, John and {Hessel}, Kaitlyn and {Heyl}, Jeremy and {Heymans}, Catherine and {Hezaveh}, Yashar and {Hlozek}, Renee and {Hoenk}, Michael E. and {Holland}, Andrew and {Huff}, Eric and {Hutchinson}, Ian and {Iwata}, Ikuru and {Jewell}, April D. and {Johnstone}, Doug and {Jones}, Maia and {Jones}, Todd and {Lang}, Dustin and {Lapington}, Jon and {Larivi{\`e}re}, Justin and {Lawlor-Forsyth}, Cameron and {Laurin}, Denis and {Lee}, Charles and {Legin}, Ronan and {Li}, Ting S. and {Lim}, Sungsoon and {Ludwig}, Bethany and {Kozun}, Matt and {Vivek}, M. and {Mann}, Robert and {McConnachie}, Alan W. and {McDonough}, Evan and {Metchev}, Stanimir and {Miller}, David R. and {Moriya}, Takashi and {Morgan}, Cameron and {Navarro}, Julio and {Naz{\'e}}, Ya{\"e}l. and {Nikzad}, Shouleh and {Oad}, Vivek and {Ouellette}, Nathalie and {Pass}, Emily K. and {Percival}, Will J. and {Levasseur}, Laurence Perreault and {Postma}, Joe and {Raza}, Nayyer and {Richards}, Gordon T. and {Richer}, Harvey and {Robert}, Carmelle and {Rosolowsky}, Erik and {Ruan}, John J. and {Rugheimer}, Sarah and {Safi-Harb}, Samar and {Saha}, Kanak and {Scowcroft}, Vicky and {Sestito}, Federico and {Sharma}, Himanshu and {Sikora}, James and {Sivakoff}, Gregory R. and {Sivarani}, Thirupathi and {Smith}, Patrick and {Soh}, Warren and {Sorba}, Robert and {Subramanian}, Smitha and {Teimoorinia}, Hossen and {Teplitz}, Harry I. and {Thadani}, Shaylin and {Thadani}, Shavon and {Tohuvavohu}, Aaron and {Venn}, Kim A. and {Vieira}, Nicholas and {Webb}, Jeremy J. and {Wiegert}, Paul and {Wierckx}, Ryan and {Wu}, Yanqin and {Yeung}, Jade and {Yi}, Sukyoung K.},
        title = "{The CASTOR mission}",
      journal = {Journal of Astronomical Telescopes, Instruments, and Systems},
         year = 2025,
        month = oct,
       volume = {11},
          eid = {042202},
        pages = {042202},
          doi = {10.1117/1.JATIS.11.4.042202},
       adsurl = {https://ui.adsabs.harvard.edu/abs/2025JATIS..11d2202C}
}

@BOOK{nrc2003,
       author = {{Committee On The Physics Of The Universe}, Board On Physics and {Astronomy}, Division On Engineering and {Physical Sciences}, National Research Council Of The National Academies},
        title = "{Connecting quarks with the cosmos : eleven science questions for the new century}",
         year = 2003,
       adsurl = {https://ui.adsabs.harvard.edu/abs/2003cqwc.book.....C}
}

@ARTICLE{cowan_et_al2009,
       author = {{Cowan}, Nicolas B. and {Agol}, Eric and {Meadows}, Victoria S. and {Robinson}, Tyler and {Livengood}, Timothy A. and {Deming}, Drake and {Lisse}, Carey M. and {A'Hearn}, Michael F. and {Wellnitz}, Dennis D. and {Seager}, Sara and {Charbonneau}, David and {EPOXI Team}},
        title = "{Alien Maps of an Ocean-bearing World}",
      journal = {\apj},
         year = 2009,
        month = aug,
       volume = {700},
       number = {2},
        pages = {915-923},
          doi = {10.1088/0004-637X/700/2/915},
archivePrefix = {arXiv},
       eprint = {0905.3742},
 primaryClass = {astro-ph.EP},
       adsurl = {https://ui.adsabs.harvard.edu/abs/2009ApJ...700..915C}
}

@ARTICLE{cowan_et_al2011,
       author = {{Cowan}, Nicolas B. and {Robinson}, Tyler and {Livengood}, Timothy A. and {Deming}, Drake and {Agol}, Eric and {A'Hearn}, Michael F. and {Charbonneau}, David and {Lisse}, Carey M. and {Meadows}, Victoria S. and {Seager}, Sara and {Shields}, Aomawa L. and {Wellnitz}, Dennis D.},
        title = "{Rotational Variability of Earth's Polar Regions: Implications for Detecting Snowball Planets}",
      journal = {\apj},
         year = 2011,
        month = apr,
       volume = {731},
       number = {1},
          eid = {76},
        pages = {76},
          doi = {10.1088/0004-637X/731/1/76},
archivePrefix = {arXiv},
       eprint = {1102.4345},
 primaryClass = {astro-ph.EP},
       adsurl = {https://ui.adsabs.harvard.edu/abs/2011ApJ...731...76C}
}

@ARTICLE{cranmer_et_al2023,
       author = {{Cranmer}, Steven R. and {Chhiber}, Rohit and {Gilly}, Chris R. and {Cairns}, Iver H. and {Colaninno}, Robin C. and {McComas}, David J. and {Raouafi}, Nour E. and {Usmanov}, Arcadi V. and {Gibson}, Sarah E. and {DeForest}, Craig E.},
        title = "{The Sun's Alfv{\'e}n Surface: Recent Insights and Prospects for the Polarimeter to Unify the Corona and Heliosphere (PUNCH)}",
      journal = {\solphys},
         year = 2023,
        month = nov,
       volume = {298},
       number = {11},
          eid = {126},
        pages = {126},
          doi = {10.1007/s11207-023-02218-2},
archivePrefix = {arXiv},
       eprint = {2310.05887},
 primaryClass = {astro-ph.SR},
       adsurl = {https://ui.adsabs.harvard.edu/abs/2023SoPh..298..126C}
}

@techreport{crill2022,
  author      = {Crill, Brendan P.},
  title       = {Progress in Technology for Exoplanet Missions: An Appendix to the {NASA} Exoplanet Exploration Program Technology Plan},
  institution = {National Aeronautics and Space Administration (NASA), Jet Propulsion Laboratory, California Institute of Technology},
  year        = {2022},
  type        = {JPL Document},
  number      = {D-108825},
  url         = {https://assets.science.nasa.gov/content/dam/science/astro/programs/exep/technology/files/Progress_in_Technology_for_Exoplanet_Missions.pdf},
  note        = {Approved for release January 3, 2023. JPL CL \#22-6678}
}

@ARTICLE{crowther+dessart1998,
       author = {{Crowther}, P.~A. and {Dessart}, Luc},
        title = "{Quantitative spectroscopy of Wolf-Rayet stars in HD97950 and R136a - the cores of giant HII regions}",
      journal = {\mnras},
         year = 1998,
        month = may,
       volume = {296},
       number = {3},
        pages = {622-642},
          doi = {10.1046/j.1365-8711.1998.01400.x},
       adsurl = {https://ui.adsabs.harvard.edu/abs/1998MNRAS.296..622C}
}

@ARTICLE{crowther+hadfield2006,
       author = {{Crowther}, P.~A. and {Hadfield}, L.~J.},
        title = "{Reduced Wolf-Rayet line luminosities at low metallicity}",
      journal = {\aap},
         year = 2006,
        month = apr,
       volume = {449},
       number = {2},
        pages = {711-722},
          doi = {10.1051/0004-6361:20054298},
archivePrefix = {arXiv},
       eprint = {astro-ph/0512183},
 primaryClass = {astro-ph},
       adsurl = {https://ui.adsabs.harvard.edu/abs/2006A&A...449..711C}
}

@ARTICLE{crowther_et_al2010,
       author = {{Crowther}, Paul A. and {Schnurr}, Olivier and {Hirschi}, Raphael and {Yusof}, Norhasliza and {Parker}, Richard J. and {Goodwin}, Simon P. and {Kassim}, Hasan Abu},
        title = "{The R136 star cluster hosts several stars whose individual masses greatly exceed the accepted 150M$_{solar}$ stellar mass limit}",
      journal = {\mnras},
         year = 2010,
        month = oct,
       volume = {408},
       number = {2},
        pages = {731-751},
          doi = {10.1111/j.1365-2966.2010.17167.x},
archivePrefix = {arXiv},
       eprint = {1007.3284},
 primaryClass = {astro-ph.SR},
       adsurl = {https://ui.adsabs.harvard.edu/abs/2010MNRAS.408..731C}
}

@ARTICLE{crowther_et_al2016,
       author = {{Crowther}, Paul A. and {Caballero-Nieves}, S.~M. and {Bostroem}, K.~A. and {Ma{\'\i}z Apell{\'a}niz}, J. and {Schneider}, F.~R.~N. and {Walborn}, N.~R. and {Angus}, C.~R. and {Brott}, I. and {Bonanos}, A. and {de Koter}, A. and {de Mink}, S.~E. and {Evans}, C.~J. and {Gr{\"a}fener}, G. and {Herrero}, A. and {Howarth}, I.~D. and {Langer}, N. and {Lennon}, D.~J. and {Puls}, J. and {Sana}, H. and {Vink}, J.~S.},
        title = "{The R136 star cluster dissected with Hubble Space Telescope/STIS. I. Far-ultraviolet spectroscopic census and the origin of He II {\ensuremath{\lambda}}1640 in young star clusters}",
      journal = {\mnras},
         year = 2016,
        month = may,
       volume = {458},
       number = {1},
        pages = {624-659},
          doi = {10.1093/mnras/stw273},
archivePrefix = {arXiv},
       eprint = {1603.04994},
 primaryClass = {astro-ph.SR},
       adsurl = {https://ui.adsabs.harvard.edu/abs/2016MNRAS.458..624C}
}

@ARTICLE{cubillos_et_al2025,
       author = {{Cubillos}, Patricio E. and {Brogi}, Matteo and {Garc{\'\i}a Mu{\~n}oz}, Antonio and {Fossati}, Luca and {Boro Saikia}, Sudeshna and {Bourrier}, Vincent and {Caballero}, Jose A. and {Cabrera}, Juan and {Chiavassa}, Andrea and {Fludra}, Andrzej and {Gkouvelis}, Leonardos and {Grenfell}, John Lee and {Guedel}, Manuel and {Labiano}, Alvaro and {Lendl}, Monika and {Rodgers-Lee}, Donna and {Salvador}, Arnaud and {Schroetter}, Ilane and {Strugarek}, Antoine and {Taysum}, Benjamin and {Vidotto}, Aline and {Wilson}, Thomas G.},
        title = "{High-resolution Ultraviolet-to-nearinfrared Characterization of Exoplanet Atmospheres}",
      journal = {arXiv e-prints},
         year = 2025,
        month = jul,
          eid = {arXiv:2507.03060},
        pages = {arXiv:2507.03060},
          doi = {10.48550/arXiv.2507.03060},
archivePrefix = {arXiv},
       eprint = {2507.03060},
 primaryClass = {astro-ph.IM},
       adsurl = {https://ui.adsabs.harvard.edu/abs/2025arXiv250703060C}
}

@ARTICLE{cuntz_et_al2000,
       author = {{Cuntz}, Manfred and {Saar}, Steven H. and {Musielak}, Zdzislaw E.},
        title = "{On Stellar Activity Enhancement Due to Interactions with Extrasolar Giant Planets}",
      journal = {\apjl},
         year = 2000,
        month = apr,
       volume = {533},
       number = {2},
        pages = {L151-L154},
          doi = {10.1086/312609},
       adsurl = {https://ui.adsabs.harvard.edu/abs/2000ApJ...533L.151C}
}

@ARTICLE{dalcanton_et_al2015,
       author = {{Dalcanton}, Julianne J. and {Fouesneau}, Morgan and {Hogg}, David W. and {Lang}, Dustin and {Leroy}, Adam K. and {Gordon}, Karl D. and {Sandstrom}, Karin and {Weisz}, Daniel R. and {Williams}, Benjamin F. and {Bell}, Eric F. and {Dong}, Hui and {Gilbert}, Karoline M. and {Gouliermis}, Dimitrios A. and {Guhathakurta}, Puragra and {Lauer}, Tod R. and {Schruba}, Andreas and {Seth}, Anil C. and {Skillman}, Evan D.},
        title = "{The Panchromatic Hubble Andromeda Treasury. VIII. A Wide-area, High-resolution Map of Dust Extinction in M31}",
      journal = {\apj},
         year = 2015,
        month = nov,
       volume = {814},
       number = {1},
          eid = {3},
        pages = {3},
          doi = {10.1088/0004-637X/814/1/3},
archivePrefix = {arXiv},
       eprint = {1509.06988},
 primaryClass = {astro-ph.GA},
       adsurl = {https://ui.adsabs.harvard.edu/abs/2015ApJ...814....3D}
}

@ARTICLE{dalcanton_et_al2023,
       author = {{Dalcanton}, Julianne J. and {Bell}, Eric F. and {Choi}, Yumi and {Dolphin}, Andrew E. and {Fouesneau}, Morgan and {Girardi}, L{\'e}o and {Hogg}, David W. and {Seth}, Anil C. and {Williams}, Benjamin F.},
        title = "{The Panchromatic Hubble Andromeda Treasury. XX. The Disk of M31 is Thick}",
      journal = {\aj},
         year = 2023,
        month = aug,
       volume = {166},
       number = {2},
          eid = {80},
        pages = {80},
          doi = {10.3847/1538-3881/accc83},
archivePrefix = {arXiv},
       eprint = {2304.08613},
 primaryClass = {astro-ph.GA},
       adsurl = {https://ui.adsabs.harvard.edu/abs/2023AJ....166...80D}
}

@ARTICLE{damiano_et_al2025,
       author = {{Damiano}, Mario and {Burr}, Zachary and {Hu}, Renyu and {Burt}, Jennifer and {Kataria}, Tiffany},
        title = "{Effects of Planetary Mass Uncertainties on the Interpretation of the Reflectance Spectra of Earth-like Exoplanets}",
      journal = {\aj},
         year = 2025,
        month = feb,
       volume = {169},
       number = {2},
          eid = {97},
        pages = {97},
          doi = {10.3847/1538-3881/ada610},
archivePrefix = {arXiv},
       eprint = {2502.01513},
 primaryClass = {astro-ph.EP},
       adsurl = {https://ui.adsabs.harvard.edu/abs/2025AJ....169...97D}
}

@article{damiano+hu2022,
  title={Reflected spectroscopy of small exoplanets II: characterization of terrestrial exoplanets},
  author={Damiano, Mario and Hu, Renyu},
  journal={The Astronomical Journal},
  volume={163},
  number={6},
  pages={299},
  year={2022},
  publisher={IOP Publishing}
}

@ARTICLE{dassarma+schwieterman2021,
       author = {{DasSarma}, Shiladitya and {Schwieterman}, Edward W.},
        title = "{Early evolution of purple retinal pigments on Earth and implications for exoplanet biosignatures}",
      journal = {International Journal of Astrobiology},
         year = 2021,
        month = jun,
       volume = {20},
       number = {3},
        pages = {241-250},
          doi = {10.1017/S1473550418000423},
       adsurl = {https://ui.adsabs.harvard.edu/abs/2021IJAsB..20..241D}
}

@ARTICLE{das_et_al2017,
       author = {{Das}, Anindita and {Singh}, Tanya and {LokaBharathi}, P.~A. and {Dhakephalkar}, Prashant K. and {Mallik}, Sweta and {Kshirsagar}, Pranav R. and {Khadge}, N.~H. and {Nath}, B. Nagender and {Bhattacharya}, Satadru and {Dagar}, Aditya Kumar and {Kaur}, Prabhjot and {Ray}, Dwijesh and {Shukla}, Anil D. and {Fernandes}, Christabelle E.~G. and {Fernandes}, Sheryl O. and {Thomas}, Tresa Remya A. and {Mamatha}, S.~S. and {Mourya}, Babu Shashikant and {Meena}, Ram Murti},
        title = "{Astrobiological implications of dim light phototrophy in deep-sea red clays}",
      journal = {Life Sciences in Space Research},
         year = 2017,
        month = feb,
       volume = {12},
        pages = {39-50},
          doi = {10.1016/j.lssr.2017.01.002},
       adsurl = {https://ui.adsabs.harvard.edu/abs/2017LSSR...12...39D}
}

@ARTICLE{dawson_et_al2016,
       author = {{Dawson}, Kyle S. and {Kneib}, Jean-Paul and {Percival}, Will J. and {Alam}, Shadab and {Albareti}, Franco D. and {Anderson}, Scott F. and {Armengaud}, Eric and {Aubourg}, {\'E}ric and {Bailey}, Stephen and {Bautista}, Julian E. and {Berlind}, Andreas A. and {Bershady}, Matthew A. and {Beutler}, Florian and {Bizyaev}, Dmitry and {Blanton}, Michael R. and {Blomqvist}, Michael and {Bolton}, Adam S. and {Bovy}, Jo and {Brandt}, W.~N. and {Brinkmann}, Jon and {Brownstein}, Joel R. and {Burtin}, Etienne and {Busca}, N.~G. and {Cai}, Zheng and {Chuang}, Chia-Hsun and {Clerc}, Nicolas and {Comparat}, Johan and {Cope}, Frances and {Croft}, Rupert A.~C. and {Cruz-Gonzalez}, Irene and {da Costa}, Luiz N. and {Cousinou}, Marie-Claude and {Darling}, Jeremy and {de la Macorra}, Axel and {de la Torre}, Sylvain and {Delubac}, Timoth{\'e}e and {du Mas des Bourboux}, H{\'e}lion and {Dwelly}, Tom and {Ealet}, Anne and {Eisenstein}, Daniel J. and {Eracleous}, Michael and {Escoffier}, S. and {Fan}, Xiaohui and {Finoguenov}, Alexis and {Font-Ribera}, Andreu and {Frinchaboy}, Peter and {Gaulme}, Patrick and {Georgakakis}, Antonis and {Green}, Paul and {Guo}, Hong and {Guy}, Julien and {Ho}, Shirley and {Holder}, Diana and {Huehnerhoff}, Joe and {Hutchinson}, Timothy and {Jing}, Yipeng and {Jullo}, Eric and {Kamble}, Vikrant and {Kinemuchi}, Karen and {Kirkby}, David and {Kitaura}, Francisco-Shu and {Klaene}, Mark A. and {Laher}, Russ R. and {Lang}, Dustin and {Laurent}, Pierre and {Le Goff}, Jean-Marc and {Li}, Cheng and {Liang}, Yu and {Lima}, Marcos and {Lin}, Qiufan and {Lin}, Weipeng and {Lin}, Yen-Ting and {Long}, Daniel C. and {Lundgren}, Britt and {MacDonald}, Nicholas and {Geimba Maia}, Marcio Antonio and {Malanushenko}, Elena and {Malanushenko}, Viktor and {Mariappan}, Vivek and {McBride}, Cameron K. and {McGreer}, Ian D. and {M{\'e}nard}, Brice and {Merloni}, Andrea and {Meza}, Andres and {Montero-Dorta}, Antonio D. and {Muna}, Demitri and {Myers}, Adam D. and {Nandra}, Kirpal and {Naugle}, Tracy and {Newman}, Jeffrey A. and {Noterdaeme}, Pasquier and {Nugent}, Peter and {Ogando}, Ricardo and {Olmstead}, Matthew D. and {Oravetz}, Audrey and {Oravetz}, Daniel J. and {Padmanabhan}, Nikhil and {Palanque-Delabrouille}, Nathalie and {Pan}, Kaike and {Parejko}, John K. and {P{\^a}ris}, Isabelle and {Peacock}, John A. and {Petitjean}, Patrick and {Pieri}, Matthew M. and {Pisani}, Alice and {Prada}, Francisco and {Prakash}, Abhishek and {Raichoor}, Anand and {Reid}, Beth and {Rich}, James and {Ridl}, Jethro and {Rodriguez-Torres}, Sergio and {Carnero Rosell}, Aurelio and {Ross}, Ashley J. and {Rossi}, Graziano and {Ruan}, John and {Salvato}, Mara and {Sayres}, Conor and {Schneider}, Donald P. and {Schlegel}, David J. and {Seljak}, Uros and {Seo}, Hee-Jong and {Sesar}, Branimir and {Shandera}, Sarah and {Shu}, Yiping and {Slosar}, An{\v{z}}e and {Sobreira}, Flavia and {Streblyanska}, Alina and {Suzuki}, Nao and {Taylor}, Donna and {Tao}, Charling and {Tinker}, Jeremy L. and {Tojeiro}, Rita and {Vargas-Maga{\~n}a}, Mariana and {Wang}, Yuting and {Weaver}, Benjamin A. and {Weinberg}, David H. and {White}, Martin and {Wood-Vasey}, W.~M. and {Yeche}, Christophe and {Zhai}, Zhongxu and {Zhao}, Cheng and {Zhao}, Gong-bo and {Zheng}, Zheng and {Ben Zhu}, Guangtun and {Zou}, Hu},
        title = "{The SDSS-IV Extended Baryon Oscillation Spectroscopic Survey: Overview and Early Data}",
      journal = {\aj},
         year = 2016,
        month = feb,
       volume = {151},
       number = {2},
          eid = {44},
        pages = {44},
          doi = {10.3847/0004-6256/151/2/44},
archivePrefix = {arXiv},
       eprint = {1508.04473},
 primaryClass = {astro-ph.CO},
       adsurl = {https://ui.adsabs.harvard.edu/abs/2016AJ....151...44D}
}

@ARTICLE{debes_et_al2008,
       author = {{Debes}, John H. and {Weinberger}, Alycia J. and {Schneider}, Glenn},
        title = "{Complex Organic Materials in the Circumstellar Disk of HR 4796A}",
      journal = {\apjl},
         year = 2008,
        month = feb,
       volume = {673},
       number = {2},
        pages = {L191},
          doi = {10.1086/527546},
archivePrefix = {arXiv},
       eprint = {0712.3283},
 primaryClass = {astro-ph},
       adsurl = {https://ui.adsabs.harvard.edu/abs/2008ApJ...673L.191D}
}

@ARTICLE{de_cia_et_al2024,
       author = {{De Cia}, Annalisa and {Roman-Duval}, Julia and {Konstantopoulou}, Christina and {Noterdaeme}, Pasquier and {Ramburuth-Hurt}, Tanita and {Velichko}, Anna and {Fox}, Andrew J. and {Ledoux}, C{\'e}dric and {Petitjean}, Patrick and {Jermann}, Iris and {Krogager}, Jens-Kristian},
        title = "{{\ensuremath{\alpha}}-element enhancements in the ISM of the LMC and SMC: Evidence of recent star formation}",
      journal = {\aap},
         year = 2024,
        month = mar,
       volume = {683},
          eid = {A216},
        pages = {A216},
          doi = {10.1051/0004-6361/202346611},
       adsurl = {https://ui.adsabs.harvard.edu/abs/2024A&A...683A.216D}
}

@ARTICLE{defrere_et_al2015,
       author = {{Defr{\`e}re}, D. and {Hinz}, P.~M. and {Skemer}, A.~J. and {Kennedy}, G.~M. and {Bailey}, V.~P. and {Hoffmann}, W.~F. and {Mennesson}, B. and {Millan-Gabet}, R. and {Danchi}, W.~C. and {Absil}, O. and {Arbo}, P. and {Beichman}, C. and {Brusa}, G. and {Bryden}, G. and {Downey}, E.~C. and {Durney}, O. and {Esposito}, S. and {Gaspar}, A. and {Grenz}, P. and {Haniff}, C. and {Hill}, J.~M. and {Lebreton}, J. and {Leisenring}, J.~M. and {Males}, J.~R. and {Marion}, L. and {McMahon}, T.~J. and {Montoya}, M. and {Morzinski}, K.~M. and {Pinna}, E. and {Puglisi}, A. and {Rieke}, G. and {Roberge}, A. and {Serabyn}, E. and {Sosa}, R. and {Stapeldfeldt}, K. and {Su}, K. and {Vaitheeswaran}, V. and {Vaz}, A. and {Weinberger}, A.~J. and {Wyatt}, M.~C.},
        title = "{First-light LBT Nulling Interferometric Observations: Warm Exozodiacal Dust Resolved within a Few AU of {\ensuremath{\eta}} Crv}",
      journal = {\apj},
         year = 2015,
        month = jan,
       volume = {799},
       number = {1},
          eid = {42},
        pages = {42},
          doi = {10.1088/0004-637X/799/1/42},
archivePrefix = {arXiv},
       eprint = {1501.04144},
 primaryClass = {astro-ph.EP},
       adsurl = {https://ui.adsabs.harvard.edu/abs/2015ApJ...799...42D}
}

@ARTICLE{defrere_et_al2021,
       author = {{Defr{\`e}re}, D. and {Hinz}, P.~M. and {Kennedy}, G.~M. and {Stone}, J. and {Rigley}, J. and {Ertel}, S. and {Gaspar}, A. and {Bailey}, V.~P. and {Hoffmann}, W.~F. and {Mennesson}, B. and {Millan-Gabet}, R. and {Danchi}, W.~C. and {Absil}, O. and {Arbo}, P. and {Beichman}, C. and {Bonavita}, M. and {Brusa}, G. and {Bryden}, G. and {Downey}, E.~C. and {Esposito}, S. and {Grenz}, P. and {Haniff}, C. and {Hill}, J.~M. and {Leisenring}, J.~M. and {Males}, J.~R. and {McMahon}, T.~J. and {Montoya}, M. and {Morzinski}, K.~M. and {Pinna}, E. and {Puglisi}, A. and {Rieke}, G. and {Roberge}, A. and {Rousseau}, H. and {Serabyn}, E. and {Spalding}, E. and {Skemer}, A.~J. and {Stapelfeldt}, K. and {Su}, K. and {Vaz}, A. and {Weinberger}, A.~J. and {Wyatt}, M.~C.},
        title = "{The HOSTS Survey: Evidence for an Extended Dust Disk and Constraints on the Presence of Giant Planets in the Habitable Zone of {\ensuremath{\beta}} Leo}",
      journal = {\aj},
         year = 2021,
        month = apr,
       volume = {161},
       number = {4},
          eid = {186},
        pages = {186},
          doi = {10.3847/1538-3881/abe3ff},
archivePrefix = {arXiv},
       eprint = {2103.03268},
 primaryClass = {astro-ph.EP},
       adsurl = {https://ui.adsabs.harvard.edu/abs/2021AJ....161..186D}
}

@INPROCEEDINGS{delacroix_et_al2016,
       author = {{Delacroix}, Christian and {Savransky}, Dmitry and {Garrett}, Daniel and {Lowrance}, Patrick and {Morgan}, Rhonda},
        title = "{Science yield modeling with the Exoplanet Open-Source Imaging Mission Simulator (EXOSIMS)}",
    booktitle = {Modeling, Systems Engineering, and Project Management for Astronomy VI},
         year = 2016,
       editor = {{Angeli}, George Z. and {Dierickx}, Philippe},
       series = {Society of Photo-Optical Instrumentation Engineers (SPIE) Conference Series},
       volume = {9911},
        month = aug,
          eid = {991119},
        pages = {991119},
          doi = {10.1117/12.2233913},
       adsurl = {https://ui.adsabs.harvard.edu/abs/2016SPIE.9911E..19D}
}

@ARTICLE{della_bruna_et_al2022,
       author = {{Della Bruna}, Lorenza and {Adamo}, Angela and {McLeod}, Anna F. and {Smith}, Linda J. and {Savard}, Gabriel and {Robert}, Carmelle and {Sun}, Jiayi and {Amram}, Philippe and {Bik}, Arjan and {Blair}, William P. and {Long}, Knox S. and {Renaud}, Florent and {Walterbos}, Rene and {Usher}, Christopher},
        title = "{Stellar feedback in M 83 as observed with MUSE. II. Analysis of the H II region population: Ionisation budget and pre-SN feedback}",
      journal = {\aap},
         year = 2022,
        month = oct,
       volume = {666},
          eid = {A29},
        pages = {A29},
          doi = {10.1051/0004-6361/202243395},
archivePrefix = {arXiv},
       eprint = {2206.09741},
 primaryClass = {astro-ph.GA},
       adsurl = {https://ui.adsabs.harvard.edu/abs/2022A&A...666A..29D}
}

@INPROCEEDINGS{demyk2011,
       author = {{Demyk}, K.},
        title = "{Interstellar dust within the life cycle of the interstellar medium}",
    booktitle = {European Physical Journal Web of Conferences},
         year = 2011,
       series = {European Physical Journal Web of Conferences},
       volume = {18},
        month = jan,
          eid = {03001},
        pages = {03001},
          doi = {10.1051/epjconf/20111803001},
       adsurl = {https://ui.adsabs.harvard.edu/abs/2011EPJWC..1803001D}
}

@ARTICLE{desi_et_al2016,
       author = {{DESI Collaboration} and {Aghamousa}, Amir and {Aguilar}, Jessica and {Ahlen}, Steve and {Alam}, Shadab and {Allen}, Lori E. and {Allende Prieto}, Carlos and {Annis}, James and {Bailey}, Stephen and {Balland}, Christophe and {Ballester}, Otger and {Baltay}, Charles and {Beaufore}, Lucas and {Bebek}, Chris and {Beers}, Timothy C. and {Bell}, Eric F. and {Bernal}, Jos{\'e} Luis and {Besuner}, Robert and {Beutler}, Florian and {Blake}, Chris and {Bleuler}, Hannes and {Blomqvist}, Michael and {Blum}, Robert and {Bolton}, Adam S. and {Briceno}, Cesar and {Brooks}, David and {Brownstein}, Joel R. and {Buckley-Geer}, Elizabeth and {Burden}, Angela and {Burtin}, Etienne and {Busca}, Nicolas G. and {Cahn}, Robert N. and {Cai}, Yan-Chuan and {Cardiel-Sas}, Laia and {Carlberg}, Raymond G. and {Carton}, Pierre-Henri and {Casas}, Ricard and {Castander}, Francisco J. and {Cervantes-Cota}, Jorge L. and {Claybaugh}, Todd M. and {Close}, Madeline and {Coker}, Carl T. and {Cole}, Shaun and {Comparat}, Johan and {Cooper}, Andrew P. and {Cousinou}, M. -C. and {Crocce}, Martin and {Cuby}, Jean-Gabriel and {Cunningham}, Daniel P. and {Davis}, Tamara M. and {Dawson}, Kyle S. and {de la Macorra}, Axel and {De Vicente}, Juan and {Delubac}, Timoth{\'e}e and {Derwent}, Mark and {Dey}, Arjun and {Dhungana}, Govinda and {Ding}, Zhejie and {Doel}, Peter and {Duan}, Yutong T. and {Ealet}, Anne and {Edelstein}, Jerry and {Eftekharzadeh}, Sarah and {Eisenstein}, Daniel J. and {Elliott}, Ann and {Escoffier}, St{\'e}phanie and {Evatt}, Matthew and {Fagrelius}, Parker and {Fan}, Xiaohui and {Fanning}, Kevin and {Farahi}, Arya and {Farihi}, Jay and {Favole}, Ginevra and {Feng}, Yu and {Fernandez}, Enrique and {Findlay}, Joseph R. and {Finkbeiner}, Douglas P. and {Fitzpatrick}, Michael J. and {Flaugher}, Brenna and {Flender}, Samuel and {Font-Ribera}, Andreu and {Forero-Romero}, Jaime E. and {Fosalba}, Pablo and {Frenk}, Carlos S. and {Fumagalli}, Michele and {Gaensicke}, Boris T. and {Gallo}, Giuseppe and {Garcia-Bellido}, Juan and {Gaztanaga}, Enrique and {Pietro Gentile Fusillo}, Nicola and {Gerard}, Terry and {Gershkovich}, Irena and {Giannantonio}, Tommaso and {Gillet}, Denis and {Gonzalez-de-Rivera}, Guillermo and {Gonzalez-Perez}, Violeta and {Gott}, Shelby and {Graur}, Or and {Gutierrez}, Gaston and {Guy}, Julien and {Habib}, Salman and {Heetderks}, Henry and {Heetderks}, Ian and {Heitmann}, Katrin and {Hellwing}, Wojciech A. and {Herrera}, David A. and {Ho}, Shirley and {Holland}, Stephen and {Honscheid}, Klaus and {Huff}, Eric and {Hutchinson}, Timothy A. and {Huterer}, Dragan and {Hwang}, Ho Seong and {Illa Laguna}, Joseph Maria and {Ishikawa}, Yuzo and {Jacobs}, Dianna and {Jeffrey}, Niall and {Jelinsky}, Patrick and {Jennings}, Elise and {Jiang}, Linhua and {Jimenez}, Jorge and {Johnson}, Jennifer and {Joyce}, Richard and {Jullo}, Eric and {Juneau}, St{\'e}phanie and {Kama}, Sami and {Karcher}, Armin and {Karkar}, Sonia and {Kehoe}, Robert and {Kennamer}, Noble and {Kent}, Stephen and {Kilbinger}, Martin and {Kim}, Alex G. and {Kirkby}, David and {Kisner}, Theodore and {Kitanidis}, Ellie and {Kneib}, Jean-Paul and {Koposov}, Sergey and {Kovacs}, Eve and {Koyama}, Kazuya and {Kremin}, Anthony and {Kron}, Richard and {Kronig}, Luzius and {Kueter-Young}, Andrea and {Lacey}, Cedric G. and {Lafever}, Robin and {Lahav}, Ofer and {Lambert}, Andrew and {Lampton}, Michael and {Landriau}, Martin and {Lang}, Dustin and {Lauer}, Tod R. and {Le Goff}, Jean-Marc and {Le Guillou}, Laurent and {Le Van Suu}, Auguste and {Lee}, Jae Hyeon and {Lee}, Su-Jeong and {Leitner}, Daniela and {Lesser}, Michael and {Levi}, Michael E. and {L'Huillier}, Benjamin and {Li}, Baojiu and {Liang}, Ming and {Lin}, Huan and {Linder}, Eric and {Loebman}, Sarah R. and {Luki{\'c}}, Zarija and {Ma}, Jun and {MacCrann}, Niall and {Magneville}, Christophe and {Makarem}, Laleh and {Manera}, Marc and {Manser}, Christopher J. and {Marshall}, Robert and {Martini}, Paul and {Massey}, Richard and {Matheson}, Thomas and {McCauley}, Jeremy and {McDonald}, Patrick and {McGreer}, Ian D. and {Meisner}, Aaron and {Metcalfe}, Nigel and {Miller}, Timothy N. and {Miquel}, Ramon and {Moustakas}, John and {Myers}, Adam and {Naik}, Milind and {Newman}, Jeffrey A. and {Nichol}, Robert C. and {Nicola}, Andrina and {Nicolati da Costa}, Luiz and {Nie}, Jundan and {Niz}, Gustavo and {Norberg}, Peder and {Nord}, Brian and {Norman}, Dara and {Nugent}, Peter and {O'Brien}, Thomas and {Oh}, Minji and {Olsen}, Knut A.~G.},
        title = "{The DESI Experiment Part I: Science,Targeting, and Survey Design}",
      journal = {arXiv e-prints},
         year = 2016,
        month = oct,
          eid = {arXiv:1611.00036},
        pages = {arXiv:1611.00036},
          doi = {10.48550/arXiv.1611.00036},
archivePrefix = {arXiv},
       eprint = {1611.00036},
 primaryClass = {astro-ph.IM},
       adsurl = {https://ui.adsabs.harvard.edu/abs/2016arXiv161100036D}
}

@ARTICLE{devecchi+volonteri2009,
       author = {{Devecchi}, B. and {Volonteri}, M.},
        title = "{Formation of the First Nuclear Clusters and Massive Black Holes at High Redshift}",
      journal = {\apj},
         year = 2009,
        month = mar,
       volume = {694},
       number = {1},
        pages = {302-313},
          doi = {10.1088/0004-637X/694/1/302},
archivePrefix = {arXiv},
       eprint = {0810.1057},
 primaryClass = {astro-ph},
       adsurl = {https://ui.adsabs.harvard.edu/abs/2009ApJ...694..302D}
}

@ARTICLE{diamond-stanic_et_al2016,
       author = {{Diamond-Stanic}, Aleksandar M. and {Coil}, Alison L. and {Moustakas}, John and {Tremonti}, Christy A. and {Sell}, Paul H. and {Mendez}, Alexander J. and {Hickox}, Ryan C. and {Rudnick}, Greg H.},
        title = "{Galaxies Probing Galaxies at High Resolution: Co-rotating Gas Associated with a Milky Way Analog at z=0.4}",
      journal = {\apj},
         year = 2016,
        month = jun,
       volume = {824},
       number = {1},
          eid = {24},
        pages = {24},
          doi = {10.3847/0004-637X/824/1/24},
archivePrefix = {arXiv},
       eprint = {1507.01945},
 primaryClass = {astro-ph.GA},
       adsurl = {https://ui.adsabs.harvard.edu/abs/2016ApJ...824...24D}
}

@ARTICLE{diana+dini2024,
       author = {{Diana}, Lorenzo and {Dini}, Pierpaolo},
        title = "{Review on Hardware Devices and Software Techniques Enabling Neural Network Inference Onboard Satellites}",
      journal = {Remote Sensing},
         year = 2024,
        month = oct,
       volume = {16},
       number = {21},
          eid = {3957},
        pages = {3957},
          doi = {10.3390/rs16213957},
       adsurl = {https://ui.adsabs.harvard.edu/abs/2024RemS...16.3957D}
}

@ARTICLE{dihingia_et_al2021,
       author = {{Dihingia}, Indu K. and {Vaidya}, Bhargav and {Fendt}, Christian},
        title = "{Jets, disc-winds, and oscillations in general relativistic, magnetically driven flows around black hole}",
      journal = {\mnras},
         year = 2021,
        month = aug,
       volume = {505},
       number = {3},
        pages = {3596-3615},
          doi = {10.1093/mnras/stab1512},
archivePrefix = {arXiv},
       eprint = {2105.11468},
 primaryClass = {astro-ph.HE},
       adsurl = {https://ui.adsabs.harvard.edu/abs/2021MNRAS.505.3596D}
}

@ARTICLE{dos_santos_et_al2019,
       author = {{dos Santos}, Leonardo A. and {Bourrier}, Vincent and {Ehrenreich}, David and {Kameda}, Shingo},
        title = "{Observability of hydrogen-rich exospheres in Earth-like exoplanets}",
      journal = {\aap},
         year = 2019,
        month = feb,
       volume = {622},
          eid = {A46},
        pages = {A46},
          doi = {10.1051/0004-6361/201833392},
archivePrefix = {arXiv},
       eprint = {1812.02145},
 primaryClass = {astro-ph.EP},
       adsurl = {https://ui.adsabs.harvard.edu/abs/2019A&A...622A..46D}
}

@INPROCEEDINGS{dos_santos_et_al2023,
       author = {{Dos Santos}, Leonardo A.},
        title = "{Observations of planetary winds and outflows}",
    booktitle = {Winds of Stars and Exoplanets},
         year = 2023,
       editor = {{Vidotto}, Aline A. and {Fossati}, Luca and {Vink}, Jorick S.},
       series = {IAU Symposium},
       volume = {370},
        month = jan,
        pages = {56-71},
          doi = {10.1017/S1743921322004239},
archivePrefix = {arXiv},
       eprint = {2211.16243},
 primaryClass = {astro-ph.EP},
       adsurl = {https://ui.adsabs.harvard.edu/abs/2023IAUS..370...56D}
}

@ARTICLE{dotti_et_al2022,
       author = {{Dotti}, Massimo and {Bonetti}, Matteo and {D'Orazio}, Daniel J. and {Haiman}, Zolt{\'a}n and {Ho}, Luis C.},
        title = "{Binary black hole signatures in polarized light curves}",
      journal = {\mnras},
         year = 2022,
        month = jan,
       volume = {509},
       number = {1},
        pages = {212-223},
          doi = {10.1093/mnras/stab2893},
archivePrefix = {arXiv},
       eprint = {2103.14652},
 primaryClass = {astro-ph.HE},
       adsurl = {https://ui.adsabs.harvard.edu/abs/2022MNRAS.509..212D}
}

@ARTICLE{douglas_et_al2022,
       author = {{Douglas}, Ewan S. and {Debes}, John and {Mennesson}, Bertrand and {Nemati}, Bijan and {Ashcraft}, Jaren and {Ren}, Bin and {Stapelfeldt}, Karl R. and {Savransky}, Dmitry and {Lewis}, Nikole K. and {Macintosh}, Bruce},
        title = "{Sensitivity of the Roman Coronagraph Instrument to Exozodiacal Dust}",
      journal = {\pasp},
         year = 2022,
        month = feb,
       volume = {134},
       number = {1032},
          eid = {024402},
        pages = {024402},
          doi = {10.1088/1538-3873/ac3f7b},
archivePrefix = {arXiv},
       eprint = {2112.12804},
 primaryClass = {astro-ph.EP},
       adsurl = {https://ui.adsabs.harvard.edu/abs/2022PASP..134b4402D}
}

@ARTICLE{droser_et_al2017,
       author = {{Droser}, Mary L. and {Tarhan}, Lidya G. and {Gehling}, James G.},
        title = "{The Rise of Animals in a Changing Environment: Global Ecological Innovation in the Late Ediacaran}",
      journal = {Annual Review of Earth and Planetary Sciences},
         year = 2017,
        month = aug,
       volume = {45},
       number = {1},
        pages = {593-617},
          doi = {10.1146/annurev-earth-063016-015645},
       adsurl = {https://ui.adsabs.harvard.edu/abs/2017AREPS..45..593D}
}

@ARTICLE{edge_et_al1997,
       author = {{Edge}, R. and {McGarvey}, D.~J. and {Truscott}, T.~G.},
        title = "{The carotenoids as anti-oxidants {\textemdash} a review}",
      journal = {Journal of Photochemistry and Photobiology B: Biology},
         year = 1997,
        month = jan,
       volume = {41},
       number = {3},
        pages = {189-200},
          doi = {10.1016/S1011-1344(97)00092-4},
       adsurl = {https://ui.adsabs.harvard.edu/abs/1997JPPB...41..189E}
}

@ARTICLE{ehrenreich2012,
       author = {{Ehrenreich}, D. and {Vidal-Madjar}, A. and {Widemann}, T. and
         {Gronoff}, G. and {Tanga}, P. and {Barth{\'e}lemy}, M. and
         {Lilensten}, J. and {Lecavelier Des Etangs}, A. and {Arnold}, L.},
        title = "{Transmission spectrum of Venus as a transiting exoplanet}",
      journal = {\aap},
         year = "2012",
        month = "Jan",
       volume = {537},
          eid = {L2},
        pages = {L2},
          doi = {10.1051/0004-6361/201118400},
archivePrefix = {arXiv},
       eprint = {1112.0572},
 primaryClass = {astro-ph.EP},
       adsurl = {https://ui.adsabs.harvard.edu/abs/2012A&A...537L...2E}
}

@ARTICLE{ekstrom_et_al2012,
       author = {{Ekstr{\"o}m}, S. and {Georgy}, C. and {Eggenberger}, P. and {Meynet}, G. and {Mowlavi}, N. and {Wyttenbach}, A. and {Granada}, A. and {Decressin}, T. and {Hirschi}, R. and {Frischknecht}, U. and {Charbonnel}, C. and {Maeder}, A.},
        title = "{Grids of stellar models with rotation. I. Models from 0.8 to 120 M$_{{\ensuremath{\odot}}}$ at solar metallicity (Z = 0.014)}",
      journal = {\aap},
         year = 2012,
        month = jan,
       volume = {537},
          eid = {A146},
        pages = {A146},
          doi = {10.1051/0004-6361/201117751},
archivePrefix = {arXiv},
       eprint = {1110.5049},
 primaryClass = {astro-ph.SR},
       adsurl = {https://ui.adsabs.harvard.edu/abs/2012A&A...537A.146E}
}

@ARTICLE{elmegreen+efremov1997,
       author = {{Elmegreen}, Bruce G. and {Efremov}, Yuri N.},
        title = "{A Universal Formation Mechanism for Open and Globular Clusters in Turbulent Gas}",
      journal = {\apj},
         year = 1997,
        month = may,
       volume = {480},
       number = {1},
        pages = {235-245},
          doi = {10.1086/303966},
       adsurl = {https://ui.adsabs.harvard.edu/abs/1997ApJ...480..235E}
}

@ARTICLE{eldridge+stanway2009,
       author = {{Eldridge}, John J. and {Stanway}, Elizabeth R.},
        title = "{Spectral population synthesis including massive binaries}",
      journal = {\mnras},
         year = 2009,
        month = dec,
       volume = {400},
       number = {2},
        pages = {1019-1028},
          doi = {10.1111/j.1365-2966.2009.15514.x},
archivePrefix = {arXiv},
       eprint = {0908.1386},
 primaryClass = {astro-ph.CO},
       adsurl = {https://ui.adsabs.harvard.edu/abs/2009MNRAS.400.1019E}
}

@INPROCEEDINGS{enya_et_al2026,
       author = {{Enya}, Keigo and {Yoneta}, Kenta and {Murakami}, Naoshi and {Nishikawa}, Jun and {Itoh}, Satoshi and {Matsuo}, Taro and {Kojima}, Reiki and {Kotani}, Takayuki and {Guyon}, Olivier and {Sumi}, Takahiro and {Miyazaki}, Satoshi and {Yamada}, Toru and {Takahashi}, Aoi and {Kawahara}, Hajime and {Miyazaki}, Shota and {Kondo}, Iona and {Higashio}, Nana and {Yamasaki}, Noriko and {Kuzuhara}, Masayuki and {Tamura}, Motohide and {Ikoma}, Masahiro and {Narita}, Norio and {Lozi}, Julien},
        title = "{Japan's Possible Contributions for Coronagraph of the Habitable Worlds Observatory (HWO)}",
    booktitle = {Astronomical Society of the Pacific Conference Series},
         year = 2026,
       editor = {{Lee}, Janice C. and {Noviello}, Jessica and {LaMassa}, Stephanie and {Postman}, Marc},
       series = {Astronomical Society of the Pacific Conference Series},
       volume = {543},
        month = nov,
        pages = {409},
       adsurl = {https://ui.adsabs.harvard.edu/abs/2026ASPC..543..409E}
}

@ARTICLE{ertel_et_al2014,
       author = {{Ertel}, S. and {Absil}, O. and {Defr{\`e}re}, D. and {Le Bouquin}, J. -B. and {Augereau}, J. -C. and {Marion}, L. and {Blind}, N. and {Bonsor}, A. and {Bryden}, G. and {Lebreton}, J. and {Milli}, J.},
        title = "{A near-infrared interferometric survey of debris-disk stars. IV. An unbiased sample of 92 southern stars observed in H band with VLTI/PIONIER}",
      journal = {\aap},
         year = 2014,
        month = oct,
       volume = {570},
          eid = {A128},
        pages = {A128},
          doi = {10.1051/0004-6361/201424438},
archivePrefix = {arXiv},
       eprint = {1409.6143},
 primaryClass = {astro-ph.EP},
       adsurl = {https://ui.adsabs.harvard.edu/abs/2014A&A...570A.128E}
}

@ARTICLE{ertel_et_al2016,
       author = {{Ertel}, S. and {Defr{\`e}re}, D. and {Absil}, O. and {Le Bouquin}, J. -B. and {Augereau}, J. -C. and {Berger}, J. -P. and {Blind}, N. and {Bonsor}, A. and {Lagrange}, A. -M. and {Lebreton}, J. and {Marion}, L. and {Milli}, J. and {Olofsson}, J.},
        title = "{A near-infrared interferometric survey of debris-disc stars. V. PIONIER search for variability}",
      journal = {\aap},
         year = 2016,
        month = oct,
       volume = {595},
          eid = {A44},
        pages = {A44},
          doi = {10.1051/0004-6361/201527721},
archivePrefix = {arXiv},
       eprint = {1608.05731},
 primaryClass = {astro-ph.SR},
       adsurl = {https://ui.adsabs.harvard.edu/abs/2016A&A...595A..44E}
}

@ARTICLE{ertel_et_al2018,
       author = {{Ertel}, S. and {Defr{\`e}re}, D. and {Hinz}, P. and {Mennesson}, B. and {Kennedy}, G.~M. and {Danchi}, W.~C. and {Gelino}, C. and {Hill}, J.~M. and {Hoffmann}, W.~F. and {Rieke}, G. and {Shannon}, A. and {Spalding}, E. and {Stone}, J.~M. and {Vaz}, A. and {Weinberger}, A.~J. and {Willems}, P. and {Absil}, O. and {Arbo}, P. and {Bailey}, V.~P. and {Beichman}, C. and {Bryden}, G. and {Downey}, E.~C. and {Durney}, O. and {Esposito}, S. and {Gaspar}, A. and {Grenz}, P. and {Haniff}, C.~A. and {Leisenring}, J.~M. and {Marion}, L. and {McMahon}, T.~J. and {Millan-Gabet}, R. and {Montoya}, M. and {Morzinski}, K.~M. and {Pinna}, E. and {Power}, J. and {Puglisi}, A. and {Roberge}, A. and {Serabyn}, E. and {Skemer}, A.~J. and {Stapelfeldt}, K. and {Su}, K.~Y.~L. and {Vaitheeswaran}, V. and {Wyatt}, M.~C.},
        title = "{The HOSTS Survey{\textemdash}Exozodiacal Dust Measurements for 30 Stars}",
      journal = {\aj},
         year = 2018,
        month = may,
       volume = {155},
       number = {5},
          eid = {194},
        pages = {194},
          doi = {10.3847/1538-3881/aab717},
archivePrefix = {arXiv},
       eprint = {1803.11265},
 primaryClass = {astro-ph.SR},
       adsurl = {https://ui.adsabs.harvard.edu/abs/2018AJ....155..194E}
}

@ARTICLE{ertel_et_al2020,
       author = {{Ertel}, S. and {Defr{\`e}re}, D. and {Hinz}, P. and {Mennesson}, B. and {Kennedy}, G.~M. and {Danchi}, W.~C. and {Gelino}, C. and {Hill}, J.~M. and {Hoffmann}, W.~F. and {Mazoyer}, J. and {Rieke}, G. and {Shannon}, A. and {Stapelfeldt}, K. and {Spalding}, E. and {Stone}, J.~M. and {Vaz}, A. and {Weinberger}, A.~J. and {Willems}, P. and {Absil}, O. and {Arbo}, P. and {Bailey}, V.~P. and {Beichman}, C. and {Bryden}, G. and {Downey}, E.~C. and {Durney}, O. and {Esposito}, S. and {Gaspar}, A. and {Grenz}, P. and {Haniff}, C.~A. and {Leisenring}, J.~M. and {Marion}, L. and {McMahon}, T.~J. and {Millan-Gabet}, R. and {Montoya}, M. and {Morzinski}, K.~M. and {Perera}, S. and {Pinna}, E. and {Pott}, J. -U. and {Power}, J. and {Puglisi}, A. and {Roberge}, A. and {Serabyn}, E. and {Skemer}, A.~J. and {Su}, K.~Y.~L. and {Vaitheeswaran}, V. and {Wyatt}, M.~C.},
        title = "{The HOSTS Survey for Exozodiacal Dust: Observational Results from the Complete Survey}",
      journal = {\aj},
         year = 2020,
        month = apr,
       volume = {159},
       number = {4},
          eid = {177},
        pages = {177},
          doi = {10.3847/1538-3881/ab7817},
archivePrefix = {arXiv},
       eprint = {2003.03499},
 primaryClass = {astro-ph.SR},
       adsurl = {https://ui.adsabs.harvard.edu/abs/2020AJ....159..177E}
}

@ARTICLE{ertel_et_al2025,
       author = {{Ertel}, Steve and {Pearce}, Tim D. and {Debes}, John H. and {Faramaz}, Virginie C. and {Danchi}, William C. and {Anche}, Ramya M. and {Defr{\`e}re}, Denis and {Hasegawa}, Yasuhiro and {Hom}, Justin and {Kirchschlager}, Florian and {Rebollido}, Isabel and {Rousseau}, H{\'e}l{\`e}ne and {Scott}, Jeremy and {Stapelfeldt}, Karl and {Stuber}, Thomas A.},
        title = "{Review and Prospects of Hot Exozodiacal Dust Research For Future Exo-Earth Direct Imaging Missions}",
      journal = {\pasp},
         year = 2025,
        month = mar,
       volume = {137},
       number = {3},
          eid = {031001},
        pages = {031001},
          doi = {10.1088/1538-3873/adb6d5},
archivePrefix = {arXiv},
       eprint = {2504.00295},
 primaryClass = {astro-ph.EP},
       adsurl = {https://ui.adsabs.harvard.edu/abs/2025PASP..137c1001E}
}

@techreport{esa_voyage2050,
  author      = {Tacconi, Linda and Arridge, Christopher and {Voyage 2050 Senior Committee}},
  title       = {Voyage 2050: Final Recommendations from the Senior Committee to the ESA Director of Science},
  institution = {European Space Agency (ESA)},
  year        = {2021},
  month       = jun,
  type        = {Report},
  url         = {https://www.cosmos.esa.int/documents/1866264/1866292/Voyage2050-Senior-Committee-report-public.pdf},
  note        = {Available at ESA Cosmos}
}

@ARTICLE{evans_et_al2018,
       author = {{Evans}, Nancy Remage and {Proffitt}, Charles and {Carpenter}, Kenneth G. and {Winston}, Elaine M. and {Kober}, Gladys V. and {G{\"u}nther}, H. Moritz and {Gorynya}, Natalia and {Rastorguev}, Alexey and {Inno}, L.},
        title = "{The Mass of the Cepheid V350 Sgr}",
      journal = {\apj},
         year = 2018,
        month = oct,
       volume = {866},
       number = {1},
          eid = {30},
        pages = {30},
          doi = {10.3847/1538-4357/aade03},
archivePrefix = {arXiv},
       eprint = {1808.10472},
 primaryClass = {astro-ph.SR},
       adsurl = {https://ui.adsabs.harvard.edu/abs/2018ApJ...866...30E}
}

@ARTICLE{evans_et_al2019,
       author = {{Evans}, C.~J. and {Castro}, N. and {Gonzalez}, O.~A. and {Garcia}, M. and {Bastian}, N. and {Cioni}, M. -R.~L. and {Clark}, J.~S. and {Davies}, B. and {Ferguson}, A.~M.~N. and {Kamann}, S. and {Lennon}, D.~J. and {Patrick}, L.~R. and {Vink}, J.~S. and {Weisz}, D.~R.},
        title = "{First stellar spectroscopy in Leo P}",
      journal = {\aap},
         year = 2019,
        month = feb,
       volume = {622},
          eid = {A129},
        pages = {A129},
          doi = {10.1051/0004-6361/201834145},
archivePrefix = {arXiv},
       eprint = {1901.01295},
 primaryClass = {astro-ph.SR},
       adsurl = {https://ui.adsabs.harvard.edu/abs/2019A&A...622A.129E}
}

@ARTICLE{evans_et_al2023,
       author = {{Evans}, Nancy Remage and {Ferrari}, Mckenzie G. and {Kuraszkiewicz}, Joanna and {Silverberg}, Steven and {Nichols}, Joy and {Torres}, Guillermo and {Fischbach}, Makenzi},
        title = "{The Mass-Temperature Relation for B and Early A Stars Based on International Ultraviolet Explorer Spectra of Detached Eclipsing Binaries}",
      journal = {\aj},
         year = 2023,
        month = sep,
       volume = {166},
       number = {3},
          eid = {109},
        pages = {109},
          doi = {10.3847/1538-3881/ace89b},
archivePrefix = {arXiv},
       eprint = {2308.01374},
 primaryClass = {astro-ph.SR},
       adsurl = {https://ui.adsabs.harvard.edu/abs/2023AJ....166..109E}
}

@BOOK{exoplanet_science_strategy2018,
  author    = "{National Academies of Sciences, Engineering, and Medicine}",
  title     = "Exoplanet Science Strategy",
  isbn      = "978-0-309-47941-7",
  doi       = "10.17226/25187",
  url       = "https://nap.nationalacademies.org/catalog/25187/exoplanet-science-strategy",
  year      = 2018,
  publisher = "The National Academies Press",
  address   = "Washington, DC"
}

@ARTICLE{faramaz_et_al2017,
       author = {{Faramaz}, V. and {Ertel}, S. and {Booth}, M. and {Cuadra}, J. and {Simmonds}, C.},
        title = "{Inner mean-motion resonances with eccentric planets: a possible origin for exozodiacal dust clouds}",
      journal = {\mnras},
         year = 2017,
        month = feb,
       volume = {465},
       number = {2},
        pages = {2352-2365},
          doi = {10.1093/mnras/stw2846},
archivePrefix = {arXiv},
       eprint = {1611.02196},
 primaryClass = {astro-ph.EP},
       adsurl = {https://ui.adsabs.harvard.edu/abs/2017MNRAS.465.2352F}
}

@ARTICLE{feinberg_et_al2026,
       author = {{Feinberg}, Lee D. and {Sitarski}, Breann N. and {McElwain}, Michael W. and {Arney}, Giada and {Baker}, Caleb and {Bolcar}, Matthew R. and {Levine}, Marie and {Liu}, Alice and {Mennesson}, Bertrand and {Roberge}, Aki and {Smith}, J. Scott and {Zhao}, Feng and {Ziemer}, John},
        title = "{Habitable Worlds Observatory's Concept and Technology Maturation: Initial Feasibility and Trade Space Exploration}",
      journal = {arXiv e-prints},
         year = 2026,
        month = jan,
          eid = {arXiv:2601.11803},
        pages = {arXiv:2601.11803},
          doi = {10.48550/arXiv.2601.11803},
archivePrefix = {arXiv},
       eprint = {2601.11803},
 primaryClass = {astro-ph.IM},
       adsurl = {https://ui.adsabs.harvard.edu/abs/2026arXiv260111803F}
}

@ARTICLE{fetherolf_et_al2026,
       author = {{Fetherolf}, Tara and {Gupta}, Arvind F. and {Newton}, Elisabeth R. and {Buccino}, Andrea P. and {Burt}, Jennifer A. and {Caballero}, Jose A. and {Carrazco-Gaxiola}, Sebastian and {Vieytes}, Mariela C. and {Hinkel}, Natalie R. and {Mamajek}, Eric E.},
        title = "{HWO Target Stars and Systems: Activity and Rotation Catalog (ARC) of Potential Target Stars for the Habitable Worlds Observatory}",
      journal = {arXiv e-prints},
         year = 2026,
        month = may,
          eid = {arXiv:2605.22618},
        pages = {arXiv:2605.22618},
          doi = {10.48550/arXiv.2605.22618},
archivePrefix = {arXiv},
       eprint = {2605.22618},
 primaryClass = {astro-ph.SR},
       adsurl = {https://ui.adsabs.harvard.edu/abs/2026arXiv260522618F}
}

@ARTICLE{feruglio_et_al2010,
       author = {{Feruglio}, C. and {Maiolino}, R. and {Piconcelli}, E. and {Menci}, N. and {Aussel}, H. and {Lamastra}, A. and {Fiore}, F.},
        title = "{Quasar feedback revealed by giant molecular outflows}",
      journal = {\aap},
         year = 2010,
        month = jul,
       volume = {518},
          eid = {L155},
        pages = {L155},
          doi = {10.1051/0004-6361/201015164},
archivePrefix = {arXiv},
       eprint = {1006.1655},
 primaryClass = {astro-ph.CO},
       adsurl = {https://ui.adsabs.harvard.edu/abs/2010A&A...518L.155F}
}

@ARTICLE{finkelstein_et_al2019,
       author = {{Finkelstein}, Steven L. and {D'Aloisio}, Anson and {Paardekooper}, Jan-Pieter and {Ryan}, Jr., Russell and {Behroozi}, Peter and {Finlator}, Kristian and {Livermore}, Rachael and {Upton Sanderbeck}, Phoebe R. and {Dalla Vecchia}, Claudio and {Khochfar}, Sadegh},
        title = "{Conditions for Reionizing the Universe with a Low Galaxy Ionizing Photon Escape Fraction}",
      journal = {\apj},
         year = 2019,
        month = jul,
       volume = {879},
       number = {1},
          eid = {36},
        pages = {36},
          doi = {10.3847/1538-4357/ab1ea8},
archivePrefix = {arXiv},
       eprint = {1902.02792},
 primaryClass = {astro-ph.CO},
       adsurl = {https://ui.adsabs.harvard.edu/abs/2019ApJ...879...36F}
}

@ARTICLE{fischer_et_al2016,
       author = {{Fischer}, Woodward W. and {Hemp}, James and {Johnson}, Jena E.},
        title = "{Evolution of Oxygenic Photosynthesis}",
      journal = {Annual Review of Earth and Planetary Sciences},
         year = 2016,
        month = jun,
       volume = {44},
        pages = {647-683},
          doi = {10.1146/annurev-earth-060313-054810},
       adsurl = {https://ui.adsabs.harvard.edu/abs/2016AREPS..44..647F}
}

@ARTICLE{fitzpatrick+massa2007,
       author = {{Fitzpatrick}, E.~L. and {Massa}, D.},
        title = "{An Analysis of the Shapes of Interstellar Extinction Curves. V. The IR-through-UV Curve Morphology}",
      journal = {\apj},
         year = 2007,
        month = jul,
       volume = {663},
       number = {1},
        pages = {320-341},
          doi = {10.1086/518158},
archivePrefix = {arXiv},
       eprint = {0705.0154},
 primaryClass = {astro-ph},
       adsurl = {https://ui.adsabs.harvard.edu/abs/2007ApJ...663..320F}
}

@INPROCEEDINGS{flambaum+shuryak2008,
       author = {{Flambaum}, V.~V. and {Shuryak}, E.~V.},
        title = "{How changing physical constants and violation of local position invariance may occur?}",
    booktitle = {Nuclei and Mesoscopic Physic - WNMP 2007},
         year = 2008,
       editor = {{Danielewicz}, Pawel and {Piecuch}, Piotr and {Zelevinsky}, Vladimir},
       series = {American Institute of Physics Conference Series},
       volume = {995},
        month = apr,
    publisher = {AIP},
        pages = {1-11},
          doi = {10.1063/1.2915601},
archivePrefix = {arXiv},
       eprint = {physics/0701220},
 primaryClass = {physics.atom-ph},
       adsurl = {https://ui.adsabs.harvard.edu/abs/2008AIPC..995....1F}
}

@ARTICLE{flury_et_al2022a,
       author = {{Flury}, Sophia R. and {Jaskot}, Anne E. and {Ferguson}, Harry C. and {Worseck}, G{\'a}bor and {Makan}, Kirill and {Chisholm}, John and {Saldana-Lopez}, Alberto and {Schaerer}, Daniel and {McCandliss}, Stephan and {Wang}, Bingjie and {Ford}, N.~M. and {Heckman}, Timothy and {Ji}, Zhiyuan and {Giavalisco}, Mauro and {Amorin}, Ricardo and {Atek}, Hakim and {Blaizot}, Jeremy and {Borthakur}, Sanchayeeta and {Carr}, Cody and {Castellano}, Marco and {Cristiani}, Stefano and {De Barros}, Stephane and {Dickinson}, Mark and {Finkelstein}, Steven L. and {Fleming}, Brian and {Fontanot}, Fabio and {Garel}, Thibault and {Grazian}, Andrea and {Hayes}, Matthew and {Henry}, Alaina and {Mauerhofer}, Valentin and {Micheva}, Genoveva and {Oey}, M.~S. and {Ostlin}, Goran and {Papovich}, Casey and {Pentericci}, Laura and {Ravindranath}, Swara and {Rosdahl}, Joakim and {Rutkowski}, Michael and {Santini}, Paola and {Scarlata}, Claudia and {Teplitz}, Harry and {Thuan}, Trinh and {Trebitsch}, Maxime and {Vanzella}, Eros and {Verhamme}, Anne and {Xu}, Xinfeng},
        title = "{The Low-redshift Lyman Continuum Survey. I. New, Diverse Local Lyman Continuum Emitters}",
      journal = {\apjs},
         year = 2022,
        month = may,
       volume = {260},
       number = {1},
          eid = {1},
        pages = {1},
          doi = {10.3847/1538-4365/ac5331},
archivePrefix = {arXiv},
       eprint = {2201.11716},
 primaryClass = {astro-ph.GA},
       adsurl = {https://ui.adsabs.harvard.edu/abs/2022ApJS..260....1F}
}

@ARTICLE{flury_et_al2022b,
       author = {{Flury}, Sophia R. and {Jaskot}, Anne E. and {Ferguson}, Harry C. and {Worseck}, G{\'a}bor and {Makan}, Kirill and {Chisholm}, John and {Saldana-Lopez}, Alberto and {Schaerer}, Daniel and {McCandliss}, Stephan R. and {Xu}, Xinfeng and {Wang}, Bingjie and {Oey}, M.~S. and {Ford}, N.~M. and {Heckman}, Timothy and {Ji}, Zhiyuan and {Giavalisco}, Mauro and {Amor{\'\i}n}, Ricardo and {Atek}, Hakim and {Blaizot}, Jeremy and {Borthakur}, Sanchayeeta and {Carr}, Cody and {Castellano}, Marco and {De Barros}, Stephane and {Dickinson}, Mark and {Finkelstein}, Steven L. and {Fleming}, Brian and {Fontanot}, Fabio and {Garel}, Thibault and {Grazian}, Andrea and {Hayes}, Matthew and {Henry}, Alaina and {Mauerhofer}, Valentin and {Micheva}, Genoveva and {Ostlin}, Goran and {Papovich}, Casey and {Pentericci}, Laura and {Ravindranath}, Swara and {Rosdahl}, Joakim and {Rutkowski}, Michael and {Santini}, Paola and {Scarlata}, Claudia and {Teplitz}, Harry and {Thuan}, Trinh and {Trebitsch}, Maxime and {Vanzella}, Eros and {Verhamme}, Anne},
        title = "{The Low-redshift Lyman Continuum Survey. II. New Insights into LyC Diagnostics}",
      journal = {\apj},
         year = 2022,
        month = may,
       volume = {930},
       number = {2},
          eid = {126},
        pages = {126},
          doi = {10.3847/1538-4357/ac61e4},
archivePrefix = {arXiv},
       eprint = {2203.15649},
 primaryClass = {astro-ph.GA},
       adsurl = {https://ui.adsabs.harvard.edu/abs/2022ApJ...930..126F}
}

@ARTICLE{ford_et_al2001,
       author = {{Ford}, E.~B. and {Seager}, S. and {Turner}, E.~L.},
        title = "{Characterization of extrasolar terrestrial planets from diurnal photometric variability}",
      journal = {\nat},
         year = 2001,
        month = aug,
       volume = {412},
       number = {6850},
        pages = {885-887},
          doi = {10.1038/35091009},
archivePrefix = {arXiv},
       eprint = {astro-ph/0109054},
 primaryClass = {astro-ph},
       adsurl = {https://ui.adsabs.harvard.edu/abs/2001Natur.412..885F}
}

@ARTICLE{fossati_et_al2010,
       author = {{Fossati}, L. and {Haswell}, C.~A. and {Froning}, C.~S. and {Hebb}, L. and {Holmes}, S. and {Kolb}, U. and {Helling}, Ch. and {Carter}, A. and {Wheatley}, P. and {Collier Cameron}, A. and {Loeillet}, B. and {Pollacco}, D. and {Street}, R. and {Stempels}, H.~C. and {Simpson}, E. and {Udry}, S. and {Joshi}, Y.~C. and {West}, R.~G. and {Skillen}, I. and {Wilson}, D.},
        title = "{Metals in the Exosphere of the Highly Irradiated Planet WASP-12b}",
      journal = {\apjl},
         year = 2010,
        month = may,
       volume = {714},
       number = {2},
        pages = {L222-L227},
          doi = {10.1088/2041-8205/714/2/L222},
archivePrefix = {arXiv},
       eprint = {1005.3656},
 primaryClass = {astro-ph.SR},
       adsurl = {https://ui.adsabs.harvard.edu/abs/2010ApJ...714L.222F}
}

@INPROCEEDINGS{fossati_et_al2015,
       author = {{Fossati}, Luca and {Haswell}, Carole A. and {Linsky}, Jeffrey L. and {Kislyakova}, Kristina G.},
        title = "{Observations of Exoplanet Atmospheres and Surrounding Environments}",
    booktitle = {Characterizing Stellar and Exoplanetary Environments},
         year = 2015,
       editor = {{Lammer}, Helmut and {Khodachenko}, Maxim},
       series = {Astrophysics and Space Science Library},
       volume = {411},
        month = jan,
        pages = {59},
          doi = {10.1007/978-3-319-09749-7_4},
       adsurl = {https://ui.adsabs.harvard.edu/abs/2015ASSL..411...59F}
}

@ARTICLE{flury_et_al2024,
       author = {{Flury}, Sophia R. and {Jaskot}, Anne E. and {Saldana-Lopez}, Alberto and {Oey}, M.~S. and {Chisholm}, John and {Amor{\'\i}n}, Ricardo and {Bait}, Omkar and {Borthakur}, Sanchayeeta and {Carr}, Cody and {Ferguson}, Henry C. and {Giavalisco}, Mauro and {Hayes}, Matthew and {Heckman}, Timothy and {Henry}, Alaina and {Ji}, Zhiyuan and {Komarova}, Lena and {Leclercq}, Floriane and {Le Reste}, Alexandra and {McCandliss}, Stephan and {Marques-Chaves}, Rui and {{\"O}stlin}, G{\"o}ran and {Pentericci}, Laura and {Ravindranath}, Swara and {Rutkowski}, Michael and {Scarlata}, Claudia and {Schaerer}, Daniel and {Thuan}, Trinh and {Trebitsch}, Maxime and {Vanzella}, Eros and {Verhamme}, Anne and {Wang}, Bingjie and {Worseck}, G{\'a}bor and {Xu}, Xinfeng},
        title = "{The Low-Redshift Lyman Continuum Survey: The Roles of Stellar Feedback and ISM Geometry in LyC Escape}",
      journal = {arXiv e-prints},
         year = 2024,
        month = sep,
          eid = {arXiv:2409.12118},
        pages = {arXiv:2409.12118},
          doi = {10.48550/arXiv.2409.12118},
archivePrefix = {arXiv},
       eprint = {2409.12118},
 primaryClass = {astro-ph.GA},
       adsurl = {https://ui.adsabs.harvard.edu/abs/2024arXiv240912118F}
}

@ARTICLE{frebel2010,
       author = {{Frebel}, A.},
        title = "{Stellar archaeology: Exploring the Universe with metal-poor stars}",
      journal = {Astronomische Nachrichten},
         year = 2010,
        month = may,
       volume = {331},
       number = {5},
        pages = {474-488},
          doi = {10.1002/asna.201011362},
archivePrefix = {arXiv},
       eprint = {1006.2419},
 primaryClass = {astro-ph.GA},
       adsurl = {https://ui.adsabs.harvard.edu/abs/2010AN....331..474F}
}

@ARTICLE{freedman_et_al2025,
       author = {{Freedman}, Wendy L. and {Madore}, Barry F. and {Hoyt}, Taylor J. and {Jang}, In Sung and {Lee}, Abigail J. and {Owens}, Kayla A.},
        title = "{Status Report on the Chicago-Carnegie Hubble Program (CCHP): Measurement of the Hubble Constant Using the Hubble and James Webb Space Telescopes}",
      journal = {\apj},
         year = 2025,
        month = jun,
       volume = {985},
       number = {2},
          eid = {203},
        pages = {203},
          doi = {10.3847/1538-4357/adce78},
archivePrefix = {arXiv},
       eprint = {2408.06153},
 primaryClass = {astro-ph.CO},
       adsurl = {https://ui.adsabs.harvard.edu/abs/2025ApJ...985..203F}
}

@ARTICLE{fryer_et_al2001,
       author = {{Fryer}, C.~L. and {Woosley}, S.~E. and {Heger}, A.},
        title = "{Pair-Instability Supernovae, Gravity Waves, and Gamma-Ray Transients}",
      journal = {\apj},
         year = 2001,
        month = mar,
       volume = {550},
       number = {1},
        pages = {372-382},
          doi = {10.1086/319719},
archivePrefix = {arXiv},
       eprint = {astro-ph/0007176},
 primaryClass = {astro-ph},
       adsurl = {https://ui.adsabs.harvard.edu/abs/2001ApJ...550..372F}
}

@ARTICLE{fujii_et_al2018,
       author = {{Fujii}, Yuka and {Angerhausen}, Daniel and {Deitrick}, Russell and {Domagal-Goldman}, Shawn and {Grenfell}, John Lee and {Hori}, Yasunori and {Kane}, Stephen R. and {Pall{\'e}}, Enric and {Rauer}, Heike and {Siegler}, Nicholas and {Stapelfeldt}, Karl and {Stevenson}, Kevin B.},
        title = "{Exoplanet Biosignatures: Observational Prospects}",
      journal = {Astrobiology},
         year = 2018,
        month = jun,
       volume = {18},
       number = {6},
        pages = {739-778},
          doi = {10.1089/ast.2017.1733},
archivePrefix = {arXiv},
       eprint = {1705.07098},
 primaryClass = {astro-ph.EP},
       adsurl = {https://ui.adsabs.harvard.edu/abs/2018AsBio..18..739F}
}

@ARTICLE{fukumura_et_al2010,
       author = {{Fukumura}, Keigo and {Kazanas}, Demosthenes and {Contopoulos}, Ioannis and {Behar}, Ehud},
        title = "{Magnetohydrodynamic Accretion Disk Winds as X-ray Absorbers in Active Galactic Nuclei}",
      journal = {\apj},
         year = 2010,
        month = may,
       volume = {715},
       number = {1},
        pages = {636-650},
          doi = {10.1088/0004-637X/715/1/636},
archivePrefix = {arXiv},
       eprint = {0910.3001},
 primaryClass = {astro-ph.HE},
       adsurl = {https://ui.adsabs.harvard.edu/abs/2010ApJ...715..636F}
}

@ARTICLE{fukumura_et_al2014,
       author = {{Fukumura}, Keigo and {Tombesi}, Francesco and {Kazanas}, Demosthenes and {Shrader}, Chris and {Behar}, Ehud and {Contopoulos}, Ioannis},
        title = "{Stratified Magnetically Driven Accretion-disk Winds and Their Relations to Jets}",
      journal = {\apj},
         year = 2014,
        month = jan,
       volume = {780},
       number = {2},
          eid = {120},
        pages = {120},
          doi = {10.1088/0004-637X/780/2/120},
archivePrefix = {arXiv},
       eprint = {1311.0077},
 primaryClass = {astro-ph.HE},
       adsurl = {https://ui.adsabs.harvard.edu/abs/2014ApJ...780..120F}
}

@ARTICLE{fuller2017,
       author = {{Fuller}, Jim},
        title = "{Heartbeat stars, tidally excited oscillations and resonance locking}",
      journal = {\mnras},
         year = 2017,
        month = dec,
       volume = {472},
       number = {2},
        pages = {1538-1564},
          doi = {10.1093/mnras/stx2135},
archivePrefix = {arXiv},
       eprint = {1706.05054},
 primaryClass = {astro-ph.SR},
       adsurl = {https://ui.adsabs.harvard.edu/abs/2017MNRAS.472.1538F}
}

@ARTICLE{fuller+tsuna2024,
       author = {{Fuller}, Jim and {Tsuna}, Daichi},
        title = "{Boil-off of red supergiants: mass loss and type II-P supernovae}",
      journal = {The Open Journal of Astrophysics},
         year = 2024,
        month = jun,
       volume = {7},
          eid = {47},
        pages = {47},
          doi = {10.33232/001c.120130},
archivePrefix = {arXiv},
       eprint = {2405.21049},
 primaryClass = {astro-ph.SR},
       adsurl = {https://ui.adsabs.harvard.edu/abs/2024OJAp....7E..47F}
}

@ARTICLE{fulton_et_al2017,
       author = {{Fulton}, Benjamin J. and {Petigura}, Erik A. and {Howard}, Andrew W. and {Isaacson}, Howard and {Marcy}, Geoffrey W. and {Cargile}, Phillip A. and {Hebb}, Leslie and {Weiss}, Lauren M. and {Johnson}, John Asher and {Morton}, Timothy D. and {Sinukoff}, Evan and {Crossfield}, Ian J.~M. and {Hirsch}, Lea A.},
        title = "{The California-Kepler Survey. III. A Gap in the Radius Distribution of Small Planets}",
      journal = {\aj},
         year = 2017,
        month = sep,
       volume = {154},
       number = {3},
          eid = {109},
        pages = {109},
          doi = {10.3847/1538-3881/aa80eb},
archivePrefix = {arXiv},
       eprint = {1703.10375},
 primaryClass = {astro-ph.EP},
       adsurl = {https://ui.adsabs.harvard.edu/abs/2017AJ....154..109F}
}

@ARTICLE{furano_et_al2020,
       author = {{Furano}, Gianluca and {Meoni}, Gabriele and {Dunne}, Aubrey and {Moloney}, David and {Ferlet-Cavrois}, Veronique and {Tavoularis}, Antonis and {Byrne}, Jonathan and {Buckley}, Leonie and {Psarakis}, Mihalis and {Voss}, Kay-Obbe and {Fanucci}, Luca},
        title = "{Towards the Use of Artificial Intelligence on the Edge in Space Systems: Challenges and Opportunities}",
      journal = {IEEE Aerospace and Electronic Systems Magazine},
         year = 2020,
        month = dec,
       volume = {35},
       number = {12},
        pages = {44-56},
          doi = {10.1109/MAES.2020.3008468},
       adsurl = {https://ui.adsabs.harvard.edu/abs/2020IAESM..35l..44F}
}

@ARTICLE{gaillard+scaillet2014,
       author = {{Gaillard}, Fabrice and {Scaillet}, Bruno},
        title = "{A theoretical framework for volcanic degassing chemistry in a comparative planetology perspective and implications for planetary atmospheres}",
      journal = {Earth and Planetary Science Letters},
         year = 2014,
        month = oct,
       volume = {403},
        pages = {307-316},
          doi = {10.1016/j.epsl.2014.07.009},
       adsurl = {https://ui.adsabs.harvard.edu/abs/2014E&PSL.403..307G}
}

@ARTICLE{gaudi_et_al2020,
       author = {{Gaudi}, B. Scott and {Seager}, Sara and {Mennesson}, Bertrand and {Kiessling}, Alina and {Warfield}, Keith and {Cahoy}, Kerri and {Clarke}, John T. and {Domagal-Goldman}, Shawn and {Feinberg}, Lee and {Guyon}, Olivier and {Kasdin}, Jeremy and {Mawet}, Dimitri and {Plavchan}, Peter and {Robinson}, Tyler and {Rogers}, Leslie and {Scowen}, Paul and {Somerville}, Rachel and {Stapelfeldt}, Karl and {Stark}, Christopher and {Stern}, Daniel and {Turnbull}, Margaret and {Amini}, Rashied and {Kuan}, Gary and {Martin}, Stefan and {Morgan}, Rhonda and {Redding}, David and {Stahl}, H. Philip and {Webb}, Ryan and {Alvarez-Salazar}, Oscar and {Arnold}, William L. and {Arya}, Manan and {Balasubramanian}, Bala and {Baysinger}, Mike and {Bell}, Ray and {Below}, Chris and {Benson}, Jonathan and {Blais}, Lindsey and {Booth}, Jeff and {Bourgeois}, Robert and {Bradford}, Case and {Brewer}, Alden and {Brooks}, Thomas and {Cady}, Eric and {Caldwell}, Mary and {Calvet}, Rob and {Carr}, Steven and {Chan}, Derek and {Cormarkovic}, Velibor and {Coste}, Keith and {Cox}, Charlie and {Danner}, Rolf and {Davis}, Jacqueline and {Dewell}, Larry and {Dorsett}, Lisa and {Dunn}, Daniel and {East}, Matthew and {Effinger}, Michael and {Eng}, Ron and {Freebury}, Greg and {Garcia}, Jay and {Gaskin}, Jonathan and {Greene}, Suzan and {Hennessy}, John and {Hilgemann}, Evan and {Hood}, Brad and {Holota}, Wolfgang and {Howe}, Scott and {Huang}, Pei and {Hull}, Tony and {Hunt}, Ron and {Hurd}, Kevin and {Johnson}, Sandra and {Kissil}, Andrew and {Knight}, Brent and {Kolenz}, Daniel and {Kraus}, Oliver and {Krist}, John and {Li}, Mary and {Lisman}, Doug and {Mandic}, Milan and {Mann}, John and {Marchen}, Luis and {Marrese-Reading}, Colleen and {McCready}, Jonathan and {McGown}, Jim and {Missun}, Jessica and {Miyaguchi}, Andrew and {Moore}, Bradley and {Nemati}, Bijan and {Nikzad}, Shouleh and {Nissen}, Joel and {Novicki}, Megan and {Perrine}, Todd and {Pineda}, Claudia and {Polanco}, Otto and {Putnam}, Dustin and {Qureshi}, Atif and {Richards}, Michael and {Eldorado Riggs}, A.~J. and {Rodgers}, Michael and {Rud}, Mike and {Saini}, Navtej and {Scalisi}, Dan and {Scharf}, Dan and {Schulz}, Kevin and {Serabyn}, Gene and {Sigrist}, Norbert and {Sikkia}, Glory and {Singleton}, Andrew and {Shaklan}, Stuart and {Smith}, Scott and {Southerd}, Bart and {Stahl}, Mark and {Steeves}, John and {Sturges}, Brian and {Sullivan}, Chris and {Tang}, Hao and {Taras}, Neil and {Tesch}, Jonathan and {Therrell}, Melissa and {Tseng}, Howard and {Valente}, Marty and {Van Buren}, David and {Villalvazo}, Juan and {Warwick}, Steve and {Webb}, David and {Westerhoff}, Thomas and {Wofford}, Rush and {Wu}, Gordon and {Woo}, Jahning and {Wood}, Milana and {Ziemer}, John and {Arney}, Giada and {Anderson}, Jay and {Ma{\'\i}z-Apell{\'a}niz}, Jes{\'u}s and {Bartlett}, James and {Belikov}, Ruslan and {Bendek}, Eduardo and {Cenko}, Brad and {Douglas}, Ewan and {Dulz}, Shannon and {Evans}, Chris and {Faramaz}, Virginie and {Feng}, Y. Katherina and {Ferguson}, Harry and {Follette}, Kate and {Ford}, Saavik and {Garc{\'\i}a}, Miriam and {Geha}, Marla and {Gelino}, Dawn and {G{\"o}tberg}, Ylva and {Hildebrandt}, Sergi and {Hu}, Renyu and {Jahnke}, Knud and {Kennedy}, Grant and {Kreidberg}, Laura and {Isella}, Andrea and {Lopez}, Eric and {Marchis}, Franck and {Macri}, Lucas and {Marley}, Mark and {Matzko}, William and {Mazoyer}, Johan and {McCandliss}, Stephan and {Meshkat}, Tiffany and {Mordasini}, Christoph and {Morris}, Patrick and {Nielsen}, Eric and {Newman}, Patrick and {Petigura}, Erik and {Postman}, Marc and {Reines}, Amy and {Roberge}, Aki and {Roederer}, Ian and {Ruane}, Garreth and {Schwieterman}, Edouard and {Sirbu}, Dan and {Spalding}, Christopher and {Teplitz}, Harry and {Tumlinson}, Jason and {Turner}, Neal and {Werk}, Jessica and {Wofford}, Aida and {Wyatt}, Mark and {Young}, Amber and {Zellem}, Rob},
        title = "{The Habitable Exoplanet Observatory (HabEx) Mission Concept Study Final Report}",
      journal = {arXiv e-prints},
         year = 2020,
        month = jan,
          eid = {arXiv:2001.06683},
        pages = {arXiv:2001.06683},
          doi = {10.48550/arXiv.2001.06683},
archivePrefix = {arXiv},
       eprint = {2001.06683},
 primaryClass = {astro-ph.IM},
       adsurl = {https://ui.adsabs.harvard.edu/abs/2020arXiv200106683G}
}

@ARTICLE{gazagnes_et_al2020,
       author = {{Gazagnes}, S. and {Chisholm}, J. and {Schaerer}, D. and {Verhamme}, A. and {Izotov}, Y.},
        title = "{The origin of the escape of Lyman {\ensuremath{\alpha}} and ionizing photons in Lyman continuum emitters}",
      journal = {\aap},
         year = 2020,
        month = jul,
       volume = {639},
          eid = {A85},
        pages = {A85},
          doi = {10.1051/0004-6361/202038096},
archivePrefix = {arXiv},
       eprint = {2005.07215},
 primaryClass = {astro-ph.GA},
       adsurl = {https://ui.adsabs.harvard.edu/abs/2020A&A...639A..85G}
}

@ARTICLE{garcia_et_al2018,
       author = {{Garcia}, Miriam},
        title = "{Massive stars in the Sagittarius Dwarf Irregular Galaxy}",
      journal = {\mnras},
         year = 2018,
        month = feb,
       volume = {474},
       number = {1},
        pages = {L66-L70},
          doi = {10.1093/mnrasl/slx194},
archivePrefix = {arXiv},
       eprint = {1711.11299},
 primaryClass = {astro-ph.SR},
       adsurl = {https://ui.adsabs.harvard.edu/abs/2018MNRAS.474L..66G}
}

@ARTICLE{garcia_et_al2021,
       author = {{Garcia}, Miriam and {Evans}, Christopher J. and {Bestenlehner}, Joachim M. and {Bouret}, Jean Claude and {Castro}, Norberto and {Cervi{\~n}o}, Miguel and {Fullerton}, Alexander W. and {Gieles}, Mark and {Herrero}, Artemio and {de Koter}, Alexander and {Lennon}, Daniel J. and {van Loon}, Jacco Th. and {Martins}, Fabrice and {de Mink}, Selma E. and {Najarro}, Francisco and {Negueruela}, Ignacio and {Sana}, Hugues and {Sim{\'o}n-D{\'\i}az}, Sergio and {Sz{\'e}csi}, Dorottya and {Tramper}, Frank and {Vink}, Jorick S. and {Wofford}, Aida},
        title = "{Massive stars in extremely metal-poor galaxies: a window into the past}",
      journal = {Experimental Astronomy},
         year = 2021,
        month = jun,
       volume = {51},
       number = {3},
        pages = {887-911},
          doi = {10.1007/s10686-021-09785-x},
       adsurl = {https://ui.adsabs.harvard.edu/abs/2021ExA....51..887G}
}

@ARTICLE{garcia-burillo_et_al2021,
       author = {{Garc{\'\i}a-Burillo}, S. and {Alonso-Herrero}, A. and {Ramos Almeida}, C. and {Gonz{\'a}lez-Mart{\'\i}n}, O. and {Combes}, F. and {Usero}, A. and {H{\"o}nig}, S. and {Querejeta}, M. and {Hicks}, E.~K.~S. and {Hunt}, L.~K. and {Rosario}, D. and {Davies}, R. and {Boorman}, P.~G. and {Bunker}, A.~J. and {Burtscher}, L. and {Colina}, L. and {D{\'\i}az-Santos}, T. and {Gandhi}, P. and {Garc{\'\i}a-Bernete}, I. and {Garc{\'\i}a-Lorenzo}, B. and {Ichikawa}, K. and {Imanishi}, M. and {Izumi}, T. and {Labiano}, A. and {Levenson}, N.~A. and {L{\'o}pez-Rodr{\'\i}guez}, E. and {Packham}, C. and {Pereira-Santaella}, M. and {Ricci}, C. and {Rigopoulou}, D. and {Rouan}, D. and {Shimizu}, T. and {Stalevski}, M. and {Wada}, K. and {Williamson}, D.},
        title = "{The Galaxy Activity, Torus, and Outflow Survey (GATOS). I. ALMA images of dusty molecular tori in Seyfert galaxies}",
      journal = {\aap},
         year = 2021,
        month = aug,
       volume = {652},
          eid = {A98},
        pages = {A98},
          doi = {10.1051/0004-6361/202141075},
archivePrefix = {arXiv},
       eprint = {2104.10227},
 primaryClass = {astro-ph.GA},
       adsurl = {https://ui.adsabs.harvard.edu/abs/2021A&A...652A..98G}
}

@ARTICLE{garcia_munoz2018,
       author = {{Garc{\'\i}a Mu{\~n}oz}, A.},
        title = "{On Mapping Exoplanet Atmospheres with High-dispersion Spectro-polarimetry: Some Model Predictions}",
      journal = {\apj},
         year = 2018,
        month = feb,
       volume = {854},
       number = {2},
          eid = {108},
        pages = {108},
          doi = {10.3847/1538-4357/aaaa1f},
archivePrefix = {arXiv},
       eprint = {1802.01024},
 primaryClass = {astro-ph.EP},
       adsurl = {https://ui.adsabs.harvard.edu/abs/2018ApJ...854..108G}
}

@ARTICLE{gardner_et_al2006,
       author = {{Gardner}, Jonathan P. and {Mather}, John C. and {Clampin}, Mark and {Doyon}, Rene and {Greenhouse}, Matthew A. and {Hammel}, Heidi B. and {Hutchings}, John B. and {Jakobsen}, Peter and {Lilly}, Simon J. and {Long}, Knox S. and {Lunine}, Jonathan I. and {McCaughrean}, Mark J. and {Mountain}, Matt and {Nella}, John and {Rieke}, George H. and {Rieke}, Marcia J. and {Rix}, Hans-Walter and {Smith}, Eric P. and {Sonneborn}, George and {Stiavelli}, Massimo and {Stockman}, H.~S. and {Windhorst}, Rogier A. and {Wright}, Gillian S.},
        title = "{The James Webb Space Telescope}",
      journal = {\ssr},
         year = 2006,
        month = apr,
       volume = {123},
       number = {4},
        pages = {485-606},
          doi = {10.1007/s11214-006-8315-7},
archivePrefix = {arXiv},
       eprint = {astro-ph/0606175},
 primaryClass = {astro-ph},
       adsurl = {https://ui.adsabs.harvard.edu/abs/2006SSRv..123..485G}
}

@ARTICLE{garnett_et_al1995,
       author = {{Garnett}, D.~R. and {Dufour}, R.~J. and {Peimbert}, M. and {Torres-Peimbert}, S. and {Shields}, G.~A. and {Skillman}, E.~D. and {Terlevich}, E. and {Terlevich}, R.~J.},
        title = "{Si/O Abundance Ratios in Extragalactic H II Regions from Hubble Space Telescope UV Spectroscopy}",
      journal = {\apjl},
         year = 1995,
        month = aug,
       volume = {449},
        pages = {L77},
          doi = {10.1086/309620},
       adsurl = {https://ui.adsabs.harvard.edu/abs/1995ApJ...449L..77G}
}

@ARTICLE{garreau_et_al2025,
       author = {{Garreau}, G. and {Defr{\`e}re}, D. and {Ertel}, S. and {Faramaz-Gorka}, V. and {Bryden}, G. and {Sommer}, M. and {Mesa}, D. and {Wagner}, K. and {De Prins}, T. and {Laugier}, R. and {Weinberger}, A. and {Farinato}, J. and {Haniff}, C. and {Hinz}, P.~M. and {Isbell}, J.~W. and {Kennedy}, G.~M. and {Lorenzetto}, A. and {Maier}, E.~R. and {Marafatto}, L. and {Marino}, S. and {Martinod}, M.~A. and {Mennesson}, B. and {Rousseau}, H. and {Spalding}, E. and {Vassallo}, D. and {Wyatt}, M.~C.},
        title = "{The HOSTS survey: Suspected variable dust emission and constraints on companions around {\ensuremath{\theta}} Boo}",
      journal = {\aap},
         year = 2025,
        month = jul,
       volume = {699},
          eid = {A107},
        pages = {A107},
          doi = {10.1051/0004-6361/202452653},
archivePrefix = {arXiv},
       eprint = {2505.07585},
 primaryClass = {astro-ph.EP},
       adsurl = {https://ui.adsabs.harvard.edu/abs/2025A&A...699A.107G}
}

@ARTICLE{garvin2022,
       author = {{Garvin}, James B. and {Getty}, Stephanie A. and {Arney}, Giada N. and {Johnson}, Natasha M. and {Kohler}, Erika and {Schwer}, Kenneth O. and {Sekerak}, Michael and {Bartels}, Arlin and {Saylor}, Richard S. and {Elliott}, Vincent E. and {Goodloe}, Colby S. and {Garrison}, Matthew B. and {Cottini}, Valeria and {Izenberg}, Noam and {Lorenz}, Ralph and {Malespin}, Charles A. and {Ravine}, Michael and {Webster}, Christopher R. and {Atkinson}, David H. and {Aslam}, Shahid and {Atreya}, Sushil and {Bos}, Brent J. and {Brinckerhoff}, William B. and {Campbell}, Bruce and {Crisp}, David and {Filiberto}, Justin R. and {Forget}, Francois and {Gilmore}, Martha and {Gorius}, Nicolas and {Grinspoon}, David and {Hofmann}, Amy E. and {Kane}, Stephen R. and {Kiefer}, Walter and {Lebonnois}, Sebastien and {Mahaffy}, Paul R. and {Pavlov}, Alexander and {Trainer}, Melissa and {Zahnle}, Kevin J. and {Zolotov}, Mikhail},
        title = "{Revealing the Mysteries of Venus: The DAVINCI Mission}",
      journal = {\psj},
         year = 2022,
        month = may,
       volume = {3},
       number = {5},
          eid = {117},
        pages = {117},
          doi = {10.3847/PSJ/ac63c2},
archivePrefix = {arXiv},
       eprint = {2206.07211},
 primaryClass = {astro-ph.EP},
       adsurl = {https://ui.adsabs.harvard.edu/abs/2022PSJ.....3..117G}
}

@ARTICLE{gebek+oza2020,
       author = {{Gebek}, Andrea and {Oza}, Apurva V.},
        title = "{Alkaline exospheres of exoplanet systems: evaporative transmission spectra}",
      journal = {\mnras},
         year = 2020,
        month = oct,
       volume = {497},
       number = {4},
        pages = {5271-5291},
          doi = {10.1093/mnras/staa2193},
archivePrefix = {arXiv},
       eprint = {2005.02536},
 primaryClass = {astro-ph.EP},
       adsurl = {https://ui.adsabs.harvard.edu/abs/2020MNRAS.497.5271G}
}

@INPROCEEDINGS{girardot_et_al2024,
       author = {{Girardot}, Adrien and {Neiner}, Coralie and {Reess}, Jean-Michel},
        title = "{Design of a FUV polarimeter for Pollux aboard HWO}",
    booktitle = {Space Telescopes and Instrumentation 2024: Ultraviolet to Gamma Ray},
         year = 2024,
       editor = {{den Herder}, Jan-Willem A. and {Nikzad}, Shouleh and {Nakazawa}, Kazuhiro},
       series = {Society of Photo-Optical Instrumentation Engineers (SPIE) Conference Series},
       volume = {13093},
        month = aug,
          eid = {130933V},
        pages = {130933V},
          doi = {10.1117/12.3017994},
       adsurl = {https://ui.adsabs.harvard.edu/abs/2024SPIE13093E..3VG}
}

@ARTICLE{giuffrida_et_al2022,
       author = {{Giuffrida}, Gianluca and {Fanucci}, Luca and {Meoni}, Gabriele and {Bati{\v{c}}}, Matej and {Buckley}, L{\'e}onie and {Dunne}, Aubrey and {van Dijk}, Chris and {Esposito}, Marco and {Hefele}, John and {Vercruyssen}, Nathan and {Furano}, Gianluca and {Pastena}, Massimiliano and {Aschbacher}, Josef},
        title = "{The {\ensuremath{\Phi}}-Sat-1 Mission: The First On-Board Deep Neural Network Demonstrator for Satellite Earth Observation}",
      journal = {IEEE Transactions on Geoscience and Remote Sensing},
         year = 2022,
        month = jan,
       volume = {60},
          eid = {TGRS.2021},
        pages = {TGRS.2021},
          doi = {10.1109/TGRS.2021.3125567},
       adsurl = {https://ui.adsabs.harvard.edu/abs/2022ITGRS..6025567G}
}

@ARTICLE{gohin_et_al2002,
       author = {{Gohin}, F. and {Druon}, J.~N. and {Lampert}, L.},
        title = "{A five channel chlorophyll concentration algorithm applied to SeaWiFS data processed by SeaDAS in coastal waters}",
      journal = {International Journal of Remote Sensing},
         year = 2002,
        month = jan,
       volume = {23},
       number = {8},
        pages = {1639-1661},
          doi = {10.1080/01431160110071879},
       adsurl = {https://ui.adsabs.harvard.edu/abs/2002IJRS...23.1639G}
}

@ARTICLE{gomez-gonzalez_et_al2021,
       author = {{G{\'o}mez-Gonz{\'a}lez}, V.~M.~A. and {Mayya}, Y.~D. and {Toal{\'a}}, J.~A. and {Arthur}, S.~J. and {Zaragoza-Cardiel}, J. and {Guerrero}, M.~A.},
        title = "{Wolf-Rayet stars in the Antennae unveiled by MUSE}",
      journal = {\mnras},
         year = 2021,
        month = jan,
       volume = {500},
       number = {2},
        pages = {2076-2095},
          doi = {10.1093/mnras/staa3304},
archivePrefix = {arXiv},
       eprint = {2010.09781},
 primaryClass = {astro-ph.GA},
       adsurl = {https://ui.adsabs.harvard.edu/abs/2021MNRAS.500.2076G}
}

@ARTICLE{goodis_gordon_et_al2025,
       author = {{Goodis Gordon}, Kenneth E. and {Karalidi}, Theodora and {Bott}, Kimberly M. and {Wogan}, Nicholas F. and {Arney}, Giada N. and {Parenteau}, Mary N. and {Kataria}, Tiffany and {Meadows}, Victoria S.},
        title = "{Polarized Signatures of the Earth Through Time: An Outlook for the Habitable Worlds Observatory}",
      journal = {\apj},
         year = 2025,
        month = apr,
       volume = {983},
       number = {2},
          eid = {168},
        pages = {168},
          doi = {10.3847/1538-4357/adc09c},
archivePrefix = {arXiv},
       eprint = {2410.02194},
 primaryClass = {astro-ph.EP},
       adsurl = {https://ui.adsabs.harvard.edu/abs/2025ApJ...983..168G}
}

@ARTICLE{gordon_et_al2003,
       author = {{Gordon}, Karl D. and {Clayton}, Geoffrey C. and {Misselt}, K.~A. and {Landolt}, Arlo U. and {Wolff}, Michael J.},
        title = "{A Quantitative Comparison of the Small Magellanic Cloud, Large Magellanic Cloud, and Milky Way Ultraviolet to Near-Infrared Extinction Curves}",
      journal = {\apj},
         year = 2003,
        month = sep,
       volume = {594},
       number = {1},
        pages = {279-293},
          doi = {10.1086/376774},
archivePrefix = {arXiv},
       eprint = {astro-ph/0305257},
 primaryClass = {astro-ph},
       adsurl = {https://ui.adsabs.harvard.edu/abs/2003ApJ...594..279G}
}

@ARTICLE{gordon_et_al2009,
       author = {{Gordon}, Karl D. and {Cartledge}, Stefan and {Clayton}, Geoffrey C.},
        title = "{FUSE Measurements of Far-Ultraviolet Extinction. III. The Dependence on R(V) and Discrete Feature Limits from 75 Galactic Sightlines}",
      journal = {\apj},
         year = 2009,
        month = nov,
       volume = {705},
       number = {2},
        pages = {1320-1335},
          doi = {10.1088/0004-637X/705/2/1320},
archivePrefix = {arXiv},
       eprint = {0909.3087},
 primaryClass = {astro-ph.GA},
       adsurl = {https://ui.adsabs.harvard.edu/abs/2009ApJ...705.1320G}
}

@ARTICLE{gordon_et_al2024,
       author = {{Gordon}, Karl D. and {Fitzpatrick}, E.~L. and {Massa}, Derck and {Bohlin}, Ralph and {Chastenet}, J{\'e}r{\'e}my and {Murray}, Claire E. and {Clayton}, Geoffrey C. and {Lennon}, Daniel J. and {Misselt}, Karl A. and {Sandstrom}, Karin},
        title = "{Expanded Sample of Small Magellanic Cloud Ultraviolet Dust Extinction Curves: Correlations between the 2175 {\r{A}} Bump, q $_{PAH}$, Ultraviolet Extinction Shape, and N(H I)/A(V)}",
      journal = {\apj},
         year = 2024,
        month = jul,
       volume = {970},
       number = {1},
          eid = {51},
        pages = {51},
          doi = {10.3847/1538-4357/ad4be1},
archivePrefix = {arXiv},
       eprint = {2405.12792},
 primaryClass = {astro-ph.GA},
       adsurl = {https://ui.adsabs.harvard.edu/abs/2024ApJ...970...51G}
}

@ARTICLE{gotberg_et_al2017,
       author = {{G{\"o}tberg}, Y. and {de Mink}, S.~E. and {Groh}, J.~H.},
        title = "{Ionizing spectra of stars that lose their envelope through interaction with a binary companion: role of metallicity}",
      journal = {\aap},
         year = 2017,
        month = nov,
       volume = {608},
          eid = {A11},
        pages = {A11},
          doi = {10.1051/0004-6361/201730472},
archivePrefix = {arXiv},
       eprint = {1701.07439},
 primaryClass = {astro-ph.SR},
       adsurl = {https://ui.adsabs.harvard.edu/abs/2017A&A...608A..11G}
}

@ARTICLE{gotberg_et_al2019,
       author = {{G{\"o}tberg}, Y. and {de Mink}, S.~E. and {Groh}, J.~H. and {Leitherer}, C. and {Norman}, C.},
        title = "{The impact of stars stripped in binaries on the integrated spectra of stellar populations}",
      journal = {\aap},
         year = 2019,
        month = sep,
       volume = {629},
          eid = {A134},
        pages = {A134},
          doi = {10.1051/0004-6361/201834525},
archivePrefix = {arXiv},
       eprint = {1908.06102},
 primaryClass = {astro-ph.GA},
       adsurl = {https://ui.adsabs.harvard.edu/abs/2019A&A...629A.134G}
}

@ARTICLE{gottweis_et_al2025,
       author = {{Gottweis}, Juraj and {Weng}, Wei-Hung and {Daryin}, Alexander and {Tu}, Tao and {Sirkovic}, Petar and {Myaskovsky}, Artiom and {Glowaty}, Grzegorz and {Weissenberger}, Felix and {Orlandi}, Alessio and {Popovici}, Dan and {Palepu}, Anil and {Rong}, Keran and {Tanno}, Ryutaro and {Saab}, Khaled and {Zhang}, Fan and {Blum}, Jacob and {Carroll}, Andrew and {Kulkarni}, Kavita and {Tomasev}, Nenad and {Zverinski}, Dina and {Rendulic}, Ivor and {Vedadi}, Elahe and {Hasler}, Florian and {Rimanic}, Luka and {Boia}, Marina and {Budiselic}, Ivan and {Feinstein}, Ben and {Bellaiche}, Mathias and {Sheffer}, Tom and {Freyberg}, Jan and {Ratcliff}, Jeremy and {Bertolli}, Ottavia and {Chou}, Katherine and {Hassidim}, Avinatan and {Gokturk}, Burak and {Vahdat}, Amin and {Guan}, Yuan and {Dhillon}, Vikram and {Dhaval Vaishnav}, Eeshit and {Lee}, Byron and {Costa}, Tiago R D and {Penad{\'e}s}, Jos{\'e} R and {Peltz}, Gary and {Matias}, Yossi and {Manyika}, James and {Hassabis}, Demis and {Xu}, Yunhan and {Kohli}, Pushmeet and {Pawlosky}, Annalisa and {Karthikesalingam}, Alan and {Natarajan}, Vivek},
        title = "{Accelerating scientific discovery with Co-Scientist}",
      journal = {arXiv e-prints},
         year = 2025,
        month = feb,
          eid = {arXiv:2502.18864},
        pages = {arXiv:2502.18864},
          doi = {10.48550/arXiv.2502.18864},
archivePrefix = {arXiv},
       eprint = {2502.18864},
 primaryClass = {cs.AI},
       adsurl = {https://ui.adsabs.harvard.edu/abs/2025arXiv250218864G}
}

@ARTICLE{grafener+hamann2008,
       author = {{Gr{\"a}fener}, G. and {Hamann}, W. -R.},
        title = "{Mass loss from late-type WN stars and its Z-dependence. Very massive stars approaching the Eddington limit}",
      journal = {\aap},
         year = 2008,
        month = may,
       volume = {482},
       number = {3},
        pages = {945-960},
          doi = {10.1051/0004-6361:20066176},
archivePrefix = {arXiv},
       eprint = {0803.0866},
 primaryClass = {astro-ph},
       adsurl = {https://ui.adsabs.harvard.edu/abs/2008A&A...482..945G}
}

@ARTICLE{grafener_et_al2021,
       author = {{Gr{\"a}fener}, G{\"o}tz},
        title = "{Physics and evolution of the most massive stars in 30 Doradus. Mass loss, envelope inflation, and a variable upper stellar mass limit}",
      journal = {\aap},
         year = 2021,
        month = mar,
       volume = {647},
          eid = {A13},
        pages = {A13},
          doi = {10.1051/0004-6361/202040037},
archivePrefix = {arXiv},
       eprint = {2101.03837},
 primaryClass = {astro-ph.SR},
       adsurl = {https://ui.adsabs.harvard.edu/abs/2021A&A...647A..13G}
}

@ARTICLE{greene_et_al2020,
       author = {{Greene}, Jenny E. and {Strader}, Jay and {Ho}, Luis C.},
        title = "{Intermediate-Mass Black Holes}",
      journal = {\araa},
         year = 2020,
        month = aug,
       volume = {58},
        pages = {257-312},
          doi = {10.1146/annurev-astro-032620-021835},
archivePrefix = {arXiv},
       eprint = {1911.09678},
 primaryClass = {astro-ph.GA},
       adsurl = {https://ui.adsabs.harvard.edu/abs/2020ARA&A..58..257G}
}

@ARTICLE{greene2023,
       author = {{Greene}, Thomas P. and {Bell}, Taylor J. and {Ducrot}, Elsa and {Dyrek}, Achr{\`e}ne and {Lagage}, Pierre-Olivier and {Fortney}, Jonathan J.},
        title = "{Thermal emission from the Earth-sized exoplanet TRAPPIST-1 b using JWST}",
      journal = {\nat},
         year = 2023,
        month = jun,
       volume = {618},
       number = {7963},
        pages = {39-42},
          doi = {10.1038/s41586-023-05951-7},
archivePrefix = {arXiv},
       eprint = {2303.14849},
 primaryClass = {astro-ph.EP},
       adsurl = {https://ui.adsabs.harvard.edu/abs/2023Natur.618...39G}
}

@ARTICLE{guilluy_et_al2020,
       author = {{Guilluy}, G. and {Andretta}, V. and {Borsa}, F. and {Giacobbe}, P. and {Sozzetti}, A. and {Covino}, E. and {Bourrier}, V. and {Fossati}, L. and {Bonomo}, A.~S. and {Esposito}, M. and {Giampapa}, M.~S. and {Harutyunyan}, A. and {Rainer}, M. and {Brogi}, M. and {Bruno}, G. and {Claudi}, R. and {Frustagli}, G. and {Lanza}, A.~F. and {Mancini}, L. and {Pino}, L. and {Poretti}, E. and {Scandariato}, G. and {Affer}, L. and {Baffa}, C. and {Baruffolo}, A. and {Benatti}, S. and {Biazzo}, K. and {Bignamini}, A. and {Boschin}, W. and {Carleo}, I. and {Cecconi}, M. and {Cosentino}, R. and {Damasso}, M. and {Desidera}, S. and {Falcini}, G. and {Martinez Fiorenzano}, A.~F. and {Ghedina}, A. and {Gonz{\'a}lez-{\'A}lvarez}, E. and {Guerra}, J. and {Hernandez}, N. and {Leto}, G. and {Maggio}, A. and {Malavolta}, L. and {Maldonado}, J. and {Micela}, G. and {Molinari}, E. and {Nascimbeni}, V. and {Pagano}, I. and {Pedani}, M. and {Piotto}, G. and {Reiners}, A.},
        title = "{The GAPS programme at TNG. XXII. The GIARPS view of the extended helium atmosphere of HD 189733 b accounting for stellar activity}",
      journal = {\aap},
         year = 2020,
        month = jul,
       volume = {639},
          eid = {A49},
        pages = {A49},
          doi = {10.1051/0004-6361/202037644},
archivePrefix = {arXiv},
       eprint = {2005.05676},
 primaryClass = {astro-ph.EP},
       adsurl = {https://ui.adsabs.harvard.edu/abs/2020A&A...639A..49G}
}

@ARTICLE{gull_et_al2022,
       author = {{Gull}, Maude and {Weisz}, Daniel R. and {Senchyna}, Peter and {Sandford}, Nathan R. and {Choi}, Yumi and {McLeod}, Anna F. and {El-Badry}, Kareem and {G{\"o}tberg}, Ylva and {Gilbert}, Karoline M. and {Boyer}, Martha and {Dalcanton}, Julianne J. and {GuhaThakurta}, Puragra and {Goldman}, Steven and {Marigo}, Paola and {McQuinn}, Kristen B.~W. and {Pastorelli}, Giada and {Stark}, Daniel P. and {Skillman}, Evan and {Ting}, Yuan-sen and {Williams}, Benjamin F.},
        title = "{A Panchromatic Study of Massive Stars in the Extremely Metal-poor Local Group Dwarf Galaxy Leo A}",
      journal = {\apj},
         year = 2022,
        month = dec,
       volume = {941},
       number = {2},
          eid = {206},
        pages = {206},
          doi = {10.3847/1538-4357/aca295},
archivePrefix = {arXiv},
       eprint = {2211.14349},
 primaryClass = {astro-ph.SR},
       adsurl = {https://ui.adsabs.harvard.edu/abs/2022ApJ...941..206G}
}

@INPROCEEDINGS{guyon_et_al2012,
       author = {{Guyon}, Olivier and {Martinache}, Frantz and {Cady}, Eric J. and {Belikov}, Ruslan and {Balasubramanian}, Kunjithapatham and {Wilson}, Daniel and {Clergeon}, Christophe S. and {Mateen}, Mala},
        title = "{How ELTs will acquire the first spectra of rocky habitable planets}",
    booktitle = {Adaptive Optics Systems III},
         year = 2012,
       editor = {{Ellerbroek}, Brent L. and {Marchetti}, Enrico and {V{\'e}ran}, Jean-Pierre},
       series = {Society of Photo-Optical Instrumentation Engineers (SPIE) Conference Series},
       volume = {8447},
        month = jul,
          eid = {84471X},
        pages = {84471X},
          doi = {10.1117/12.927181},
       adsurl = {https://ui.adsabs.harvard.edu/abs/2012SPIE.8447E..1XG}
}

@ARTICLE{haddock_et_al2010,
       author = {{Haddock}, Steven H.~D. and {Moline}, Mark A. and {Case}, James F.},
        title = "{Bioluminescence in the Sea}",
      journal = {Annual Review of Marine Science},
         year = 2010,
        month = jan,
       volume = {2},
        pages = {443-493},
          doi = {10.1146/annurev-marine-120308-081028},
       adsurl = {https://ui.adsabs.harvard.edu/abs/2010ARMS....2..443H}
}

@ARTICLE{hadfield_et_al2005,
       author = {{Hadfield}, L.~J. and {Crowther}, P.~A. and {Schild}, H. and {Schmutz}, W.},
        title = "{A spectroscopic search for the non-nuclear Wolf-Rayet population of the metal-rich spiral galaxy M 83}",
      journal = {\aap},
         year = 2005,
        month = aug,
       volume = {439},
       number = {1},
        pages = {265-277},
          doi = {10.1051/0004-6361:20042262},
archivePrefix = {arXiv},
       eprint = {astro-ph/0506343},
 primaryClass = {astro-ph},
       adsurl = {https://ui.adsabs.harvard.edu/abs/2005A&A...439..265H}
}

@ARTICLE{haiman+loeb2001,
       author = {{Haiman}, Zolt{\'a}n and {Loeb}, Abraham},
        title = "{What Is the Highest Plausible Redshift of Luminous Quasars?}",
      journal = {\apj},
         year = 2001,
        month = may,
       volume = {552},
       number = {2},
        pages = {459-463},
          doi = {10.1086/320586},
archivePrefix = {arXiv},
       eprint = {astro-ph/0011529},
 primaryClass = {astro-ph},
       adsurl = {https://ui.adsabs.harvard.edu/abs/2001ApJ...552..459H}
}

@ARTICLE{hall_et_al2023,
       author = {{Hall}, Sawyer and {Krissansen-Totton}, Joshua and {Robinson}, Tyler and {Salvador}, Arnaud and {Fortney}, Jonathan J.},
        title = "{Constraining Background N$_{2}$ Inventories on Directly Imaged Terrestrial Exoplanets to Rule Out O$_{2}$ False Positives}",
      journal = {\aj},
         year = 2023,
        month = dec,
       volume = {166},
       number = {6},
          eid = {254},
        pages = {254},
          doi = {10.3847/1538-3881/ad03e9},
archivePrefix = {arXiv},
       eprint = {2311.13001},
 primaryClass = {astro-ph.EP},
       adsurl = {https://ui.adsabs.harvard.edu/abs/2023AJ....166..254H}
}

@ARTICLE{harada_et_al2024a,
       author = {{Harada}, Caleb K. and {Dressing}, Courtney D. and {Kane}, Stephen R. and {Ardestani}, Bahareh Adami},
        title = "{Setting the Stage for the Search for Life with the Habitable Worlds Observatory: Properties of 164 Promising Planet-survey Targets}",
      journal = {\apjs},
         year = 2024,
        month = jun,
       volume = {272},
       number = {2},
          eid = {30},
        pages = {30},
          doi = {10.3847/1538-4365/ad3e81},
archivePrefix = {arXiv},
       eprint = {2401.03047},
 primaryClass = {astro-ph.EP},
       adsurl = {https://ui.adsabs.harvard.edu/abs/2024ApJS..272...30H}
}

@ARTICLE{harada_et_al2024b,
       author = {{Harada}, Caleb K. and {Dressing}, Courtney D. and {Kane}, Stephen R. and {Blunt}, Sarah and {Dietrich}, Jamie and {Hinkel}, Natalie R. and {Li}, Zhexing and {Mamajek}, Eric and {Rice}, Malena and {Tuchow}, Noah W. and {Turtelboom}, Emma V. and {Wittenmyer}, Robert A.},
        title = "{SPORES-HWO. II. Limits on Planetary Companions of Future High-contrast Imaging Targets from $>$20 Years of HIRES and HARPS Radial Velocities}",
      journal = {arXiv e-prints},
         year = 2024,
        month = sep,
          eid = {arXiv:2409.10679},
        pages = {arXiv:2409.10679},
          doi = {10.48550/arXiv.2409.10679},
archivePrefix = {arXiv},
       eprint = {2409.10679},
 primaryClass = {astro-ph.EP},
       adsurl = {https://ui.adsabs.harvard.edu/abs/2024arXiv240910679H}
}

@ARTICLE{hartman_et_al2026,
       author = {{Hartman}, Zachary D. and {Clark}, Catherine A. and {Lund}, Michael B. and {Lester}, Kathryn V. and {Caballero}, Jos{\'e} A. and {Howell}, Steve B. and {Ciardi}, David and {Deveny}, Sarah and {Everett}, Mark E. and {Furlan}, Elise and {Kalari}, Venu and {Littlefield}, Colin and {Stephens}, Andrew W. and {Burt}, Jennifer A. and {Huber}, Guillaume and {Matson}, Rachel and {Mamajek}, Eric E. and {Tuchow}, Noah},
        title = "{Paving the Road to the Habitable Worlds Observatory with High-resolution Imaging. I. New and Archival Speckle Observations of Potential HWO Target Stars}",
      journal = {\aj},
         year = 2026,
        month = mar,
       volume = {171},
       number = {3},
          eid = {174},
        pages = {174},
          doi = {10.3847/1538-3881/ae2fe7},
archivePrefix = {arXiv},
       eprint = {2601.05387},
 primaryClass = {astro-ph.SR},
       adsurl = {https://ui.adsabs.harvard.edu/abs/2026AJ....171..174H}
}

@ARTICLE{hartwig_et_al2018,
       author = {{Hartwig}, Tilman and {Yoshida}, Naoki and {Magg}, Mattis and {Frebel}, Anna and {Glover}, Simon C.~O. and {G{\'o}mez}, Facundo A. and {Griffen}, Brendan and {Ishigaki}, Miho N. and {Ji}, Alexander P. and {Klessen}, Ralf S. and {O'Shea}, Brian W. and {Tominaga}, Nozomu},
        title = "{Descendants of the first stars: the distinct chemical signature of second-generation stars}",
      journal = {\mnras},
         year = 2018,
        month = aug,
       volume = {478},
       number = {2},
        pages = {1795-1810},
          doi = {10.1093/mnras/sty1176},
archivePrefix = {arXiv},
       eprint = {1801.05044},
 primaryClass = {astro-ph.GA},
       adsurl = {https://ui.adsabs.harvard.edu/abs/2018MNRAS.478.1795H}
}

@ARTICLE{haswell_et_al2012,
       author = {{Haswell}, C.~A. and {Fossati}, L. and {Ayres}, T. and {France}, K. and {Froning}, C.~S. and {Holmes}, S. and {Kolb}, U.~C. and {Busuttil}, R. and {Street}, R.~A. and {Hebb}, L. and {Collier Cameron}, A. and {Enoch}, B. and {Burwitz}, V. and {Rodriguez}, J. and {West}, R.~G. and {Pollacco}, D. and {Wheatley}, P.~J. and {Carter}, A.},
        title = "{Near-ultraviolet Absorption, Chromospheric Activity, and Star-Planet Interactions in the WASP-12 system}",
      journal = {\apj},
         year = 2012,
        month = nov,
       volume = {760},
       number = {1},
          eid = {79},
        pages = {79},
          doi = {10.1088/0004-637X/760/1/79},
archivePrefix = {arXiv},
       eprint = {1301.1860},
 primaryClass = {astro-ph.EP},
       adsurl = {https://ui.adsabs.harvard.edu/abs/2012ApJ...760...79H}
}

@ARTICLE{hatt_et_al2018,
       author = {{Hatt}, Dylan and {Freedman}, Wendy L. and {Madore}, Barry F. and {Beaton}, Rachael L. and {Hoyt}, Taylor J. and {Jang}, In Sung and {Lee}, Myung Gyoon and {Monson}, Andrew J. and {Rich}, Jeffrey A. and {Scowcroft}, Victoria and {Seibert}, Mark},
        title = "{The Carnegie-Chicago Hubble Program. IV. The Distance to NGC 4424, NGC 4526, and NGC 4356 via the Tip of the Red Giant Branch}",
      journal = {\apj},
         year = 2018,
        month = jul,
       volume = {861},
       number = {2},
          eid = {104},
        pages = {104},
          doi = {10.3847/1538-4357/aac9cc},
archivePrefix = {arXiv},
       eprint = {1806.02900},
 primaryClass = {astro-ph.CO},
       adsurl = {https://ui.adsabs.harvard.edu/abs/2018ApJ...861..104H}
}

@ARTICLE{heckman_et_al2011,
       author = {{Heckman}, Timothy M. and {Borthakur}, Sanchayeeta and {Overzier}, Roderik and {Kauffmann}, Guinevere and {Basu-Zych}, Antara and {Leitherer}, Claus and {Sembach}, Ken and {Martin}, D. Chris and {Rich}, R. Michael and {Schiminovich}, David and {Seibert}, Mark},
        title = "{Extreme Feedback and the Epoch of Reionization: Clues in the Local Universe}",
      journal = {\apj},
         year = 2011,
        month = mar,
       volume = {730},
       number = {1},
          eid = {5},
        pages = {5},
          doi = {10.1088/0004-637X/730/1/5},
archivePrefix = {arXiv},
       eprint = {1101.4219},
 primaryClass = {astro-ph.CO},
       adsurl = {https://ui.adsabs.harvard.edu/abs/2011ApJ...730....5H}
}

@ARTICLE{heller_et_al2014,
       author = {{Heller}, Ren{\'e} and {Williams}, Darren and {Kipping}, David and {Limbach}, Mary Anne and {Turner}, Edwin and {Greenberg}, Richard and {Sasaki}, Takanori and {Bolmont}, {\'E}meline and {Grasset}, Olivier and {Lewis}, Karen and {Barnes}, Rory and {Zuluaga}, Jorge I.},
        title = "{Formation, Habitability, and Detection of Extrasolar Moons}",
      journal = {Astrobiology},
         year = 2014,
        month = sep,
       volume = {14},
       number = {9},
        pages = {798-835},
          doi = {10.1089/ast.2014.1147},
archivePrefix = {arXiv},
       eprint = {1408.6164},
 primaryClass = {astro-ph.EP},
       adsurl = {https://ui.adsabs.harvard.edu/abs/2014AsBio..14..798H}
}

@ARTICLE{hinkel_et_al2014,
       author = {{Hinkel}, Natalie R. and {Timmes}, F.~X. and {Young}, Patrick A. and {Pagano}, Michael D. and {Turnbull}, Margaret C.},
        title = "{Stellar Abundances in the Solar Neighborhood: The Hypatia Catalog}",
      journal = {\aj},
         year = 2014,
        month = sep,
       volume = {148},
       number = {3},
          eid = {54},
        pages = {54},
          doi = {10.1088/0004-6256/148/3/54},
archivePrefix = {arXiv},
       eprint = {1405.6719},
 primaryClass = {astro-ph.SR},
       adsurl = {https://ui.adsabs.harvard.edu/abs/2014AJ....148...54H}
}

@ARTICLE{hinkel_et_al2018,
       author = {{Hinkel}, Natalie R. and {Unterborn}, Cayman T.},
        title = "{The Star-Planet Connection. I. Using Stellar Composition to Observationally Constrain Planetary Mineralogy for the 10 Closest Stars}",
      journal = {\apj},
         year = 2018,
        month = jan,
       volume = {853},
       number = {1},
          eid = {83},
        pages = {83},
          doi = {10.3847/1538-4357/aaa5b4},
archivePrefix = {arXiv},
       eprint = {1709.08630},
 primaryClass = {astro-ph.EP},
       adsurl = {https://ui.adsabs.harvard.edu/abs/2018ApJ...853...83H}
}

@ARTICLE{hinkel_et_al2024,
       author = {{Hinkel}, Natalie R. and {Youngblood}, Allison and {Soares-Furtado}, Melinda},
        title = "{Host Stars and How Their Compositions Influence Exoplanets}",
      journal = {Reviews in Mineralogy and Geochemistry},
         year = 2024,
        month = jul,
       volume = {90},
       number = {1},
        pages = {1-26},
          doi = {10.2138/rmg.2024.90.01},
archivePrefix = {arXiv},
       eprint = {2404.15422},
 primaryClass = {astro-ph.EP},
       adsurl = {https://ui.adsabs.harvard.edu/abs/2024RvMG...90....1H}
}

@article{hitchcock+lovelock1967,
title = {Life detection by atmospheric analysis},
journal = {Icarus},
volume = {7},
number = {1},
pages = {149-159},
year = {1967},
issn = {0019-1035},
doi = {https://doi.org/10.1016/0019-1035(67)90059-0},
url = {https://www.sciencedirect.com/science/article/pii/0019103567900590},
author = {Dian R. Hitchcock and James E. Lovelock}
}

@ARTICLE{ho_et_al2017,
       author = {{Ho}, Stephanie H. and {Martin}, Crystal L. and {Kacprzak}, Glenn G. and {Churchill}, Christopher W.},
        title = "{Quasars Probing Galaxies. I. Signatures of Gas Accretion at Redshift Approximately 0.2}",
      journal = {\apj},
         year = 2017,
        month = feb,
       volume = {835},
       number = {2},
          eid = {267},
        pages = {267},
          doi = {10.3847/1538-4357/835/2/267},
archivePrefix = {arXiv},
       eprint = {1611.04579},
 primaryClass = {astro-ph.GA},
       adsurl = {https://ui.adsabs.harvard.edu/abs/2017ApJ...835..267H}
}

@ARTICLE{merrill1926,
       author = {{Merrill}, Paul W.},
        title = "{Notes on Lines in the Spectra of Red Stars}",
      journal = {\apj},
         year = 1926,
        month = jan,
       volume = {63},
        pages = {13},
          doi = {10.1086/142946},
       adsurl = {https://ui.adsabs.harvard.edu/abs/1926ApJ....63...13M}
}
